\documentclass[12pt,openany]{book}

\usepackage{etoolbox}  
\newtoggle{UNT}
\togglefalse{UNT}

\usepackage{graphicx}
\usepackage{amsmath, fullpage, mathrsfs}
\usepackage[Bjornstrup]{fncychap}
\usepackage{srcltx, tabularx, pdfsync, multirow, float}
\usepackage{fancyhdr}
\usepackage[table]{xcolor}  
\usepackage{color, colortbl, rotate, listings, rotating}
\usepackage{caption, booktabs, tocvsec2, longtable}
\usepackage[numbers,sort]{natbib}
\usepackage{tocbibind}
\usepackage{subfig}
\usepackage{bm}  
\usepackage{lettrine}  
\usepackage{cool}  
\usepackage{braket}
\usepackage[pagebackref=true, plainpages=false, bookmarks, bookmarksnumbered, colorlinks, linktocpage=true, linkcolor=blue, citecolor=black, filecolor=black,urlcolor=black, %
			pdflang={en}, pdftitle={Variational Calculations of Positronium Scattering with Hydrogen}, pdfauthor={Denton Woods}, %
			pdfsubject={Denton Woods's Dissertation at the University of North Texas}, pdfkeywords={positronium, hydrogen, scattering, positronium hydride, Denton Woods}]{hyperref}
\usepackage{siunitx}  
\usepackage{wasysym}  
\usepackage{gensymb}  
\usepackage{pdflscape}  
\usepackage{dcolumn}
\usepackage[makeroom]{cancel}  
\usepackage{tocstyle}  
\usepackage{bigints}
\usepackage{dirtree}

\renewcommand*{\backref}[1]{}
\renewcommand*{\backrefalt}[4]{%
    \ifcase #1 (Not cited)%
    \or        (Cited on page~#2)%
    \else      (Cited on pages~#2)%
    \fi}

\usepackage{cleveref}
\crefname{equation}{Equation}{Equations}
\crefname{table}{Table}{Tables}
\crefname{figure}{Figure}{Figures}
\crefname{chapter}{Chapter}{Chapters}
\crefname{section}{Section}{Sections}
\crefname{appendix}{Appendix}{Appendices}
\crefname{listing}{Listing}{Listings}

\newcolumntype{d}[1]{D{.}{\cdot}{#1}}
\newcolumntype{.}{D{.}{.}{-1}}

\newcommand*{\thead}[1]{\multicolumn{1}{c}{#1}}

\usepackage{mathpazo}

\usepackage{setspace}

\newcommand{\ee} {\,\text{e}}
\newcommand{\beq}{\begin{equation}}
\newcommand{\eeq}{\end{equation}}
\newcommand{\beqs}{\begin{equation*}}
\newcommand{\eeqs}{\end{equation*}}
\newcommand{\barr}{\begin{array}}
\newcommand{\earr}{\end{array}}
\newcommand{\bce}{\begin{center}}
\newcommand{\ece}{\end{center}}

\newcommand{\kr} {\kappa\rho}
\newcommand{\krp} {\kappa\rhop}
\newcommand{\mr} {\mu\rho}
\newcommand{\mrp} {\mu\rho'}
\newcommand{\rhop} {\rho'}
\newcommand{\ii}{{\textrm{i}}}

\newcommand{\noop}[1]{}

\def\biblio{\bibliographystyle{UNTamsplain-modded}\bibliography{Dissertation}}

\DeclareSymbolFont{euler}{U}{eur}{m}{n}
\DeclareMathSymbol \uppi \mathalpha {euler} {"19}

\makeatletter 
\renewcommand{\maketitle} 
{ \begingroup \vskip 10pt \begin{center} \large {\bf \@title}
	\vskip 10pt \large \@author \hskip 20pt \@date \end{center}
  \vskip 10pt \endgroup \setcounter{footnote}{0} }
\makeatother 
\newcommand{\gv}[1]{\ensuremath{\mbox{\boldmath$ #1 $}}} 
\newcommand{\abs}[1]{\left| #1 \right|} 
 
\newcommand{\grad}[1]{\gv{\nabla} #1} 

\let\baraccent=\= 
\renewcommand{\=}[1]{\stackrel{#1}{=}} 

\definecolor{Gray}{gray}{0.9}
\definecolor{LightCyan}{rgb}{0.88,1,1}
\definecolor{startcolor}{HTML}{B32018}

\graphicspath{{IPython/}{Images/}}

\makeatletter
\patchcmd{\@makechapterhead}{\vspace*{50\p@}}{}{}{}
\patchcmd{\@makeschapterhead}{\vspace*{50\p@}}{}{}{}
\makeatother

\allowdisplaybreaks  
\begin{document}

\def\biblio{}

\pdfbookmark[1]{Title Page}{title} 

\pagenumbering{roman}


\begin{titlepage}
    \begin{center}
        \vspace*{1cm}
        
		\iftoggle{UNT}{
			VARIATIONAL CALCULATIONS OF POSITRONIUM SCATTERING WITH HYDROGEN
		}{
			\LARGE
			\textbf{Variational Calculations of Positronium Scattering with Hydrogen}
		}
        \vspace{0.5cm}
		
		\iftoggle{UNT}{
			Denton Woods
		}{
			\Large
			\textbf{Denton Woods}
		}

		\iftoggle{UNT}{
			\vspace{2.5cm}
		}{
			\vspace{4cm}
		}
        
		\normalsize
        Dissertation Prepared for the Degree of\\
        DOCTOR OF PHILOSOPHY
        
		\iftoggle{UNT}{
			\vspace{2.5cm}
		}{
			\vspace{4cm}
		}
		
        \normalsize
		UNIVERSITY OF NORTH TEXAS\\
		August 2015
    \end{center}

	\vfill
  
	\hspace{8cm} APPROVED:
	\vspace{0.1cm}
	
	\singlespacing
	
	\hspace{8cm} Sandra J. Quintanilla (Ward),
	
	\hspace{9cm} Major Professor

	\hspace{8cm} Peter Van Reeth, Minor Professor

	\hspace{8cm} Duncan Weathers, Committee Member

	\hspace{8cm} David Shiner, Committee Member

	\hspace{8cm} Carlos Ordonez, Committee Member

	\hspace{8cm} David Schultz, Chair of the Department
	
	\hspace{9cm} 	of Physics
	
	\hspace{8cm} Mark Wardell, Dean of the Toulouse
	
	\hspace{9cm} 	Graduate School

\end{titlepage}

\newpage
\vspace*{\fill}
\begingroup
\centering
Copyright 2015

by

Denton Woods

\endgroup
\vspace*{\fill}
\newpage

\iftoggle{UNT}{
}{
	\pdfbookmark[1]{Abstract}{abstract}

\thispagestyle{plain}
\begin{center}
    \large
    \textbf{Abstract}
\end{center}

Positronium-hydrogen (Ps-H) scattering is of interest, as it is a fundamental 
four-body Coulomb problem. We have investigated low-energy Ps-H scattering 
below the Ps(n=2) excitation threshold using the Kohn variational method and 
variants of the method with a trial wavefunction that includes highly 
correlated Hylleraas-type short-range terms.
We give an elegant formalism that combines all Kohn-type variational
methods into a single form. Along with this, we have also developed a general
formalism for Kohn-type matrix elements that allows us to evaluate arbitrary 
partial waves with a single codebase. Computational strategies we have 
developed and use in this work are also discussed.

With these methods, we have computed 
phase shifts for the first six partial waves for both the singlet and triplet 
states. The $^{1,3}$S and $^{1,3}$P phase shifts are highly accurate results
and could potentially be viewed as benchmark results.
Resonance positions and widths for the $^1$S-, $^1$P-, $^1$D-, and
$^1$F-waves have been calculated.

We present elastic integrated, elastic differential, and momentum transfer 
cross sections using all six partial waves and note interesting features of 
each. We use multiple effective range theories, including several that 
explicitly take into account the long-range van der Waals interaction, to 
investigate scattering lengths and effective ranges.

}

\clearpage
\pdfbookmark[1]{Acknowledgements}{acknowledgements}
\thispagestyle{plain}
\begin{center}
	\iftoggle{UNT}{
		ACKNOWLEDGEMENTS
	}{
		\large
		\textbf{Acknowledgements}
	}
\end{center}

Many people helped me through this project. Special thanks goes to Ryan 
Bosca, who helped me at times and let me bounce ideas off of him. Keri Ward,
Cassie Whitmire, Robert Whitmire, 
and Lauren Murphy have been part of a great support network that helped me
survive this process, along with all of my friends in the physics department.

Great thanks goes to my advisor, Dr. S.~J.~Ward and to our collaborator, Dr. Peter 
Van Reeth, who helped immensely with this project through discussions, plenty 
of notes, and codes. I also appreciate discussions with Drs. Y.~K.~Ho, J.~W.~
Humberston, K.~Pachucki, Z.-C.~Yan, Edward Armour, James Cooper, Martin 
Plummer, and Gillian Peach during the course of this project.

I received several travel grants to help me go to and present at conferences, 
including from DAMOP, UNT's Student Government Association, and UNT's College 
of Arts and Sciences. Dr. Ward received a grant from the National Science 
Foundation under grant no. PHYS-0968638 and another from UNT through the UNT 
faculty research grant GA9150, both of which supported my research.

Computational resources were provided by UNT's High Performance Computing 
Services, a project of Academic Computing and User Services division of the 
University Information Technology with additional support from UNT Office of 
Research and Economic Development.

\hypersetup{linktocpage}
\hypersetup{
    colorlinks,
    citecolor=black,
    filecolor=black,
    linkcolor=black,
    urlcolor=black
}
\hypersetup{breaklinks=true}

\tableofcontents
\newpage
\listoftables
\listoffigures

\newpage

\chapter*{List of Abbreviations}
\label{chp:nomenclature}
\addcontentsline{toc}{chapter}{List of Abbreviations}

\begin{table}[H]
	\begin{tabular}{l l}
short-short & short-range--short-range \\
long-short & long-range--short-range \\
long-long & long-range--long-range \\
BO & Born-Oppenheimer \\
CC & close coupling                \\
CI & configuration interaction \\
CVM & confined variational method \\
DMC & diffusion Monte Carlo \\
Ps & positronium \\
PsH & positronium hydride \\
SVM & stochastic variational method  \\
SE & static-exchange \\
vdW & van der Waals \\
	\end{tabular}
\end{table}

\newpage
\clearpage
\pagenumbering{arabic}


\chapter{Introduction}
\label{sec:Introduction}

\section{Positronium}
\label{sec:Positronium}

\nocite{GitHub,figshare,Plotly,Wiki,Conferences1,Conferences2,Conferences3,Conferences4,Conferences5}

Positronium (Ps) is an exotic atom formed from the bound state of a positron
(antielectron, e$^+$) and an electron (e$^-$). Stjepan Mohorovi\u{c}i\'{c} 
theorized the existence of Ps in 1934 \cite{Mohorovicic1934}, 
but it was not created until 1951 by Martin Deutsch \cite{Deutsch1951}.
This atom is similar in some ways to hydrogen (H) but also differs 
in some key aspects. Namely that Ps annihilates, emitting two or three $\gamma$
rays, depending on the spin \cite{Charlton2001}. In the singlet state, also 
known as parapositronium (p-Ps), the lifetime is $\SI{125}{ps}$ \cite{Czarnecki1999}.
The triplet state, or orthopositronium (o-Ps), lasts approximately 1000 times
longer with a lifetime of $\SI{142}{ns}$ \cite{Vallery2003}. Due to the short 
lifetime of p-Ps, the majority of experimental data of Ps-atom and
Ps-molecule scattering comes from o-Ps.

Working in atomic units (see \cref{sec:Units}), the ground state wavefunctions
of H and that of Ps are
\begin{equation}
\Phi_{\rm{H}}\left(r_3\right) = \frac{1}{\sqrt{\pi}} \ee^{-r_3}
\label{eq:HWave}
\end{equation}
and
\begin{equation}
\Phi_{\rm{Ps}}\left(r_{12}\right) = \frac{1}{\sqrt{8 \pi}} \ee^{-r_{12}/2} \;.
\label{eq:PsWave}
\end{equation}
When the Schr\"{o}dinger equation is solved for the hydrogen atom, the energy
is seen to be (neglecting higher-order effects)
\beq
\label{eq:HEnergy}
E_{n,\rm{H}} = -\frac{R_\infty}{n^2}.
\eeq
In Hartree atomic units, $R_\infty = \frac{1}{2}$, giving a ground state energy of
\beq
\label{eq:HEnergyAU}
E_{\rm{H}} = -\frac{1}{2}.
\eeq
Due to the reduced mass of half that of hydrogen, the ground state energy of Ps is
\beq
\label{eq:PsEnergyAU}
E_{\rm{Ps}} = -\frac{1}{4}.
\eeq

\section{Motivation}
\label{sec:Motivation}

Ps formation is important in the galactic core \cite{Kinzer1996}, and Ps-atom 
scattering is of interest in the study of solar processes \cite{Crannell1976}.
As well as the basic interest of Ps-atom scattering in atomic physics, Ps 
is also important in material science. As Ps is neutral, it penetrates deeper 
into material than a charged particle, such as a positron. Ps scattering also 
has applications in other areas of physics such as biophysics and 
astrophysics \cite{Laricchia2012}. A brief overview of the state of the art
in antimatter atomic physics is Ref.~\cite{Walters2010}, and a more in-depth
review of Ps collisions is Ref.~\cite{Laricchia2012}.

With the increased interest in antihydrogen ($\bar{\rm{H}}$) production at CERN
\cite{ALPHACollaboration2011}, there have been investigations by groups
exploring alternate mechanisms other than dumping antiprotons ($\bar{\rm{p}}$)
into a cloud of e$^+$ and relying on the reaction of
$\bar{\rm{p}} + \rm{e}^+ + \rm{e}^+ \to \bar{\rm{H}} + \rm{e}^+$.
Refs.~\cite{Kadyrov2013,Elkilany2014} explore the inelastic low-energy
reaction of $\bar{\rm{p}} + \rm{Ps} \to \bar{\rm{H}} + \rm{e}^-$, where the Ps
is in its ground state. A recent paper \cite{Kadyrov2015} (also mentioned
in the popular science press \cite{Kadyrov2015b}) found that if the target Ps
is in an excited state ($1 < n \leq 3$), the $\bar{\rm{H}}$ production rate is
improved by several orders of magnitude.

The original
motivation for this research was a proposed experiment to measure 
low-energy Ps scattering from alkali metals by Jason Engbrecht of the Positron 
Research Group at St. Olaf College. There has not been much theoretical work
on these systems so far. Unfortunately, it appears that this project
is put on hold indefinitely, and the group's website \cite{Engbrecht2013} no 
longer exists. We started investigating Ps-H scattering and plan to extend
this work into Ps scattering from the alkali metals.

However, there is still interesting ongoing experimental work on Ps scattering,
though with different targets. The University College London (UCL) Positron
Group \cite{UCL2015} has developed energy-tunable o-Ps beams
\cite{Brown1985,Laricchia1987,Zafar1996,Garner1996,Laricchia2008} over the
course of many years. This group has been able to study Ps scattering from
He, Ne, Ar, Kr, and Xe
\cite{Garner1996,Garner2000,Armitage2002,Laricchia2004,Armitage2006,Laricchia2008,Engbrecht2008,Brawley2010a}
and the H$_2$, N$_2$, O$_2$, CO$_2$, H$_2$O, and SF$_6$ molecules
\cite{Garner1996,Garner1998,Garner2000,Laricchia2004,Armitage2006,Beale2006,Brawley2010a}.

\begin{figure}
	\centering
	\resizebox{1.0\textwidth}{!}{\includegraphics{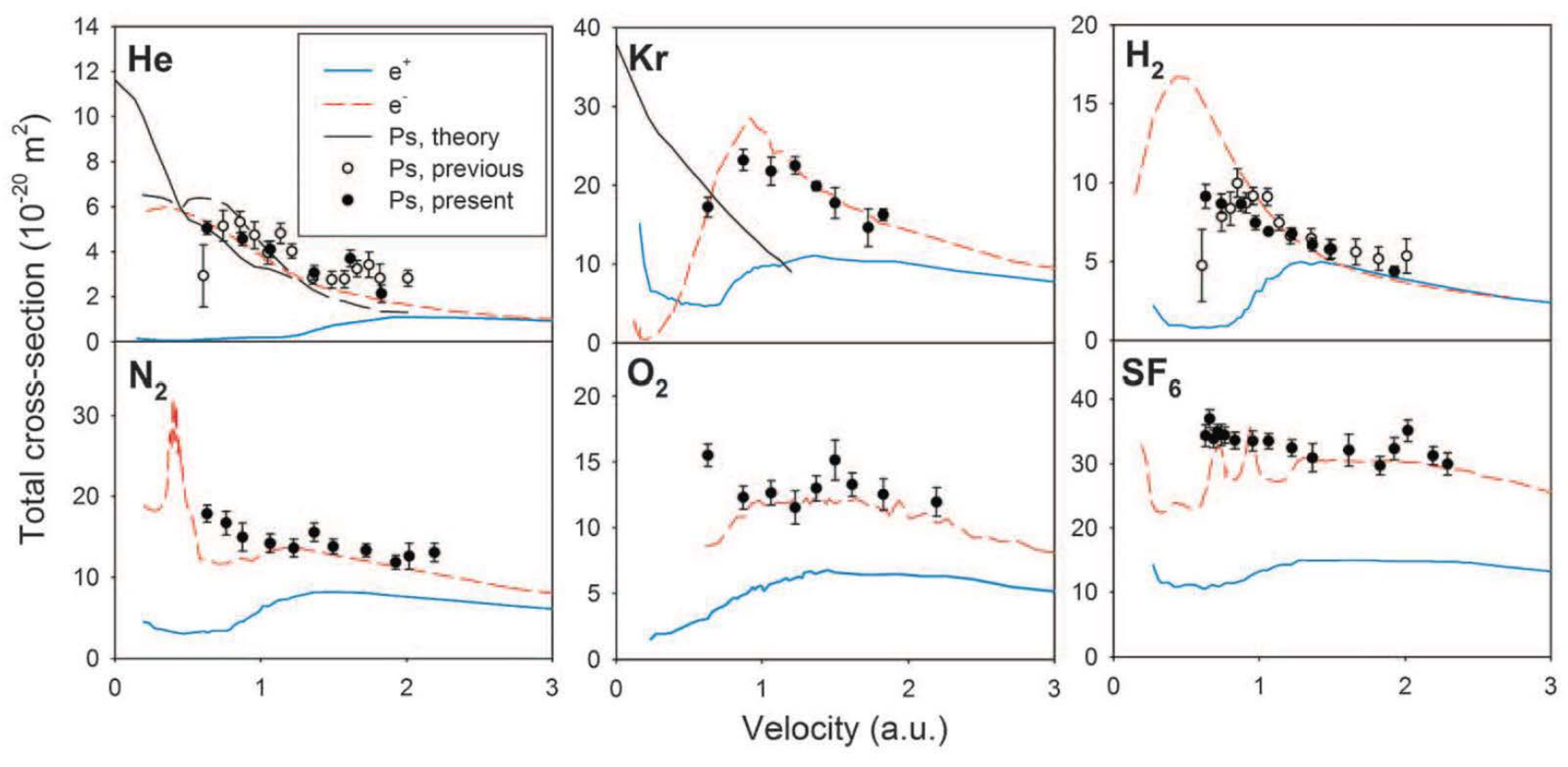}}
	\caption[Comparisons of e$^-$, e$^+$, and Ps scattering]{Comparisons of
e$^-$, e$^+$, and Ps scattering from different atomic and molecular targets 
from Ref.~\cite{Brawley2010a}. Reprinted with permission from AAAS.}
	\label{fig:ScienceBrawley}
\end{figure}

A recent development in the field is the surprising discovery that Ps 
scattering is electron-like \cite{Brawley2010,Brawley2010a}, which was also reported in the
popular science literature \cite{NewScientist2015}. If the cross sections are 
plotted with respect to the velocity of the incoming projectile, not momentum 
like typical, e$^-$ and Ps scattering look similar when the target is the same.
This is despite the Ps projectile having twice the mass of e$^-$ and being 
electrically neutral versus negatively charged. It can be seen in
\cref{fig:ScienceBrawley} that e$^+$ scattering looks different when compared
to these two. Fabrikant and Gribakin
\cite{Fabrikant2014,Fabrikant2014a} compare low-energy e$^-$ and Ps scattering
from Kr and Ar targets, also finding that the cross sections are similar for
e$^-$ and Ps projectiles. The tentative conclusion is that the
e$^+$ plays a much smaller role in the scattering process than the e$^-$
in Ps-atom and Ps-molecule scattering. This shows that there is still
plenty of work to do to understand Ps scattering more fully, but it
does suggest that for certain cases, a decent first approximation to Ps
scattering can be made by using e$^-$ scattering data.

\section{Partial Waves and Kohn-Type Variational Methods}
\label{sec:KohnIntro}

The most common way of approaching scattering problems is to use the complete
set of Legendre polynomials to expand the scattering wavefunction. For a
central potential, this can be written as \cite{Bransden2003}
\beq
\label{eq:PartialWave}
\Psi(k,r,\theta) = \sum_{\ell=0}^\infty R_\ell(k,r) P_\ell(\cos\theta).
\eeq
The method of partial waves evaluates each term in this summation separately,
with each referred to as a partial wave, and each has a different angular 
momentum. The typical naming of each partial wave starting from $\ell = 0$
is the S-, P-, D-, F-, G-, H-wave, etc., which is similar to the usual 
spectroscopic notation. For low energies, usually only a few terms in this
expansion are required, and for very low energies, the S-wave ($\ell = 0$) is 
typically the only significant contribution.

The Kohn variational method \cite{Kohn1948} and its variants, derived and 
described in \cref{chp:WaveKohn}, have been used successfully in many 
scattering problems, such as e$^-$-H \cite{Schwartz1961}, e$^-$-methane
\cite{McCurdy1989}, H-$\bar{\rm{H}}$ \cite{Armour2002}, e$^+$-H$_2$
\cite{Cooper2008}, e$^-$-Ps \cite{Ward1987}, e$^+$-He \cite{VanReeth1997},
and nucleon-nucleon scattering \cite{Tomio1995,Kievsky1997}.
The Kohn variational method and its variants suffer from well-known spurious
singularities (see \cref{eq:SchwartzSing}), so they are often used in conjunction
with each other to identify these. To avoid cumbersome wording in this document,
the Kohn variational method and variants of the method are simply referred to as
``Kohn-type methods''. Complex Kohn methods that use spherical Hankel functions 
instead of the spherical Bessel and Neumann functions are often used due to a 
smaller, but nonzero, chance of the Schwartz singularities
\cite{McCurdy1989,Lucchese1989,Cooper2010}. This work uses the Kohn,
inverse Kohn, generalized Kohn \cite{Armour1991}, $S$-matrix complex Kohn,
and $T$-matrix complex Kohn variational methods.

\section{Ps-H Scattering}
\label{sec:ScatIntro}

For this work \cite{Woods2015,Conferences1,Conferences2,Conferences3,Conferences4,Conferences5},
we have computed phase shifts for the first six partial waves of Ps-H
scattering and resonance parameters for $^1$S through $^1$F
(\cref{chp:SWave,chp:PWave,chp:DWave,chp:HigherWaves}).
We also calculate scattering lengths and effective ranges for $^{1,3}$S
(\cref{sec:ScatteringLength}), scattering lengths for $^{1,3}$P
(\cref{sec:PWaveScatLen}), and multiple cross sections
(\cref{chp:CrossSections}). Each of these is compared to previously published
research where possible.

This work is an extension of the earlier work on Ps-H collisions using the 
Kohn and inverse Kohn variational methods by Van Reeth and Humberston
\cite{VanReeth2003,VanReeth2004}. In these papers, they calculated $^{1,3}$S
and $^{1,3}$P phase shifts and obtained resonance parameters, scattering lengths
and effective ranges from these. The most important difference from this earlier 
work is that we have increased the number of partial waves examined from two 
to six, requiring developing a general formalism and code that works for 
arbitrary partial waves. This allows us to calculate elastic integrated, elastic
differential, and momentum transfer cross sections (\cref{chp:CrossSections}),
which would not be possible with only the
$^{1,3}$S- and $^{1,3}$P-waves. We also develop a very general form
(\cref{chp:WaveKohn}) of the scattering wavefunction and codes that allows us to
calculate phase shifts with not only the Kohn and inverse Kohn variational methods
but also with the generalized Kohn, $S$-matrix complex Kohn, $T$-matrix complex
Kohn, generalized $S$-matrix complex Kohn, and generalized $T$-matrix complex
Kohn variational methods. Over the previous work, we also perform a thorough
analysis of the van der Waals contribution to the $^{1,3}$S scattering lengths
and effective ranges and investigate the $^{1,3}$P scattering lengths
(\cref{sec:ScatteringLength}).

We have increased the number of short-range terms (\cref{eq:PhiDef}) used 
over prior work \cite{VanReeth2003,VanReeth2004}. This is enabled by several
changes. The largest improvement has been the introduction of the Todd method
(\cref{sec:ToddBound}), which selects the ``best'' set of short-range terms
from the full set determined by $\omega$ in \cref{eq:GeneralWaveTrial}. Some
short-range terms contribute more to linear dependence than others, and this
method removes those in a systematic manner. This is often an improvement 
over the restriction in powers that Van Reeth and Humberston
\cite{VanReeth2003} did, though we still use that restricted basis set for
the $^{1,3}$F-wave (\cref{sec:FWave}). We also implement the asymptotic
expansion \cite{Drake1995,Yan1997} for the short-range--short-range integrals
instead of only doing a direction summation (\cref{sec:AsymptoticExpansion}).
This gives much more accurate short-range--short-range integrals, allowing us
to use more short-range terms and solve larger matrices in
\cref{eq:GeneralKohnMatrix}. We specifically noted a threefold increase in the
number of terms we could use for the $^3$S state when the asymptotic expansion
was included. We use approximately seven times as many
integration points as this previous work (\cref{sec:SelQuadPoints}).
For $\ell \geq 1$ especially, an increase in the number of
integration points for the long-range--long-range and long-range--short-range
integrals (\cref{sec:CompLong}) lead to more stable results and the ability to
use more short-range terms. For $\ell \geq 2$ (not investigated by the prior
work), we also introduced extra exponentials in several coordinates to the
Gauss-Laguerre quadratures (\cref{sec:ExtraExp}) that are subsequently removed,
increasing the convergence rate of the long-range--short-range integrals.

The Ps-H collision problem has been treated by multiple groups with different
methods over the years.
Our paper \cite{Woods2015} also has discussion of the different methods used
for low-energy Ps-H scattering. Various properties of this system have been
calculated using the first Born approximation \cite{Massey1954,McAlinden1996},
diffusion Monte Carlo (DMC)
\cite{Chiesa2002}, the SVM \cite{Ivanov2001,Ivanov2002}, CC
\cite{Sinha1997,Campbell1998,Adhikari1999,Sinha2000,Blackwood2002,Blackwood2002b,Walters2004},
static exchange \cite{Hara1975,Ray1996,Ray1997}, Kohn variational
\cite{Page1976,VanReeth2003,VanReeth2004}, and inverse Kohn
variational \cite{VanReeth2003,VanReeth2004} methods.

The CC method has been used in multiple papers for Ps-H scattering by the
\href{http://web.am.qub.ac.uk/ctamop/people.php}{Belfast group}. Campbell et
al. \cite{Campbell1998} used 22 pseudostates and eigenstates of Ps with the
ground state of H in an approximation they denote as 22Ps1H for singlet and
triplet Ps-H scattering. Blackwood et al. \cite{Blackwood2002b} later included
the e$^+$-H$^-$ channel, using a 22Ps1H + H$^-$ approximation, finding that it
improved the convergence of the binding energy and allowed them to calculate
resonance parameters for the $^1$F-wave and $^1$G-wave. In another paper by
Blackwood \cite{Blackwood2002}, they use a 14Ps14H approximation with
14 Ps and 14 H pseudostates and eigenstates to compute
$^{1,3}$S phase shifts and a 9Ps9H approximation to compute $^{1,3}$P and
$^{1,3}$D phase shifts, finding that this performed better than the
earlier 22Ps1H calculation \cite{Campbell1998}. Walters et al.
\cite{Walters2004} also included the e$^+$-H$^-$ channel in a 14Ps14H + H$^-$
approximation for the $^1$S-wave phase shifts and a 9Ps9H + H$^-$ approximation
for the $^1$S-, $^1$P, and $^1$D-wave phase shifts and resonances through the
$^1$F-wave.

Page \cite{Page1976} calculated $^{1,3}$S Ps-H scattering lengths using the
Kohn variational method with 35 short-range terms. The Kohn/inverse Kohn
variational methods \cite{VanReeth2003,VanReeth2004} have been used to calculate
$^{1,3}$S phase shifts, scattering lengths, and effective ranges, along with
$^1$S resonance parameters. The Kohn/inverse Kohn variational methods
\cite{VanReeth2004} have also been used to calculate $^{1,3}$P phase shifts
and parameters for the first $^1$P resonance. The Kohn-type variational methods
give empirical bounds on the phase shifts, and adding short-range terms to the
wavefunction allows us to improve the phase shift convergence in a rigorous way.
The Kohn-type methods can generate very accurate 
phase shifts, but the choice of trial wavefunction can make computation very 
difficult. Then there are the spurious Schwartz singularities, but these can 
often be mitigated by using complex Kohn methods.

\section{Positronium Hydride}
\label{sec:PsH}
Positronium hydride (PsH) is a bound state comprised of 
a hydrogen atom and a positronium atom. After Wheeler \cite{Wheeler1946} 
showed that positrons could be part of what he called a polyelectronic 
compound, Ore shortly thereafter predicted PsH in 1951 \cite{Ore1951}. PsH 
was not experimentally verified until 1992 by Schrader \cite{Schrader1992}
using the reaction e$^+$ + CH$_4$ $\to$ CH$_3^+$ + PsH.

We first investigated the bound state of PsH instead of Ps-H scattering, as 
it is a simpler problem and has been studied extensively in the literature
(see \cref{tab:BoundEnergyOther} on page \pageref{tab:BoundEnergyOther}).
The purpose of first studying PsH was not to try to contribute more accurate
results but to develop the experience with the short-range Hylleraas-type 
correlation terms that we used in both the $^1$S PsH and $^1$S Ps-H
scattering problems. 
The full discussion of our work on PsH is found in \cref{chp:PsHBound}.
There are dozens of calculations of the ground state
or binding energy of PsH given in \cref{tab:BoundEnergyOther}.
The binding energy of \SI{1.066 406}{eV} compares well with the most accurate
value from Ref.~\cite{Bubin2006} of \SI{1.066 598}{eV}, which gives confidence
in the short-range part of the scattering wavefunction in
\cref{sec:GeneralWave}.

\section{Positronium Hydride Structure}
\label{sec:PsHStructure}
There has been some discussion in the literature about whether PsH is more 
like an atomic structure or more like a molecule with Ps and H. We did not 
attempt an analysis of this problem, since the PsH system is not the goal of 
this work. Bressanini and Morosi \cite{Bressanini2003} give a good overview 
of the lack of consensus on this problem. 

Frolov and Smith \cite{Frolov1997c} note that they expect PsH to be a cluster 
consisting of a Ps atom and an H atom. Then from their calculations, they 
conclude that it acts as some kind of sum of H$^-$ with Ps$^-$.

Saito \cite{Saito2000} attempted to answer this question using Ho's
\cite{Ho1986} Hylleraas basis set by plotting e$^+$ and e$^-$ densities. Saito's 
conclusion was that PsH has an atomic structure but also has a diatomic 
molecular structure, or Ps with H. Bromley and Mitroy \cite{Bromley2001} also 
state that PsH has a molecular structure, comparing it ``to a light isotope 
of the H$_2$ molecule.''

Biswas and Darewych \cite{Biswas2002} find that the difference between the
S(1) resonance and the binding energy (\cref{sec:BoundSinglet}) for various 
calculations of differing accuracy is roughly constant. They suggest that 
this means that PsH is less like e$^+$ orbiting H$^-$ and more like a 
diatomic molecule.

Bressanini and Morosi \cite{Bressanini2003} perform calculations on PsH with 
a highly optimized one term wavefunction to determine the structure and 
conclude that PsH cannot be viewed as Ps+H or e$^+$ orbiting around H$^-$. 
They state, ``Keeping in mind the quantum nature of the leptons and so the 
impossibility of defining a structure, we suggest to look at PsH as a 
hydrogen negative ion with the positron that, staying more distant from the 
nucleus than the electrons, correlates its motion with those of both the 
electrons. Its attraction on the electrons squeezes them nearer to each other 
and nearer to the nucleus.''

Heyrovska \cite{Heyrovska2011} treats PsH as a molecule and calculates bond
lengths to try to gain a better understanding of its structure. However, this
preprint does not settle the debate.

\section{Final Notes}
\label{sec:Units}

Unless otherwise stated, values throughout are given in atomic units, i.e.
$\hbar = m_{\rm{e}} = e = 4\pi\epsilon_0 = 1$ \cite{Hartree1928}. Energies are given
in hartrees, with $\SI{1}{\hartree} = \SI{27.211 385 05(60)}{\electronvolt}$ 
\cite{Mohr2012,NISTConversions}. Momentum is given as units of $a_0^{-1}$,
where $a_0$ is the Bohr radius. Cross sections are given in units of
$\pi a_0^2$, and differential cross sections are given in units of
$a_0^2 / \rm{sr}$, unless otherwise noted.

Some of the figures in this document are adapted from our paper submitted to
Physical Review A \cite{Woods2015}. Some figures are available as interactive
plots on the plotly page at \url{http://plot.ly/~Denton}.
Additional notes for derivations are
available at \url{http://figshare.com/authors/Denton_Woods/581638}, and
codes discussed are available at \url{https://github.com/DentonW/Ps-H-Scattering}.
These notes and codes are also linked at \url{http://www.dentonwoods.com} and
on the Research Wiki \cite{Wiki}.

%


\chapter{Positronium Hydride and Short-Range Terms}
\label{chp:PsHBound}

\begin{figure}
	\centering
	\includegraphics[width=4in]{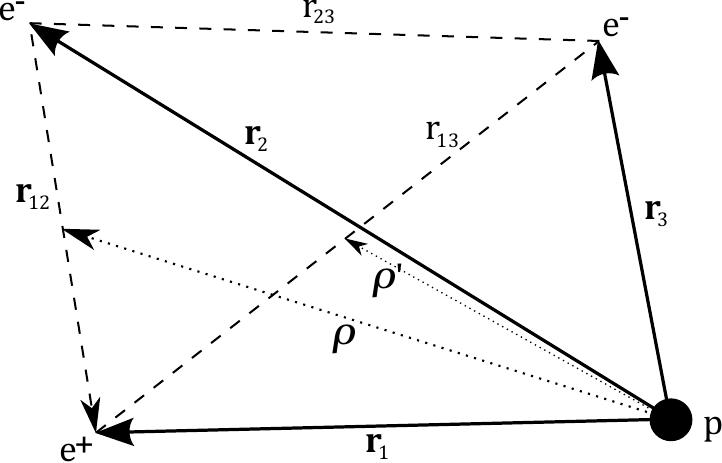}
	\caption{Positronium hydride coordinate system}
	\label{fig:PsHCoords}
\end{figure}

\iftoggle{UNT}{As}{\lettrine{\textcolor{startcolor}{A}}{s}}
discussed in \cref{sec:PsH}, PsH 
consists of one atom of both Ps and of H.
\Cref{fig:PsHCoords} shows the PsH coordinate system. There are 6 interparticle 
coordinates, given by $r_1$, $r_2$, $r_3$, $r_{12}$, $r_{13}$, and $r_{23}$. 
The proton is considered infinitely heavy in this treatment.
Armour et al.~\cite{Armour2005} point out that positronium antihydride is an
equivalent system, assuming CPT symmetry. Another related system is e$^+$PsH,
which is stable and can be thought of as e$^+$ orbiting around PsH \cite{Armour2005}.

\section{PsH Wavefunction}
\label{sec:BoundWavefn}

The wavefunction we use for the bound state has a Hylleraas-style
\cite{Yan1999,VanReeth2003} set of terms, given by
\begin{subequations}
\label{eq:BoundWavefn}
\begin{align}
 \Psi^\pm &= \sum_{i=1}^{N(\omega)} c_i \bar{\phi}_i^\pm \label{eq:BoundWavefn_psi} \\
 \bar{\phi}_i^\pm &= (1 \pm P_{23}) \phi_i \label{eq:BoundWavefn_phibar} \\
 \phi_i &= \ee^{-(\alpha r_1 + \beta r_2 + \gamma r_3)} r_1^{k_i} r_2^{l_i} r_{12}^{m_i} r_3^{n_i} r_{13}^{p_i} r_{23}^{q_i} \label{eq:BoundWavefn_phi} \, ,
\end{align}
\end{subequations}
where the plus sign indicates the spatially symmetric singlet case, and the 
minus sign indicates the spatially antisymmetric triplet case. The 
permutation operator $P_{23}$ is needed, as the two electrons are 
indistinguishable. The $\frac{1}{\sqrt{2}}$ needed to normalize $\Psi^\pm$
(to cancel out the 2 in \cref{eq:RRfinal}) is absorbed into the $c_i$ constant 
in \cref{eq:BoundWavefn_psi}. The constant
$\SphericalHarmY{0}{0}{\theta}{\phi} = \frac{1}{\sqrt{4\pi}}$ is also absorbed
into $c_i$. The Hylleraas-type basis set satisfies the Kato cusp condition
\cite{Kato1957} well \cite{Armour1991}.

\begin{figure}
	\centering
	\includegraphics[width=4in]{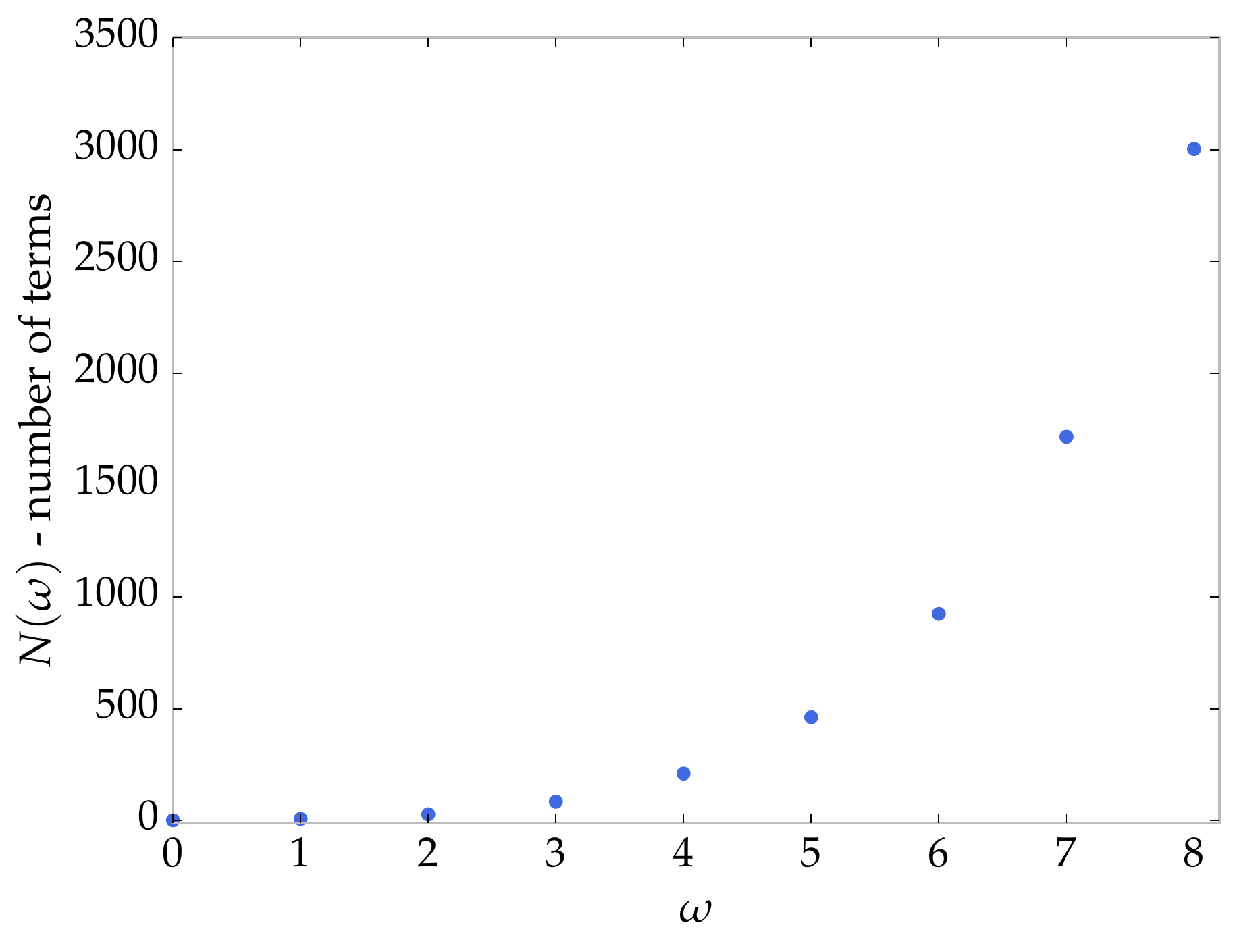}
	\caption{$N(\omega)$ versus $\omega$}
	\label{fig:num-omega}
\end{figure}

The variable $\omega$ is an integer $\geq 0$ that determines the number of 
terms in the basis set. For a chosen value of $\omega$, the integer powers of 
$r_i$ and $r_{ij}$ are constructed in such a way that \cite{VanReeth2003}
\beq
\label{eq:OmegaDef}
k_i + l_i + m_i + n_i + p_i + q_i \leq \omega,
\eeq
with all $k_i$, $l_i$, $m_i$, $n_i$, $q_i$ and $p_i$ $\geq 0$. 
Using combination with repetition, an explicit formula for $N(\omega)$ is 
given as
\beq
\label{eq:NumberTermsOmega}
N(\omega) = \Binomial{\omega+6}{6} \, ,
\eeq
where the 6 comes from the 6 coordinates of $r_i$ and $r_{ij}$. A plot of
$N(\omega)$ versus $\omega$ is given in \cref{fig:num-omega}.

\section{Rayleigh-Ritz Variational Method}
\label{sec:RayleighRitz}
The Rayleigh-Ritz variational method is given as the functional \cite{Bransden2003}
\beq
\label{eq:RayleighRitz}
E[\Psi] = \frac{\Braket{\Psi | H | \Psi}}{\Braket{\Psi | \Psi}}.
\eeq
This provides an upper bound to the ground-state energy, $E_0$. In other words,
\beq
E_0 \leq E[\Psi].
\eeq
\Cref{eq:RayleighRitz} can be rewritten in matrix notation as a generalized eigenvalue problem
\cite{RayleighRitz}
\beq
\label{eq:BoundGenEig}
\textbf{Hc} = E\textbf{Sc},
\eeq
where
\beq
\label{eq:HijSij}
H_{ij} = \left< \bar{\phi}_i \left| H \right| \bar{\phi}_j \right>\!, \, S_{ij} = \left< \bar{\phi}_i \left| \right.\! \bar{\phi}_j \right>, 
\eeq
and \textbf{c} is the vector of coefficients for the wavefunction $\Psi$. The
normalization here is unimportant due to the division in \cref{eq:RayleighRitz}
and the form of \cref{eq:BoundGenEig}.

For PsH, the non-relativistic Hamiltonian is
\beq
\label{eq:BoundHamiltonian}
H = -\frac{1}{2} \Laplacian_{r_1} - \frac{1}{2} \Laplacian_{r_2} - \frac{1}{2} \Laplacian_{r_3} + \frac {1}{r_1}-\frac {1}{r_2}-\frac {1}{r_3}-\frac {1}{r_{12}}-\frac {1}{r_{13}}+\frac {1}{r_{23}}.
\eeq
The Laplacians in \cref{eq:BoundHamiltonian} are complicated when applied to the $\phi_j$ function.  We exploit the short-range nature of $\phi_i$ and $\phi_j$ by using integration by parts, similar to equation (3.21) of Armour and Humberston \cite{Armour1991}.
\beq
\label{eq:BoundGradient}
-\Int{ \phi_i \left(\Laplacian_{r_1} + \Laplacian_{r_2} + \Laplacian_{r_3} \right) \phi_j }{\tau} = \int \sum_{l=1}^3 \grad_{\!\bm{r}_l} \phi_i \bm{\cdot} \grad_{\!\bm{r}_l} \phi_j \, d\tau
\eeq
The differential $d\tau$ represents the 9-dimensional configuration 
space given by $r_1$, $r_2$, and $r_3$ (see \cref{chp:AngularInt}).
This expression is simpler than 
applying the Laplacian operators directly to $\phi_j$, and the summation is 
given by the following expression (a full derivation is given on the research 
Wiki \cite{Wiki}):
\begin{align}
\label{eq:GradGradShort}
\nonumber \sum_{l=1}^3 \grad_{\!\mathbf{r}_l} \phi_i \bm{\cdot} \grad_{\!\mathbf{r}_l} \phi_j = &\phi_i \phi_j \Bigg\{(\alpha^2 + \beta^2 + \gamma^2) - \frac{\alpha}{r_1}(k_i + k_j) - \frac{\beta}{r_2}(l_i + l_j) + \frac{\gamma}{r_3}(n_i + n_j) \\
\nonumber  &+ \frac{k_i k_j}{r_1^2} + \frac{l_i l_j}{r_2^2} + \frac{n_i n_j}{r_3^2} + \frac{2 m_i m_j}{r_{12}^2} + \frac{2 p_i p_j}{r_{13}^2} + \frac{2 q_i q_j}{r_{23}^2} \\
\nonumber  &+ \frac{r_1^2 + r_{12}^2 - r_2^2}{2 r_1^2 r_{12}^2} \left[-\alpha r_1(m_i+m_j) + (k_i m_j + k_j m_i)\right] \\
\nonumber  &+ \frac{r_1^2 + r_{13}^2 - r_3^2}{2 r_1^2 r_{13}^2} \left[-\alpha r_1(p_i+p_j) + (k_i p_j + k_j p_i)\right] \\
\nonumber  &+ \frac{r_{12}^2 + r_{13}^2 - r_{23}^2}{2 r_{12}^2 r_{13}^2} \left[m_i p_j + m_j p_i\right] \\
\nonumber  &+ \frac{r_2^2 + r_{12}^2 - r_1^2}{2 r_2^2 r_{12}^2} \left[-\beta r_2(m_i+m_j) + (m_i l_j + m_j l_i)\right] \\
\nonumber  &+ \frac{r_2^2 + r_{23}^2 - r_3^2}{2 r_2^2 r_{23}^2} \left[-\beta r_2(q_i+q_j) + (l_i q_j + l_j q_i)\right] \\
\nonumber  &+ \frac{r_{12}^2 + r_{23}^2 - r_{13}^2}{2 r_{12}^2 r_{23}^2} \left[m_i q_j + m_j q_i\right] \\
\nonumber  &+ \frac{r_3^2 + r_{13}^2 - r_1^2}{2 r_3^2 r_{13}^2} \left[-\gamma r_3(p_i+p_j) + (n_i p_j + n_j p_i)\right] \\
\nonumber  &+ \frac{r_3^2 + r_{23}^2 - r_2^2}{2 r_3^2 r_{23}^2} \left[-\gamma r_3(q_i+q_j) + (n_i q_j + n_j q_i)\right] \\
		   &+ \frac{r_{13}^2 + r_{23}^2 - r_{12}^2}{2 r_{13}^2 r_{23}^2} \left[p_i q_j + p_j q_i\right] \Bigg\}.
\end{align}
This is similar to the forms given in Refs.~\cite{Armour1991,VanReethThesis}.
The S-wave code also has an alternate formalism using the Laplacian, as given in
Ref.~\cite{Yan1997}.

The full expression for $H_{ij}$ from \cref{eq:HijSij} for real-valued $\phi$
after using \cref{eq:BoundHamiltonian,eq:BoundGradient} is then
\beq
\label{eq:BoundHFull}
\left< \bar{\phi}_i \left| H \right| \bar{\phi}_j \right> = \bigintsss{ \left[ \frac{1}{2}\sum_{l=1}^3 \grad_{\!\mathbf{r}_l} \bar{\phi}_i \boldsymbol{\cdot} \grad_{\!\mathbf{r}_l} \bar{\phi}_j + \left( \frac {1}{r_1}-\frac {1}{r_2}-\frac {1}{r_3}-\frac {1}{r_{12}}-\frac {1}{r_{13}}+\frac {1}{r_{23}} \right) \bar{\phi}_i \bar{\phi}_j \right]}{d\tau}.
\eeq
To reduce the number of integrations needed by half, we use a property of the
permutation operator. Since 
\beq
\left< \phi_i \left| H \right| \phi_j \right> = \left< P_{23} \phi_i \left| H \right| P_{23} \phi_j \right>
\eeq
and
\beq
\left< \phi_i \left| H \right| P_{23} \phi_j \right> = \left< P_{23} \phi_i \left| H \right| \phi_j \right>,
\eeq
\cref{eq:BoundHFull} becomes
\begin{subequations}
\label{eq:RRfinal}
\begin{align}
\nonumber \left< (1 \pm P_{23}) \phi_i \left| H \right| (1 \pm P_{23}) \phi_j \right> =& \left< \phi_i \left| H \right| \phi_j \right> \pm \left< P_{23} \phi_i \left| H \right| \phi_j \right> \\
&\pm \left< \phi_i \left| H \right| P_{23} \phi_j \right> + \left< P_{23} \phi_i \left| H \right| P_{23} \phi_j \right> \\
=& 2 \left[ \left< \phi_i \left| H \right| \phi_j \right> \pm \left< P_{23} \phi_i \left| H \right| \phi_j \right> \right] \\
=& 2 \left[ \left< \phi_i \left| H \right| \phi_j \right> \pm \left< \phi_i \left| H \right| P_{23} \phi_j \right> \right].
\end{align}
\end{subequations}

\section{Results}
\label{sec:BoundResults}

The binding energy (also known as the dissociation energy) is given by Ref.~\cite{Page1974} as
\beq
\label{eq:DissociationE}
E_d = -E_1 - \frac{3}{4} \text{ a.u.}
\eeq

The $-\frac{3}{4}$ comes from adding the ground state energies of Ps and H. 
If the PsH system has a lower energy than $-\frac{3}{4}$, the system is bound. 
Using the accurate Bubin and Adamowicz energy in \cref{tab:BoundEnergyOther}, 
this gives that $^1$S PsH is stable against dissociation into Ps and H by
$\SI{0.039 196 765 251}{au}$ or $\SI{1.066 598 271 959}{eV}$. 

\subsection{Bound State: Singlet}
\label{sec:BoundSinglet}

\setlength{\abovecaptionskip}{6pt}   
\setlength{\belowcaptionskip}{6pt}   
\begin{table}
\small
\centering
\begin{tabular}{c c c c c c c d{1.12}}
\toprule
$\omega$ & Terms & $\alpha$ & $\beta$ & $\gamma$ & Total Energy (a.u.) & Binding Energy (eV) & \multicolumn{1}{c}{$\Delta E$ (a.u.)} \\ [0.5ex]
\midrule
0 & 1 &     0.60 & 0.60 & 1.00 & $-$0.541 492 378 889 & --- & \multicolumn{1}{c}{---} \\
1 & 7 &     0.60 & 0.60 & 1.00 & $-$0.744 334 244 165 & ---               &  0.202 841 865 276 \\
2 & 28 &    0.60 & 0.60 & 1.00 & $-$0.778 357 106 972 & 0.771 636 156 726 &  0.034 022 862 807 \\
3 & 84 &    0.60 & 0.60 & 1.00 & $-$0.786 807 448 395 & 1.001 581 651 009 &  0.008 450 341 423 \\
4 & 210 &   0.60 & 0.60 & 1.00 & $-$0.788 685 563 109 & 1.052 687 753 648 &  0.001 878 114 714 \\
5 & 462 &   0.60 & 0.60 & 1.00 & $-$0.789 082 645 582 & 1.063 492 917 716 &  0.000 397 082 473 \\
6 & 916 &   0.60 & 0.60 & 1.00 & $-$0.789 169 509 836 & 1.065 856 614 384 &  0.000 086 864 254 \\
7 & 1585 &  0.60 & 0.60 & 1.00 & $-$0.789 189 568 390 & 1.066 402 435 425 &  0.000 020 058 554 \\
8 & 1925 &  0.60 & 0.60 & 1.00 & $-$0.789 194 559 324 & 1.066 538 245 640 &  0.000 004 990 934 \\
9 & 2166 &  0.60 & 0.60 & 1.00 & $-$0.789 195 830 870 & 1.066 572 846 182 &  0.000 001 271 546 \\
10 & 2205 & 0.60 & 0.60 & 1.00 & $-$0.789 196 323 586 & 1.066 586 253 647 &  0.000 000 492 716 \\
11 & 1674 & 0.58 & 0.60 & 1.00 & $-$0.789 196 284 600 & 1.066 585 192 793 & -0.000 000 038 986 \\
\bottomrule
\end{tabular}
\caption{Ground state energy of PsH}
\label{tab:BoundEnergyOld}
\end{table}

The original double precision PsH code was run for a simple choice of the nonlinear parameters $\alpha$, $\beta$ and $\gamma$. During these initial runs, LAPACK returned valid energies through $\omega = 5$. With the $\omega = 6$ runs, it had trouble using the full 924 terms, with \texttt{dsygv} giving an error. Using Todd's algorithm (\cref{sec:ToddBound}), this code returned a usable 916 terms for $\omega = 6$, as given in \cref{tab:BoundEnergyOld}. The original code worked well through $\omega = 10$, but going from $\omega = 9$ to $\omega = 10$ only added an additional 39 terms. The run for $\omega = 11$ was obviously a problem, since it gave less terms than $\omega = 10$ and a higher energy, even when changing the $\alpha$ parameter slightly.

\setlength{\abovecaptionskip}{6pt}   
\setlength{\belowcaptionskip}{6pt}   
\begin{table}
\small
\centering
\centerline{
\begin{tabular}{c c c c c c c c}
\toprule
$\omega$ & Terms & $\alpha$ & $\beta$ & $\gamma$ & Total Energy (a.u.) & Binding Energy (eV) & $\Delta E$ (a.u.) \\ [0.5ex]
\midrule
0 & 1    & 0.586 & 0.580 & 1.093 & -0.558 977 058 051 & --- & --- \\
1 & 7    & 0.586 & 0.580 & 1.093 & -0.744 698 936 920 & ---               & 0.185 721 878 869 \\
2 & 28   & 0.586 & 0.580 & 1.093 & -0.778 246 602 473 & 0.768 629 176 247 & 0.033 547 665 553 \\
3 & 84   & 0.586 & 0.580 & 1.093 & -0.786 743 703 126 & 0.999 847 053 924 & 0.008 497 100 653 \\
4 & 210  & 0.586 & 0.580 & 1.093 & -0.788 672 801 036 & 1.052 340 479 962 & 0.001 929 097 910 \\
5 & 462  & 0.586 & 0.580 & 1.093 & -0.789 082 645 582 & 1.063 460 197 197 & 0.000 409 844 546 \\
6 & 924  & 0.586 & 0.580 & 1.093 & -0.789 169 509 836 & 1.065 861 354 038 & 0.000 086 864 254 \\
7 & 1716 & 0.586 & 0.580 & 1.093 & -0.789 189 730 694 & 1.066 406 851 931 & 0.000 020 220 858 \\
\bottomrule
\end{tabular}
}
\caption{Ground state energy of $^1$S PsH with full set of terms and original ordering}
\label{tab:BoundEnergy1}
\end{table}

The next and current version of the code uses quadruple precision and is able 
to do full runs through $\omega = 8$ without omitting terms (not shown in 
\cref{tab:BoundEnergy1}). Linear dependence is also decreased when the 
nonlinear parameters are different (\cref{sec:BoundOptimization} for 
parameter optimization). The Ps-H scattering problem (\cref{chp:SWave}) is 
more difficult, so a run with $\omega = 7$ is all that is needed.
\Cref{tab:BoundEnergy1} shows the PsH energies through $\omega = 7$ for the full 
set of terms described by \cref{eq:OmegaDef}.


\setlength{\abovecaptionskip}{6pt}   
\setlength{\belowcaptionskip}{6pt}   
\begin{table}
\small
\centering
\centerline{
\begin{tabular}{c c c c c c c c}
\toprule
$\omega$ & Terms & $\alpha$ & $\beta$ & $\gamma$ & Total Energy (a.u.) & Binding Energy (eV) & $\Delta E$ (a.u.) \\ [0.5ex]
\midrule
0 & 1    & 0.586 & 0.580 & 1.093 & -0.558 977 058 051 & --- & --- \\
1 & 5    & 0.586 & 0.580 & 1.093 & -0.718 445 865 883 & ---               & 0.159 468 807 832 \\
2 & 25   & 0.586 & 0.580 & 1.093 & -0.776 355 701 568 & 0.717 175 143 631 & 0.057 909 835 685 \\
3 & 77   & 0.586 & 0.580 & 1.093 & -0.786 645 720 870 & 0.997 180 821 018 & 0.010 290 019 302 \\
4 & 199  & 0.586 & 0.580 & 1.093 & -0.788 665 304 510 & 1.052 136 489 091 & 0.002 019 583 640 \\
5 & 436  & 0.586 & 0.580 & 1.093 & -0.789 080 739 334 & 1.063 441 046 057 & 0.000 415 434 824 \\
6 & 856  & 0.586 & 0.580 & 1.093 & -0.789 169 644 174 & 1.065 860 269 913 & 0.000 088 904 841 \\
7 & 1505 & 0.586 & 0.580 & 1.093 & -0.789 189 725 050 & 1.066 406 698 333 & 0.000 020 080 875 \\
\bottomrule
\end{tabular}
}
\caption{Ground state energy of $^1$S PsH with Todd set of terms and original ordering}
\label{tab:BoundEnergyTodd1}
\end{table}

As described later in \cref{sec:CompPhase}, we cannot use the full 1716 terms 
for the Ps-H scattering problem. \Cref{tab:BoundEnergyTodd1} gives the ground 
state energies using the restricted set of terms using Todd's method with the 
original ordering. The cutoffs in $\omega$ are easily seen using the
ViewOmegaCutoffs.py script (\cref{chp:Programs}).

{
	\iftoggle{UNT}{\singlespacing}

	\begin{center}
	\setlength{\aboverulesep}{0pt}
	\setlength{\belowrulesep}{0pt}
	\setlength{\extrarowheight}{.75ex}
	\footnotesize
	\rowcolors{2}{gray!15}{white}
	\begin{longtable}{l c l l}
	\rowcolors{2}{gray!15}{white}
	\label{tab:BoundEnergyOther} \\
	\toprule
	\rowcolor{gray!25} \multicolumn{1}{c}{} & \multicolumn{1}{c}{} & \multicolumn{1}{c}{Total} & \multicolumn{1}{c}{Binding} \\
	\rowcolor{gray!25} \multicolumn{1}{c}{Group / Method} & \multicolumn{1}{c}{Terms} & \multicolumn{1}{c}{Energy (au)} & \multicolumn{1}{c}{Energy (eV)} \\
	\midrule
	\endfirsthead

	\rowcolor{white}\multicolumn{4}{r}{{  }} \\
	\toprule
	\rowcolor{gray!25} \multicolumn{1}{c}{} & \multicolumn{1}{c}{} & \multicolumn{1}{c}{Dissociation} & \multicolumn{1}{c}{Binding} \\
	\rowcolor{gray!25} \multicolumn{1}{c}{Group / Method} & \multicolumn{1}{c}{Terms} & \multicolumn{1}{c}{Energy (au)} & \multicolumn{1}{c}{Energy (eV)} \\
	\midrule
	\endhead

	\hline \multicolumn{4}{r}{{Continued on next page}} \\ \hline
	\rowcolor{white} \caption[Positronium hydride energy values]{Positronium hydride energy values. Starred values are the reported values. Unstarred values are obtained by using the conversion factor given in \cref{sec:Units}. Results marked by $^a$ take into account the finite mass correction.} \\
	\endfoot

	\caption[Positronium hydride energy values]{Continued from previous page. Positronium hydride energy values. Starred values are the reported values. Unstarred values are obtained by using the conversion factor given in \cref{sec:Units}. Results marked by $^a$ take into account the finite mass correction.}
	\endlastfoot
	\rowcolors{2}{gray!15}{white}
	Current work / Variational Hylleraas $(\omega = 7)$ & 1505 & -0.789 189 725 & 1.066 406 705 \\
	Frolov (2010) \cite{Frolov2010} / Semi-exponential & 84 & -0.788 516 419$^\star$ & 1.048 085 11 \\
	Bubin (2006) \cite{Bubin2006} / ECGs variational & 5000 & -0.789 196 765 251$^\star$ & 1.066 598 271 959 \\
	Bubin (2006) \cite{Bubin2006} / ECGs variational$^a$ & 5000 & -0.788 870 710 444$^\star$ & 1.057 725 869 06 \\
	Mitroy (2006) \cite{Mitroy2006} / ECGs with SVM & 1800 & -0.789 196 740$^\star$ & 1.066 597 58 \\
	Chiesa (2004) \cite{Chiesa2004} / Quantum Monte Carlo & --- & -0.784 620$^\star$ & 0.942 058 \\
	Bubin (2004) \cite{Bubin2004} / ECGs$^a$ & 3200 & -0.788 870 706 6$^\star$ & 1.057 725 764 \\
	Van Reeth (2003) \cite{VanReeth2003} / Variational Hylleraas $(\omega = 6)$ & 721 & -0.789 156 & 1.065 5$^\star$ \\
	Bressanini (2003) \cite{Bressanini2003} / Variational Monte Carlo & 1 & -0.786 073$^\star$ & 0.981 596 \\
	Saito (2003) \cite{Saito2003a} / CI & 13230 & -0.786 793$^\star$ & 1.001 19 \\
	Bromley (2001) \cite{Bromley2001} / CI & 95324 & -0.786 776 1$^\star$ & 1.000 729 \\
	Saito (2000) \cite{Saito2000} / Hylleraas & 396 & -0.788 951$^\star$ & 1.059 91 \\
	Bromley (2000) \cite{Bromley2000} / CI & --- & -0.784 301 8$^\star$ & 0.933 399 5 \\
	Yan (1999) \cite{Yan1999} / Variational Hylleraas $(\omega = 12)$ & 5741 & -0.789 196 705 1$^\star$ & 1.066 596 635 \\
	Yan (1999) \cite{Yan1999} / Variational Hylleraas $(\omega \rightarrow \infty)$ & --- & -0.789 196 714 7$^\star$ & 1.066 596 896 \\
	Yan (1999) \cite{Yan1999a} / Variational Hylleraas$^a$ & 4705 & -0.788 853 107$^\star$ & 1.057 246 85 \\
	Ryzhikh (1999) \cite{Ryzhikh1999} / ECGs & 750 & -0.789 196 0$^\star$ & 1.066 577 \\
	Adhikari (1999) \cite{Adhikari1999} / Five-state CC & --- & -0.788 6 & 1.05$^\star$ \\
	Mella (1999) \cite{Mella1999} / DMC & --- & -0.789 15$^\star$ & 1.065 3 \\
	Ryzhikh (1998) \cite{Ryzhikh1998} / ECGs with SVM & 500 & -0.789 194 4$^\star$ & 1.066 534 \\
	Strasburger (1998) \cite{Strasburger1998} / ECGs & 332 & -0.789 185$^\star$ & 1.066 278 \\
	Bressanini (1998) \cite{Bressanini1998} / DMC & --- & -0.789 175$^\star$ & 1.066 01 \\
	Jiang (1998) \cite{Jiang1998} / DMC & --- & -0.789 18$^\star$ & 1.066 1 \\
	Jiang (1998) \cite{Jiang1998} / Variational Monte Carlo & 1 & -0.777 4$^\star$ & 0.745 6 \\
	Le Sech (1998) \cite{LeSech1998} / Variational Monte Carlo & 1 & -0.772 3$^\star$ & 0.606 8 \\
	Usukura (1998) \cite{Usukura1998} / ECGs with SVM & 1600 & -0.789 196 553 6$^\star$ & 1.066 592 513 \\
	Frolov (1997) \cite{Frolov1997a} / James-Coolidge variational & 924 & -0.789 136 9$^\star$ & 1.064 969 \\
	Frolov (1997) \cite{Frolov1997a} / James-Coolidge variational & --- & -0.789 181 8$^\star$ & 1.066 191 \\
	Frolov (1997) \cite{Frolov1997c} / Kolesnikov-Tarasov variational & --- & -0.789 179 4$^\star$ & 1.066 126 \\
	Frolov (1997) \cite{Frolov1997c} / Kolesnikov-Tarasov variational$^a$ & --- & -0.788 853 4$^\star$ & 1.057 254 \\
	Ryzhikh (1997) \cite{Ryzhikh1997} / SVM & 400 & -0.789 183$^\star$ & 1.066 22 \\
	Yoshida (1996) \cite{Yoshida1996} / DMC & --- & -0.789 1$^\star$ & 1.06$^\star$ \\
	Saito (1995) \cite{Saito1995a} / Hylleraas & --- & -0.774 71$^\star$ & 0.672 39 \\*
	Saito (1995) \cite{Saito1995} / Restricted Hartree-Fock & --- & -0.776 & 0.70$^\star$ \\*
	Strasburger (1995) \cite{Strasburger1995} / CI & --- & -0.763 7$^\star$ & 0.372 8 \\*
	Strasburger (1995) \cite{Strasburger1995} / SCF & --- & -0.666 9$^\star$ & -2.261 \\*
	Schrader (1992) \cite{Schrader1992} / Experiment & --- & -0.790 & $1.1 \pm 0.2$ \\*
	Ho (1986) \cite{Ho1986} / Variational with Hylleraas & 396 & -0.788 945$^\star$ & 1.059 75 \\*
	Maruyama (1985) \cite{Maruyama1985,Saito2003a} / Hylleraas & --- & -0.788 211$^\star$ & 1.039 77 \\*
	Ho (1978) \cite{Ho1978} / Variational with Hylleraas & 210 & -0.787 525$^\star$ & 1.021 12$^\star$ \\*
	Clary (1976) \cite{Clary1976} / Variational & 67 & -0.784 161$^\star$ & 0.929 568 \\*
	Page (1974) \cite{Page1974} / Variational & 70 & -0.786 79$^\star$ & 1.00 11 \\*
	Navin (1974) \cite{Navin1974} / Variational & 17 & -0.779 2$^\star$ & 0.794 6 \\*
	Houston (1973) \cite{Houston1973} / Variational & 56 & -0.774 7$^\star$ & 0.672 5 \\*
	Lebeda (1969) \cite{Lebeda1969} / Variational & 12 & -0.774 2$^\star$ & 0.658 5 \\*
	Ludwig (1966) \cite{Ludwig1966} / Configuration interaction & 9 & -0.759 0$^\star$ & 0.244 9 \\*
	Goldanskii (1964) \cite{Goldanskii1964,Clary1976} / --- & --- & -0.667 7$^\star$ & -2.2395 \\*
	Neamtan (1962) \cite{Neamtan1962} / Variational exponential & 2 & -0.758 4$^\star$ & 0.228 6 \\*
	Ore (1951) \cite{Ore1951} / Variational exponential & 2 & -0.752 51$^\star$ & 0.068 301 \\*
	Walters (2004) \cite{Walters2004} / CC 14Ps14H + H$^-$ & --- & -0.787 9 & 1.03$^\star$\\*
	Walters (2004) \cite{Walters2004} / CC 9Ps9H + H$^-$ & --- & -0.787 5 & 1.02$^\star$\\*
	Blackwood (2002) \cite{Blackwood2002} / CC 14Ps14H & --- & -0.786 5  & 0.994$^\star$ \\*
	Blackwood (2002) \cite{Blackwood2002} / CC 9Ps9H & --- & -0.785 4 & 0.963$^\star$ \\*
	Blackwood (2002) \cite{Blackwood2002b} / CC 22Ps1H + H$^-$ & --- & -0.781 2 & 0.850$^\star$ \\*
	Campbell (1998) \cite{Campbell1998} / CC 22Ps1H & --- & -0.773 3 & 0.634$^\star$ \\*
	\bottomrule
	\end{longtable}
	\end{center}

}

There have been a large number of calculations of the PsH binding energy over 
the years, starting with Ore's prediction in 1951 that PsH could exist
\cite{Ore1951}. As computing power increased, using a Hylleraas-type basis 
set with hundreds or even thousands of terms became possible
\cite{Ho1978,Ho1986,Yan1999,VanReeth2003}. The Hylleraas-type results were 
the most accurate until very accurate ECG results from Mitroy
\cite{Mitroy2006} and Bubin and Adamowicz \cite{Bubin2004,Bubin2006}. The 
ECGs do not satisfy the Kato cusp condition \cite{Kato1957}, but good
optimization of the basis set can give good results \cite{Mitroy2013}.

Another paper by Frolov \cite{Frolov2010} is more of an introduction of a 
modified basis set to this problem, showing how only 84 terms gives a better 
energy than a much larger Hylleraas set. Yan and Ho \cite{Yan1999} use a 
Hylleraas basis set with 5 sectors that each have different nonlinear 
parameters. Frolov's work essentially uses the same basis set but gives each 
term its own set of nonlinear parameters, optimizing all of them 
simultaneously.

The last six entries of this table give the CC results of the Belfast group. 
These are grouped together to show how different states and pseudostates in 
the CC calculations can give better approximations to the binding energy. It 
is particularly clear that adding the H$^-$ channel vastly improves the 
accuracy.

The Hylleraas binding energy from our work in the first line of the table compares
well with the accurate energies of Refs.~\cite{Bubin2006, Mitroy2006, Yan1999},
but it is not as accurate of a calculation. The purpose of doing the PsH 
calculation was not to get the best result but to test how well the short-
range terms for Ps-H scattering represent the short-range interactions. Based 
on \cref{tab:BoundEnergyOther}, the short-range interactions are described 
well. This also gave experience working with the Hylleraas basis set, and 
finding the PsH energy is a simpler problem than Ps-H scattering.

\subsection{Triplet Energy Eigenvalues}
\label{sec:TripletEigenvalues}

Our code does not predict a triplet bound state. Mitroy and Bromley have 
published a paper \cite{Mitroy2007} claiming a stable triplet bound state, 
but our code does not have the appropriate type of wavefunction to see this, 
since it is for a H($2p$) + Ps($2p$) state. They also use a very large 
configuration interaction basis, and this bound state is very shallow. 

Despite not predicting a bound state, we run the energy eigenvalue code for 
the triplet so that we can use this for the short-range terms for the 
scattering programs. The triplet case was more sensitive to the accuracy of 
the matrix elements. Changing the energy eigenvalue code to quadruple 
precision let us use more short-range terms in our calculations.


\setlength{\abovecaptionskip}{6pt}   
\setlength{\belowcaptionskip}{6pt}   
\begin{table}
\centering
\begin{tabular}{c c c c c c c}
\toprule
$\omega$ & Terms & $\alpha$ & $\beta$ & $\gamma$ & Total Energy (a.u.) & $\Delta E$ (a.u.) \\ [0.5ex]
\midrule
0 & 1    & 0.323 & 0.334 & 0.975 & -0.500 031 146 247 & --- \\
1 & 7    & 0.323 & 0.334 & 0.975 & -0.623 725 636 031 & 0.123 694 489 784 \\
2 & 28   & 0.323 & 0.334 & 0.975 & -0.679 176 805 541 & 0.055 451 169 510 \\
3 & 84   & 0.323 & 0.334 & 0.975 & -0.710 407 818 871 & 0.031 231 013 330 \\
4 & 210  & 0.323 & 0.334 & 0.975 & -0.727 102 619 903 & 0.016 694 801 032 \\
5 & 462  & 0.323 & 0.334 & 0.975 & -0.735 860 693 040 & 0.008 758 073 137 \\
6 & 924  & 0.323 & 0.334 & 0.975 & -0.740 622 381 908 & 0.004 761 688 868 \\
7 & 1716 & 0.323 & 0.334 & 0.975 & -0.743 386 825 704 & 0.002 764 443 796 \\
\bottomrule
\end{tabular}
\caption{Eigenvalues of $^3$S with full set of terms and original ordering}
\label{tab:BoundEnergy3}
\end{table}

Similar to the singlet, we cannot use the full 1716 
terms for the Ps-H scattering problem. \Cref{tab:BoundEnergyTodd3} gives the 
energy eigenvalues using the restricted set of terms using Todd's method
with original ordering. 

\setlength{\abovecaptionskip}{6pt}   
\setlength{\belowcaptionskip}{6pt}   
\begin{table}
\centering
\begin{tabular}{c c c c c c c}
\toprule
$\omega$ & Terms & $\alpha$ & $\beta$ & $\gamma$ & Total Energy (a.u.) & $\Delta E$ (a.u.) \\ [0.5ex]
\midrule
0 & 1    & 0.323 & 0.334 & 0.975 & -0.500 031 146 247 & --- \\
1 & 7    & 0.323 & 0.334 & 0.975 & -0.623 725 636 031 & 0.123 694 489 784 \\
2 & 27   & 0.323 & 0.334 & 0.975 & -0.679 173 632 477 & 0.055 447 996 446 \\
3 & 81   & 0.323 & 0.334 & 0.975 & -0.710 405 860 558 & 0.031 232 228 081 \\
4 & 201  & 0.323 & 0.334 & 0.975 & -0.727 100 988 826 & 0.016 695 128 267 \\
5 & 432  & 0.323 & 0.334 & 0.975 & -0.735 860 470 972 & 0.008 759 482 146 \\
6 & 854  & 0.323 & 0.334 & 0.975 & -0.740 622 201 811 & 0.004 761 730 840 \\
7 & 1633 & 0.323 & 0.334 & 0.975 & -0.743 386 893 723 & 0.002 764 691 911 \\
\bottomrule
\end{tabular}
\caption{Eigenvalues of $^3$S with Todd set of terms and original ordering}
\label{tab:BoundEnergyTodd3}
\end{table}

\section{Stabilization}

Hazi and Taylor \cite{Hazi1970} introduced the concept of stabilization plots 
to PsH bound state calculations.
Specifically, they plotted 
the four lowest eigenvalues with respect to $N$ and looked for avoided 
crossings. A resonance only exists if there is an avoided crossing between 
eigenvalues. The avoided crossing that they found at $\SI{0.428 985}{a.u.}$ (or 
$\SI{5.8366}{eV}$ as shown in \cref{tab:SWaveResonancesOther}) corresponds 
roughly with the first $^1$S resonance. Van Reeth and Humberston \cite{
VanReeth2004} also did an analysis of stabilized eigenvalues but with the 
same type of basis set that we use.

\begin{figure}
	\centering
	\includegraphics[width=4.5in]{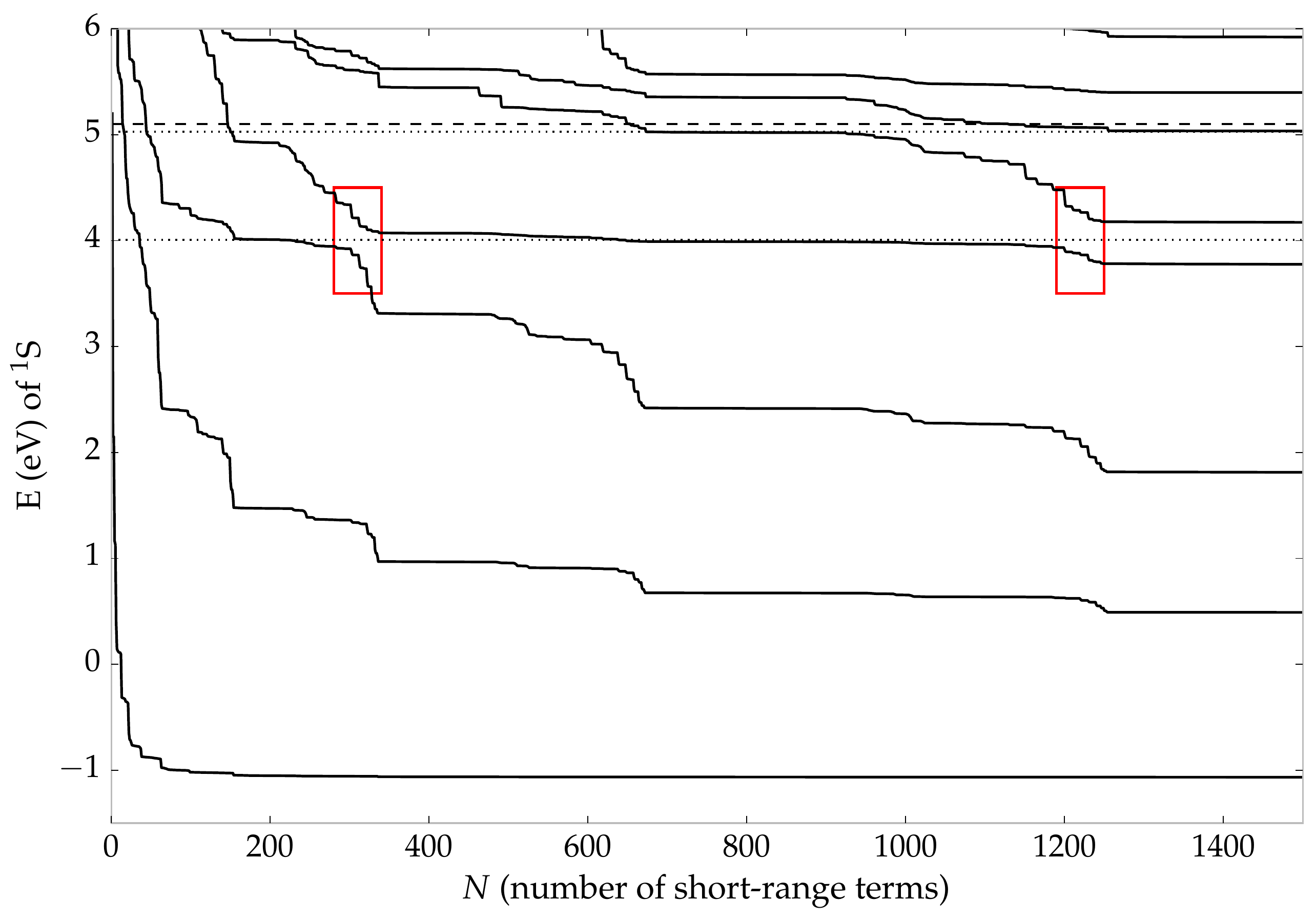}
	\caption[$^1$S eigenvalues]{$^1$S eigenvalues. The dashed line represents the Ps(n=2) threshold. Dotted lines represent complex Kohn $^1$S resonance positions. The red rectangles are potential avoided crossings.}
	\label{fig:swavesinglet-eigen}
\end{figure}

Plotting the first 10 eigenvalues with respect to $N$ for $^1$S gives
\cref{fig:swavesinglet-eigen}. As noted, the dashed line represents the Ps(n=2) 
inelastic threshold at $\SI{5.102}{eV}$. The dotted lines correspond to the 
first two complex Kohn $^1$S resonance positions given in
\cref{tab:SWaveResonancesOther}. The wavefunction we use is not optimized for this type of 
analysis, but there is evidence of avoided crossings given by the red 
rectangles in \cref{fig:swavesinglet-eigen}. The position of the plateau in 
between these rectangles corresponds relatively well with the first resonance 
position, $^1E_R$, we find in \cref{tab:SWaveResonancesOther} from the full 
scattering calculations in \cref{chp:SWave}. The line for the second 
resonance lines up with the 5th eigenvalue from roughly 700 to 900 terms and 
with the 6th eigenvalue starting at about 1250 terms. 
The estimates of the two resonance positions are $\SI{3.99}{eV}$ 
and $\SI{5.03}{eV}$.

Van Reeth and Humberston \cite{VanReeth2004} had difficulty doing the same 
type of stabilization with these Hylleraas-type terms. Other wavefunctions or 
stabilization methods may work better for this system. For instance, Yan and 
Ho \cite{Yan2003} vary a scaling factor and plot the eigenvalue energies
with respect to this.

\begin{figure}
	\centering
	\includegraphics[width=4.5in]{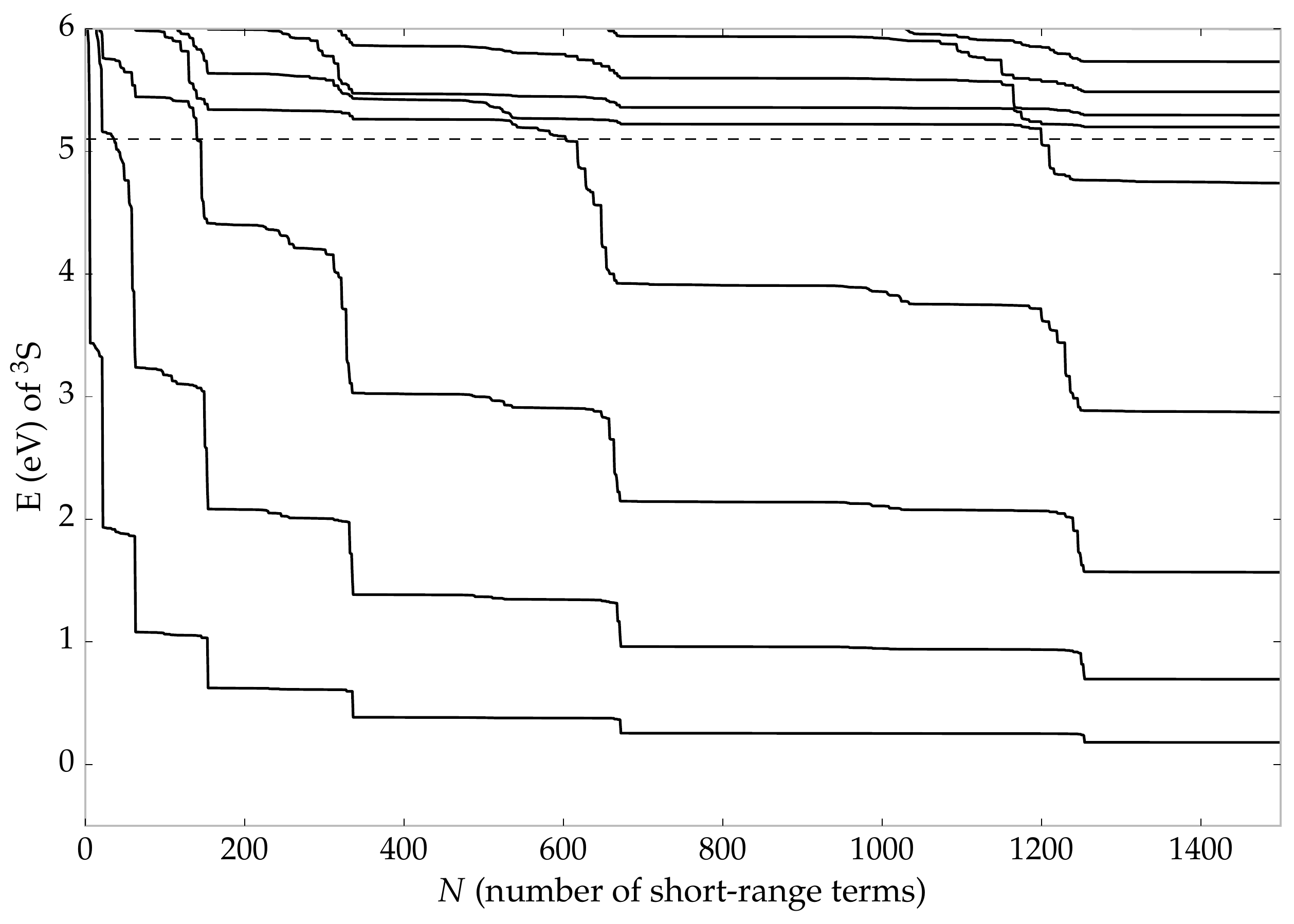}
	\caption[$^3$S eigenvalues]{$^3$S eigenvalues. The dashed line represents the Ps(n=2) threshold.}
	\label{fig:swavetriplet-eigen}
\end{figure}

\Cref{fig:swavetriplet-eigen} shows the same type of stabilization plot for
$^3$S. It is clear that below the Ps(n=2) threshold, there are no avoided 
crossings, meaning that there are no resonances in the S-wave triplet for 
this energy range. 

\begin{figure}
	\centering
	\includegraphics[width=4.5in]{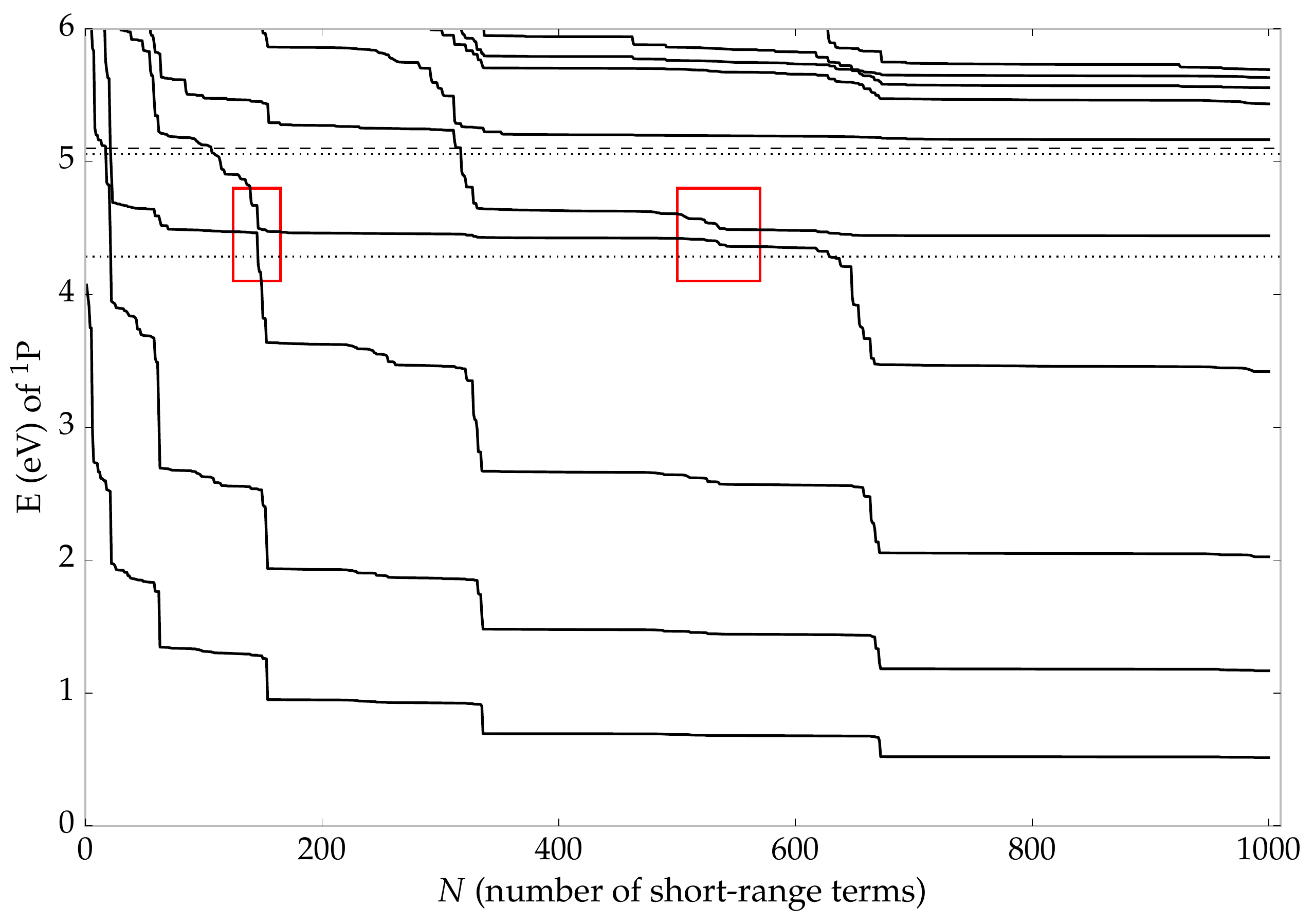}
	\caption[$^1$P eigenvalues for first symmetry only]{$^1$P eigenvalues for first symmetry only. The dashed line represents the Ps(n=2) threshold. Dotted lines represent complex Kohn $^1$P resonance positions.}
	\label{fig:pwavesinglet-eigen-unpaired}
\end{figure}

\begin{figure}
	\centering
	\includegraphics[width=4.5in]{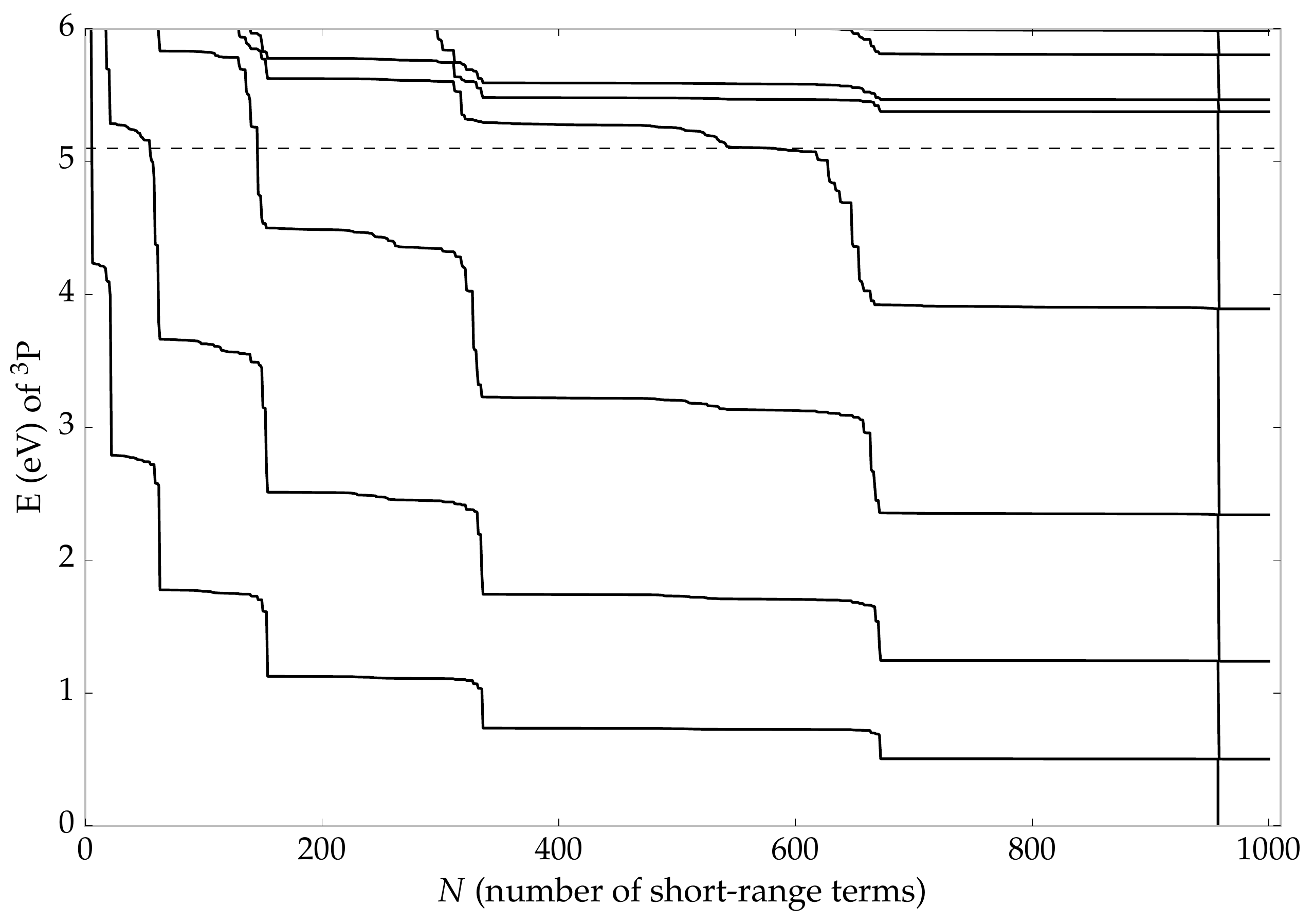}
	\caption[$^3$P eigenvalues for first symmetry only]{$^3$P eigenvalues for first symmetry only. The dashed line represents the Ps(n=2) threshold.}
	\label{fig:pwavetriplet-eigen-unpaired}
\end{figure}

\Cref{fig:pwavesinglet-eigen-unpaired,fig:pwavetriplet-eigen-unpaired} show 
the stabilization plots for the P-wave eigenvalues when only the first 
symmetry (see \cref{sec:PWaveFn}) is evaluated. 
In \cref{fig:pwavetriplet-eigen-unpaired}, linear dependence becomes a 
problem at 957 terms; hence there is an extra eigenvalue below
zero, which is not actually indicative of a bound state.

In \cref{fig:pwavesinglet-eigen-unpaired}, there is evidence of avoided 
crossings for the first resonance marked by the red rectangles. The second 
avoided crossing is barely noticeable. Taking the fifth eigenvalue at 800 
terms gives $\SI{4.44}{eV}$, which is not very much in line with the accurate 
calculations shown in \cref{tab:PWaveResonancesOther}, including the complex 
Kohn. The second resonance is above the Ps(n=2) threshold here, making it 
physically different from the actual resonance.

\begin{figure}
	\centering
	\includegraphics[width=4.5in]{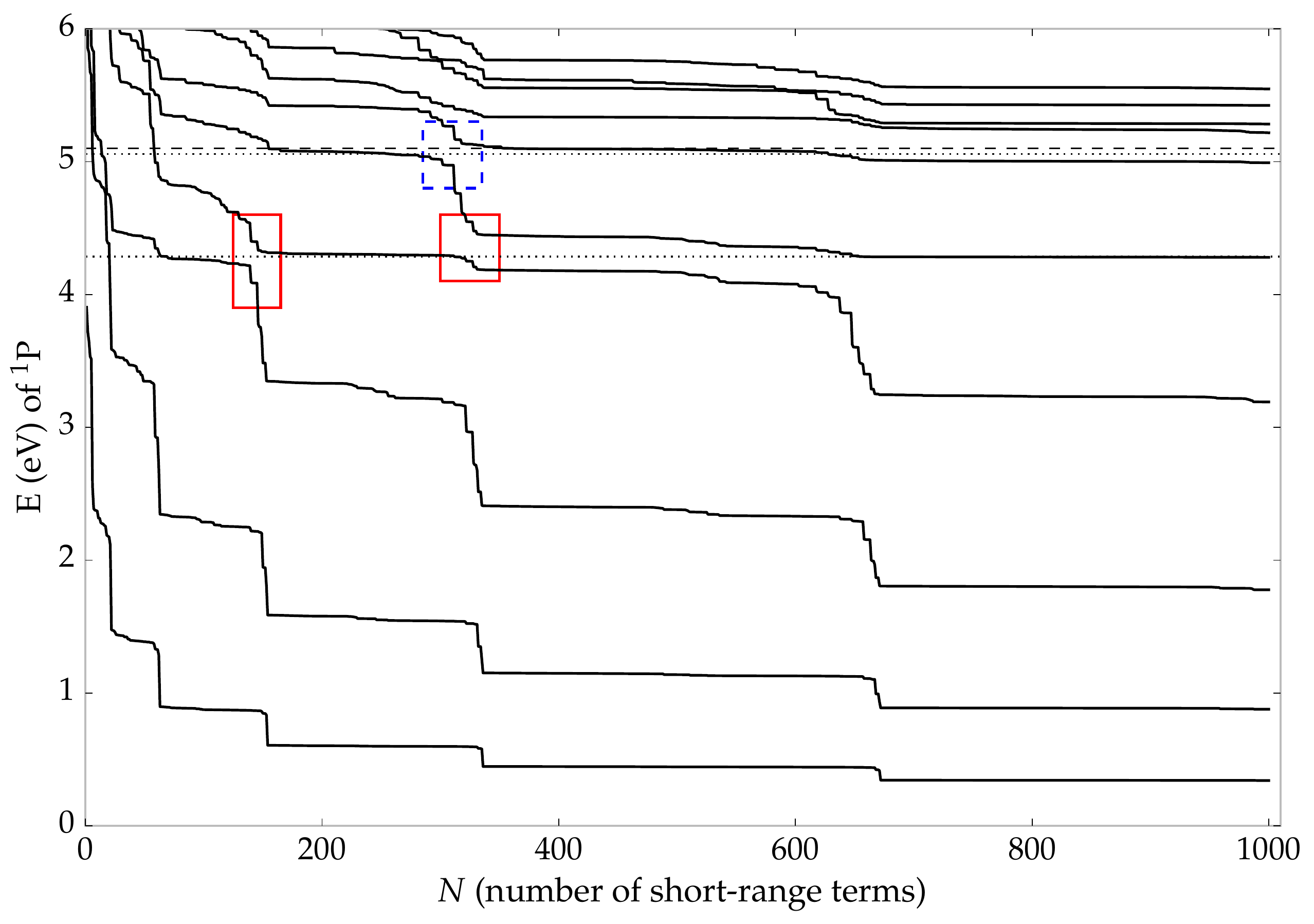}
	\caption[$^1$P eigenvalues for both symmetries]{$^1$P eigenvalues for both symmetries. The dashed line represents the Ps(n=2) threshold. Dotted horizontal lines represent complex Kohn $^1$P resonance positions. The approximate avoided crossings are denoted by rectangles.}
	\label{fig:pwavesinglet-eigen-paired}
\end{figure}

\begin{figure}
	\centering
	\includegraphics[width=4.5in]{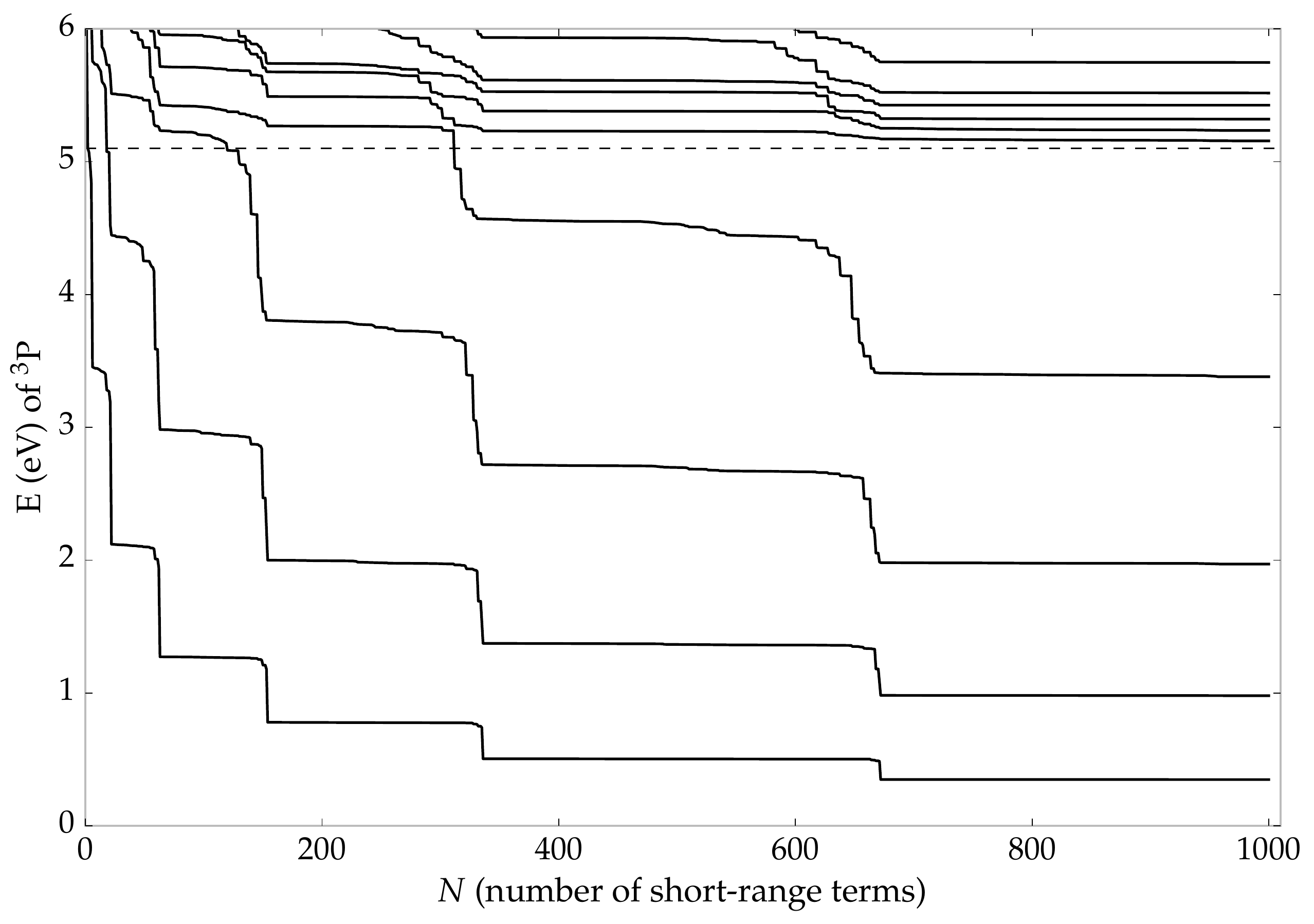}
	\caption[$^3$P eigenvalues for both symmetries]{$^3$P eigenvalues for both symmetries. The dashed line represents the Ps(n=2) threshold.}
	\label{fig:pwavetriplet-eigen-paired}
\end{figure}

\Cref{fig:pwavesinglet-eigen-paired,fig:pwavetriplet-eigen-paired} are also 
for the P-wave, but they use both the first and second symmetries paired. So 
each term number given on the x-axis is actually two terms, i.e. $N = 200$ is 
a total of 400 short-range terms, 200 of the first symmetry and 200 of the 
second symmetry. 

In \cref{fig:pwavesinglet-eigen-paired}, the first resonance position lines 
up relatively well with the full scattering calculations. The second 
resonance is a narrow resonance, and there is an avoided crossing enclosed in 
the dashed blue rectangle. The approximate position of the resonances are 
taken as the fifth and sixth eigenvalues at 1000 terms, which are tabulated 
in \cref{tab:Stabilization} and can be compared to the full complex Kohn 
calculation in \cref{tab:PWaveResonancesOther}.

\begin{figure}
	\centering
	\includegraphics[width=4.5in]{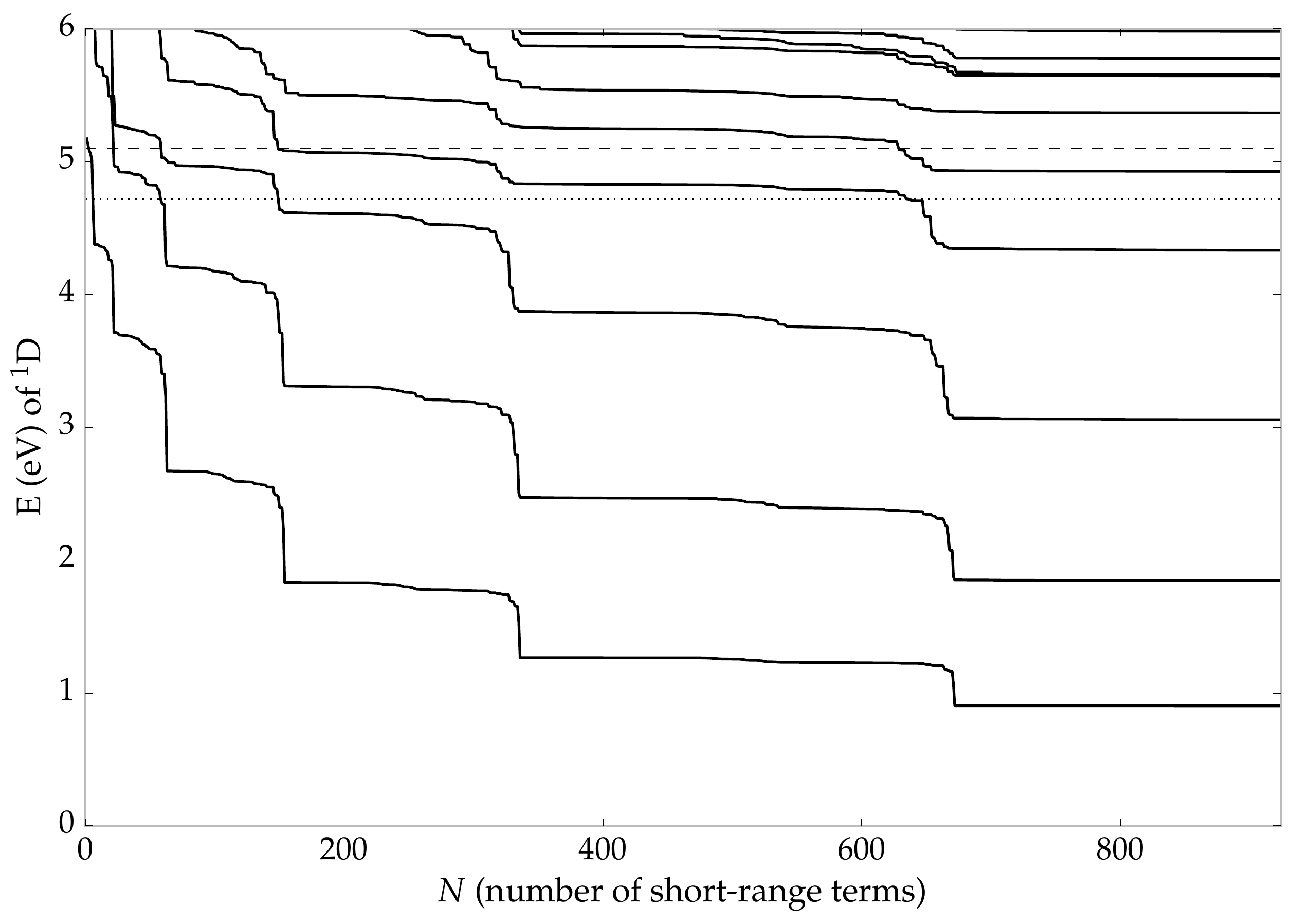}
	\caption[$^1$D eigenvalues for first symmetry only]{$^1$D eigenvalues for first symmetry only. The dashed line represents the Ps(n=2) threshold. The dotted line represents the complex Kohn $^1$D resonance position.}
	\label{fig:dwavesinglet-eigen-unpaired}
\end{figure}

\begin{figure}
	\centering
	\includegraphics[width=4.5in]{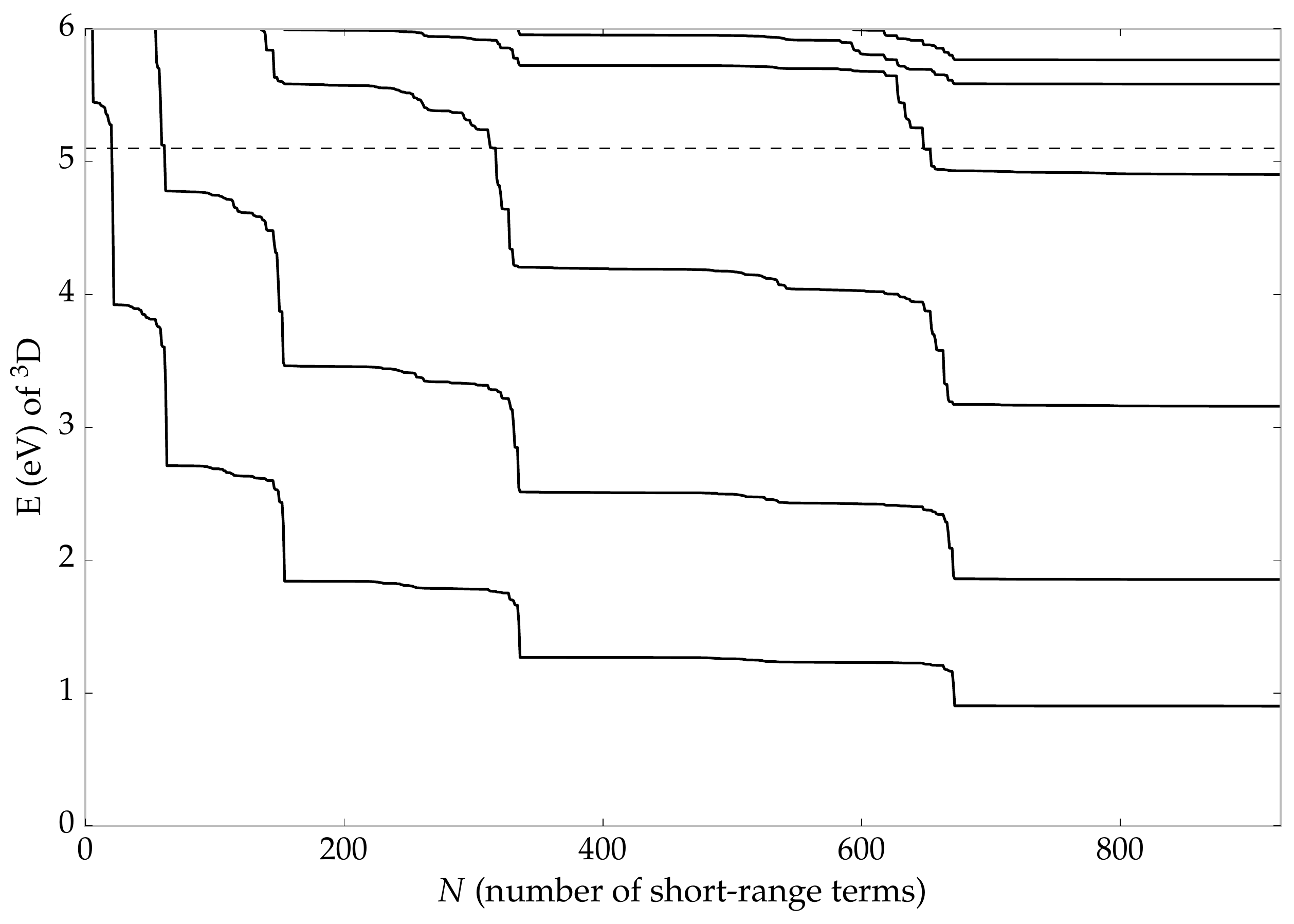}
	\caption[$^3$D eigenvalues for first symmetry only]{$^3$D eigenvalues for first symmetry only. The dashed line represents the Ps(n=2) threshold.}
	\label{fig:dwavetriplet-eigen-unpaired}
\end{figure}

\begin{figure}
	\centering
	\includegraphics[width=4.5in]{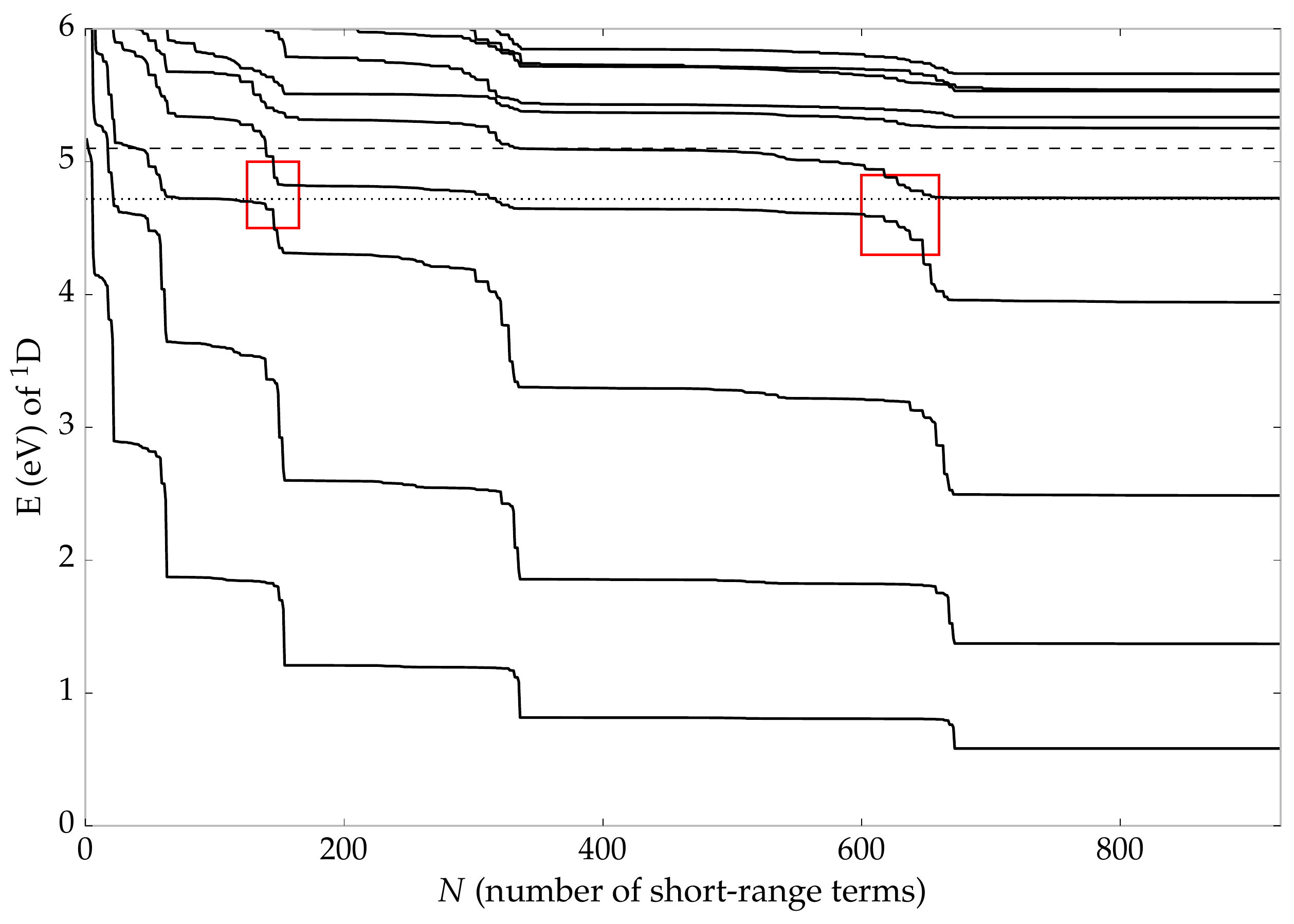}
	\caption[$^1$D eigenvalues for both symmetries]{$^1$D eigenvalues for both symmetries. The dashed line represents the Ps(n=2) threshold. The dotted line represents the complex Kohn $^1$D resonance position.}
	\label{fig:dwavesinglet-eigen-paired}
\end{figure}

\begin{figure}
	\centering
	\includegraphics[width=4.5in]{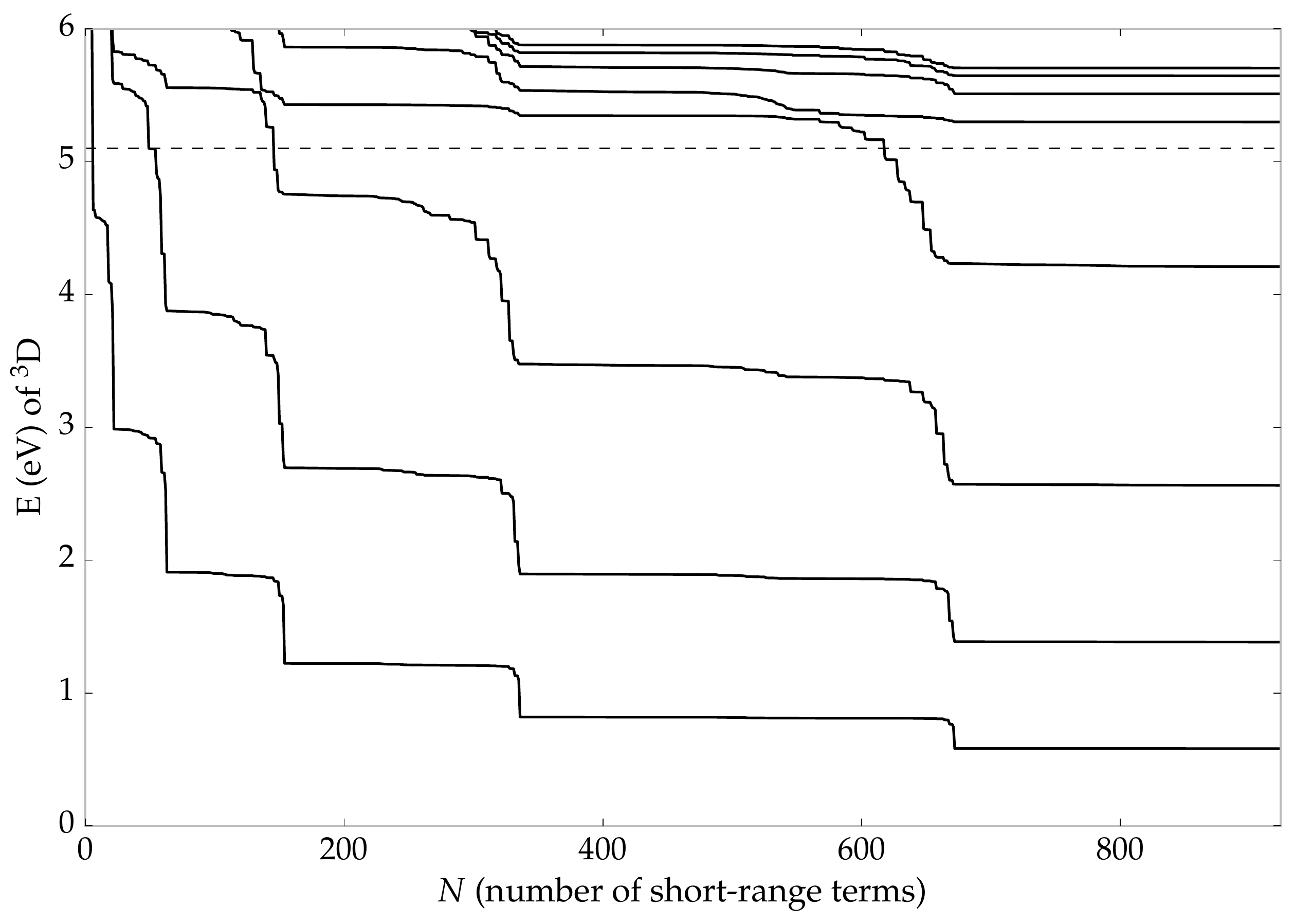}
	\caption[$^3$D eigenvalues for both symmetries]{$^3$D eigenvalues for both symmetries. The dashed line represents the Ps(n=2) threshold.}
	\label{fig:dwavetriplet-eigen-paired}
\end{figure}

To complete the discussion for resonances below the inelastic threshold, the
D-wave stabilization plots for the first symmetry only are in
\cref{fig:dwavesinglet-eigen-unpaired,fig:dwavetriplet-eigen-unpaired}. Compared to 
\cref{fig:dwavesinglet-eigen-paired,fig:dwavetriplet-eigen-paired}, the first 
symmetry obviously is not enough to adequately describe the system. There is 
only one resonance before the threshold, and the fifth eigenvalue is taken at 
924 paired terms in \cref{tab:Stabilization}.

\begin{table}
\centering
\begin{tabular}{c c c}
\toprule
Partial wave & $^1E_R$ & $^2E_R$ \\
\midrule
$^1$S & 3.99 & 5.03  \\
$^1$P & 4.28 & 4.99  \\
$^1$D & 4.73 & ---   \\
\bottomrule
\end{tabular}
\caption{Approximate resonance positions found using stabilization method}
\label{tab:Stabilization}
\end{table}

\Cref{tab:Stabilization} gives the resonance positions found using the 
stabilization method. Despite the fact that this stabilization method does 
not give very accurate resonance positions in this work, we can learn some 
things from these. First, we can determine the number of resonances before 
the Ps(n=2) threshold and where to look for them in the full complex Kohn 
calculations. Secondly, the triplet states do not have resonances before this 
threshold for any of these partial waves. Lastly, for $\ell > 0$, both the 
first and second symmetries are needed to accurately describe the system. The 
first symmetry alone cannot adequately describe the P-wave and D-wave.



\chapter{Scattering Theory}

\label{chp:WaveKohn}
\iftoggle{UNT}{As}{\lettrine{\textcolor{startcolor}{A}}{s}}
mentioned in the introduction, the 
Kohn variational method and its variants have been used for many types of 
systems. This chapter discusses the trial wavefunction used, derives the
Kohn-type variational methods, and applies them to this system.

\section{General Wavefunction}
\label{sec:GeneralWave}

The applications of the Kohn, inverse Kohn, complex Kohn for the
$S$-matrix and $T$-matrix, generalized Kohn, and generalized complex Kohn
for the $S$-matrix and $T$-matrix to the trial wavefunctions for each of the 
partial waves through the H-wave are all very similar in form. The trial 
wavefunctions for the partial waves
(\cref{eq:SWaveTrial,eq:PWaveTrial,eq:DWaveTrial}) can be written in a general
form as
\beq
\Psi_\ell^{\pm,t} = \widetilde{S}_\ell + L_\ell^{\pm,t} \, \widetilde{C}_\ell + \sum_{i=1}^{N(\omega)} c_i \bar{\phi}_i.
\label{eq:GeneralWaveTrial}
\eeq
We only consider the Ps(1s)+H(1s) system for energies up to the excitation
threshold of Ps(n=2)+H(1s), which is at an energy of $\tfrac{3}{16}$ a.u.
($5.102$ eV) \cite{Woods2015}. The coordinate system used is the same as
in \cref{fig:PsHCoords}.

To avoid confusion with $L_\ell^{\pm,t}$ here and $\mathcal{L}$ in
\cref{eq:LDef}, since the orbital angular momentum $\ell$ of the incoming Ps is
the same as the total angular momentum $L$, we use $\ell$ to indicate the partial 
wave.

The short-range $\bar{\phi}_i$ terms can represent terms of different 
symmetries, such as the $\bar{\phi}_{1i}$ and $\bar{\phi}_{2j}$ of the P-wave 
in \cref{eq:PWaveTrial}. The only requirement in this derivation is that 
these are Hylleraas-type short-range terms. In addition to letting the
$\widetilde{S}_\ell$ and $\widetilde{C}_\ell$ represent
the $\bar{S}_\ell$ and $\bar{C}_\ell$ for 
the different partial waves, we can define them in such a way as to use 
multiple Kohn methods (Kohn, inverse Kohn, etc.). We begin by defining a
matrix $\textbf{u}$ which satisfies
\beq
\label{eq:GenSCMatrix}
\begin{bmatrix}
\widetilde{S}_\ell \\
\widetilde{C}_\ell
\end{bmatrix}
=
\textbf{u}
\begin{bmatrix}
\bar{S}_\ell \\
\bar{C}_\ell
\end{bmatrix}
=
\begin{bmatrix}
u_{00} & u_{01} \\
u_{10} & u_{11}
\end{bmatrix}
\begin{bmatrix}
\bar{S}_\ell \\
\bar{C}_\ell
\end{bmatrix}.
\eeq

\noindent This notation is similar to that of Lucchese \cite{Lucchese1989}
and Cooper et al.~\cite{Cooper2010}. From this, it can easily be seen that
\begin{subequations}
\label{eq:TildeSCDef}
\begin{align}
\widetilde{S}_\ell &= u_{00} \bar{S}_\ell + u_{01} \bar{C}_\ell  \label{eq:TildeSDef} \\
\widetilde{C}_\ell &= u_{10} \bar{S}_\ell + u_{11} \bar{C}_\ell. \label{eq:TildeCDef}
\end{align}
\end{subequations}

We define
\beq
\label{eq:SCBarDef}
\bar{S}_\ell = \frac{1}{\sqrt{2}}(S_\ell \pm S_\ell^\prime) \text{ and } \bar{C}_\ell = \frac{1}{\sqrt{2}}(C_\ell \pm C_\ell^\prime),
\eeq
where
\beq
\label{eq:SCPrime}
S_\ell^\prime = P_{23}S_\ell \text{ and } C_\ell^\prime = P_{23}C_\ell.
\eeq
The general form for the long-range terms $S_\ell$ and $C_\ell$ is
\begin{subequations}
\label{eq:GenSandC}
\begin{align}
S_\ell = \,&\SphericalHarmonicY{\ell}{0}{\theta_\rho}{\varphi_\rho} \Phi_{Ps}\!\left(r_{12}\right) \Phi_H\!\left(r_3\right) \sqrt{2\kappa} \,j_\ell\!\left(\kappa\rho\right) \label{eq:GenSDef} \\
C_\ell = -&\SphericalHarmonicY{\ell}{0}{\theta_\rho}{\varphi_\rho} \Phi_{Ps}\!\left(r_{12}\right) \Phi_H\!\left(r_3\right) \sqrt{2\kappa} \,n_\ell\!\left(\kappa\rho\right) f_\ell(\rho) \label{eq:GenCDef}.
\end{align}
\end{subequations}
The $\SphericalHarmonicY{\ell}{0}{\theta_\rho}{\varphi_\rho}$ are the 
spherical harmonics, $j_\ell(\kappa\rho)$ are the spherical Bessel functions,
and $n_\ell(\kappa\rho)$ are the spherical Neumann functions. These are all
given in \cref{sec:SphericalFunc} through $\ell = 5$.
$\Phi_{\rm{Ps}}\!\left(r_{12}\right)$ and $\Phi_{\rm{H}}\!\left(r_3\right)$ are the Ps 
and H ground state wavefunctions given in \cref{eq:PsWave,eq:HWave}.

The shielding function, $f_\ell$, removes the singularity at the origin 
due to the spherical Neumann function, $n_\ell$. The form that we have chosen 
for this is
\begin{equation}
  \label{eq:PartialWaveShielding}
  f_\ell(\rho) = \left[1 - \ee^{-\mu \rho} \left(1+\frac{\mu}{2}\rho\right)
  \right]^{m_\ell}.
\end{equation}
At a minimum, $m_\ell$ is chosen so that $C_\ell$ behaves like $S_\ell$ as
$\rho \to 0$. For more discussion of this, see \Cref{sec:ShieldingFunc}. The
values used for the different partial waves are given in \cref{tab:Nonlinear}.
Prior work \cite{VanReeth2003} used a slightly simpler shielding function 
for the S-wave of
\begin{equation}
\label{eq:OldShielding}
f(\rho) = (1 - \ee^{-\lambda \rho})^3.
\end{equation}
Note that their paper is missing the negative sign in the exponential.

The Hylleraas-type short-range terms are similar to that used
in \cref{eq:BoundWavefn}, again with $k_i + l_i + m_i + n_i + p_i + q_i \leq \omega$
(\cref{eq:OmegaDef}).
These are chosen to have two symmetries, one 
with a prefactor of $r_1^\ell$ and the other with a prefactor of $r_2^\ell$.
The prefactors are included so that the correct asymptotic form of
$\Psi_\ell^{\pm,t} \sim r_k^\ell$ at the origin follows \cite[p.87]{BrownThesis}.
The first and second symmetries are given respectively by
\begin{subequations}
\label{eq:PhiDef}
\begin{align}
  \bar{\phi}_{1i} &= \left(1 \pm P_{23}\right) \SphericalHarmonicY{\ell}{0}{\theta_1}{\varphi_1}
    \ee^{-(\alpha r_1 + \beta r_2 + \gamma r_3)}
    r_1^{\ell} r_1^{k_i} r_2^{l_i} r_{12}^{m_i} r_3^{n_i} r_{13}^{p_i} r_{23}^{q_i} \label{eq:PhiDef1} \\
  \bar{\phi}_{2j} &= \left(1 \pm P_{23}\right) \SphericalHarmonicY{\ell}{0}{\theta_2}{\varphi_2}
    \ee^{-(\alpha r_1 + \beta r_2 + \gamma r_3)}
    r_2^{\ell} r_1^{k_j} r_2^{l_j} r_{12}^{m_j} r_3^{n_j} r_{13}^{p_j} r_{23}^{q_j}. \label{eq:PhiDef2}
\end{align}
\end{subequations}

From Refs.~\cite{Schwartz1961a,VanReethThesis},
the D-wave and higher can 
have additional symmetries where the angular momentum is shared between the 
Ps and H. From these references, we see that there are a possible
$\ell+1$ sets of short-range terms for each partial wave. We do not consider 
these mixed terms in this work (see \cref{sec:MixedTerms}). 

The S-wave has only a single symmetry, so $\bar{\phi}_i$ is a single set of
terms. Similar to \cref{sec:BoundWavefn}, the $\frac{1}{\sqrt{2}}$ is absorbed
into $c_{i0}$. The full S-wave trial wavefunction can be written as
\begin{equation}
  \label{eq:TrialWave}
  \Psi_0^{\pm,t} = \widetilde{S}_0 + L_0^{\pm,t} \, \widetilde{C}_0 + \sum_{i=1}^{N(\omega)} c_{i0} \bar{\phi}_{i1}.
\end{equation}
For the P-wave and higher ($\ell > 0$), 
\begin{equation}
  \label{eq:TrialWaveHigher}
  \Psi_\ell^{\pm,t} = \widetilde{S}_\ell + L^{\pm,t}_\ell \, \widetilde{C}_\ell
  + \sum_{i=1}^{N(\omega)} c_{i \ell} \bar{\phi}_{i1}
  + \!\!\!\sum_{i=N(\omega)+1}^{2N(\omega)} \!\! d_{i \ell} \bar{\phi}_{i2}.
\end{equation}

The symbols $\rho$ and $\rho'$ are defined as (refer to \cref{fig:PsHCoords,sec:RhoDef})
\begin{subequations}
\begin{align}
\bm{\rho} &= \frac{1}{2}\left(\bm{r}_1 + \bm{r}_2\right) \label{eq:RhoDef}\\
\bm{\rho}^\prime &= \frac{1}{2}\left(\bm{r}_1 + \bm{r}_3\right) \label{eq:RhopDef}.
\end{align}
\end{subequations}

\section{General Kohn Principle Derivation}
\label{sec:KohnDerivation}

Much of this derivation is similar to that in Peter Van Reeth's thesis \cite{
VanReethThesis} but is for single channel scattering and also generalized to 
a variety of Kohn-type variational methods. His thesis covers the Kohn and inverse 
Kohn methods for two channel e$^+$-He scattering. For this derivation, I will 
use \cref{eq:GeneralWaveTrial} but drop the short-range $\bar{\phi}_i^t$ 
terms. The derivation follows through the same with these terms, but it is 
clearer to ignore them here. Likewise, we only consider the direct terms 
here, unless otherwise specified. The final result of this section applies 
equally well to both the direct and exchanged terms.

The kinetic energy for Ps is
\beq
E_{\bm \kappa} = \frac{\hbar^2 \kappa^2}{2 m} = \frac{\kappa^2}{2 m} = \frac{1}{4} \kappa^2,
\label{eq:Wavenumber}
\eeq
where $\kappa$ is the momentum of the Ps atom.
Including this in the total energy with the ground-state energies of H and Ps, $E_H$ and $E_{Ps}$, gives
\beq
\label{eq:TotalEnergy}
E = E_H + E_{Ps} + E_{\bm \kappa} = -\frac{1}{2} - \frac{1}{4} + \frac{1}{4} \kappa^2 = -\frac{3}{4} + \frac{1}{4} \kappa^2.
\eeq

The functional $I_\ell$ is defined as \cite{Adhikari1998}
\begin{equation}
I[\Psi_\ell^t]\equiv \left<{\Psi_\ell^t}^\star | \mathcal{L} | \Psi_\ell^t \right> = \left(\Psi_\ell^t, \mathcal{L} \Psi_\ell^t \right) = \int \Psi_\ell^t \mathcal{L}
  \Psi_\ell^t \,d\tau,
\label{eq:IlDefPsi}
\end{equation}
where the operator $\mathcal{L}$ is given by
\beq
\label{eq:LDef}
\mathcal{L} = 2(H-E).
\eeq
Note that the exact wavefunction $\Psi_\ell$ solves the Schr\"{o}dinger equation, giving
\beq
\label{eq:Il0}
I[\Psi_\ell] = 0.
\eeq
\label{BraNote}Normally, the bra in bra-ket notation is conjugated,
but as noted by Refs.~\cite{Cooper2010,Lucchese1989,Zhang1988},
the bra is not conjugated for the Kohn-type variational methods.

The trial wavefunction is related to the exact solution by
\beq
\label{eq:PsilTrialRelation}
\Psi_\ell^t = \Psi_\ell + \delta \Psi_\ell.
\eeq
The variation of $I_\ell$ is
\begin{align}
\label{eq:IlPsiVariation1}
\nonumber \delta I_\ell &= I_\ell[\Psi_\ell^t] - I_\ell[\Psi_\ell] \\
\nonumber &= I_\ell[\Psi_\ell + \delta \Psi_\ell] - I_\ell[\Psi_\ell] \\
&= (\Psi_\ell, \mathcal{L} \Psi_\ell) + (\Psi_\ell, \mathcal{L} \,\delta\Psi_\ell) + (\delta\Psi_\ell, \mathcal{L} \Psi_\ell) + (\delta\Psi_\ell, \mathcal{L} \,\delta\Psi_\ell) - (\Psi_\ell, \mathcal{L} \Psi_\ell).
\end{align}
The first and last terms are equal to 0, by virtue of \cref{eq:Il0}.

We define $\delta I_\ell^\prime$ as
\beq
\label{eq:IlPrimeDef}
\delta I_\ell^\prime = \delta I_\ell - (\delta\Psi_\ell, \mathcal{L} \,\delta\Psi_\ell).
\eeq
Since $\mathcal{L}\Psi_\ell = 0$,
\beq
(\delta\Psi_\ell, \mathcal{L} \Psi_\ell) = -(\delta\Psi_\ell, \mathcal{L} \Psi_\ell),
\eeq
which combined with the definition of $\mathcal{L}$ from \cref{eq:LDef}, allows us to write the above equation as
\beq
\delta I_\ell^\prime = 2 \left(\Psi_\ell, (H\!-\!E) \delta\Psi_\ell \right) - 2 \left(\delta\Psi_\ell, (H\!-\!E) \Psi_\ell \right).
\label{eq:IlPsiVariation2}
\eeq

The Hamiltonian for the fundamental Coulombic system is
\begin{align}
\label{eq:Hamiltonian1}
H = -\frac{1}{2} \Laplacian_{\bm{r}_1} - \frac{1}{2} \Laplacian_{\bm{r}_2} - \frac{1}{2} \Laplacian_{\bm{r}_3} + \frac{1}{r_1} - \frac{1}{r_2} - \frac{1}{r_3} - \frac{1}{r_{12}} -\frac {1}{r_{13}} + \frac{1}{r_{23}}.
\end{align}

\noindent The Hamiltonian can also be expressed in terms of other variables in Jacobi coordinates as
\begin{align}
H = -\frac{1}{4} \Laplacian_{\bm{\rho}} - \frac{1}{2} \Laplacian_{\bm{r}_3} - \Laplacian_{\bm{r}_{12}} + \frac{1}{r_1} - \frac{1}{r_2} - \frac{1}{r_3} - \frac{1}{r_{12}} - \frac{1}{r_{13}} + \frac{1}{r_{23}}
\label{eq:Hamiltonian2}
\end{align}
and for the permuted version,
\begin{align}
H = -\frac{1}{4} \Laplacian_{\bm{\rho}^\prime} - \frac{1}{2} \Laplacian_{\bm{r}_2} - \Laplacian_{\bm{r}_{13}} + \frac{1}{r_1} - \frac{1}{r_2} - \frac{1}{r_3} - \frac{1}{r_{12}} - \frac{1}{r_{13}} + \frac{1}{r_{23}}.
\label{eq:Hamiltonian3}
\end{align}

\noindent Substituting the second form of $H$ in \cref{eq:IlPsiVariation2} and using the total energy from \cref{eq:TotalEnergy} gives
\begin{align}
\label{eq:IlPsiVariation3}
\delta I_\ell^\prime = 2 \int\limits_{V_{12}} \int\limits_{V_3} \int\limits_{V_\rho} \Psi_\ell &\left[-\frac{1}{4} \Laplacian_{\bm{\rho}} - \frac{1}{2} \Laplacian_{\bm{r}_3} - \Laplacian_{\bm{r}_{12}} + \frac{1}{r_1} - \frac{1}{r_2} - \frac{1}{r_3} \right. \nonumber \\
  &- \left. \frac{1}{r_{12}} - \frac{1}{r_{13}} + \frac{1}{r_{23}} - E_H - E_{Ps} - \frac{1}{4}\kappa^2 \right] \delta \Psi_\ell \,d\tau_\rho d\tau_{r_3} d\tau_{r_{12}} \nonumber \\
 - 2 \int\limits_{V_{12}} \int\limits_{V_3} \int\limits_{V_\rho} \delta \Psi_\ell &\left[-\frac{1}{4} \Laplacian_{\bm{\rho}} - \frac{1}{2} \Laplacian_{\bm{r}_3} - \Laplacian_{\bm{r}_{12}} + \frac{1}{r_1} - \frac{1}{r_2} - \frac{1}{r_3} \right. \nonumber \\
  &- \left. \frac{1}{r_{12}} - \frac{1}{r_{13}} + \frac{1}{r_{23}} - E_H - E_{Ps} - \frac{1}{4}\kappa^2 \right] \Psi_\ell \,d\tau_\rho d\tau_{r_3} d\tau_{r_{12}}.
\end{align}

The H and Ps equations are respectively (for large values of $\rho$)
\begin{subequations}
\label{eq:HPsEqn}
\begin{align}
\left(-\frac{1}{2} \Laplacian_{\bm{r}_3} - \frac{1}{r_3}\right) \Phi_H(r_3) &= E_H \Phi_H(r_3) \label{eq:HEqn} \\
\left(-\Laplacian_{\bm{r}_{12}} - \frac{1}{r_{12}}\right) \Phi_{Ps}(r_{12}) &= E_{Ps} \Phi_{Ps}(r_{12}). \label{eq:PsEqn}
\end{align}
\end{subequations}
Realizing then that the Hamiltonians for H and Ps are given by
\begin{subequations}
\label{eq:HPsHamil}
\begin{align}
H_H =& -\frac{1}{2} \Laplacian_{\bm{r}_3} - \frac{1}{r_3} \label{eq:HHamil} \\
H_{Ps} =& -\Laplacian_{\bm{r}_{12}} - \frac{1}{r_{12}}, \label{eq:PsHamil}
\end{align}
\end{subequations}
and rearranging terms, \cref{eq:IlPsiVariation3} becomes
\begin{align}
\label{eq:IlPsiVariation4}
\delta I_\ell^\prime = -&\frac{1}{2} \int\limits_{V_{12}} \int\limits_{V_3} \int\limits_{V_\rho} \left[\Psi_\ell \Laplacian_{\bm{\rho}} \,\delta\Psi_\ell - \delta\Psi_\ell \Laplacian_{\bm{\rho}} \Psi_\ell \right] \,d\tau_\rho d\tau_{r_3} d\tau_{r_{12}} \nonumber \\
+ &2 \int\limits_{V_{12}} \int\limits_{V_3} \int\limits_{V_\rho} \Psi_\ell \left[H_H + H_{Ps} + \frac{1}{r_1} - \frac{1}{r_2} - \frac{1}{r_{13}} + \frac{1}{r_{23}} - E_H - E_{Ps} - \frac{1}{4}\kappa^2 \right] \delta\Psi_\ell \,d\tau_\rho d\tau_{r_3} d\tau_{r_{12}} \nonumber \\
- &2 \int\limits_{V_{12}} \int\limits_{V_3} \int\limits_{V_\rho} \delta\Psi_\ell \left[H_H + H_{Ps} + \frac{1}{r_1} - \frac{1}{r_2} - \frac{1}{r_{13}} + \frac{1}{r_{23}} - E_H - E_{Ps} - \frac{1}{4}\kappa^2 \right] \Psi_\ell \,d\tau_\rho d\tau_{r_3} d\tau_{r_{12}}.
\end{align}
From Green's theorem,
\begin{align}
\label{eq:GreensThm}
\nonumber \Braket{\Psi_\ell^\star | \Laplacian_{\bm{\rho}} | \delta\Psi_\ell } - \Braket{\delta\Psi_\ell^\star | \Laplacian_{\bm{\rho}} | \Psi_\ell}
&= \int\limits_{V_3} \int\limits_{V_{12}} \int\limits_{V_\rho} \left[ \Psi_\ell \Laplacian_{\bm{\rho}} \,\delta\Psi_\ell - \delta\Psi_\ell \Laplacian_{\bm{\rho}} \Psi_\ell \right] d\tau_\rho \, d\tau_{12} d\tau_3 \\
&= \int\limits_{V_3} \int\limits_{V_{12}} \int\limits_{S_\rho} \left[ \Psi_\ell \grad_\rho \delta\Psi - \delta\Psi_\ell \grad_\rho \Psi_\ell \right] \cdot d\bm{\sigma}_\rho d\tau_{12} d\tau_3.
\end{align}
$S_\rho$ is the surface at $\rho \rightarrow \infty$.
Due to the exponential form of $\Phi_{Ps}(r_{12})$ and $\Phi_H(r_3)$ given in
\cref{eq:PsWave,eq:HWave}, the last two lines of \cref{eq:IlPsiVariation4} cancel each other.

Since we are only considering the direct terms in \cref{eq:GeneralWaveTrial} so
far, let us define a direct term only version of \cref{eq:TildeSCDef} with
\begin{subequations}
\label{eq:TildeSCDefDir}
\begin{align}
\widetilde{S}_d &= u_{00} S_\ell + u_{01} C_\ell \\
\widetilde{C}_d &= u_{10} S_\ell + u_{11} C_\ell.
\end{align}
\end{subequations}

From \cref{eq:PsilTrialRelation} and \cref{eq:GeneralWaveTrial},
\beq
\label{eq:DeltaPsi}
\delta \Psi = \Psi_\ell^t - \Psi_\ell = (\widetilde{S}_d + L_\ell^t \, \widetilde{C}_d) - (\widetilde{S}_d + L_\ell \, \widetilde{C}_d) = (L_\ell^t - L_\ell) \widetilde{C}_d \,.
\eeq
Substituting this into \cref{eq:IlPsiVariation4} with \cref{eq:GreensThm},
\beq
\label{eq:IlPsiVariation5}
\delta I_\ell^\prime = -\frac{1}{2} (L_\ell^t - L_\ell) \int\limits_{V_{12}} \int\limits_{V_3} \int\limits_{S_\rho} \left[(\widetilde{S}_d + L_\ell \, \widetilde{C}_d) \grad_\rho \widetilde{C}_d - \widetilde{C}_d \grad_\rho (\widetilde{S}_d + L_\ell \, \widetilde{C}_d) \right] \cdot d\bm{\sigma}_\rho d\tau_{12} d\tau_3 \;.
\eeq

From \cref{eq:SphBesDerRel,eq:GenSandC}, to first order, the gradient acting
on $\widetilde{S}_d$ and $\widetilde{C}_d$ in \cref{eq:TildeSCDef} gives
\begin{subequations}
\label{eq:GradSC}
\begin{align}
\grad_\rho \widetilde{S}_d &\sim \kappa \left[u_{00} C_\ell - u_{01} S_\ell \right] \hat{\bm\rho} \\
\grad_\rho \widetilde{C}_d &\sim \kappa \left[u_{10} C_\ell - u_{11} S_\ell \right] \hat{\bm\rho} \,.
\end{align}
\end{subequations}
Substituting this into \cref{eq:IlPsiVariation5} and dropping the dot product,
since the surface elements are in the same direction as $\hat{\bm\rho}$,
this becomes
\begin{align}
\label{eq:IlPsiVariation6a}
\delta I_\ell^\prime \sim -\frac{1}{2} (L_\ell^t - L_\ell) & \int\limits_{V_{12}} \int\limits_{V_3} \int\limits_{S_\rho}  \left\{(\widetilde{S}_d + L_\ell \, \widetilde{C}_d) \kappa (u_{10} C_\ell - u_{11} S_\ell) \right. \nonumber \\
 &- \left. \widetilde{C}_d \kappa \left[(u_{00} C_\ell - u_{01} S_\ell) + L_\ell (u_{10} C_\ell - u_{11} S_\ell) \right] \right\} d\sigma_\rho d\tau_{12} d\tau_3.
\end{align}
Omitting terms quadratic in $L_\ell$ or $L_\ell^t$, including $L_\ell^t L_\ell$,
\begin{align}
\label{eq:IlPsiVariation6b}
\delta I_\ell^\prime \sim -\frac{1}{2} \kappa (L_\ell^t - L_\ell) & \int\limits_{V_{12}} \int\limits_{V_3} \int\limits_{S_\rho} \left[\widetilde{S}_d (u_{10} C_\ell - u_{11} S_\ell) - \widetilde{C}_d (u_{00} C_\ell - u_{01} S_\ell) \right] d\sigma_\rho d\tau_{12} d\tau_3 \nonumber \\
= -\frac{1}{2} \kappa (L_\ell^t - L_\ell) & \int\limits_{V_{12}} \int\limits_{V_3} \int\limits_{S_\rho} \left[\widetilde{S}_d u_{10} C_\ell - \widetilde{S}_d u_{11} S_\ell - \widetilde{C}_d u_{00} C_\ell + \widetilde{C}_d u_{01} S_\ell \right] d\sigma_\rho d\tau_{12} d\tau_3 \nonumber \\
= -\frac{1}{2} \kappa (L_\ell^t - L_\ell) & \int\limits_{V_{12}} \int\limits_{V_3} \int\limits_{S_\rho} \left[u_{00} u_{10} S_\ell C_\ell + u_{01} u_{10} C_\ell^2 - u_{00} u_{11} S_\ell^2 - u_{01} u_{11} C_\ell S_\ell\right. \nonumber \\
& \left. - u_{10} u_{00} S_\ell C_\ell - u_{11} u_{00} C_\ell^2 + u_{10} u_{01} S_\ell^2 + u_{11} u_{01} C_\ell S_\ell \right] d\sigma_\rho d\tau_{12} d\tau_3 \nonumber \\
= -\frac{1}{2} \kappa (L_\ell^t - L_\ell) & \int\limits_{V_{12}} \int\limits_{V_3} \int\limits_{S_\rho} (u_{10} u_{01} - u_{00} u_{11}) \left( S_\ell^2 + C_\ell^2 \right) d\sigma_\rho d\tau_{12} d\tau_3 \nonumber \\
= \frac{1}{2} \kappa (L_\ell^t - L_\ell) & \Det{\textbf{u}} \int\limits_{V_{12}} \int\limits_{V_3} \int\limits_{S_\rho} \left( S_\ell^2 + C_\ell^2 \right) d\sigma_\rho d\tau_{12} d\tau_3.
\end{align}

The rest of this derivation considers only the direct terms. The final result
applies as well when the exchanged terms are included. Since we are considering
the surface as $\rho \rightarrow \infty$, $f_\ell(\rho)$ in \cref{eq:GenCDef}
becomes 1. Then from \cref{eq:GenSandC},
\beq
S_\ell^2 + C_\ell^2 = \SphericalHarmonicY{\ell}{0}{\theta_\rho}{\varphi_\rho}^2 \Phi_{Ps}\left(r_{12}\right)^2 \Phi_H\left(r_3\right)^2 (2 \kappa) \left[j_\ell\!\left(\kappa\rho\right)^2 + n_\ell\!\left(\kappa\rho\right)^2\right].
\eeq
The asymptotic forms of $j_\ell$ and $n_\ell$ as $\rho \to \infty$ are given by \cite[p.729]{Arfken2005}
\begin{subequations}
\label{eq:AsymSphBes}
\begin{align}
j_\ell\!\left(\kappa\rho\right) &\sim \frac{1}{\kappa\rho} \sin\left(\kappa\rho - \frac{n \pi}{2}\right) \label{eq:AsymJl} \\
n_\ell\!\left(\kappa\rho\right) &\sim \frac{1}{\kappa\rho} \cos\left(\kappa\rho - \frac{n \pi}{2}\right). \label{eq:AsymNl}
\end{align}
\end{subequations}
As $\rho \to \infty$,
\beq
S_\ell^2 + C_\ell^2 \sim \SphericalHarmonicY{\ell}{0}{\theta_\rho}{\varphi_\rho}^2 \Phi_{Ps}\left(r_{12}\right)^2 \Phi_H\left(r_3\right)^2 (2 \kappa) \frac{1}{\kappa^2 \rho^2}.
\eeq

Substituting this in \cref{eq:IlPsiVariation6b} and expanding the
$d\sigma_\rho$ volume element,
\begin{align}
\label{eq:IlPsiVariation7}
\delta I_\ell^\prime \sim& \; \kappa (L_\ell^t - L_\ell) \Det{\textbf{u}} \int\limits_{V_{12}} \int\limits_{V_3} \int\limits_{S_\rho} \SphericalHarmonicY{\ell}{0}{\theta_\rho}{\varphi_\rho}^2 \Phi_{Ps}\left(r_{12}\right)^2 \Phi_H\left(r_3\right)^2 \frac{1}{\kappa \rho^2} \rho^2 \sin\theta_\rho d\theta_\rho d\varphi_\rho d\tau_{12} d\tau_3 \nonumber \\
=& (L_\ell^t - L_\ell) \Det{\textbf{u}} \int\limits_{V_{12}} \int\limits_{V_3} \int\limits_{S_\rho} \SphericalHarmonicY{\ell}{0}{\theta_\rho}{\varphi_\rho}^2 \Phi_{Ps}\left(r_{12}\right)^2 \Phi_H\left(r_3\right)^2 \sin\theta_\rho d\theta_\rho d\varphi_\rho d\tau_{12} d\tau_3.
\end{align}
Since the Ps and H eigenfunctions are normalized, i.e.
\beq
\int\limits_{V_3}\! \left| \Phi_H(r_3) \right|^2 d\tau_3 = 1 \text{ and } \int\limits_{V_{12}}\! \left|\Phi_{Ps}(r_{12})\right|^2 d\tau_{12} = 1,
\label{eq:PsHNormalization}
\eeq
we now have
\beq
\label{eq:IlPsiVariation8}
\delta I_\ell^\prime = (L_\ell^t - L_\ell) \Det{\textbf{u}} \int\limits_{S_\rho} \SphericalHarmonicY{\ell}{0}{\theta_\rho}{\varphi_\rho}^2 \sin\theta_\rho d\theta_\rho d\varphi_\rho.
\eeq
The spherical harmonics are normalized so that \cite[p.788]{Arfken2005}
\beq
\label{eq:SphHarmNorm}
\int\limits_{S_\rho} \left| \SphericalHarmonicY{\ell}{0}{\theta_\rho}{\varphi_\rho} \right|^2 d\Omega = 1.
\eeq
This gives that
\beq
\label{eq:IlPsiVariation9}
\delta I_\ell^\prime = (L_\ell^t - L_\ell) \Det{\textbf{u}}.
\eeq

From \cref{eq:IlPrimeDef,eq:IlPsiVariation9},
\beq
\label{eq:KatoIdent}
\delta I_\ell^\prime = (L_\ell^t - L_\ell) \Det{\textbf{u}} + (\delta\Psi_\ell, \mathcal{L} \,\delta\Psi_\ell).
\eeq
This is the Kato identity \cite{Kato1951a}. For the Kohn-type variational methods,
the last term is neglected, since it is second order in $\delta\Psi_\ell$. Using the
approximation $\delta I_\ell^\prime \approx \delta I_\ell$, we have
\beq
\delta I_\ell = I_\ell[\Psi_\ell^t] - I_\ell[\Psi_\ell] \approx (L_\ell^t - L_\ell) \Det{\textbf{u}}.
\eeq
Replacing the exact $L_\ell$ by the variational $L_\ell^v$ and rearranging,
we finally get the general Kohn variational method of
\beq
\label{eq:GenKohn}
L_\ell^v = L_\ell^t - I_\ell[\Psi_\ell^t] / \! \Det{\textbf{u}},
\eeq
which is correct to second-order.
This was only derived using the direct terms, but the exchange terms follow 
the same steps with $\rho^\prime$ instead of $\rho$. This was also only shown 
for the long-range terms, but it applies equally as well to the full
wavefunction with the short-range terms.

\section{Application of the Kohn-Type Variational Methods}
\label{sec:KohnApplied}

We use the general Kohn variational method (\cref{eq:GenKohn}) with the full
trial wavefunction to get
\beq
\label{eq:GenKohnApplied}
L_\ell^v = L_\ell^t - \tfrac{1}{\Det{\textbf{u}}} \Big((\widetilde{S}_\ell + L_\ell^t \, \widetilde{C}_\ell + \sum_i c_i \bar{\phi}_i^t), \mathcal{L} (\widetilde{S}_\ell + L_\ell^t \, \widetilde{C}_\ell + \sum_j c_j \bar{\phi}_j^t )\Big).
\eeq
The property of the Kohn functional that it is stationary with respect to
variations in the linear parameters \cite{Joachain1979} can be written in
this case as
\beq
\frac{\partial L_\ell^v}{\partial L_\ell^t} = 0  \text{ and } \frac{\partial L_\ell^v}{\partial c_i} = 0 \text{, where $i = 1,\ldots,N$}.
\label{eq:KohnStationary}
\eeq

Performing the first variation gives
\begin{align}
\nonumber 0 &= \pderiv{L_\ell^v}{L_\ell^t} \\
&= \Det{\textbf{u}} - \! \left[(\widetilde{S}_\ell,\mathcal{L}\widetilde{C}_\ell) + (\widetilde{C}_\ell,\mathcal{L}\widetilde{S}_\ell) + \frac{\partial}{\partial L_\ell^t}(L_\ell^t \widetilde{C}_\ell,\mathcal{L} L_\ell^t \widetilde{C}_\ell) + (\widetilde{C}_\ell, \mathcal{L} \sum_i c_i \bar{\phi}_i) + (\sum_i c_i \bar{\phi}_i, \mathcal{L} \widetilde{C}_\ell) \right]\!.
\label{eq:PdLambda1}
\end{align}

\noindent The third term in brackets becomes
\beq
\pderiv{}{L_\ell^t} (L_\ell^t \widetilde{C}_\ell,\mathcal{L} L_\ell^t \widetilde{C}_\ell) = (\widetilde{C}_\ell,\mathcal{L} \widetilde{C}_\ell) \frac{\partial}{\partial L_\ell^t} {L_\ell^t}^2 = 2(\widetilde{C}_\ell,\mathcal{L}\widetilde{C}_\ell) L_\ell^t.
\eeq

\noindent The last two terms of \cref{eq:PdLambda1} are equal to each other, and we can use \cref{eq:GenSLCandCLS} to rewrite this as
\beq
0 = -2 (\widetilde{C}_\ell,\mathcal{L}\widetilde{S}_\ell) - 2 L_\ell^t (\widetilde{C}_\ell,\mathcal{L}\widetilde{C}_\ell) - 2 \sum_i c_i (\widetilde{C}_\ell,\mathcal{L}\bar{\phi}_i).
\eeq

\noindent Rearranging gives
\beq
\label{eq:PdLambda}
-(\widetilde{C}_\ell,\mathcal{L}\widetilde{S}_\ell) = L_\ell^t (\widetilde{C}_\ell,\mathcal{L}\widetilde{C}_\ell) + \sum_i c_i (\widetilde{C}_\ell,\mathcal{L}\bar{\phi}_i).
\eeq

Now we perform the variation with respect to a general $c_k$ as in \cref{eq:KohnStationary}.
\beq
0 = \frac{\partial \mathcal{L}_v}{\partial c_k} = -\!\!\left[ (\widetilde{S}_\ell,\mathcal{L} \bar{\phi}_k) + L_\ell^t (\widetilde{C}_\ell,\mathcal{L} \bar{\phi}_k) + (\bar{\phi}_k,\mathcal{L} \widetilde{S}_\ell) + L_\ell^t (\bar{\phi}_k,\mathcal{L} \widetilde{C}_\ell) + \frac{\partial}{\partial c_k} (\sum_i c_i \bar{\phi}_i, \mathcal{L} \sum_j c_j \bar{\phi}_j) \right]
\label{eq:PdCk1}
\eeq
Rearranging gives
\beq
-\left( \bar{\phi}_k, \mathcal{L} \widetilde{S}_\ell \right) = L_\ell^t \left( \bar{\phi}_k, \mathcal{L} \widetilde{C}_\ell \right) + \sum_i \left( \bar{\phi}_k, \mathcal{L} c_i \bar{\phi}_i \right).
\label{eq:PdCk}
\eeq

The set of linear equations in \cref{eq:PdLambda,eq:PdCk} can be written in matrix form as
\begin{equation}
\label{eq:GeneralKohnMatrix}
\begin{bmatrix} 
 (\widetilde{C}_\ell,\mathcal{L}\widetilde{C}_\ell) & (\widetilde{C}_\ell,\mathcal{L}\bar{\phi}_1) & \cdots & (\widetilde{C}_\ell,\mathcal{L}\bar{\phi}_j) & \cdots\\
 (\bar{\phi}_1,\mathcal{L}\widetilde{C}_\ell) & (\bar{\phi}_1,\mathcal{L}\bar{\phi}_1) & \cdots & (\bar{\phi}_1,\mathcal{L}\bar{\phi}_j) & \cdots\\
 \vdots & \vdots & \ddots & \vdots \\
 (\bar{\phi}_i,\mathcal{L}\widetilde{C}_\ell) & (\bar{\phi}_i,\mathcal{L}\bar{\phi}_1) & \cdots & (\bar{\phi}_i,\mathcal{L}\bar{\phi}_j) & \cdots\\
 \vdots & \vdots & & \vdots & \\
\end{bmatrix}
\begin{bmatrix}
L_\ell^t\\
c_1\\
\vdots\\
c_i\\
\vdots
\end{bmatrix}
= -
\begin{bmatrix}
(\widetilde{C}_\ell,\mathcal{L}\widetilde{S}_\ell) \\
(\bar{\phi}_1,\mathcal{L}\widetilde{S}_\ell) \\
\vdots \\
(\bar{\phi}_i,\mathcal{L}\widetilde{S}_\ell) \\
\vdots
\end{bmatrix}.
\end{equation}

\noindent This matrix equation can be rewritten as
\beq
\label{eq:GenKohnMatrixAXB}
\textbf{\emph{AX}} = -\textbf{\emph{B}}.
\eeq

\noindent Solving this for $\textbf{\emph{X}}$ gives
\beq
\textbf{\emph{X}} = -\textbf{\emph{A}}^{-1}\textbf{\emph{B}}.
\eeq

To obtain $L_\ell^v$ from this matrix equation, we must next expand \cref{eq:GenKohnApplied}.

\begin{align}
\nonumber L_\ell^v = L_\ell^t - & \tfrac{1}{\Det{\textbf{u}}} \left[ (\widetilde{S}_\ell,\mathcal{L}\widetilde{S}_\ell) + L_\ell^t (\widetilde{S}_\ell,\mathcal{L}\widetilde{C}_\ell) + \sum_i c_i (\widetilde{S}_\ell,\mathcal{L} \bar{\phi}_i) + L_\ell^t (\widetilde{C}_\ell,\mathcal{L}\widetilde{S}_\ell) + {L_\ell^t}^2 (\widetilde{C}_\ell,\mathcal{L}\widetilde{C}_\ell)  \right. \\
& + \left. L_\ell^t \sum_i c_i (\widetilde{C}_\ell,\mathcal{L} \bar{\phi}_i) + \sum_i c_i (\bar{\phi}_i, \mathcal{L} \widetilde{S}_\ell) + L_\ell^t \sum_i c_i (\bar{\phi}_i, \mathcal{L} \widetilde{C}_\ell) + \sum_i \sum_j c_i c_j (\bar{\phi}_i, \mathcal{L} \bar{\phi}_j) \right]
\end{align}

\noindent By substituting \cref{eq:GenSLCandCLS} in for $(\widetilde{S}_\ell,\mathcal{L}\widetilde{C}_\ell)$, the first $L_\ell^t$ above is canceled, leaving

\begin{align}
\label{eq:GenKohnApplied2}
L_\ell^v = - & \tfrac{1}{\Det{\textbf{u}}} \left[ (\widetilde{S}_\ell,\mathcal{L}\widetilde{S}_\ell) + L_\ell^t (\widetilde{C}_\ell,\mathcal{L}\widetilde{S}_\ell) + \sum_i c_i (\widetilde{S}_\ell, \mathcal{L} \bar{\phi}_i) + L_\ell^t (\widetilde{C}_\ell,\mathcal{L}\widetilde{S}_\ell) + {L_\ell^t}^2 (\widetilde{C}_\ell,\mathcal{L}\widetilde{C}_\ell) \right.  \nonumber \\
& + \left. L_\ell^t \sum_i c_i (\widetilde{C}_\ell,\mathcal{L} \bar{\phi}_i)
+ \sum_i c_i (\bar{\phi}_i, \mathcal{L} \widetilde{S}_\ell) + L_\ell^t \sum_i c_i (\bar{\phi}_i, \mathcal{L} \widetilde{C}_\ell) + \sum_i \sum_j c_i c_j (\bar{\phi}_i, \mathcal{L} \bar{\phi}_j) \right].
\end{align}

Using the following definitions of
\beq
D = 
\begin{bmatrix}
L_\ell^t & c_1 & \cdots & c_N & 1
\end{bmatrix}
\text{ and}
\eeq
\beq
\label{eq:GenFandD}
F =
\begin{bmatrix}
(\boldsymbol{\widetilde{C}_\ell,\mathcal{L}\widetilde{C}_\ell}) & (\boldsymbol{\widetilde{C}_\ell,\mathcal{L}\bar{\phi}}) & (\boldsymbol{\widetilde{C}_\ell,\mathcal{L}\widetilde{S}_\ell}) \\
(\boldsymbol{\bar{\phi},\mathcal{L}\widetilde{C}_\ell}) & (\boldsymbol{\bar{\phi},\mathcal{L}\bar{\phi}}) & (\boldsymbol{\bar{\phi},\mathcal{L}\widetilde{S}_\ell}) \\
(\boldsymbol{\widetilde{C}_\ell,\mathcal{L}\widetilde{S}_\ell}) & (\boldsymbol{\widetilde{S}_\ell,\mathcal{L}\bar{\phi}}) & (\boldsymbol{\widetilde{S}_\ell,\mathcal{L}\widetilde{S}_\ell})
\end{bmatrix},
\eeq
\cref{eq:GenKohnApplied2} can be rewritten as the following matrix equation:
\beq
\label{eq:GenDFDT}
L_\ell^v = - \tfrac{1}{\Det{\textbf{u}}} D F D^T.
\eeq

\noindent Using \cref{eq:GenKohnMatrixAXB} in \cref{eq:GenDFDT} and expanding gives
\begin{align}
\label{eq:GenDFDT2}
\nonumber \Det{\textbf{u}} \: L_\ell^v &= - 
\begin{bmatrix}
\boldsymbol{X^T} & 1 
\end{bmatrix}
\begin{bmatrix}
\boldsymbol{A} & \boldsymbol{B} \\
\boldsymbol{B^T} & \boldsymbol{(\widetilde{S}_\ell,\mathcal{L}\widetilde{S}_\ell)}
\end{bmatrix}
\begin{bmatrix}
\boldsymbol{X} \\
1
\end{bmatrix}
= -
\begin{bmatrix}
\boldsymbol{X^T} & 1 
\end{bmatrix}
\begin{bmatrix}
0 \\
\boldsymbol{B^T X} + (\widetilde{S}_\ell,\mathcal{L}\widetilde{S}_\ell)
\end{bmatrix} \\
&= -\boldsymbol{B^T X} - (\widetilde{S}_\ell,\mathcal{L}\widetilde{S}_\ell),
\end{align}
where
\beq
\boldsymbol{B^T X} = L_\ell^t (\widetilde{C}_\ell,\mathcal{L}\widetilde{S}_\ell) + \sum_i c_i (\bar{\phi}_i, \mathcal{L} \widetilde{S}_\ell).
\eeq

\noindent A more compact way of writing \cref{eq:GenDFDT2} is by
\beq
L_\ell^v = -\tfrac{1}{\Det{\textbf{u}}} \left( \Psi^{t,0},\mathcal{L} \widetilde{S}_\ell \right),
\eeq
where $\Psi^{t,0}$ is the full general wavefunction in \cref{eq:GeneralWaveTrial} with its nonlinear parameters optimized.
Finally, to obtain the phase shifts, we use the relation given by Ref.~\cite{Lucchese1989} as
\begin{equation}
\label{eq:GenKohnL}
K_\ell = \tan \delta_\ell = (u_{01} + u_{11} L_\ell)(u_{00} + u_{10} L_\ell)^{-1}.
\end{equation}

The $\textbf{u}$ and $L^{\pm,t}_\ell$ for the various Kohn methods are 
described now. Note that for each of these, $\Det{\textbf{u}} = 1$, except
for the ones describing the $S$-matrix complex Kohn and generalized
$S$-matrix complex Kohn. When we 
create the matrix in \cref{eq:GeneralKohnMatrix}, we only calculate the 
matrix elements for the Kohn, along with $(\bar{S}_\ell,\mathcal{L}\bar{S}_\ell)$
and $(\bar{S}_\ell,\mathcal{L}\bar{C}_\ell)$. Then from the definitions in
\cref{eq:TildeSCDef}, this matrix can be changed to any of the other Kohn 
methods without recomputing any of the integrals. 

\subsubsection*{Kohn}
\label{sec:Kohn}
\beq
\textbf{u} =
\begin{bmatrix}
1 & 0 \\
0 & 1 
\end{bmatrix}
\label{eq:uKohn}
\eeq

\beq
L^{\pm,t}_\ell = \lambda_t = K_t
\label{eq:LKohn}
\eeq

\subsubsection*{Inverse Kohn}
\label{sec:InvKohn}
\beq
\textbf{u} =
\begin{bmatrix}
0 & 1 \\
-1 & 0 
\end{bmatrix}
\label{eq:uInvKohn}
\eeq

\beq
L^{\pm,t}_\ell = -\mu_t = -K^{-1}_t = -\bar{K}_t
\label{eq:LInvKohn}
\eeq

\subsubsection*{Generalized Kohn}
\label{sec:GenKohn}
\beq
\textbf{u} =
\begin{bmatrix}
\cos\tau & \sin\tau \\
-\sin\tau & \cos\tau 
\end{bmatrix}
\label{eq:uGenKohn}
\eeq
The generalized Kohn method is described by Cooper et al.\ \cite{
Cooper2009, Cooper2010}.  When $\tau = 0$ is substituted in \cref{eq:uGenKohn}
, the $\textbf{u}$-matrix for the Kohn method is generated (\cref{eq:uKohn}). 
Similarly, when $\tau = \frac{\pi}{2}$, the $\textbf{u}$-matrix for the 
inverse Kohn method is generated (\cref{eq:uInvKohn}).

\subsubsection*{$T$-matrix Complex Kohn}
\label{sec:ComplexTKohn}
\beq
\textbf{u} =
\begin{bmatrix}
1 & 0 \\
\ii & 1
\end{bmatrix}
\label{eq:uCompTKohn}
\eeq

\beq
L^{\pm,t}_\ell = T_\ell
\label{eq:LCompTKohn}
\eeq
Lucchese \cite{Lucchese1989} denotes this as $\mathcal{L}_\ell = -\pi T$,
but we use the definition of the $T$-matrix from Bransden \cite{Bransden1970}:
\begin{equation}
\label{eq:TMatrix}
K_\ell = \frac{T_\ell}{1 + \ii T_\ell}
\end{equation}

\subsubsection*{$S$-matrix Complex Kohn}
\label{sec:ComplexSKohn}
\beq
\textbf{u} =
\begin{bmatrix}
-\ii & 1 \\
\ii & 1
\end{bmatrix}
\label{eq:uCompSKohn}
\eeq

\beq
L^{\pm,t}_\ell = -S_\ell
\label{eq:LCompSKohn}
\eeq

The Lucchese \cite{Lucchese1989} version of $\textbf{u}$ differs from this, 
since he uses a different definition for the $S$-matrix. The form of the
$S$-matrix we are using is related to the $K$-matrix by
\cite{Bransden2003}
\begin{equation}
\label{eq:SMatrix}
K_\ell = \frac{\ii(1-S_\ell)}{1+S_\ell},
\end{equation}
which is satisfied by the above $\textbf{u}$-matrix. Also of note is that
$\det \textbf{u} = -2\ii$ instead of 1 like most of the other Kohn methods
presented here. Cooper et al.~\cite{Cooper2010} use the $T$-matrix but also
provide a relation between the two.

\subsubsection*{Generalized $T$-matrix Complex Kohn}
\label{sec:GenComplexTKohn}
\beq
\textbf{u} =
\begin{bmatrix}
\cos\tau & \sin\tau \\
-\sin\tau + \ii \cos\tau & \cos\tau + \ii \sin\tau
\end{bmatrix}
\label{eq:uGenTKohn}
\eeq
This is a generalized form of the $T$-matrix complex Kohn, similar to how the 
generalized Kohn works. When $\tau = 0$, this reduces to the $T$-matrix complex
Kohn. This is also a slightly different form than that of Cooper
et al.~\cite{Cooper2010}, who have the real and imaginary parts of $\widetilde{C}_\ell$
swapped.

\subsubsection*{Generalized $S$-matrix Complex Kohn}
\label{sec:GenComplexSKohn}
\beq
\textbf{u} =
\begin{bmatrix}
-\ii \cos\tau - \sin\tau & -\ii \sin\tau + \cos\tau \\
\ii \cos\tau - \sin\tau & \ii \sin\tau + \cos\tau
\end{bmatrix}
\label{eq:uGenSKohn}
\eeq
This is a generalized form of the $S$-matrix complex Kohn. When $\tau = 0$, 
this reduces to the $S$-matrix complex Kohn.

\section{Matrix Elements}
In this section, we examine the matrix elements of \cref{eq:GeneralKohnMatrix}.
The three types of matrix elements are short-range--short-range (short-short),
short-range--long-range (short-long) and long-range--long-range (long-long).
The short-long and long-long matrix elements have a similar analysis. For 
these, the effect of the $\mathcal{L} = 2(H-E)$ operator on the long-range 
terms must be considered, and then integrations over the external angles (see 
\cref{chp:AngularInt}) are performed. The remaining 6-dimensional integral is 
then numerically integrated as described in \cref{sec:LongLongInt,sec:ShortLongInt}.

For all of these matrix elements, looking at the barred terms, we have 4
integrations to perform. Using a property of the $P_{23}$ permutation operator,
for a general $f$ and $g$,
\beq
\label{eq:PermProp}
(f,\mathcal{L}g) = (f^\prime,\mathcal{L}g^\prime) \text{ and } (f,\mathcal{L}g^\prime) = (f^\prime,\mathcal{L}g).
\eeq
The functions $f$ and $g$ are any of $\widetilde{S}_\ell$, $\widetilde{C}_\ell$
and $\bar{\phi}_i$. This relation allows us to reduce the number of integrations
needed by half by doing \cite{VanReethThesis}
\beq
\label{eq:PermPropFull}
(\bar{f},\mathcal{L}\bar{g}) = (f,\mathcal{L}g) \pm (f,\mathcal{L}g^\prime) \pm (f^\prime,\mathcal{L}g) + (f^\prime,\mathcal{L}g^\prime) = 2\left[(f,\mathcal{L}g) \pm (f,\mathcal{L}g^\prime)\right].
\eeq

\subsection{Matrix Element Symmetries}
\label{sec:Symmetries}

Not all matrix elements in \cref{eq:GeneralKohnMatrix} have to be calculated 
as presented. Some matrix elements are identical to other matrix elements, 
such as $(\widetilde{C}_\ell,\mathcal{L}\bar{\phi}_i) = (\bar{\phi}_i,\mathcal{L}\widetilde{C}_\ell)$.
In this particular case, it is much easier to 
calculate $(\bar{\phi}_i,\mathcal{L}\widetilde{C}_\ell)$ instead of
$(\widetilde{C}_\ell,\mathcal{L}\bar{\phi}_i)$, due to the complexity of 
operating $\mathcal{L}$ on $\bar{\phi}_i$ (see \cref{eq:GradGradShort}).
We prove these claims in this section.

These arguments follow that of Appendix A of Van Reeth's thesis \cite{VanReethThesis}.
We start with the functional
\begin{align}
	\label{eq:Fsymm}
	F \equiv \left(g,\mathcal{L}f \right)-\left(f,\mathcal{L}g \right) \, ,
\end{align}
with $\mathcal{L}$ given by \cref{eq:LDef}.
Using the Hamiltonian given by \cref{eq:Hamiltonian1}, only the first
three terms of the above functional have to be evaluated, as the other terms go to 0 with the subtraction.
\begin{align}
	F=&\left({-g,\frac{1}{2} \Laplacian_{\bm{r}_1} f}\right)+\left({f,\frac{1}{2} \Laplacian_{\bm{r}_1} g}\right)+
	\left({-g,\frac{1}{2} \Laplacian_{\bm{r}_2} f}\right)+\left({f,\frac{1}{2} \Laplacian_{\bm{r}_2} g}\right) \nonumber \\
	&+\left({-g,\frac{1}{2} \Laplacian_{\bm{r}_3} f}\right)+\left({f,\frac{1}{2} \Laplacian_{\bm{r}_3} g}\right)  \nonumber \\
=& \int\limits_{V_3} \int\limits_{V_2} \int\limits_{V_1} \left[
	-g  \Laplacian_{\bm{r}_1} f + f \Laplacian_{\bm{r}_1} g
	-g \Laplacian_{\bm{r}_2} f + f \Laplacian_{\bm{r}_2} g - g \Laplacian_{\bm{r}_3} f + f \Laplacian_{\bm{r}_3} g \right] \,d\tau_1 d\tau_2 d\tau_3
\end{align}
Using Green's theorem on each pair of terms,
\begin{align}
F = \int\limits_{V_3} \int\limits_{V_2} \int\limits_{S_1} & \left[-g \grad_{r_1} f + f \grad_{r_1} g \right] \cdot d\bm{\sigma}_1 d\tau_2 d\tau_3 + \int\limits_{V_3} \int\limits_{V_1} \int\limits_{S_2} \left[ -g \grad_{r_2} f + f \grad_{r_2} g \right] \cdot d\bm{\sigma}_2 d\tau_1 d\tau_3 \nonumber \\
	& + \int\limits_{V_1} \int\limits_{V_2} \int\limits_{S_3} \left[ -g \grad_{r_3} f + f \grad_{r_3} g \right] \cdot d\bm{\sigma}_3 d\tau_2 d\tau_1 \;.
\end{align}
For the first pair of terms, the differential surface element contains $r_1^2$.
The surface we are integrating over is at $r_1 \rightarrow \infty$, so if 
the integrand falls off faster than $r_1^{-2}$, the integral vanishes, the 
same as the argument in \cref{eq:IlPsiVariation4}. The same argument applies 
for $r_2^2$ and $r_3^2$ in the second and third pairs of terms, respectively. 
The short-range Hylleraas-type terms in \cref{eq:PhiDef} fulfill this 
requirement for $r_1$, $r_2$, and $r_3$, so if the matrix elements contain
$\bar{\phi}_i$, the right hand side is equal to 0. From \cref{eq:Fsymm}, this 
gives that matrix elements with short-range terms are symmetric, or that
\begin{subequations}
\label{eq:ShortElemSymm}
\begin{align}
\left(\bar{\phi}_i, \mathcal{L} \bar{\phi}_j \right) &= \left(\bar{\phi}_j, \mathcal{L} \bar{\phi}_i \right) \\
\left(\bar{\phi}_i, \mathcal{L} \widetilde{S}_\ell \right) &= \left(\widetilde{S}_\ell, \mathcal{L} \bar{\phi}_i \right) \label{eq:SLPhi} \\
\left(\bar{\phi}_i, \mathcal{L} \widetilde{C}_\ell \right) &= \left(\widetilde{C}_\ell, \mathcal{L} \bar{\phi}_i \right) \label{eq:CLPhi}.
\end{align}
\end{subequations}
From \cref{eq:LDef,eq:BoundHFull,eq:GradGradShort}, the form of $\mathcal{L}\phi_i$ is very complicated, but \cref{eq:SLPhi,eq:CLPhi} allow us to avoid having to operate $\mathcal{L}$ on the $\bar{\phi}_i$ terms.

Now if we let $g = \bar{S}_\ell$ and $f = \bar{C}_\ell$ in \cref{eq:Fsymm},
\begin{align}
F &= \left(\frac{1}{\sqrt{2}} \left[S_\ell \pm S_\ell^\prime \right], \mathcal{L} \frac{1}{\sqrt{2}}\left[C_\ell \pm C_\ell^\prime \right] \right) -
    \left(\frac{1}{\sqrt{2}} \left[C_\ell \pm C_\ell^\prime \right], \mathcal{L} \frac{1}{\sqrt{2}}\left[S_\ell \pm S_\ell^\prime \right] \right) \nonumber \\
& = \frac{1}{2}\left[(S_\ell,\mathcal{L}C_\ell) \pm (S_\ell,\mathcal{L}C_\ell^\prime) \pm (S_\ell^\prime,\mathcal{L}C_\ell) + (S_\ell^\prime,\mathcal{L}C_\ell^\prime) \right. \nonumber\\
& \left. \phantom{move} - (C_\ell,\mathcal{L}S_\ell) \mp (C_\ell,\mathcal{L}S_\ell^\prime) \mp (C_\ell^\prime,\mathcal{L}S_\ell) - (C_\ell^\prime,\mathcal{L}S_\ell^\prime)\right]
\end{align}
From the property of the permutation operators, $(S_\ell,\mathcal{L}C_\ell) = 
(S_\ell^\prime,\mathcal{L}C_\ell^\prime)$, $(S_\ell^\prime,\mathcal{L}C_\ell) 
= (S_\ell,\mathcal{L}C_\ell^\prime)$, $(C_\ell,\mathcal{L}S_\ell) = (S_\ell^
\prime,\mathcal{L}C_\ell^\prime)$ and
$(C_\ell^\prime,\mathcal{L}S_\ell) = (C_\ell,\mathcal{L}S_\ell^\prime)$,
causing the above to reduce to
\begin{align}
F = &\left[ (S_\ell,\mathcal{L}C_\ell) - (C_\ell,\mathcal{L}S_\ell)\right] \pm \left[ (S_\ell,\mathcal{L}C_\ell^\prime) - (C_\ell^\prime,\mathcal{L}S_\ell)\right] \nonumber \\
&\equiv G \pm G^\prime.
\label{GBarDef}
\end{align}
Using \cref{eq:Hamiltonian2} and Green's theorem,
\begin{align}
G &= \frac{1}{2} \int\limits_{V_3} \int\limits_{V_{12}} \int\limits_{S_\rho} \left[ S_\ell \grad_\rho C_\ell - C_\ell \grad_\rho S_\ell \right] \cdot d\bm{\sigma}_\rho d\tau_{12} d\tau_3  \nonumber \\
  &+ \int\limits_{V_\rho} \int\limits_{V_{12}} \int\limits_{S_3} \left[ S_\ell \grad_{r_3} C_\ell - C_\ell \grad_{r_3} S_\ell\right] \cdot d\bm{\sigma}_3 d\tau_{12} d\tau_{\rho} \nonumber \\
  &+ 2 \int\limits_{V_\rho} \int\limits_{V_{13}}\int\limits_{S_{12}} \left[ S_\ell \grad_{r_{12}} C_\ell - C_\ell \grad_{r_{12}} S_\ell \right] \cdot d\bm{\sigma}_{12} d\tau_{13} d\tau_\rho
  \label{eq:GDef}
\end{align}
As before, the Ps and H functions have an exponential dependence on $r_3$ and 
$r_{12}$, respectively, so the second and third terms go to 0. The surface 
elements under consideration at $\rho \to \infty$ are normal to $\hat{\rho}$, 
so we can ignore the angular dependence in $\grad_\rho$. Then \cref{eq:GDef} 
becomes
\beq
G = \frac{1}{2} \int\limits_{V_3} \int\limits_{V_{12}} \left[\int\limits_{S_\rho} \left(S_\ell \frac{\partial C_\ell}{\partial \rho} - C_\ell \frac{\partial S}{\partial \rho} \right) \rho^2 \sin \theta_\rho d\theta_\rho d\varphi_\rho \right] d\tau_{12} d\tau_3.
\label{eq:GDef2}
\eeq

Using \cref{eq:GenSandC} and realizing that $f_\ell \rightarrow 1$ as
$\rho \to \infty$, from the \emph{Mathematica} notebook ``SLC - CLS Proof.nb''
\cite{Wiki}, we have
\begin{equation}
\label{eq:CSderdiff}
\left(S_\ell \frac{\partial C_\ell}{\partial \rho} - C_\ell \frac{\partial S_\ell}{\partial \rho} \right) = \frac{2}{\rho^2} \SphericalHarmonicY{\ell}{0}{\theta_\rho}{\varphi_\rho}^2 \Phi_{Ps}\left(r_{12}\right)^2 \Phi_H\left(r_3\right)^2 .
\end{equation}
Substituting \cref{eq:CSderdiff} into \cref{eq:GDef2} yields
\begin{align}
G &= \cancel{\frac{1}{2}} \int\limits_{V_3} \int\limits_{V_{12}} \left[\int\limits_{S_\rho} \cancel{\frac{2}{\rho^2}} \SphericalHarmonicY{\ell}{0}{\theta_\rho}{\varphi_\rho}^2 \Phi_{Ps}\left(r_{12}\right)^2 \Phi_H\left(r_3\right)^2 \cancel{\rho^2} \sin \theta_\rho d\theta_\rho d\varphi_\rho \right] d\tau_{12} d\tau_3  \nonumber \\
&= \int\limits_{V_3} \int\limits_{V_{12}} \Phi_{Ps}(r_{12})^2 \Phi_{H}(r_{3})^2 \left[ \int\limits_{S_\rho} \SphericalHarmonicY{\ell}{0}{\theta_\rho}{\varphi_\rho}^2 \sin \theta_\rho d\theta_\rho d\varphi_\rho \right] d\tau_{12} d\tau_3  \nonumber \\
&= \int\limits_{V_3} \int\limits_{V_{12}} \Phi_{Ps}(r_{12})^2 \Phi_{H}(r_{3})^2 d\tau_{12} d\tau_3 = 1,
\label{eq:GDef3}
\end{align}
which follows from the orthonormality of the spherical harmonics and the
normalization of the Ps and H wavefunctions. These, when combined in
\cref{GBarDef}, give that
\beq
(S_\ell,\mathcal{L}C_\ell) = (C_\ell,\mathcal{L}S_\ell) + 1.
\label{eq:SLCandCLSdirect}
\eeq
As this also applies to the permuted versions, writing this in terms of $\bar{S}_\ell$ and $\bar{C}_\ell$ gives the final relation of
\beq
\label{eq:SLCandCLS}
\left(\bar{S}_\ell,\mathcal{L}\bar{C}_\ell\right) = \left(\bar{C}_\ell,\mathcal{L}\bar{S}_\ell\right) + 1.
\eeq

This can also be shown more generally for the $(\widetilde{S}_\ell,\mathcal{L}\widetilde{C}_\ell)$
and $(\widetilde{C}_\ell,\mathcal{L}\widetilde{S}_\ell)$ matrix elements.
From the definitions of $\widetilde{S}_\ell$ and $\widetilde{C}_\ell$ in \cref{eq:TildeSCDef},
\begin{align}
\nonumber (\widetilde{S}_\ell,\mathcal{L}\widetilde{C}_\ell) &= \left((u_{00}\bar{S}_\ell + u_{01}\bar{C}_\ell),\mathcal{L}(u_{10}\bar{S}_\ell + u_{11}\bar{C}_\ell)\right) \\
&= u_{00} u_{10} (\bar{S}_\ell,\mathcal{L}\bar{S}_\ell) + u_{00} u_{11} (\bar{S}_\ell,\mathcal{L}\bar{C}_\ell) + u_{01} u_{10} (\bar{C}_\ell,\mathcal{L}\bar{S}_\ell) + u_{01} u_{11} (\bar{C}_\ell,\mathcal{L}\bar{C}_\ell) .
\label{eq:GenSLC}
\end{align}
Likewise,
\begin{align}
\nonumber (\widetilde{C}_\ell,\mathcal{L}\widetilde{S}_\ell) &= \left((u_{10}\bar{S}_\ell + u_{11}\bar{C}_\ell),\mathcal{L}(u_{00}\bar{S}_\ell + u_{01}\bar{C}_\ell)\right) \\
&= u_{10} u_{00} (\bar{S}_\ell,\mathcal{L}\bar{S}_\ell) + u_{10} u_{01} (\bar{S}_\ell,\mathcal{L}\bar{C}_\ell) + u_{11} u_{00} (\bar{C}_\ell,\mathcal{L}\bar{S}_\ell) + u_{11} u_{01} (\bar{C}_\ell,\mathcal{L}\bar{C}_\ell).
\label{eq:GenCLS}
\end{align}

\noindent Combining \cref{eq:GenSLC,eq:GenCLS} gives
\begin{align}
\nonumber (\widetilde{S}_\ell,\mathcal{L}\widetilde{C}_\ell) - (\widetilde{C}_\ell,\mathcal{L}\widetilde{S}_\ell) = \,\, &[u_{00} u_{10} - u_{10} u_{00}] (\bar{S}_\ell,\mathcal{L}\bar{S}_\ell) + [u_{00} u_{11} - u_{10} u_{01}] (\bar{S}_\ell,\mathcal{L}\bar{C}_\ell) \\
\nonumber + &[u_{01} u_{10} - u_{11} u_{00}] (\bar{C}_\ell,\mathcal{L}\bar{S}_\ell) + [u_{01} u_{11} - u_{11} u_{01}] (\bar{C}_\ell,\mathcal{L}\bar{C}_\ell) \\
\nonumber = \,\, &[u_{00} u_{11} - u_{10} u_{01}] [(\bar{S}_\ell,\mathcal{L}\bar{C}_\ell) - (\bar{C}_\ell,\mathcal{L}\bar{S}_\ell)] \\
\nonumber = \,\, & \Det{\textbf{u}} [(\bar{S}_\ell,\mathcal{L}\bar{C}_\ell) - (\bar{C}_\ell,\mathcal{L}\bar{S}_\ell)]
\end{align}
This is finally written as the general form of \cref{eq:SLCandCLS}, giving
\beq
(\widetilde{S}_\ell,\mathcal{L}\widetilde{C}_\ell) = (\widetilde{C}_\ell,\mathcal{L}\widetilde{S}_\ell) + \Det{\textbf{u}}.
\label{eq:GenSLCandCLS}
\eeq

This relation can let us obtain the
$(\widetilde{S}_\ell,\mathcal{L}\widetilde{C}_\ell)$ matrix element from
$(\widetilde{C}_\ell,\mathcal{L}\widetilde{S}_\ell)$, but it also gives us a 
nice numerical check. I calculate these two matrix elements separately in the 
long-long code and calculate their difference for the Kohn variational method.
\label{GenSLCandCLS}If it is close to 1, this
gives us confidence that the long-long integrations are accurate. This also
allows us to check that the difficult formulation of $\mathcal{L}C_\ell$
is correct.

\subsection{Matrix Elements Involving Long-Range Terms}
\label{sec:MatrixLong}
The short-long and long-long matrix elements have a similar analysis. For all 
of these, the effect of the $\mathcal{L} = 2(H-E)$ operator on the long-range 
terms must be considered, and then integrations over the external angles (see 
\cref{chp:AngularInt}) are performed. The remaining 6-dimensional integral is 
then numerically integrated as described in \cref{sec:LongLongInt,sec:ShortLongInt}.

\subsubsection{\texorpdfstring{$\mathcal{L}\bar{S}_\ell$}{LS} Terms}
\label{sec:LSTerms}
The matrix elements in \cref{eq:GeneralKohnMatrix} require us to first determine
$\mathcal{L}\bar{S}_\ell$. We start with examining $\mathcal{L}S_\ell$ first.
Using \cref{eq:GenSDef,eq:LDef}, 
\begin{align}
\label{eq:LS1}
\mathcal{L}S_\ell = &\left(-\frac{1}{2} \Laplacian_{\bm{\rho}} - \Laplacian_{\bm{r}_3} - 2 \Laplacian_{\bm{r}_{12}} + \frac{2}{r_1} - \frac{2}{r_2} - \frac{2}{r_3} - \frac{2}{r_{12}} - \frac{2}{r_{13}} + \frac{2}{r_{23}} - 2 E_H - 2 E_{Ps} - \frac{1}{2} \kappa^2\right) \nonumber \\
& \times \SphericalHarmonicY{\ell}{0}{\theta_\rho}{\varphi_\rho} \Phi_{Ps}\left(r_{12}\right) \Phi_H\left(r_3\right) \sqrt{2\kappa} \,j_\ell\!\left(\kappa\rho\right).
\end{align}
Since $S_\ell$ is independent of $r_3$ and $r_{12}$ except for the $\Phi_H$ and $\Phi_{Ps}$ functions, respectively, using \cref{eq:HPsEqn,eq:HPsHamil} simplifies this to
\begin{align}
\label{eq:LS2}
\mathcal{L}S_\ell = \left(-\frac{1}{2} \Laplacian_{\bm{\rho}} + \frac{2}{r_1} - \frac{2}{r_2} - \frac{2}{r_{13}} + \frac{2}{r_{23}} - \frac{1}{2} \kappa^2\right) \SphericalHarmonicY{\ell}{0}{\theta_\rho}{\varphi_\rho} \Phi_{Ps}\left(r_{12}\right) \Phi_H\left(r_3\right) \sqrt{2\kappa} \,j_\ell\!\left(\kappa\rho\right).
\end{align}
From \cref{sec:SphBess2}, we find that $\SphericalHarmonicY{\ell}{0}{\theta_\rho}{\varphi_\rho} j_\ell(\kappa\rho)$ is an eigenfunction of $\Laplacian_{\bm{\rho}}$ with eigenvalue $-\kappa^2$:
\begin{equation}
\Laplacian_{\bm{\rho}} \left[\SphericalHarmonicY{\ell}{0}{\theta_\rho}{\varphi_\rho} j_\ell(\kappa\rho) \right] = \frac{(-\kappa^2 \rho^2) \LegendreP{\ell, \cos\theta} } {\rho^2 \LegendreP{\ell, \cos\theta}}
= -\kappa^2 \, \SphericalHarmonicY{\ell}{0}{\theta_\rho}{\varphi_\rho} j_\ell(\kappa\rho).
\end{equation}
Then \cref{eq:LS2} reduces down to
\begin{equation}
\label{eq:LS3}
\mathcal{L}S_\ell = \left(\frac{2}{r_1} - \frac{2}{r_2} - \frac{2}{r_{13}} + \frac{2}{r_{23}}\right) \SphericalHarmonicY{\ell}{0}{\theta_\rho}{\varphi_\rho} \Phi_{Ps}\left(r_{12}\right) \Phi_H\left(r_3\right) \sqrt{2\kappa} \,j_\ell\!\left(\kappa\rho\right)
\end{equation}
or
\begin{equation}
\label{eq:LSFinal}
\mathcal{L}S_\ell = \left(\frac{2}{r_1} - \frac{2}{r_2} - \frac{2}{r_{13}} + \frac{2}{r_{23}}\right) S_\ell.
\end{equation}

$\mathcal{L}S_\ell^\prime$ is simply the same as \cref{eq:LSFinal} but with $2 \leftrightarrow 3$ due to the permutation operator, or
\begin{equation}
\label{eq:LSPrimeFinal}
\mathcal{L}S_\ell^\prime = \left(\frac{2}{r_1} - \frac{2}{r_3} - \frac{2}{r_{12}} + \frac{2}{r_{23}}\right) S_\ell^\prime.
\end{equation}

\subsubsection{\texorpdfstring{$\mathcal{L}\bar{C}_\ell$}{LC} Terms}
\label{sec:LCTerms}
\begin{figure}
	\centering
	\includegraphics[width=6.5in]{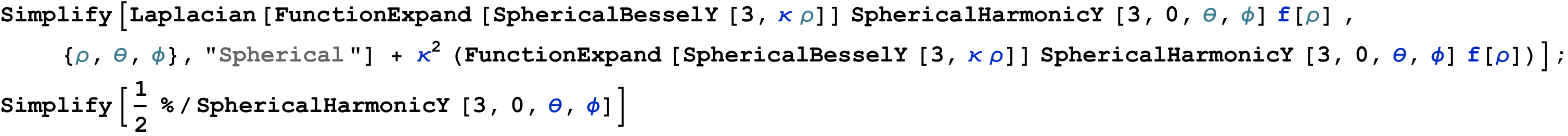}
	\caption[\emph{Mathematica} code to calculate part of $\mathcal{L}C_\ell$]{Listing of \emph{Mathematica} code in ``First Partial Waves LC.nb'' to calculate part of $\mathcal{L}C_\ell$ for the F-wave}
	\label{fig:LCMath}
\end{figure}

To calculate the matrix elements in \cref{eq:GeneralKohnMatrix} that include $\bar{C}_\ell$ in the ket, we start by writing a general form of $\mathcal{L}C_\ell$ using \cref{eq:LDef,eq:GenCDef}:
\begin{align}
\label{eq:LC1}
\mathcal{L}C_\ell = -&\left(-\frac{1}{2} \Laplacian_{\bm{\rho}} - \Laplacian_{\bm{r}_3} - 2 \Laplacian_{\bm{r}_{12}} + \frac{2}{r_1} - \frac{2}{r_2} - \frac{2}{r_3} - \frac{2}{r_{12}} - \frac{2}{r_{13}} + \frac{2}{r_{23}} - 2 E_H - 2 E_{Ps} - \frac{1}{2} \kappa^2\right) \nonumber \\
& \times \SphericalHarmonicY{\ell}{0}{\theta_\rho}{\varphi_\rho} \Phi_{Ps}\left(r_{12}\right) \Phi_H\left(r_3\right) \sqrt{2\kappa} \,n_\ell\!\left(\kappa\rho\right) f_\ell(\rho) 
\end{align}
Similar to \cref{eq:LS2}, using \cref{eq:HPsEqn,eq:HPsHamil} reduces this to
\begin{align}
\label{eq:LC2}
\mathcal{L}C_\ell = -&\left(-\frac{1}{2} \Laplacian_{\bm{\rho}} + \frac{2}{r_1} - \frac{2}{r_2} - \frac{2}{r_{13}} + \frac{2}{r_{23}} - 2 E_H - 2 E_{Ps} - \frac{1}{2} \kappa^2\right) \nonumber \\
& \times \SphericalHarmonicY{\ell}{0}{\theta_\rho}{\varphi_\rho} \Phi_{Ps}\left(r_{12}\right) \Phi_H\left(r_3\right) \sqrt{2\kappa} \,n_\ell\!\left(\kappa\rho\right) f_\ell(\rho).
\end{align}
Unlike with $\mathcal{L}S_\ell$ in \cref{sec:LSTerms}, there is not a direct cancellation 
with the $\Laplacian_{\bm{\rho}}$ and $\kappa^2$ terms, as these also operate on the
shielding function $f_\ell$. The combination of these 
terms was calculated in the ``First Partial Waves LC.nb'' \emph{Mathematica} 
notebook \cite{GitHub,Wiki} using the code given in \cref{fig:LCMath}. This is for the F-wave, 
and replacing the $\ell$-value of 3 in \texttt{SphericalBesselY} allows this 
to be used for any partial wave. The results of these derivations are given in 
each partial wave chapter through the D-wave. The full $\mathcal{L}C_\ell$ for the
S-wave is shown on page~\pageref{pg:SWaveLC} by substituting the
$\frac{1}{2} \left(\Laplacian_{\bm{\rho}} + \kappa^2\right) \SphericalHarmonicY{\ell}{0}{\theta_\rho}{\varphi_\rho} n_\ell(\kappa\rho) f_\ell(\rho)$
in \cref{eq:LC2}.

\noindent Matrix elements involving $\mathcal{L}C_\ell^\prime$ look similar but have the 2 and 3 coordinates swapped. In other words, $\rho \leftrightarrow \rho^\prime$.

From \cref{eq:PartialWaveShielding}, the general shielding function for $\widetilde{C}_\ell$ to keep it regular at the origin is given by
\begin{equation}
  f_\ell(\rho) = \left[1 - \ee^{-\mu \rho} \left(1+\frac{\mu}{2}\rho\right)
  \right]^{m_\ell}.
\end{equation}
In \cref{fig:LCMath}, the derivatives $f_\ell^\prime(\rho)$ and $f_\ell^{\prime\prime}(\rho)$ are needed. In the \emph{Mathematica} notebook ``Shielding Factor.nb'' \cite{GitHub,Wiki}, I found out that the derivatives can be written generally as
\beq
\label{eq:Shielding1Der}
f_\ell^\prime(\rho) = -\frac{\mu m_\ell (\mu  \rho +1) \left[1-\frac{1}{2} e^{-\mu  \rho } (\mu  \rho +2)\right]^{m_\ell}}{\mu  \rho -2 e^{\mu  \rho }+2}
\eeq
and
\beq
\label{eq:Shielding2Der}
f_\ell^{\prime\prime}(\rho) = \frac{\mu^2 m_\ell \left[-2 \mu  \rho  e^{\mu  \rho }+ m_\ell (\mu  \rho +1)^2-1\right] \left[1-\frac{1}{2} e^{-\mu  \rho } (\mu  \rho +2)\right]^{m_\ell}} {\left(\mu  \rho -2 e^{\mu  \rho }+2\right)^2}.
\eeq

\subsubsection{\texorpdfstring{$(\bar{S}_\ell,\mathcal{L}\bar{S}_\ell)$}{SLS} and \texorpdfstring{$(\bar{C}_\ell,\mathcal{L}\bar{S}_\ell)$}{CLS} Matrix Elements}
\label{sec:SLSandCLS}

From \cref{eq:TildeSCDef}, we see that, in general, any matrix element in 
\cref{eq:GeneralKohnMatrix} containing only long-range terms will contain 
both $(\bar{S}_\ell,\mathcal{L}\bar{S}_\ell)$ and
$(\bar{C}_\ell,\mathcal{L}\bar{S}_\ell)$, along with the other two combinations
given in \cref{sec:LCTerms}.
When $(\bar{S}_\ell,\mathcal{L}\bar{S}_\ell)$ is expanded,

\beq
(\bar{S}_\ell,\mathcal{L}\bar{S}_\ell) = \frac{1}{2} \left[ (S_\ell,\mathcal{L}S_\ell) \pm (S_\ell^\prime,\mathcal{L}S_\ell) \pm (S_\ell,\mathcal{L}S_\ell^\prime) \pm (S_\ell^\prime,\mathcal{L}S_\ell^\prime) \right].
\eeq
The properties of the permutation operator give that
\beq
(S_\ell,\mathcal{L}S_\ell) = (S_\ell^\prime,\mathcal{L}S_\ell^\prime) \text{ and } (S_\ell^\prime,\mathcal{L}S_\ell) = (S_\ell,\mathcal{L}S_\ell^\prime),
\eeq
so this becomes
\beq
(\bar{S}_\ell,\mathcal{L}\bar{S}_\ell) = (S_\ell,\mathcal{L}S_\ell) \pm (S_\ell^\prime,\mathcal{L}S_\ell).
\eeq

From \cref{eq:LSFinal,eq:LSPrimeFinal}, the Laplacian on $S_\ell$ and
$S_\ell^\prime$ leaves only some of the potential terms. The potential terms in
\cref{eq:LSFinal} are antisymmetric upon the $1 \leftrightarrow 2$ swap, and the 
terms in \cref{eq:LSPrimeFinal} are antisymmetric upon the $1 \leftrightarrow 3$
swap. $S_\ell$ and $S_\ell^\prime$ are symmetric with the $1 \leftrightarrow 2$
and $1 \leftrightarrow 3$ swaps, respectively. When these 
are integrated over these coordinates, the combination of symmetric with 
antisymmetric functions causes the integral to be 0:
\beq
\label{eq:SLS0Test}
(S_\ell,\mathcal{L}S_\ell) = (S_\ell^\prime,\mathcal{L}S_\ell^\prime) = 0.
\eeq
Therefore, using \cref{eq:LSFinal,eq:LSPrimeFinal},
\begin{subequations}
\label{eq:SbarLSbar}
\begin{align}
(\bar{S}_\ell,\mathcal{L}\bar{S}_\ell) &= \pm (S_\ell^\prime,\mathcal{L}S_\ell) = \pm \left(S_\ell^\prime, \left[ \frac{2}{r_1} - \frac{2}{r_2} - \frac{2}{r_{13}} + \frac{2}{r_{23}} \right] S_\ell\right)  \label{eq:SbarLSbar1} \\
& = \pm (S_\ell,\mathcal{L}S_\ell^\prime) = \pm \left(S_\ell, \left[ \frac{2}{r_1} - \frac{2}{r_3} - \frac{2}{r_{12}} + \frac{2}{r_{23}} \right] S_\ell^\prime \right) . \label{eq:SbarLSbar2}
\end{align}
\end{subequations}
Either form can be used to calculate $(\bar{S}_\ell,\mathcal{L}\bar{S}_\ell)$ in the long-range code.

Since $C_\ell$ and $C_\ell^\prime$ are also symmetric with their respective swaps, the $(\bar{C}_\ell,\mathcal{L}\bar{S}_\ell)$ matrix element is a similar form given by
\begin{subequations}
\label{eq:CbarLSbar}
\begin{align}
(\bar{C}_\ell,\mathcal{L}\bar{S}_\ell) &= \pm (C_\ell^\prime,\mathcal{L}S_\ell) = \pm \left(C_\ell^\prime, \left[ \frac{2}{r_1} - \frac{2}{r_2} - \frac{2}{r_{13}} + \frac{2}{r_{23}} \right] S_\ell\right)  \label{eq:CbarLSbar1} \\
& = \pm (C_\ell,\mathcal{L}S_\ell^\prime) = \pm \left(C_\ell, \left[ \frac{2}{r_1} - \frac{2}{r_3} - \frac{2}{r_{12}} + \frac{2}{r_{23}} \right] S_\ell^\prime \right) . \label{eq:CbarLSbar2}
\end{align}
\end{subequations}

\subsubsection{\texorpdfstring{$(\bar{\phi}_i, \mathcal{L}\bar{S}_\ell)$}{phiLS} and \texorpdfstring{$(\bar{\phi}_i, \mathcal{L}\bar{C}_\ell)$}{phiLC} Matrix Elements}
\label{sec:phiLSC}

The $(\bar{\phi}_i,\mathcal{L}\widetilde{S}_\ell)$ and $(\bar{\phi}_i,\mathcal{L}\widetilde{C}_\ell)$ in \cref{eq:GeneralKohnMatrix} have combinations of $(\bar{\phi}_i, \mathcal{L}\bar{S})$ and $(\bar{\phi}_i, \mathcal{L}\bar{C})$, as seen in \cref{eq:TildeSCDef}.

Let us investigate $(\bar{\phi}_i, \mathcal{L}\bar{S}_\ell)$ first.
\beq
\label{eq:PhiBarLSBar1}
(\bar{\phi}_i, \mathcal{L}\bar{S}_\ell) = \left( (\phi_i \pm \phi_i'), \mathcal{L} \frac{(S_\ell \pm S_\ell^\prime)}{\sqrt{2}}\right)
= \frac{1}{\sqrt{2}} \left[ (\phi_i, \mathcal{L} S_\ell) \pm (\phi_i, \mathcal{L} S_\ell^\prime) \pm (\phi_i^\prime, \mathcal{L} S_\ell) + (\phi_i^\prime, \mathcal{L} S_\ell^\prime) \right]
\eeq
Again, from the properties of the $P_{23}$ permutation operator,
\beq
(\phi_i,\mathcal{L}S_\ell) = (\phi_i^\prime,\mathcal{L}S_\ell^\prime) \text{ and } (\phi_i,\mathcal{L}S_\ell^\prime) = (\phi_i^\prime,\mathcal{L}S_\ell).
\eeq

\noindent \Cref{eq:PhiBarLSBar1} becomes
\begin{subequations}
\label{eq:PhiBarLSBar2}
\begin{align}
(\bar{\phi}_i, \mathcal{L}\bar{S}_\ell) = \frac{1}{\sqrt{2}} \left[2(\phi_i,\mathcal{L}S_\ell) \pm 2(\phi_i^\prime,\mathcal{L}S_\ell)\right] &= \frac{2}{\sqrt{2}} \left[(\phi_i,\mathcal{L}S_\ell) \pm (\phi_i',\mathcal{L}S_\ell)\right] \label{eq:PhiBarLSBar2a} \\
 &= \frac{2}{\sqrt{2}} \left[(\phi_i,\mathcal{L}S_\ell) \pm (\phi_i,\mathcal{L}S_\ell^\prime)\right]  \label{eq:PhiBarLSBar2b}
\end{align}
\end{subequations}

\noindent Notice that \cref{eq:PhiBarLSBar2a,eq:PhiBarLSBar2b}) are equivalent ways of writing this expression. Either could be used, depending on the form desired for the computation.
From \cref{eq:LSFinal,eq:LSPrimeFinal},
\begin{align}
(\bar{\phi}_i, \mathcal{L}\bar{S}_\ell) &= \frac{2}{\sqrt{2}} \left[\left( \phi_i \left( \frac{2}{r_1} - \frac{2}{r_2} - \frac{2}{r_{13}} + \frac{2}{r_{23}} \right) S_\ell\right) \pm \left( \phi_i^\prime \left( \frac{2}{r_1} - \frac{2}{r_2} - \frac{2}{r_{13}} + \frac{2}{r_{23}} \right) S_\ell\right)\right] \\
 &= \frac{2}{\sqrt{2}} \left[\left( \phi_i \left( \frac{2}{r_1} - \frac{2}{r_2} - \frac{2}{r_{13}} + \frac{2}{r_{23}} \right) S_\ell\right) \pm \left( \phi_i \left( \frac{2}{r_1} - \frac{2}{r_3} - \frac{2}{r_{12}} + \frac{2}{r_{23}} \right) S_\ell^\prime \right)\right]
\end{align}
Unlike the long-long matrix elements in \cref{sec:SLSandCLS}, $\phi_i$ is
neither symmetric nor anti-symmetric in the $1 \leftrightarrow 2$ swap, so
the direct-direct and exchange-exchange terms are nonzero.

\subsection{Matrix Elements Involving Only Short-Range Terms}
\label{sec:MatrixShort}
Using the short-range terms given by \cref{eq:PhiDef} in \cref{eq:BoundHFull}
with \cref{eq:LDef} and realizing that the bra is
not conjugated in the Kohn-type variational methods \cite{Cooper2010,Lucchese1989},
the short-short integrals are of the form
\beq
\label{eq:SWaveShortShort}
\left(\bar{\phi}_i, \mathcal{L} \bar{\phi}_j\right) = \bigintsss \left[ \sum_{l=1}^3 \boldsymbol{\nabla}_{\!\mathbf{r}_l} \bar{\phi}_i \boldsymbol{\cdot} \boldsymbol{\nabla}_{\!\mathbf{r}_l} \bar{\phi}_j + \left( \frac{2}{r_1} - \frac{2}{r_2} - \frac{2}{r_3} - \frac{2}{r_{12}} - \frac{2}{r_{13}} + \frac{2}{r_{23}} - 2 E \right) \bar{\phi}_i \bar{\phi}_j \right] d\tau.
\eeq
Again, the $\bar{\phi}$ represents any of the short-range terms given in
\cref{eq:PhiDef} and could also represent any other Hylleraas-type terms, such as 
the mixed symmetry terms (see \cref{sec:MixedTerms}) or the second formalism of the
P-wave (see \cref{sec:PWave2Formalism}). These matrix elements are numerically 
integrated using the methods described in \cref{sec:ShortInt}.

\section{Schwartz Singularities}
\label{eq:SchwartzSing}

A disadvantage of the Kohn-type variational methods is the presence of spurious 
singularities in the phase shifts. Looking at \cref{eq:GeneralKohnMatrix}, if 
$\textbf{\emph{A}}$ becomes near-singular
(i.e. $\Det{\textbf{\emph{A}}} \approx 0$), solving this matrix equation
will yield incorrect phase shifts. 
These ``Schwartz singularities'' were described first by Schwartz
\cite{Schwartz1961} and analyzed by others \cite{Nesbet1968,Nesbet1969}.
These singularities do not make the Kohn-type variational methods 
unusable, however. These singularities are often easily noticeable, because 
they do not follow the pattern of other phase shifts, and they do not
agree with the results of the other Kohn-type variational methods.

\begin{figure}
	\centering
	\includegraphics[width=\textwidth]{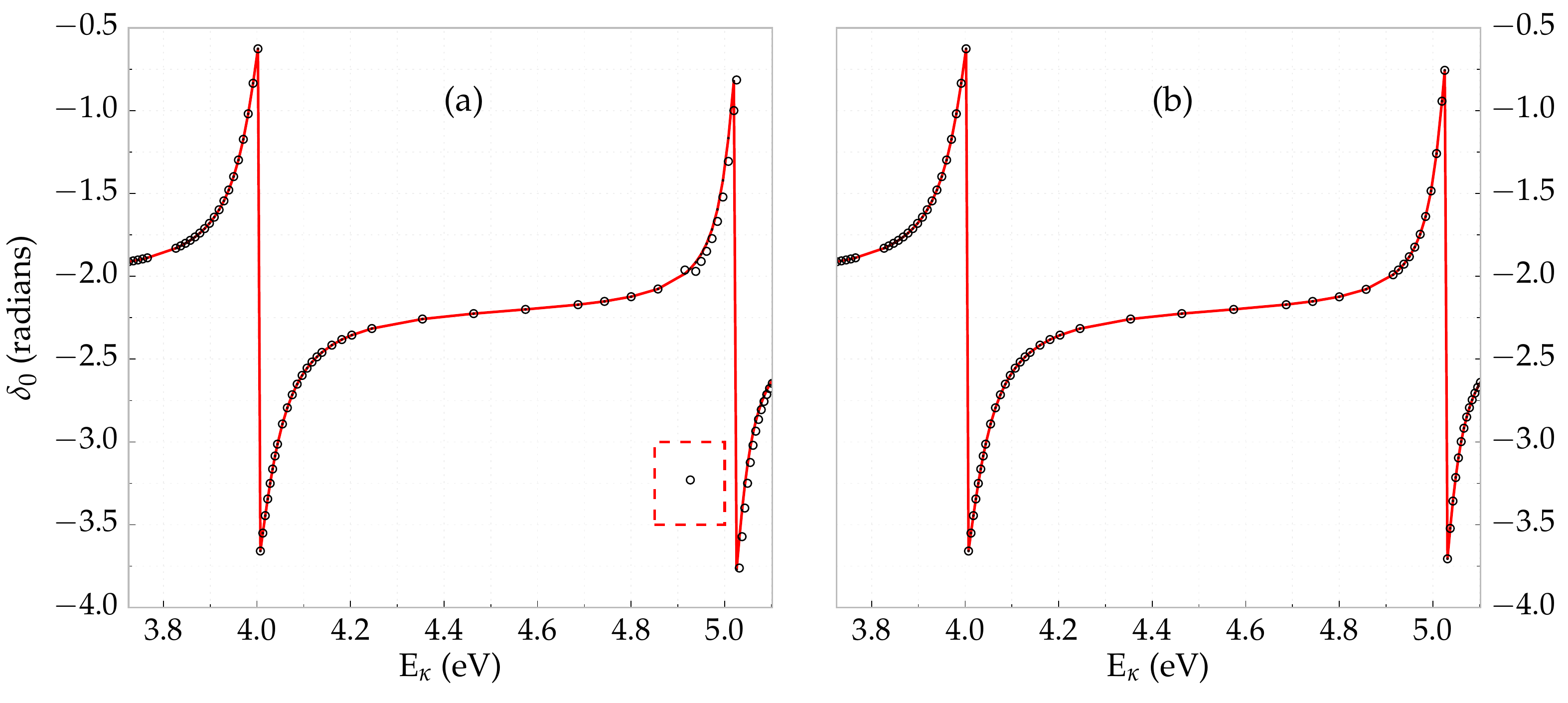}
	\caption[Example of a Schwartz singularity for $^1$S]{Example of Schwartz singularity for $^1$S at $\omega = 7$.
The generalized Kohn method with $\tau = 1.4$ is shown in (a), and the dashed rectangle
is surrounding the Schwartz singularity. The corresponding $S$-matrix complex Kohn phase shifts are shown in (b),
and no Schwartz singularities are present.}
	\label{fig:schwartz-singularity}
\end{figure}

As an example, refer to \cref{fig:schwartz-singularity}(a). This shows an 
example of Schwartz singularities for the generalized Kohn method with
$\tau = 1.4$. A clear Schwartz singularity exists at $\kappa = 0.851$ or $E = \SI{4.927}{eV}$,
which can be seen as a point on the graph that is far away from 
the red fitting curve. There is also a Schwartz singularity at $\kappa = 0.85$
or $E = \SI{4.915}{eV}$, seen as a slight deviation from the fitting curve.

The same calculation as in \cref{fig:schwartz-singularity}(a) is performed 
using the $S$-matrix complex Kohn in \cref{fig:schwartz-singularity}(b), but 
no Schwartz singularities are evident. Normally, if one Kohn-type
method described in the previous section (\cref{sec:KohnApplied}) has a 
Schwartz singularity, other Kohn-type methods will not. This gives us a strategy 
of simply rejecting Kohn methods that have obvious Schwartz singularities.

Additionally, the complex Kohn methods in
\cref{eq:uCompTKohn,eq:uCompSKohn,eq:uGenTKohn,eq:uGenSKohn} are far less likely to 
suffer from these singularities. A complex-valued Kohn variational method was 
first proposed by Miller et al. \cite{Miller1987} and used the same year by 
McCurdy et al.~\cite{McCurdy1987}. Subsequent work by
Lucchese~\cite{Lucchese1989} showed that the complex Kohn methods can indeed have
Schwartz singularities, but they are not likely to show up in practice.
Cooper et al.~\cite{Cooper2010} also showed that the phase shifts obtained using the 
complex Kohn variational method can be obtained exactly from the real-valued 
generalized Kohn method in \cref{eq:uGenKohn}, but we do not use this method.

There is also much less variability in the results of the different complex 
Kohn-type variational methods, and even for different values of $\tau$ in
\cref{eq:uGenTKohn,eq:uGenSKohn}, the phase shifts generally agree to great 
precision (greater than the accuracy that we quote for the phase shifts). Due 
to the stability and agreement of the complex Kohn variational methods, the results quoted 
throughout this paper are for the $S$-matrix complex Kohn unless noted 
otherwise.

\section{Resonances}
\label{sec:Resonances}
Resonances are where the phase shifts rapidly change by $\pi$. We find that 
there are resonances in each singlet partial wave for Ps-H scattering. Some 
papers refer to these resonances as Breit-Wigner resonances
\cite{Tennyson1984,Stibbe1998}, but according to Bransden and Joachain
\cite[p.596]{Bransden2003}, Breit-Wigner resonances are a special case of Fano
resonances. 

The resonance positions and widths can be calculated to high 
accuracy by fitting the phase shift data to the following curve \cite{VanReeth2004}:
\beq
\label{eq:ResonanceCurve}
\delta(E) = A + BE + CE^2 + \arctan\left[ \frac{^1\Gamma}{2\left(^1E_R-E\right)} \right]
  + \arctan\left[ \frac{^2\Gamma}{2\left(^2E_R-E\right)} \right].
\eeq
The polynomial part of the above equation corresponds to hard sphere 
scattering. The arctangent parts correspond to the first and second Fano 
resonances \cite{Fano1961,Macek1970,Hazi1979}, with $^1E_R$ and $^2E_R$ as 
the positions of the resonances and $^1\Gamma$ and $^2\Gamma$ as the widths 
of the respective resonances. The $^1$S and $^1$P partial waves use this 
fitting, and the $^1$D and $^1$F partial waves omit the second term, since we 
consider only a single resonance. See
\cref{sec:SWaveResonances,sec:PWaveResonances,sec:DWaveResonance,sec:FWaveResonance}
for further discussion of the resonances for each partial wave.

\begin{figure}
	\centering
	\includegraphics[width=5.25in]{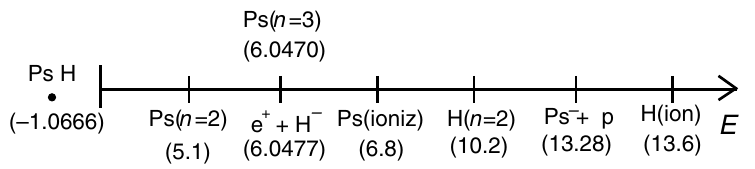}
	\caption[Event line for singlet Ps-H scattering]{Event line for singlet Ps-H scattering
from Ref.~\cite{Walters2004}. Reprinted with permission from Elsevier Limited.}
	\label{fig:WaltersEventLine}
\end{figure}

As Blackwood et al.\ \cite{Blackwood2002} mention, there are an infinite 
number of Rydberg resonances in each partial wave converging on the e$^+$+H$^-$
threshold at \SI{6.05}{eV}. Walters et al.\ \cite{Walters2004} gives this as a
a more accurate value of \SI{6.0477}{eV}, as shown in \cref{fig:WaltersEventLine}
from their paper. They also note that the Ps(n=3)
threshold is at \SI{6.0470}{eV}, making these two thresholds nearly the same
energy. This work only considers the single channel problem up to the Ps(n=2)
threshold at \SI{5.102}{eV}.

For Ps-H scattering, Drachman predicted that these resonances correspond to 
the metastable state of e$^+$ with the H$^-$ ion \cite{Drachman1979}. A set 
of close coupling papers \cite{Blackwood2002,Blackwood2002b,Walters2004} 
confirms that the H$^-$ channel is important for the resonances, indicating 
that Drachman's prediction was correct. Biswas \cite{Biswas2001} also showed 
that H$^-$ formation is important for describing this system.

The first $^1$S resonance is associated with the $2s$ state \cite{DiRienzi2002b} 
and was first calculated by Hazi and Taylor using a stabilization method
\cite{Hazi1970}. The first $^1$P Rydberg resonance is associated with the $3p$ 
state, not the $2p$ state \cite{DiRienzi2002b}, while the $^1$D resonance 
corresponds with the $3d$ state \cite{DiRienzi2002a}. No such analysis exists
in the current literature for higher partial waves or resonances for the
S-, P- and D-waves other than the first resonance of each.


\chapter{Computation}
\label{chp:Computation}

\iftoggle{UNT}{This}{\lettrine{\textcolor{startcolor}{T}}{his}}
chapter covers some of the major
computational details of this work. For additional details, such as resonance
fittings, determination of nonlinear parameters and the use of Gaussian
quadratures, refer to \cref{chp:ExtraNumerics}.

\section{Short-Range Terms}
\label{sec:CompShort}

Matrix elements in \cref{eq:GeneralKohnMatrix} involving only short-range 
terms are handled differently than those that involve long-range terms
(short-long and long-long). The integrals resulting from using
\Cref{eq:BoundHFull,eq:GradGradShort} in \cref{eq:GeneralKohnMatrix} are 
handled in this section.

The bound state problem is a generalized eigenvalue problem (see
\cref{eq:BoundGenEig}), but to simplify the following discussion, I will refer
to the set of matrices \textbf{H} and \textbf{S} as a single matrix. The 
diagrams in this section show a single matrix, but it is in actuality a pair 
of matrices forming the generalized eigenvalue problem.

\subsection{Short-Range -- Short-Range Integrations}
\label{sec:ShortInt}
The short-range--short-range (short-short) matrix elements make up the
bulk of the $\textbf{\emph{A}}$ matrix
(\cref{eq:GenKohnMatrixAXB}). The PsH bound state problem in 
\cref{chp:PsHBound} consists of only these types of matrix elements
(see \cref{eq:BoundWavefn}). The form of these short-range integrals is
\beq
\label{eq:FourBody}
I = \int e^{-(\bar{\alpha} r_1 + \bar{\beta} r_2 + \bar{\gamma} r_3)} r_1^{k_i} r_2^{l_i} r_{12}^{m_i} r_3^{n_i} r_{13}^{p_i} r_{23}^{q_i} d\textbf{r}_1 d\textbf{r}_2 d\textbf{r}_3,
\eeq
with real-valued $\bar{\alpha}, \bar{\beta}, \bar{\gamma} > 0$. The $\bar{\alpha}$ is related
to $\alpha$ through \cref{eq:GeneralKohnMatrix}, as are the relations of the other
nonlinear parameters $\bar{\beta}$, $\beta$, $\bar{\gamma}$, and $\gamma$.

This class of integrals, the Hylleraas three-electron or four-body 
integrals, has been studied extensively. See
Refs.~\cite{Drake1995,Frolov2003,Pelzl1998,Ruiz2009,Pachucki2004} for just some of 
the papers detailing strategies on how to compute these integrals.
Refs.~\cite{Drake1995,Frolov2003,Pelzl1998} use the same infinite summation to numerically solve this 
integral but use different techniques to accelerate the convergence, as the 
summation converges slowly for some arguments. The paper by Pachucki et 
al. \cite{Pachucki2004} uses a very different approach with recursion relations,
described in \cref{sec:RecursionRelations}.

Each of these methods has some restriction on how singular these integrals 
can be. For Drake and Yan's asymptotic expansion \cite{Drake1995,Yan1997},
$k_i, l_i, n_i \geq -2$ and $m_i, p_i, q_i \geq -1$. Yan extends these to
$k_i, l_i, m_i, n_i, p_i, q_i \geq -3$ in an additional paper \cite{Yan2000a}. For 
the recursion relations of Pachucki et al.\ \cite{Pachucki2004},
$k_i, l_i, m_i, n_i, p_i, q_i \geq -1$. Pachucki and Puchalski have two other papers,
one extending this to $k_i, l_i, m_i, n_i, p_i, q_i \geq -2$ \cite{Pachucki2005} 
and another even extending these to $k_i, l_i, m_i, n_i, p_i, q_i \geq -3$ 
\cite{Pachucki2008}. Our work for the D-wave has some terms with $k_i$, $l_i$,
or $n_i = -2$, restricting what methods we can use. The biggest need for 
integrals more singular than $r_i^{-1}$ or $r_{ij}^{-1}$ in most 
work (such as lithium energies \cite{Yan1997a,Puchalski2010}) is for 
relativistic or quantum electrodynamics effects
\cite{Yan1997,Pachucki2008,Puchalski2010}.

{\"O}hrn and Nordling \cite{Ohrn1963} were the first to give a method to 
solve this type of integral by splitting it into summations over $W$ 
functions. To deal with odd powers of $r_{ij}$, these terms are usually 
expanded in a Laplace expansion using Perkins's expression \cite{Perkins1968}.
A related expression is given by Sack \cite{Sack1964}. Porras and King
\cite{Porras1994} also use another expansion with Gegenbauer polynomials.

From Drake and Yan's paper \cite{Drake1995}, splitting into $W$ functions gives
\begin{align}
\label{eq:FourBodyExpansion}
I &= (4\pi)^3 \sum_{q=0}^\infty \sum_{k_{12} = 0}^{L_{12}} \sum_{k_{12} = 0}^{L_{23}} \sum_{k_{12} = 0}^{L_{13}} \frac{1}{(2q+1)^2} C_{j_{12} q k_{12}} C_{j_{23} q k_{23}} C_{j_{13} q k_{13}} \nonumber \\
& \times [W(\tilde{k}_i + 2q + 2 k_{12} + 2 k_{13}, \tilde{l}_i + m_i - 2 k_{12} + 2 k_{23}, \tilde{n}_i + q_i - 2 q - 2 k_{23} + p_i - 2 k_{13}; \alpha, \beta, \gamma) \nonumber \\
      & + W(\tilde{k}_i + 2q + 2 k_{12} + 2 k_{13}, \tilde{n}_i + p_i - 2 k_{13} + 2 k_{23}, \tilde{l}_i + m_i - 2 q - 2 k_{23} + q_i - 2 k_{13}; \alpha, \gamma, \beta) \nonumber \\
      & + W(\tilde{l}_i + 2q + 2 k_{12} + 2 k_{23}, \tilde{k}_i + m_i - 2 k_{12} + 2 k_{13}, \tilde{n}_i + q_i - 2 q - 2 k_{23} + p_i - 2 k_{13}; \beta, \alpha, \gamma) \nonumber \\
      & + W(\tilde{l}_i + 2q + 2 k_{12} + 2 k_{23}, \tilde{n}_i + q_i - 2 k_{23} + 2 k_{13}, \tilde{k}_i + m_i - 2 q - 2 k_{23} + p_i - 2 k_{13}; \beta, \gamma, \alpha) \nonumber \\
      & + W(\tilde{n}_i + 2q + 2 k_{23} + 2 k_{13}, \tilde{k}_i + p_i - 2 k_{13} + 2 k_{12}, \tilde{l}_i + m_i - 2 q - 2 k_{23} + q_i - 2 k_{13}; \gamma, \alpha, \beta) \nonumber \\
      & + W(\tilde{n}_i + 2q + 2 k_{23} + 2 k_{13}, \tilde{l}_i + q_i - 2 k_{23} + 2 k_{12}, \tilde{k}_i + m_i - 2 q - 2 k_{23} + p_i - 2 k_{13}; \gamma, \beta, \alpha)].
\end{align}
The $C_{jqk}$ coefficients are given by Perkins \cite{Perkins1968} as
\beq
\label{eq:Ccoeff}
C_{jqk} = \frac{2q+1}{j+2} \Binomial{j+2}{2k+1} \prod_{t=0}^{\min{(q-1),\frac{1}{2}(j+1)}} \frac{2k+2t-j}{2k+2q-2t+1}.
\eeq
The $W$ functions are expressed as an infinite summation of the $_2F_1$ hypergeometric functions \cite{Drake1995}:
\begin{align}
\label{eq:Wfunc}
W(l,m,n;\alpha,\beta,\gamma) = &\frac{\Factorial{l}}{(\alpha+\beta+\gamma)^{l+m+n+3}} \sum_{p=0}^\infty \frac{\Factorial{(l+m+n+p+2)}}{\Factorial{(l+1+p)} (l+m+2+p)} \left( \frac{\alpha}{\alpha+\beta+\gamma} \right)^p  \nonumber \\
& \times \Hypergeometric{2}{1}{1,l+m+n+p+3}{l+m+p+3}{\frac{\alpha+\beta}{\alpha+\beta+\gamma}}.
\end{align}
To reduce computation time, we also use the recursion relation in their paper of
\beq
\Hypergeometric{2}{1}{1,a}{c}{z} = 1 + \left( \frac{a}{c} \right) z \,\, \Hypergeometric{2}{1}{1,a+1}{c+1}{z}.
\eeq
A derivation of this is given in \cref{sec:Hypergeometric}.

The $k_{ij}$ summations in \cref{eq:FourBodyExpansion} are all finite. If all 
powers of $r_{ij}$ are odd, the $q$ summation is infinite, and if any of
$r_{ij}$ is even, the $q$ summation becomes finite. For the finite $q$ sums, 
these direct sums are solved very accurately, but for the infinite sums, some 
integrals converge slowly. Particularly when all three $r_{ij}$ powers are 
odd and at least one is $-1$, the integrals converge the slowest. It is 
possible to restrict the basis set described by \cref{eq:OmegaDef} so that at 
least one of the $r_{ij}$ powers is even, making solving the integrals 
easier. However, this leads to slow convergence in the energy \cite{Drake1995}.

As Frolov and Bailey \cite{Frolov2003} note, these four-body integrals only 
work for systems with one infinitely heavy particle, such as PsH or Ps-H 
scattering. For systems with arbitrary masses, such as Ps$_2$ or Ps-Ps 
scattering, this has to be generalized to
\beq
\label{eq:FourBodyIntGen}
I = \int e^{-(\bar{\alpha} r_1 + \bar{\beta} r_2 + \bar{\gamma} r_3 + a_{12} r_{12} + a_{13} r_{13} + a_{23} r_{23})} r_1^{k_i} r_2^{l_i} r_{12}^{m_i} r_3^{n_i} r_{13}^{p_i} r_{23}^{q_i} d\textbf{r}_1 d\textbf{r}_2 d\textbf{r}_3.
\eeq
Fromm and Hill give an analytic solution to this \cite{Fromm1987}, but it is extremely difficult to work with. Harris more recently solved this problem analytically using recursion relations \cite{Harris2009}, using a similar method to the recursion relations of Pachucki et al.\ \cite{Pachucki2004}. Both of these solutions restrict the powers of $r_i$ and $r_{ij}$ to $k_i, l_i, m_i, n_i, p_i, q_i \geq -1$. We have also looked at a subset of these integrals.
As another extension of the Hylleraas basis set, some four-electron integrals can be reduced down to the three-electron integrals \cite{King1993,Pelzl2002}.

\subsubsection{Asymptotic Expansion}
\label{sec:AsymptoticExpansion}
For cases of \cref{eq:FourBodyExpansion} with all odd powers of $r_{ij}$, the $q$ summation is infinite, and the summation converges slowly. The convergence accelerator approach of Pelzl and King \cite{Pelzl1998,Pelzl2002} is one way to deal with this. I did limited testing with this approach but chose to instead use the asymptotic expansion method of Drake and Yan \cite{Drake1995}. In my testing, it was more numerically stable, and it has been generalized to arbitrary angular momenta \cite{Yan1997}.

As an example in Drake and Yan's paper \cite{Drake1995}, direct calculation of the $q$ summation for $I(0,0,0,-1,-1,-1; 1,1,1)$ only reaches an accuracy of $1.5 x 10^{-13}$ after 6860 terms. With their asymptotic expansion method, the integral has converged to approximately $2.2 x 10^{-16}$ after only 21 terms. The summation converges monotonically and asymptotically. Drake and Yan use this knowledge to speed up the convergence of the integration. Details of this method can be found in their paper \cite{Drake1995}.

I use this asymptotic expansion method in all calculations of the short--short integrals through the H-wave, and it performed very well. In fact, my quadruple precision code often calculates matrix elements to better than $1$ part in $10^{20}$ for the S-wave. \Cref{tab:AsympExpan} gives an example of the convergence of a single integral by only calculating the direct sum of \cref{eq:FourBodyExpansion} in the $S_d(N)$ column and with the asymptotic expansion in the $S_a(N)$ column. These four-body integrals are relatively quick to calculate, and most of the runs to compute them for $\ell \leq 2$ complete in a matter of hours on a typical desktop computer.

\begin{table}
\centering
\begin{tabular}{c c c c}
\toprule
$N$ & $S_d(N)$ & $\Delta S_d(N)$ & $S_a(N)$ \\
\midrule
17 & 684.106 432 091 &               & 684.113 411 842 629 912 374 349 \\
18 & 684.107 475 306 & 0.001 043 214 & 684.113 411 842 629 911 836 645 \\
19 & 684.108 320 637 & 0.000 845 331 & 684.113 411 842 629 911 829 661 \\
20 & 684.109 012 851 & 0.000 692 213 & 684.113 411 842 629 911 835 071 \\
21 & 684.109 585 093 & 0.000 572 241 & 684.113 411 842 629 911 836 095 \\
22 & 684.110 062 261 & 0.000 477 167 & 684.113 411 842 629 911 836 195 \\
23 & 684.110 463 302 & 0.000 401 041 & 684.113 411 842 629 911 836 186 \\
24 & 684.110 802 809 & 0.000 339 506 & 684.113 411 842 629 911 836 178 \\
25 & 684.111 092 142 & 0.000 289 333 & 684.113 411 842 629 911 836 174 \\
26 & 684.111 340 236 & 0.000 248 094 & 684.113 411 842 629 911 836 173 \\
27 & 684.111 554 183 & 0.000 213 946 & 684.113 411 842 629 911 836 172 \\
28 & 684.111 739 660 & 0.000 185 477 & 684.113 411 842 629 911 836 172 \\
29 & 684.111 901 249 & 0.000 161 588 & 684.113 411 842 629 911 836 172 \\
30 & 684.112 042 674 & 0.000 141 425 & 684.113 411 842 629 911 836 172 \\
\bottomrule
\end{tabular}
\caption[Convergence of direct sum against the asymptotic expansion]
{Convergence of direct sum, $S_d(N)$, against the 
asymptotic expansion, $S_a(N)$. This is an extension of Table I in Drake and 
Yan's work \cite{Drake1995} and uses $\Lambda = 15$. $\Delta S_d(N)$ gives 
the difference between successive direct sums.}
\label{tab:AsympExpan}
\end{table}

\subsubsection{Recursion Relations}
\label{sec:RecursionRelations}
After the S-wave calculations were completed, we learned of an analytic, 
instead of numerical, solution to these three-electron integrals, derived by 
Pachucki et al.\ \cite{Pachucki2004}. They
were not the first to derive an analytic solution to the three-electron 
integrals, but the first by Fromm and Hill \cite{Fromm1987} is not very 
practical to use, considering its very complicated form. Each set of $r_i$ 
and $r_{ij}$ powers requires a new rederivation from the Fromm and Hill 
result. The recursion relations from Pachucki et al.\ are complicated but also general.

Like many other types of recursion relations, these recursion relations may 
not be stable for higher $\omega$ values, depending on the calculated 
precision. I have tested through $\omega = 8$, and this method produced 
stable results under quadruple precision. In some of their work on Li and
Be$^+$, their group uses sextuple precision, as quadruple precision becomes 
insufficient near $\omega = 10$ \cite{Puchalski2006}, which is much higher 
than we can use in the Ps-H scattering calculations.

These are used as a check on the accuracy of the asymptotic expansion method in
\cref{sec:AsymptoticExpansion} for the S-wave and P-wave. The D-wave short-range
integrals cannot be evaluated using the recursion relations in
Ref.~\cite{Pachucki2004} due to the $r_i^{-2}$ terms that appear. Solving these
using the recursion relations would require implementing the extended method in
Ref.~\cite{Pachucki2005}. However, the asymptotic expansion method has proven
to be stable and accurate so far.

\subsubsection{W Functions}
\label{sec:WFunctions}

While evaluating the integrals in the PsH short-range code, we noticed that the
$W$ functions in \cref{eq:FourBodyExpansion} could be called more than once by
multiple integrals. For the S-wave, there are 34 terms in \cref{eq:GradGradShort},
and each matrix element in \textbf{H} in \cref{eq:BoundGenEig} requires these 34
integrals to be evaluated. The overall powers of $r_1$, $r_2$, $r_{12}$, etc.\ in
\cref{eq:FourBody} can be the same for a set of $\phi_i$ with $\phi_j$. Also for a
set of integrals without the overall powers the same, there is the possibility of
two different integrals calling the same $W$ function.

To speed up the partial wave specific programs for the short-short
S-, P-, and D-wave matrix elements significantly, we precompute the $W$ functions
and store the results in a look up table stored in a 4-dimensional matrix.
Even though there is a possibility for the $W$ function arguments to be
anywhere in this space, not all of the $W$ matrix elements will be used.

The current S-, P-, and D-wave short-range codes first do a dummy run
where the integrals are not actually calculated, but any $W$ matrix elements
that need to be computed are marked. Only these are computed, and then the code
runs through again using this $W$ matrix look up table. \Cref{tab:WFuncUnusedS}
shows that only $10.6\%$ of the $W$ matrix elements are actually needed for the
S-wave at a relatively high value of $\omega$. Similarly for the D-wave,
\cref{tab:WFuncUnusedD} shows that only $10.5\%$ are needed.

\begin{table}
\centering
\begin{tabular}{c c c c}
\toprule
$\omega$ & Used Terms & Total Terms & Percentage Used \\
\midrule
1 & 	115   &  20,000 & 	0.575\% \\
2 & 	921   &  27,040 & 	3.41\% \\
3 & 	1,939 &  34,992 & 	5.54\% \\
4 & 	3,140 &  43,904 & 	7.15\% \\
5 & 	4,573 &  53,824 &	8.49\% \\
6 & 	6,246 &  64,800 &	9.64\% \\
7 & 	8,147 &  76,880 &	10.6\% \\
\bottomrule
\end{tabular}
\caption{S-wave W function terms used}
\label{tab:WFuncUnusedS}
\end{table}

\begin{table}
\centering
\begin{tabular}{c c c c}
\toprule
$\omega$ & Used Terms & Total Terms & Percentage Used \\
\midrule
0 & 150		&	196,290 &	0.0764\% \\
1 & 1,670	&	259,932 &	0.642\% \\
2 & 14,076	&	332,262 &	4.24\% \\
3 & 23,384	&	413,520 &	5.65\% \\
4 & 36,418	&	503,946 &	7.23\% \\
5 & 51,420  &	603,780 &	8.52\% \\
6 & 68,438	&	713,262 &	9.60\% \\
7 & 87,520	&	832,632 &  10.5\% \\
\bottomrule
\end{tabular}
\caption{D-wave W function terms used}
\label{tab:WFuncUnusedD}
\end{table}
Another more sophisticated method that we have started using is
looking instead at the overall integrations instead of the $W$ function
arguments using prime factorization to avoid calculating integrals
more than once \cite{VanReethPrivate}.

\subsection{Linear Dependence in the Bound State Calculation}
With infinite precision in calculations, all terms from the basis set could 
be used. However, due to the limited precision inherent in computer 
calculations, when using large basis sets, near linear dependences will exist 
in the matrices. The goal is to identify and eliminate terms that exhibit 
near linear dependence with other terms. These terms are not exactly linearly 
dependent, if infinite precision was possible, but they are linearly 
dependent to computer precision.

To calculate the eigenvalues of the generalized eigenproblem, the LAPACK 
routine \texttt{dsygv} is used \cite{dsygv}. For the basis set consisting of 
terms from $\omega$ = 5, LAPACK computes the eigenvalues without errors. 
Adding in terms corresponding to $\omega = 6$ for some sets of nonlinear 
parameters causes \texttt{dsygv} to fail with an error set in the last 
parameter, info. This error is always an integer greater than the number of 
terms, indicating from the library documentation that ``the leading minor of 
order i of B is not positive definite'' \cite{dsygv}. L\"uchow and 
Kleindienst \cite{Luchow1993} also encountered a similar problem using
other libraries.

This error does suggest that one approach to identifying problematic terms is 
to check for positive definiteness of the overlap matrix
$\left\langle \phi_i | \phi_j \right\rangle$. This is one method that
Yan and Ho \cite{Yan1999} use to isolate 
problematic terms. They test the eigenvalues of the overlap 
matrix to see if any are small or negative, though there is no mention in 
their paper of what value of ``small'' is used. We attempted to use this fact 
to remove problematic terms, but too many terms were removed, leading to an 
energy that converged too slowly. In several papers by Yan and others
\cite{Yan1998,Yan1998a,Yan1999,Drake1995,Yan1997a}, terms with $j1 > j2$ are 
omitted if $l_1 = l_2$ and $\bar{\alpha} \approx \bar{\beta}$, along with $j_1 = j_2$ if
$j_{23} > j_{31}$. 

Another technique Yan and Ho \cite{Yan1999} used was to partition the basis set into five 
sectors, each with a different set of nonlinear parameters and maximum
$\omega$. The sectors also have restrictions on the interparticle
$r_{ij}$ terms, mainly limiting the power of $r_{23}$ and $r_{31}$, which are the 
electron-positron coordinates in their paper (corresponding to $r_{12}$ and
$r_{13}$ in our work). These techniques used for restricting 
the set of terms are not used in our work.

\subsection{Todd's Method}
\label{sec:ToddBound}

In trying to determine the energy eigenvalues, we noticed that the ordering 
of the terms could determine whether there was linear dependence in the 
matrices. Todd's method \cite{Todd2007,Armour2008} was attractive,
because it reorders the matrices to obtain the best possible energy, and it 
is a purely computational approach. We have not seen 
any physical reason why certain terms should introduce a near linear 
dependence. Todd's method is very similar to the algorithm in
Ref.~\cite{Luchow1992}
that we also considered, but it did not work as well. A description of his
algorithm as implemented in this work follows.

The total number of terms to look at is $N = N(\omega)$ (see \cref{eq:NumberTermsOmega}).
N matrices of size 1x1 are created for each term. This 
is done for the overlap and the
$\left\langle \phi_i \left| \,H \right| \phi_j \right\rangle$
matrices together. The LAPACK \texttt{dsygv} routine is used to 
determine the lowest eigenvalue for each of these N sets. These energy 
eigenvalues are compared against one another, and the term with the lowest 
energy is chosen. In the next step, the first basis function from the 
previous step is combined with each unused term to create N-1 matrices of 
size 2x2. Again, the energy eigenvalues for each of the N-1 matrices are 
compared against each other, and the term yielding the lowest energy is 
chosen as the second basis function. This is done again with 3x3 matrices for 
each of the N-2 remaining terms combined with the basis functions chosen in 
the first two steps. This procedure is repeated until all terms have been 
used or the remaining terms are problematic.

In his original algorithm, Todd looked at the eigenvalues computed from 
the upper and lower triangular matrices. Normally, the overlap and H matrices 
are symmetric, but this is not true to machine precision due to truncation 
and rounding. If the energy eigenvalues from the upper and lower triangles 
differ by more than $10^{-6}$ (in atomic units), the last added term is 
considered problematic and discarded.

In our testing for $\omega = 6$ for the S-wave, no terms were omitted due to the reordering. 
As noted earlier, before implementing this algorithm, LAPACK would fail when 
trying to calculate the eigenvalues, so the ordering is important for getting 
the best possible energy. For $\omega$ = 7, 116 terms were omitted, out of a 
total of 1716 terms. The criteria that the eigenvalues for the upper and 
lower matrices differs by no more than a certain amount was not needed in 
this case. The info parameter of the LAPACK \texttt{dsygv} function is checked for 
both the upper and lower matrix eigenvalue calculations, and the last added 
term is discarded if it causes an error due to linear dependence. When only 
the 116 problematic terms were left, every one of them caused LAPACK to 
error. If a term was problematic at any stage, it continued to be problematic 
in all further stages, so computation time can be decreased by immediately 
discarding it.

For larger basis sets, this algorithm becomes extremely slow, as determining 
the eigenvalues is an $O(N^3)$ operation. It can easily be parallelized, 
since we are computing the eigenvalues for a large number of matrices. Our 
program has been parallelized using OpenMP \cite{OpenMP} for intranode
communications and MPI \cite{MPI} for internode communications. Todd's
algorithm provides the best converged energy for a set of terms, albeit
at a cost of computational speed.

\begin{figure}
	\centering
	{\includegraphics[height=1.5in]{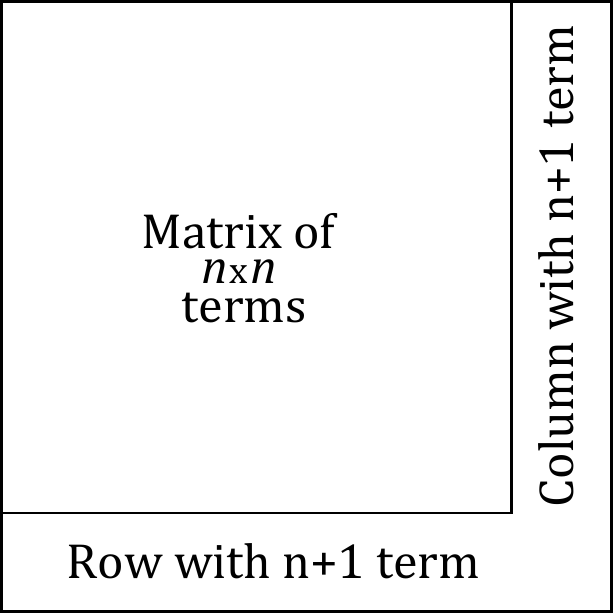}}
	\caption{Diagram of Todd's procedure}
	\label{fig:Todd}
\end{figure}

\subsection{Restricted Set}
\label{sec:Restricted}
Van Reeth and Humberston \cite{VanReeth2003} found that restricting the power 
of the $r_3$ coordinate could significantly improve their numerics, allowing 
them to use more short-range terms. Specifically, we restrict the $r_3$ power 
(see \cref{eq:PhiDef}) so that $n_i \leq 2$ if $\omega \geq 3$, and we refer 
to this as the restricted set.

We normally use Todd's method applied to the scattering problem,
described in \cref{sec:CompPhase}, or we 
use the full unrestricted set when possible. For the cases of $^1$F and $^3$F 
for low $\kappa$, described in \cref{sec:FNonlinear}, we use the restricted 
set. This allowed us to keep the convergence ratios less than 1 for these
cases.

\section{Long-Range Terms}
\label{sec:CompLong}

To minimize verbosity, the long-long and long-short integrations are 
collectively referred to as the long-range integrations. Each of the
long-range integrations are performed using Gaussian quadratures, which
are described fully in \cref{sec:GaussQuad}.

\subsection{Long-Range -- Long-Range}
\label{sec:LongLongInt}

The scattering program calculates only the short-long and long-long matrix 
elements. The volume element in \cref{eq:dTauS23} has an internal angle 
of $\varphi_{23}$ to integrate over. When a term has a negative power of
$r_{23}$ due to the Hamiltonian in
\cref{eq:Hamiltonian1,eq:Hamiltonian2,eq:Hamiltonian3}, a large
number of integration points must be used for reasonable 
accuracy. Instead, we split the integration such that one part is missing 
the $r_{23}^{-1}$ term and the other contains only the $r_{23}^{-1}$ term
\cite{VanReethThesis,VanReethPrivate}.

The first integration excluding the $r_{23}^{-1}$ term has negative powers of 
$r_i$ and $r_{ij}$ canceled by the corresponding terms in the volume element 
given by \cref{eq:dTauS23}. For the integration over the $r_{23}^{-1}$ term, 
we use an alternative volume element, namely that given by equation
\cref{eq:dTauS12}.
The $r_{23}^{-1}$ is then canceled by the $r_{23}$ in this volume element.

\subsubsection{Integration without the \texorpdfstring{$r_{23}^{-1}$} {1/r23} term}
\label{sec:LongLongNoR23}
The simplest long-long matrix element to evaluate is
$(\bar{S}_\ell,\mathcal{L} \bar{S}_\ell)$. From \cref{eq:SbarLSbar}, not
including its $r_{23}^{-1}$ term, this is
\beq
(\bar{S}_\ell,\mathcal{L} \bar{S}_\ell)_A = \pm \left(S_\ell^\prime,\mathcal{L} S_\ell\right)_A = \pm \left(S_\ell^\prime, \left[ \frac{2}{r_1} - \frac{2}{r_2} - \frac{2}{r_{13}}\right] S_\ell\right).
\eeq

For this type of integration, we use perimetric coordinates as described in
\cref{sec:PerimetricCoords}.
\begin{align}
\label{eq:SBarSBarInt}
(\bar{S}_\ell,\mathcal{L} \bar{S}_\ell)_A = \pm &2\pi^2 \int_0^\infty \int_0^\infty \int_0^\infty \int_0^\infty \int_{|r_1 - r_3|}^{|r_1 + r_3|} \int_0^{2\pi}  S_\ell^\prime S_\ell \left[ \frac{2}{r_1} - \frac{2}{r_2} - \frac{2}{r_{13}}\right] \\
&\times r_2 r_3 r_{12} r_{13}\, d\varphi_{23}\, dr_{13}\, dr_3\, dz\, dy\, dx
\end{align}

The $\varphi_{23}$ integration is done analytically. Since $S_\ell$ and
$S_\ell^\prime$ have no $r_{23}$ dependence and there is no $r_{23}$ term in
the brackets, the integration over $\varphi_{23}$ is simply $2\pi$. The
$r_{13}$ integration uses the Gauss-Laguerre quadrature from
\cref{sec:GaussQuad}. The $x$, $y$ and $z$ integrations use Gauss-Laguerre 
quadrature, since they are semi-infinite.

The $r_3$ integration could also be performed using just the Gauss-Laguerre 
quadrature. However, the integrand for the $r_3$ integration has a 
discontinuity in its slope at $r_3=r_1$, creating a cusp
(see \cref{sec:Cusps}), so the accuracy is improved greatly if we split the
integration interval into two parts and employ different quadratures for each.
The integration is split to use Gauss-Legendre on the interval $(0,r_1)$ and
Gauss-Laguerre on the interval $(r_1,\infty)$.

\subsubsection{Integration over the \texorpdfstring{$r_{23}^{-1}$} {1/r23} term}
\label{sec:LongLongR23}
The other part of the $(\bar{S}_\ell,\mathcal{L} \bar{S}_\ell)$ integral
contains the $r_{23}^{-1}$ term.

\beq
(\bar{S}_\ell,\mathcal{L} \bar{S}_\ell)_B = \pm \left(S_\ell^\prime, \left[ \frac{2}{r_{23}}\right] S_\ell\right)
\eeq
The volume element for this integral is $d\tau$ from \cref{eq:dTauS12}.
The integration also does not need to be converted to perimetric coordinates,
so its form is
\begin{align}
(\bar{S}_\ell,\mathcal{L} \bar{S}_\ell)_B = \pm 8\pi^2 \int_0^\infty & \int_0^\infty \int_0^\infty \int_{|r_1 - r_3|}^{|r_1 + r_3|} \int_{|r_2 - r_3|}^{|r_2 + r_3|} \int_0^{2\pi}  S_\ell^\prime S_\ell \frac{2}{r_{23}} r_1 r_2 r_{13} r_{23} \nonumber \\
& \times d\varphi_{12}\, dr_{23}\, dr_{13}\, dr_2\, dr_3\, dr_1.
\end{align}

The $r_{13}$ and $r_{23}$ integrals have finite limits, so here we use Gauss-
Legendre quadrature. Again, for the internal angular integration, this time 
over $\varphi_{12}$, we use Chebyshev-Gauss quadrature. The cusp in the $r_3$ 
integration is at $r_3 = r_1$, and the cusp in the $r_2$ integration is at
$r_2 = r_3$. Similar to before, we split up these integrations by using Gauss-
Legendre before the cusp and Gauss-Laguerre after the cusp.

The $(\bar{C}_\ell,\mathcal{L} \bar{S}_\ell)$ and
$(\bar{C}_\ell,\mathcal{L} \bar{C}_\ell)$ terms
are integrated in the same manner as the
$(\bar{S}_\ell,\mathcal{L} \bar{S}_\ell)$ integral just described. With all
four matrix elements calculated, as mentioned on \pageref{GenSLCandCLS}, if
the difference of $(\bar{S}_\ell,\mathcal{L} \bar{C}_\ell)$ and
$(\bar{C}_\ell,\mathcal{L} \bar{S}_\ell)$ is close to 1, we can have reasonable
confidence in the long-long integrations. Another check is that if we
explicitly calculate $(S_\ell,\mathcal{L} S_\ell)$, we should get 0
(\cref{eq:SLS0Test}). We find that the choice of innermost integration,
over $\varphi_{12}$ or $\varphi_{13}$, can cause this to be nonzero, so this
gives us another check on the accuracy of the long-long integrations.

\subsection{Short-Range -- Long-Range}
\label{sec:ShortLongInt}
We will consider only the integrations for $(\bar{\phi}_i,\mathcal{L} \bar{S}_\ell)$ 
here, as the integrations for $(\bar{\phi}_i,\mathcal{L} \bar{C}_\ell)$
are evaluated in the same manner. As in the case of long-long 
integrations in \cref{sec:LongLongInt}, we split up the integration 
into two parts -- one containing the $r_{23}^{-1}$ term and another 
containing the rest of the terms. The short-range terms have the added 
benefit of the possibility of the polynomial $r_{23}^{\,q_i}$ being present, 
which cancels the $r_{23}^{-1}$ term or gives it an overall positive power.

From \cref{eq:PhiBarLSBar2b,eq:LSFinal,eq:LSPrimeFinal},
\begin{align}
\label{eq:PhiLSBarInt}
\nonumber (\bar{\phi}_i, \mathcal{L} \bar{S}_\ell) &= \frac{2}{\sqrt{2}} \left(\phi_i,\mathcal{L} \bar{S}_\ell\right) \\
 &= \frac{2}{\sqrt{2}} \int \phi_i \left[ \left( \frac{2}{r_1} - \frac{2}{r_2} - \frac{2}{r_{13}} + \frac{2}{r_{23}} \right)S_\ell \pm \left( \frac{2}{r_1} - \frac{2}{r_3} - \frac{2}{r_{12}} + \frac{2}{r_{23}} \right) S_\ell^\prime \right]  d\tau.
\end{align}

\subsubsection{Case I: \texorpdfstring{$q_i > 0$}{qi > 0}}
\label{sec:Swaveqigt0}
When $q_i > 0$ in $\phi_i$ (\cref{eq:PhiDef}), the power of $r_{23}$ is
equal to or greater than 0. Gaussian quadratures can safely integrate this type
of term, so we integrate the full expression in \cref{eq:PhiLSBarInt}.
\begin{align}
\label{eq:PhiLSBarIntFull}
\nonumber (\bar{\phi}_i, \mathcal{L} \bar{S}_\ell) =& \, \frac{2}{\sqrt{2}} \cdot 8\pi^2  \int_0^\infty \int_0^\infty \int_0^\infty \int_{|r_1 - r_2|}^{|r_1 + r_2|} \int_{|r_1 - r_3|}^{|r_1 + r_3|} \int_0^{2\pi} \phi_i \\
&\times \left[ \left( \frac{2}{r_1} - \frac{2}{r_2} - \frac{2}{r_{13}} + \frac{2}{r_{23}} \right)S_\ell \pm \left( \frac{2}{r_1} - \frac{2}{r_3} - \frac{2}{r_{12}} + \frac{2}{r_{23}} \right) S_\ell^\prime \right] \\
&\times r_2 r_3 r_{12} r_{13}\, d\varphi_{23}\, dr_{13}\, dr_{12}\, dr_3\, dr_2\, dr_1
\end{align}

Similar to the long-long integrations from \cref{sec:LongLongInt}, the 
$r_1$ integration is performed using the Gauss-Laguerre quadrature. The $r_2$
integral is broken into two parts at the cusp of $r_2 = r_1$, with the
Gauss-Legendre quadrature before the cusp and the Gauss-Laguerre quadrature 
after the cusp. In the $r_3$ coordinate, there is a cusp at $r_3 = r_2$, so 
the integration is also split up into Gauss-Legendre before the cusp and
Gauss-Laguerre after the cusp. The finite intervals for $r_{12}$ and $r_{13}$ 
ensure that we can use Gauss-Legendre quadratures for these coordinates. The 
$\varphi_{23}$ integration uses the Chebyshev-Gauss quadrature.

\subsubsection{Case II: \texorpdfstring{$q_i = 0$}{qi = 0}}
When $q_i = 0$, the overall power of the $r_{23}$ term is $-1$, so we 
cannot use the Gaussian quadratures in the form of \cref{eq:PhiLSBarIntFull}.  
Similar to the long-long integrations, the $r_{23}^{-1}$ term is integrated 
separately, using the same type of integrations as \cref{eq:PhiLSBarIntFull}.
Refer to the previous section for the description of the quadratures used.
\begin{align}
\label{eq:PhiLSBarIntNoR23}
\nonumber (\bar{\phi}_i, \mathcal{L} \bar{S}_\ell)_A =& \,\frac{2}{\sqrt{2}} \bigintsss \phi_i \left[ \left( \frac{2}{r_1} - \frac{2}{r_2} - \frac{2}{r_{13}} \right)S_\ell \pm \left( \frac{2}{r_1} - \frac{2}{r_3} - \frac{2}{r_{12}} \right) S_\ell^\prime \right]  d\tau \\
\nonumber =&\, \frac{2}{\sqrt{2}} \cdot 8\pi^2  \int_0^\infty \int_0^\infty \int_0^\infty \int_{|r_1 - r_2|}^{|r_1 + r_2|} \int_{|r_1 - r_3|}^{|r_1 + r_3|} \int_0^{2\pi} \phi_i r_2 r_3 r_{12} r_{13} \\
&\times \left[ \left( \frac{2}{r_1} - \frac{2}{r_2} - \frac{2}{r_{13}} \right)S_\ell \pm \left( \frac{2}{r_1} - \frac{2}{r_3} - \frac{2}{r_{12}} \right) S_\ell^\prime \right]  d\varphi_{23}\, dr_{13}\, dr_{12}\, dr_3\, dr_2\, dr_1
\end{align}

The integration over the $r_{23}^{-1}$ term is done the same way as the 
second integration of the long-long matrix elements in \cref{sec:LongLongInt}.
The $r_{23}$ in the $d\tau$ volume element cancels the $r_{23}^{-1}$ 
term. Refer to \cref{sec:LongLongR23} for a description of 
the quadratures used here.
\begin{align}
\label{eq:PhiLSBarIntR23}
(\bar{\phi}_i, \mathcal{L} \bar{S}_\ell)_B =& \,\frac{2}{\sqrt{2}} \bigintsss \phi_i \left[ \frac{2}{r_{23}}\left(S_\ell \pm S_\ell^\prime\right) \right] d\tau^\prime  \nonumber \\
=&\, \frac{2}{\sqrt{2}} \cdot 8\pi^2  \int_0^\infty \int_0^\infty \int_0^\infty \int_{|r_1 - r_3|}^{|r_1 + r_3|} \int_{|r_2 - r_3|}^{|r_2 + r_3|} \int_0^{2\pi} \phi_i \left[ \frac{2}{r_{23}}\left(S_\ell \pm S_\ell^\prime\right) \right]  \nonumber \\
&\times  r_1 r_2 r_{13} r_{23}\, d\varphi_{12}\, dr_{23}\, dr_{13}\, dr_2\, dr_3\, dr_1
\end{align}

For a discussion on the final quadrature points used, refer to \cref{sec:QuadraturePoints}.

\subsection{Cusp Behavior}
\label{sec:Cusps}

The short-long and long-long matrix element integrals have cusps in the
integrands that must be dealt with \cite{VanReethThesis,Armour1991}.
As an example, consider$(\bar{\phi}_i, L\bar{C}_0)$ from the S-wave. Using the
notation of \cref{sec:LongLongR23}, where we are only computing the
$r_{23}^{-1}$ terms (given fully later in \cref{eq:LCBar} on \pageref{eq:LCBar}),
\begin{align}
(\bar{\phi}_i, L\bar{C}_0)_B = 8\pi^2 & \int_0^\infty \int_0^\infty \int_0^\infty \int_{|r_1 - r_2|}^{|r_1 + r_2|} \int_{|r_2 - r_3|}^{|r_2 + r_3|} \int_0^{2\pi} \bar{\phi}_i (\mathcal{L} \bar{C})_B  \nonumber \\
& \times r_1 r_3 r_{13} r_{23}\, d\varphi_{13}\, dr_{23}\, dr_{12}\, dr_3\, dr_2\, dr_1.
\end{align}
Due to the integration limits of the $r_{13}$ and $r_{23}$ integrations, the
$r_2$ integrand has a cusp at $r_2 = r_1$, and the $r_3$ integrand has a cusp
at $r_3 = r_2$.

\begin{figure}
	\centering
	\includegraphics[width=5in]{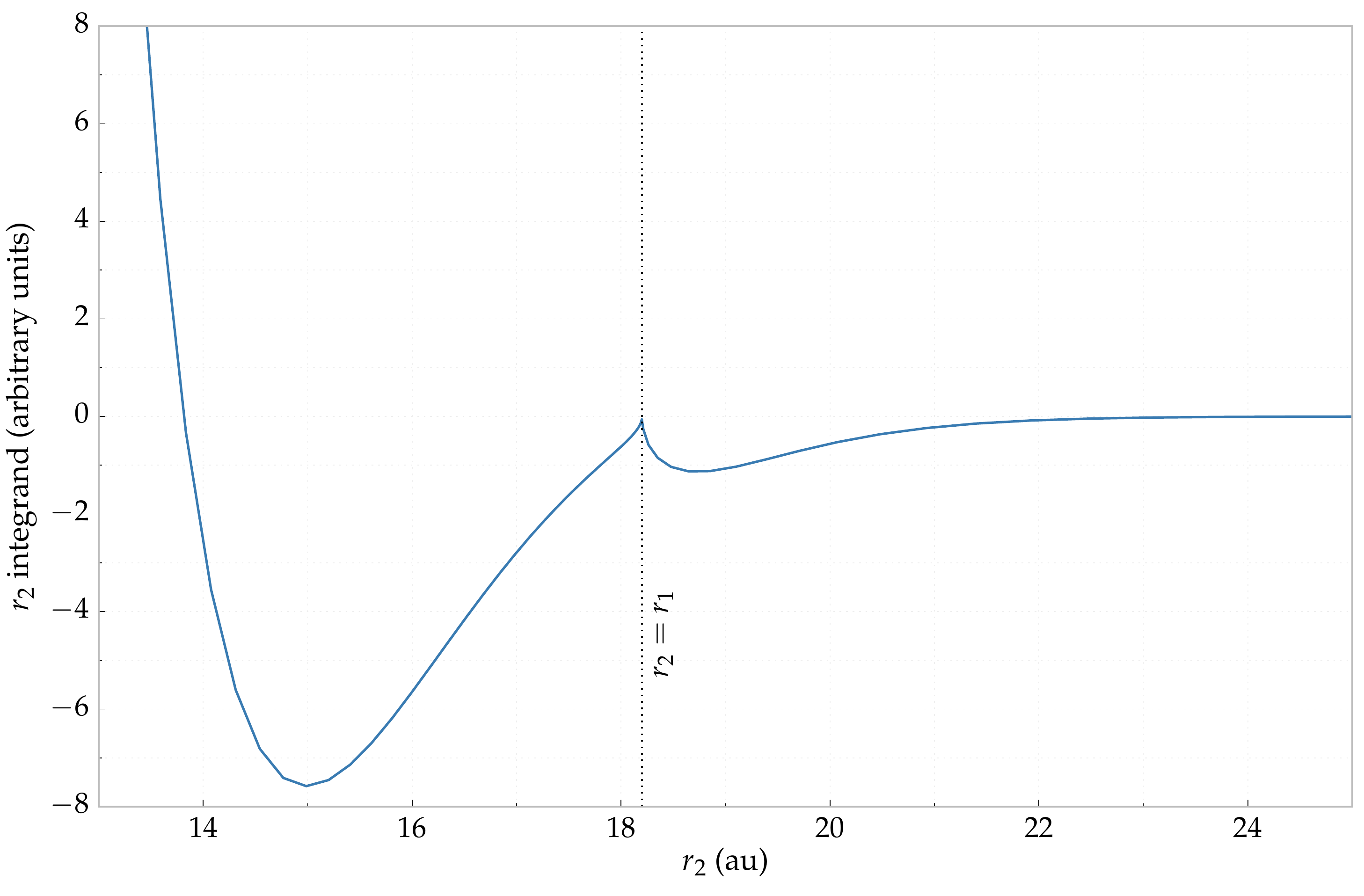}
	\caption[Example of cusp in S-wave short-long integration]{Example of cusp
in S-wave short-long integration for $1/r_{23}$ term of
$(\bar{\phi}_3, \mathcal{L}\bar{C}_0)$ with \mbox{$r_1 = 18.201$}}
	\label{fig:cusp}
\end{figure}

\Cref{fig:cusp} shows one such example of this cusp behavior for the $r_2$ 
integrand. All inner integrations ($\varphi_{13}$, $r_{23}$, $r_{12}$, and $r_3$)
are performed with a full set of integration points as described in
\cref{tab:OptimalEffectiveCoords}.

The $r_2$ integration could be performed using just the Gauss-Laguerre 
quadrature, since we are integrating over $[0,\infty)$. However, the cusp 
makes this integration slowly convergent. We instead split the integration 
interval into two parts and employ different quadratures for each. The 
integration is split to use Gauss-Legendre on the 
interval $[0,r_1]$ and Gauss-Laguerre (\cref{sec:GaussQuad}) on the interval 
$[r_1,\infty)$. Likewise, the $r_3$ integration is split into the intervals
$[0,r_2]$ and $[r_2,\infty)$.

Doing this splitting requires many more function evaluations, so we use an 
approximation whenever $r_1$ is large enough. Specifically when $r_1$ is 
greater than a chosen distance, we use Gauss-Laguerre over the entire
$[0,\infty)$ range. 
In the example given in \cref{fig:cusp}, at $r_2 = 100$, the $r_2$ integrand 
is approximately $10^{-57}$, while at $r_2 = 25$, the integrand is 
approximately $10^{-9}$.

For all partial waves, runs were performed with the cusp parameters set at
$r_1 = 100$. When $r_1 > 100$, the $r_2$ and $r_3$ integrations are done using only 
Gauss-Laguerre, since the cusp is considered unimportant at that distance. 
When $r_1 \leq 100$, we use Gauss-Legendre before the cusp and Gauss-Laguerre 
after the cusp, as described in \cref{sec:LongLongNoR23,sec:LongLongR23}.

\subsection{Extra Exponential}
\label{sec:ExtraExp}
To further improve the convergence of the short-long matrix elements in 
equation \cref{eq:GeneralKohnMatrix}, we investigated the integrands
\cite{VanReethPrivate}. The biggest source of 
difficulty in converging these results comes through the Gauss-Laguerre 
quadratures in the $r_1$, $r_2$ and $r_3$ integrations. Specifically, the 
region near the origin is not adequately represented. The integrands fall off 
quickly due to the exponential falloffs in $r_1$, $r_2$ and $r_3$, so it is 
not as important to have abscissae far away from the origin.
We are using approximately 7 times as many integration points total as the
earlier Kohn and inverse Kohn work \cite{VanReeth2003,VanReeth2004}
(see \cref{sec:QuadraturePoints}), but this brute force approach of adding 
quadrature points can increase the computational time greatly. We took 
another approach to further increase the accuracy. For each of the Gauss-
Laguerre quadratures, we introduce an extra $e^{-\lambda r_i}$ and remove it 
with $e^{\lambda r_i}$ after the quadrature, bringing the abscissae closer to 
the origin without increasing the number of integration points.

The basic form of the Gauss-Laguerre quadrature is given in \cref{eq:GaussLag} as
\beq
\int_0^\infty e^{-x} f(x) dx \approx \sum_{i=1}^n w_i f(x_i).
\eeq
Introducing an extra $e^{-\lambda x}$ and removing it with $e^{\lambda x}$, we can bring the abscissae closer in by
\beq
\label{eq:GaussLagLambda}
\int_0^\infty e^{-x} f(x) dx \approx \sum_{i=1}^n \frac{w_i}{\lambda} f\! \left(\frac{x_i}{\lambda}\right) \ee^{\lambda x_i}.
\eeq

As an example, the full expression for the $r_2$ Gauss-Laguerre quadrature is given by
\beq
\int_a^\infty e^{-\beta r_2} f(r_2) dr_2 \approx \frac{e^{-\beta a}}{\beta} \sum_{i=1}^n w_i f\left(\frac{y_i}{\beta}+a\right),
\eeq
where $y_i = \beta x_i$, and $a$ is an arbitrary lower limit starting at the
cusp described in \cref{sec:Cusps}. With the exponential in $\lambda$, this becomes
\beq
\int_a^\infty e^{-\beta r_2} f(r_2) dr_2 \approx \frac{e^{-a (\beta + \lambda)}}{\beta + \lambda} \sum_{i=1}^n w_i f\left(\frac{r_2 + a(\beta + \lambda)}{\beta + \lambda} \right) \ee^{\lambda r_2}.
\eeq

\begin{figure}
	\centering
	\includegraphics[width=5in]{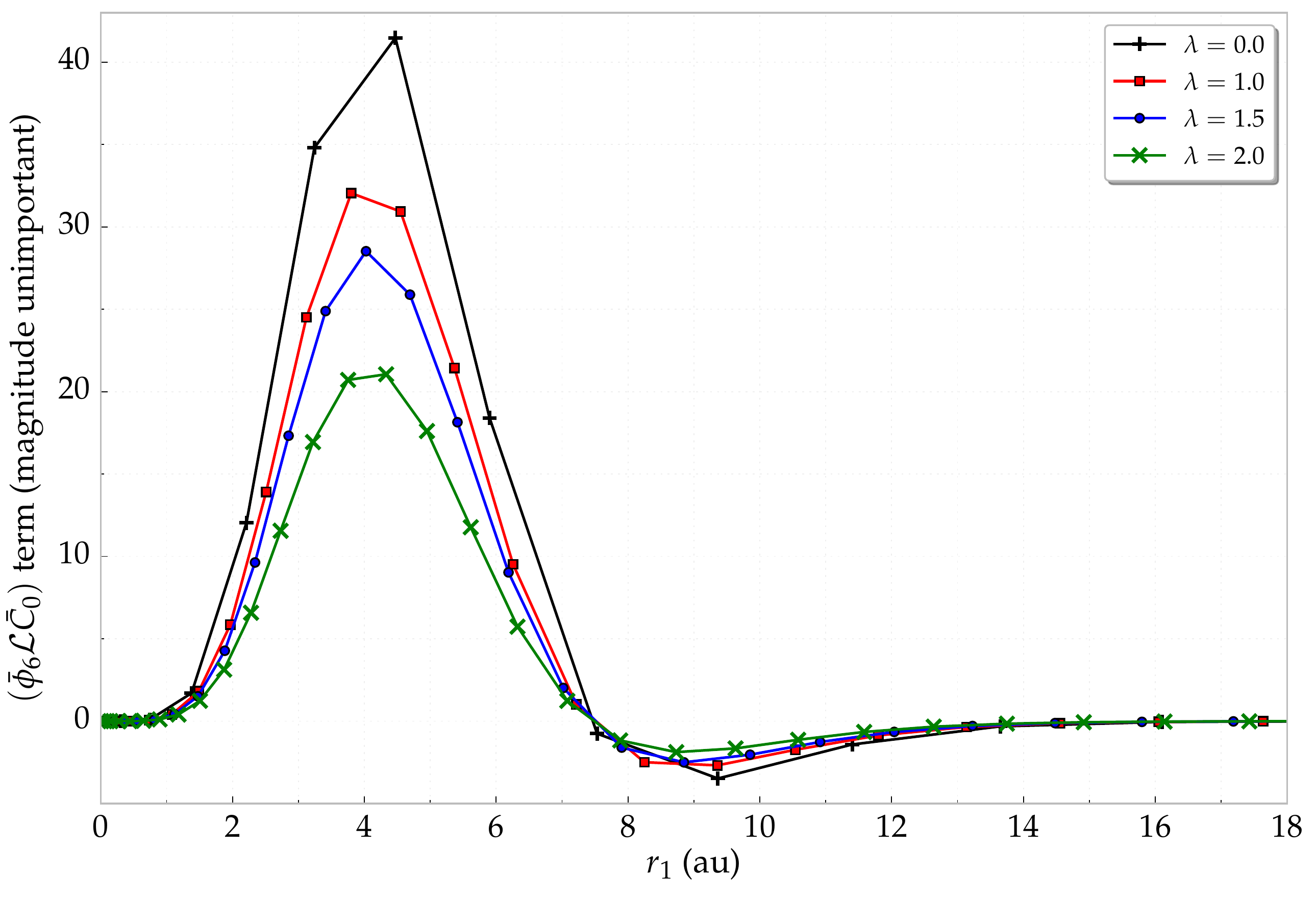}
	\caption{Effect of introducing $\lambda$ in $r_1$ exponential}
	\label{fig:lambda}
\end{figure}

\Cref{fig:lambda} shows the effect of introducing this $\lambda$ into the $r_1$
exponential. We use 50 quadrature points for each curve, meaning that there 
are points represented past the cutoff of 18 au in the graph. For the black 
curve with $\lambda = 0$, the curve is very jagged around the peak of 
approximately 4 au. With $\lambda = 1.0$, the curve is much smoother, and as
$\lambda$ is increased to 2.0, the peak is represented even better. If $\lambda$
is too large, the area near the origin may be overrepresented and the area 
farther out may be underrepresented. We have chosen $\lambda = 1.0$ for all 
of our runs for the $r_1$, $r_2$ and $r_3$ coordinates, which gives better
representation near the origin but does not  run the risk of neglecting the
small contribution of the curve for large $r_i$ values. Our $D$-wave and
general long-range codes could use different $\lambda$ for each of the
$r_1$, $r_2$ and $r_3$ integrations, but we set them equal here. The matrix
elements converge better for $\lambda = 1.0$ than for $\lambda = 0$.

\section{Phase Shifts}
\label{sec:CompPhase}

\begin{figure}
	\centering
	\includegraphics[width=5in]{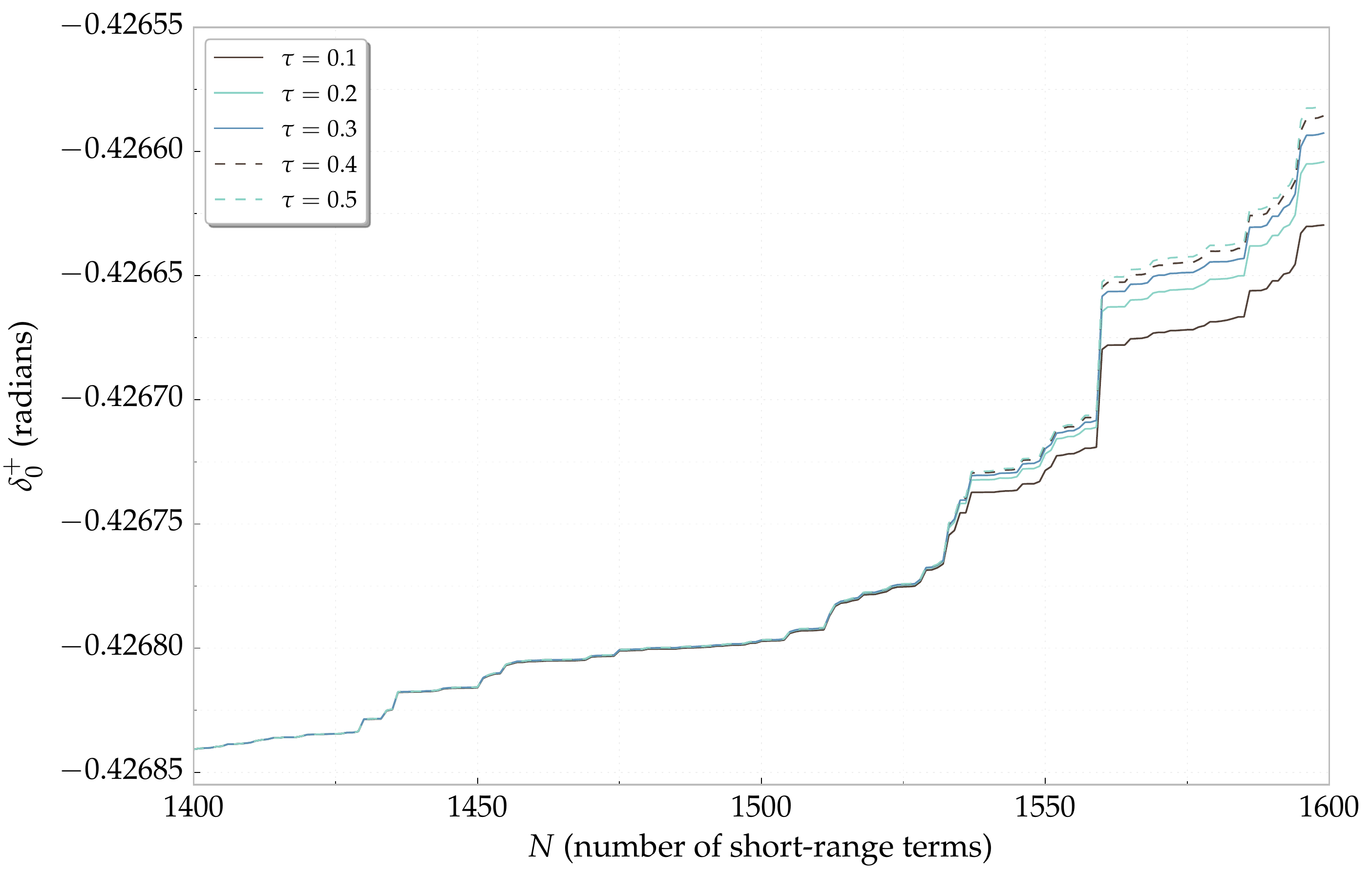}
	\caption[Breakdown in convergence of the $^1S$ phase shifts]
{Breakdown in convergence of the $^1S$ phase
shifts with respect to number of short-range terms for different $\tau$
values for the generalized Kohn variational method}
	\label{fig:swave-phase-divergence}
\end{figure}

We solve for the phase shifts by solving \cref{eq:GeneralKohnMatrix} for
$\boldsymbol{X}$ and then using \cref{eq:GenDFDT2,eq:GenKohnL}. Note that
as mentioned in \cref{sec:Kohn}, we only have to calculate one set of integrals
for the Kohn variational method, and the other Kohn-type methods are
used by rearrangement of these integrals in the matrix equation. The different
Kohn-type variational methods typically agree well when linear dependence in the
matrix equation does not occur. We use this fact to determine how many
short-range terms we are able to use in our final calculations. If we plot the
$^1$S phase shifts with respect to the number of terms, as in \cref{fig:swave-phase-divergence},
it is clear that around 1530 terms, the phase shifts from the different
generalized Kohn methods begin to diverge. In this particular example, we
chose the cutoff as a more conservative 1505 terms--before the ``jump'' that 
precedes the clear divergence. No Schwartz singularities (\cref{eq:SchwartzSing})
are evident here, but if any are present, we discard the generalized Kohn
variational methods that contain spurious singularities.

Using the short-range terms chosen by Todd's method (\cref{sec:ToddBound})
normally allows us to increase the number of terms used over using the terms
chosen just by \cref{eq:OmegaDef} for the Ps-H scattering problem.
Since Todd's method chooses an ordering with the most ``important'' terms 
at the beginning, we obtain well-converged phase shifts using this method with
the above-mentioned cutoffs for the first two partial waves (which do not 
have omitted symmetries). So we perform one truncation of the basis set by
using Todd's method for just the short-range terms and then an additional 
truncation by using plots as in \cref{fig:swave-phase-divergence} to reduce
linear dependence.

The number of terms used for each partial wave are given in \cref{sec:NonlinParam},
denoted as $N'(\omega)$. For $\ell \geq 3$, we either use the full set of
terms described by \cref{eq:OmegaDef} or the restricted set described in
\cref{sec:Restricted}.

The phase shifts presented in this work are found using the $S$-matrix 
complex Kohn, unless otherwise indicated. We normally only use the Kohn, 
inverse Kohn, and generalized Kohn variational method results to determine 
the number of short-range terms to use as described in this section. Then the 
phase shifts for this set of terms with the $S$-matrix complex Kohn method 
are determined.

\section{Convergence and Extrapolations}
\label{sec:Extrapolations}

For comparing quantities such as the convergence of the matrix elements or 
the convergence of the differential cross sections, we use the percent
difference:
\begin{equation}
\label{eq:PercentDiff}
\% \text{ Diff} = \left| \frac{a - b}{(a + b) / 2} \right| \times 100\%,
\end{equation}
where $a$ and $b$ can represent any of these quantities.
When we can compare convergence with respect to $\omega$, we define a
convergence ratio as
\begin{equation}
\label{eq:ConvRatio}
R'(\omega) = \frac{\delta_\ell^\pm(\omega)-\delta_\ell^\pm(\omega-1)}
  {\delta_\ell^\pm(\omega-1)-\delta_\ell^\pm(\omega-2)}.
\end{equation}
This is similar to the inverse of the ratio for the energy eigenvalues given in
Ref.~\cite{Yan1999}. If $R'(\omega) \geq 1$, there is no convergence
pattern. A ratio of less than 1 shows convergence but does not guarantee a
reliable extrapolation. We find that $R'(\omega) \lesssim 0.5$ is
needed to properly extrapolate phase shifts. We also find that when
$\delta_\ell \lesssim 10^{-4}$, the convergence often becomes poor
(see \cref{sec:DWavePhase}).

The tangent of the phase shifts is fitted to the function
\beq
\label{eq:PhaseExtrap}
\tan \delta_\ell^\pm(\omega) = \tan \delta_\ell^\pm(\omega \to \infty) + \frac{c}{\omega^p}.
\eeq
The $c$ and $p$ in this equation are fitting parameters and depend on each extrapolation. When plotted with respect to $\omega^{-p}$, the tangents of the phase shifts form nearly a straight line, with the y-intercept being the tangent of the extrapolated value of the phase shift, $\tan \delta^\pm(\omega \to \infty)$.

Van Reeth and Humberston \cite{VanReeth2003} do the extrapolation using $\omega = 3$ 
through $\omega = 6$.
We have completed the extrapolation using the sets $\omega = 3-6$,
$\omega = 3-7$ and $\omega = 4-7$. The smallest residuals are normally found
with the set $\omega = 4-7$. The values of the extrapolated phase shifts using
this method are in \cref{tab:SWavePhase}.

\begin{figure}
	\centering
	\includegraphics[width=\textwidth]{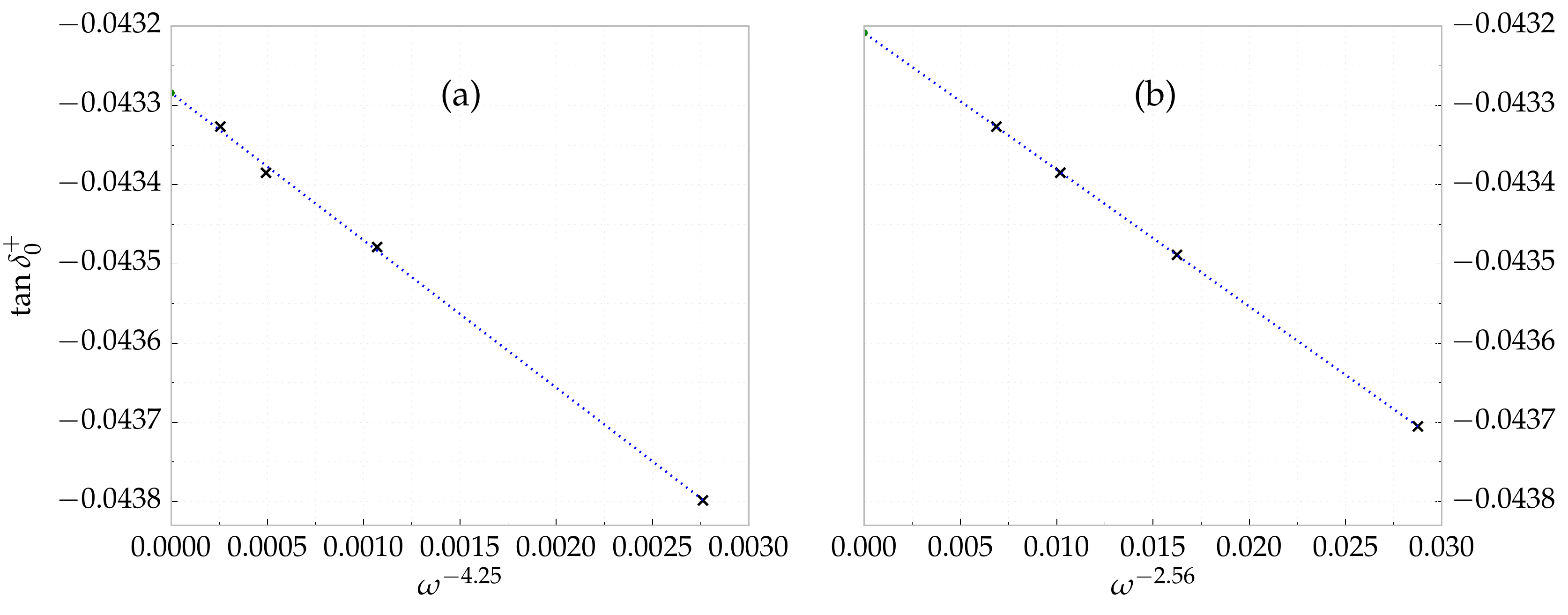}
	\caption[Extrapolation for $^1$S at $\kappa = 0.01$]{Extrapolation for $^1$S at $\kappa = 0.01$. The extrapolation without resorting the short-range terms into increasing $\omega$ is given in (a), and the resorted version is in (b).}
	\label{fig:extrap-phase}
\end{figure}

As described in \cref{sec:ToddBound}, we omit certain terms using 
Todd's method. The output of this method from the bound state program 
described in \cref{sec:ToddBound} is not ordered in terms of 
increasing $\omega$. This leads to difficulties when attempting to do the 
extrapolation in \cref{eq:PhaseExtrap}. The tangent of the phase 
shifts cannot be fitted to a straight line in this order, as seen in
\cref{fig:extrap-phase}(a). If we reorder the 
short-range terms back into their original order while still omitting terms, 
the tangent of the phase shifts can now be fitted to this straight line, as 
can be seen in \cref{fig:extrap-phase}(b). The reordering does not affect 
the phase shifts in any case that we tested, since more terms are normally 
omitted in the scattering problem than just the bound state problem.

We also extrapolate scattering lengths as shown in \cref{eq:ScatLenExtrap}.
There is no clear convergence pattern for the effective ranges, which are
described in \cref{sec:ScatteringLength}.


\chapter{S-Wave}
\label{chp:SWave}

\iftoggle{UNT}{The}{\lettrine[lines=2, lhang=0.33, loversize=0.1]{\textcolor{startcolor}{T}}{he}}
S-wave is the simplest of the 
partial waves to consider, since the spherical harmonic
$\SphericalHarmonicY{0}{0}{\theta_\rho}{\varphi_\rho}$ is constant. The form
of the Kohn matrix elements was derived in \cref{chp:WaveKohn} for general
$\ell$, but when we started this project, we went through the S-wave derivation
first and wrote code to compute this partial wave only. This chapter shows the
matrix elements specifically for the S-wave, which can be shown from the
general results in \cref{chp:WaveKohn}. The code described in
\cref{chp:General} works for any partial wave but is slower than the S-wave
specific code (\cref{chp:Programs}).

Van Reeth and Humberston \cite{VanReeth2003} previously performed $^{1,3}$S-wave
Ps-H calculations using the Kohn and inverse Kohn. We extend their work here
and used Van Reeth's notes and code \cite{VanReethPrivate} for guidance, 
though we rederived everything, and I wrote my own codes.

\section{Wavefunction}
\label{sec:SWaveFn}

The general wavefunction for any $\ell$ is given in \cref{eq:GeneralWaveTrial}, and the S-wave version is
\beq
\Psi_0^{\pm,t} = \widetilde{S}_0 + L_0^{\pm,t} \, \widetilde{C}_0 + \sum_{i=1}^{N'(\omega)} c_i \bar{\phi}_i.
\label{eq:SWaveTrial}
\eeq
The S-wave has only one set of short-range terms. $\widetilde{S}_0$,
$\widetilde{C}_0$, and $\bar{\phi}_i$ are given by
\cref{eq:TildeSCDef,eq:PhiDef}. The forms of the short-range terms in
\cref{eq:PhiDef} are equivalent for the S-wave ($\ell = 0$).
There is one notable difference about what we did here for the S-wave versus the
general code, namely that we absorb the spherical harmonic on the short-range
terms into the $c_i$ coefficients, similar to what we did with the $\frac{1}{\sqrt{2}}$.

\section{Matrix Elements with \texorpdfstring{$\mathcal{L}\bar{C}_0$}{LC}}
\label{sec:LCElements}

These can be calculated using the results in \cref{sec:LCTerms}, but the S-wave
code was written using these derivations.
The analysis for $\mathcal{L}C_0$ is more difficult than that of
$\mathcal{L}S_0$ in \cref{eq:LSFinal} and \cref{eq:LSPrimeFinal}.
The shielding factor, $f_0(\rho)$, complicates the derivatives slightly.
\begin{align}
\mathcal{L}C_0 = & \left(-\frac{1}{2}\Laplacian_{\bm{\rho}} - \Laplacian_{\mathbf{r}_3} - 2\Laplacian_{\mathbf{r}_{12}} + \frac{2}{r_1} - \frac{2}{r_2} - \frac{2}{r_3} - \frac{2}{r_{12}} - \frac{2}{r_{13}} + \frac{2}{r_{23}} - 2 E_H - 2 E_{Ps} - \frac{1}{2}\kappa^2 \right) \nonumber \\
 & \times \Phi_{Ps}(r_{12}) \Phi_H(r_3) \frac{\cos(\kappa\rho)}{\kappa\rho} \sqrt{\frac{2\kappa}{4\pi}} \left[1 + e^{-\mu\rho} \left(1 + \frac{\mu}{2} \rho \right) \right]
\label{eq:LC1S}
\end{align}
Again, we use \cref{eq:HEqn,eq:PsEqn}) to simplify this expression.
\begin{align}
\mathcal{L}C_0 = & \left(-\frac{1}{2}\Laplacian_{\bm{\rho}} + \frac{2}{r_1} - \frac{2}{r_2} - \frac{2}{r_{13}} + \frac{2}{r_{23}}  - \frac{1}{2}\kappa^2\right) \nonumber \\
 & \times \Phi_{Ps}(r_{12}) \Phi_H(r_3) \frac{\cos(\kappa\rho)}{\kappa\rho} \sqrt{\frac{2\kappa}{4\pi}} \left[1 + e^{-\mu\rho} \left(1 + \frac{\mu}{2} \rho \right) \right]
\label{eq:LC2S}
\end{align}

\label{pg:SWaveLC}Similar to \cref{sec:SphBess2}, $n_0(\kr)$ is an eigenfunction of $\Laplacian_{\bm{\rho}}$ with eigenvalue $-\kappa^2$.
To properly take into account the shielding function, we use the code in
\cref{fig:LCMath} for the S-wave, yielding
\beq
\label{eq:LCMathS}
\frac{1}{2} \left(\Laplacian_{\bm{\rho}} + \kappa^2\right) \SphericalHarmonicY{0}{0}{\theta_\rho}{\varphi_\rho} n_0(\kappa\rho) f_0(\rho) = \frac{2 \kappa  f^\prime(\rho ) \sin (\kappa \rho )-f^{\prime\prime}(\rho) \cos (\kappa \rho )}{2 \kappa \rho}.
\eeq
The $f^\prime(\rho)$ and $f^{\prime\prime}(\rho)$ are given in \cref{eq:Shielding1Der}.
Combining \cref{eq:LCMathS,eq:LC2S} and the permuted versions gives
\begin{align}
\mathcal{L}\bar{C}_0 = \:\: &\frac{1}{\sqrt{2}} \mathcal{L}(C_0 \pm C_0^\prime) = \frac{1}{\sqrt{2}} (\mathcal{L}C_0 \pm \mathcal{L}C_0^\prime) \nonumber \\
= \:\: &\frac{1}{\sqrt{8\pi}} \Phi_{Ps}(r_{12}) \Phi_H(r_3) \sqrt{2\kappa} \nonumber  \\
&\times \left\{ \frac{\kappa\mu}{2} e^{-\mr} (1+\mr) \frac{\sin(\kr)}{\kr} + \frac{\mu^3 \rho}{4} e^{-\mr} \frac{\cos(\kr)}{\kr} \right. \nonumber \\
&+ \left. \left(\frac{2}{r_1} - \frac{2}{r_2} - \frac{2}{r_{13}} + \frac{2}{r_{23}}\right) \frac{\cos(\kr)}{\kr} \left[1 - e^{-\mr} \left(1 + \frac{\mu}{2}\rho\right)\right] \right\} \nonumber \\
\pm &\frac{1}{\sqrt{8\pi}} \Phi_{Ps}(r_{13}) \Phi_H(r_2) \sqrt{2\kappa} \nonumber  \\
&\times \left\{ \frac{\kappa\mu}{2} e^{-\mrp} (1+\mrp) \frac{\sin(\krp)}{\krp} + \frac{\mu^3 \rhop}{4} e^{-\mrp} \frac{\cos(\krp)}{\krp} \right. \nonumber \\
&+ \left. \left(\frac{2}{r_1} - \frac{2}{r_3} - \frac{2}{r_{12}} + \frac{2}{r_{23}}\right) \frac{\cos(\krp)}{\krp} \left[1 - e^{-\mrp} \left(1 + \frac{\mu}{2}\rhop\right)\right] \right\}.
\label{eq:LCBar}
\end{align}

\subsection{\texorpdfstring{$(\bar{C}_0,\mathcal{L}\bar{C}_0)$}{CLC} Matrix Element}
Using \cref{eq:PermPropFull,eq:SCBarDef},
\begin{align}
\left(\bar{C}_0,\mathcal{L}\bar{C}_0\right) = \frac{1}{2}\left[2(C_0,\mathcal{L}C_0) \pm 2(C_0',\mathcal{L}C_0)\right] 
 = (C_0,\mathcal{L}C_0) \pm (C_0',\mathcal{L}C_0).
 \label{eq:CLC1}
\end{align}

Substitute \cref{eq:GenCDef,eq:LCBar} in \cref{eq:CLC1} and simplify to get
\begin{align}
\left(\bar{C}_0,\mathcal{L}\bar{C}_0\right) = &\frac{\kappa\mu}{2} \left((C_0 \pm C_0^\prime) e^{-\mr}(1+\mr) S_0 \right) \nonumber\\
 &+ \sqrt{\frac{2\kappa}{4\pi}} \frac{\mu^3}{4} \left((C_0 \pm C_0^\prime) e^{-\mr} \frac{\cos(\kr)}{\kr} \Phi_{Ps}(r_{12}) \Phi_H(r_3) \right) \nonumber\\
 &+ \left((C_0 \pm C_0^\prime) \left(\frac{2}{r_1} - \frac{2}{r_3} - \frac{2}{r_{12}} + \frac{2}{r_{23}}\right) C_0 \right).
 \label{eq:CBarLCBar2}
\end{align}

Looking at just the last term:
\begin{align}
&\left((C_0 \pm C_0') \left(\frac{2}{r_1} - \frac{2}{r_3} - \frac{2}{r_{12}} + \frac{2}{r_{23}}\right) C_0 \right) \nonumber \\
 &\phantom{Space} = \left( \left(\frac{2}{r_1} - \frac{2}{r_3} - \frac{2}{r_{12}} + \frac{2}{r_{23}}\right) C_0^2 \right) \pm \left( \left(\frac{2}{r_1} - \frac{2}{r_3} - \frac{2}{r_{12}} + \frac{2}{r_{23}}\right) C_0' C_0 \right).
\end{align}

\noindent The first set of parentheses has the same form as \cref{eq:SbarLSbar}. These terms in parentheses are antisymmetric with respect to the $1 \leftrightarrow 2$ permutation. Also, $C_0$ is symmetric with respect to this permutation. So the first set of parentheses is 0.  Thus,
\beq
\left((C_0 \pm C_0') \left(\frac{2}{r_1} - \frac{2}{r_3} - \frac{2}{r_{12}} + \frac{2}{r_{23}}\right) C_0 \right) = \pm \left(C_0' \left(\frac{2}{r_1} - \frac{2}{r_3} - \frac{2}{r_{12}} + \frac{2}{r_{23}}\right) C_0 \right)
\eeq

\noindent We finally have that
\begin{align}
\left(\bar{C}_0,\mathcal{L}\bar{C}_0\right) = \: &\frac{\kappa\mu}{2} \left((C_0 \pm C_0') e^{-\mr}(1+\mr) S_0 \right) \nonumber\\
 &+ \sqrt{\frac{2\kappa}{4\pi}} \frac{\mu^3}{4} \left((C_0 \pm C_0') e^{-\mr} \frac{\cos(\kr)}{\kr} \Phi_{Ps}(r_{12}) \Phi_H(r_3) \right) \nonumber\\
 &\pm \left(C_0' \left(\frac{2}{r_1} - \frac{2}{r_3} - \frac{2}{r_{12}} + \frac{2}{r_{23}}\right) C_0 \right).
 \label{CBarLCBar}
\end{align}
This is the form that is used in the S-wave long-range code (\cref{chp:Programs}).

\subsection{\texorpdfstring{$(\bar{\phi_i},\mathcal{L}\bar{C}_0)$}{PhiLC} Matrix Elements}
For the S-wave, following the work of Van Reeth \cite{VanReethThesis}, we 
chose to absorb both the $\frac{1}{\sqrt{2}}$ and
$\SphericalHarmonicY{0}{0}{\theta_\rho}{\varphi_\rho} = \frac{1}{\sqrt{4\pi}}$ into the $c_i$ 
constants of the short-range terms in \cref{eq:PhiDef}.

From \cref{eq:PermPropFull},
\begin{subequations}
\begin{align}
(\bar{\phi}_i, \mathcal{L}\bar{C}_0) &= \frac{2}{\sqrt{2}} \left[(\phi_i,\mathcal{L}C_0) \pm (\phi_i',\mathcal{L}C_0)\right] \label{PhiBarLCBar2a} \\
 &= \frac{2}{\sqrt{2}} \left[(\phi_i,\mathcal{L}C_0) \pm (\phi_i,\mathcal{L}C_0')\right].  \label{PhiBarLCBar2b}
\end{align}
\end{subequations}

\noindent Using \cref{eq:LCBar} in the above gives the results for $(\bar{\phi}_i, \mathcal{L}\bar{C}_0)$.
\begin{subequations}
\begin{align}
(\bar{\phi}_i, \mathcal{L}\bar{C}_0) = &\sqrt{\frac{\kappa}{\pi}} \left( (\phi_i \pm \phi_i') \Phi_{Ps}(r_{12}) \Phi_H(r_3) \left\{ \left(\frac{2}{r_1} - \frac{2}{r_2} - \frac{2}{r_{13}} + \frac{2}{r_{23}}\right) \frac{\cos(\kr)}{\kr} \right.\right. \nonumber\\
&\times \left[1 - e^{-\mr}\left(1 + \frac{\mu}{2}\rho\right)\right] + \left.\left.\frac{e^{-\mr}\mu^3\rho}{4} \frac{\cos(\kr)}{\kr} + \frac{e^{-\mr}}{2} \kappa\mu (1+\mr) \frac{\sin(\kr)}{\kr}  \right\}\right) \\
= \sqrt{\frac{\kappa}{\pi}} &\left( \phi_i \Phi_{Ps}(r_{12}) \Phi_H(r_3) \left\{ \left(\frac{2}{r_1} - \frac{2}{r_2} - \frac{2}{r_{13}} + \frac{2}{r_{23}}\right) \frac{\cos(\kr)}{\kr} \left[1 - e^{-\mr}\left(1 + \frac{\mu}{2}\rho\right)\right] \right.\right. \nonumber\\
&+ \left.\left.\frac{e^{-\mr}\mu^3\rho}{4} \frac{\cos(\kr)}{\kr} + \frac{e^{-\mr}}{2} \kappa\mu (1+\mr) \frac{\sin(\kr)}{\kr}  \right\}\right) \nonumber\\
\pm \sqrt{\frac{\kappa}{\pi}} &\left( \phi_i \Phi_{Ps}(r_{12}) \Phi_H(r_3) \left\{ \left(\frac{2}{r_1} - \frac{2}{r_3} - \frac{2}{r_{12}} + \frac{2}{r_{23}}\right) \frac{\cos(\krp)}{\krp} \left[1 - e^{-\mrp}\left(1 + \frac{\mu}{2}\rhop\right)\right] \right.\right. \nonumber\\
&+ \left.\left.\frac{e^{-\mr}\mu^3\rhop}{4} \frac{\cos(\krp)}{\krp} + \frac{e^{-\mrp}}{2} \kappa\mu (1+\mrp) \frac{\sin(\krp)}{\krp}  \right\}\right)
\end{align}
\end{subequations}
Either of these forms can be used. We use the second form, since this only has
$\phi_i$, not $\phi_i$ and $\phi_i'$.

\section{Results}
\label{sec:SWaveResults}

\subsection{Phase Shifts}
\label{sec:SWavePhase}

All runs here were performed using the set of integration points described in 
\cref{sec:QuadraturePoints}. The number of terms used for $^1$S was determined
using the procedure in \cref{sec:MaxMu}, and for $^3$S, the method in
\cref{sec:CompPhase}. The phase shifts are calculated with the programs 
described in \cref{chp:Programs}. \Cref{tab:SWavePhase} shows the $^{1,3}$S phase shifts 
calculated using the $S$-matrix complex Kohn at regular intervals of $\kappa$, which we 
compare to the results from other groups in \cref{tab:SWaveComparisons}. 
The extrapolations in the fourth and fifth columns are performed using the 
technique described in \cref{sec:Extrapolations}. The last two columns show 
the percent difference given by \cref{eq:PercentDiff}.

\begin{table}
\centering
\begin{tabular}{c c c c c c c c}
\toprule
$\kappa$ & $\delta^+ (\omega = 7)$ & $\delta^- (\omega = 7)$ & $\delta^+ (\omega \rightarrow \infty)$ & $\delta^- (\omega \rightarrow \infty)$ & \% Diff$^+$ & \% Diff$^-$ \\
\midrule
0.1 & $-0.427$ & $-0.215$ & $-0.426$ & $-0.214$ & $0.223\%$ & $0.120\%$ \\
0.2 & $-0.820$ & $-0.431$ & $-0.819$ & $-0.431$ & $0.010\%$ & $0.063\%$ \\
0.3 & $-1.161$ & $-0.645$ & $-1.161$ & $-0.645$ & $0.040\%$ & $0.094\%$ \\
0.4 & $-1.446$ & $-0.850$ & $-1.446$ & $-0.849$ & $0.022\%$ & $0.130\%$ \\
0.5 & $-1.678$ & $-1.041$ & $-1.677$ & $-1.040$ & $0.031\%$ & $0.166\%$ \\
0.6 & $-1.858$ & $-1.217$ & $-1.857$ & $-1.214$ & $0.040\%$ & $0.273\%$ \\
0.7 & $-1.964$ & $-1.375$ & $-1.963$ & $-1.372$ & $0.045\%$ & $0.250\%$ \\
\bottomrule
\end{tabular}
\caption[$^{1,3}$S complex Kohn phase shifts]{$^{1,3}$S phase shifts using the $S$-matrix complex Kohn. \% Diff$^+$ and \% Diff$^-$ are the percent differences between the
 current complex Kohn $\omega = 7$ and $\omega \rightarrow \infty$ results.}
\label{tab:SWavePhase}
\end{table}

\begin{figure}
	\centering
	\includegraphics[width=\textwidth]{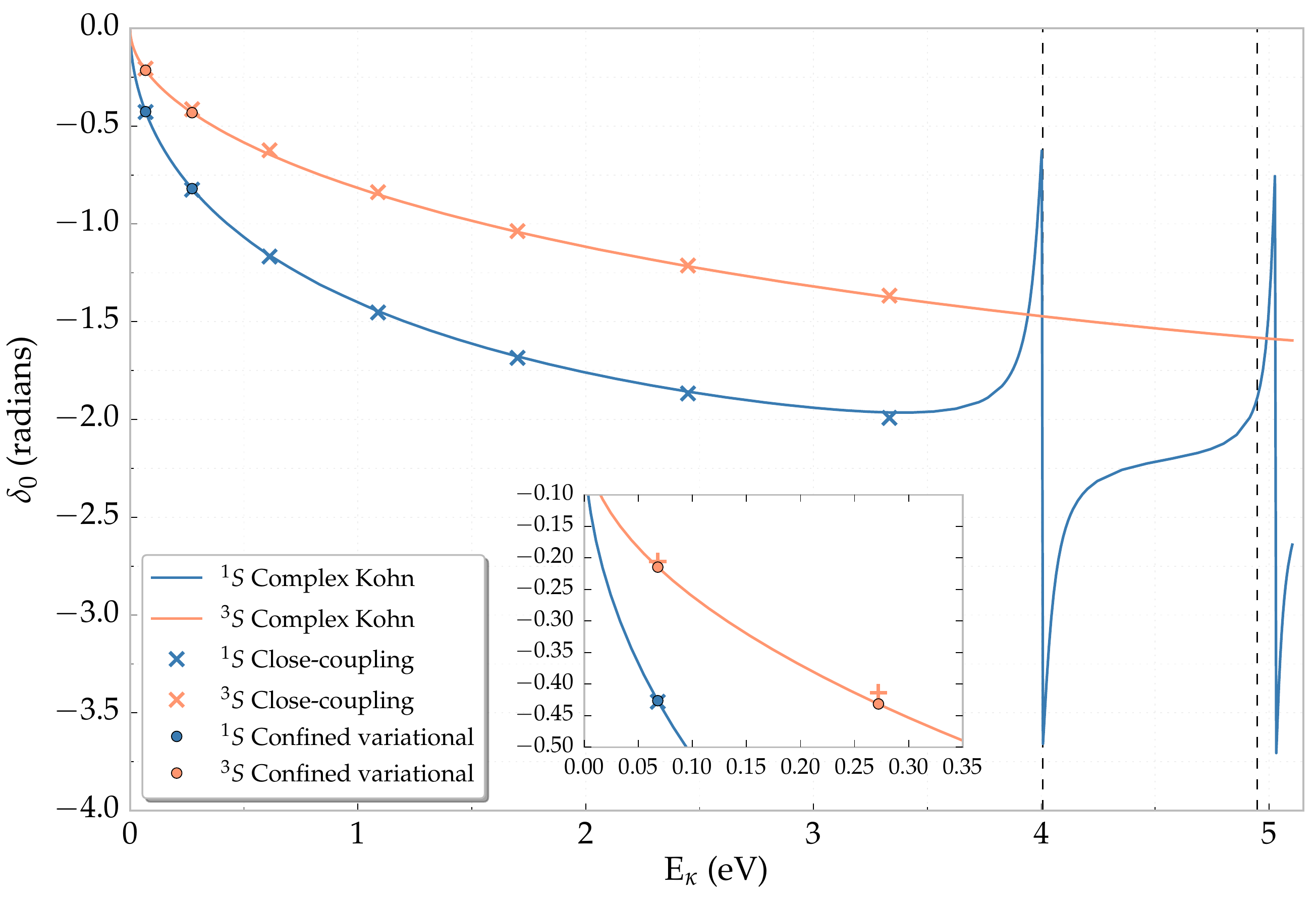}
	\caption[$^{1,3}$S phase shifts]{$^{1,3}$S complex Kohn phase shifts. The $^1S$ CC phase shifts
\cite{Walters2004} are given by \mbox{\textcolor{blue}{$\times$}}, and the
$^3S$ CC phase shifts \cite{Blackwood2002} are given by
\mbox{\textcolor{red}{\textbf{+}}}. The CVM $^1S$ and $^3S$ phase shifts
\cite{Zhang2012} are blue and red circles,
respectively. Vertical dashed lines denote the complex rotation resonance
positions \cite{Yan1999,Yan1998a,Ho1998}. An interactive version of this
figure is available online \cite{Plotly} at
\url{https://plot.ly/~Denton/3/s-wave-ps-h-scattering/}.}
	\label{fig:SWavePhase}
\end{figure}

\Cref{tab:SWaveComparisons} gives comparisons between the complex Kohn phase 
shifts and phase shifts calculated elsewhere in the literature. The $\omega = 
7$ phase shifts from this table are the same as Van Reeth and 
Humberston's results for $\omega = 6$ \cite{VanReeth2003}, with some 
exceptions in the last digit. This indicates that the prior Kohn variational 
method S-wave phase shifts were well converged, despite only using 721
short-range terms. 
We use a larger basis set here, which brings the phase shifts up slightly,
but the larger set of integration points for the long-range terms (see
\cref{sec:SelQuadPoints}) tended to bring the phase shifts down slightly.
\Cref{fig:SWavePhase} has the more complete set of phase shifts plotted with 
respect to the incoming Ps energy, $E_{\bm \kappa}$. This compares the 
complex Kohn phase shifts to that of the $^1$S CC \cite{Walters2004} and the $
^3$S CC \cite{Blackwood2002}, along with the $^{1,3}$S CVM \cite{Zhang2012}. 

As seen in \cref{tab:SWaveComparisons,fig:SWavePhase}, the $^1$S CC phase 
shifts are slightly below the complex Kohn phase shifts, with a larger 
difference at higher $\kappa$. The complex Kohn $^3$S results are almost 
exactly the same as the prior Kohn \cite{VanReeth2003}, and Van Reeth and 
Humberston noted then that the CC $^3$S phase shifts were higher. For 
scattering problems, there is no rigorous bound, but the phase shifts are
typically empirically bound. In the inset in
\cref{fig:SWavePhase} however, we see that the recent CVM results line up 
well with the complex Kohn results for both $^1$S and $^3$S, potentially indicating 
that complex Kohn phase shifts are more accurate than the CC.

\begin{figure}
	\centering
	\includegraphics[width=5.25in]{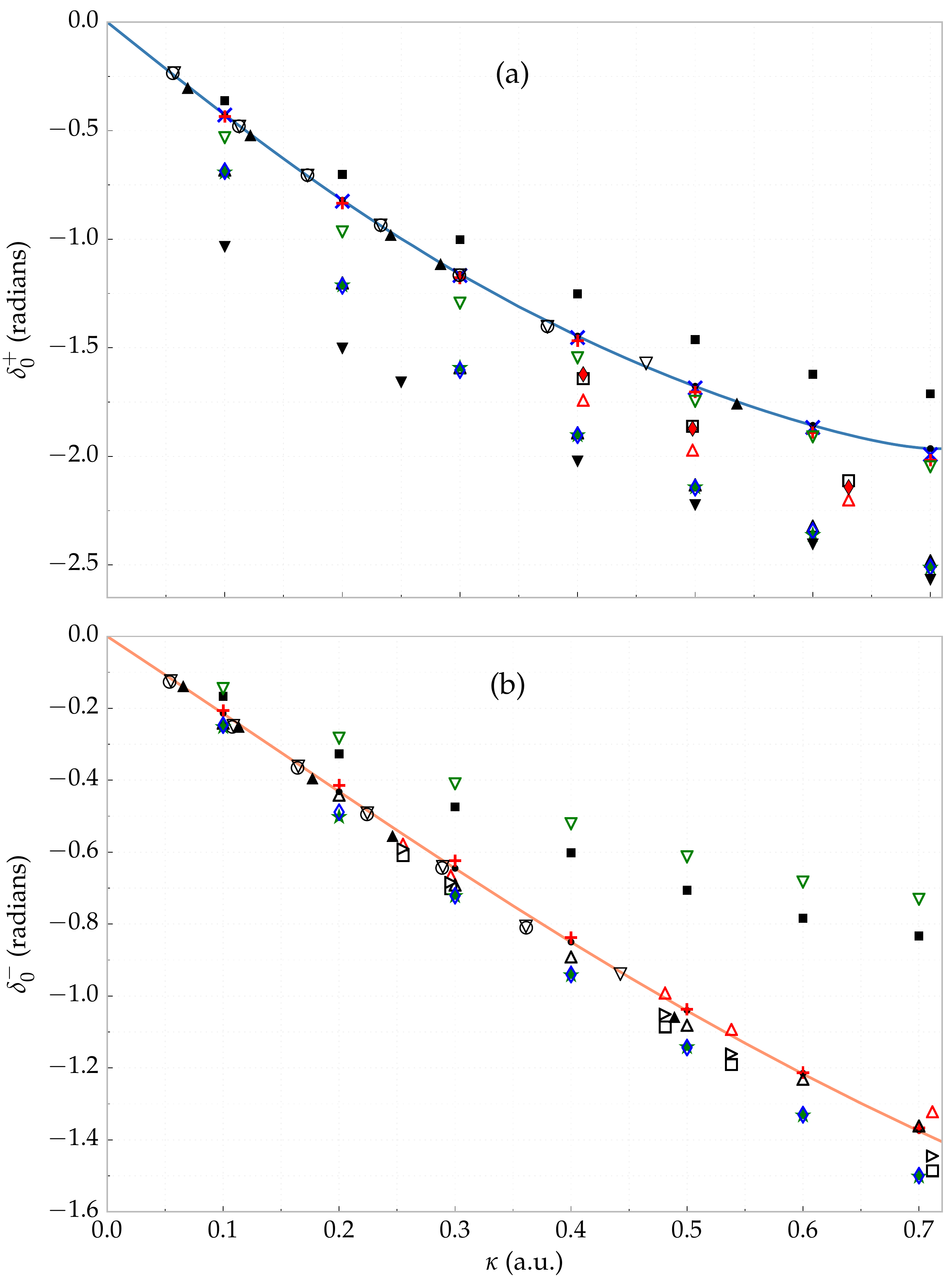}
	\caption[Comparison of $^1S$ and $^3S$ phase shifts]{Comparison of $^1S$ (a) and $^3S$ (b) phase shifts
with results from other groups. Results are ordered according to year of
publication. Solid curves -- this work;
\mbox{\textcolor{blue}{$\times$} -- CC \cite{Walters2004};}
\mbox{$\CIRCLE$ -- Kohn \cite{VanReeth2003};}
\mbox{\textcolor{red}{\textbf{+}} -- CC \cite{Blackwood2002};}
\mbox{$\blacktriangle$ -- DMC \cite{Chiesa2002};} 
\mbox{$\triangledown$ -- SVM 2002 \cite{Ivanov2002};} 
\mbox{\textcolor[RGB]{0,127,0}{$\blacktriangle$} -- T-matrix \cite{Biswas2002a};} 
\mbox{$\Circle$ -- SVM 2001 \cite{Ivanov2001};} 
\mbox{\textcolor[RGB]{0,127,0}{$\triangledown$} -- 2 channel / static exchange with model exchange \cite{Biswas2001};} 
\mbox{\textcolor{red}{$\vartriangle$} -- 6-state CC \cite{Sinha2000};} 
\mbox{$\blacksquare$ -- 5-state CC \cite{Adhikari1999};} 
\mbox{$\square$ -- Coupled-pseudostate \cite{Campbell1998};} 
\mbox{$\vartriangle$ -- 3-state CC \cite{Sinha1997};} 
\mbox{\textcolor[RGB]{0,127,0}{$\bigstar$} -- Static-exchange \cite{Ray1997};} 
\mbox{$\triangleright$ -- Stabilization \cite{Drachman1976};} 
\mbox{\textcolor{red}{$\blacklozenge$} -- Stabilization \cite{Drachman1975};}
\mbox{\textcolor{blue}{$\lozenge$} -- Static-exchange \cite{Hara1975};}
\mbox{$\blacktriangledown$ -- Static-exchange \cite{Fraser1961}.}}
	\label{fig:SWaveComparisons}
\end{figure}

\Cref{fig:SWaveComparisons} shows comparisons of the complex Kohn phase 
shifts to that of other groups for calculations of $^1$S and $^3$S. The
different Kohn-type methods agree to the accuracy given after methods with
Schwartz singularities are removed. The current $S$-matrix complex Kohn results are extremely close to Van
Reeth and Humberston's results \cite{VanReeth2003}, so they 
follow along the solid line as well. Several groups have results that cluster 
very closely to the current $S$-matrix complex Kohn phase shifts, namely Blackwood et al. \cite{Blackwood2002}, Walters 
et al. \cite{Walters2004}, Chiesa et al. \cite{Chiesa2002} and Ivanov et al. \cite{Ivanov2002}. 
Ivanov et al.\ \cite{Ivanov2002} discuss that the higher phase shifts of
Adhikari and Biswas \cite{Adhikari1999} are likely to be in error. Likewise, 
the further $^1$S calculations by Biswas et al. \cite{Biswas2002a} are near 
that of Adhikari and Biswas \cite{Adhikari1999}. The Biswas et al. \cite{Biswas2001} phase shifts 
agree relatively well with the current complex Kohn and that of the accurate 
Refs.~\cite{Blackwood2002,VanReeth2003,Walters2004} for $^1$S but seem to 
overestimate the $^3$S phase shifts.

\begin{table}
\centering
\setlength{\tabcolsep}{-2pt}
\footnotesize
\begin{tabular}{@{\hskip 0.1cm}l . . . . . . .}
\toprule
Method & \multicolumn{1}{c}{\phantom{1}0.1} & \multicolumn{1}{c}{\phantom{1}0.2} & \multicolumn{1}{c}{\phantom{1}0.3} & \multicolumn{1}{c}{\phantom{1}0.4} & \multicolumn{1}{c}{\phantom{1}0.5} & \multicolumn{1}{c}{\phantom{1}0.6} & \multicolumn{1}{c}{\phantom{1}0.7} \\
\midrule
This work $\omega = 7$ $\delta_0^+$									& -0.427   & -0.820   & -1.161   & -1.446  & -1.678  & -1.858  & -1.964 \\
This work $\omega \to \infty$ $\delta_0^+$							& -0.426   & -0.819   & -1.161   & -1.446  & -1.677  & -1.857  & -1.963 \\
Kohn $\omega = 6$ \cite{VanReeth2003} $\delta_0^+$					& -0.427   & -0.820   & -1.161   & -1.446  & -1.677  & -1.857  & -1.964 \\
Kohn $\omega \rightarrow \infty$ \cite{VanReeth2003} $\delta_0^+$	& -0.425   & -0.817   & -1.158   & -1.443  & -1.674  & -1.852  & -1.959 \\
CVM \cite{Zhang2012} $\delta_0^+$									& -0.42636 & -0.81973 & $---$    & $---$   & $---$   & $---$   & $---$  \\
CC 14Ps14H+H$^-$ \cite{Walters2004} $\delta_0^+$					& -0.428   & -0.825   & -1.167   & -1.453  & -1.685  & -1.867  & -1.992 \\
CC 14Ps14H \cite{Blackwood2002} $\delta_0^+$						& -0.434   & -0.834   & -1.178   & -1.467  & -1.704  & -1.890  & -2.018 \\
T-matrix \cite{Biswas2002a} $\delta_0^+$							& -0.38269 & -0.73419 & -1.03799 & -1.2924 & -1.5014 & -1.6667 & $---$  \\
2 channel ME \cite{Biswas2001} $\delta_0^+$							& -0.532   & -0.966   & -1.294   & -1.546  & -1.746  & -1.910  & -2.048 \\
3-state CC \cite{Sinha1997} $\delta_0^+$							& -0.68    & -1.20    & -1.59    & -1.89   & -2.13   & -2.32   & -2.48  \\
SE \cite{Ray1997} $\delta_0^+$										& -0.692   & -1.212   & -1.592   & -1.902  & -2.142  & -2.362  & -2.512 \\
5-state CC \cite{Adhikari1999} $\delta_0^+$							& -0.362   & -0.702   & -1.002   & -1.252  & -1.462  & -1.622  & -1.712 \\
SE \cite{Hara1975} $\delta_0^+$										& -0.68649 & -1.2147  & -1.6029  & -1.9026 & -2.144  & -2.344  & -2.511 \\
\midrule                                                            
This work $\omega = 7$ $\delta_0^-$									& -0.215   & -0.431   & -0.645   & -0.850  & -1.041  & -1.217  & -1.375  \\
This work $\omega \to \infty$ $\delta_0^-$							& -0.214   & -0.431   & -0.645   & -0.849  & -1.040  & -1.214  & -1.372  \\
Kohn $\omega = 6$ \cite{VanReeth2003} $\delta_0^-$ 					& -0.215   & -0.432   & -0.645   & -0.850  & -1.040  & -1.215  & -1.373  \\
Kohn $\omega \rightarrow \infty$ \cite{VanReeth2003} $\delta_0^-$	& -0.214   & -0.431   & -0.645   & -0.849  & -1.038  & -1.211  & -1.366  \\
CVM \cite{Zhang2012} $\delta_0^-$ 									& -0.21464 & -0.43159 & $---$    & $---$   & $---$   & $---$   & $---$   \\
CC 14Ps14H \cite{Blackwood2002} $\delta_0^-$ 						& -0.206   & -0.414   & -0.624   & -0.838  & -1.037  & -1.213  & -1.367  \\
SE ME \cite{Biswas2001} $\delta_0^-$ 								& -0.145   & -0.283   & -0.410   & -0.521  & -0.613  & -0.683  & -0.731  \\
3-state CC \cite{Sinha1997} $\delta_0^+$							& -0.24    & -0.44    & -0.69    & -0.89   & -1.08   & -1.23   & -1.362  \\
SE \cite{Ray1997} $\delta_0^-$ 										& -0.252   & -0.502   & -0.722   & -0.942  & -1.142  & -1.332  & -1.502  \\
5-state CC \cite{Adhikari1999} $\delta_0^-$ 						& -0.167   & -0.327   & -0.474   & -0.602  & -0.706  & -0.784  & -0.833  \\
SE \cite{Hara1975} $\delta_0^-$										& -0.2469  & -0.4888  & -0.7211  & -0.9402 & -1.1435 & -1.3300 & -1.4996 \\
\bottomrule
\end{tabular}
\caption[Comparison of $^{1,3}$S phase shifts]{Comparison of $^{1,3}$S phase shifts between complex Kohn results and those from other groups. Values in the header are $\kappa$ in a.u.}
\label{tab:SWaveComparisons}
\end{table}

An extensive comparison of the S-wave phase shifts with calculations from other
groups is also shown in \cref{tab:SWaveComparisons}. Fewer groups have attempted this
problem than the PsH bound state problem in \cref{tab:BoundEnergyOther}. We see
that the accurate complex Kohn results are very similar to the prior Kohn
\cite{VanReeth2003} and agree extremely well with the CVM results
\cite{Zhang2012}. The CC phase shifts \cite{Walters2004,Blackwood2002} also
agree relatively well, with closer agreement for $^1$S than $^3$S.

\subsection{Resonance Parameters}
\label{sec:SWaveResonances}

Before the Ps(n=2) threshold at \SI{5.102}{eV}, 
there are two very clear resonances for $^1$S scattering, which can be seen 
in \cref{fig:SWavePhase}. 
The trial wavefunction we use cannot be 
extended past this threshold without modifications to take into account the
Ps(n=2) channel. We use the numerical methods described in
\cref{sec:ResonanceFit} to accurately determine the resonance parameters.

\setlength{\abovecaptionskip}{6pt}   
\setlength{\belowcaptionskip}{6pt}   
\begin{table}
\footnotesize
\centering
\begin{tabular}{l l l l l}
\toprule
Method & \thead{$^1E_R \text{ (eV)}$} & \thead{$^1\Gamma \text{ (eV)}$} & \thead{$^2E_R \text{ (eV)}$} & \thead{$^2\Gamma \text{ (eV)}$} \\
\midrule
Current work: &  &  &  & \\
\ Average $\pm$ standard deviation & $4.0065 \pm 0.0001$ & $0.0955 \pm 0.0001$ & $5.0272 \pm 0.0029$ & $0.0608 \pm 0.0007$ \\
Current work: &  &  &  & \\
\ $S$-matrix complex Kohn & $4.0065$ & $0.0955$ & $5.0278$ & $0.0608$ \\
CC (9Ps9H + H$^-$) \cite{Walters2004} & $4.149$ & $0.103$ & $4.877$ & $0.0164$ \\
Kohn variational \cite{VanReeth2004} & $4.0072 \pm 0.0020$ & $0.0956 \pm 0.010$ & $5.0267 \pm 0.0020$ & $0.0597 \pm 0.0010$ \\
Stabilization \cite{Yan2003} & $4.007$ & $0.0969$ & $4.953$ & $0.0574$ \\
CC (22Ps1H + H$^-$) \cite{Blackwood2002b} & $4.141$ & $0.071$ & $4.963$ & $0.033$ \\
CC (9Ps9H) \cite{Blackwood2002} & $4.37$ & $0.10$ & --- & --- \\
Optical potential \cite{DiRienzi2002b} & $4.021$ & $0.0259$ & --- & --- \\
T-matrix \cite{Biswas2002a} & $4.06$ & --- & --- & --- \\
CC \cite{Biswas2002} & $4.04$ & --- & --- & --- \\
Five-state CC \cite{Adhikari2001e} & $4.01$ & $0.15$ & --- & --- \\
Complex rotation \cite{Yan1999} & $4.0058 \pm 0.0005$ & $0.0952 \pm 0.0011$ & $4.9479 \pm 0.0014$ & $0.0585 \pm 0.0027$ \\
Coupled-pseudostate \cite{Campbell1998} & $4.55$ & $0.084$ & --- & --- \\
Complex rotation \cite{Ho1978} & $4.013 \pm 0.014$ & $0.075 \pm 0.027$ & --- & --- \\
Complex rotation \cite{Drachman1975} & $4.455 \pm 0.010$ & $0.062 \pm 0.015$ & --- & --- \\
Stabilization \cite{Hazi1970} & $5.8366$ & $0.2693$ & --- & --- \\
\bottomrule
\end{tabular}
\caption{S-wave resonance parameters}
\label{tab:SWaveResonancesOther}
\end{table}

The current complex Kohn resonance parameters in
\cref{tab:SWaveResonancesOther} are very similar to that of the previous Kohn 
calculations \cite{VanReeth2004}, but the first resonance position matches 
better with the complex rotation of Yan and Ho \cite{Yan1999}. The complex
rotation can be considered one of the best calculations for these 
resonances, and the stabilization method in Ref.~\cite{Yan2003} from the same 
authors agrees well with the complex rotation. Both the complex Kohn and 
prior Kohn calculations give a second resonance position that is at a higher 
energy than the complex rotation, but the current results agree relatively 
well with the complex rotation for most parameters.

The CC calculations of Ref.~\cite{Walters2004} tabulate resonance parameters 
through the F-wave singlet. Their resonance parameters are close to the 
complex Kohn and CC results, but there is a significant difference in the 
second resonance width, which is much smaller than the other calculations. 
Other calculations shown in this table agree approximately on the position 
and width of the resonances, with the exception of the much earlier 
stabilization by Hazi and Taylor \cite{Hazi1970}.

\begin{figure}
	\centering
	\includegraphics[width=5in]{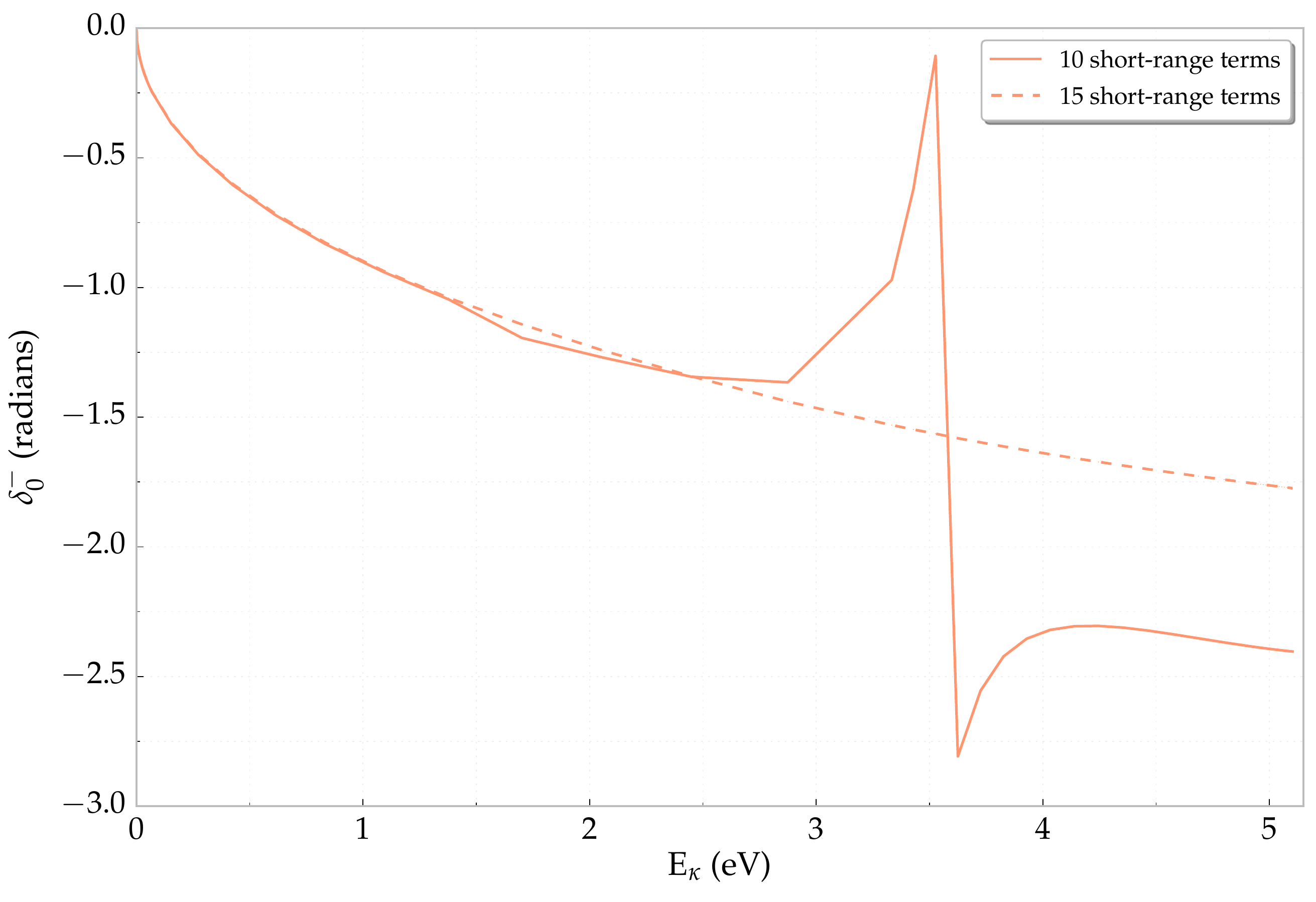}
	\caption{$^3$S plot showing Schwartz singularity at low $N$}
	\label{fig:triplet-false-resonance}
\end{figure}

Ray \cite{Ray2006} reports a $^3$S resonance in a 3-state CC approximation. 
This has not been reported by any other groups, and it appears that this may 
not be a true resonance. The Belfast group notes this in Ref.~\cite{Campbell1998}, 
stating that ``Since H$^-$ is in a spin singlet state, the resonances must 
have total electronic spin zero. Accordingly, we find no resonances in our 
triplet partial waves.'' \Cref{fig:triplet-false-resonance} shows that for a 
small number of terms, we get a Schwartz singularity in the triplet. As the 
number of terms increases, this singularity disappears, which can illustrate 
how critical it is to have basis sets with a large number of terms for Ps-H 
scattering.

\section{Summary}
\label{sec:SummaryS}

The Kohn-type variational methods have provided highly accurate phase shifts 
and reliable resonance parameters for the $^{1,3}$S-wave. The $S$-matrix
complex Kohn phase shifts and resonance parameters compare well with those
of accurate calculations from other groups \cite{Blackwood2002,Walters2004,Zhang2012}.


\chapter{P-Wave}
\label{chp:PWave}

\iftoggle{UNT}{Like}{\lettrine{\textcolor{startcolor}{L}}{ike}}
the S-wave in \cref{chp:SWave}, the general formalism in
\cref{chp:WaveKohn} was developed much later than the P-wave derivations and code 
were developed. So though the analysis in \cref{chp:WaveKohn} will work 
properly for the P-wave, it is worthwhile to show the exact formulas that are 
used in the P-wave long-range code. Similar to the S-wave (\cref{chp:SWave}),
we used Van Reeth's \cite{VanReethPrivate} P-wave notes and code for guidance
in performing our derivations and writing code.

\section{Wavefunction}
\label{sec:PWaveFn}
We include both short-range symmetries for the P-wave, so that \cref{eq:GeneralWaveTrial} becomes
\beq
\label{eq:PWaveTrial}
\Psi_1^{\pm,t} = \widetilde{S}_1 + L_1^{\pm,t} \, \widetilde{C}_1 + \sum_{i=1}^{N'(\omega)} c_i \bar{\phi}_{1i} + \sum_{j=1}^{N'(\omega)} d_j \bar{\phi}_{2j}.
\eeq
The forms of the long-range terms $\widetilde{S}_1$ and $\widetilde{C}_1$ are
found in \cref{sec:GeneralWave}.
Like for the S-wave and as mentioned in \cref{sec:KohnApplied}, we compute only
matrix elements with the barred $\bar{S}_1$ and $\bar{C}_1$ for the Kohn, then
use these to construct the matrices (\cref{eq:GeneralKohnMatrix}) to compute
the other Kohn-type variational methods.

The short-range terms are given by \cref{eq:PhiDef}, rewritten as
\begin{subequations}
\label{eq:PWavePhiBar}
\begin{align}
\bar{\phi}_{1i} &= \left(1 \pm P_{23}\right) Y_{10}(\theta_1) r_1 \phi_i \label{eq:PWavePhi1i}\\
\bar{\phi}_{2j} &= \left(1 \pm P_{23}\right) Y_{10}(\theta_2) r_2 \phi_j \label{eq:PWavePhi2j},
\end{align}
\end{subequations}
where $\phi_i$ and $\phi_j$ are given by \cref{eq:BoundWavefn_phi}. We also use the shortcuts
\begin{equation}
\label{eq:PWavePhi}
\phi_{1i} = r_1 \phi_i \text{ and } \phi_{2j} = r_2 \phi_j.
\end{equation}
We refer to the short-range Hylleraas-type terms $\phi_{1i}$ and $\phi_{2j}$ as
the first and second symmetries, respectively.

To derive the form of the matrix elements involving $\widetilde{C}_1$, we used
the \emph{Mathematica} code in \cref{fig:LCMath} to find that
\begin{align}
\frac{1}{2} & \left(\Laplacian_\rho + \kappa^2\right) \SphericalHarmonicY{1}{0}{\theta_\rho}{\varphi_\rho} n_1(\kappa\rho) f_1(\rho)  \nonumber \\
&= -\frac{\rho  f^{\prime\prime}(\rho ) \left[\kappa  \rho  \sin (\kappa  \rho )+\cos (\kappa  \rho )\right]+2 f^\prime(\rho ) \left[\left(\kappa ^2 \rho ^2-1\right) \cos (\kappa  \rho )-\kappa  \rho  \sin (\kappa \rho )\right]}{2 \kappa ^2 \rho ^3}.
\end{align}
Derivations are not given in the following sections -- just the final
results for each matrix element. The notes on derivations are available on
my \href{http://figshare.com/authors/Denton_Woods/581638}{figshare page} \cite{figshare}.

\section{First Formalism}

We used two different formalisms for the P-wave. The first places the orbital angular
momentum mainly on the positron ($\bm{r}_1$) and the electron of Ps ($\bm{r}_2$,
and $\bm{r}_3$ with exchange). \Cref{sec:PWave2Formalism} covers the second
formalism that we tried.

\subsection{Matrix Equation}
\label{sec:PWaveMatrix}

Since the P-wave trial wavefunction has two sets of short-range terms, the matrix equation can be seen from \cref{eq:GeneralKohnMatrix} to be
{
\begin{equation}
\label{eq:PWaveKohnMatrix}
\resizebox{\textwidth}{!}{$
	\begin{bmatrix} 
	 (\widetilde{C},\mathcal{L}\widetilde{C}) & (\widetilde{C},\mathcal{L}\bar{\phi}_{11}) & \cdots & (\widetilde{C},\mathcal{L}\bar{\phi}_{1N}) & (\widetilde{C},\mathcal{L}\bar{\phi}_{21}) & \cdots & (\widetilde{C},\mathcal{L}\bar{\phi}_{2N})\\
	 (\bar{\phi}_{11},\mathcal{L}\widetilde{C}) & (\bar{\phi}_{11},\mathcal{L}\bar{\phi}_{11}) & \cdots & (\bar{\phi}_{11},\mathcal{L}\bar{\phi}_{1N}) & (\bar{\phi}_{11},\mathcal{L}\bar{\phi}_{21}) & \cdots & (\bar{\phi}_{11},\mathcal{L}\bar{\phi}_{2N})\\
	 \vdots & \vdots & \ddots & \vdots & \vdots & \ddots & \vdots \\
	 (\bar{\phi}_{1N},\mathcal{L}\widetilde{C}) & (\bar{\phi}_{1N},\mathcal{L}\bar{\phi}_{11}) & \cdots & (\bar{\phi}_{1N},\mathcal{L}\bar{\phi}_{1N}) & (\bar{\phi}_{1N},\mathcal{L}\bar{\phi}_{21}) & \cdots & (\bar{\phi}_{1N},\mathcal{L}\bar{\phi}_{2N})\\
	 (\bar{\phi}_{21},\mathcal{L}\widetilde{C}) & (\bar{\phi}_{21},\mathcal{L}\bar{\phi}_{11}) & \cdots & (\bar{\phi}_{21},\mathcal{L}\bar{\phi}_{1N}) & (\bar{\phi}_{21},\mathcal{L}\bar{\phi}_{21}) & \cdots & (\bar{\phi}_{21},\mathcal{L}\bar{\phi}_{2N})\\
	 \vdots & \vdots & \ddots & \vdots & \vdots & \ddots & \vdots \\
	 (\bar{\phi}_{2N},\mathcal{L}\widetilde{C}) & (\bar{\phi}_{2N},\mathcal{L}\bar{\phi}_{11}) & \cdots & (\bar{\phi}_{2N},\mathcal{L}\bar{\phi}_{1N}) & (\bar{\phi}_{2N},\mathcal{L}\bar{\phi}_{21}) & \cdots & (\bar{\phi}_{2N},\mathcal{L}\bar{\phi}_{2N})\\
	\end{bmatrix}
	\begin{bmatrix}
  L_\ell^t\\
	c_1\\
	\vdots\\
	c_N\\
	d_1\\
	\vdots\\
	d_N\\
	\end{bmatrix}
	= -
	\begin{bmatrix}
	(\widetilde{C},\mathcal{L}\widetilde{S}) \\
	(\bar{\phi}_{11},\mathcal{L}\widetilde{S}) \\
	\vdots \\
	(\bar{\phi}_{1N},\mathcal{L}\widetilde{S}) \\
	(\bar{\phi}_{21},\mathcal{L}\widetilde{S}) \\
	\vdots \\
	(\bar{\phi}_{2N},\mathcal{L}\widetilde{S}) \\
	\end{bmatrix}.$
	}
\end{equation}
}
The analysis in \cref{sec:KohnApplied} applies directly to the P-wave.

\subsection{Short-Range -- Short-Range Matrix Elements}
\label{sec:PWaveShortShort}

Due to the angular dependence of the $\SphericalHarmonicY{1}{0}{\theta}{\phi}$
spherical harmonic, the P-wave short-short integrals
are more complicated than the S-wave. Using the results in \cref{chp:AngularInt},
these integrals with both symmetries can be written as shown in
\cref{eq:PWavePhi1Phi1,eq:PWavePhi2Phi2,eq:PWavePhi1Phi2,eq:PWavePhi2Phi1}.
Each of these equations has re-expressed the Laplacian as the gradient-gradient,
allowing us to use \cref{eq:GradGradShort}.
Full derivations are available from the Research Wiki \cite{Wiki} and on
figshare at \url{http://figshare.com/s/82e47cbe0d9b11e58ed806ec4b8d1f61}.

The short-short matrix elements involving all four combinations of the two
symmetries are given below:
\begin{align}
\label{eq:PWavePhi1Phi1}
\left(\bar{\phi}_{1i},\mathcal{L} \bar{\phi}_{1j}\right) = &2 \cdot 2\pi \int{ \Bigg\{ \sum_{k=1}^3 \left[ \boldsymbol{\nabla}_{\!\mathbf{r}_k} \nonumber \phi_{1i} \boldsymbol{\cdot} \boldsymbol{\nabla}_{\!\mathbf{r}_k} \phi_{1j} \pm \boldsymbol{\nabla}_{\!\mathbf{r}_k} \phi_{1i} \boldsymbol{\cdot} \boldsymbol{\nabla}_{\!\mathbf{r}_k} \phi_{1j}^\prime \right] } \\
\nonumber  &+ \left. \left[\frac{2}{r_1} - \frac{2}{r_2} - \frac{2}{r_3} - \frac{2}{r_{12}} - \frac{2}{r_{13}} + \frac{2}{r_{23}} - 2 E_H - 2 E_{Ps} - \frac{1}{2}\kappa^2 + \frac{2}{r_1^2} \right] \right. \\
 &\;\;\;\;\; \times \left(\phi_{1i} \phi_{1j} \pm \phi_{1i} \phi_{1j}^\prime \right) \Bigg\} d\tau_{int}
\end{align}

\begin{align}
\label{eq:PWavePhi2Phi2}
\left(\bar{\phi}_{2i},\mathcal{L} \bar{\phi}_{2j}\right) = 2 & \cdot 2\pi \int \Bigg\{ \sum_{k=1}^3 \left[ \boldsymbol{\nabla}_{\!\mathbf{r}_k} \nonumber \phi_{2i} \boldsymbol{\cdot} \boldsymbol{\nabla}_{\!\mathbf{r}_k} \phi_{2j} \pm \cos\theta_{23} \boldsymbol{\nabla}_{\!\mathbf{r}_k} \phi_{2i} \boldsymbol{\cdot} \boldsymbol{\nabla}_{\!\mathbf{r}_k} \phi_{2j}^\prime \right]  + \frac{2}{r_2^2}\phi_{2i}\phi_{2j} \\
 \nonumber &\mp \phi_{2i} \phi_{2j}^\prime \left[p_i \frac{r_1}{r_3 r_{13}^2} (\cos\theta_{12} - \cos\theta_{23} \cos\theta_{13}) + m_j^\prime \frac{r_1}{r_2 r_{12}^2}(\cos\theta_{13}\right.\\
 \nonumber & \left. \;\;\;\;\; -\cos\theta_{23} \cos\theta_{12}) + \sin^2\theta_{23} \left(q_i \frac{r_2}{r_3 r_{23}^2} + q_j^\prime \frac{r_3}{r_2 r_{23}^2} \right) \right] \\
 \nonumber &+ \left. \left[\frac{2}{r_1} - \frac{2}{r_2} - \frac{2}{r_3} - \frac{2}{r_{12}} - \frac{2}{r_{13}} + \frac{2}{r_{23}} - 2 E_H - 2 E_{Ps} - \frac{1}{2}\kappa^2 \right] \right. \\
 &\;\;\;\;\; \times \left(\phi_{2i} \phi_{2j} \pm \cos\theta_{23} \phi_{2i} \phi_{2j}^\prime \right) \Bigg\} d\tau_{int}
\end{align}

\begin{align}
\label{eq:PWavePhi1Phi2}
\left(\bar{\phi}_{1i},\mathcal{L} \bar{\phi}_{2j}\right) = 2 & \cdot 2\pi \int \Bigg\{ \sum_{k=1}^3 \left[ \cos\theta_{12} \boldsymbol{\nabla}_{\!\mathbf{r}_k} \nonumber \phi_{1i} \boldsymbol{\cdot} \boldsymbol{\nabla}_{\!\mathbf{r}_k} \phi_{2j} \pm \cos\theta_{13} \boldsymbol{\nabla}_{\!\mathbf{r}_k} \phi_{1i} \boldsymbol{\cdot} \boldsymbol{\nabla}_{\!\mathbf{r}_k} \phi_{2j}^\prime \right] \\
 \nonumber &\mp \phi_{1i} \phi_{2j} \left[q_i \frac{r_3}{r_2 r_{23}^2} (\cos\theta_{13} - \cos\theta_{12} \cos\theta_{23}) + p_j \frac{r_3}{r_1 r_{13}^2}(\cos\theta_{23}\right.\\
 \nonumber & \left. \;\;\;\;\; -\cos\theta_{12} \cos\theta_{13}) + \sin^2\theta_{12} \left(m_i \frac{r_1}{r_2 r_{12}^2} + m_j \frac{r_2}{r_1 r_{12}^2} \right) \right] \\
 \nonumber &\mp \phi_{1i} \phi_{2j}^\prime \left[q_i \frac{r_2}{r_3 r_{23}^2} (\cos\theta_{12} - \cos\theta_{13} \cos\theta_{23}) + m_j^\prime \frac{r_2}{r_1 r_{12}^2}(\cos\theta_{23}\right.\\
 \nonumber & \left. \;\;\;\;\; -\cos\theta_{12} \cos\theta_{13}) + \sin^2\theta_{13} \left(p_i \frac{r_1}{r_3 r_{13}^2} + p_j^\prime \frac{r_3}{r_1 r_{13}^2} \right) \right] \\
 \nonumber &+ \left. \left[\frac{2}{r_1} - \frac{2}{r_2} - \frac{2}{r_3} - \frac{2}{r_{12}} - \frac{2}{r_{13}} + \frac{2}{r_{23}} - 2 E_H - 2 E_{Ps} - \frac{1}{2}\kappa^2 \right] \right. \\
 &\;\;\;\;\; \times \left(\cos\theta_{12} \phi_{1i} \phi_{2j} \pm \cos\theta_{13} \phi_{1i} \phi_{2j}^\prime \right) \Bigg\} d\tau_{int}
\end{align}

\begin{align}
\label{eq:PWavePhi2Phi1}
\left(\bar{\phi}_{2i},\mathcal{L} \bar{\phi}_{1j}\right) = 2 & \cdot 2\pi \int \Bigg\{ \sum_{k=1}^3 \cos\theta_{12} \left[ \boldsymbol{\nabla}_{\!\mathbf{r}_k} \nonumber \phi_{2i} \boldsymbol{\cdot} \boldsymbol{\nabla}_{\!\mathbf{r}_k} \phi_{1j} \pm \boldsymbol{\nabla}_{\!\mathbf{r}_k} \phi_{2i} \boldsymbol{\cdot} \boldsymbol{\nabla}_{\!\mathbf{r}_k} \phi_{1j}^\prime \right] \\
 \nonumber &\mp \phi_{2i} \phi_{1j} \left[p_i \frac{r_3}{r_1 {r_{13}}^2} (\cos\theta_{23} - \cos\theta_{12} \cos\theta_{13}) + q_j \frac{r_3}{r_2 {r_{23}}^2}(\cos\theta_{13}\right.\\
 \nonumber & \left. \;\;\;\;\; -\cos\theta_{12} \cos\theta_{23}) + \sin^2\theta_{12} \left(m_i \frac{r_2}{r_1 {r_{12}}^2} + m_j \frac{r_1}{r_2 {r_{12}}^2} \right) \right] \\
 \nonumber &\mp \phi_{2i} \phi_{1j}^\prime \left[p_i \frac{r_3}{r_1 r_{13}^2} (\cos\theta_{23} - \cos\theta_{12} \cos\theta_{13}) + q_j^\prime \frac{r_3}{r_2 {r_{23}}^2}(\cos\theta_{13}\right.\\
 \nonumber & \left. \;\;\;\;\; -\cos\theta_{12} \cos\theta_{23}) + \sin^2\theta_{12} \left(m_i \frac{r_2}{r_1 {r_{12}}^2} + m_j^\prime \frac{r_1}{r_2 {r_{12}}^2} \right) \right] \\
 \nonumber &+ \left. \left[\frac{2}{r_1} - \frac{2}{r_2} - \frac{2}{r_3} - \frac{2}{r_{12}} - \frac{2}{r_{13}} + \frac{2}{r_{23}} - 2 E_H - 2 E_{Ps} - \frac{1}{2}\kappa^2 \right] \right. \\
 &\;\;\;\;\; \times \cos\theta_{12} \left(\phi_{2i} \phi_{12j} \pm \phi_{2i} \phi_{1j}^\prime \right) \Bigg\} d\tau_{int}
\end{align}

The $\cos\theta_{ij}$ that are present in these equations have to be rewritten in
terms of $r_i$ and $r_{ij}$ using \cref{eq:Cosines} so that the short-short
code in \cref{sec:CompShort} can be used with these.
I wrote a \emph{Mathematica} notebook
(see \cref{chp:Programs}) that transforms
these equations into a form that allows us to copy and paste them directly
into Fortran code.

\subsection{Short-Range -- Long-Range Matrix Elements}
\label{sec:PWaveShortLong}

Like the S-wave in \cref{sec:LCElements}, we derived these and wrote code 
using these results based on Van Reeth's \cite{VanReethPrivate} notes and 
codes before we developed a general formalism described in \cref{chp:WaveKohn}.
They can easily be shown to be equivalent, but it is easier to compare these
forms to the P-wave long-range code.

\begin{align}
\label{eq:PWavePhi1SBar}
\nonumber \left(\bar{\phi}_{1i},\mathcal{L} \bar{S}_1\right) = \sqrt{2} \pi & \int \phi_{1i} \left[ \frac{r_1 + r_2 \cos\theta_{12}}{\rho} \left( \frac{2}{r_1} - \frac{2}{r_2} - \frac{2}{r_{13}} + \frac{2}{r_{23}} \right) S_{22} \right. \\
& \pm \left. \frac{r_1 + r_3 \cos\theta_{13}}{\rho^\prime} \left( \frac{2}{r_1} - \frac{2}{r_3} - \frac{2}{r_{12}} + \frac{2}{r_{23}} \right) S_{23} \right]  d\tau_{int}
\end{align}

\begin{align}
\label{eq:PWavePhi1CBar}
\nonumber (\bar{\phi}_{1i},&\mathcal{L} \bar{C}_1) = -2 \pi \sqrt{\kappa} \int \phi_{1i} \\
\nonumber & \times \left\{ \frac{r_1 + r_2 \cos\theta_{12}}{\rho} \Phi_{Ps}(r_{12}) \Phi_H(r_3) \left[ \left( \frac{2}{r_1} - \frac{2}{r_2} - \frac{2}{r_{13}} + \frac{2}{r_{23}} \right) n_1(\kappa\rho) f_1(\rho) \right. \right. \\
\nonumber & + \left.\left. \left[f_1^\prime(\rho) \frac{1}{\rho} \left( n_1(\kappa\rho) + \cos(\kappa\rho) \right) - \frac{1}{2} f_1^{\prime\prime}(\rho) n_1(\kappa\rho) \right]\right] \pm \frac{r_1 + r_3 \cos\theta_{13}}{\rho^\prime}  \Phi_{Ps}(r_{13}) \Phi_H(r_2) \right. \\
\nonumber & \times \left[ \left( \frac{2}{r_1} - \frac{2}{r_3} - \frac{2}{r_{12}} + \frac{2}{r_{23}} \right) n_1(\kappa\rho^\prime) f_1(\rho^\prime) \right. \\ 
& \left.\left. + \ \left[f_1^\prime(\rho^\prime) \frac{1}{\rho^\prime} \left( n_1(\kappa\rho^\prime) + \cos(\kappa\rho^\prime) \right) - \frac{1}{2} f_1^{\prime\prime}(\rho^\prime) n_1(\kappa\rho^\prime) \right]\right]\right\} d\tau_{int}
\end{align}

\begin{align}
\label{eq:PWavePhi2SBar}
\nonumber \left(\bar{\phi}_{2j},\mathcal{L} \bar{S}_1\right) = \sqrt{2} \pi & \int \phi_{2j} \left[ \frac{1}{\rho} \left(r_2 + r_1 \cos\theta_{12}\right) \left( \frac{2}{r_1} - \frac{2}{r_2} - \frac{2}{r_{13}} + \frac{2}{r_{23}} \right) S_{22} \right. \\
& \pm \left. \frac{1}{\rho^\prime} \left(r_1 \cos\theta_{12} + r_3 \cos\theta_{23}\right) \left( \frac{2}{r_1} - \frac{2}{r_3} - \frac{2}{r_{12}} + \frac{2}{r_{23}} \right) S_{23} \right]  d\tau_{int}
\end{align}

\begin{align}
\label{eq:PWavePhi2CBar}
\nonumber (\bar{\phi}_{2j},&\mathcal{L} \bar{C}_1) = -2 \uppi \sqrt{\kappa} \int \phi_{2j} \\
&\times \left\{ \frac{r_2 + r_1 \cos\theta_{12}}{\rho} \Phi_{Ps}(r_{12}) \Phi_H(r_3) \left[ \left( \frac{2}{r_1} - \frac{2}{r_2} - \frac{2}{r_{13}} + \frac{2}{r_{23}} \right) n_1(\kappa\rho) f_1(\rho) \right. \right. \\
\nonumber & + \left.\left. \left[f_1^\prime(\rho) \frac{1}{\rho} \left( n_1(\kappa\rho) + \cos(\kappa\rho) \right) - \frac{1}{2} f_1^{\prime\prime}(\rho) n_1(\kappa\rho) \right]\right] \right. \\
\nonumber & \pm \frac{r_1 \cos\theta_{12} + r_3 \cos\theta_{23}}{\rho^\prime}  \Phi_{Ps}(r_{13}) \Phi_H(r_2) \\
& \times \left. \left[ \left( \frac{2}{r_1} - \frac{2}{r_3} - \frac{2}{r_{12}} + \frac{2}{r_{23}} \right) n_1(\kappa\rho^\prime) f_1(\rho^\prime) \right.\right. \\
\nonumber & + \left.\left. \left[f_1^\prime(\rho^\prime) \frac{1}{\rho^\prime} \left( n_1(\kappa\rho^\prime) + \cos(\kappa\rho^\prime) \right) - \frac{1}{2} f_1^{\prime\prime}(\rho^\prime) n_1(\kappa\rho^\prime) \right]\right]\right\}
\end{align}

\subsection{Long-Range -- Long-Range Matrix Elements}
\label{sec:PWaveLongLong}

The long-long matrix elements are a straightforward application of the
$\mathcal{L}$ operator on $S_1$ and $C_1$. The resulting integrals after
external angular integration (see \cref{chp:AngularInt}) are given below: 

\begin{align}
\label{eq:PWaveSBarSBar}
\left(\bar{S}_1,\mathcal{L}\bar{S}_1\right) = \pm \frac{\uppi}{2} \int \frac{1}{\rho\rho^\prime} & \left[S_1' S_1 \left(\frac{2}{r_1} - \frac{2}{r_2} - \frac{2}{r_{13}} + \frac{2}{r_{23}} \right) \right.  \nonumber \\
& \left. \times \left(r_1^2 + r_1 r_2 \cos\theta_{12} + r_1 r_3 \cos\theta_{13} + r_2 r_3 \cos\theta_{23} \right) \right] d\tau_{int}
\end{align}

\begin{align}
\label{eq:PWaveCBarSBar}
\left(\bar{C}_1,\mathcal{L}\bar{S}_1\right) = \pm \frac{\uppi}{2} \int \frac{1}{\rho\rho^\prime} & \left[C_1' S_1 \left(\frac{2}{r_1} - \frac{2}{r_2} - \frac{2}{r_{13}} + \frac{2}{r_{23}} \right) \right.  \nonumber \\
& \left. \times \left(r_1^2 + r_1 r_2 \cos\theta_{12} + r_1 r_3 \cos\theta_{13} + r_2 r_3 \cos\theta_{23} \right) \right] d\tau_{int}
\end{align}

\begin{align}
\label{eq:PWaveCBarCBar}
\nonumber \left(\bar{C}_1,\mathcal{L}\bar{C}_1\right) = 2 \uppi \kappa & \int \Phi_{Ps}(r_{12}) \Phi_H(r_3) \Bigg\{ 2 \Phi_{Ps}(r_{12}) \Phi_H(r_3) n_1(\kappa\rho) f_1(\rho) \\
\nonumber & \times \left(\frac{1}{\rho} f_1^\prime(\rho) \left(n_1(\kappa\rho) + \cos(\kappa\rho)\right) - \frac{1}{2} f_1^{\prime\prime}(\rho) n_1(\kappa\rho)\right) \\
\nonumber \pm & \frac{1}{2\rho\rho^\prime} (r_1^2 + r_1 r_2 \cos\theta_{12} + r_1 r_3 \cos\theta_{13} + r_2 r_3 \cos\theta_{23}) \\
\nonumber & \times \Phi_{Ps}(r_{13}) \Phi_H(r_2) n_1(\kappa\rho^\prime) f_1(\rho^\prime) \\
\nonumber & \times \left[ n_1(\kappa\rho) f_1(\kappa\rho) \left(\frac{2}{r_1} - \frac{2}{r_2} - \frac{2}{r_{13}} + \frac{2}{r_{23}} \right) \right. \\
& \; \; + \left. \left(\frac{1}{\rho} f_1^\prime(\rho) \left(n_1(\kappa\rho) + \cos(\kappa\rho)\right) - \frac{1}{2} f_1^{\prime\prime}(\rho)  n_1(\kappa\rho)\right) \right]\Bigg\} d\tau_{int}
\end{align}

\section{Second Formalism}
\label{sec:PWave2Formalism}

Van Reeth and Humberston \cite{VanReeth2004} had difficulty with convergence 
of the $^3$P phase shifts. They specifically state:
\begin{quote}
``The present triplet p-wave phase shifts are not yet fully converged; at very 
low energies we estimate them to be $\approx 20\%$ below the fully converged 
values and $\approx 3\%$ in the higher energy range. This relatively poor 
convergence is due to the fact that we have not yet included in the 
wavefunction a set of terms for which the unit of angular momentum is on the 
electron in the H atom.''
\end{quote}
Due to this, we looked at a second formalism that includes these short-range 
terms instead of what we call the first formalism in \cref{sec:PWaveFn}. This
second formalism places the orbital angular momentum mainly on the Ps
($\bm{\rho}$) and the electron of H ($\bm{r}_3$, and $\bm{r}_2$ with exchange).

\subsection{Wavefunction}
The trial wavefunction that we use for the second formalism is similar to the
first formalism in \cref{sec:PWaveFn}.
\begin{equation}
\Psi_1^{\pm,t} = \widetilde{S}_1 + L_1^{\pm,t} \, \widetilde{C}_1 + \sum_{i=1}^{N'(\omega)} c_i \bar{\phi}_{\rho i} + \sum_{j=1}^{N'(\omega)} d_j \bar{\phi}_{3j}
\label{eq:PWave2ndWavefn}
\end{equation}

\noindent The new short-range terms are given by
\begin{subequations}
\label{eq:PWave2ndPhiBar}
\begin{align}
\bar{\phi}_{\rho i} &= \left(1 \pm P_{23}\right) \SphericalHarmonicY{1}{0}{\theta_\rho}{\varphi_\rho} \rho \, \phi_i \label{eq:PWave2ndPhi1i}\\
\bar{\phi}_{3j} &= \left(1 \pm P_{23}\right) \SphericalHarmonicY{1}{0}{\theta_3}{\varphi_3} r_3 \phi_j \label{eq:PWave2ndPhi2j}.
\end{align}
\end{subequations}
These place the angular momentum on the Ps and on the electron of H, instead of
on the positron and the electron in the Ps. In the first formalism, some amount
of the angular momentum is placed on the second electron through the $P_{23}$
permutation operator.

The short-long terms can be directly computed by changing the short-range terms
in the code, but the integration routines (see \cref{sec:ShortInt}) for the
short-short terms cannot handle the $\rho$ factor of $\bar{\phi}_{\rho 1}$
directly. The following relations are used:
\begin{subequations}
\label{eq:P2rhoY10}
\begin{align}
\rho \SphericalHarmonicY{1}{0}{\theta_\rho}{\varphi_\rho} &= \frac{1}{2}\left[ r_1 \SphericalHarmonicY{1}{0}{\theta_1}{\varphi_1} + r_2 \SphericalHarmonicY{1}{0}{\theta_2}{\varphi_2} \right] \\
\rho^\prime \SphericalHarmonicY{1}{0}{\theta_{\rho'}}{\varphi_{\rho'}} &= \frac{1}{2}\left[ r_1 \SphericalHarmonicY{1}{0}{\theta_1}{\varphi_1} + r_3 \SphericalHarmonicY{1}{0}{\theta_3}{\varphi_3} \right].
\end{align}
\end{subequations}
The first of these is obtained by substituting
\begin{equation}
\label{eq:CosRho}
\cos\theta_\rho = \frac{r_1 \cos\theta_1 + r_2 \cos\theta_2}{2\rho}
\end{equation}
in the $\SphericalHarmonicY{1}{0}{\theta_\rho}{\varphi_\rho}$ spherical
harmonic. The second is found by using the $P_{23}$ permutation of this.

\subsection{Short-Short Matrix Elements}
\label{sec:PWave2ndShortShort}

The $\left(\bar{\phi}_{\rho i},\mathcal{L} \bar{\phi}_{\rho j}\right)$ matrix
elements of the second formalism can easily be rewritten in terms of the first
formalism by using \cref{eq:P2rhoY10}. In terms of the first formalism (denoted
by $1st$ subscripts),
\beq
\left(\bar{\phi}_{\rho i},\mathcal{L} \bar{\phi}_{\rho j}\right) = \frac{1}{4} \left[ \left(\bar{\phi}_{1i},\mathcal{L} \bar{\phi}_{1j}\right)_{1st} + \left(\bar{\phi}_{1i},\mathcal{L} \bar{\phi}_{2j}\right)_{1st} + \left(\bar{\phi}_{2i},\mathcal{L} \bar{\phi}_{1j}\right)_{1st} + \left(\bar{\phi}_{2i},\mathcal{L} \bar{\phi}_{2j}\right)_{1st} \right].
\eeq
The other types of terms cannot be expressed as combinations of the first
formalism results, but these calculations are similar to the first formalism.
\begin{align}
\nonumber \left(\bar{\phi}_{\rho i},\mathcal{L} \bar{\phi}_{3j}\right) = &(Y_{10}(\theta_1) r_1 \phi_i, \mathcal{L} Y_{10}(\theta_3) r_3 \phi_j) + (Y_{10}(\theta_2) r_2 \phi_i, \mathcal{L} Y_{10}(\theta_3) r_3 \phi_j) \\
\pm &(Y_{10}(\theta_1) r_1 \phi_i, \mathcal{L} Y_{10}(\theta_2) r_2 \phi_j^\prime) \pm (Y_{10}(\theta_2) r_2 \phi_i, \mathcal{L} Y_{10}(\theta_2) r_2 \phi_j^\prime)
\end{align}
\begin{align}
\nonumber \left(\bar{\phi}_{\rho i},\mathcal{L} \bar{\phi}_{3j}\right) = \pm &(Y_{10}(\theta_2) r_2 \phi_i^\prime, \mathcal{L} Y_{10}(\theta_1) r_1 \phi_j) \pm (Y_{10}(\theta_2) r_2 \phi_i^\prime, \mathcal{L} Y_{10}(\theta_2) r_2 \phi_j) \\
+ &(Y_{10}(\theta_2) r_2 \phi_i^\prime, \mathcal{L} Y_{10}(\theta_1) r_1 \phi_j^\prime) + (Y_{10}(\theta_2) r_2 \phi_i^\prime, \mathcal{L} Y_{10}(\theta_3) r_3 \phi_j^\prime)
\end{align}
\beq
\nonumber \left(\bar{\phi}_{\rho i},\mathcal{L} \bar{\phi}_{3j}\right) = 2 \left[ \pm(Y_{10}(\theta_2) r_2 \phi_i, \mathcal{L} Y_{10}(\theta_3) r_3 \phi_j) + (Y_{10}(\theta_2) r_2 \phi_i^\prime, \mathcal{L} Y_{10}(\theta_2) r_2 \phi_j^\prime) \right]
\eeq
Derivations for all four of these are found on figshare \cite{figshare}.

\subsection{Short-Long Matrix Elements}
\label{sec:PWave2ndShortLong}

The short-long second formalism matrix elements are easily written as
combinations of matrix elements from the first formalism. In terms of the first
formalism (denoted by $1st$ subscripts),
\begin{subequations}
\begin{align}
\left(\bar{\phi}_{\rho i},\mathcal{L} \bar{S}_1\right) &= \frac{1}{2} \left[ \left(\bar{\phi}_{1i},\mathcal{L} \bar{S}_1\right)_{1st} + \left(\bar{\phi}_{2i},\mathcal{L} \bar{S}_1\right)_{1st} \right] \\
\left(\bar{\phi}_{\rho i},\mathcal{L} \bar{C}_1\right) &= \frac{1}{2} \left[ \left(\bar{\phi}_{1i},\mathcal{L} \bar{C}_1\right)_{1st} + \left(\bar{\phi}_{2i},\mathcal{L} \bar{C}_1\right)_{1st} \right].
\end{align}
\end{subequations}
Since these are both linear combinations of the first formalism results,
the code is a straightforward adaptation of the first formalism code.

\subsection{Phase Shifts}
\label{sec:PWave2ndPhase}

\begin{table}
\centering
\begin{tabular}{c | c c | c c}
\toprule
\multicolumn{1}{c}{ } & \multicolumn{2}{c}{$\delta_1^+$} & \multicolumn{2}{c}{$\delta_1^-$} \\
$\kappa$ & $1^{st}$ Formalism & $2^{nd}$ Formalism & $1^{st}$ Formalism & $2^{nd}$ Formalism \\
\midrule
0.1 & $2.26^{-2}$ & $2.27^{-2}$ & $-1.79^{-3}$ & $-1.80^{-3}$ \\
0.2 & $1.91^{-1}$ & $1.91^{-1}$ & $-1.68^{-2}$ & $-1.67^{-2}$ \\
0.3 & $6.08^{-1}$ & $6.08^{-1}$ & $-5.54^{-2}$ & $-5.52^{-2}$ \\
0.4 & $9.93^{-1}$ & $9.93^{-1}$ & $-1.15^{-1}$ & $-1.15^{-1}$ \\
0.5 & $1.14$      & $1.14$      & $-1.84^{-1}$ & $-1.84^{-1}$ \\
0.6 & $1.16$      & $1.16$      & $-2.49^{-1}$ & $-2.48^{-3}$ \\
0.7 & $1.15$      & $1.15$      & $-2.93^{-1}$ & $-2.93^{-1}$ \\
\bottomrule
\end{tabular}
\caption[Comparison of $^{1,3}$P phase shifts for the first and second formalisms]{Comparison of $^{1,3}$P phase shifts for the first and second formalisms $(\omega = 6)$. The $2^{nd}$ formalism for $\delta_1^-$ uses 889 terms, while the others use 924 terms.}
\label{tab:PWaveFormalPhase}
\end{table}

The phase shifts for $^1$P and $^3$P in \cref{tab:PWaveFormalPhase} were not 
improved in general by using the second formalism and were even lower in
the $\kappa = 0.1$ triplet case.
One exception is $\kappa = 0.3$ for $^3$P, where the second formalism phase shift
is slightly higher. We note that this is the same result as the $\omega = 7$
case with 1000 terms as presented in \cref{tab:PWaveComparisons}.
The difficulty with the second formalism is that linear dependence becomes more
of an issue, forcing us to only use 889 terms for $\omega = 6$ and unable to
do an $\omega = 7$ run. I tried an $\omega = 7$ run with the Todd energy
program, but it could only select 767 terms, leading to worse phase shifts than
the $\omega = 6$ runs.

The energy eigenvalues in \cref{tab:SimplexPWaveSingOpt,tab:SimplexPWaveTripOpt}
were also lower for the first formalism, so we did not pursue using the second
formalism any further. The convergence of the phase shifts for the first
formalism, even for $^3$P, appears to be good. The second formalism was not
tried for the D-wave due to its difficulty and because it did not generally
improve the results for the P-wave.

\section{Results}

Results in this section were computed using the nonlinear parameters $\alpha$,
$\beta$, and $\gamma$ described in \cref{sec:PWaveOpt}, which has discussion
on how these were chosen.

\subsection{Phase Shifts}

\Cref{tab:PWaveComparisons} compares the current $S$-matrix complex Kohn phase shifts for the 
first P-wave formalism with that of other groups. We use $\omega = 4 - 7$ 
phase shifts to do the extrapolations for $\omega \rightarrow \infty$. The
\% Diff entries comparing the $\omega = 7$ and $\omega \to \infty$ extrapolated
phases shifts give an estimate of the error for the complex Kohn results. The singlet 
results are converged very well. The triplet results are also converged well, 
but there is more possible error at $\kappa = 0.1$ and $\kappa = 0.7$. Even 
with these slightly larger percentages, the $^3$P phase shifts appear to be 
well converged.

\begin{table}
\centering
\setlength{\tabcolsep}{-2pt}
\footnotesize
\begin{tabular}{@{\hskip 0.1cm}l . . . . . . .}
\toprule
Method & \multicolumn{1}{c}{\phantom{1}0.1} & \multicolumn{1}{c}{\phantom{1}0.2} & \multicolumn{1}{c}{\phantom{1}0.3} & \multicolumn{1}{c}{\phantom{1}0.4} & \multicolumn{1}{c}{\phantom{1}0.5} & \multicolumn{1}{c}{\phantom{1}0.6} & \multicolumn{1}{c}{\phantom{1}0.7} \\
\midrule
This work $(\omega = 7)$ $\delta_1^+$ 					& 0.226^{-1} & 0.191 & 0.609 & 0.994 & 1.140 & 1.162 & 1.152 \\
This work $(\omega \to \infty)$ $\delta_1^+$			& 0.227^{-1} & 0.192 & 0.611 & 0.996 & 1.142 & 1.163 & 1.154 \\
\% Diff$^+$												& 0.465\% & 0.306\% & 0.314\% & 0.205\% & 0.140\% & 0.137\% & 0.181\% \\
\arrayrulecolor[RGB]{220,220,220}\midrule\arrayrulecolor{black}
Kohn $(\omega = 6)$ \cite{VanReethPrivate} $\delta_1^+$	& 0.226^{-1} & 0.192 & 0.612 & 0.997 & 1.143 & 1.165 & 1.155 \\
CC 9Ps9H+H$^-$ \cite{Walters2004} $\delta_1^+$			& 0.221^{-1} & 0.183 & 0.580 & 0.956 & 1.106 & 1.134 & 1.133 \\
CC 9Ps9H \cite{Blackwood2002} $\delta_1^+$				& 0.213^{-1} & 0.175 & 0.545 & 0.908 & 1.068 & 1.103 & 1.099 \\
3-state CC \cite{Sinha1997} $\delta_1^+$				& 8.14^{-3} & 6.27^{-2} & 0.190 & 0.358 & 0.489 & 0.551 & 0.556 \\
SE \cite{Ray1997} $\delta_1^+$ 							& 0.798^{-2} & 0.614^{-1} & 0.186 & 0.349 & 0.477 & 0.536 & 0.538 \\
5-state CC \cite{Adhikari1999} $\delta_1^+$				& 0.477^{-2} & 0.370^{-1} & 0.116 & 0.239 & 0.372 & 0.478 & 0.541 \\
SE \cite{Hara1975} $\delta_1^+$							& 0.79^{-2}  & 0.611^{-1} & 0.1853 & 0.3487 & 0.4772 & 0.5361 & 0.5388 \\
\midrule                                                
This work $(\omega = 7)$ $\delta_1^-$					& -0.178^{-2} & -0.167^{-1} & -0.552^{-1} & -0.115 & -0.183 & -0.248 & -0.292 \\
This work $(\omega \to \infty)$ $\delta_1^-$			& -0.172^{-2} & -0.165^{-1} & -0.540^{-1} & -0.114 & -0.182 & -0.246 & -0.288 \\
\% Diff$^-$												& 3.176\% & 0.993\% & 0.749\% & 0.698\% & 0.749\% & 0.896\% & 1.237\% \\
\arrayrulecolor[RGB]{220,220,220}\midrule\arrayrulecolor{black}
CC 9Ps9H \cite{Blackwood2002} $\delta_1^-$				& -0.953^{-3} & -0.122^{-1} & -0.456^{-1} & -0.104 & -0.178 & -0.247 & -0.295 \\
3-state CC \cite{Sinha1997} $\delta_1^-$				& -4.43^{-3} & -3.08^{-2} & -8.51^{-2} & -0.159 & -0.236 & -0.302 & -0.332 \\
SE \cite{Ray1997} $\delta_1^-$							& -0.503^{-2} & -0.352^{-1} & -0.980^{-1} & -0.186 & -0.287 & -0.390 & -0.488 \\
5-state CC \cite{Adhikari1999} $\delta_1^-$				& -0.233^{-2} & -0.167^{-1} & -0.476^{-1} & -0.918^{-1} & -0.142 & -0.190 & -0.228 \\
SE \cite{Hara1975} $\delta_1^-$							& -0.50^{-2}  & -0.350^{-1} & -0.978^{-1} & -0.1860 & -0.2872 & -0.3906 & -0.4882 \\
\bottomrule
\end{tabular}
\caption[$^{1,3}$P comparisons]{Comparison of the $S$-matrix complex Kohn
$^{1,3}$P phase shifts with results from other groups. \% Diff$^\pm$ is the percent difference
between the current complex Kohn $\omega = 7$ and $\omega \rightarrow \infty$
results. Values in the header are $\kappa$ in au. Exponents denote powers of 10.}
\label{tab:PWaveComparisons}
\end{table}

The different Kohn-type variational methods generate phase shifts that agree
with the $S$-matrix complex Kohn results presented in \cref{tab:PWaveComparisons}
when methods with Schwartz singularities are removed. The previous Kohn / inverse
Kohn $^1$P phase shifts \cite{VanReeth2004}, published only in graph form in
their paper, agree well with the current $S$-matrix complex Kohn phase shifts
though generally are slightly higher.

\begin{figure}
	\centering
	\includegraphics[width=\textwidth]{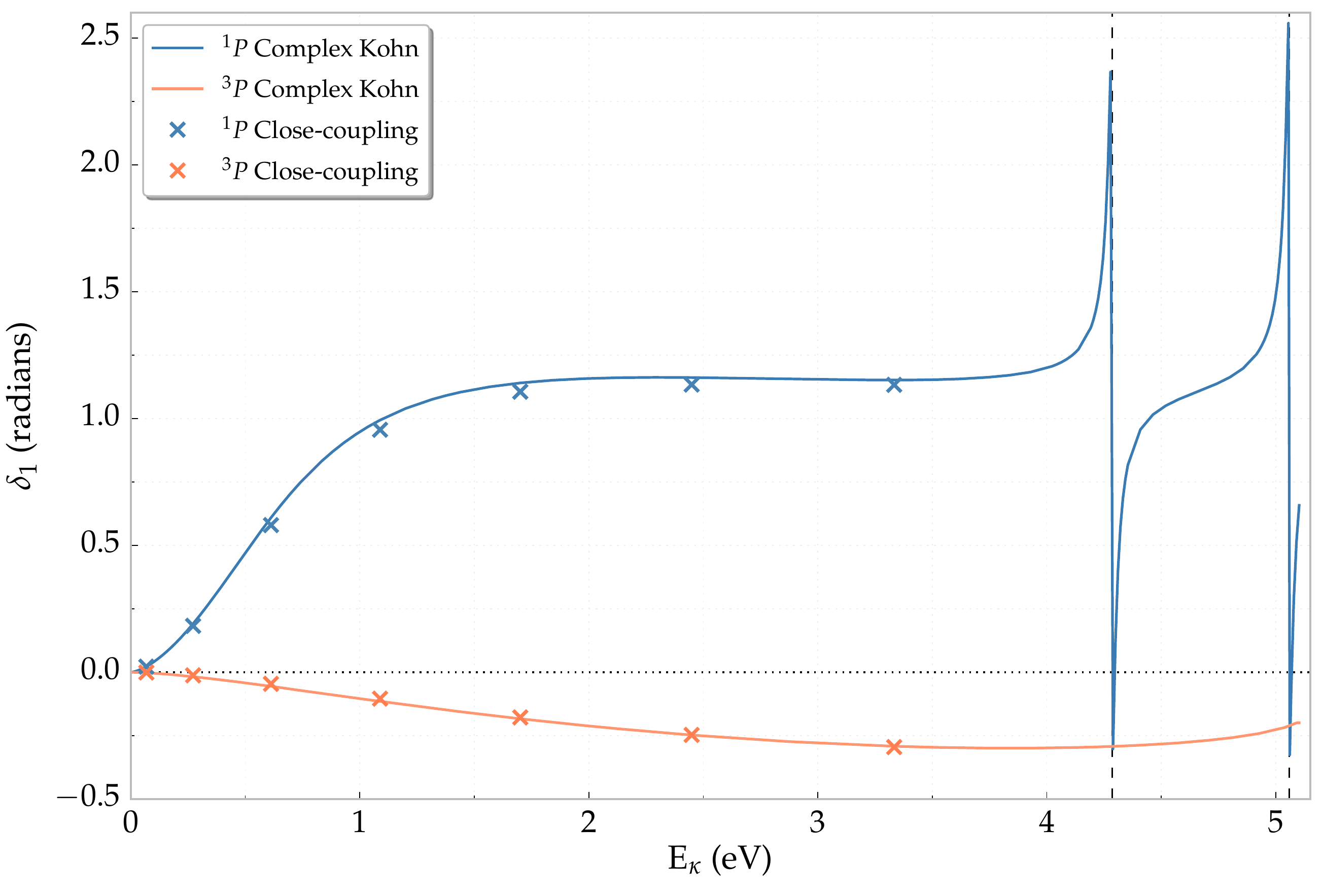}
	\caption[$^{1,3}$P phase shifts]{$^{1,3}$P complex Kohn phase shifts. The $^1$P CC phase shifts
\cite{Walters2004} are given by \mbox{\textcolor{blue}{$\times$}}, and the
$^3$P CC phase shifts \cite{Blackwood2002} are given by
\mbox{\textcolor{red}{\textbf{+}}}. Vertical dashed lines denote the complex rotation resonance
positions \cite{Yan1999}. An interactive version of this figure is available online \cite{Plotly}
at \url{https://plot.ly/~Denton/4/p-wave-ps-h-scattering/}.}
	\label{fig:PWavePhase}
\end{figure}

\Cref{tab:PWaveComparisons,fig:PWavePhase} give good comparisons with the CC 
results of the Belfast group \cite{Blackwood2002,Walters2004}. The complex 
Kohn phase shifts are higher for $^1$P, but they are slightly lower for much 
of the range for $^3$P. We note that this was also the case for the S-wave
(see \cref{tab:SWaveComparisons} on \pageref{tab:SWaveComparisons}), but the 
CVM results seem to confirm the complex Kohn $^3$S phase shifts. Both the 
$S$-matrix complex Kohn and the CC phase shifts of the Belfast group
appear to give the most accurate sets of phase shifts, with the current complex
Kohn giving highly accurate results.

\begin{figure}
	\centering
	\includegraphics[width=5.25in]{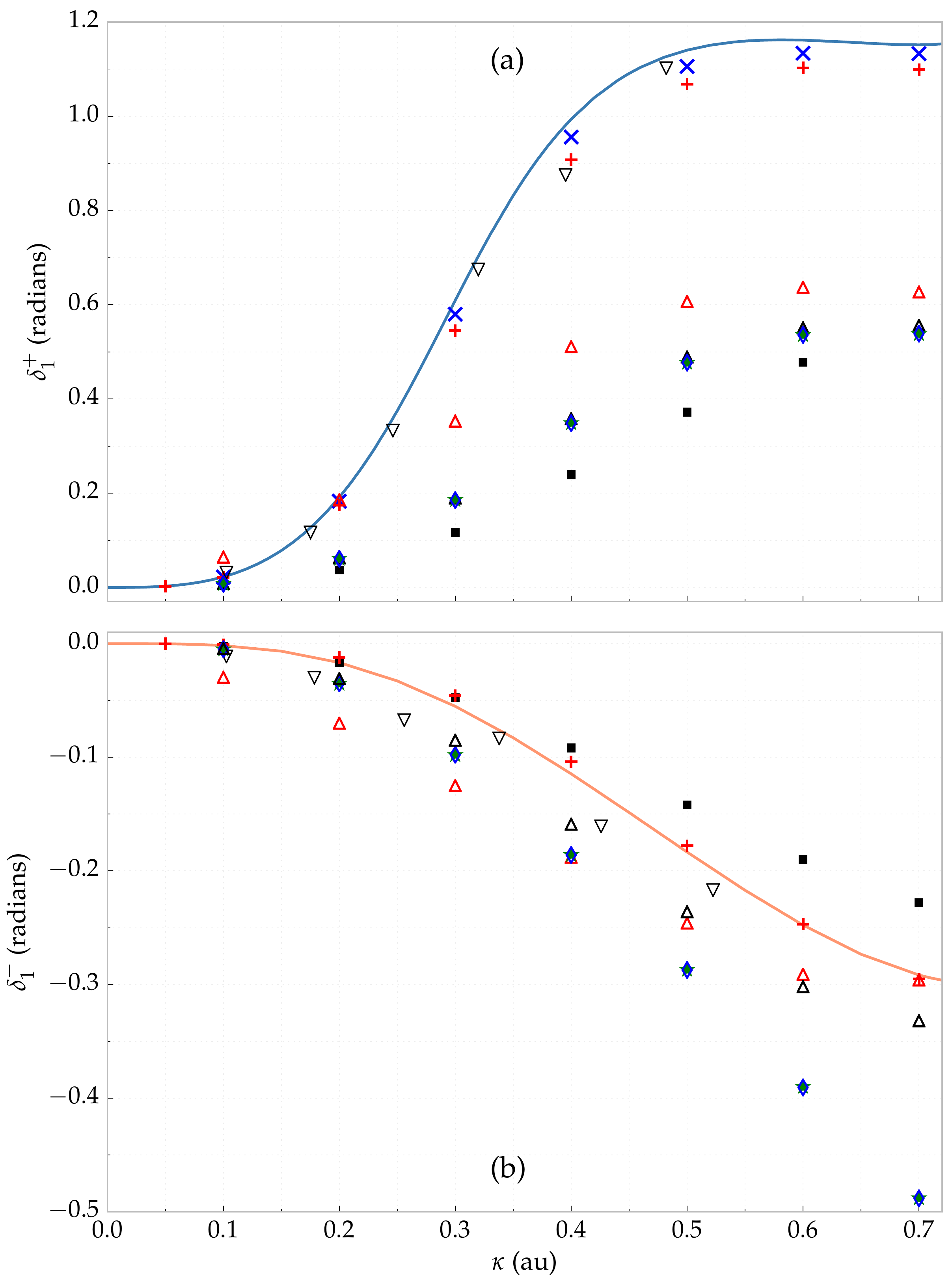}
	\caption[Comparison of P-wave phase shifts]{Comparison of $^1$P (a) and $^3$P (b) phase shifts with results from other groups. Results are ordered according to year of publication. Solid curves -- this work; \mbox{\textcolor{blue}{$\times$} -- CC \cite{Walters2004};} \mbox{$\CIRCLE$ -- Kohn \cite{VanReeth2003};} \mbox{\textcolor{red}{\textbf{+}} -- CC \cite{Blackwood2002};}\mbox{$\triangledown$ -- SVM 2002 \cite{Ivanov2002};} \mbox{\textcolor{red}{$\vartriangle$} -- 6-state CC \cite{Sinha2000};} \mbox{$\blacksquare$ -- 5-state CC \cite{Adhikari1999};} \mbox{$\vartriangle$ -- 3-state CC \cite{Sinha1997};} \mbox{\textcolor[RGB]{0,127,0}{$\bigstar$} -- CC \cite{Ray1997};} \mbox{\textcolor{blue}{$\lozenge$} -- Static-exchange \cite{Hara1975}.}}
	\label{fig:PWaveComparisons}
\end{figure}

\Cref{fig:PWaveComparisons} gives a detailed comparison to phase shifts from 
many other groups, similar to the S-wave comparisons in
\cref{fig:SWaveComparisons}. The 5-state CC method of Adhikari and Biswas
\cite{Adhikari1999} again gives phase shifts that do not agree well with other 
methods, and Ivanov et al. \cite{Ivanov2002} has a detailed discussion of 
this. The SVM results of Ivanov et al. \cite{Ivanov2002} follow along well 
with the current complex Kohn and the CC \cite{Blackwood2002,Walters2004} but 
are below throughout the energy range. The SE
\cite{Hara1975,Ray1997} and smaller CC calculations \cite{Sinha1997} do not
match the complex Kohn phase shifts well.

\subsection{Resonance Parameters}
\label{sec:PWaveResonances}

The P-wave has two 
resonances before the inelastic threshold, as seen in \cref{fig:PWavePhase}. 
These resonances are shifted to higher energies than the
S-wave, and we use the numerical fittings described in \cref{sec:ResonanceFit}
to determine the resonance parameters given in \cref{tab:PWaveResonancesOther}.

\setlength{\abovecaptionskip}{6pt}   
\setlength{\belowcaptionskip}{6pt}   
\begin{table}
\footnotesize
\centering
\begin{tabular}{l l l l l}
\toprule
Method & \thead{$^1E_R \text{ (eV)}$} & \thead{$^1\Gamma \text{ (eV)}$} & \thead{$^2E_R \text{ (eV)}$} & \thead{$^2\Gamma \text{ (eV)}$} \\
\midrule
Current work: &  &  &  &  \\
\ Average $\pm$ standard deviation & $4.2856 \pm 0.0001$ & $0.0445 \pm 0.0001$ & $5.0577 \pm 0.0004$ & $0.0459 \pm 0.0005$ \\
Current work: &  &  &  &  \\
\ $S$-matrix complex Kohn & $4.2856$ & $0.0445$ & $5.0579$ & $0.0459$ \\
Kohn variational \cite{VanReeth2004} & $4.29 \pm 0.01$ & $0.042 \pm 0.005$ & --- & --- \\
CC (9Ps9H + H$^-$) \cite{Walters2004} & $4.475$ & $0.0827$ & $4.905$ & $0.0043$ \\
Stabilization \cite{Yan2003} & $4.287$ & $0.0446$ & $5.062$ & $0.0563$ \\
CC (22Ps1H + H$^-$) \cite{Blackwood2002b} & $4.401$ & $0.029$ & $5.108$ & $0.017$ \\
Optical potential \cite{DiRienzi2002b} & $4.472$ & $0.082$ & --- & --- \\
Five-state CC \cite{Adhikari2001e} & $5.08$ & $0.004$ & --- & --- \\
CC (9Ps9H) \cite{Blackwood2002} & $4.66$ & $0.084$ & --- & --- \\
Complex rotation \cite{Yan1999} & $4.2850 \pm 0.0014$ & $0.0435 \pm 0.0027$ & $5.0540 \pm 0.0027$ & $0.0585 \pm 0.0054$ \\
Coupled-pseudostate \cite{Campbell1998} & $4.88$ & $0.058$ & --- & --- \\
\bottomrule
\end{tabular}
\caption{$^1$P-wave resonance parameters}
\label{tab:PWaveResonancesOther}
\end{table}

\Cref{tab:PWaveResonancesOther} has the $^1$P-wave resonances calculated in this 
work and compared to that of calculations from other groups. The complex 
rotation result of Yan and Ho \cite{Yan1999} is one of the most accurate 
calculations of these resonance parameters. The positions of the first and 
resonances using the complex Kohn agree very well with the complex rotation, 
as does the width of the first resonance. We find a width for the second 
resonance about half that of the complex rotation. A more recent result of 
Yan and Ho uses a stabilization method \cite{Yan2003}, which has a result for the
second resonance closer to the complex Kohn result.

The CC results of the Belfast group \cite{Blackwood2002,Blackwood2002b,Walters2004}
show again that the H$^-$ channel is important for the resonances. 
The $^1E_R$ that they calculate is still higher 
than this work and that of Yan and Ho \cite{Yan1999,Yan2003}, but it is roughly comparable. 
The $^2E_R$ they calculate is lower than the complex 
Kohn, stabilization and complex rotation, although it is also close. The complex
Kohn and complex rotation second resonance widths are much larger than the CC.

\section{Summary}
\label{sec:SummaryP}

We have investigated the P-wave by using two different formalisms but found little 
difference in the resulting phase shifts between them. Using the first formalism
$S$-matrix results, we have obtained accurate $^{1,3}$P-wave phase shifts that
converge well and compare against the results from multiple groups. We also
obtain reliable $^1$P resonance parameters that compare well with those of other
groups.


\chapter{D-Wave}
\label{chp:DWave}

\iftoggle{UNT}{Van}{\lettrine{\textcolor{startcolor}{V}}{an}}
Reeth \cite{VanReethPrivate}
had D-wave notes that I used as guidance for
the D-wave derivations. He also had code for the $^1$D-wave that we used to
compare our independent code against after it was written. His D-wave work was
unpublished, unlike that of the S-wave \cite{VanReeth2003,VanReeth2004}
and P-wave \cite{VanReeth2004}.

\section{Wavefunction}
\label{sec:DWaveFn}

Similar to the discussions for the S-wave (\cref{chp:SWave}) and the P-wave
(\cref{chp:PWave}), here we present the D-wave wavefunction for multiple Kohn-type
variational methods in \cref{eq:DWaveTrial}, but throughout this chapter, we
only consider the Kohn variational method. The wavefunction for each of the 
variants of the Kohn can be easily constructed from this by using the general
formalism in \cref{sec:GeneralWave,sec:KohnApplied}. The general D-wave
wavefunction is given by
\begin{equation}
\Psi_t^\pm = \widetilde{S}_2 + L_2^{\pm,t} \, \widetilde{C}_2 + \sum_{i=1}^{N'(\omega)} c_i \bar{\phi}_{1i} + \sum_{j=1}^{N'(\omega)} d_j \bar{\phi}_{2j} + \sum_{j=1}^{N'(\omega)} f_k \bar{\phi}_{12k}.
\label{eq:DWaveTrial}
\end{equation}
The long-range $\widetilde{S}_\ell$ and $\widetilde{C}_\ell$ are given by
\cref{eq:TildeSCDef}.
As noted in \cref{sec:GeneralWave}, the full wavefunction has $(\ell+1) = 3$
short-range symmetries. The short-range terms are given by
\begin{subequations}
\label{eq:DWavePhiBar}
\begin{align}
\bar{\phi}_{1i} &= \left(1 \pm P_{23}\right) Y_{20}(\theta_1) r_1^2 \phi_i \label{eq:DWavePhi1i}\\
\bar{\phi}_{2j} &= \left(1 \pm P_{23}\right) Y_{20}(\theta_2) r_2^2 \phi_j \label{eq:DWavePhi2j}\\
\bar{\phi}_{12k} &= \left(1 \pm P_{23}\right) \psi_{(1,1,2,0)}(\theta_1,\theta_2) r_1 r_2 \phi_k, \label{eq:DWavePhi12k}
\end{align}
\end{subequations}
where $\phi_i$, $\phi_j$ and $\phi_k$ are given by \cref{eq:BoundWavefn_phi}.
We also use the shortcuts
\begin{equation}
\label{eq:DWavePhi}
\phi_{1i} = r_1^2 \phi_i \text{, }
\phi_{2j} = r_2^2 \phi_j \text{, and }
\phi_{12k} = r_1 r_2 \phi_k.
\end{equation}
The mixed symmetry terms, given as $\bar{\phi}_{12k}$,
are discussed in \cref{sec:MixedTerms}. There is also a second formalism,
similar to that for the P-wave (\cref{sec:PWave2Formalism}), which we did not
use.

Only the results of the derivations of the matrix elements are shown here, but
full derivations are found on figshare \cite{figshare}.

\section{Short-Range -- Short-Range Matrix Elements}
\label{sec:DWaveShortShort}

The D-wave short-short integrals are generally more complicated than those 
for the S-wave and P-wave.
\Cref{eq:DWavePhi1Phi1,eq:DWavePhi2Phi2,eq:DWavePhi1Phi2,eq:DWavePhi2Phi1}
give the short-short integrals needed evaluate the 
matrix in \cref{eq:GeneralKohnMatrix}. Full derivations for each of these are 
given in separate notes available on the Wiki \cite{Wiki} and figshare
\cite{figshare}.

\begin{align}
\label{eq:DWavePhi1Phi1}
\left(\bar{\phi}_{1i},\mathcal{L} \bar{\phi}_{1j}\right) = &2 \cdot 2\pi \bigintsss \Bigg\{ \sum_{k=1}^3 \left[ \boldsymbol{\nabla}_{\!\mathbf{r}_k} \nonumber \phi_{1i} \boldsymbol{\cdot} \boldsymbol{\nabla}_{\!\mathbf{r}_k} \phi_{1j} \pm \boldsymbol{\nabla}_{\!\mathbf{r}_k} \phi_{1i} \boldsymbol{\cdot} \boldsymbol{\nabla}_{\!\mathbf{r}_k} \phi_{1j}^\prime \right]  \\
\nonumber  &+ \left. \left[\frac{2}{r_1} - \frac{2}{r_2} - \frac{2}{r_3} - \frac{2}{r_{12}} - \frac{2}{r_{13}} + \frac{2}{r_{23}} - 2 E_H - 2 E_{Ps} - \frac{1}{2}\kappa^2 + \frac{6}{r_1^2} \right] \right. \\
 &\;\;\;\;\; \times \left(\phi_{1i} \phi_{1j} \pm \phi_{1i} \phi_{1j}^\prime \right) \Bigg\} d\tau_{int}
\end{align}
\begin{align}
\label{eq:DWavePhi2Phi2}
\left(\bar{\phi}_{2i},\mathcal{L} \bar{\phi}_{2j}\right) = 2 & \cdot 2\pi \bigintsss \Bigg\{ \sum_{k=1}^3 \left[ \boldsymbol{\nabla}_{\!\mathbf{r}_k} \nonumber \phi_{2i} \boldsymbol{\cdot} \boldsymbol{\nabla}_{\!\mathbf{r}_k} \phi_{2j} \pm \left(1-\tfrac{3}{2}\sin^2\theta_{23}\right) \boldsymbol{\nabla}_{\!\mathbf{r}_k} \phi_{2i} \boldsymbol{\cdot} \boldsymbol{\nabla}_{\!\mathbf{r}_k} \phi_{2j}^\prime \right]  + \frac{6}{r_2^2}\phi_{2i}\phi_{2j} \\
 \nonumber &\mp 3 \phi_{2i} \phi_{2j}^\prime \cos\theta_{23} \left[p_i \frac{r_1}{r_3 r_{13}^2} (\cos\theta_{12} - \cos\theta_{23} \cos\theta_{13}) + m_j^\prime \frac{r_1}{r_2 r_{12}^2}(\cos\theta_{13} \right.\\
 \nonumber & \left. \;\;\;\;\; - \cos\theta_{23} \cos\theta_{12}) + \sin^2\theta_{23} \left(q_i \frac{r_2}{r_3 r_{23}^2} + q_j^\prime \frac{r_3}{r_2 r_{23}^2} \right) \right] \\
 \nonumber &+ \left. \left[\frac{2}{r_1} - \frac{2}{r_2} - \frac{2}{r_3} - \frac{2}{r_{12}} - \frac{2}{r_{13}} + \frac{2}{r_{23}} - 2 E_H - 2 E_{Ps} - \frac{1}{2}\kappa^2 \right] \right. \\
 &\;\;\;\;\; \times \left[\phi_{2i} \phi_{2j} \pm \left(1-\tfrac{3}{2}\sin^2\theta_{23}\right) \phi_{2i} \phi_{2j}^\prime \right] \Bigg\} d\tau_{int}
\end{align}
\begin{align}
\label{eq:DWavePhi1Phi2}
\left(\bar{\phi}_{1i},\mathcal{L} \bar{\phi}_{2j}\right) = 2 & \cdot 2\pi \bigintsss \Bigg\{ \sum_{k=1}^3 \left[ \left(1-\tfrac{3}{2}\sin^2\theta_{12}\right) \boldsymbol{\nabla}_{\!\mathbf{r}_k} \nonumber \phi_{1i} \boldsymbol{\cdot} \boldsymbol{\nabla}_{\!\mathbf{r}_k} \phi_{2j} \right. \\
\nonumber & \left. \;\;\;\;\; \pm \left(1-\tfrac{3}{2}\sin^2\theta_{13}\right) \boldsymbol{\nabla}_{\!\mathbf{r}_k} \phi_{1i} \boldsymbol{\cdot} \boldsymbol{\nabla}_{\!\mathbf{r}_k} \phi_{2j}^\prime \right] \\
 \nonumber &\mp 3 \phi_{1i} \phi_{2j} \cos\theta_{12} \left[q_i \frac{r_3}{r_2 r_{23}^2} (\cos\theta_{13} - \cos\theta_{12} \cos\theta_{23}) + p_j \frac{r_3}{r_1 r_{13}^2}(\cos\theta_{23} \right.\\
 \nonumber & \left. \;\;\;\;\;  - \cos\theta_{12} \cos\theta_{13}) + \sin^2\theta_{12} \left(m_i \frac{r_1}{r_2 r_{12}^2} + m_j \frac{r_2}{r_1 r_{12}^2} \right) \right] \\
 \nonumber &\mp 3 \phi_{1i} \phi_{2j}^\prime \cos\theta_{13} \left[q_i \frac{r_2}{r_3 r_{23}^2} (\cos\theta_{12} - \cos\theta_{13} \cos\theta_{23}) + m_j^\prime \frac{r_2}{r_1 r_{12}^2}(\cos\theta_{23} \right.\\
 \nonumber & \left. \;\;\;\;\;  - \cos\theta_{12} \cos\theta_{13}) + \sin^2\theta_{13} \left(p_i \frac{r_1}{r_3 r_{13}^2} + p_j^\prime \frac{r_3}{r_1 r_{13}^2} \right) \right] \\
 \nonumber &+ \left. \left[\frac{2}{r_1} - \frac{2}{r_2} - \frac{2}{r_3} - \frac{2}{r_{12}} - \frac{2}{r_{13}} + \frac{2}{r_{23}} - 2 E_H - 2 E_{Ps} - \frac{1}{2}\kappa^2 \right] \right. \\
 &\;\;\;\;\; \times \left[\left(1-\tfrac{3}{2}\sin^2\theta_{12}\right) \phi_{1i} \phi_{2j} \pm \left(1-\tfrac{3}{2}\sin^2\theta_{13}\right) \phi_{1i} \phi_{2j}^\prime \right] \Bigg\} d\tau_{int}
\end{align}
\begin{align}
\label{eq:DWavePhi2Phi1}
\left(\bar{\phi}_{2i},\mathcal{L} \bar{\phi}_{1j}\right) = 2 & \cdot 2\pi \bigintsss \Bigg\{ \sum_{k=1}^3 \left(1-\tfrac{3}{2}\sin^2\theta_{12}\right) \left[ \boldsymbol{\nabla}_{\!\mathbf{r}_k} \nonumber \phi_{2i} \boldsymbol{\cdot} \boldsymbol{\nabla}_{\!\mathbf{r}_k} \phi_{1j} \pm \boldsymbol{\nabla}_{\!\mathbf{r}_k} \phi_{2i} \boldsymbol{\cdot} \boldsymbol{\nabla}_{\!\mathbf{r}_k} \phi_{1j}^\prime \right] \\
 \nonumber &\mp 3 \phi_{2i} \phi_{1j} \cos\theta_{12} \left[p_i \frac{r_3}{r_1 {r_{13}}^2} (\cos\theta_{23} - \cos\theta_{12} \cos\theta_{13}) + q_j \frac{r_3}{r_2 {r_{23}}^2}(\cos\theta_{13} \right.\\
 \nonumber & \left. \;\;\;\;\; - \cos\theta_{12} \cos\theta_{23}) + \sin^2\theta_{12} \left(m_i \frac{r_2}{r_1 {r_{12}}^2} + m_j \frac{r_1}{r_2 {r_{12}}^2} \right) \right] \\
 \nonumber &\mp 3 \phi_{2i} \phi_{1j}^\prime \cos\theta_{12} \left[p_i \frac{r_3}{r_1 r_{13}^2} (\cos\theta_{23} - \cos\theta_{12} \cos\theta_{13}) + q_j^\prime \frac{r_3}{r_2 {r_{23}}^2}(\cos\theta_{13} \right.\\
 \nonumber & \left. \;\;\;\;\; - \cos\theta_{12} \cos\theta_{23}) + \sin^2\theta_{12} \left(m_i \frac{r_2}{r_1 {r_{12}}^2} + m_j^\prime \frac{r_1}{r_2 {r_{12}}^2} \right) \right] \\
 \nonumber &+ \left. \left[\frac{2}{r_1} - \frac{2}{r_2} - \frac{2}{r_3} - \frac{2}{r_{12}} - \frac{2}{r_{13}} + \frac{2}{r_{23}} - 2 E_H - 2 E_{Ps} - \frac{1}{2}\kappa^2 \right] \right. \\
 &\;\;\;\;\; \times \left(1-\tfrac{3}{2}\sin^2\theta_{12}\right) \left( \phi_{2i} \phi_{1j} \pm \phi_{2i} \phi_{1j}^\prime \right) \Bigg\} d\tau_{int}
\end{align}
The $\cos\theta_{ij}$ factors in each of these must be re-expressed in terms of $r_i$
and $r_{ij}$ by using \cref{eq:Cosines} so that these can solved using the
techniques given in \cref{sec:ShortInt}. Similar to the P-wave, I wrote a
\emph{Mathematica} notebook (see \cref{chp:Programs}) that transforms
these equations into a form that allows us to copy and paste them directly
into Fortran code.

\section{Short-Range -- Long-Range  Matrix Elements}
\label{sec:DWaveShortLong}

To calculate $\mathcal{L} C_2$, we again use the code in \cref{fig:LCMath} to get
\begin{align}
\label{eq:LCMathD}
\nonumber \frac{1}{2} \left(\Laplacian_\rho + \kappa^2\right) \SphericalHarmonicY{2}{0}{\theta_\rho}{\varphi_\rho} & n_2(\kappa\rho) f_2(\rho) = \\
\nonumber -\frac{1}{2 \kappa ^3 \rho ^4} &\left\{\rho  f_2^{\prime\prime}(\rho ) \left[\left(3-\kappa ^2 \rho ^2\right) \cos (\kappa  \rho )+3 \kappa  \rho  \sin (\kappa  \rho )\right] \right. \\
& \left.+2 f_2^\prime(\rho ) \left[\kappa  \rho  \left(\kappa ^2 \rho ^2-6\right) \sin (\kappa  \rho )+3 \left(\kappa ^2 \rho ^2-2\right) \cos (\kappa  \rho )\right] \right\}.
\end{align}

To simplify these equations, define
\begin{align}
\nonumber \mathscr{L} S_2 = &\frac{\mathcal{L} S_2}{Y_{20}(\theta_\rho)} = \left(\frac{2}{r_1} - \frac{2}{r_2} - \frac{2}{r_{13}} + \frac{2}{r_{23}} \right) \Phi_{Ps}(r_{12}) \Phi_H(r_3) \sqrt{2\kappa} \, j_2(\kappa\rho) \\
\nonumber \mathscr{L} C_2 = &\frac{\mathcal{L} C_2}{Y_{20}(\theta_\rho)} = - \left(\frac{2}{r_1} - \frac{2}{r_2} - \frac{2}{r_{13}} + \frac{2}{r_{23}} \right) \Phi_{Ps}(r_{12}) \Phi_H(r_3) \sqrt{2\kappa} \, n_2(\kappa\rho) f_{2}(\rho) \\
& - \Phi_{Ps}(r_{12}) \Phi_H(r_3) \sqrt{2\kappa} \frac{1}{2\rho} \left\{ \left[4 n_2(\kappa\rho) - 2 \kappa\rho \, n_1(\kappa\rho) \right] f_{2}^\prime(\rho) - \rho \, n_2(\kappa\rho) f_{2}^{\prime\prime}(\rho) \right\}.
\end{align}
Equivalently,
\begin{equation}
\mathscr{L} S_2^\prime = \frac{\mathcal{L} S^\prime}{Y_{20}(\theta_{\rho^\prime})} \text{ and }
\mathscr{L} C_2^\prime = \frac{\mathcal{L} C^\prime}{Y_{20}(\theta_{\rho^\prime})}.
\end{equation}

Each of the following equations has multiple forms that can be used due to the
properties of the permutation operator, but these equations show the form that
we used in our code. These 
equations are mainly straightforward applications of the external angular
integrations in \cref{chp:AngularInt} to the matrix elements of
\cref{eq:GeneralKohnMatrix}.

\begin{equation}
\label{eq:DWavePhi1SBar}
\left(\bar{\phi}_{1i},\mathcal{L} \bar{S}_2\right) 
= \sqrt{2} \cdot 2\pi \bigintsss \phi_{1i} \left[ \left(1 - \frac{3 r_2^2 \sin^2\theta_{12}}{8 \rho^2} \right) \mathscr{L}S_2 \pm \left(1 - \frac{3 r_3^2 \sin^2\theta_{13}}{8 {\rho^\prime}^2} \right) \mathscr{L}S_2^\prime \right] d\tau_{int}
\end{equation}

\begin{equation}
\label{eq:DWavePhi1CBar}
\left(\bar{\phi}_{1i},\mathcal{L} \bar{C}_2\right)
= \sqrt{2} \cdot 2\pi \bigintsss \phi_{1i} \left[ \left(1 - \frac{3 r_2^2 \sin^2\theta_{12}}{8 \rho^2} \right) \mathscr{L}C_2 \pm \left(1 - \frac{3 r_3^2 \sin^2\theta_{13}}{8 {\rho^\prime}^2} \right) \mathscr{L}C_2^\prime \right] d\tau_{int}
\end{equation}

\begin{align}
\label{eq:DWavePhi2SBar}
\nonumber \left(\bar{\phi}_{2j},\mathcal{L} \bar{S}_2\right) =& \sqrt{2} \cdot 2\pi \bigintsss \phi_{2j} \left[ \left( \frac{3(r_1 \cos\theta_{12} + r_2)^2}{8 \rho^2} - \frac{1}{2} \right) \mathscr{L}S_2 \right. \\
& \left. \pm \left( \frac{3(r_1 \cos\theta_{12} + r_3 \cos\theta_{23})^2}{8 {\rho^\prime}^2} - \frac{1}{2} \right) \mathscr{L}S_2^\prime \right] d\tau_{int}
\end{align}

\begin{align}
\label{eq:DWavePhi2CBar}
\nonumber \left(\bar{\phi}_{2j},\mathcal{L} \bar{C}_2\right) =& \sqrt{2} \cdot 2\pi \bigintsss \phi_{2j} \left[ \left( \frac{3(r_1 \cos\theta_{12} + r_2)^2}{8 \rho^2} - \frac{1}{2} \right) \mathscr{L}C_2 \right.\\
& \left. \pm \left( \frac{3(r_1 \cos\theta_{12} + r_3 \cos\theta_{23})^2}{8 {\rho^\prime}^2} - \frac{1}{2} \right) \mathscr{L}C_2^\prime \right] d\tau_{int}
\end{align}

\section{Long-Range -- Long-Range  Matrix Elements}
\label{sec:DWaveLongLong}

Similar to the short-short terms, the derivations for the long-long terms are 
available on the Wiki \cite{Wiki} and figshare \cite{figshare}. After external
angular integrations (see \cref{chp:AngularInt}), these matrix elements are:
\begin{align}
\label{eq:DWaveSBarSBar}
\left(\bar{S}_2,\mathcal{L}\bar{S}_2\right) = \pm 2\pi \bigintsss \left\{ S_2 S_2' \left(\frac{2}{r_1} - \frac{2}{r_2} - \frac{2}{r_{13}} + \frac{2}{r_{23}} \right) \left[ \frac{3}{8} \frac{(4\rho^2 + 4 {\rho^\prime}^2 - r_{23}^2)^2}{16 \rho^2 {\rho^\prime}^2} - \frac{1}{2} \right] \right\} d\tau_{int}
\end{align}
\begin{align}
\label{eq:DWaveCBarSBar}
\left(\bar{C}_2,\mathcal{L}\bar{S}_2\right) = \pm 2\pi \bigintsss \left\{ S_2 C_2' \left(\frac{2}{r_1} - \frac{2}{r_2} - \frac{2}{r_{13}} + \frac{2}{r_{23}} \right) \left[ \frac{3}{8} \frac{(4\rho^2 + 4 {\rho^\prime}^2 - r_{23}^2)^2}{16 \rho^2 {\rho^\prime}^2} - \frac{1}{2} \right] \right\} d\tau_{int}
\end{align}
\begin{align}
\label{eq:DWaveSBarCBar}
\nonumber \left(\bar{S}_2,\mathcal{L}\bar{C}_2\right) = 2\pi \bigintsss \Bigg\{ \pm & S_2' C_2 \left(\frac{2}{r_1} - \frac{2}{r_2} - \frac{2}{r_{13}} + \frac{2}{r_{23}} \right) \left[ \frac{3}{8} \frac{(4\rho^2 + 4 {\rho^\prime}^2 - r_{23}^2)^2}{16 \rho^2 {\rho^\prime}^2} - \frac{1}{2} \right] \\
\nonumber - & \left[ S_2 \pm \left( \frac{3}{8} \frac{(4\rho^2 + 4 {\rho^\prime}^2 - r_{23}^2)^2}{16 \rho^2 {\rho^\prime}^2} - \frac{1}{2} \right) S_2' \right] \sqrt{2\kappa} \, \Phi_{Ps}\left(r_{12}\right) \Phi_H\left(r_3\right) \\
& \times \frac{1}{2\rho} \left( \left[ 4 n_2(\kappa\rho) - 2 \kappa\rho \, n_1(\kappa\rho) \right] f_2^\prime(\rho) - \rho \, n_2(\kappa\rho) f_2^{\prime\prime}(\rho) \right) \Bigg\} d\tau_{int}
\end{align}
\begin{align}
\label{eq:DWaveCBarCBar}
\nonumber \left(\bar{C}_2,\mathcal{L}\bar{C}_2\right) = 2\pi \bigintsss \Bigg\{ \pm & C_2' C_2 \left(\frac{2}{r_1} - \frac{2}{r_2} - \frac{2}{r_{13}} + \frac{2}{r_{23}} \right) \left[ \frac{3}{8} \frac{(4\rho^2 + 4 {\rho^\prime}^2 - r_{23}^2)^2}{16 \rho^2 {\rho^\prime}^2} - \frac{1}{2} \right] \\
\nonumber - & \left[ C_2 \pm \left( \frac{3}{8} \frac{(4\rho^2 + 4 {\rho^\prime}^2 - r_{23}^2)^2}{16 \rho^2 {\rho^\prime}^2} - \frac{1}{2} \right) C_2' \right] \sqrt{2\kappa} \, \Phi_{Ps}\left(r_{12}\right) \Phi_H\left(r_3\right) \\
& \times \frac{1}{2\rho} \left( \left[ 4 n_2(\kappa\rho) - 2 \kappa\rho \, n_1(\kappa\rho) \right] f_2^\prime(\rho) - \rho \, n_2(\kappa\rho) f_2^{\prime\prime}(\rho) \right) \Bigg\} d\tau_{int}.
\end{align}

\section{Mixed Symmetry Terms}
\label{sec:MixedTerms}

According to Schwartz \cite{Schwartz1961a}, to have a complete description, 
each partial wave needs $\ell+1$ symmetries. As shown in
\cref{eq:SWaveTrial,eq:PWaveTrial}, we use the full sets of symmetries for the
S-wave and P-wave. There is a third symmetry for the D-wave, which we refer to
as the mixed symmetry terms, or mixed terms, given by
\begin{align}
\label{eq:MixedAng}
\psi(\ell_1,\ell_2,L,M) &= \psi_{(1,1,2,0)} = \sum_{m=-1}^{+1} Y_{1,m}(\theta_1,\varphi_1) Y_{1,m}(\theta_2,\varphi_2) \left< 1,m; 1,-m,0 | 2,0 \right> \nonumber \\
	&= Y_{1,-1}(\theta_1,\varphi_1) Y_{1,+1}(\theta_2,\varphi_2)
    \left< 1,-1,1,+1 | 2,0 \right> \nonumber \\
& \hspace{0.5cm}  + Y_{1,0}(\theta_1,\varphi_1) Y_{1,0}(\theta_2,\varphi_2)
    \left< 1,0,1,0 | 2,0 \right> \nonumber \\
& \hspace{0.5cm} + Y_{1,+1}(\theta_1,\varphi_1) Y_{1,-1}(\theta_2,\varphi_2)
   \left< 1,+1,1,-1 | 2,0 \right>,
\end{align}
where $\ell_1$ and $\ell_2$ are the angular momenta on the particles in Ps,
and $L$ and $M$ give the angular momentum of the Ps.
These three terms can be combined into a single set as
\begin{equation}
\label{eq:MixedAngSimple}
\psi_{(1,1,2,0)}(\theta_1,\theta_2) = \frac{3}{4\uppi} \frac{1}{\sqrt{6}} \left(3 \cos\theta_1 \cos\theta_2 - \cos\theta_{12} \right).
\end{equation}
This avoids the issue of dealing with complex terms in the $m = -1$ and $m = 1$ cases.
Refer to \cref{sec:MixedDerivation} for the derivation of \cref{eq:MixedAngSimple}
from \cref{eq:MixedAng}.

The short-long matrix elements that include the mixed terms are not
much more difficult to deal with than the first and second symmetry terms.
The evaluation of the short-short matrix elements involving the mixed terms is
much more difficult than those with just the first and second symmetry terms. 
I spent some time trying to derive the expressions for the short-short
matrix elements but ran into difficulty doing so. There has been some
preliminary progress on this front. 

For three-body problems, the mixed terms are substantially easier to deal
with. For e$^+$-H, Refs.~\cite{Brown1985a,BrownThesis,WattsThesis,Humberston1997,VanReeth1997}
used the mixed terms. Dunn et al. \cite{Dunn2000,DunnThesis} also treated
e$^+$-He as a three-body problem, using one-electron models of He and
including the mixed terms.

Prior work \cite{VanReeth1997,VanReethThesis} on e$^+$-He scattering (also a
four-body problem) neglected the mixed terms. The justification they used was
that for e$^+$-H scattering, adding the third symmetry changed the $K$-matrix
elements less than $1.5\%$. The e$^+$-H scattering problem also has mixed
terms, but these are much easier to deal with analytically, since it is a
three-body problem. However, this conclusion is now believed to be in error, as
described in Ref.~\cite{Woods2015}. A corrected code in a preliminary
investigation by Van Reeth and Humberston \cite{VanReeth2015} for e$^+$-H
scattering found that these mixed terms can change the $K$-matrix elements by
about $10\%$ near the Ps formation threshold. Since they believed that the
mixed terms did not contribute much, they were neglected for e$^+$-He scattering
and virtual Ps terms were added in to improve the convergence, getting the final
results to within 1 to 2\% of the corrected code from this preliminary
investigation. It is clear that these virtual Ps terms are no longer needed if
the mixed terms are included.

The same preliminary investigation \cite{VanReeth2015} also looked at e$^-$-H
scattering. This investigation found that the mixed terms change the $^1$D and
$^3$D phase shifts less than 1\%, but the $^1$D phase shifts are affected more
by the inclusion of the mixed terms. The $^1$D phase shifts are
affected less at low $\kappa$ and more at higher $\kappa$, though the
contributions are still small.

Due to the difficulty we had evaluating the analytical portion (the external
angular integrals) for the mixed terms, we only included the first two
symmetries in the D-wave calculations in this work. The preliminary investigation
\cite{VanReeth2015} took place only recently, after we had attempted other
methods to accelerate the convergence
of the D-wave phase shifts. As we discovered in \cref{sec:DWaveResults}, this
turns out to be an acceptable approximation.

\section{Results}
\label{sec:DWaveResults}

This section gives the results of the $^{1,3}$D-wave calculations using the 
nonlinear parameters determined in \cref{sec:DWaveNonlinear}.

\subsection{Phase Shifts}
\label{sec:DWavePhase}

\begin{table}
\centering
\setlength{\tabcolsep}{-2pt}
\footnotesize
\begin{tabular}{@{\hskip 0.1cm}l . . . . . . .}
\toprule
Method & \multicolumn{1}{c}{\phantom{1}0.1} & \multicolumn{1}{c}{\phantom{1}0.2} & \multicolumn{1}{c}{\phantom{1}0.3} & \multicolumn{1}{c}{\phantom{1}0.4} & \multicolumn{1}{c}{\phantom{1}0.5} & \multicolumn{1}{c}{\phantom{1}0.6} & \multicolumn{1}{c}{\phantom{1}0.7} \\
\midrule
This work $(\omega = 6)$ $\delta_2^+$ 				& 1.36^{-4}  & 2.99^{-3}  & 1.60^{-2}  & 4.98^{-2}  & 1.13^{-1}  & 2.06^{-1}  & 3.28^{-1} \\
This work $(\omega \to \infty)$ $\delta_2^+$ 		& 4.24^{-4}  & 3.18^{-3}  & 1.62^{-2}  & 5.05^{-2}  & 1.14^{-1}  & 2.09^{-1}  & 3.33^{-1} \\
\% Diff$^+$											& 103.0\%    & 6.27\%     & 1.54\%     & 1.33\%     & 1.52\%     & 1.67\%     & 1.67\% \\
\% Diff$^+$	CC										& 39.1\%     & 15.4\%     & 7.81\%     & 4.71\%     & 2.62\%     & 0.97\%     & 1.23\% \\
\arrayrulecolor[RGB]{220,220,220}\midrule\arrayrulecolor{black}
CC 9Ps9H+H$^-$ \cite{Walters2004} $\delta_2^+$		& 2.02^{-4}  & 3.49^{-3}  & 1.73^{-2}  & 5.22^{-2}  & 1.16^{-1}  & 2.08^{-1}  & 3.24^{-1} \\
CC 9Ps9H \cite{Blackwood2002} $\delta_2^+$			& 1.46^{-4}  & 3.15^{-3}  & 1.65^{-2}  & 4.95^{-2}  & 1.08^{-1}  & 1.94^{-1}  & 3.02^{-1} \\
3-state CC \cite{Sinha1997} $\delta_2^-$			& 3.22^{-5}  & 9.29^{-4}  & 5.96^{-3}  & 2.01^{-2}  & 4.63^{-2}  & 8.29^{-2}  & 1.23^{-1} \\
SE \cite{Ray1997} $\delta_2^+$ 						& 3.18^{-5}  & 9.17^{-4}  & 5.87^{-3}  & 1.97^{-2}  & 4.54^{-2}  & 8.09^{-2}  & 1.19^{-1} \\
5-state CC \cite{Adhikari1999} $\delta_2^+$			& 1.8^{-5}   & 5.3^{-4}   & 3.5^{-3}   & 1.2^{-2}   & 2.9^{-2}   & 5.5^{-2}   & 8.8^{-2} \\
SE \cite{Hara1975} $\delta_2^+$						& 0.0        & 0.0009     & 0.0058     & 0.0195     & 0.0453     & 0.0810     & 0.1194 \\
\midrule
This work $(\omega = 6)$ $\delta_2^-$ 				& 5.81^{-5}  & 7.12^{-4}  & 1.07^{-3}  & -2.00^{-3} & -1.12^{-2} & -2.65^{-2} & -4.45^{-2} \\
This work $(\omega \to \infty)$ $\delta_2^-$ 		& 3.13^{-4}  & 8.67^{-4}  & 1.41^{-3}  & -1.20^{-3} & -9.34^{-2} & -2.32^{-2} & -4.02^{-2} \\
\% Diff$^-$											& 137.4\%    & 19.6\%     & 24.4\%     & 39.8\%     & 13.5\%     & 9.10\%     & 6.28\% \\
\arrayrulecolor[RGB]{220,220,220}\midrule\arrayrulecolor{black}
CC 9Ps9H \cite{Blackwood2002} $\delta_2^-$			& 8.48^{-5}  & 1.15^{-3}  & 2.84^{-3}  & 2.37^{-3}  & -4.66^{-3} & -1.85^{-2} & -3.27^{-2} \\
3-state CC \cite{Sinha1997} $\delta_2^-$			& -2.74^{-5} & -7.77^{-4} & -4.83^{-3} & -1.55^{-2} & -3.41^{-2} & -5.83^{-2} & -8.25^{-2} \\
SE \cite{Ray1997} $\delta_2^+$ 						& -3.00^{-5} & -8.56^{-4} & -5.37^{-3} & -1.76^{-2} & -3.95^{-2} & -7.03^{-2} & -1.06^{-1} \\
5-state CC \cite{Adhikari1999} $\delta_2^+$			& -1.4^{-5}  & -4.0^{-4}  & -2.6^{-3}  & -8.6^{-3}  & -2.0^{-2}  & -3.6^{-2}  & -5.5^{-2} \\
SE \cite{Hara1975} $\delta_2^+$						& 0.0        & -8.0^{-4}  & -5.3^{-3}  & -1.74^{-2} & -3.95^{-2} & -7.04^{-2} & -1.062^{-1} \\
\bottomrule
\end{tabular}
\caption[$^{1,3}$D results and comparisons]{Comparison of $^{1,3}$D phase
shifts and comparisons with other groups. Values in the header are $\kappa$
in au. \% Diff$^\pm$ is the 
percent difference between the current complex Kohn $\omega = 6$ and 
extrapolated values. \% Diff$^+$ CC is the percent difference between the 
complex Kohn $\omega = 6$ and CC 14Ps14H+H$^-$ \cite{Walters2004} phase
shifts. Exponents denote powers of 10.}
\label{tab:DWaveComparisons}
\end{table}

\Cref{tab:DWaveComparisons,fig:DWaveComparisons} show the final results of 
the phase shifts using the $S$-matrix complex Kohn with the nonlinear 
parameters given in \cref{sec:DWaveNonlinear} and compares to multiple other 
calculations by other groups. One of the most easily noticeable problems is
the poor convergence for $\kappa = 0.1$ for both $^1$D and $^3$D. As mentioned
in \cref{sec:Extrapolations} on page~\pageref{sec:Extrapolations}, our code
does not seem to handle very small phase shifts particularly well, and we
likely would have to increase the number of integration points further. Also, the 
lack of mixed terms likely affects the convergence. Both of these effects
likely combine to give a bad extrapolation of over 100\%. However, note that
the phase shifts for $\kappa = 0.1$ are not extremely far from the CC of
Refs.~\cite{Walters2004,Blackwood2002}. We do not show this extrapolation for
$^1$D in Ref.~\cite{Woods2015}, as it is obviously not good.

For $\kappa \geq 0.2$, the extrapolations look more reasonable, and for $^1$D,
the phase shifts look relatively well converged, especially for $\kappa \geq 0.3$.
We believe that with the exception of $\kappa < 0.2$, the $^1$D phase shifts are
reasonably accurate, even with the omission of the mixed terms. The $^1$D phase
shifts for $\omega = 5$ and $\omega = 6$ differ by less than 10\%.

For $^3$D, the extrapolations are worse when the phase shift curve goes
from positive to negative, which happens around $\kappa = 0.35$, as can be seen in
\cref{fig:DWaveComparisons}(b). In Ref.~\cite{Blackwood2002}, they note \label{DWaveSwitch} that
this change in sign of the phase shifts indicates a switch from being repulsive
at low $\kappa$ and attractive at high $\kappa$. In the higher
$\kappa$ region ($\kappa \geq 0.5$),
the $^3$D phase shift convergence is much better, getting down to $\sim\!6\%$.
Due to the worse extrapolations for $^3$D than for the $^{1,3}$S, $^{1,3}$P,
and $^1$D partial waves, we do not report any $^3$D extrapolations in
Ref.~\cite{Woods2015}.

We see from \Cref{tab:DWaveComparisons,fig:DWaveComparisons} that the CC phase shifts
\cite{Blackwood2002,Walters2004} are generally above the $S$-matrix complex Kohn phase shifts. The exception is the
$\kappa = 0.7$ phase shift for $^1$D and the extrapolated phase shifts for
$\kappa = 0.6$ and $0.7$, where the complex Kohn results are slightly higher.
For $^1$D, the complex Kohn and CC \cite{Walters2004} results overall agree
very well, as seen in the fourth line of \cref{tab:DWaveComparisons}, with
percent differences $\sim\!1\%$ at higher $\kappa$.

\begin{figure}
	\centering
	\includegraphics[width=\textwidth]{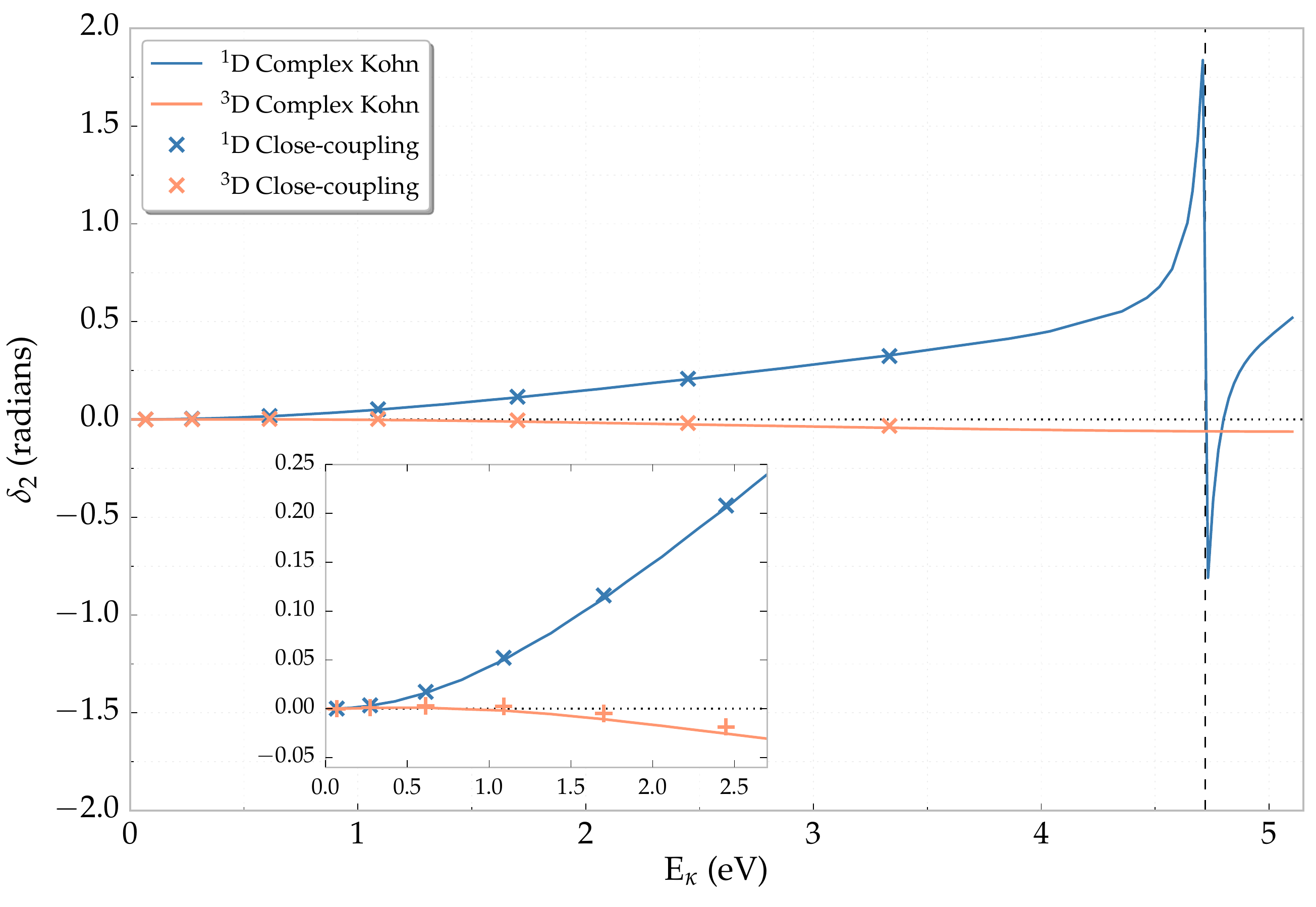}
	\caption[$^{1,3}$D phase shifts]{$^{1,3}$D complex Kohn phase shifts. The $^1$D CC phase shifts
\cite{Walters2004} are given by \mbox{\textcolor{blue}{$\times$}}, and the
$^3$D CC phase shifts \cite{Blackwood2002} are given by
\mbox{\textcolor{red}{\textbf{+}}}. Vertical dashed lines denote the complex rotation resonance
positions \cite{Ho1998}. An interactive version of this figure is available online \cite{Plotly}
at \url{https://plot.ly/~Denton/5/d-wave-ps-h-scattering/}.}
	\label{fig:DWavePhase}
\end{figure}

\begin{figure}
	\centering
	\includegraphics[width=5.25in]{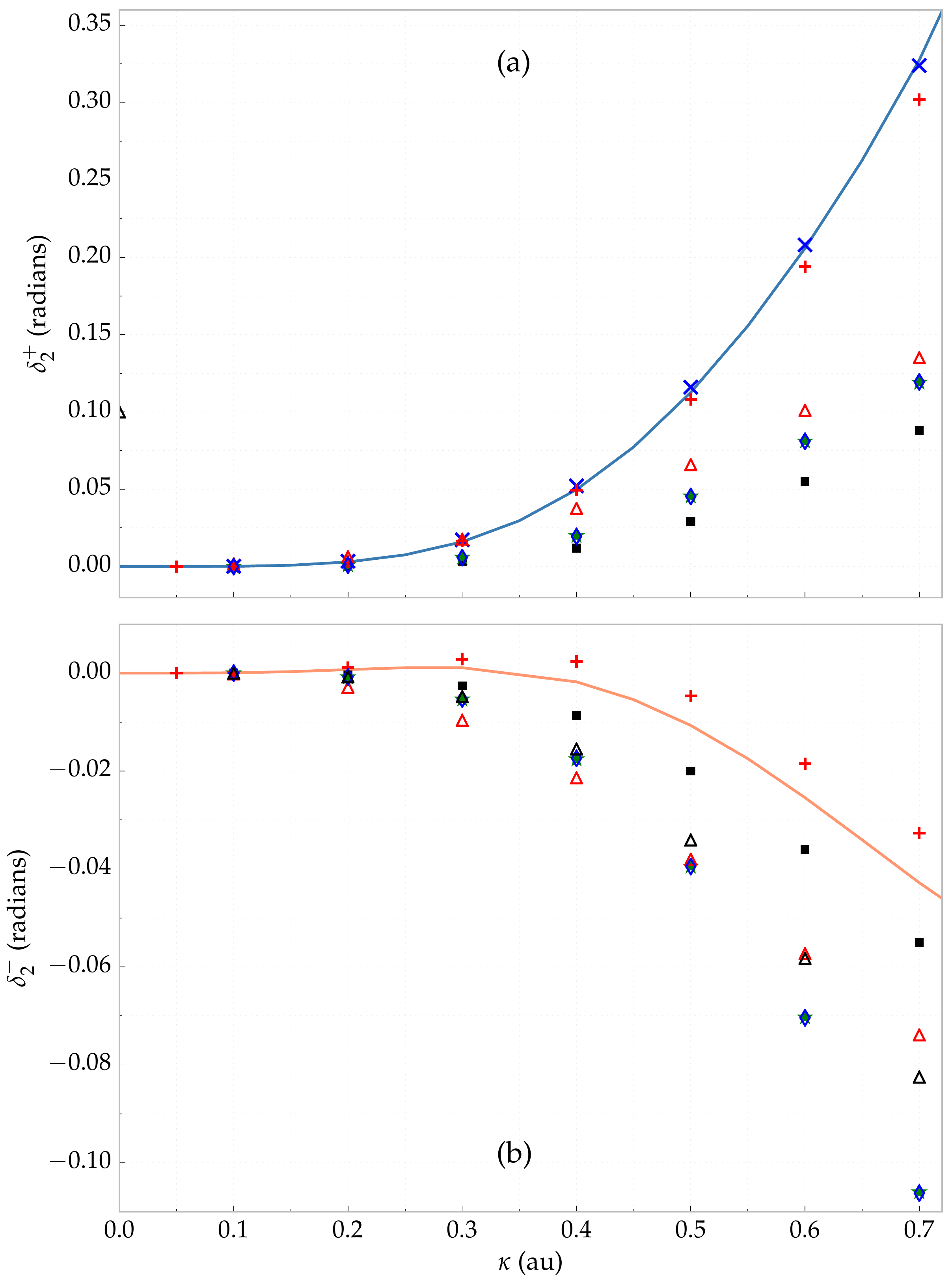}
	\caption[Comparison of D-wave phase shifts]{Comparison of $^1$D (a) and $^3$D (b) phase shifts with results from other groups. Results are ordered according to year of publication. Solid curves -- this work; \mbox{\textcolor{blue}{$\times$} -- CC \cite{Walters2004};} \mbox{$\CIRCLE$ -- Kohn \cite{VanReeth2003};} \mbox{\textcolor{red}{\textbf{+}} -- CC \cite{Blackwood2002};} \mbox{\textcolor{red}{$\vartriangle$} -- 6-state CC \cite{Sinha2000};} \mbox{$\blacksquare$ -- 5-state CC \cite{Adhikari1999};} \mbox{$\vartriangle$ -- 3-state CC \cite{Sinha1997};} \mbox{\textcolor[RGB]{0,127,0}{$\bigstar$} -- CC \cite{Ray1997};} \mbox{\textcolor{blue}{$\lozenge$} -- Static-exchange \cite{Hara1975}.}}
	\label{fig:DWaveComparisons}
\end{figure}

There is much more discrepancy between the $^3$D complex Kohn and CC phase
shifts, as can be seen easily in \cref{fig:DWaveComparisons} and the inset of
\cref{fig:DWavePhase}. This discrepancy
plus the poorer $^3$D extrapolations show that the $^3$D phase shifts are not
fully converged. It should be noted however that the CC phase shifts could be
overestimates, considering that the CVM \cite{Zhang2012} $^3$S phase shifts
agree extremely well with the complex Kohn phase shifts, but the CC $^3$S phase shifts are
slightly higher than both. The CC also differs from the complex
Kohn at low $\kappa$ for the $^3$P-wave, where we do not have a neglected
symmetry, and the phase shifts are well converged.

Unfortunately, it is not possible to know at the moment how far the complex 
Kohn phase shifts differ from the true phase shifts, and further 
investigation into the mixed terms could help resolve this discrepancy.
Thankfully, the contribution to the cross sections
in \cref{chp:CrossSections} from the $^3D$ is small, even in the differential
cross section. The $^1$D contribution to the cross sections is very small
in the low $\kappa$ region where the phase shifts are less converged. In the
resonance region, where the $^1$D resonance contributes significantly, the
phase shifts are well converged.

\subsection{Resonance Parameters}
\label{sec:DWaveResonance}

The $^1$D-wave has a single resonance before the Ps(n=2) threshold, unlike 
the $^1$S- and $^1$P-waves, which have two. We use \cref{eq:ResonanceCurve}
to fit this curve but without the second $\arctan$ term.
As mentioned in the previous section, the $^1$D phase shifts are
well converged in the higher $\kappa$ region, where the resonance
is located.

From the discussion in \cref{sec:DWaveNonlinear}, the complex Kohn resonance 
parameters presented in \cref{tab:DWaveResonancesOther} are more 
sensitive to the nonlinear parameters than the two lower partial waves. The
complex Kohn 
resonance parameters agree very well with the complex rotation \cite{Ho1998}.
The 9Ps9H CC results \cite{Blackwood2002} are brought closer to the complex
Kohn and complex rotation results by the inclusion of the H$^-$ channel for
both the position and width. The stabilization \cite{Yan2003} results are
also similar to the complex rotation results, both of which are carried out by
Yan and Ho.

This $^1$D-wave resonance has a very important contribution to the cross
sections in \cref{chp:CrossSections}. Due to the $(2\ell+1)$ dependence
for the elastic integrated and elastic differential cross sections and
the $(\ell+1)$ dependence for the momentum transfer cross section, the
$^1$D resonance has a larger contribution to these cross
sections than the $^1$S- and $^1$P-waves.

\setlength{\abovecaptionskip}{6pt}   
\setlength{\belowcaptionskip}{6pt}   
\begin{table}
\centering
\begin{tabular}{l l l}
\toprule
Method & $^1E_R \text{ (eV)}$ & $^1\Gamma \text{ (eV)}$ \\
\midrule
Current work: Average $\pm$ standard deviation & $4.720 \pm 0.001$ & $0.0908 \pm 0.0010$ \\
Current work: $S$-matrix complex Kohn & $4.720$ & $0.0909$ \\
CC (9Ps9H + H$^-$) \cite{Walters2004} & $4.899$ & $0.0872$ \\
Stabilization \cite{Yan2003} & $4.714$ & $0.0969$ \\
CC (9Ps9H) \cite{Blackwood2002} & $5.16$ & $0.15$ \\
CC (22Ps1H + H$^-$) \cite{Blackwood2002b} & $4.814$ & $0.065$ \\
Optical potential \cite{DiRienzi2002a} & $4.729$ & $0.327$ \\
Complex rotation \cite{Ho1998} & $4.710 \pm 0.0027$ & $0.0925 \pm 0.0054$  \\
Coupled-pseudostate \cite{Campbell1998} & $5.28$ & $0.47$ \\
\bottomrule
\end{tabular}
\caption{$^1$D-wave resonance parameters}
\label{tab:DWaveResonancesOther}
\end{table}

\section{Summary}
\label{sec:SummaryD}

The $^1$D-wave phase shifts obtained by the Kohn-type methods using the first 
two short-range symmetries are relatively well converged at higher
$\kappa$ but do not appear to be as well converged at low $\kappa$ values. The
$^3$D-wave phase shifts are less converged than the $^1$D phase shifts and 
show poor convergence at low $\kappa$. Including the mixed symmetry terms may
improve the convergence, but we have not included them in this work, primarily
due to the difficulty of evaluating the short-short integrations. However, we
are able to reliably calculate the $^1$D resonance, and the contribution to the
cross sections from the less accurate $^3$D phase shifts is minimal
(see \cref{chp:CrossSections}). The contribution to the cross sections from the
$^1$D low $\kappa$ phase shifts is also small.



\chapter{General Ps-H Formalism}
\label{chp:General}

\iftoggle{UNT}{To}{\lettrine{\textcolor{startcolor}{T}}{o}}
extend this work to higher partial 
waves, we could have continued as before, performing derivations and writing 
new code for each partial wave. This takes time and is error prone, so during
the course of writing the D-wave code, I investigated creating a general
formalism that works for arbitrary $\ell$. The long-range
code, covered in \cref{sec:GeneralLong}, is a straightforward 
generalization from the previous codes. The short-range code, discussed in 
\cref{sec:GeneralShort}, uses the Laplacian formalism instead of the
gradient-gradient that we used in the S-, P-, and D-wave short-short codes,
which also enabled us to compare the two different methods for the first three
partial waves.

\section{General Long-Range Matrix Elements}
\label{sec:GeneralLong}

If the mixed symmetry terms in \cref{sec:MixedTerms} are not included for the
D-wave, the P-wave and D-wave long-range codebases are very similar. The S-wave 
is not much different as well, but it only has a single symmetry. We also
only treat the first two symmetries for $\ell \geq 2$.

\subsection{Long-Long Matrix Elements}
\label{sec:GeneralLongLong}

One major difference for similar matrix elements is the angular integrations. 
As shown in \cref{chp:AngularInt}, the external angular integrations 
for each partial wave have different results, due to the different spherical 
harmonics. $\widetilde{S}_\ell$ and $\widetilde{C}_\ell$ are also of a similar form between the 
partial waves, as seen in \cref{eq:TildeSCDef,eq:SCBarDef}.
Other than the spherical harmonics, the spherical Bessel functions are 
different, as is the shielding function for $\widetilde{C}_\ell$. The spherical Bessel 
functions are easily called through the GNU Scientific Library \cite{GSL} for 
any $\ell$-value. The power of the shielding function varies as
$m_\ell \geq (2\ell + 1)$. All of this allows us to easily generalize the long-long 
integrations.

\subsection{Short-Long Matrix Elements}
\label{sec:GeneralShortLong}
The short-range functions are also easily generalizable. For the P-wave and
D-wave, these are seen in \cref{eq:PWavePhiBar,eq:DWavePhiBar}. The $\phi_i$
and $\phi_j$ parts for each are the same. This fact can only 
be easily used for the short-long calculations and not the short-short 
calculations, due to the action of the $\mathcal{L}$-operator on these terms. The 
formalism we have used for the short-long terms at every stage only operates
$\mathcal{L}$ on the long-range parts. \Cref{sec:GeneralShort} covers 
the calculations of the short-short terms using a general formalism.

\section{General Short-Range--Short-Range}
\label{sec:GeneralShort}

The short-short integrals are more complicated due to the 
gradient-gradient operator acting on the short-range terms (see \cref{eq:BoundGradient}).
As an example of the nature of these, refer to \cref{sec:DWaveShortShort}.
The derivations were previously performed by rotating 
the coordinate system and integrating over external angles first, as 
described in \cref{chp:AngularInt}. This approach works, but it 
cannot be extended to higher partial waves without completing a full 
derivation for each partial wave. 
Drake and Yan \cite{Yan1997} expand upon 
their previous work for the three-electron Hylleraas integrals to include the 
spherical harmonics using the Laplacian formalism instead, which still
has a complicated form but works for arbitrary $\ell$.
It is this approach that we take here. All equations in this
section have been rederived for the Ps-H system in my notes \cite{Wiki,figshare}.
The work of Harris \cite{Harris2005a} may also be directly
applicable to this problem, but this has not yet been explored.

\subsection{Hamiltonian}
\label{sec:GenShortHam}
As we have for the bound state problem, the Hamiltonian is given compactly
as in \cref{eq:BoundHamiltonian} by
\beq
\label{eq:GenHam1}
H = \sum_{i=1}^3 \left(-\frac{1}{2} \nabla_i^2 - \frac{1}{r_i} \right) + \sum_{i>j}^3 \frac{1}{r_{ij}}.
\eeq
As in Ref.~\cite{Yan1997}, when the Laplacians are expanded, this is given by
the form
\begin{align}
\label{eq:GenHam2}
\nonumber H = -\frac{1}{2} & \left[ \sum_{i=1}^3 \left( \frac{\partial^2}{\partial r_i^2} + \frac{2}{r_i} \frac{\partial}{\partial r_i} - \frac{\ell(\ell+1)}{r_i^2} \right) + \sum_{i>j}^3 \left( 2 \frac{\partial^2}{\partial r_i^2} + \frac{4}{r_{ij}} \frac{\partial}{\partial r_i} \right) + \sum_{i \neq j}^3 \left(\frac{r_i^2 - r_j^2 + r_{12}^2}{r_i r_j} \right) \right. \\
\nonumber &  \left. + \frac{r_{12}^2 + r_{13}^2 - r_{23}^2}{r_{12} r_{13}} \frac{\partial^2}{\partial r_{12} \partial r_{13}} + \frac{r_{12}^2 + r_{23}^2 - r_{13}^2}{r_{12} r_{23}} \frac{\partial^2}{\partial r_{12} \partial r_{23}} + \frac{r_{13}^2 + r_{23}^2 - r_{12}^2}{r_{13} r_{23}} \frac{\partial^2}{\partial r_{13} \partial r_{23}} \right. \\
& \left. + \sum_{i>j}^3 \frac{1}{r_{ij}} \frac{r_i}{r_j} \frac{\partial}{\partial r_{ij}} \left( \hat{\boldsymbol{r}_i} \cdot \hat{\nabla}_j^Y \right) + \sum_{i>j}^3 \frac{1}{r_{ji}} \frac{r_j}{r_i} \frac{\partial}{\partial r_{ji}} \left( \hat{\boldsymbol{r}_j} \cdot \hat{\nabla}_i^Y \right) \right].
\end{align}

\noindent This is the form when the mass polarization terms are unimportant
(with $\mu \rightarrow 0$). As defined in Ref. \cite{Yan1997},
$\hat{\boldsymbol{r}_j} = \boldsymbol{r}_i/r_i$ and $\hat{\nabla}_i^Y = r_i \nabla_i^Y$.
The terms involving $\nabla_i^Y$ only operate on the spherical 
harmonics, and they will be discussed in \cref{sec:GenSphHarm}.

The wavefunction we use is slightly different from the form Yan and Drake \cite{Yan1997}
use. Specifically, we handle the antisymmetrization operator differently in 
our code, and we do not include the $\Omega$ function, given by Equation (10) 
in their paper. Part of the difference here is that we are working with two 
electrons and one positron, whereas they are working with systems that have 
three electrons, such as binding energy calculations of Li.

\subsection{General Integrals}
\label{sec:GenOverlap}

Using the notation in Ref.~\cite{Yan1997}, the terms in \cref{eq:GenHam2} without the spherical harmonic operator, $\hat{\nabla}_i^Y$, will be of the general form
\begin{align}
\label{eq:ShortIntGen}
\nonumber I(\ell_1^\prime m_1^\prime, \ell_2^\prime m_2^\prime, &\ell_3^\prime m_3^\prime, \ell_1 m_1, \ell_2 m_2, \ell_3 m_3; j_1,j_2,j_3,j_{12},j_{23},j_{31}; \bar{\alpha}, \bar{\beta}, \bar{\gamma}) \\
= & \; \int d \textit{\textbf{r}}_1 d \textit{\textbf{r}}_2 d \textit{\textbf{r}}_3
r_1^{j_1} r_2^{j_2} r_3^{j_3} r_{12}^{j_{12}}
r_{23}^{j_{23}} r_{31}^{j_{31}}
e^{-(\bar{\alpha} r_1 + \bar{\beta} r_2 + \bar{\gamma} r_3)}  \nonumber \\
& \times Y_{\ell_1^\prime m_1^\prime}^* (\textit{\textbf{r}}_1) Y_{\ell_2^\prime m_2^\prime}^* (\textit{\textbf{r}}_2) Y_{\ell_3^\prime m_3^\prime}^* (\textit{\textbf{r}}_3) Y_{\ell_1 m_1} (\textit{\textbf{r}}_1) Y_{\ell_2 m_2} (\textit{\textbf{r}}_2) Y_{\ell_3 m_3} (\textit{\textbf{r}}_3)\, .
\end{align}
As noted on page~\pageref{BraNote} for the Kohn-type variational methods, these
should not be conjugated, but for this work with $m = 0$ (so excluding the mixed
symmetry terms described in \cref{sec:MixedTerms}), the real-valued short-range
terms will be the same with and without the conjugate.
The method described here is an extension of that in \cref{sec:ShortInt}.
Note that $\alpha$, $\beta$, and $\gamma$ are not necessarily the same as those
in \cref{eq:PhiDef}. These will be $2\alpha$, $2\beta$, and $2\gamma$ for the
direct-direct terms and $2\alpha$, $\beta+\gamma$, $\gamma+\beta$ for the
direct-exchange terms. After some manipulation, this integral can be written as
\cite{Yan1997}
\begin{align}
\label{eq:ShortIntGen2}
I(\ell_1^\prime m_1^\prime, &\ell_2^\prime m_2^\prime, \ell_3^\prime m_3^\prime, \ell_1 m_1, \ell_2 m_2, \ell_3 m_3; j_1,j_2,j_3,j_{12},j_{23},j_{31}; \bar{\alpha}, \bar{\beta}, \bar{\gamma})  \nonumber \\
= & \; \sum_{q_{12}=0}^{M_{12}} \sum_{q_{23}=0}^{M_{23}} \sum_{q_{31}=0}^{M_{31}} \sum_{k_{12}=0}^{L_{12}} \sum_{k_{23}=0}^{L_{23}} \sum_{k_{31}=0}^{L_{31}}  \nonumber \\
& \times I_{\rm{ang}}(\ell_1^\prime m_1^\prime, \ell_2^\prime m_2^\prime, \ell_3^\prime m_3^\prime, \ell_1 m_1, \ell_2 m_2, \ell_3 m_3; q_{12}, q_{23}, q_{31})  \nonumber \\
& \times I_{\rm{R}}(q_{12}, q_{23}, q_{31}, k_{12}, k_{23}, k_{31}; j_1, j_2, j_3, j_{12}, j_{23}, j_{31}; \bar{\alpha}, \bar{\beta}, \bar{\gamma}).
\end{align}
For even values of $j_{12}$, $M_{12} = \frac{1}{2}j_{12}$ and $L_{12} = \frac{1}{2}j_{12} - q_{12}$. For odd values of $j_{12}$, $M_{12} = \infty$ and $L_{12} = \frac{1}{2}(j_{12}+1)$. The same type of upper limits apply to the $j_{23}$ and $j_{31}$ terms.

The angular part of $I$ is given by
\begin{align}
\label{eq:DrakeGenAng}
I_{\rm{ang}}(\ell_1^\prime m_1^\prime, & \ell_2^\prime m_2^\prime, \ell_3^\prime m_3^\prime, \ell_1 m_1, \ell_2 m_2, \ell_3 m_3; q_{12}, q_{23}, q_{31})  \nonumber \\
= & \;(-1)^{m_1^\prime + m_2^\prime + m_3^\prime + q_{12} + q_{23} + q_{31}} (\ell_1^\prime, \ell_2^\prime, \ell_3^\prime, \ell_1, \ell_2, \ell_3)^{1/2} \sum_{n_1 n_2 n_3} (n_1,n_2,n_3)  \nonumber \\
& \times \SixJSymbol{n_1,n_2,n_3}{q_{23},q_{31},q_{12}} \ThreeJSymbol{n_1, m_1^\prime - m_1}{n_2, m_2^\prime - m_2}{n_3, m_3^\prime - m_3}  \nonumber \\
& \times \ThreeJSymbol{\ell_1^\prime,-m_1^\prime}{\ell_1,m_1}{n_1, m_1^\prime - m_1} \ThreeJSymbol{\ell_2^\prime,-m_2^\prime}{\ell_2,m_2}{n_2,m_2^\prime - m_2}  \nonumber \\
& \times \ThreeJSymbol{\ell_1^\prime,-m_1^\prime}{\ell_3,m_3}{n_3,m_3^\prime - m_3} \ThreeJSymbol{\ell_1^\prime,0}{\ell_1,0}{n_1,0} \ThreeJSymbol{\ell_2^\prime,0}{\ell_2,0}{n_2,0}  \nonumber \\
& \times \ThreeJSymbol{\ell_3^\prime,0}{\ell_3,0}{n_3,0} \ThreeJSymbol{q_{31},0}{q_{12},0}{n_1,0} \ThreeJSymbol{q_{12},0}{q_{23},0}{n_2,0} \ThreeJSymbol{q_{23},0}{q_{31},0}{n_3,0}.
\end{align}
This expression includes summations over both the Wigner 3-j and 6-j
coefficients \cite{Edmonds1996,Brink1993,Rose1995}, also known as the 3-j and
6-j symbols. The 3-j symbols are related to the Clebsch-Gordan coefficients by
\cite[p.46]{Edmonds1996}
\begin{equation}
\label{eq:Clebsch3J}
\ThreeJSymbol{j_1,m_1}{j_2,m_2}{j_3,m_3} = (-1)^{j_1 - j_2 + m_3} (2 j_3 + 1)^{-1/2} \ClebschGordon{j_1,m_1}{j_2,m_2}{j_3,-m_3}.
\end{equation}
The advantage of the 3-j symbols over the Clebsch-Gordan coefficients is that 
they are a more symmetric representation of angular momentum. The factor
$(2\ell+1)$ appears often in these types of derivations, so we adopt their 
shorthand notation of $(l,m,n,\ldots) = (2l+1)(2m+1)(2n+1) \cdots$.

The radial part of \cref{eq:ShortIntGen2} is 
\begin{align}
\label{eq:ShortGenIR}
I_{\rm{R}}(q_{12}, q_{23}, q_{31}&, k_{12}, k_{23}, k_{31}; j_1, j_2, j_3, j_{12}, j_{23}, j_{31}; \bar{\alpha}, \bar{\beta}, \bar{\gamma})  \nonumber \\
= & \; C_{j_{12} q_{12} k_{12}} C_{j_{23} q_{23} k_{23}} C_{j_{31} q_{31} k_{31}}  \nonumber \\
& \times W_{\rm{R}}(q_{12}, q_{23}, q_{31}, k_{12}, k_{23}, k_{31}; j_1, j_2, j_3, j_{12}, j_{23}, j_{31}; \bar{\alpha}, \bar{\beta}, \bar{\gamma}).
\end{align}
The $C_{jqk}$ coefficients are the same as in \cref{eq:Ccoeff}. The $W_{\rm{R}}$ function is built from the $W$ functions in \cref{eq:Wfunc} as
\begin{align}
\label{eq:ShortGenWR}
W_{\rm{R}}(&q_{12}, q_{23}, q_{31}, k_{12}, k_{23}, k_{31}; j_1, j_2, j_3, j_{12}, j_{23}, j_{31}; \bar{\alpha}, \bar{\beta}, \bar{\gamma})  \nonumber \\
=\; \; & W(j_1 + 2 + s_{12} + s_{31}, j_2 + 2 + j_{12} - s_{12} + s_{23}, j_3 + 2 + j_{23} - s_{23} + j_{31} - s_{31}; \bar{\alpha}, \bar{\beta}, \bar{\gamma})  \nonumber \\
+ & W(j_1 + 2 + s_{12} + s_{31}, j_3 + 2 + s_{23} + j_{31} - s_{31}, j_2 + 2 + j_{12} - s_{12} + j_{23} - s_{23}; \bar{\alpha}, \bar{\gamma}, \bar{\beta})  \nonumber \\
+ & W(j_2 + 2 + s_{12} + s_{23}, j_1 + 2 + j_{12} - s_{12} + s_{31}, j_3 + 2 + j_{23} - s_{23} + j_{31} - s_{31}; \bar{\beta}, \bar{\alpha}, \bar{\gamma})  \nonumber \\
+ & W(j_2 + 2 + s_{12} + s_{23}, j_3 + 2 + j_{23} - s_{23} + s_{31}, j_1 + 2 + j_{12} - s_{12} + j_{31} - s_{31}; \bar{\beta}, \bar{\gamma}, \bar{\alpha})  \nonumber \\
+ & W(j_3 + 2 + s_{23} + s_{31}, j_1 + 2 + s_{12} + j_{31} - s_{31}, j_2 + 2 + j_{12} - s_{12} + j_{23} - s_{23}; \bar{\gamma}, \bar{\alpha}, \bar{\beta})  \nonumber \\
+ & W(j_3 + 2 + s_{23} + s_{31}, j_2 + 2 + s_{12} + j_{23} - s_{23}, j_1 + 2 + j_{12} - s_{12} + j_{31} - s_{31}; \bar{\gamma}, \bar{\beta}, \bar{\alpha}),
\end{align}
with $s_{ij} = q_{ij} + 2 k_{ij}$.
Note that there are similarities with \cref{eq:FourBodyExpansion}.

\subsection{Spherical Harmonic Terms}
\label{sec:GenSphHarm}

The terms in \cref{eq:GenHam2} involving the spherical harmonic operator
$\hat{\nabla}_i^Y$ have to be handled differently than the other terms. In
Ref.~\cite{Yan1997}, they do different permutations for the three-electron
problem, so for the $p$ in their paper, we always use $p = 1$.

\begin{align}
\label{eq:GenY12}
I^1(\hat{\boldsymbol{r}_1} \cdot \hat{\nabla}_2^Y) = &\sum_{q_{12}=0}^{M_{12}} \sum_{q_{23}=0}^{M_{23}} \sum_{q_{31}=0}^{M_{31}} \sum_{k_{12}=0}^{L_{12}} \sum_{k_{23}=0}^{L_{23}} \sum_{k_{31}=0}^{L_{31}} \sum_{T_1 = \abs{1 - \ell_1}}^{1+\ell_1} \sum_{T_2 = \abs{1 - \ell_2}}^{1+\ell_2} b(\ell_2; T_2) C^1(\hat{\textit{\textbf{r}}}_1 \bm{\cdot} \hat{\textit{\textbf{r}}}_2)  \nonumber \\
& \times I_{\rm{R}}(q_{12}, q_{23}, q_{31}, k_{12}, k_{23}, k_{31}; j_1, j_2, j_3, j_{12}, j_{23}, j_{31}; \bar{\alpha}, \bar{\beta}, \bar{\gamma})
\end{align}
The $b$ function here is defined by
\begin{align}
\label{eq:bfunc}
b(\ell; \ell - 1) &= \ell + 1  \nonumber \\
b(\ell; \ell + 1) &= -\ell.
\end{align}
For any other values of the arguments, $b$ gives 0. $C^1$ is given by
\begin{align}
\label{eq:CP12}
C^1(\hat{\textit{\textbf{r}}}_1 & \bm{\cdot} \hat{\textit{\textbf{r}}}_2) = (\ell_1^\prime, \ell_2^\prime, \ell_3^\prime, \ell_1, \ell_2, \ell_3)^{1/2} (-1)^{q_{12} + q_{23} + q_{31}} \sum_{n_1 n_2 n_3} (n_1, n_2, n_3, T_1, T_2)  \nonumber \\
& \times \ThreeJSymbol{1,0}{\ell_1,0}{T_1,0} \ThreeJSymbol{1,0}{\ell_2,0}{T_1,0} \ThreeJSymbol{\ell_1^\prime,0}{T_1,0}{n_1,0} \ThreeJSymbol{\ell_2^\prime,0}{T_2,0}{n_2,0} \ThreeJSymbol{\ell_3^\prime,0}{\ell_3,0}{n_3,0}  \nonumber \\
& \times \ThreeJSymbol{q_{31},0}{q_{12},0}{n_1,0} \ThreeJSymbol{q_{12},0}{q_{23},0}{n_2,0} \ThreeJSymbol{q_{23},0}{q_{31},0}{n_3,0} \SixJSymbol{n_1,n_2,n_3}{q_{23},q_{31},q_{12}} \tilde{C}^1(\hat{\textit{\textbf{r}}}_1 \bm{\cdot} \hat{\textit{\textbf{r}}}_2),
\end{align}
with
\begin{align}
\label{eq:CPtilde12}
\tilde{C}^1(\hat{\textit{\textbf{r}}}_1 & \bm{\cdot} \hat{\textit{\textbf{r}}}_2) = \sum_\mu (-1)^{\mu-m_3} \ThreeJSymbol{1,\mu}{\ell_1,m_1}{T_1,-\mu-m_1} \ThreeJSymbol{1,-\mu}{\ell_2,m_2}{T_2,\mu-m_2}  \nonumber \\
& \times \ThreeJSymbol{\ell_3^\prime,-m_3^\prime}{\ell_3,m_3}{n_3,m_3^\prime-m_3} \ThreeJSymbol{\ell_1^\prime,-m_1^\prime}{T_1,\mu+m_1}{n_1,m_1^\prime-\mu-m_1}  \nonumber \\
& \times \ThreeJSymbol{\ell_2^\prime,-m_2^\prime}{T_2,-\mu+m_2}{n_2,m_2^\prime+\mu-m_2} \ThreeJSymbol{n_1,m_1^\prime-\mu-m_1}{n_2,m_2^\prime+\mu-m_2}{n_3,m_3^\prime-m_3}.
\end{align}
The values that $\mu$ can take are $-1$, $0$, and $1$.

Considering the properties of the 3-j symbols in \cref{eq:CPtilde12}, the 
limits for the $T_1$ summation in \cref{eq:GenY12} are $\abs{1 - \ell_1}$ to
$1 + \ell_1$. Similarly, for the $T_2$ summation, the limits are
$\abs{1 - \ell_2}$ to $1 + \ell_2$. However, considering that the $b$
function in \cref{eq:bfunc} can give 0, not all $T_2$ in this range are used.

Unlike the previous section, the expressions here are not exactly the same as 
that in Yan and Drake \cite{Yan1997}, since as noted, their wavefunction
for Li has an extra set of coefficients.
This addition to their wavefunction is included in the
$\tilde{C}^1(\hat{\textit{\textbf{r}}}_1 \bm{\cdot} \hat{\textit{\textbf{r}}}_2)$
and allows them to reduce these using graphical methods as in
Refs.~\cite{Lindgren2012,Brink1993}. Unfortunately, such simplifications are not possible with the 
form of \cref{eq:GeneralWaveTrial}.

We note that $C^1(\hat{\textit{\textbf{r}}}_1 \bm{\cdot} \hat{\textit{\textbf{r}}}_2) = C^1(\hat{\textit{\textbf{r}}}_2 \bm{\cdot} \hat{\textit{\textbf{r}}}_1)$, but $I^1(\hat{\boldsymbol{r}_1} \cdot \hat{\nabla}_2^Y) \neq I^1(\hat{\boldsymbol{r}_2} \cdot \hat{\nabla}_1^Y)$. Namely,
\begin{align}
\label{eq:GenY21}
I^1(\hat{\boldsymbol{r}_2} \cdot \hat{\nabla}_1^Y) = &\sum_{q_{12}=0}^{M_{12}} \sum_{q_{23}=0}^{M_{23}} \sum_{q_{31}=0}^{M_{31}} \sum_{k_{12}=0}^{L_{12}} \sum_{k_{23}=0}^{L_{23}} \sum_{k_{31}=0}^{L_{31}} \sum_{T_1 T_2} b(\ell_1; T_1) C^1(\hat{\textit{\textbf{r}}}_1 \bm{\cdot} \hat{\textit{\textbf{r}}}_2)  \nonumber \\
& \times I_{\rm{R}}(q_{12}, q_{23}, q_{31}, k_{12}, k_{23}, k_{31}; j_1, j_2, j_3, j_{12}, j_{23}, j_{31}; \bar{\alpha}, \bar{\beta}, \bar{\gamma}).
\end{align}
More generally, these integrals can be written as
\begin{align}
\label{eq:GenYij}
I^1(\hat{\boldsymbol{r}_i} \cdot \hat{\nabla}_j^Y) = &\sum_{q_{12}=0}^{M_{12}} \sum_{q_{23}=0}^{M_{23}} \sum_{q_{31}=0}^{M_{31}} \sum_{k_{12}=0}^{L_{12}} \sum_{k_{23}=0}^{L_{23}} \sum_{k_{31}=0}^{L_{31}} \sum_{T_i T_j} b(\ell_j; T_j) C^1(\hat{\textit{\textbf{r}}}_i \bm{\cdot} \hat{\textit{\textbf{r}}}_j)  \nonumber \\
& \times I_{\rm{R}}(q_{12}, q_{23}, q_{31}, k_{12}, k_{23}, k_{31}; j_1, j_2, j_3, j_{12}, j_{23}, j_{31}; \bar{\alpha}, \bar{\beta}, \bar{\gamma}),
\end{align}
and $C^1(\hat{\textit{\textbf{r}}}_i \bm{\cdot} \hat{\textit{\textbf{r}}}_j) = C^1(\hat{\textit{\textbf{r}}}_j \bm{\cdot} \hat{\textit{\textbf{r}}}_i)$. Using cyclic permutations,
\begin{align}
\label{eq:CP23}
C^1(\hat{\textit{\textbf{r}}}_2 & \bm{\cdot} \hat{\textit{\textbf{r}}}_3) = (\ell_1^\prime, \ell_2^\prime, \ell_3^\prime, \ell_1, \ell_2, \ell_3)^{1/2} (-1)^{q_{12} + q_{23} + q_{31}} \sum_{n_1 n_2 n_3} (n_1, n_2, n_3, T_2, T_3)  \nonumber \\
& \times \ThreeJSymbol{1,0}{\ell_2,0}{T_2,0} \ThreeJSymbol{1,0}{\ell_3,0}{T_3,0} \ThreeJSymbol{\ell_1^\prime,0}{\ell_1,0}{n_1,0} \ThreeJSymbol{\ell_2^\prime,0}{T_2,0}{n_2,0} \ThreeJSymbol{\ell_3^\prime,0}{T_3,0}{n_3,0}  \nonumber \\
& \times \ThreeJSymbol{q_{31},0}{q_{12},0}{n_1,0} \ThreeJSymbol{q_{12},0}{q_{23},0}{n_2,0} \ThreeJSymbol{q_{23},0}{q_{31},0}{n_3,0} \SixJSymbol{n_1,n_2,n_3}{q_{23},q_{31},q_{12}} \tilde{C}^1(\hat{\textit{\textbf{r}}}_2 \bm{\cdot} \hat{\textit{\textbf{r}}}_3)
\end{align}
with
\begin{align}
\label{eq:CPtilde23}
\tilde{C}^1(\hat{\textit{\textbf{r}}}_2 & \bm{\cdot} \hat{\textit{\textbf{r}}}_3) = \sum_\mu (-1)^{\mu-m_1} \ThreeJSymbol{1,\mu}{\ell_2,m_2}{T_2,-\mu-m_2} \ThreeJSymbol{1,-\mu}{\ell_3,m_3}{T_3,\mu-m_3}  \nonumber \\
& \times \ThreeJSymbol{\ell_1^\prime,-m_1^\prime}{\ell_1,m_1}{n_1,m_1^\prime-m_1} \ThreeJSymbol{\ell_2^\prime,-m_2^\prime}{T_2,\mu+m_2}{n_2,m_2^\prime-\mu-m_2}  \nonumber \\
& \times \ThreeJSymbol{\ell_3^\prime,-m_3^\prime}{T_3,-\mu+m_3}{n_3,m_3^\prime+\mu-m_3} \ThreeJSymbol{n_1,m_1^\prime-m_1}{n_2,m_2^\prime-\mu-m_2}{n_3,m_3^\prime+\mu-m_3},
\end{align}
and
\begin{align}
\label{eq:CP31}
C^1(\hat{\textit{\textbf{r}}}_3 & \bm{\cdot} \hat{\textit{\textbf{r}}}_1) = (\ell_1^\prime, \ell_2^\prime, \ell_3^\prime, \ell_1, \ell_2, \ell_3)^{1/2} (-1)^{q_{12} + q_{23} + q_{31}} \sum_{n_1 n_2 n_3} (n_1, n_2, n_3, T_3, T_1)  \nonumber \\
& \times \ThreeJSymbol{1,0}{\ell_3,0}{T_3,0} \ThreeJSymbol{1,0}{\ell_1,0}{T_1,0} \ThreeJSymbol{\ell_2^\prime,0}{\ell_2,0}{n_2,0} \ThreeJSymbol{\ell_3^\prime,0}{T_3,0}{n_3,0} \ThreeJSymbol{\ell_1^\prime,0}{T_1,0}{n_1,0}  \nonumber \\
& \times \ThreeJSymbol{q_{31},0}{q_{12},0}{n_1,0} \ThreeJSymbol{q_{12},0}{q_{23},0}{n_2,0} \ThreeJSymbol{q_{23},0}{q_{31},0}{n_3,0} \SixJSymbol{n_1,n_2,n_3}{q_{23},q_{31},q_{12}} \tilde{C}^1(\hat{\textit{\textbf{r}}}_3 \bm{\cdot} \hat{\textit{\textbf{r}}}_1)
\end{align}
with
\begin{align}
\label{eq:CPtilde31}
\tilde{C}^1(\hat{\textit{\textbf{r}}}_3 & \bm{\cdot} \hat{\textit{\textbf{r}}}_1) = \sum_\mu (-1)^{\mu-m_2} \ThreeJSymbol{1,\mu}{\ell_3,m_3}{T_3,-\mu-m_3} \ThreeJSymbol{1,-\mu}{\ell_1,m_1}{T_1,\mu-m_1}  \nonumber \\
& \times \ThreeJSymbol{\ell_2^\prime,-m_2^\prime}{\ell_2,m_2}{n_2,m_2^\prime-m_2} \ThreeJSymbol{\ell_3^\prime,-m_3^\prime}{T_3,\mu+m_3}{n_3,m_3^\prime-\mu-m_3}  \nonumber \\
& \times \ThreeJSymbol{\ell_1^\prime,-m_1^\prime}{T_1,-\mu+m_1}{n_1,m_1^\prime+\mu-m_1} \ThreeJSymbol{n_1,m_1^\prime+\mu-m_1}{n_2,m_2^\prime-m_2}{n_3,m_3^\prime-\mu-m_3}.
\end{align}

\section{Programs}
\label{sec:GenShortProg}
The general codes described in this chapter are available on GitHub \cite{GitHub}.

The general short-short code implementing the equations described in \cref{sec:GeneralShort}
is at least an order of magnitude slower than the corresponding S-, P-, and D-wave
short-short code developed separately for each of the first three partial waves.
On the Talon 2 \cite{Talon2} cluster, an $\omega = 5$ run of the general short-short code
takes approximately a full day running on a single node with 16 cores.

A rough analysis of the code shows that the bulk of the processing time is 
spent calculating the angular parts, i.e. the 3-j and 6-j symbols. Due to the 
symmetries inherent in the 3-j and 6-j symbols, different inputs can generate 
the same output. Also, many input values to these do not satisfy the 
selection rules for the 3-j and 6-j symbols
\cite[p.1054-1064]{Messiah1999} \cite{Edmonds1996}, giving a result of 0.

Multiple papers \cite{Luscombe1998,Wei1998,Rasch2004,Johansson2015} have
detailed strategies to exploit the symmetries to speed up calculations of the
3-j and 6-j symbols. Storing a full lookup table with the full parameter
space of the 6 input variables would be prohibitively memory-intensive.
The most recent of these by Johansson and Forss{\'{e}}n \cite{Johansson2015}
provides a very promising algorithm with full code that could be used to speed
up this general short-short code.


\chapter{Higher Partial Waves}
\label{chp:HigherWaves}

\iftoggle{UNT}{By}{\lettrine{\textcolor{startcolor}{B}}{y}}
the D-wave, the various cross sections in \cref{chp:CrossSections} have not
fully converged, so we looked at the contributions from higher partial waves.
This work includes partial waves through the H-wave ($\ell = 5$), but
there is nothing preventing us from extending this to even higher partial waves
now that we have determined general expressions for the 
external angular integrations in \cref{chp:AngularInt}.

\section{Born-Oppenheimer Approximation}
\label{sec:Born}
In an attempt to approximate the partial waves past the D-wave, we turned to 
the Born-Oppenheimer (BO) approximation \cite{Massey1954,Oppenheimer1928,Geltman1969,Mott1965}.
The BO approximation comes from using only the 
first term in \cref{eq:TrialWave,eq:TrialWaveHigher}. Specifically, this is 
done with the Kohn variational method to get an estimate for the $K$-matrix, 
giving \cite[p.720]{Bransden2003}
\begin{equation}
\label{eq:Born}
\tan\delta_\ell \approx -(\widetilde{S}_\ell,\mathcal{L}\widetilde{S}_\ell ) = -(\bar{S}_\ell,\mathcal{L}\bar{S}_\ell ) = \mp(S_\ell,\mathcal{L}S'_\ell ) \,.
\end{equation}
We have also performed Kohn variational method runs that only use the first two 
terms ($\widetilde{S}_\ell$ and $\widetilde{C}_\ell$) in \cref{eq:TrialWave,eq:TrialWaveHigher}.
This gives phase shifts that are very similar to the BO approximation and lines up
nearly exactly with the BO on most partial waves, so we normally just use the BO
approximation. We also note that the first Born approximation,
$\tan\delta_\ell \approx -(S_\ell,\mathcal{L}S_\ell )$, cannot be used
here due to \cref{eq:SLS0Test}, since this gives 0.

\begin{figure}
	\centering
	\includegraphics[width=5in]{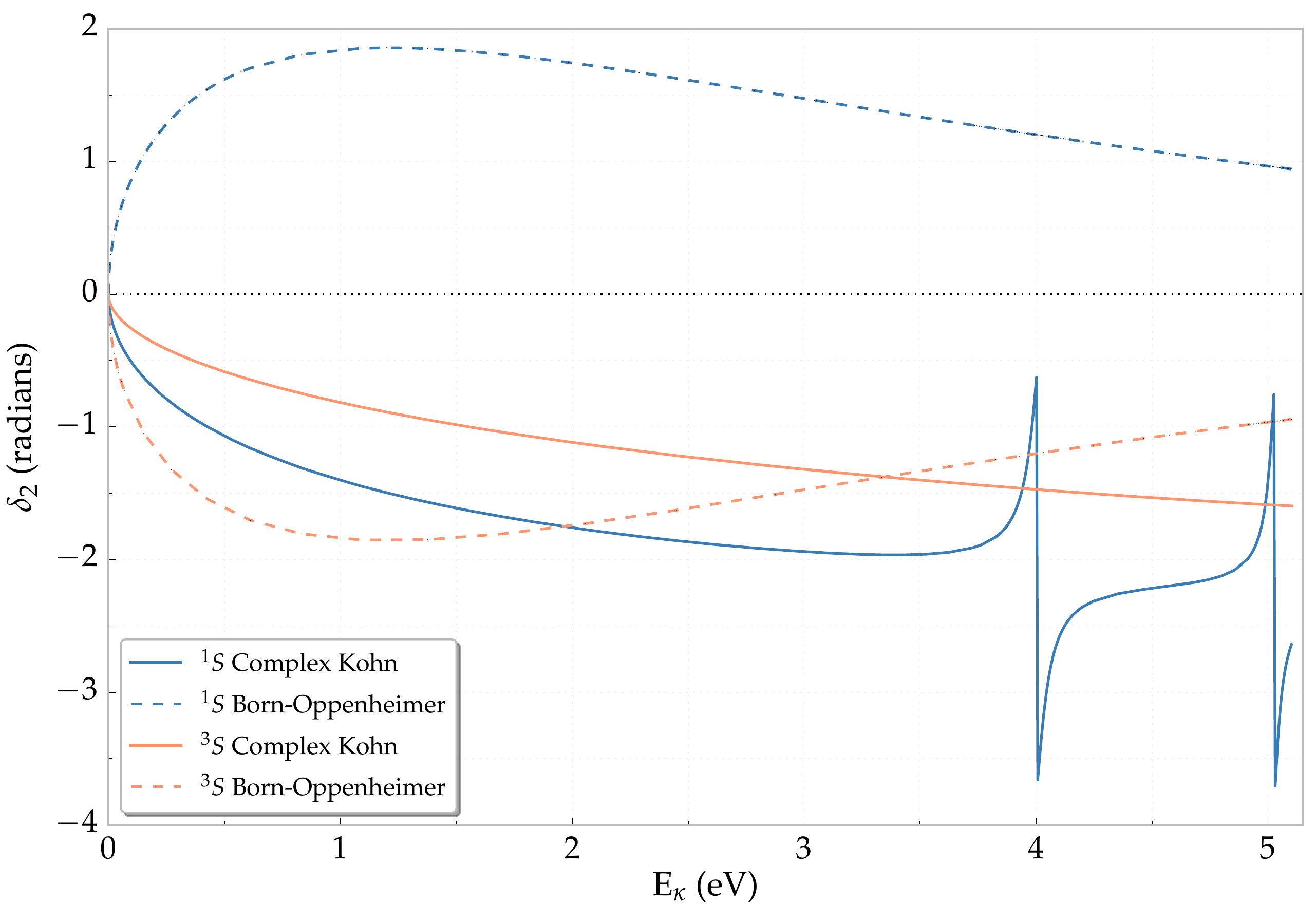}
	\caption[$^{1,3}$S complex Kohn and BO comparison]{$^{1,3}$S phase shift comparison between S-matrix complex Kohn and BO approximation}
	\label{fig:SWavePhaseBorn}
\end{figure}

\begin{figure}
	\centering
	\includegraphics[width=5in]{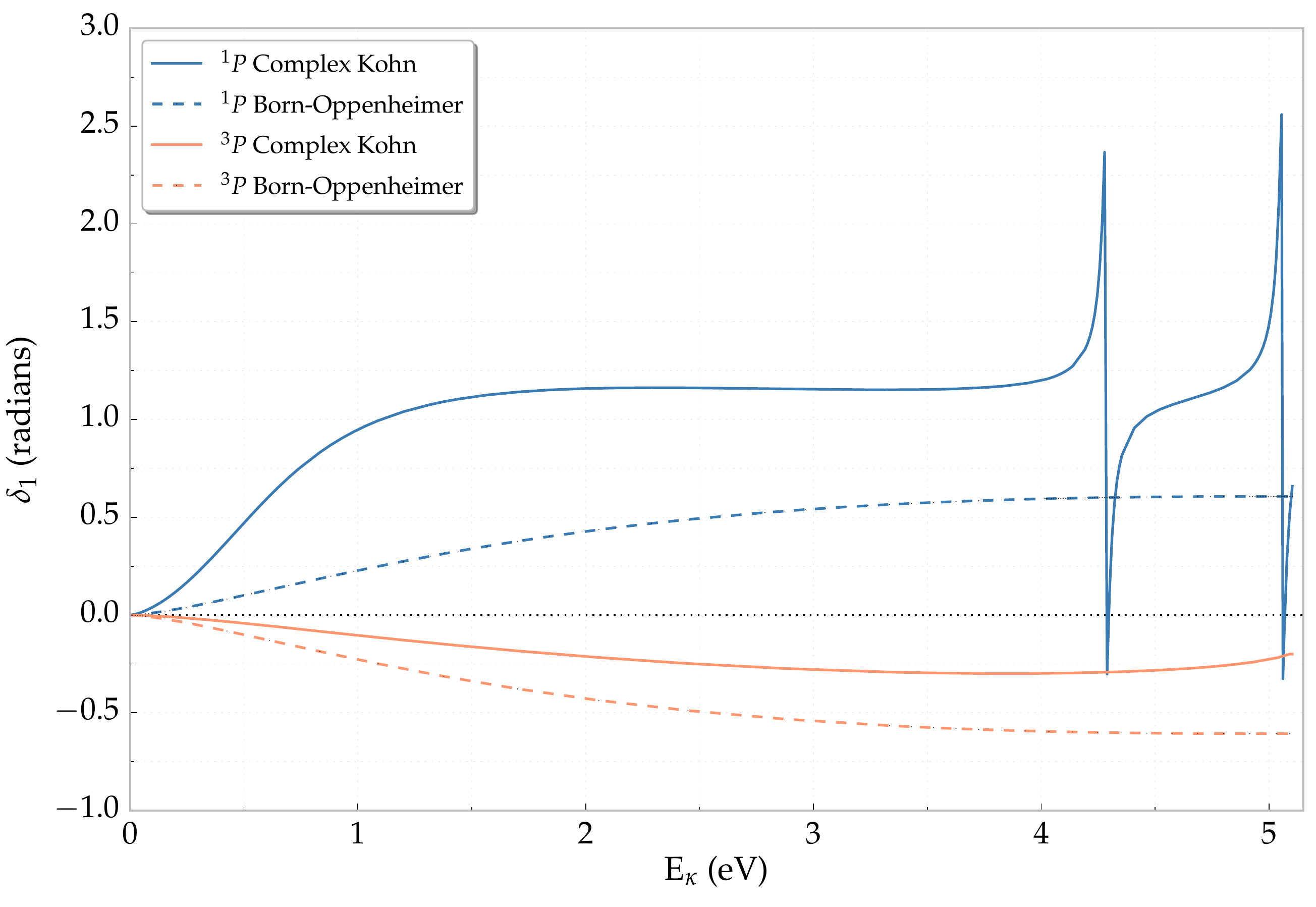}
	\caption[$^{1,3}$P complex Kohn and BO comparison]{$^{1,3}$P phase shift comparison between S-matrix complex Kohn and BO approximation}
	\label{fig:PWavePhaseBorn}
\end{figure}

\begin{figure}
	\centering
	\includegraphics[width=5in]{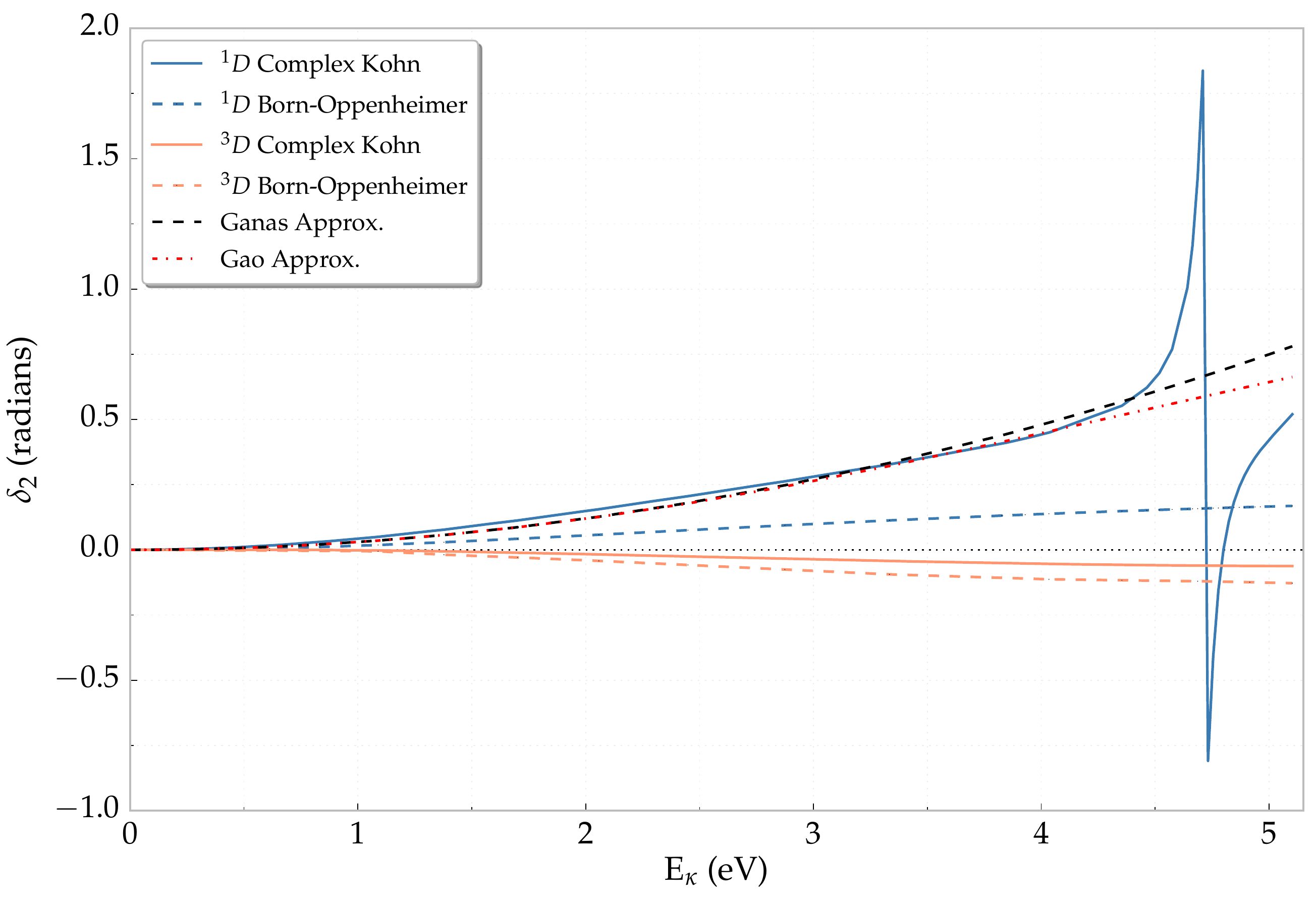}
	\caption[$^{1,3}$D complex Kohn and BO comparison]{$^{1,3}$D phase shift comparison between $S$-matrix complex Kohn, BO approximation, Ganas approximation and Gao approximation}
	\label{fig:DWavePhaseBorn}
\end{figure}

These BO approximations were calculated for the first three partial waves 
but showed huge discrepancies, especially for the S-wave, as seen in
\cref{fig:SWavePhaseBorn,fig:PWavePhaseBorn,fig:DWavePhaseBorn}. The 
$^1$S, $^1$P, and $^1$D partial waves have resonances before the Ps(n=2)
threshold, which we would not expect to be represented by the BO approximation.
The Ganas approximation described in \cref{sec:GanasPhase} is also included in
\cref{fig:DWavePhaseBorn}. This approximation agrees with the $^1$D phase
shifts much better than the BO approximation.

For the $^1$S, $^1$P, $^1$D, and $^1$F partial waves, each has at least one 
resonance before or shortly after the Ps(n=2) threshold. The BO 
approximation does not capture resonance behavior, so we can really only look 
at this for partial waves that do not contain resonances in this region. As
$\ell$ increases, the resonance positions
(in \cref{tab:SWaveResonancesOther,tab:PWaveResonancesOther,tab:DWaveResonancesOther,tab:FWaveResonanceComparisons})
increase, until they are past the threshold fully for the G-wave.
Ho and Yan \cite{Ho2000} calculate the G-wave resonance
at \SI{5.486}{eV} with a width of \SI{0.0109}{eV}.

Due to the obvious discrepancies with even the D-wave, we decided to do 
full Kohn-type calculations for the F-wave to compare, again finding that the BO 
approximation does not match up as well as we would like for either
$^1$F or $^3$F. We tried the same for the G-wave and H-wave, and the BO approximation
does not match up with the complex Kohn phase shifts.
The results of using the BO approximation for the higher partial waves are 
shown later in this chapter in
\cref{fig:FWavePhase,fig:GWavePhase,fig:HWavePhase}. The BO approximation 
unfortunately does not represent any of the partial waves through the H-wave 
well for this system.

\section{Ganas Approximation}
\label{sec:GanasPhase}

Ganas \cite{Ganas1972} gives an expression to estimate phase shifts for
$\ell \geq 2$ using a van der Waals ERT (see \cref{sec:vanderWaalsERT}),
which is given in a more convenient form by
Refs.~\cite{Fabrikant2014a,Mitroy2003a,Swann2015} as
\begin{equation}
\label{eq:vdWPhase}
\delta_\ell(\kappa) \simeq \frac{6 \uppi m C_6 \kappa^4}{(2\ell+5)(2\ell+3)(2\ell+1)(2\ell-1)(2\ell-3)}.
\end{equation}
For Ps scattering, $m = 2$.
The van der Waals coefficient for Ps-H scattering is given in
\cref{sec:vanderWaalsERT} as $C_6 = 34.78473$.

This approximation matches surprisingly well to the $^1$D phase shifts in
\cref{fig:DWavePhaseBorn}. \Cref{fig:FWavePhase,fig:GWavePhase,fig:HWavePhase}
show this approximation for the $^1$F-, $^1$G-, and $^1$H-waves, and it normally
gives a better approximation to the phase shifts than the BO
approximation. For the H-wave,
it matches relatively well but overestimates the phase shifts.

\section{Gao Approximation}
\label{sec:GaoPhase}

Gao \cite{Gao1998a} provides a QDT expansion for the van der Waals interaction,
somewhat similar to those we used for the S-wave and P-wave in
\cref{sec:GaoModel}. For $\ell \geq 2$, the $K_\ell^0$ and its derivatives
do not come into play though. This expression is given in terms of $\tan\delta_\ell$:
\begin{equation}
\label{eq:GaoPhase}
\tan\delta_{\ell \geq 2} = (3 \uppi / 32) \{(\ell+1/2) [(\ell+1/2)^2 - 4][(\ell+1/2)^2 - 1]\}^{-1} (\kappa \beta_6)^4,
\end{equation}
where $\beta_6$ is given in \cref{eq:beta6}.

This approximation matches even better to the $^1$D phase shifts than the Ganas
approximation and matches the same as the Ganas approximation for $\ell \geq 3$.
The behavior of all three of these approximations is shown in
\cref{fig:DWavePhaseBorn,fig:FWavePhase,fig:GWavePhase,fig:HWavePhase}.

\section{F-Wave}
\label{sec:FWave}

Similar to the D-wave (see \cref{sec:DWaveNonlinear}), we investigated the
dependence of the F-wave phase shifts on the nonlinear parameters.
After multiple variations of $\alpha$ and
$\beta$ with a fixed $\gamma$, we settled on using the same set of nonlinear
parameters that we used for the D-wave, as seen in \cref{tab:Nonlinear}, but
with the switchover between the two sets at $\kappa = 0.4$.
We found that to keep $R'(5) < 1$ for $^{1,3}$F, we had to use the 
restricted set described in \cref{sec:Restricted} for $\kappa < 0.4$.
We have calculated the F-wave phase shifts for $\omega = 5$ with these nonlinear
parameters for the first two short-range symmetries using the general codes
described in \cref{sec:GeneralLong,sec:GeneralShort}.

\label{sec:FNonlinear}

\subsection{Phase Shifts}
\label{sec:FWavePhase}

\begin{figure}
	\centering
	\includegraphics[width=\textwidth]{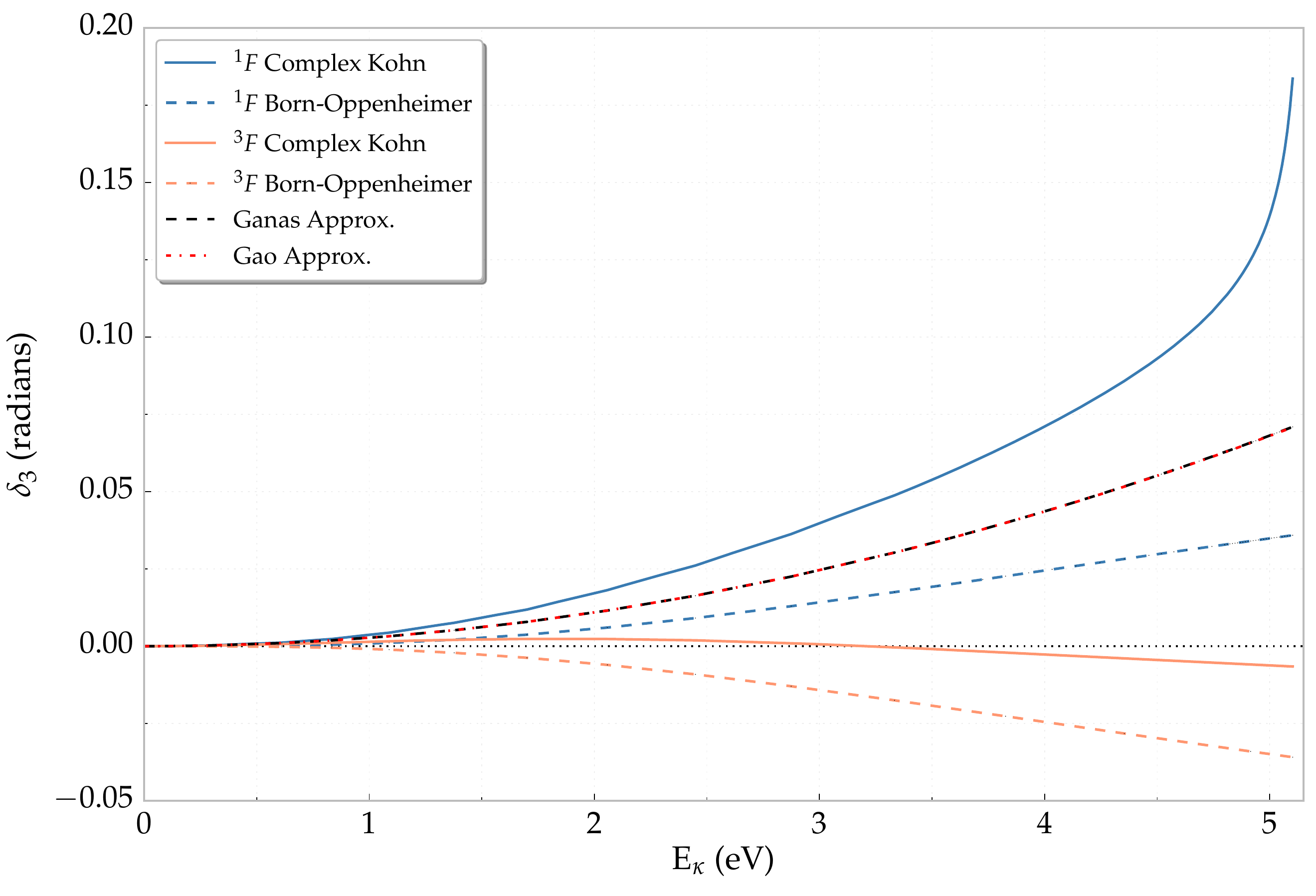}
	\caption{$^{1,3}$F phase shifts}
	\label{fig:FWavePhase}
\end{figure}

\Cref{fig:FWavePhase} shows the phase shifts for the F-wave.
Similar to the D-wave (page \pageref{DWaveSwitch}), the triplet F-wave starts
positive and becomes negative, though at a higher $\kappa$ of about 0.7
(\SI{3.3}{eV}).

We can see that the phase shifts for the (modified) BO approximation do not 
agree very well with the full Kohn calculation, though they follow roughly 
the same shape. The triplet BO is fully negative, while the Kohn only goes 
negative past about $\kappa = 0.7$. The problem with the triplet gets even 
worse for the G-wave (\cref{sec:GWave}) and H-Wave (\cref{sec:HWave}).
The Ganas \cref{sec:GanasPhase} and Gao approximations \cref{sec:GaoPhase},
which do not distinguish
between singlet and triplet states, match better with the singlet than the
BO approximation but still disagrees near the resonance.

\subsection{Resonance}
\label{sec:FWaveResonance}

\begin{figure}
	\centering
	\includegraphics[width=6in]{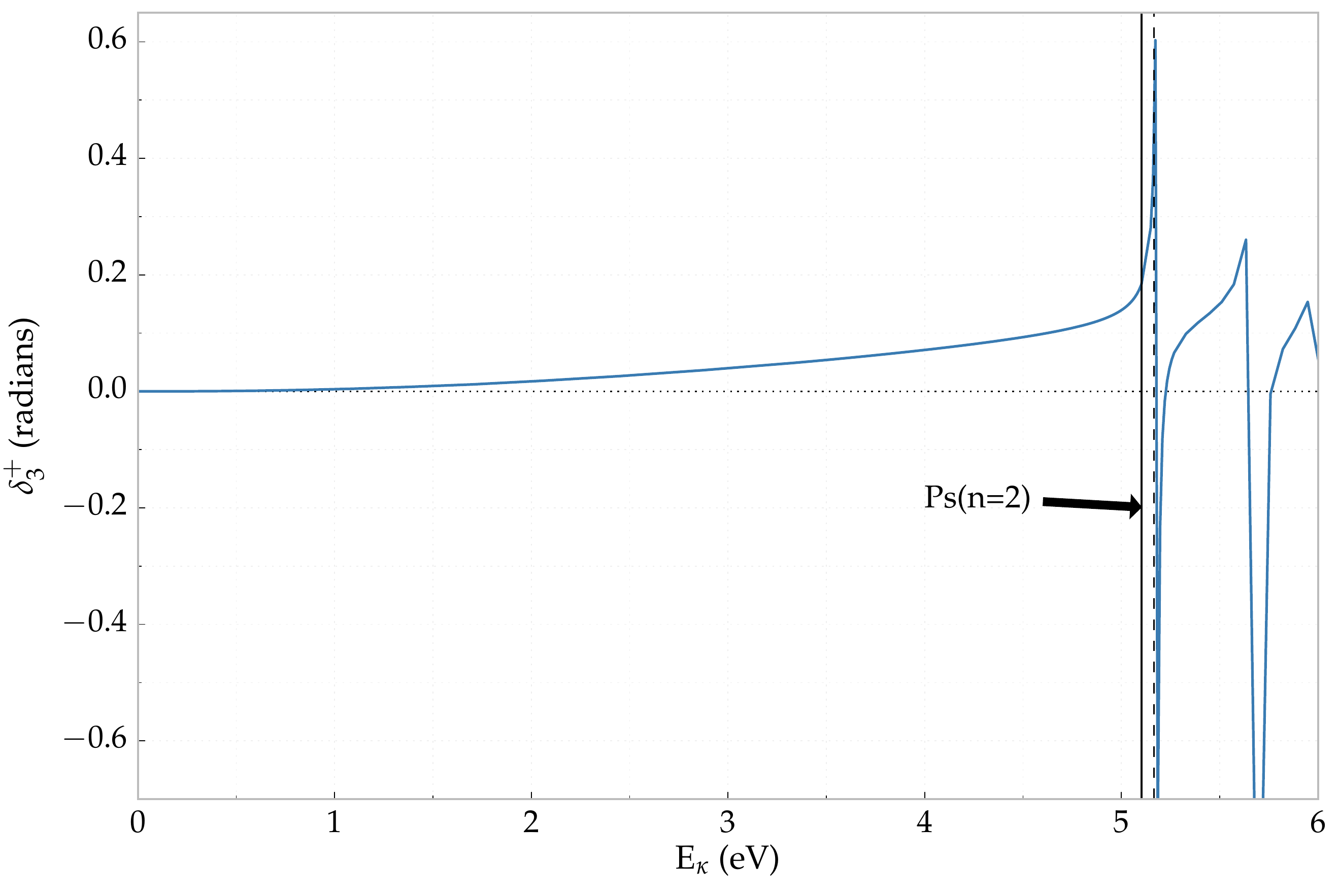}
	\caption[Full $^1$F phase shifts]{$^1$F phase shifts showing full resonance past the inelastic threshold}
	\label{fig:FWavePhaseFull}
\end{figure}

There is the start of a resonance shortly before the threshold cutoff in 
\cref{fig:FWavePhase}. \Cref{fig:FWavePhaseFull} gives a fuller
plot past this resonance. As our code does not contain the open channels 
required to extend into the region that contains the full resonance, 
we likely cannot determine the resonance parameters as accurately.
\Cref{tab:FWaveResonanceComparisons} give the resonance 
parameter fittings using a MATLAB script (\cref{sec:ResonanceFit}).
The first two lines only use data before the Ps(n=2) threshold. The next
two lines have the calculated values when we consider data on both sides of the
resonance, in the range of $\kappa = 0.74 - 0.88$. These two sets
agree well and would likely agree better if we considered the multichannel
problem above the Ps(n=2) threshold.
As with the other partial
waves, these resonance parameters compare reasonably well with the complex
rotation \cite{Ho2000}, though there is more discrepancy with this partial
wave, presumably because the resonance is past the inelastic threshold.
The CC results of Ref.~\cite{Walters2004} agree relatively well, but their
resonance position is higher than both the CC and complex Kohn results.
Interestingly, their less accurate 22Ps1H + H$^-$ calculation
\cite{Blackwood2002b} has a resonance position closer to the complex rotation
result.

\setlength{\abovecaptionskip}{6pt}   
\setlength{\belowcaptionskip}{6pt}   
\begin{table}
\centering
\begin{tabular}{l l l}
\toprule
Method & $^1E_R \text{ (eV)}$ & $^1\Gamma \text{ (eV)}$ \\
\midrule
Current work$^a$: Average $\pm$ standard deviation & $5.1867 \pm 0.0021$ & $0.0125 \pm 0.0003$ \\
Current work$^a$: $S$-matrix complex Kohn & $5.1863$ & $0.0125$ \\
Current work$^b$: Average $\pm$ standard deviation & $5.1838 \pm 0.0031$ & $0.0114 \pm 0.0015$ \\
Current work$^b$: $S$-matrix complex Kohn & $5.1857$ & $0.0145$ \\
CC (9Ps9H + H$^-$) \cite{Walters2004} & $5.200$ & $0.0095$ \\
CC (22Ps1H + H$^-$) \cite{Blackwood2002b} & $5.151$ & $0.010$ \\
Complex rotation \cite{Ho2000} & $5.1661 \pm 0.0014$ & $0.0174 \pm 0.0027$  \\
\bottomrule
\end{tabular}
\caption[F-wave resonance parameters]{F-wave resonance parameters. $^a$ denotes that the data is only taken before the Ps(n=2) threshold.
$^b$ indicates that data is used from both before and after the threshold, as described in the text.}
\label{tab:FWaveResonanceComparisons}
\end{table}

\section{G-Wave}
\label{sec:GWave}

In an effort to try to improve the convergence ratio, $R'(5)$,
of the low energy phase shifts, we
looked at the $\mu$ nonlinear parameter in the shielding function, given
in \cref{eq:PartialWaveShielding}, along with the $m_\ell$ power in the same
equation. Interestingly, the $\omega = 5$ phase shifts were very stable with
the variation of $\mu$ from 0.5 to 0.8 (with a constant $m_\ell$), agreeing to
five significant figures. Keeping $\mu$ constant and increasing $m_\ell$ from
9 to 13 yielded the same phase shifts, again agreeing to five significant
figures. The convergence ratios are greater than 1 for $^{1,3}$G when
$\kappa < 0.3$, but the phase shifts are very small in this range
($\lesssim 10^{-5}$). As for the F-wave, we also use the D-wave nonlinear
parameters, but the switchover is at $\kappa = 0.45$.

\begin{figure}
	\centering
	\includegraphics[width=\textwidth]{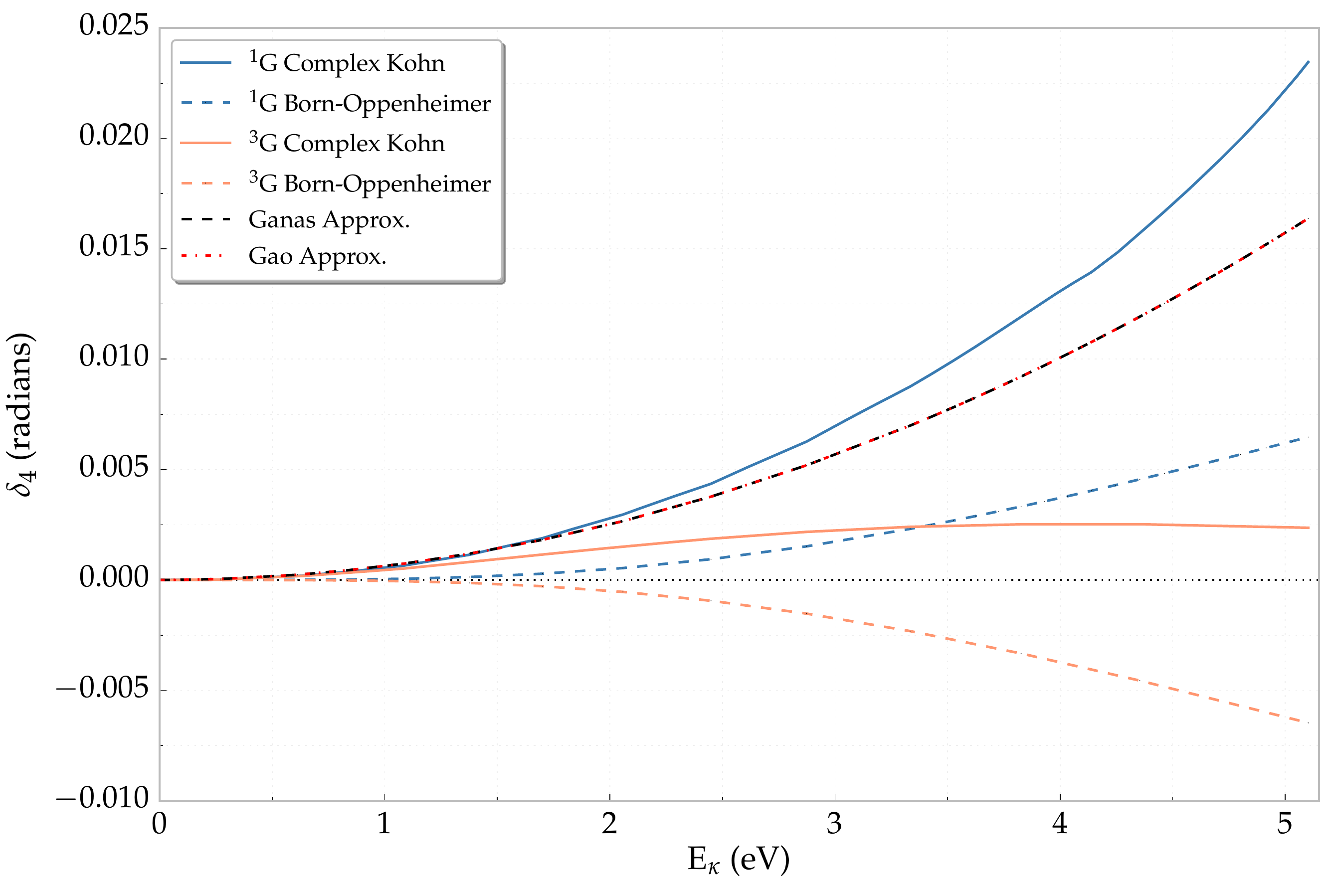}
	\caption{$^{1,3}$G phase shifts}
	\label{fig:GWavePhase}
\end{figure}

Similar to the F-wave (section \ref{sec:FWave}), the BO approximation 
does not work well for this partial wave. In fact, the G-wave triplet Kohn 
calculation is fully positive, yet the BO approximation is fully negative. 
The BO approximation gives a repulsive potential
($\delta_4^- < 0$), while the Kohn calculation gives an attractive potential
($\delta_4^- > 0$).

\section{H-Wave}
\label{sec:HWave}

As for the F-wave and G-wave, we also use the D-wave nonlinear
parameters, but the switchover is at a higher $\kappa$ of $0.45$.
The convergence ratios are greater than 1 for $^{1,3}$H when
$\kappa < 0.4$, but the phase shifts are very small in this range
($\lesssim 10^{-5}$). 

\begin{figure}
	\centering
	\includegraphics[width=\textwidth]{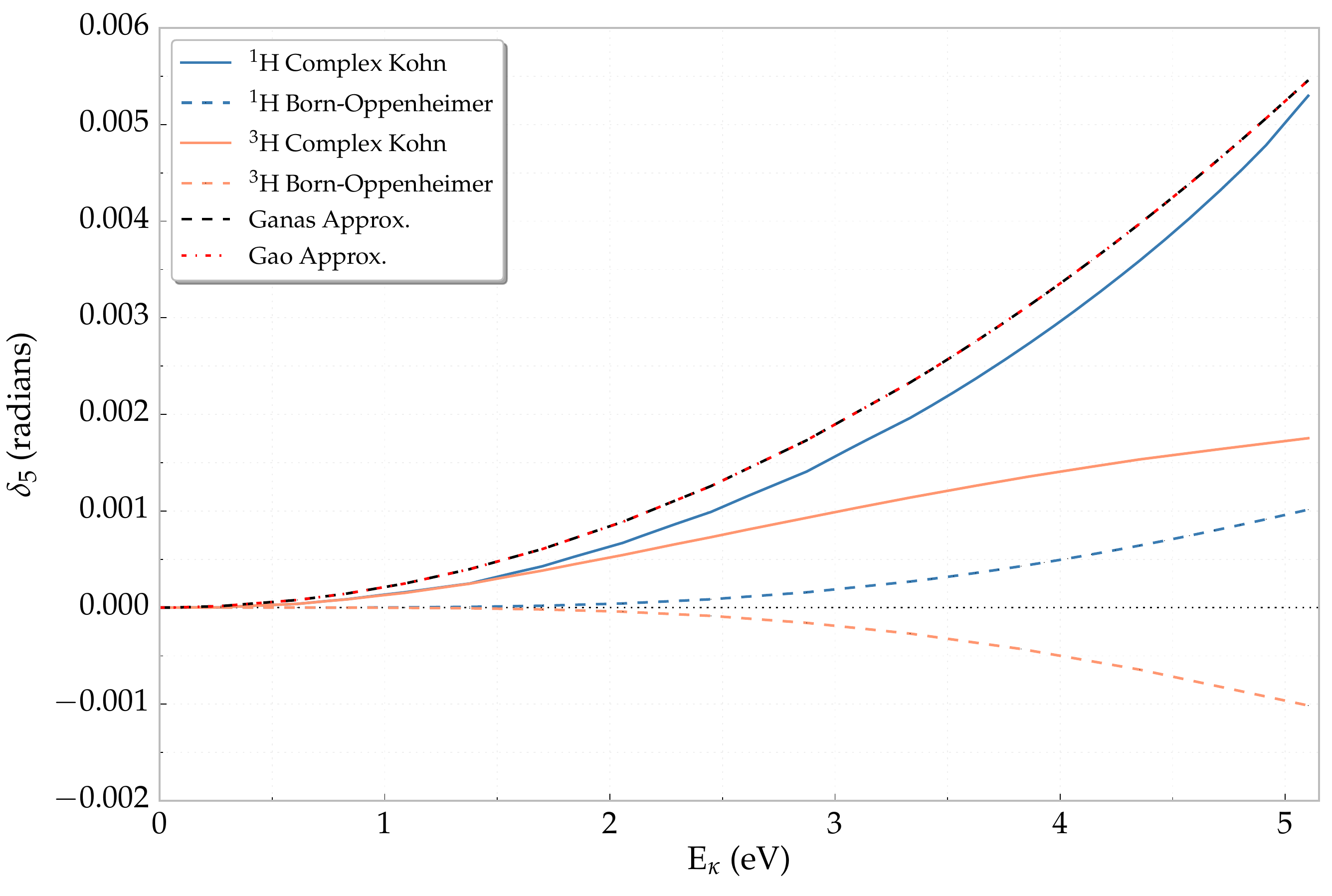}
	\caption{$^{1,3}$H phase shifts}
	\label{fig:HWavePhase}
\end{figure}

\Cref{fig:HWavePhase} shows the $^{1,3}$H phase shifts.
The BO approximation also does not work well for the H-wave. Like 
the F-wave (section \ref{sec:FWave}) and the G-wave (section \ref{sec:HWave}),
the triplet is particularly bad, giving the wrong type of potential.
The Ganas approximation agrees relatively well with the $^1$H curve. The 
phase shifts are small for this partial wave, so its contribution to the 
integrated cross section is essentially negligible, and its contribution
to the differential cross section is small (see \cref{chp:CrossSections}).

\section{Singlet/Triplet Comparisons}
\label{sec:SingTripCompare}

\begin{figure}
	\centering
	\includegraphics[width=5in]{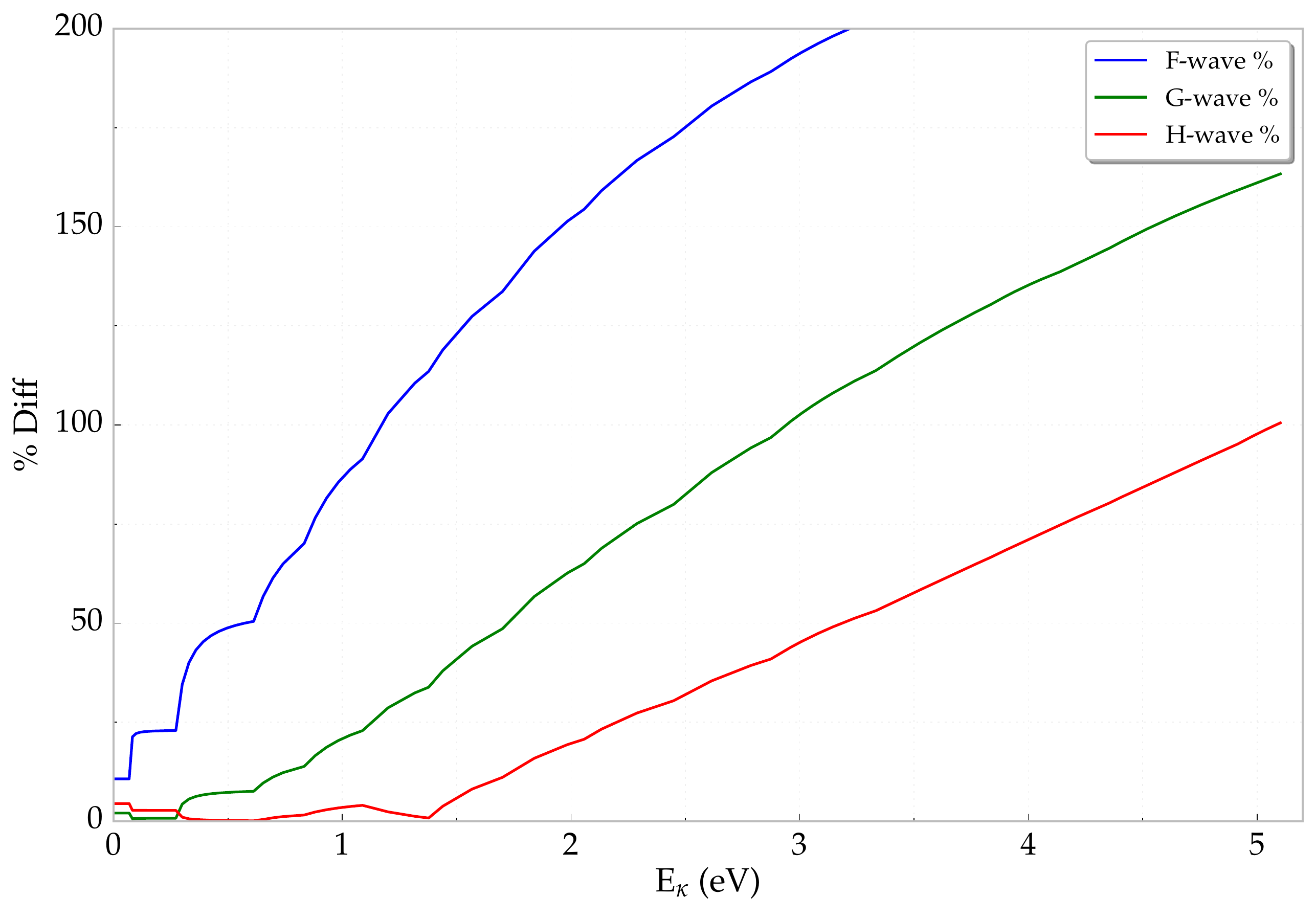}
	\caption[Singlet and triplet higher partial wave comparisons]{Percentage difference comparison between singlet and triplet higher partial waves}
	\label{fig:singlet-triplet-compare}
\end{figure}

Interestingly, there appears to be a pattern concerning the difference between
the singlet and triplet phase shifts for Ps-H scattering as $\ell$ increases.
From \cref{fig:FWavePhase,fig:GWavePhase,fig:HWavePhase}, we see that at low
energies, the singlet and triplet phase shifts are nearly the same. The energy
at which the triplet curve diverges from the singlet becomes higher as $\ell$
increases. To this end, I calculated the percent difference between the singlet
and triplet phase shifts for these three partial waves in
\cref{fig:singlet-triplet-compare}.

By \SI{0.5}{eV}, the $^1$F and $^3$F phase shifts differ by more than 50\%. The
$^1$G and $^3$G differ by more than 50\% at around \SI{1.75}{eV}, and the $^1$H
and $^3$H phase shifts differ to this degree at over \SI{3}{eV}. In fact, the
$^1$H and $^3$H phase shifts are approximately the same up to about \SI{1.5}{eV}.
This suggests that at a high enough value of $\ell$ and above, the singlet and
triplet phase shifts are close to the same in the full energy range below the
Ps(n=2) threshold.

It should be noted that for each of these, the singlet and triplet
BO phase shifts are the opposite sign of each other, but the
integrated elastic cross sections
determined from the BO (\cref{sec:Born}) will be the same due to the
$\sin^2 \! \delta_\ell^\pm$ contribution. The Ganas approximation also does
not differentiate between the singlet and triplet.

The similarity of the singlet and triplet phase shifts is also
not without precedent. An analysis of two papers on
e$^-$-H scattering \cite{Shertzer1994,Chen1997} shows that the S-wave and
P-wave singlet and triplet phase shifts do not agree, but the D-wave and
F-wave singlet and triplet phase shifts agree well. It is also interesting
that this occurs as low as $\ell = 2$, while for Ps-H scattering, these do
not agree fully for even $\ell = 5$.

\section{Summary}
\label{sec:SummaryHigh}

I was able to generalize the evaluation of the matrix elements for arbitrary
$\ell$ (see \cref{chp:General}), which enabled us to calculate phase shifts
for the F-, G-, and H-waves. The phase shifts are not fully converged, but
they are small and generally get smaller as $\ell$ increases. Including the
mixed symmetry terms may allow us to get better converged phase shifts for
these, but the very small phase shifts ($\lesssim 10^{-5}$) at low $\kappa$ 
are not likely to improve without improved numerics as well. We are able to 
calculate the F-wave resonance parameters well, even though the resonance 
lies just past the Ps(n=2) threshold.



\clearpage
\pagebreak
\newpage

\chapter{Cross Sections}
\label{chp:CrossSections}

\iftoggle{UNT}{We}{\lettrine{\textcolor{startcolor}{W}}{e}}
obtain the phase shifts directly from 
the Kohn-type variational methods, but a more relevant quantity for 
experiments is the cross section. Cross sections essentially give the 
strength of the interaction and are a quantity that can be measured 
experimentally.

If we have azimuthal symmetry, as we assume for Ps-H scattering,
the results are independent of $\varphi$.
The quantity that we are most interested in from this is the ratio
$\frac{d\sigma}{d\Omega}$, which is is the differential cross section.
The integrated cross section 
is related to the differential cross section by integrating over $d\Omega$:
\begin{equation}
\label{eq:TotalDiffCross}
\sigma = \int \frac{d\sigma}{d\Omega} \, d\Omega
 = \int_0^{2\uppi} \int_0^{\uppi} \frac{d\sigma}{d\Omega} \sin\theta \, d\theta \, d\varphi.
\end{equation}

\section{Integrated Cross Sections}
\label{sec:totalcross}

The partial wave cross sections can be 
related to the phase shifts by
\citep[p.584]{Bransden2003}
\begin{equation}
\label{eq:PartialCross}
\sigma_{el,\ell}^\pm = \frac{4}{\kappa^2} (2\ell+1) \sin^2 \delta_\ell^\pm.
\end{equation}
In addition to the relation of the integrated cross sections to the
differential cross sections in \cref{eq:TotalDiffCross},
using the partial wave expansion, the integrated cross sections can also be 
expressed as \citep[p.584]{Bransden2003}
\begin{equation}
\label{eq:TotalCross}
\sigma_{el}^\pm = \sum_{\ell=0}^\infty \sigma_{el,\ell}^\pm = \frac{4}{\kappa^2} \sum_{\ell=0}^\infty (2\ell+1) \sin^2 \delta_\ell^\pm.
\end{equation}
We consider that the H(1s) is unpolarized and the final spin states are not
determined, giving spin-weighted cross sections where the
singlet contributes $\frac{1}{4}$, and the triplet contributes $\frac{3}{4}$,
i.e. \cite{Blackwood2002,Ward1987,Joachain1979,Ray1996}
\beq
\label{eq:SpinWeightCS}
\sigma = \tfrac{1}{4} \sigma^+ + \tfrac{3}{4} \sigma^-.
\eeq

\begin{figure}
	\centering
	\includegraphics[width=5.25in]{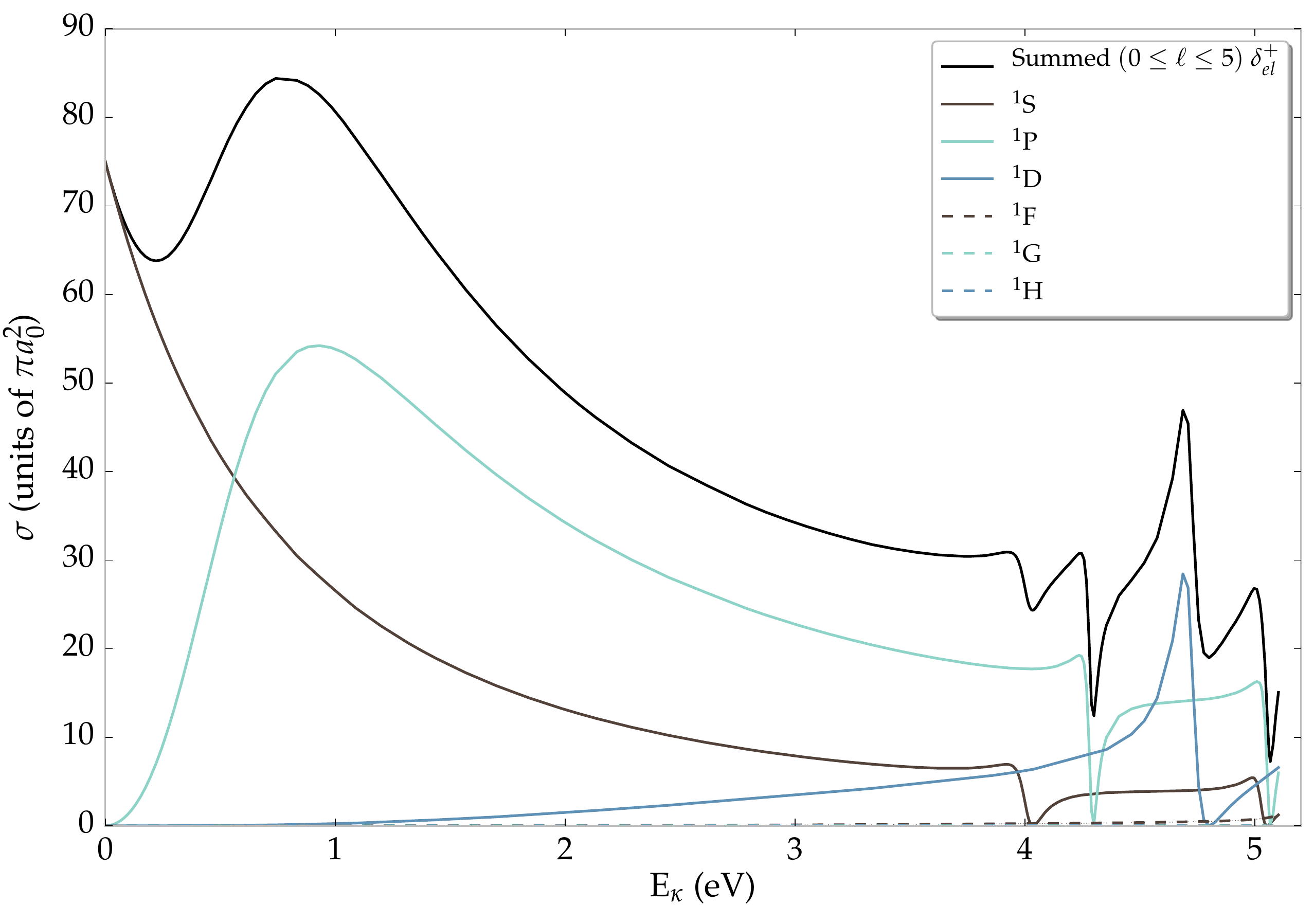}
	\caption[Singlet elastic partial wave cross sections and the sum]{$S$-matrix complex Kohn singlet elastic partial wave cross sections and the sum through $\ell = 5$. An interactive version of this
figure is available online \cite{Plotly} at
\url{https://plot.ly/~Denton/95/singlet-partial-cross-sections-ps-h-scattering/}.}
	\label{fig:singlet-cross-sections}
\end{figure}

\begin{figure}
	\centering
	\includegraphics[width=5.25in]{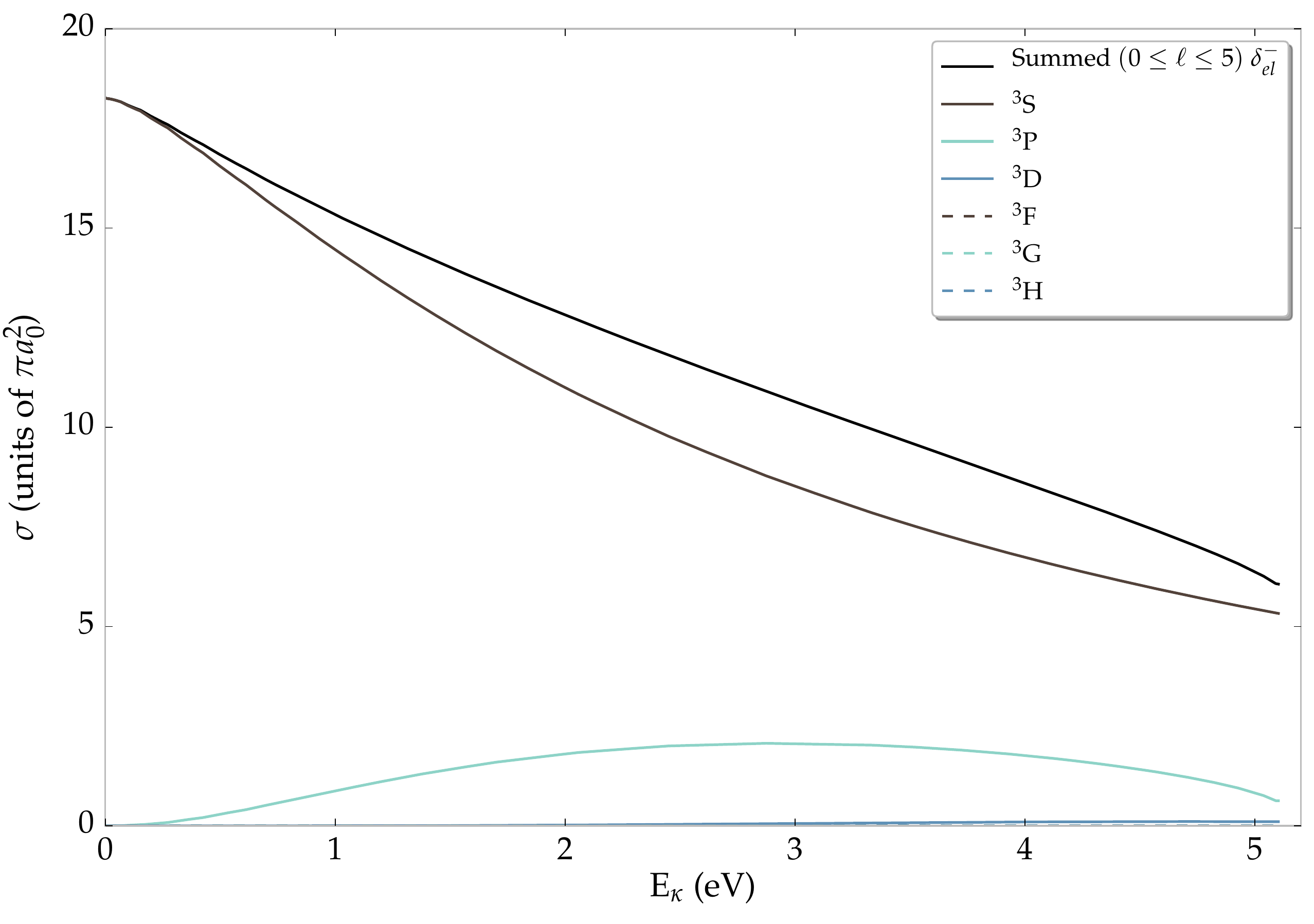}
	\caption[Triplet elastic partial wave cross sections and the sum]{$S$-matrix complex Kohn triplet elastic partial wave cross sections and the sum through $\ell = 5$. An interactive version of this
figure is available online \cite{Plotly} at
\url{https://plot.ly/~Denton/138/triplet-partial-cross-sections-ps-h-scattering/}.}
	\label{fig:triplet-cross-sections}
\end{figure}

\Cref{fig:singlet-cross-sections,fig:triplet-cross-sections} show the partial
wave cross sections for the singlet and triplet, respectively. The ``Summed''
in each is the sum of each of the singlet or triplet partial waves through the
H-wave.

The triplet cross section in \cref{fig:triplet-cross-sections} is dominated 
almost completely by the $^3$S-wave. The $^3$P-wave 
contributes less, and the $^3$D-wave barely registers on the graph. The 
higher partial waves contribute nearly negligible amounts. The 
summed cross section follows
closely with $^3$S, but the contribution from $^3$P is evident.

The singlet cross sections in \cref{fig:singlet-cross-sections} are more
interesting due to their larger partial wave cross sections and the resonances
from the first four partial waves. The ``Summed'' has peaks from each of the
first three partial waves, giving it a more complicated structure than the
summed from the triplet in \cref{fig:triplet-cross-sections}. As mentioned
in \cref{sec:Resonances}, each of these resonances goes through a phase shift
change of $\uppi$. With the $(2\ell+1)$ factor in \cref{eq:TotalCross},
higher partial wave resonances have a larger contribution to the summed
cross sections, as long as the background does not change significantly.
In \cref{fig:singlet-cross-sections}, the $^1$D resonance clearly has the
most significant contribution to the integrated elastic cross section out of
the resonances for the first three partial waves. The $^1$F resonance barely
contributes, but this is because the resonance lies past the inelastic 
threshold.

The cross section at very low energy (less than \SI{0.5}{eV}) is dominated
by the S-wave, as we would expect. At zero energy, the S-wave is the
only partial wave that has a non-zero cross section (for both $^1$S and $^3$S).
Bransden and Joachain \cite[p.589]{Bransden2003} point out that for $\ell \geq 1$,
the partial cross sections vanish as $\kappa^{4\ell}$ as $\kappa \to 0$, so
$\sigma_{el}$ = $\sigma_0$ at this limit, and the scattering is isotropic.
As can be seen in \cref{fig:singlet-cross-sections,fig:triplet-cross-sections},
the D-wave falls off more quickly than the P-wave as $\kappa \to 0$, and the F-wave
falls off even faster.

Also interesting is the minimum in the summed singlet cross section at 
approximately $\SI{0.25}{eV}$ and then the maximum at $\SI{0.74}{eV}$. The dip
here at low energy is due to the mixing of the $^1$S and $^1$P cross sections.
The $^1$S-wave cross section is decreasing rapidly while the $^1$P-wave cross
section is increasing, giving this feature. The maximum is due primarily to the
$^1$P-wave.

\begin{figure}
	\centering
	\includegraphics[width=\textwidth]{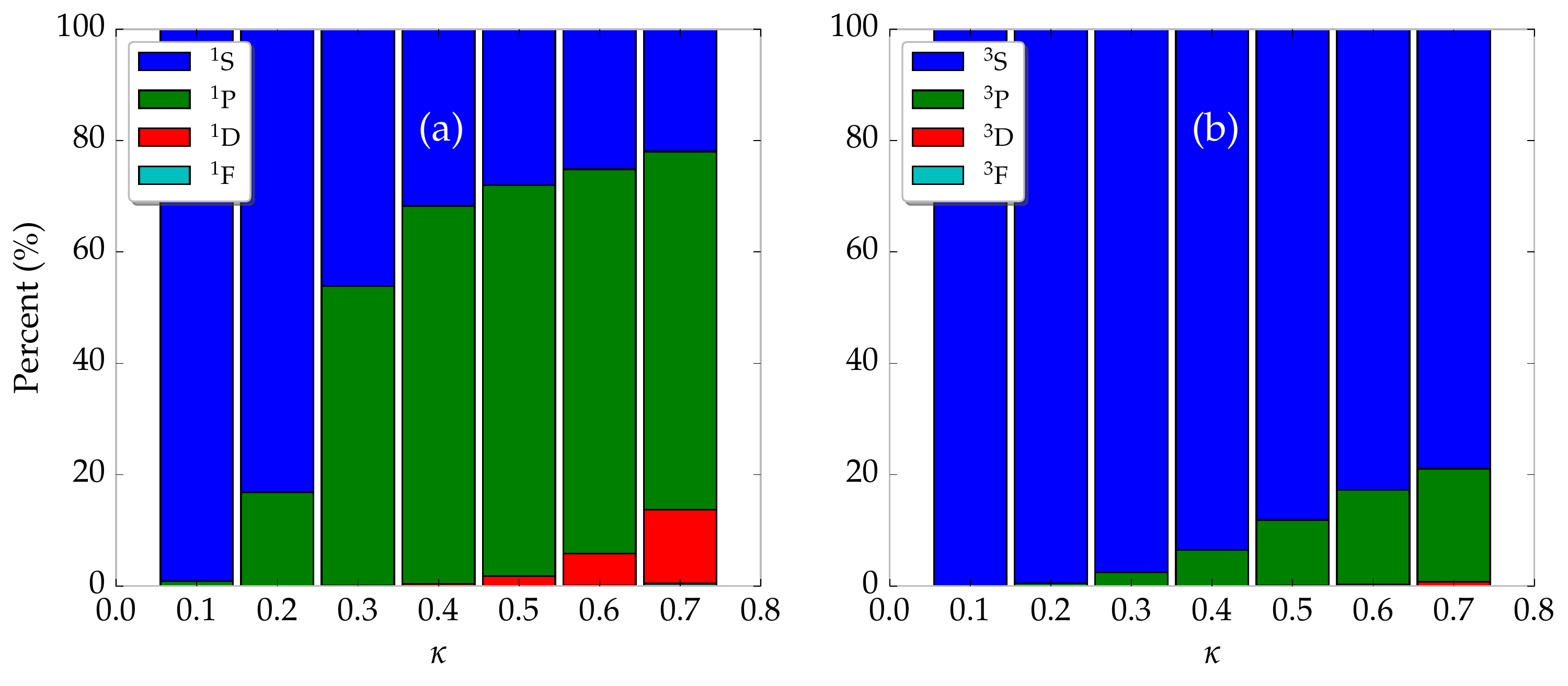}
	\caption[Percentage contribution to integrated elastic cross section for selected $\kappa$]{Percentage contribution to integrated elastic cross section from each partial wave with selected $\kappa$ for the singlet (a) and triplet (b).}
	\label{fig:percentage-cross-sections}
\end{figure}

From \cref{eq:TotalCross}, for exact cross sections, we have to do an infinite
summation. In practice, we add partial waves until the cross section no longer
changes a significant amount. In \cref{fig:percentage-cross-sections}(a), we
consider what the percentage contribution to the summed cross section for
the singlet partial waves is at the 7 ``standard'' $\kappa$ values. From this,
we can see the trend that the $^1$S-wave is by far the greatest contribution at
small $\kappa$, but the $^1$P-wave becomes the dominant contribution through
most of the rest of the energy range. When $\kappa \geq 0.5$, the $^1$D-wave is
no longer a negligible contribution. The $^1$F-wave barely contributes, even
for $\kappa = 0.7$, and the $^1$G- and $^1$H-wave are not shown due to their
insignificant contributions.

The corresponding bar chart for the triplet is in
\cref{fig:percentage-cross-sections}(b). As qualitatively described earlier,
the contribution to the elastic integrated cross section for
the triplet is mainly due to the $^3$S-wave. The $^3$P-wave contributes about
$20\%$ at $\kappa = 0.7$, and the $^3$D-wave contribution is nearly negligible.

\begin{figure}
	\centering
	\includegraphics[width=\textwidth]{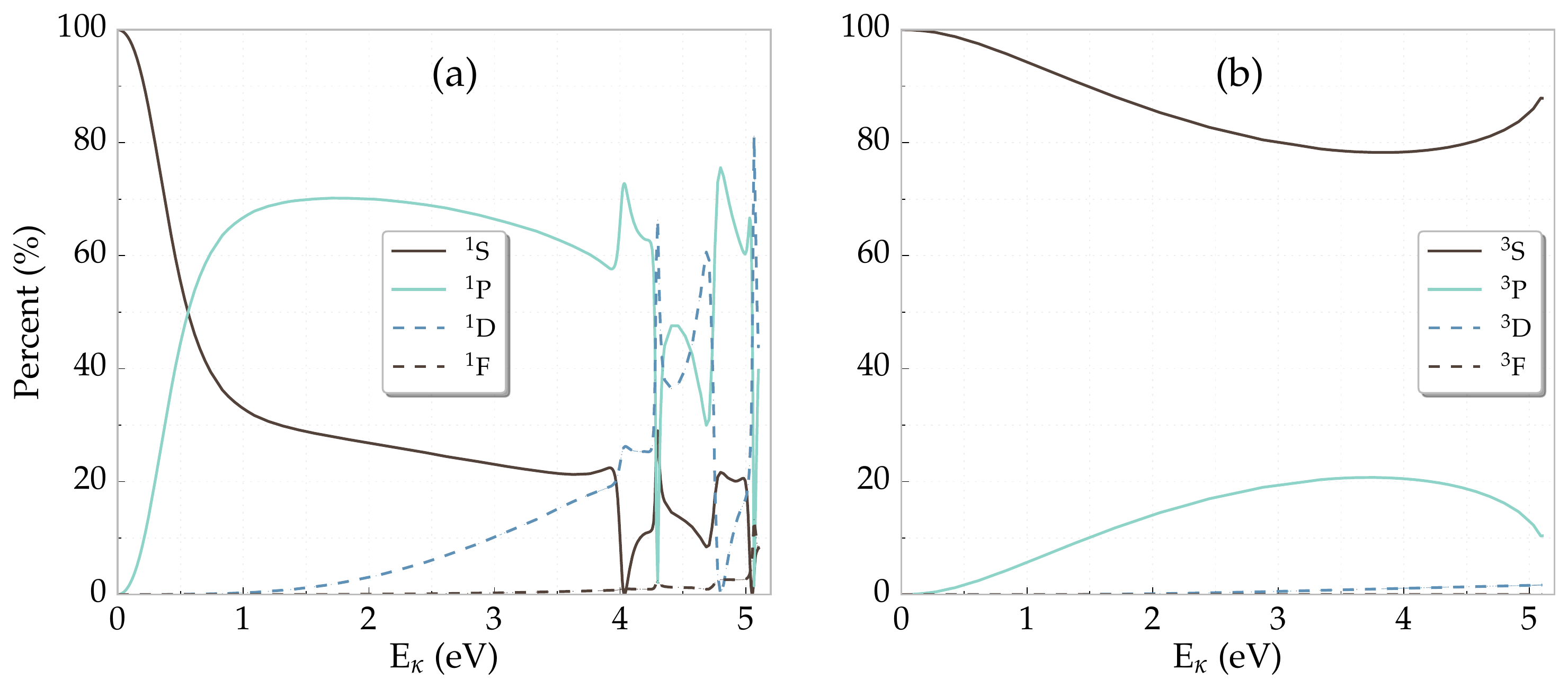}
	\caption[Percentage contribution to integrated elastic cross section]{Percentage contribution to integrated elastic cross section from each partial wave for the singlet (a) and triplet (b).}
	\label{fig:percentage-cross-sections-full}
\end{figure}

\Cref{fig:percentage-cross-sections-full} also shows the percent contributions to 
the integrated elastic cross sections from each partial wave but shows
this data for the entire energy range. The triplet graph in
\cref{fig:percentage-cross-sections-full}(b) again shows that only the $^3$S- and
$^3$P-waves give very significant contributions. The $^3$D-wave gives a
non-negligible but small contribution near the inelastic threshold

The corresponding singlet graph in \cref{fig:percentage-cross-sections-full}(a)
is more difficult to interpret due to the resonances. The $^1$S- and
$^1$P-wave dominate until about $\SI{3}{eV}$, and the $^1$P-wave is the largest 
contribution from approximately $\SI{0.5}{eV}$ to slightly over $\SI{4}{eV}$.
The $^1$D-wave starts being a more dominant contribution past $\SI{4}{eV}$, and
it has spikes in the percentage when the $^1$S and $^1$P resonances go to their
minima.

For a more quantitative approach, we calculate the percent contributions of the
different partial waves to the spin-weighted integrated elastic cross section as presented
in \cref{tab:PercentToCross}. The second and third columns give the
contribution from the singlet and triplet combined for each partial wave. From
this, it is noticeable that the F-wave should be included, though the average
is less than $0.6\%$. The G-wave and H-wave barely contribute, with the H-wave
contribution from both the singlet and triplet less than $0.002\%$ on 
average.

\begin{figure}
	\centering
	\includegraphics[width=6in]{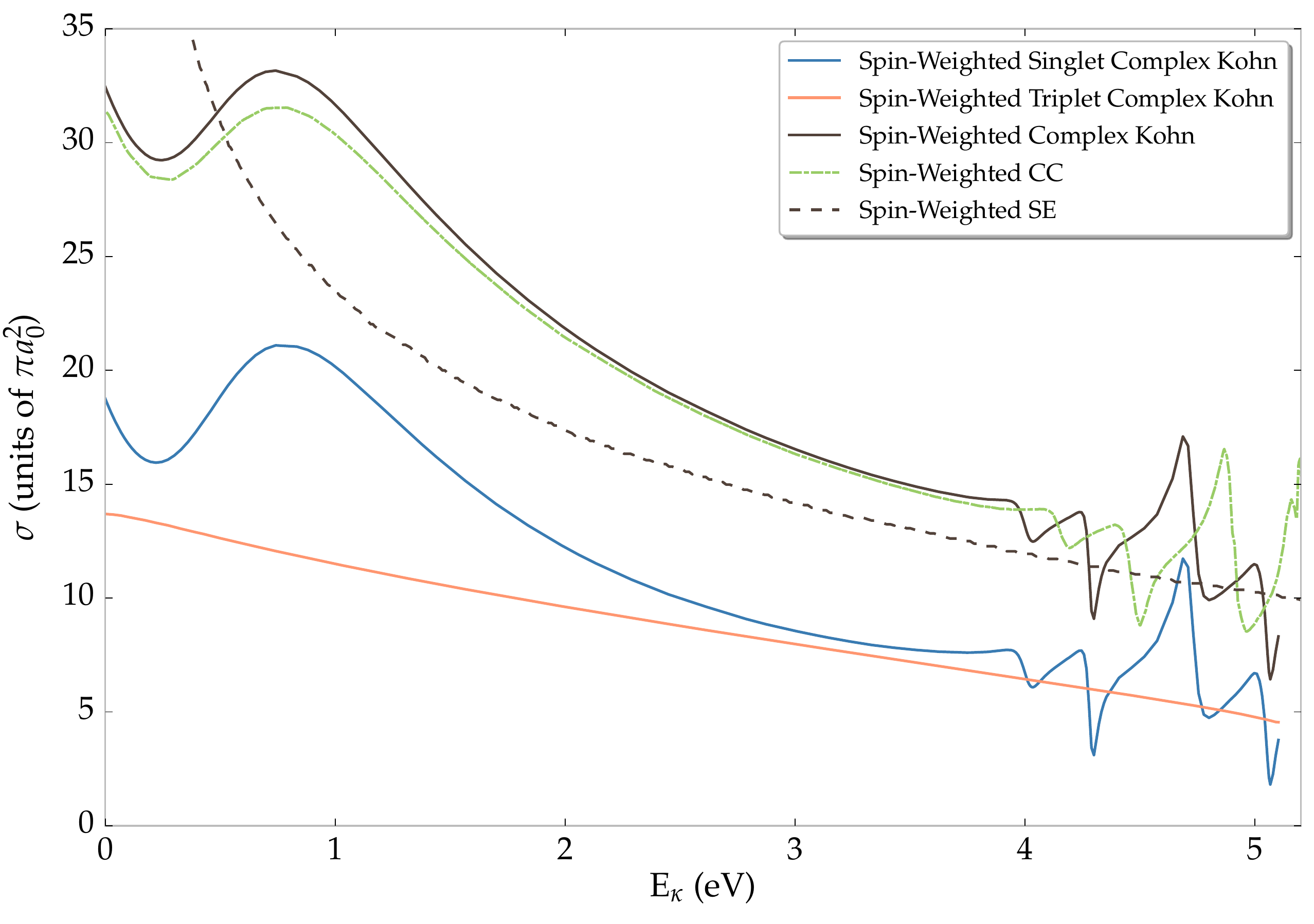}
	\caption[Elastic integrated cross sections]{Elastic integrated cross sections. CC results are from Ref.~\cite{Walters2004}, and SE results are from Ref. \cite{Hara1975}. An interactive version of this figure is available online \cite{Plotly}
at \url{https://plot.ly/~Denton/150/integrated-cross-sections-ps-h-scattering/}.}
	\label{fig:combined-cross-sections}
\end{figure}

For the last 4 columns of \cref{tab:PercentToCross}, we also compare each
partial wave's contribution to the spin-weighted integrated elastic cross section, but we
separate out the singlet and triplet. From columns 4 and 5 for the singlet,
we see that the $^1$F-wave is important, and it gives most of the combined
$^{1,3}$F-wave contribution to the elastic integrated cross section. The 
$^3$F-wave contribution is $0.0011\%$ on average, and the $^3$D contribution 
is less than $0.41\%$ through the entire energy range. We include all partial
waves through the H-wave in the final results, but these would be relatively
well converged if we stopped adding partial waves to \cref{eq:TotalCross} at
the F-wave. For comparison, the CC \cite{Walters2004} cross section shown in
\cref{fig:combined-cross-sections} uses partial waves through the G-wave, but
their graph would likely not change if they included the H-wave.

\begin{table}
\small
\centering
\begin{tabular}{cllllllll}
\toprule
& Max. & Avg. & Max. & Max. & Avg. & Max. & Max. & Avg. \\
$\ell$ & \% & \% & $^1\ell$ \% & $E_{\bm \kappa}^-$ (eV) & $\%^+$ & $\%^-$ & $E_{\bm \kappa}^+$ (eV) & $\%^-$ \\
\midrule
S-wave & 100.0\%  & 60.61\%   & 57.80\%  & 2.721$^{-7}$ & 20.80\%  & 20.86\%   & 5.067 & 13.27\% \\
P-wave & 45.97\%  & 30.21\%   & 42.89\%  & 1.200        & 24.36\%  & 4.323\%   & 4.300 & 1.947\%  \\
D-wave & 42.07\%  & 8.565\%   & 41.61\%  & 4.686        & 8.178\%  & 0.401\%   & 5.067 & 0.129\%  \\
F-wave & 3.782\%  & 0.596\%   & 3.765\%  & 5.078        & 0.593\%  & 0.00606\% & 5.067 & 0.0011\%  \\
G-wave & 0.103\%  & 0.022\%   & 0.0994\% & 5.067        & 0.0206\% & 0.00106\% & 5.067 & 0.00046\%  \\
H-wave & 0.0083\% & 0.0019\%  & 0.0062\% & 5.067        & 0.0013\% & 0.00070\% & 5.067 & 0.00021\% \\ 
\bottomrule
\end{tabular}
\caption[Cross section contribution from each partial wave]{Percent contribution to the elastic integrated cross section from each partial wave for both the maximum and average for the entire energy range. Superscripts give powers of 10.}
\label{tab:PercentToCross}
\end{table}

Finally, combining the singlet and triplet integrated elastic cross 
sections using the spin-weighting in \cref{eq:SpinWeightCS} gives the result
in \cref{fig:combined-cross-sections}. We include the spin-weighted singlet
and triplet cross sections for comparison with the combined integrated elastic cross section.
The comparison to the CC \cite{Walters2004} and SE \cite{Hara1975} results is
made possible by using the CurveSnap \cite{CurveSnap} program to extract the
curves from the respective papers. The SE curve gives a decent approximation to
the background without any resonances. We would not expect the SE to give
resonance information. 
There is
good agreement between the complex Kohn and CC cross sections, but the CC
results have resonances shifted to higher energy. The complex Kohn resonances
correspond better than the CC to the resonance positions that are given in the
complex rotation results of Yan and Ho \cite{Yan1999,Yan1998a,Ho1998,Ho2000}, as
seen in \cref{tab:SWaveResonancesOther,tab:PWaveResonancesOther,tab:DWaveResonancesOther,tab:FWaveResonanceComparisons}.

\begin{figure}
	\centering
	\includegraphics[width=\textwidth]{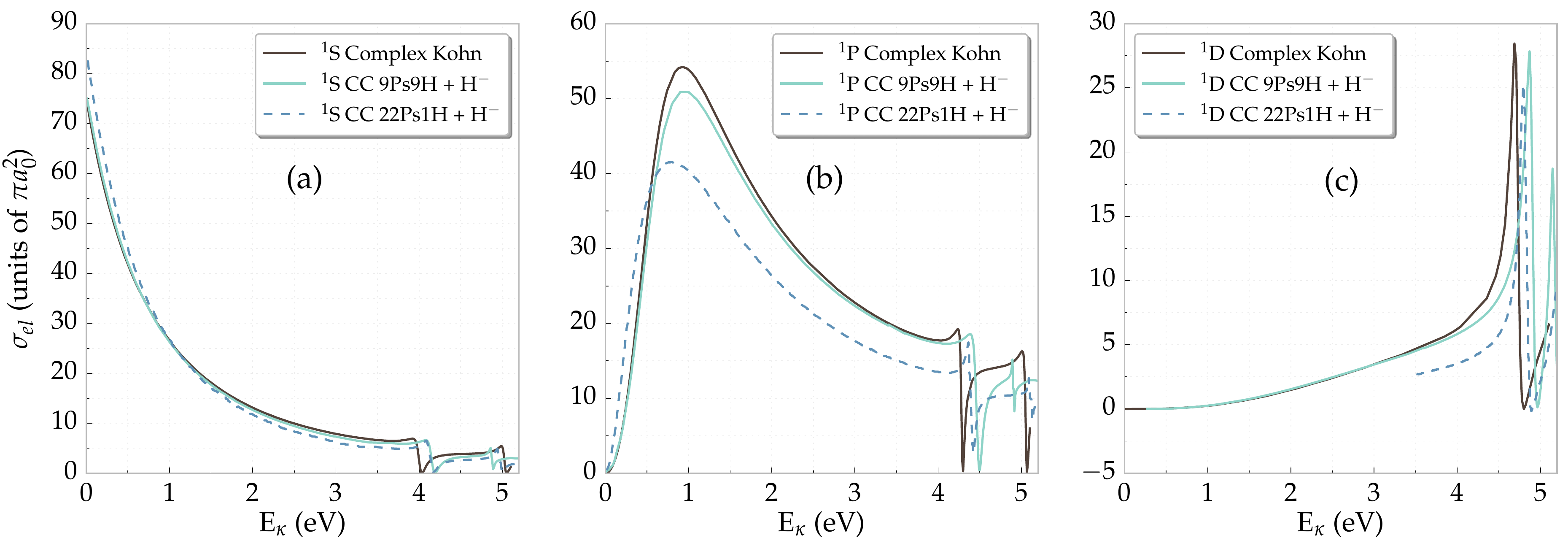}
	\caption[Comparisons of $^1$S, $^1$P, and $^1$D cross sections.]{Comparisons of $^1$S, $^1$P, and $^1$D elastic partial wave cross sections. CC 9Ps9H + H$^-$ results are from Ref.~\cite{Walters2004}, and CC 22Ps1H + H$^-$ results are from Ref.~\cite{Blackwood2002b}.}
	\label{fig:spd-singlet-cross-sections}
\end{figure}

\begin{figure}
	\centering
	\includegraphics[width=\textwidth]{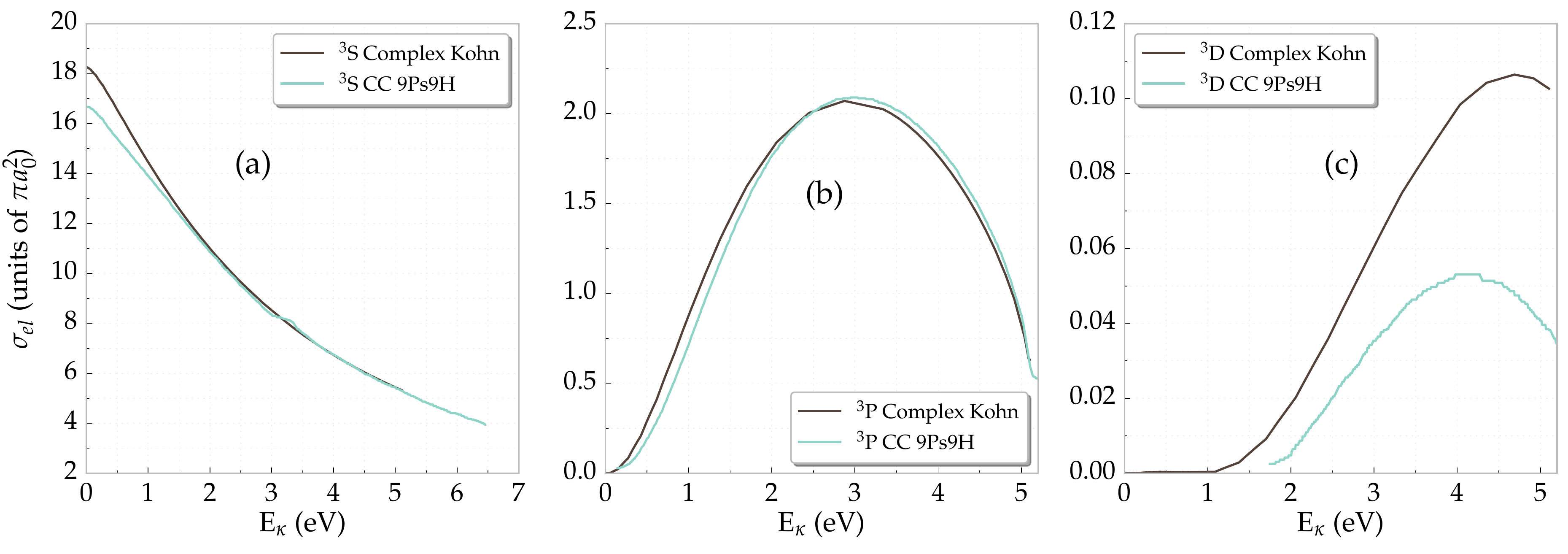}
	\caption[Comparisons of $^3$S, $^3$P, and $^3$D cross sections.]{Comparisons of $^3$S, $^3$P, and $^3$D elastic partial wave cross sections. CC results for $^3$S and $^3$P are from Ref. \cite{Walters2004}, and the CC results for $^3$D are from Ref. \cite{Blackwood2002}.}
	\label{fig:spd-triplet-cross-sections}
\end{figure}

For partial wave cross section data available from the CC papers
\cite{Walters2004,Blackwood2002,Blackwood2002b}, we compare the complex Kohn
partial wave cross sections to these in
\cref{fig:spd-singlet-cross-sections,fig:spd-triplet-cross-sections}.
The CC 9Ps9H + H$^-$ cross sections are more accurate than the CC 22Ps1H + H$^-$,
as pointed out in those papers. They also have CC data with the 9Ps9H
approximation \cite{Blackwood2002}, but that gives less accurate results than
the CC 22Ps1H + H$^-$.

Other than the shifted resonance positions, the CC results tend to match well
with the complex Kohn results. However, there are some features that are worth
noticing. For the $^1$P cross section in \cref{fig:spd-singlet-cross-sections}, the CC
9Ps9H + H$^-$ maximum is lower than the complex Kohn maximum. The CC cross section
22Ps1H + H$^-$ maximum is even lower, so we would expect that if more
eigen- and pseudo-states are included, the CC maximum would likely match up 
with the complex Kohn. Also, the $^3$D CC cross section is much smaller than
the complex Kohn. This is a reflection of the fact that in
\cref{fig:DWavePhase,tab:DWaveComparisons}, the CC results are higher than the
complex Kohn results, making them less negative and the corresponding cross sections
smaller. Noting the magnitude of the $^3$D cross section for both methods, the
contribution to the full summed elastic integrated cross section is small.
As a simple test, we tried replacing the complex Kohn $\kappa = 0.1 - 0.7$ $^3$D
phase shifts with those of the CC, and the change to the cross section is
less than 0.084\% for this range. Consequently, for the full summed
cross section, the discrepancy between the complex Kohn and the CC results
does not change things significantly.

\begin{figure}
	\centering
	\includegraphics[width=\textwidth]{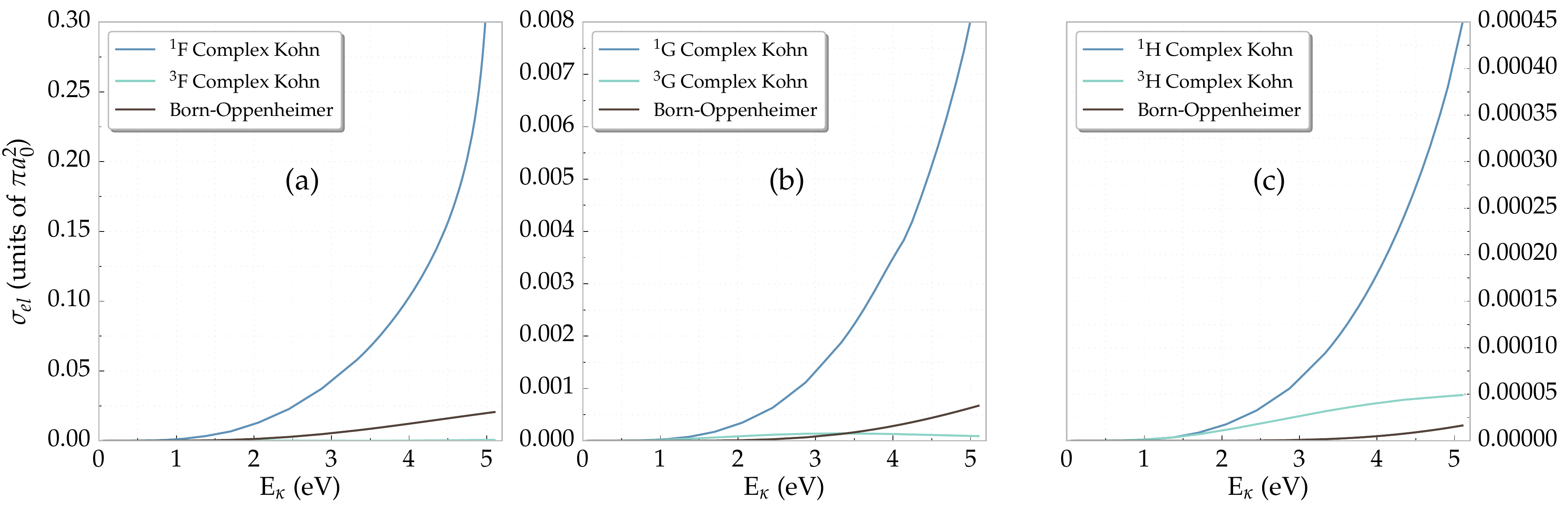}
	\caption[Comparisons of F-, G-, and H-wave cross sections]{Comparisons of F-, G-, and H-wave elastic partial wave cross sections between complex Kohn and BO approximation}
	\label{fig:fgh-born-cross-sections}
\end{figure}

In \cref{fig:fgh-born-cross-sections}, the BO approximation elastic partial
wave cross sections do not match up with either the singlet
or triplet for the F-, G-, or H-waves. Interestingly, we see that the plots
for the singlet and the BO look approximately the same in (a), (b), and (c)
but with a different vertical scale. Similar to what we
noticed in \cref{sec:SingTripCompare}, we also note that the singlet and
triplet phase shifts match for higher $\kappa$ as $\ell$ increases.

\section{Differential Cross Sections}
\label{sec:diffcross}

The differential cross section is important through \cref{eq:TotalDiffCross},
but this also gives information about the angular dependence ($\theta$) of the
system. The differential cross sections can be calculated from the phase shifts by
\cite[p.584]{Bransden2003}
\begin{align}
\label{eq:DiffCross}
\nonumber \frac{d\sigma_{el}^\pm}{d\Omega} = \frac{1}{\kappa^2} & \sum_{\ell=0}^\infty \sum_{\ell^\prime=0}^\infty (2\ell+1)(2\ell^\prime+1) \exp\left\{\ii \left[\delta_\ell(\kappa) - \delta_{\ell^\prime}(\kappa) \right] \right\} \\
& \times \sin\delta_\ell^\pm(\kappa) \sin\delta_{\ell^\prime}^\pm(\kappa) P_\ell(\cos\theta) P_{\ell^\prime}(\cos\theta)\,.
\end{align}
This expression has the complex-valued exponential, for which we can use the
well-known Euler formula of
\begin{equation}
\label{eq:ComplexExp}
\ee^{\ii x} = \cos x + \ii \sin x
\end{equation}
to split this into real-valued and imaginary-valued parts. As long as the
finite truncation of the upper limits of the summations are the same, the
imaginary part becomes 0 to within numerical accuracy. So we can use the
approximation of
\begin{align}
\label{eq:DiffCross1}
\nonumber \frac{d\sigma_{el}^\pm}{d\Omega} \approx \frac{1}{\kappa^2} & \sum_{\ell=0}^{\ell_{max}} \sum_{\ell^\prime=0}^{\ell_{max}} (2\ell+1)(2\ell^\prime+1) \cos \left[\delta_\ell(\kappa) - \delta_{\ell^\prime}(\kappa) \right] \\
& \times \sin\delta_\ell^\pm(\kappa) \sin\delta_{\ell^\prime}^\pm(\kappa) P_\ell(\cos\theta) P_{\ell^\prime}(\cos\theta)\,.
\end{align}

\begin{figure}
	\centering
	\includegraphics[width=5.25in]{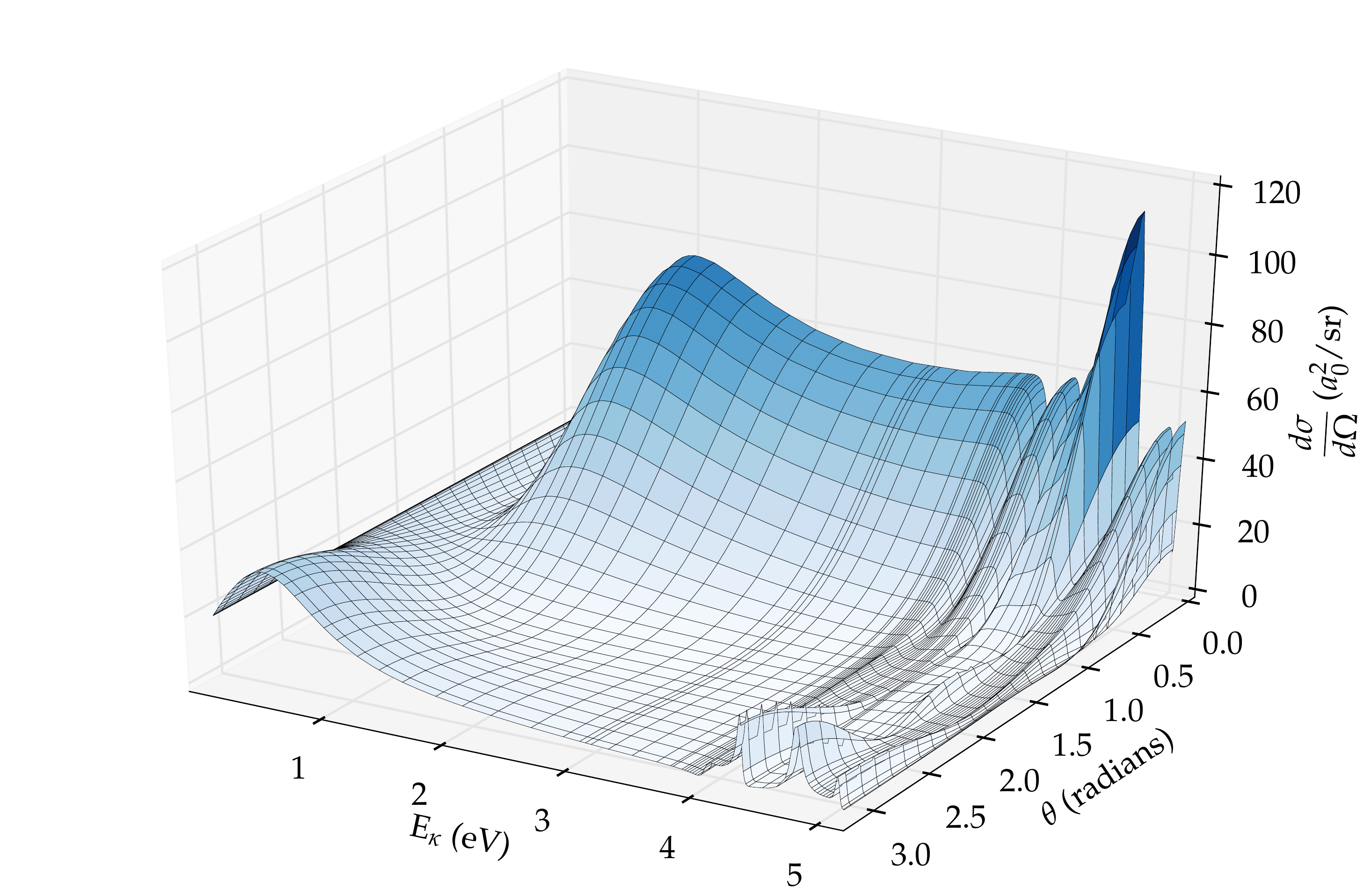}
	\caption{Singlet differential cross section}
	\label{fig:diff-cross-sections-singlet}
\end{figure}

\begin{figure}
	\centering
	\includegraphics[width=5.25in]{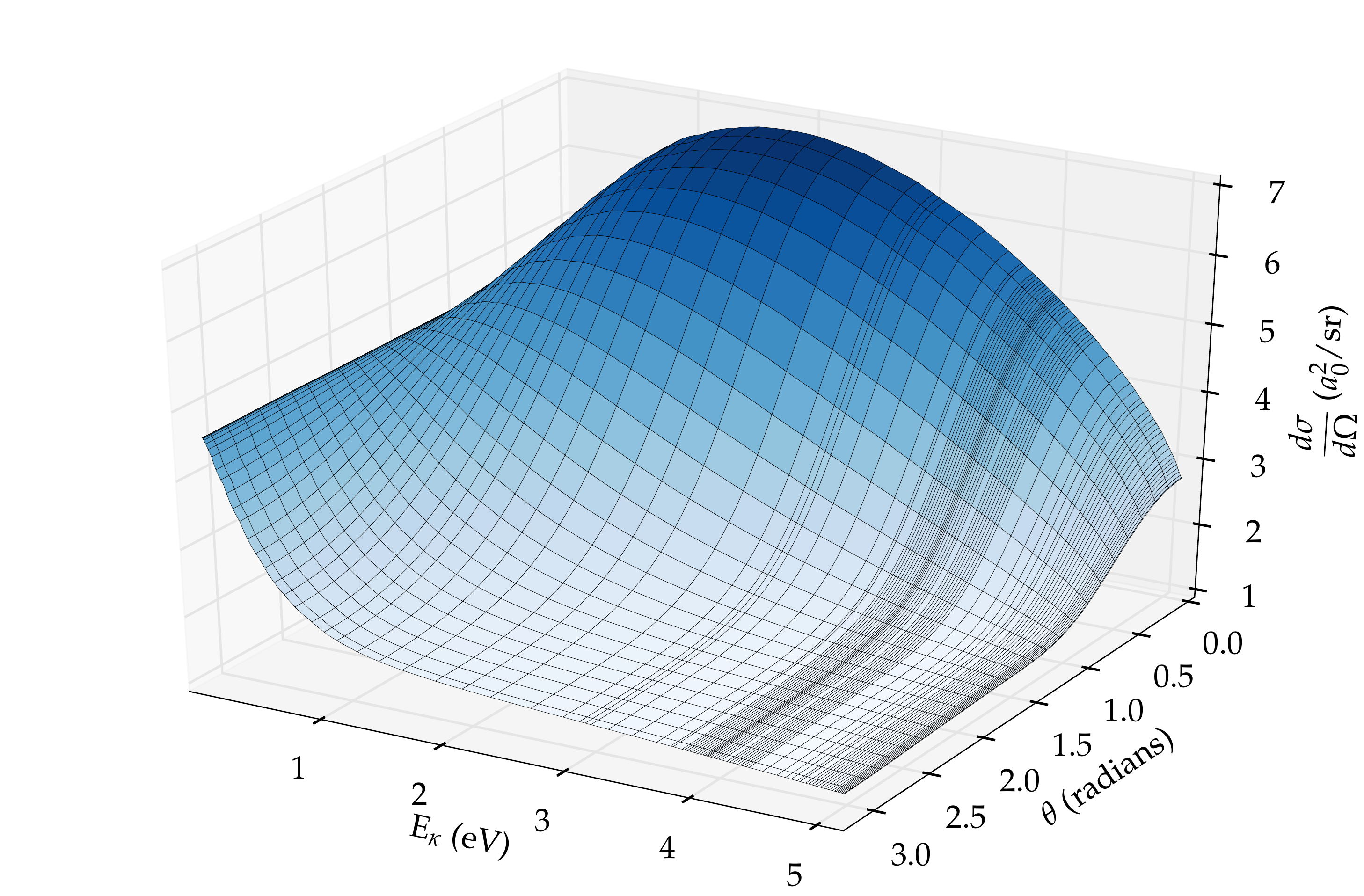}
	\caption{Triplet differential cross section}
	\label{fig:diff-cross-sections-triplet}
\end{figure}

Graphs of the differential cross sections for the singlet and triplet are found
in \cref{fig:diff-cross-sections-singlet,fig:diff-cross-sections-triplet},
respectively. Note that the $\theta$ axis is plotting backwards so that the
features are visible instead of being obscured by the higher value of the differential
cross section in the front of the graph.
The triplet $\frac{d\sigma_{el}^-}{d\Omega}$ is not particularly
remarkable, being smooth and having a maximum at intermediate energies for the
forward direction $(\theta = 0)$. The triplet differential cross section in the
forward direction is less than 7 $a_0^2/\textrm{sr}$ for all energies considered.
It is also of note that in the backward scattering direction $(\theta = \uppi)$,
the triplet differential cross section quickly becomes very small.

In contrast, the singlet differential cross section, $\frac{d\sigma_{el}^+}{d\Omega}$,
shown in \cref{fig:diff-cross-sections-singlet} has a much more complicated
structure. The resonances are clearly visible in the plot, and the highest peak
extending to 110.5 $a_0^2/\textrm{sr}$ is due to the D-wave resonance, which is
also the dominant resonance in the integrated cross section shown in
\cref{fig:singlet-cross-sections,fig:combined-cross-sections}. The features in
the integrated cross section graphs can easily be matched up to features in the
differential cross section, including the rest of the resonances. The maximum
and dip described in \cref{sec:totalcross} for the integrated cross section can
also be seen in \cref{fig:diff-cross-sections-singlet}, especially in the
forward direction.

\begin{figure}
    \centering
    \subfloat{{\includegraphics[width=6in]{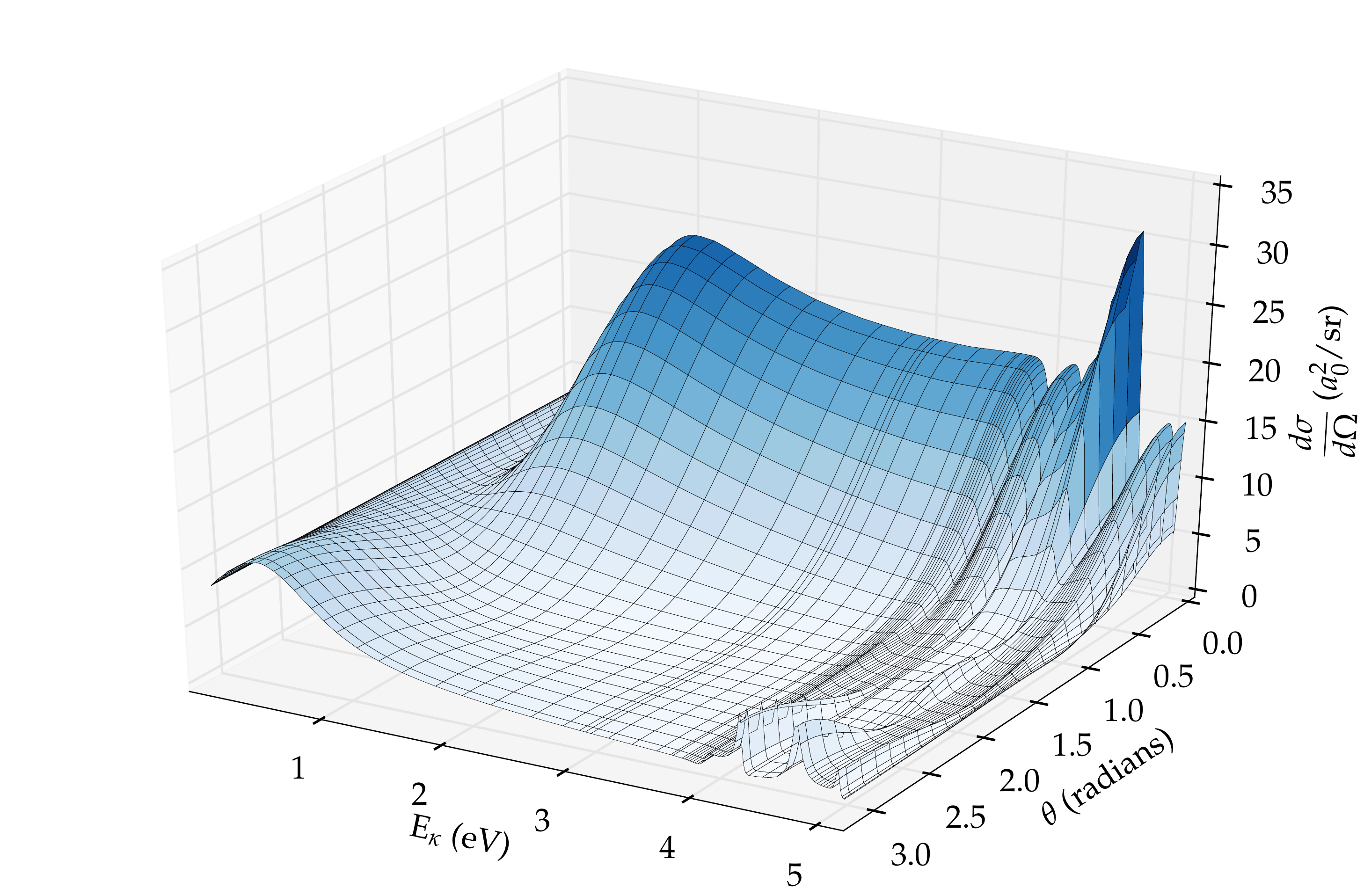} }} \\%
    \subfloat{{\includegraphics[width=6in]{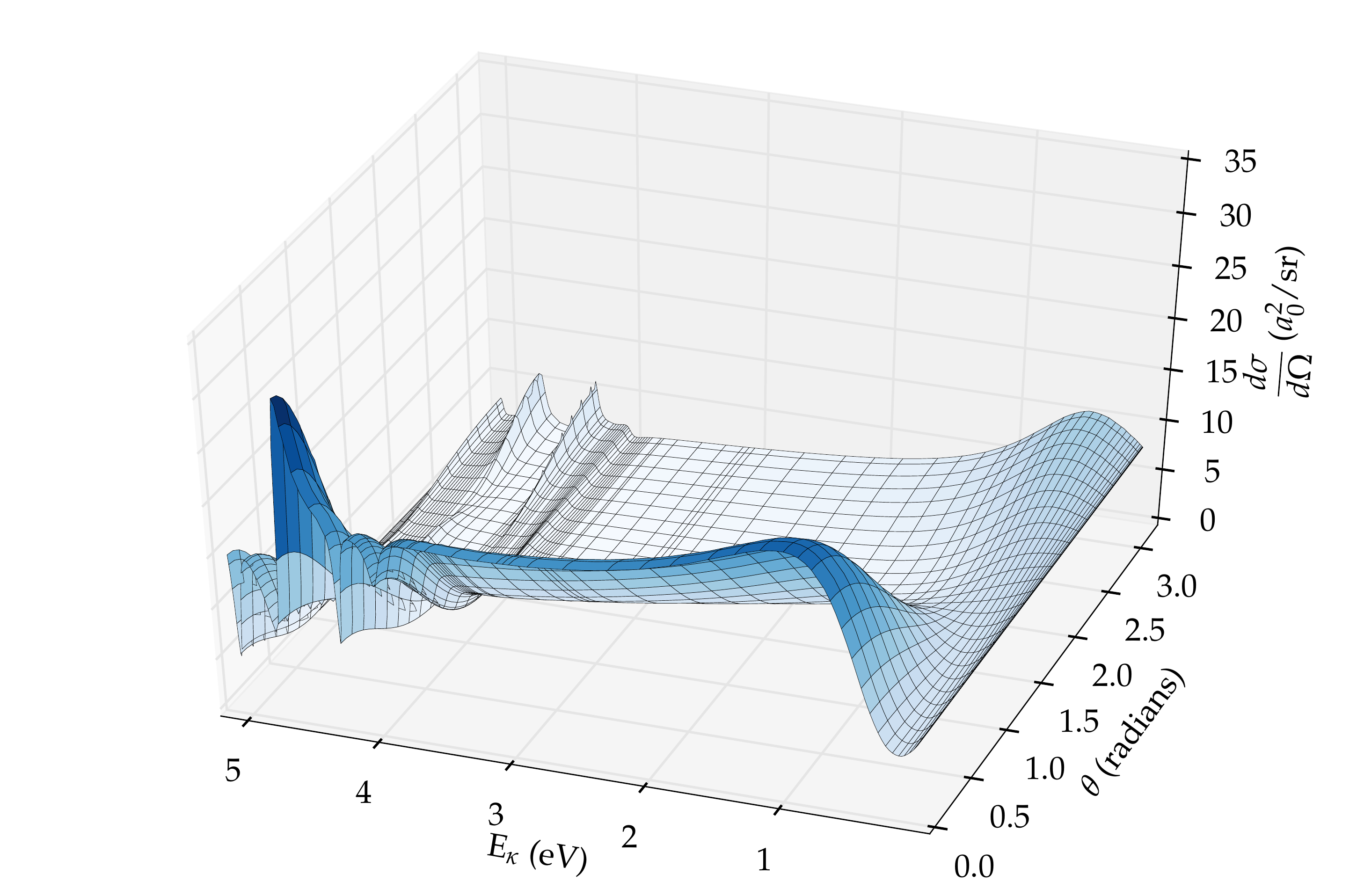} }}%
    \caption[The combined spin-weighted elastic differential cross section]{The combined spin-weighted elastic differential cross section for Ps-H scattering at two different viewing angles}%
    \label{fig:combined-diff-cross-sections}%
\end{figure}

Similar to the integrated cross section, we combine these using the same type
of spin-weighting given in \cref{eq:SpinWeightCS}. Due to the nearly featureless
nature of the triplet in \cref{fig:diff-cross-sections-triplet},
the combined differential cross section in 
\cref{fig:combined-diff-cross-sections} looks very similar to 
\cref{fig:diff-cross-sections-singlet}, but the 1/4 weighting of the singlet
brings the vertical scale down. The forward direction is enhanced slightly by
the maximum in the triplet in \cref{fig:diff-cross-sections-triplet}.

\begin{figure}
	\centering
	\includegraphics[width=5.25in]{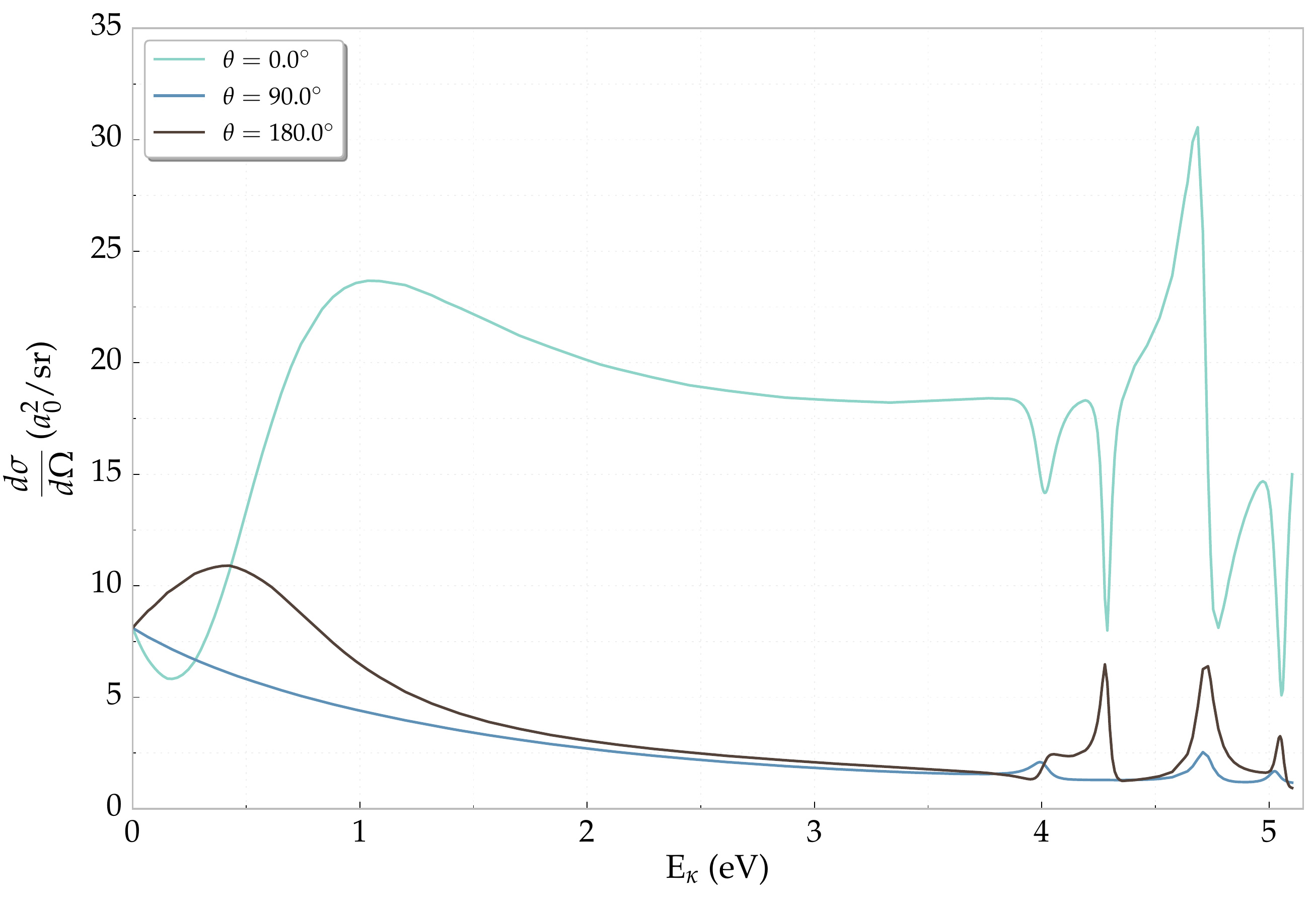}
	\caption{Differential cross section for selected $\theta$}
	\label{fig:diff-cross-section-2D-theta}
\end{figure}

With the data in \cref{fig:combined-diff-cross-sections}, it is illustrative 
to plot 2-dimensional versions to see trends more clearly. In
\cref{fig:diff-cross-section-2D-theta}, we restrict $\theta$ and vary
$E_{\bm \kappa}$ to see the the energy-dependence for several $\theta$ values.
From this figure, it is clear that scattering in the forward direction is
dominant past approximately $\SI{0.46}{eV}$. The maximum for forward scattering
is due to the $^1$D resonance. Particularly interesting is the contribution
from backward scattering for low $E_{\bm \kappa}$ and the dip in the forward
scattering direction, which corresponds to the dip in
\cref{fig:combined-cross-sections}. All angles give essentially the same value
at very low energy, as we would expect. For nearly zero energy at
$\kappa = 0.0001$ (\SI{6.8e-8}{eV}), $\frac{d\sigma_{el}^\pm}{d\Omega}$ at
$0\degree$, $90\degree$, and $180\degree$ is 8.112, 8.113, and 8.108
$a_0^2/\textrm{sr}$, showing that the differential cross section is
essentially isotropic. \label{pg:diffcross0}
At exactly \SI{0}{eV}, it will be exactly isotropic. This is due to the S-wave
having the only non-zero cross section at zero energy.

\begin{figure}
	\centering
	\includegraphics[width=5.25in]{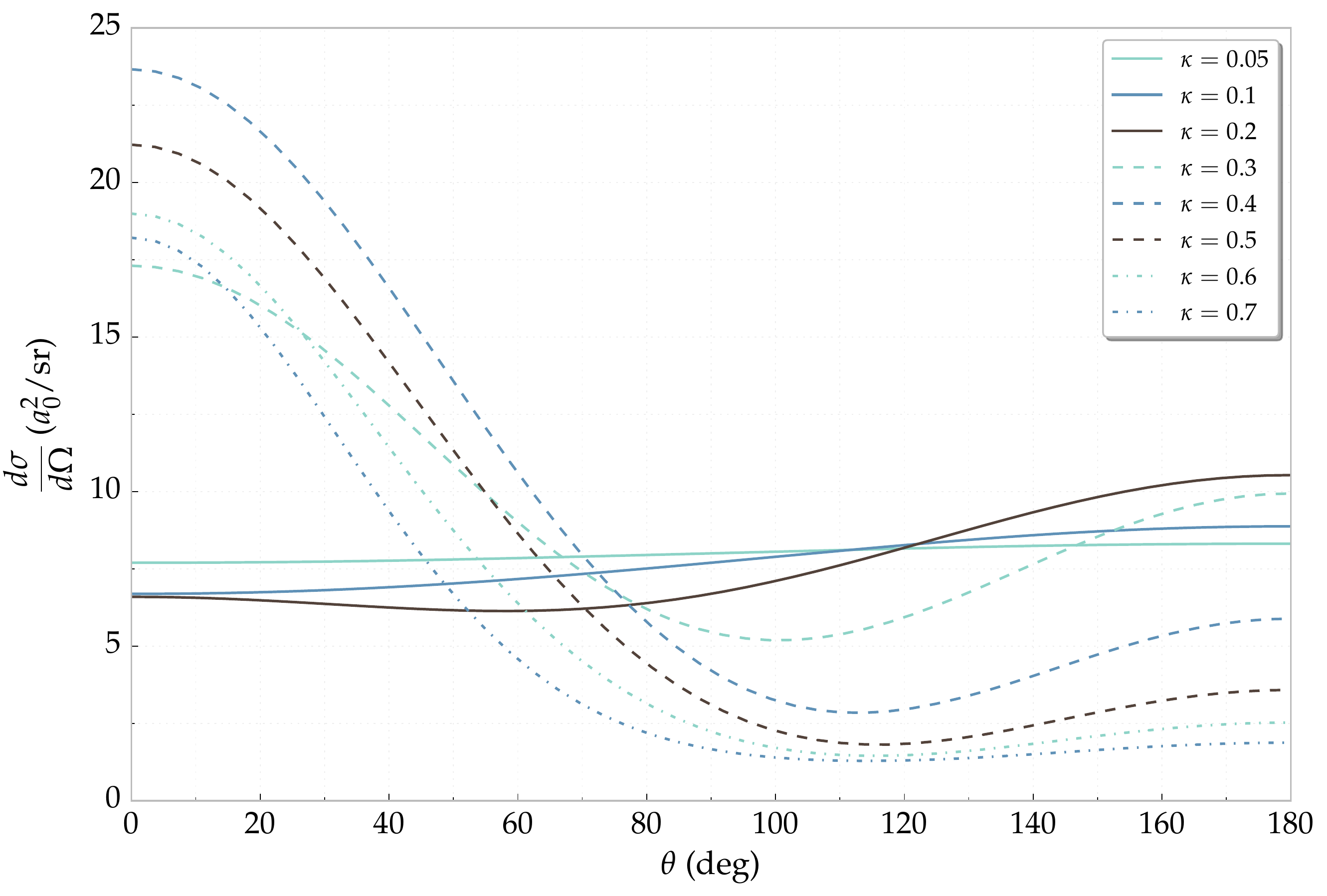}
	\caption{Differential cross section for selected $\kappa$}
	\label{fig:diff-cross-section-2D-kappa}
\end{figure}

In \cref{fig:diff-cross-section-2D-kappa}, we instead fix values of
$E_{\bm \kappa}$ and plot with respect to $\theta$. The legend gives the $\kappa$ 
value instead, so it is clear what specific values we are plotting. At low $
\kappa$ (0.05 in the plot), the differential cross section is nearly 
isotropic, with a slight bias toward backward scattering. As $\kappa$ is 
increased to 0.2, backward scattering is more prominent. Between $\kappa = 0.2$
and 0.3, there is an abrupt change in the differential cross section, where 
it becomes much more forward peaked, with a decreasing contribution to the 
backward direction, and a minimum at approximately $100\degree$. As 
$\kappa$ is increased further, the differential cross section becomes very 
strongly forward peaked, with even further decreases at larger angles and a 
nearly constant value from about $100\degree$.  We see from
\cref{fig:combined-diff-cross-sections} that the majority of the scattering
takes place between 0 and 1 radians.

We find that the elastic differential cross section converges slower with 
respect to $\ell$ than the elastic integrated cross section in
\cref{sec:totalcross}. The convergence of the differential cross section
is more difficult to quantitatively evaluate than that of the
integrated cross section, since it has mixing between all of the included
partial waves.
We calculate $\frac{d\sigma_{el}}{d\Omega}$ for two subsequent
values of $\ell_{max}$, then find the percent difference between these.

\begin{figure}
	\centering
	\includegraphics[width=5.5in]{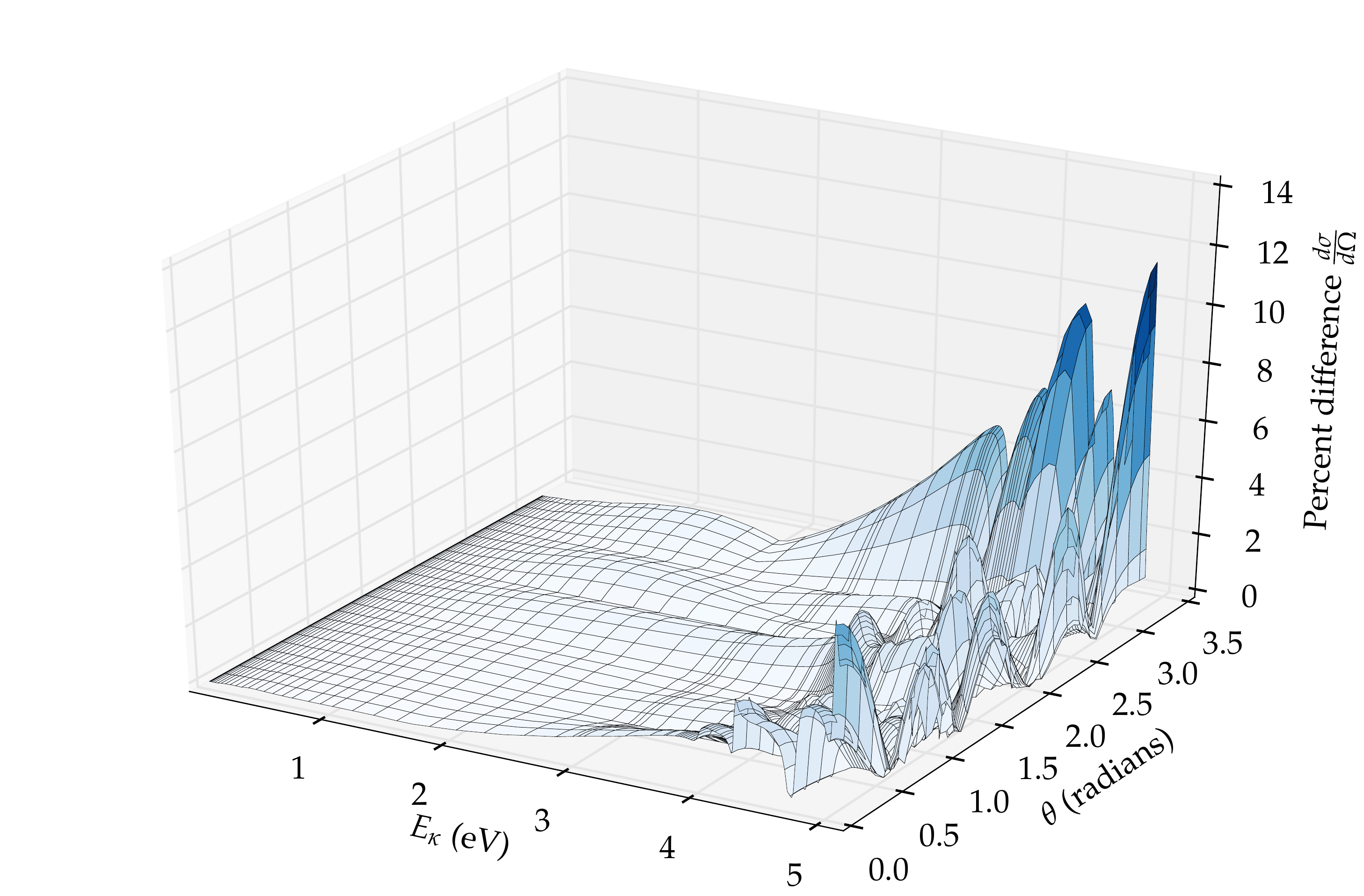}
	\caption[Percent difference of differential cross section at all angles]{Percent difference of $\frac{d\sigma_{el}}{d\Omega}$ for upper limit of summations in \cref{eq:DiffCross} as $\ell_{max} = 3$ versus $\ell_{max} = 4$ for all angles and energies}
	\label{fig:percent-diff-cross-sections-full-g}
\end{figure}

\begin{figure}
	\centering
	\includegraphics[width=5.5in]{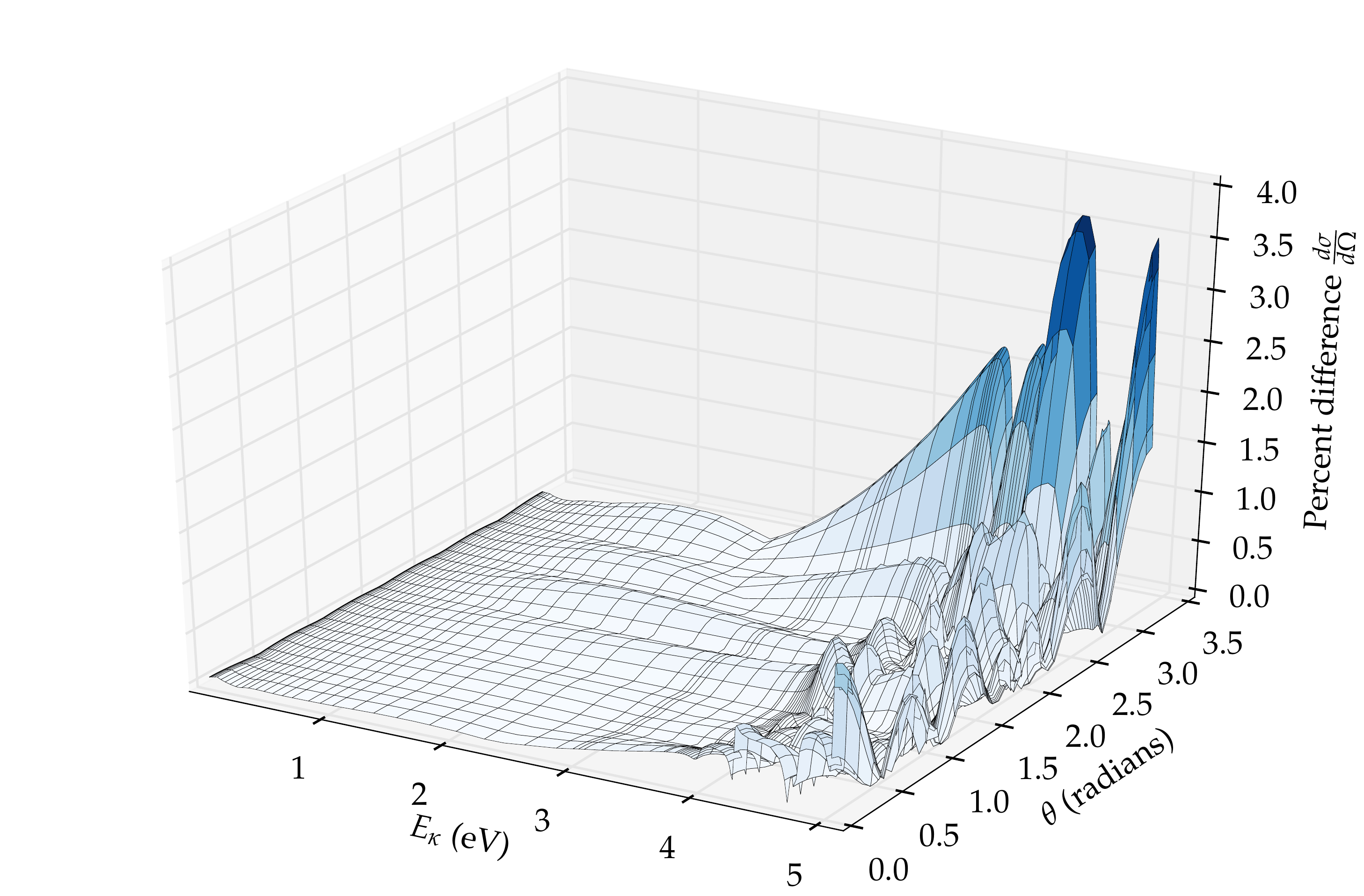}
	\caption[Percent difference of differential cross section at all angles]{Percent difference of $\frac{d\sigma_{el}}{d\Omega}$ for upper limit of summations in \cref{eq:DiffCross} as $\ell_{max} = 4$ versus $\ell_{max} = 5$ for all angles and energies}
	\label{fig:percent-diff-cross-sections-full-h}
\end{figure}

\Cref{fig:percent-diff-cross-sections-full-g,fig:percent-diff-cross-sections-full-h}
shows this percentage difference for all angles and energies. These two figures
look similar but have different vertical scales, as we would hope for if
there was convergence. There are two important trends here. One is that the 
differential cross section is well converged at low energies, and the other 
is that it is less converged in the backward scattering direction than in the 
forward direction. In \cref{fig:percent-diff-cross-sections-full-h}, which
includes all partial waves through the H-wave, the percentage differences are
all below 4\%, and most of the $E_{\bm \kappa}$ and $\theta$ range is 
much less than this. This indicates that the differential cross section is
relatively well converged.

\begin{figure}
	\centering
	\includegraphics[width=\textwidth]{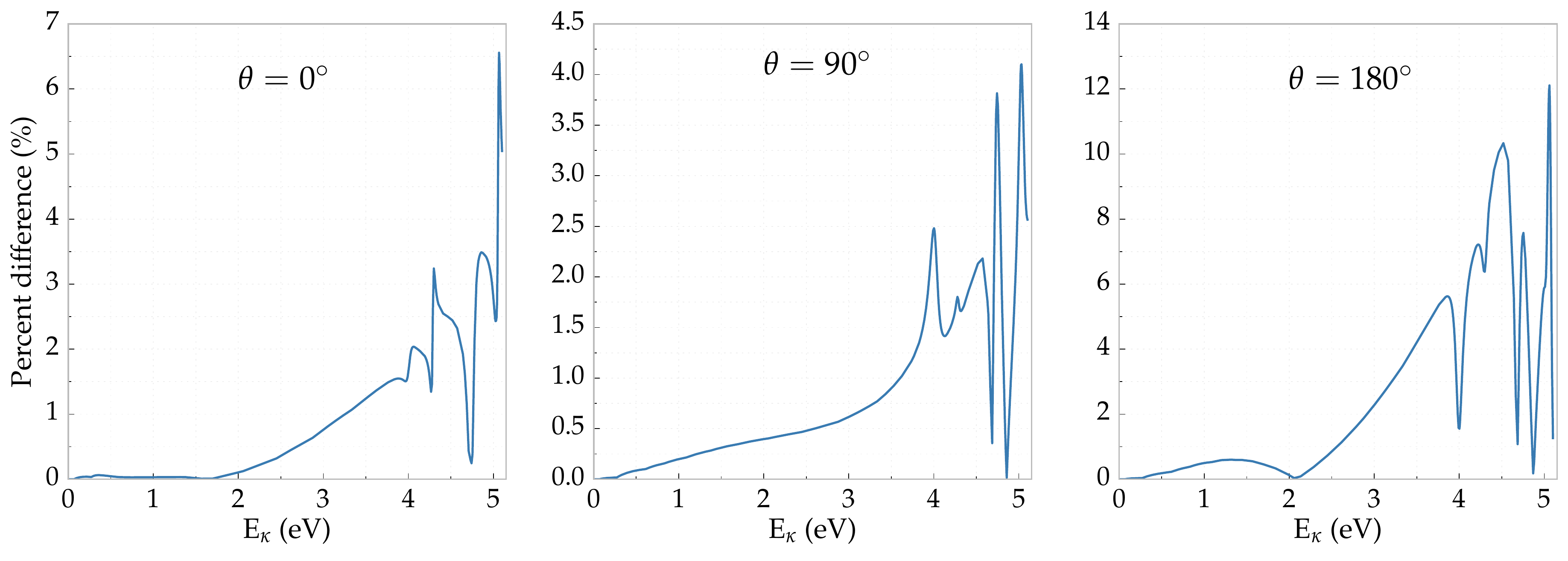}
	\caption[Percent difference of differential cross section at selected angles]{Percent difference of $\frac{d\sigma_{el}}{d\Omega}$ for upper limit of summations in \cref{eq:DiffCross} as $\ell_{max} = 3$ versus $\ell_{max} = 4$ for selected angles}
	\label{fig:percent-diff-cross-sections-g}
\end{figure}

\begin{figure}
	\centering
	\includegraphics[width=\textwidth]{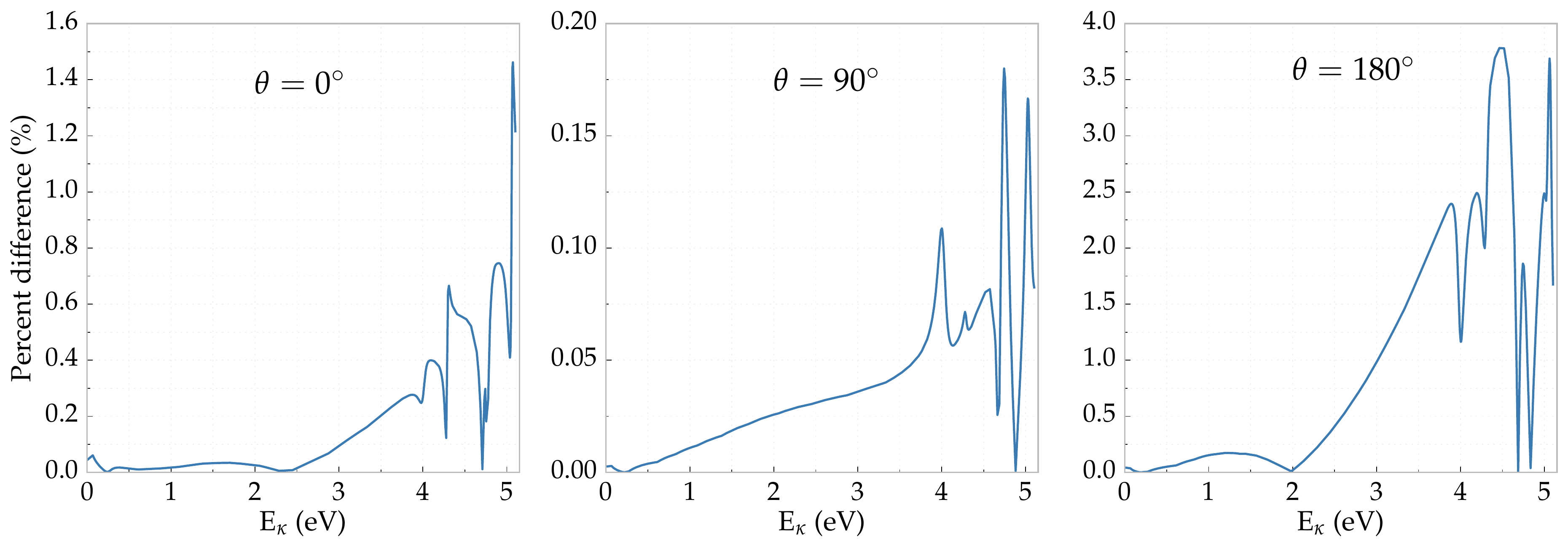}
	\caption[Percent difference of differential cross section at selected angles]{Percent difference of $\frac{d\sigma_{el}}{d\Omega}$ for upper limit of summations in \cref{eq:DiffCross} as $\ell_{max} = 4$ versus $\ell_{max} = 5$ for selected angles}
	\label{fig:percent-diff-cross-sections-h}
\end{figure}

\Cref{fig:percent-diff-cross-sections-g,fig:percent-diff-cross-sections-h}
show the same data but for selected angles. We see here that
$\theta = 90\degree$ is the best converged angle out of the three, and
the backward scattering direction of $\theta = 180\degree$ is the worst
converged. Again, as seen in \cref{fig:percent-diff-cross-sections-h}, the
differential cross section is relatively well converged if we include the
H-wave.

We would expect that as $\ell$ increases, $\theta = 180\degree$ 
will be the most sensitive to adding terms to the differential cross section. 
The partial wave expansion \cite[p.583]{Bransden2003} has a $\LegendreP{\ell,\cos\theta}$ for 
each term. If we set $\theta = 0\degree$, each of the Legendre polynomials
equals 1, meaning that each term added is positive. If $\theta = 180\degree$,
the Legendre polynomial alternates between 1 and -1, i.e. $(-1)^\ell$. This is
then an alternating series, which we would expect to converge more slowly.
The minimum for $\theta = 90\degree$ can be explained by every $\ell$ odd term
equaling 0 from the Legendre polynomial.

Similar to \cref{tab:PercentToCross}, \cref{tab:PercentDiffCrossFull} gives 
the average and maximum percent differences for adding partial waves to the
differential cross section (both the singlet and triplet). The average 
percent difference for adding the G-wave is less than 1\% but has a fairly
large maximum percent difference. Adding the H-wave is a much less
significant contribution, and the differential cross section looks to be
relatively well converged by the H-wave.

\begin{table}
\centering
\begin{tabular}{c...c}
\toprule
$\ell$  &  \multicolumn{1}{c}{Avg. \% Diff.}  &  \multicolumn{1}{c}{Max. \% Diff.} \\
\midrule
1& 49.11\% & 140.8\%  \\
2& 18.54\% & 122.6\%  \\
3& 4.50\%  & 58.1\%   \\
4& 0.86\%  & 12.1\%   \\
5& 0.26\%  & 3.8\%    \\
\bottomrule
\end{tabular}
\caption[Convergence of the full differential cross section]{Percent difference of the elastic differential cross section for each partial wave $\ell$ with respect to $\ell - 1$ for both the maximum and average for the entire $E_{\bm \kappa}$ and $\theta$ range}
\label{tab:PercentDiffCrossFull}
\end{table}

\Cref{tab:PercentDiffCrossSingTrip} separates the singlet and triplet
contributions to the differential cross section. For this table, when the
singlet contributions are analyzed, the triplet summations have
$\ell_{max} = 5$. Likewise, when the singlet contributions are analyzed,
$\ell_{max} = 5$ for the triplet. We see that with respect to $\ell$, the
triplet differential cross section converges much quicker for both the average
and maximum. Columns 5 and 6 give the values of $E_{\bm \kappa}$ and $\theta$
where the maximum percent difference is located for the singlet, and columns
8 and 9 give the maximum percent difference location for the triplet.
Unsurprisingly, for $\ell \geq 1$, the angles that are the most
sensitive are 0 and $\pi$ radians. The most sensitive energies are in the
resonance region, and most of these are near the inelastic threshold of
$\SI{5.102}{eV}$.

\begin{table}
\centering
\centerline{
\begin{tabular}{c....c..c}
\toprule
 & \multicolumn{1}{c}{Avg.}  & \multicolumn{1}{c}{Avg.}  & \multicolumn{1}{c}{Max.}  & \multicolumn{1}{c}{Max.} & Max. & \multicolumn{1}{c}{Max.}  & \multicolumn{1}{c}{Max.} & Max. \\
$\ell$ & \multicolumn{1}{c}{$\%^+$ Diff.}  & \multicolumn{1}{c}{$\%^-$ Diff.}  & \multicolumn{1}{c}{$\%^+$ Diff.}  & \multicolumn{1}{c}{$E_{\bm \kappa}^+$} (eV) & $\theta^+$ (rad) & \multicolumn{1}{c}{$\%^-$ Diff.}  & \multicolumn{1}{c}{$E_{\bm \kappa}^-$} (eV) & $\theta^-$ (rad) \\
\midrule
1 & 40.34\% & 9.84\% & 162.28\% & 4.686 & 0       & 50.11\% & 4.289 & 0.705   \\
2 & 17.55\% & 3.31\% & 121.89\% & 4.686 & 0       & 24.72\% & 4.354 & $\uppi$ \\
3 & 4.47\%  & 0.39\% & 51.89\%  & 5.072 & $\uppi$ & 3.31\%  & 5.102 & $\uppi$ \\
4 & 0.80\%  & 0.20\% & 9.22\%   & 5.067 & 0       & 1.48\%  & 5.055 & 0       \\
5 & 0.21\%  & 0.15\% & 3.15\%   & 5.061 & $\uppi$ & 1.64\%  & 4.354 & $\uppi$ \\
\bottomrule
\end{tabular}
}
\caption[Convergence of the singlet and triplet differential cross sections]{Percent difference of the elastic differential cross section for each partial wave $\ell$ with respect to $\ell - 1$ for both the maximum and average for the entire $E_{\bm \kappa}$ and $\theta$ range. The values of $E_{\bm \kappa}^\pm$ and $\theta^\pm$ given are where $\frac{d\sigma_{el}^\pm}{d\Omega}$ is at its maximum value given in columns 4 and 7.}
\label{tab:PercentDiffCrossSingTrip}
\end{table}

\section{Momentum Transfer Cross Section and Comparisons}
\label{sec:OtherCross}

Measurements of the momentum transfer cross section, $\sigma_m$,
have been made for Ps scattering with atomic and molecular targets
\cite{Engbrecht2008,Nagashima1998,Saito2003,Skalsey1998}, and calculations of
Ps scattering by inert gases have been performed \cite{Blackwood2002c}.
Calculations of momentum transfer cross sections for other systems
have been made in Refs.~\cite{Wang2014, McEachran2014}.
The momentum transfer cross section is
similar to \cref{eq:TotalDiffCross} but with a weighting factor of
$(1 - \cos\theta)$ \cite{Walters2004}:
\begin{equation}
\label{eq:MomentumCrossInt}
\sigma_m = \int (1 - \cos\theta) \frac{d\sigma_{el}}{d\Omega} \, d\Omega.
\end{equation}
The momentum transfer cross sections can also be written in terms of the
phase shifts as \citep[p.589]{Bransden2003}
\begin{equation}
\label{eq:MomentumCross}
\sigma_{m}^\pm = \frac{4}{\kappa^2} \sum_{\ell=0}^\infty (\ell+1) \sin^2 (\delta_\ell^\pm - \delta_{\ell+1}^\pm) .
\end{equation}
This is the expression used in this work.

\begin{figure}
	\centering
	\includegraphics[width=5.25in]{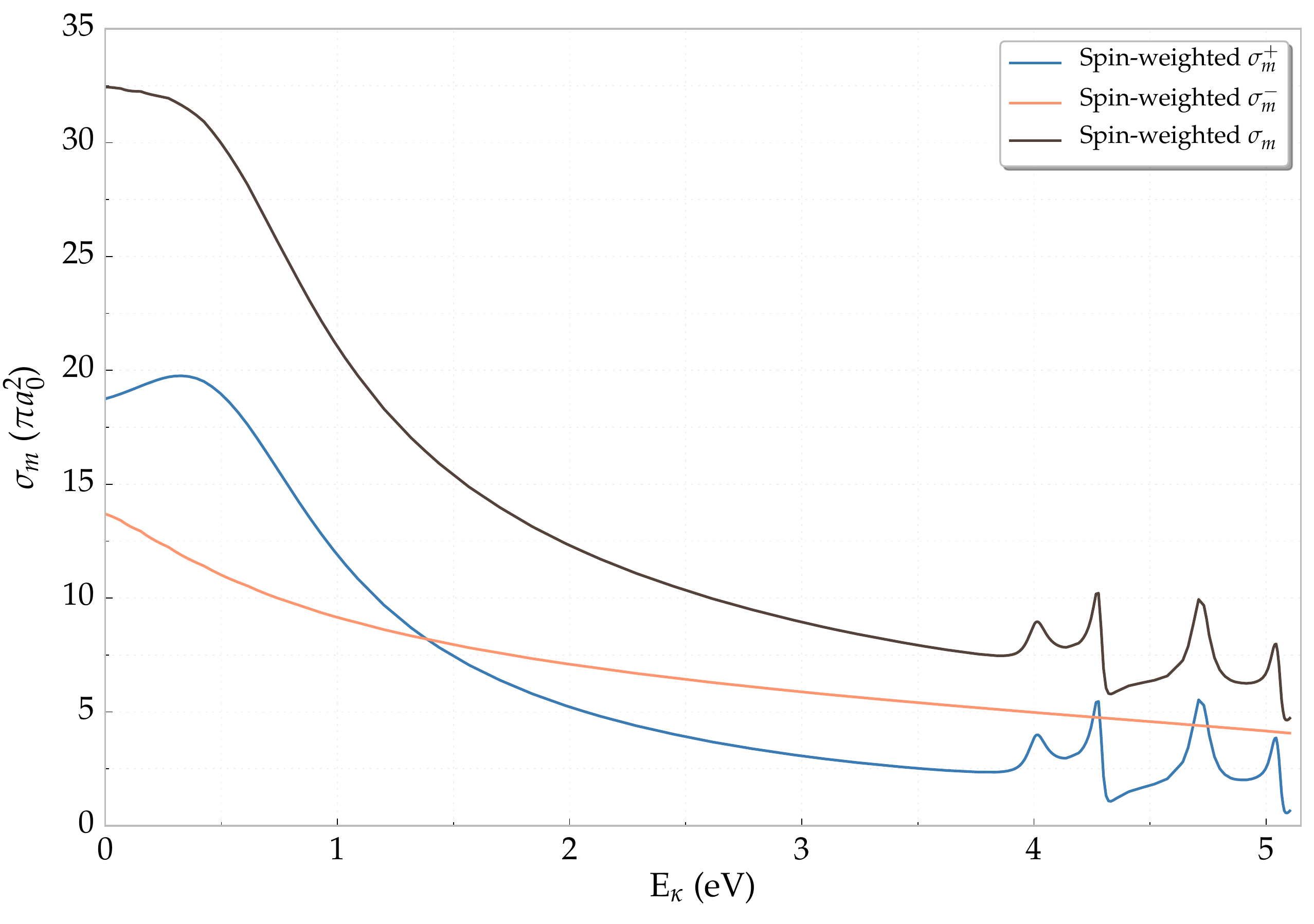}
	\caption[Momentum transfer cross sections]{$S$-matrix complex Kohn momentum transfer cross sections}
	\label{fig:momentum-cross-sections}
\end{figure}

\Cref{fig:momentum-cross-sections} shows the momentum transfer cross section
using the complex Kohn phase shifts. The spin-weighting is from \cref{eq:SpinWeightCS}.
The triplet is a nearly featureless curve that gives a nearly constant
contribution to the spin-weighted momentum transfer cross section.

\begin{figure}
	\centering
	\includegraphics[width=5.25in]{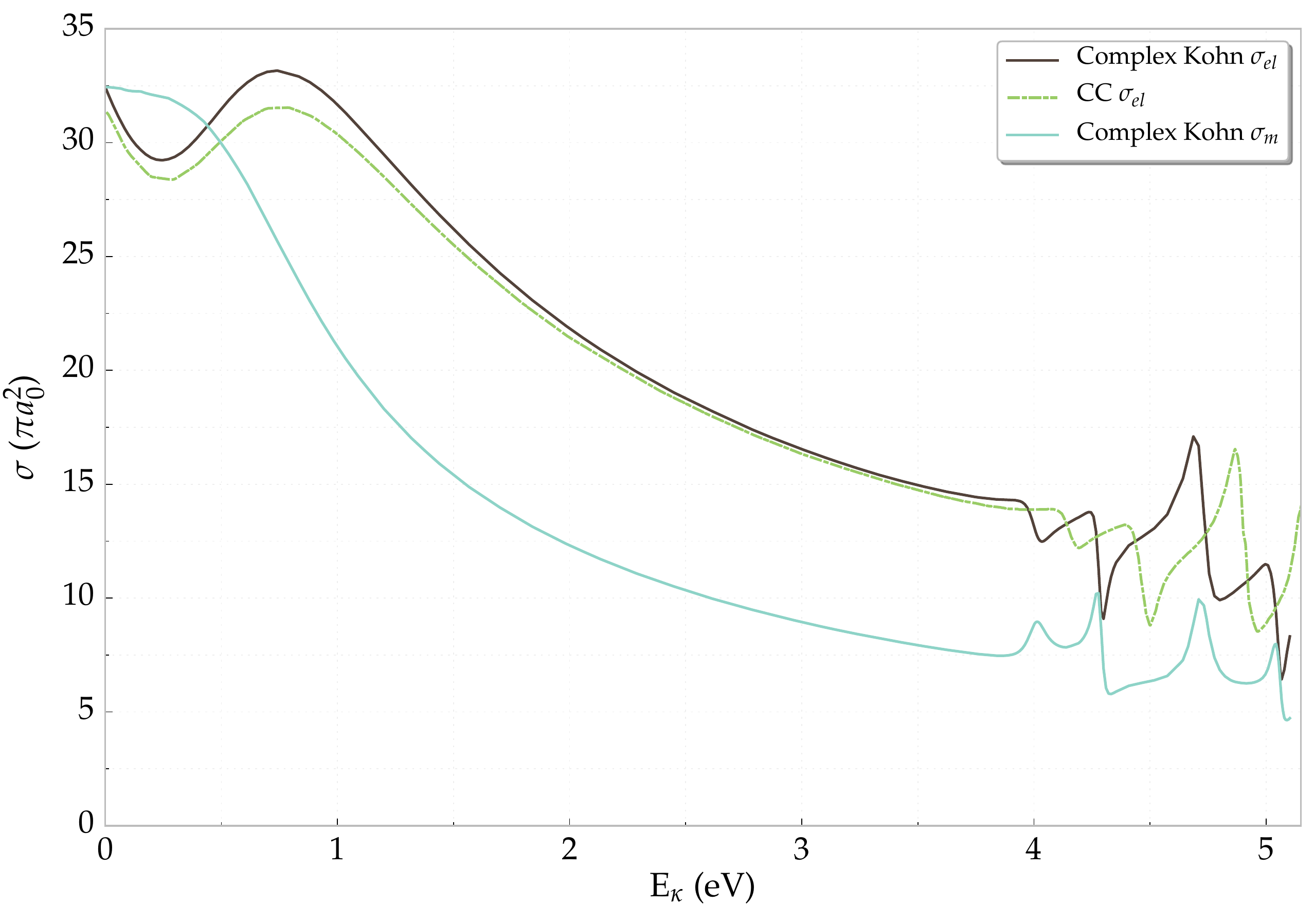}
	\caption[Comparison of cross sections.]{Comparison of cross sections. CC data is from Ref.~\cite{Walters2004}.}
	\label{fig:cross-section-comparisons}
\end{figure}

\Cref{fig:cross-section-comparisons} shows the elastic integrated and momentum
transfer cross sections.
$\sigma_m \approx \sigma_{el}$ at very low energy. 
At zero energy, $\sigma_m = \sigma_{el}$ should hold, and for Ps-H, we find that
$E_{\bm \kappa} < 10^{-6}$ eV, $\sigma_m = \sigma_{el} = 32.45$ $\pi a_0^2$. 
As Blackwood et al.\ \cite{Blackwood2002c} note for Ps-Ne scattering,
this is due to the differential cross section being essentially isotropic
at low energy (and exactly isotropic at zero energy), as seen in
\cref{fig:diff-cross-section-2D-kappa,fig:combined-diff-cross-sections}.
If $\sigma_m < \sigma_{el}$, the scattering is mainly forward peaked, and if
$\sigma_m > \sigma_{el}$, the scattering is mainly backward peaked \cite{Thumm1993}.
Past zero energy, where $\sigma_m = \sigma_{el}$, we see that $\sigma_m > \sigma_{el}$
until approximately $\SI{0.46}{eV}$ or $\kappa = 0.26$. This is backward peaked
(concentrated in the $\theta = \pi$ direction), which corresponds to the
findings from the differential cross section in \cref{sec:diffcross}. Beyond
$\SI{0.46}{eV}$, $\sigma_m < \sigma_{el}$ for the rest of the energy range,
indicating that the scattering is primarily forward peaked. The $\sigma_m$ curve,
as seen in \cref{fig:cross-section-comparisons}, gets close to the $\sigma_{el}$
curve for the dips in some of the resonances at approximately $\SI{4.3}{eV}$
and $\SI{5.1}{eV}$, showing that the scattering is concentrated nearly equally
in the forward and backward directions at these energy values.
This effect can slightly be seen in \cref{fig:combined-diff-cross-sections},
but it is most clearly shown in \cref{fig:diff-cross-section-2D-theta}.

\section{Summary}
\label{sec:SummaryCross}

Computing the phase shifts for the first three partial waves allowed us to
calculate the elastic integrated cross section for for Ps-H scattering. This
cross section compares well to the CC \cite{Walters2004} but with some 
significant differences at both low and high $\kappa$ (below the Ps(n=2) 
threshold). The $\ell = 3 - 5$ partial waves are used in this calculation but
affect the integrated cross section little. An interesting feature is the
dip and maximum under \SI{1}{eV}, which is caused by the interference of the
$^1$S and $^1$P partial wave cross sections.

The differential cross section converges more slowly than the integrated cross
section with respect to $\ell$. The differential cross section
exhibits a complicated behavior, with the singlet resonances contributing
greatly near the Ps(n=2) threshold. At very low energies, the differential
cross section is slightly backward peaked and becomes strongly forward peaked
as energy increases. The momentum transfer cross section is identical to the
elastic integrated cross section at nearly zero energy but becomes larger as
energy increases, up to approximately \SI{0.46}{eV}, which corresponds with the
differential cross section being backward peaked. Past this energy, the 
momentum transfer cross section is less than the elastic integrated cross
section, indicating that scattering is more forward peaked.



\chapter{Effective Range Theories}
\label{chp:ERT}

\iftoggle{UNT}{The}{\lettrine{\textcolor{startcolor}{T}}{he}}
scattering length and effective range give information about very low energy
scattering ($\kappa \to 0$).
The Kohn-type variational methods can give an exact upper bound on the 
scattering length \cite{Spruch1959,Joachain1979}, though we do not do a Kohn-type 
variation on the scattering length in this work. The scattering length can 
also give information about whether there is a bound state in the system.

\section{S-Wave Scattering Length and Effective Range}
\label{sec:ScatteringLength}

There are multiple methods for calculating the scattering length and effective
range. I describe the methods and results that we used in this section.

\subsection{Scattering Length Definition}
The scattering length \citep[p.589]{Bransden2003} is defined as
\beq
\label{eq:ScatLen}
a_\ell^\pm = -\lim_{\kappa \to 0} \frac{\tan{\delta_\ell^\pm}}{\kappa^{2\ell+1}}.
\eeq
We approximate this with very small $\kappa$ as
\beq
\label{eq:ScatLenApprox}
a_\ell^\pm \approx -\frac{\tan{\delta_\ell^\pm}}{\kappa^{2\ell+1}}.
\eeq
To avoid confusion with the Bohr radius, $a_0$, we define $a = a_{\ell=0}$.
If this definition is used for decreasing (but positive) values of $\kappa$, 
there is a clear convergence. \Cref{tab:ScatLenDef} has the scattering 
length calculated for $\kappa$ values of $0.001$ and $0.0001$. To the quoted 
accuracy, the values of the scattering length agree exactly for the two 
different values of $\kappa$ for $\omega = 7$. This is the method used by Van Reeth and 
Humberston \cite{VanReeth2003} for finding the scattering length (but for
$\omega = 6$ only).

\begin{table}
\centering
\begin{tabular}{c c c c c}
\toprule
$\omega$ & $a^+ (\kappa = 0.001)$ & $a^+ (\kappa = 0.0001)$ & $a^- (\kappa = 0.001)$ & $a^- (\kappa = 0.0001)$ \\
\midrule
6 & 4.3364 & 4.3364 & 2.1415 & 2.1415 \\
7 & 4.3306 & 4.3306 & 2.1363 & 2.1363 \\
\bottomrule
\end{tabular}
\caption{Scattering length from approximation to definition}
\label{tab:ScatLenDef}
\end{table}

\begin{figure}
	\centering
	\includegraphics[width=\textwidth]{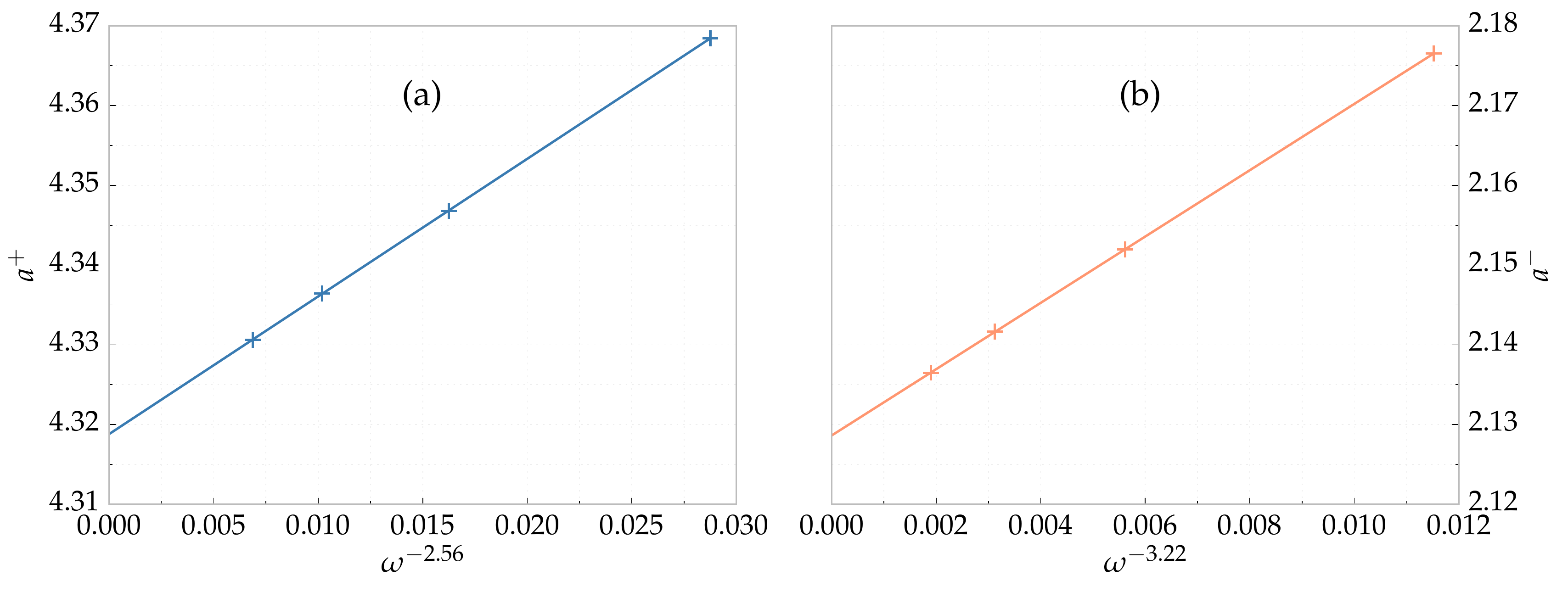}
	\caption[Convergence of S-wave scattering lengths]{Convergence of (a) $^1$S
and (b) $^3$S scattering lengths using \cref{eq:ScatLenExtrap}.}
	\label{fig:scatlen-convergence}
\end{figure}

Van Reeth and Humberston \cite{VanReeth2003} also extrapolate their 
scattering length to
$\omega \rightarrow \infty$ using
\beq
\label{eq:ScatLenExtrap}
a^\pm(\omega) = a^\pm(\omega \rightarrow \infty) + \frac{c}{\omega^p},
\eeq
where $c$ and $p$ depend on each extrapolation.
Performing this extrapolation, we obtain the fits shown in
\cref{fig:scatlen-convergence}. The extrapolated $^1$S and $^3$S scattering 
lengths shown in \cref{tab:SWaveScatLenOther} are 4.319 and 2.129, 
respectively.

At zero energy \cite{Buckman1989},
\begin{equation}
\label{eq:ScatLenCross}
\sigma_m^\pm = \sigma_{el}^\pm = 4 (a^\pm)^2.
\end{equation}
So we can compare the scattering lengths to 
the cross sections at zero energy (taken as $E_{\bm \kappa} = 10^{-7}$ eV
here) with (in units of $\pi a_0^2$)
\begin{subequations}
\label{eq:CrossScatLen}
\begin{align}
\sigma_{el}^+ = \sigma_m^+ = 75.03 & \approx 4 (a^+)^2 = 75.02 \\
\sigma_{el}^- = \sigma_m^- = 18.27 & \approx 4 (a^-)^2 = 18.26 \, .
\end{align}
\end{subequations}

Also at zero energy \cite[p.590]{Bransden2003}, 
\begin{equation}
\label{eq:diffcross0}
\frac{d\sigma_{el}}{d\Omega} = a^2
\end{equation}
Using the values for $a^+$ and $a^-$ for $\omega = 7$ from \cref{tab:ScatLenDef}
and using the appropriate spin-weighting from \cref{eq:SpinWeightCS}, this gives
$(a^+)^2 \times 1/4 + (a^-)^2 \times 3/4 = 8.11$, and as seen on
page~\pageref{pg:diffcross0}, this is equal to the differential cross section
at nearly zero energy.

\subsection{Short-Range Interactions}
\label{sec:EffectiveRangeShort}
For short-range interactions, the effective range is given by \cite{Bethe1949,Blatt1949,Drake2006}
\beq
\label{eq:EffectiveRangeShort}
\kappa \cot\delta_0^\pm = -\frac{1}{a^\pm} + \frac{1}{2} r_0^\pm \kappa^2 + \mathcal{O}(\kappa^4).
\eeq
This equation is referred to here as the short-range expansion. Phase shifts for 
low values of $\kappa$ are fitted to this equation to determine the effective 
range and scattering length, and the results are shown in
\cref{tab:ScatLenStandard}. This is a subset of \cref{tab:vdwFlannery}.
We used 10 equidistant values of $\kappa$ for all $\kappa$ ranges other than
0.1 - 0.5, where we used 5 equidistant points. This fitting is not carried 
out to $\kappa$ higher than 0.5, due to the resonance structure described in
\cref{sec:SWaveResonances}.

\begin{table}
\centering
\begin{tabular}{l c c c c}
\toprule
$\kappa$ Range & $a^+$ & $r_0^+$ & $a^-$ & $r_0^-$ \\
\midrule
$0.1 - 0.5$     & 4.3080 & 2.2816 & 2.1623 & 1.3729 \\
$0.01 - 0.09$   & 4.3306 & 2.2012 & 2.1367 & 1.9354 \\
$0.001 - 0.009$ & 4.3306 & 2.1972 & 2.1365 & 2.0354 \\
\bottomrule
\end{tabular}
\caption{Scattering length and effective range for short-range expansion}
\label{tab:ScatLenStandard}
\end{table}

\begin{figure}
	\centering
	\includegraphics[width=5.25in]{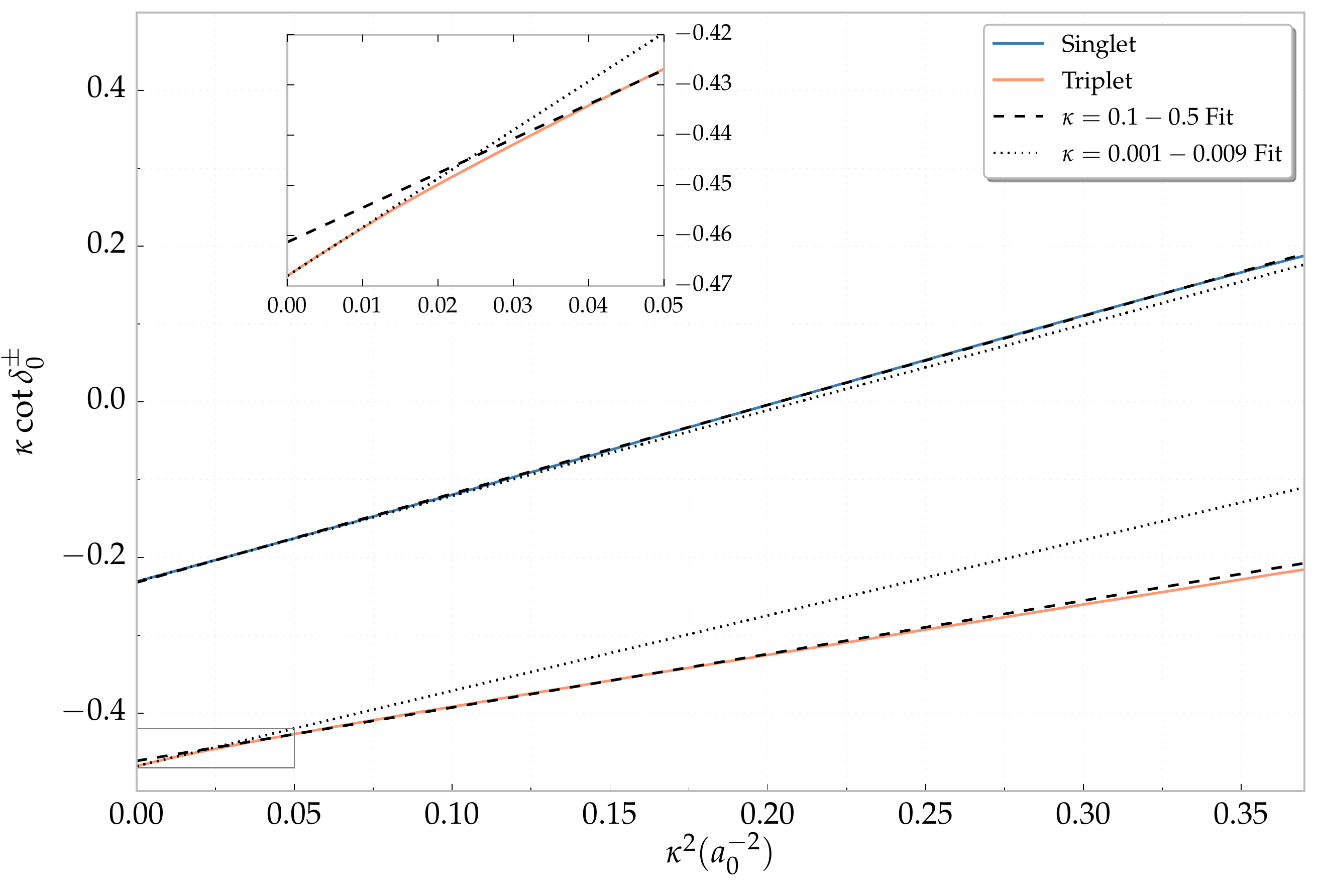}
	\caption[S-wave short-range ERT fitting]{$^1S$ and $^3S$ phase shifts, plotted as
$\kappa \cot \delta_0^\pm$ versus $\kappa^2$. The inset shows a magnified
portion of the same data as denoted by the gray box in the lower left.}
	\label{fig:swave-ERT-short}
\end{figure}

This fitting works well enough for the singlet, but the triplet data does not 
fit exactly. This can be seen in \cref{fig:swave-ERT-short}, which is similar 
to Figure 5 of Ref.~\cite{VanReeth2003}. Fraser \cite{Fraser1961} has also done
a similar graph using the SE method but found a relatively straight line for
both $^1$S and $^3$S.

For very small values of $\kappa$, the scattering length agrees with the 
definition in \cref{eq:ScatLen}. These values also agree well with 
the scattering length found by other recent calculations, shown in
\cref{tab:SWaveScatLenOther}. The effective range for the $\kappa = 0.1 - 0.5$ 
entry for the short-range expansion also agrees relatively well with the 
results from other groups. As can easily be seen, when smaller $\kappa$ 
values are used however, the value of $r_0^\pm$ changes drastically for the 
triplet and a much smaller amount for the singlet. Previous work using the 
Kohn~/~inverse Kohn variational methods found $r_0^- = 1.39$ \cite{VanReeth2003}, which is 
close to the value of $r_0^- = 1.3438$ using $\kappa = 0.1 - 0.5$.

\subsection{van der Waals Interaction}
\label{sec:vanderWaalsERT}

The dominant long-range interaction for Ps-H scattering is given by
the van der Waals potential \cite{Fabrikant2014a,VanReeth2003,Au1986}.
The van der Waals potential is given as
\beq
\label{eq:VanderWaals}
V(R) = -\frac{C_6}{R^6}.
\eeq
For the Ps-H system, $C_6$ has been calculated in atomic units by 
Martin and Fraser to be $34.78473$ \cite{Martin1980}. Mitroy and Bromley
\cite{Mitroy2003a} also calculate this to be $34.785$ in a paper
calculating $C_6$ for multiple Ps-atom problems. Ray \cite{Raye} also
investigates a modified static exchange model that explicitly includes
the van der Waals interaction.

\subsubsection{Flannery Expansion}
\label{sec:Flannery}

When the van der Waals potential is taken into account, the effective range
equation from Flannery \cite[p.669]{Drake2006} is (dropping the $\pm$ for brevity)
\beq
\label{eq:EffectiveRangeLong}
\kappa \cot\delta_0 = -\frac{1}{a} + \frac{1}{2} r_0 \kappa^2 - \frac{\pi}{15 a^2} \left(\frac{2 M C_6}{\hbar^2}\right) \kappa^3 - \frac{4}{15 a} \left(\frac{2 M C_6}{\hbar^2}\right) \kappa^4 \ln \left(\kappa a_0 \right) + \mathcal{O}(\kappa^4).
\eeq
In atomic units (see \cref{sec:Units}), $M = 2$ (mass of Ps),
$\hbar = 1$ (Planck's constant) and $a_0 = 1$ (Bohr radius), simplifying
this equation as
\beq
\label{eq:EffectiveRangeLongAu}
\kappa \cot\delta_0 = -\frac{1}{a} + \frac{1}{2} r_0 \kappa^2 - \frac{4 \pi C_6}{15 a^2} \kappa^3 - \frac{16 C_6}{15 a} \kappa^4 \ln \left(\kappa \right) + \mathcal{O}(\kappa^4).
\eeq

In \cref{tab:vdwFlannery}, the $\kappa^2$ column fits 
only to the first two terms of \cref{eq:EffectiveRangeLongAu}, which makes 
this the same as the short-range expansion, \cref{eq:EffectiveRangeShort}.  
The $\kappa^3$ column fits to the first three terms, and the $\kappa^4 \ln$ 
column fits to all four terms. Entries in these tables are given as
$a^\pm / r_0^\pm$. The scattering length, $a^\pm$, is used as a fitting parameter,
not a fixed value determined by \cref{eq:ScatLen} in \cref{tab:ScatLenDef}.

\begin{table}
\centering
\begin{tabular}{c l c c c}
\toprule
Partial wave & $\kappa$ Range & $\kappa^2$ & $\kappa^3$ & $\kappa^4 \ln$ \\
\midrule
$^1$S & $0.1 - 0.5$       & 4.3080/2.2816 & 3.9879/4.1474 & 3.1029/7.9026 \\
      & $0.01 - 0.09$     & 4.3306/2.2012 & 4.3288/2.4807 & 4.3297/2.3490 \\
      & $0.001 - 0.009$   & 4.3306/2.1972 & 4.3306/2.2251 & 4.3306/2.2207 \\
\midrule
$^3$S & $0.1 - 0.5$       & 2.1623/1.3729 & 1.9356/9.0125 & 1.7116/9.7221 \\
      & $0.01 - 0.09$     & 2.1367/1.9354 & 2.1350/3.0845 & 2.1359/2.6251 \\
      & $0.001 - 0.009$   & 2.1365/2.0354 & 2.1365/2.1502 & 2.1365/2.1394 \\
\bottomrule
\end{tabular}
\caption[Scattering length and effective range using Flannery expansion]{Scattering length and effective range using \cref{eq:EffectiveRangeLongAu}.
The column headings indicate which term this expansion is taken to.
Entries are given as $a^\pm / r_0^\pm$.}
\label{tab:vdwFlannery}
\end{table}

\subsubsection{Hinckelmann-Spruch Expansion}
\label{sec:Hinckelmann}

\Cref{eq:EffectiveRangeLong} is derived by starting with the expression given
in Hinckelmann and Spruch \cite{Hinckelmann1971}, then inverting and performing
an expansion. Hinckelmann and Spruch give this in terms of $\tan\delta_0$.

\beq
\label{eq:HinckelmannEqn}
\tan\delta_0 = -a \kappa - \frac{1}{2} r_0 a^2 \kappa^3 + \frac{1}{15} C_6^4 \kappa^4 + \frac{4}{15} C_6^4 \kappa^5 \ln|2\kappa d| + \mathcal{O}(\kappa^5)
\eeq

\begin{table}
\centering
\begin{tabular}{c c c c c}
\toprule
Partial Wave & $\kappa$ Range & $\kappa^3$ & $\kappa^4$ & $\kappa^5 \ln$ \\
\midrule
$^1$S & $0.1 - 0.5$     & 13.652/-0.7062 & 13.359/-0.6806 & 24.484/-484.65 \\
      & $0.01 - 0.09$   & 4.3286/2.3119 & 4.3276/2.3930 & 4.3306/34.834 \\
      & $0.001 - 0.009$ & 4.3306/2.1980 & 4.3306/2.2060 & 4.3306/34.622 \\
\midrule
$^3$S & $0.1 - 0.5$     & 1.7299/4.9923 & 1.4474/12.139 & 2.3287/-3.0949 \\
      & $0.01 - 0.09$   & 2.1369/1.9500 & 2.1359/2.2805 & 2.1364/2.1411 \\
      & $0.001 - 0.009$ & 2.1365/2.0382 & 2.1365/2.0711 & 2.1365/7.0068 \\
\bottomrule
\end{tabular}
\caption[Scattering length and effective range using Hinckelmann-Spruch expansion]{Scattering length and effective range using \cref{eq:HinckelmannEqn}.
The column headings indicate which term this expansion is taken to. Entries are
given as $a^\pm / r_0^\pm$.}
\label{tab:HinckERT}
\end{table}

\Cref{tab:HinckERT} has entries for the $\kappa^5 \ln$ term where the fitting 
cannot be used, since the fitting attempts to use a negative value of $d$, 
forcing the natural logarithm to return a complex value. The variable $d$ is 
given in their paper as the distance $r > d$ at which the van der Waals 
potential is dominant. We have to fit to $d$ in our problem, making this 
particular model not fit as well.

The scattering lengths for $\kappa = 0.001 - 0.009$ match well with the other 
methods, however. For the $\kappa^3$ and $\kappa^4$ terms in
\cref{eq:HinckelmannEqn} using this $\kappa$ range, the effective range matches
reasonably well with the previous methods, but the last term involving $\kappa^5$
does not match for the effective range. The effective range is even negative
when using the $\kappa^5$ term with $\kappa = 0.1 - 0.5$. The Flannery expansion,
\cref{eq:EffectiveRangeLong}, is a modified version of \cref{eq:HinckelmannEqn}
and is easier to fit to our type of problem.

\subsubsection{Arriola Expression}
Refs.~\cite{Cordon2010,Arriola2010} give an analytic solution derived using a
semiclassical approach for the effective range
with a van der Waals potential of
\beq
\label{eq:EffRangeAnalytic}
\frac{r_0}{R} = 1.395 - 1.333 \frac{R}{a_0} + 0.6373 \frac{R^2}{a_0^2} \,\,,
\eeq
where the van der Waals range, $R$, is
\beq
R = \left(\frac{M C_6}{\hbar^2}\right)^{\frac{1}{4}}.
\eeq
Using the values from \cref{tab:ScatLenDef} for $a^\pm$, this equation 
produces the results in \cref{tab:EffRangeArriola}.
\begin{table}
\centering
\begin{tabular}{c c c c c}
\toprule
$\omega$ & $r_0^+$ & $r_0^-$ \\
\midrule
6 & 2.2809 & 2.1839 \\
7 & 2.2796 & 2.1878 \\
\bottomrule
\end{tabular}
\caption{Effective range from Arriola equation}
\label{tab:EffRangeArriola}
\end{table}
Despite being derived using a semiclassical approach, it nonetheless
returns values for the effective range similar to that returned from
the other models.

\subsubsection{Gao Model}
\label{sec:GaoModel}

Gao's \cite{Gao1998} effective range theory treatment is the most complicated 
of the models tried. Gao solves the Schr\"{o}dinger equation for an 
attractive $r^{-6}$ potential to find an expression relating the phase shifts 
to a quantity he refers to as $K_\ell^0(E_{\bm \kappa})$:
\beq
\label{eq:GaoZEqn}
\tan\delta_l = [Z_{ff} - K_\ell^0(E_{\bm \kappa}) Z_{gf}]^{-1} [K_\ell^0(E_{\bm \kappa}) Z_{gg} - Z_{fg}].
\eeq
The Z functions in this equation are complicated but described in his
paper. The phase shifts are fitted to \cref{eq:GaoZEqn} to determine
$K_\ell^0(E_{\bm \kappa})$ for each $\kappa$ value. $K_\ell^0(E_{\bm \kappa})$ can
be expanded in a Taylor series as \cite{Gao1998a}
\beq
\label{eq:GaoKTaylor}
K_\ell^0(E_{\bm \kappa}) = K_\ell^0(0) + {K_\ell^0}^\prime(0) E_{\bm \kappa} + \ldots.
\eeq
We keep just the first two terms in this expression.
From this, $K_l^0(0)$ and ${K_l^0}^\prime(0)$ are determined. In another 
paper \cite{Gao1998a}, Gao performs an expansion of this expression for low
$\kappa$ to generate expressions for $a$ and $r_0$:
\beq
\label{eq:GaoScatLenS}
a_0 = \frac{2\uppi}{[\Gamma(1/4)]^2} \frac{K_{l=0}^0(0) - 1}{K_{l=0}^0(0)} \beta_6
\eeq
and
\beq
\label{eq:GaoEffRange}
r_0 = \frac{[\Gamma(1/4)]^2}{3\pi} \frac{[K_{l=0}^0(0)]^2 + 1}{[K_{l=0}^0(0) - 1]^2} \beta_6 + \frac{[\Gamma(1/4)]^2}{\pi} \frac{{K_{l=0}^0}^\prime(0)^2(\hbar^2/2\mu)(1/\beta_6)^2}{[K_{l=0}^0(0) - 1]^2} \beta_6.
\eeq
$\beta_6$ is related to $C_6$ by
\beq
\label{eq:beta6}
\beta_6 = (2\mu C_6/\hbar^2)^{1/4}.
\eeq

The results in \cref{tab:GaoResults} are computed by taking
\cref{eq:GaoKTaylor} out to the second term, which includes the $K_l^0$ derivative. 
Two values of $\kappa$ are used to solve the equations for the unknowns
$K_{\ell=0}^0(0)$ and ${K_{\ell=0}^0}^\prime(0)$. These values are in the table and are 
used to determine $a$ and $r_0$ via \cref{eq:GaoScatLenS,eq:GaoEffRange}.

\begin{table}
\centering
\begin{tabular}{c c c c c c}
\toprule
\toprule
& $\kappa$ & $K_l^0$ & $K_l^{0\prime}$ & $a$ & $r_0$ \\
\midrule
$^1S$ & $0.1, 0.2$ & -0.610967 & -1.67672 & 4.3286 & 2.3380 \\
      & $0.2, 0.3$ & -0.608704 & -1.90301 & 4.3386 & 2.3128 \\
      & $0.3, 0.4$ & --- & --- & --- & --- \\
\midrule
$^1S$ & 0.001, 0.002 & -0.61050 & -13.8371 & 4.3307 & 0.9103 \\
      & 0.002, 0.003 & -0.61052 & -2.76553 & 4.3306 & 2.2104  \\
      & 0.003, 0.004 & -0.61052 & -0.46049 & 4.3306 & 2.4811   \\
      & 0.004, 0.005 & -0.61050 & -4.51842 & 4.3307 & 2.0046   \\
      & 0.005, 0.006 & -0.61052 & -2.67368 & 4.3306 & 2.2212  \\
      & 0.006, 0.007 & -0.61052 & -2.66827 & 4.3306 & 2.2219  \\
      & 0.007, 0.008 & -0.61052 & -2.63471 & 4.3306 & 2.2258   \\
      & 0.008, 0.009 & -0.61052 & -2.59699 & 4.3306 & 2.2302  \\
\midrule
\midrule
$^3S$ & $0.1, 0.2$ & -3.33674 & -20.9648 & 2.1336 & 2.7510 \\
      & $0.2, 0.3$ & -3.30159 & -24.4797 & 2.1389 & 2.6778 \\
      & $0.3, 0.4$ & --- & --- & --- & --- \\
\midrule
$^3S$ & 0.001, 0.002 & -3.31867 & -62.3938 & 2.1363 & 2.0666  \\
      & 0.002, 0.003 & -3.31868 & -57.2099 & 2.1363 & 2.1513 \\
      & 0.003, 0.004 & -3.31868 & -57.6505 & 2.1363 & 2.1441 \\
      & 0.004, 0.005 & -3.31866 & -62.0112 & 2.1363 & 2.0728 \\
      & 0.005, 0.006 & -3.31868 & -58.7635 & 2.1363 & 2.1259 \\
      & 0.006, 0.007 & -3.31869 & -57.7088 & 2.1363 & 2.1431 \\
      & 0.007, 0.008 & -3.31871 & -55.5445 & 2.1363 & 2.1785 \\
      & 0.008, 0.009 & -3.31874 & -54.1885 & 2.1363 & 2.2006 \\
\bottomrule
\bottomrule
\end{tabular}
\caption{Full Gao model scattering length and effective range}
\label{tab:GaoResults}
\end{table}

The values for $a$ are well-converged and compare well with the values in 
\cref{tab:ScatLenDef} from the definition in \cref{eq:ScatLen}. The values of 
$r_0$ are not so well-converged, but they normally compare well with the 
fittings in \cref{sec:vanderWaalsERT}. In Ref.~\cite{Woods2015}, we used the
$\kappa = 0.002$, 0.003 values, as the smallest set of $\kappa = 0.001$, 0.002
is obviously unstable, with a much different effective range for $^1$S.

We also attempted to carry the expansion in
\cref{eq:GaoKTaylor} to the third term with the second derivative, but numerical 
inaccuracy proved to be a problem. This
term is proportional to $\kappa^4$, which is vanishingly small for 
most $\kappa$ values considered, making determining ${K_{\ell=0}^0}''$ 
impossible with the current phase shift precision.

The $K_l^0$ is not particularly sensitive, but even the $K_l^{0\prime}$ can
change drastically depending on the values of $\kappa$ used. The $K_l^{0\prime}$
values for $^3$S are also much larger than those of $^1$S. While the Gao method
gives results in line with the other ERTs presented thus far, it would likely
be suited better for systems where phase shifts could be computed to more
precision than we can do for Ps-H scattering.

Gao \cite{Gao1998a} also gives QDT expansions for the S-wave and P-wave in
another paper. Equations (6) and (7) in \cite{Gao1998a} relate the phase shifts
to $K_\ell^0$. In \cref{eq:ERTCompare}, we also see that this QDT expansion
does not do well for these two partial waves. Gao also gives an expansion for
$\ell \geq 2$, which is discussed in \cref{sec:GaoPhase}.

\subsection{Effective Range from Scattering Length and Binding Energy}
\label{sec:BlackwoodERT}
We use an expression from Ref. \cite{Blackwood2002} to get an estimate
of the $^1S$ effective range, given by
\begin{equation}
\label{eq:BlackwoodERT}
r_0^+ = \frac{a^+ \sqrt{4 E_b} - 1}{2 a^+ E_b}.
\end{equation}
This expression is also similar to one given by Page \cite{Page1976}.
Using this with the complex Kohn $a^+$ and $E_b$ at $\omega = 7$, we get
$r_0^+ = 2.106$, which only agrees somewhat with the other results in
\cref{tab:SWaveScatLenOther}. There is not an equivalent expression for $r_0^-$,
as there is not a $^3S$ PsH bound state.

\subsection{Comparison of Effective Range Theories}
\label{eq:ERTCompare}

\begin{figure}[H]%
    \centering
    \subfloat[ ]{{\includegraphics[width=3.2in]{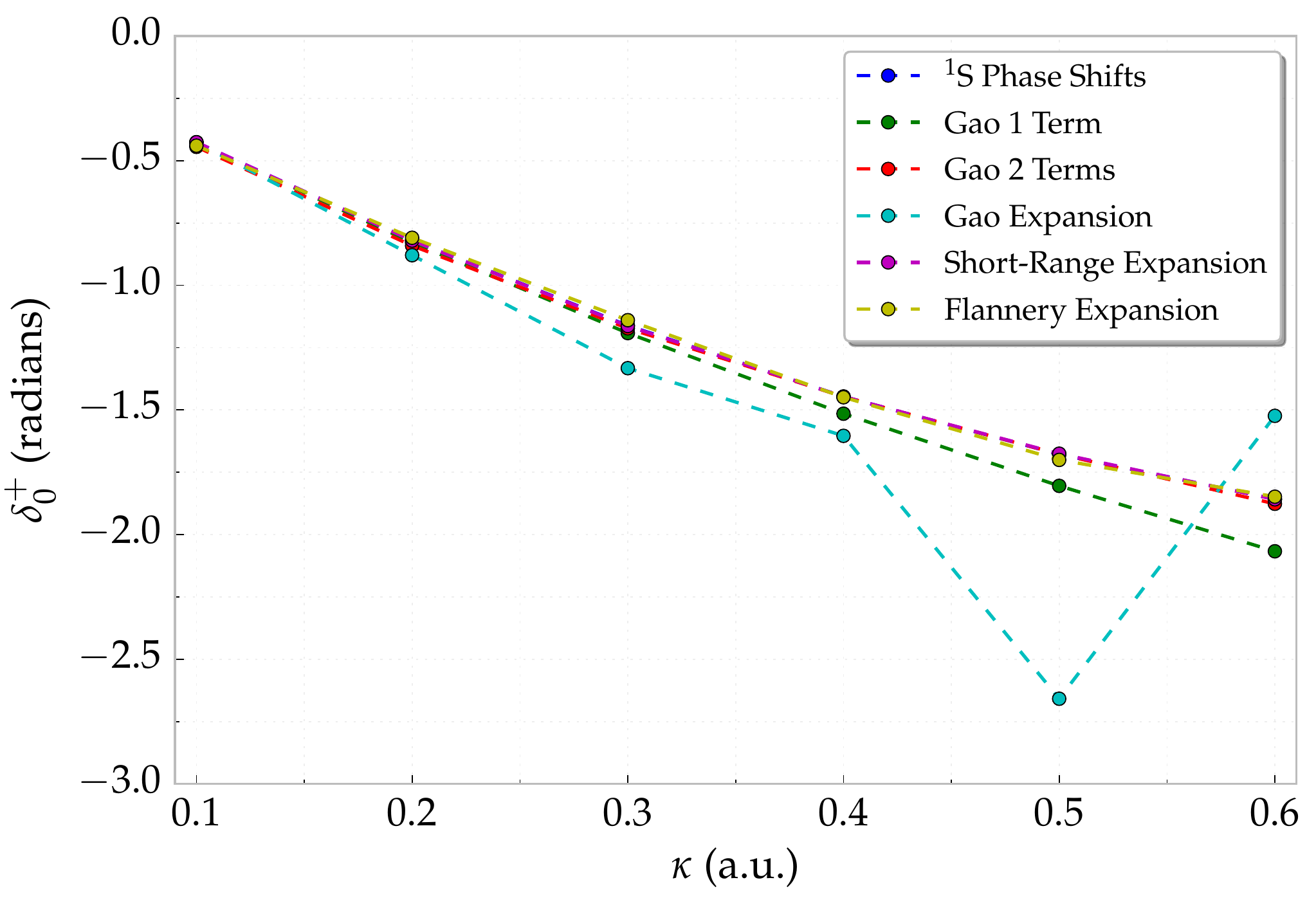} }}%
    \subfloat[ ]{{\includegraphics[width=3.2in]{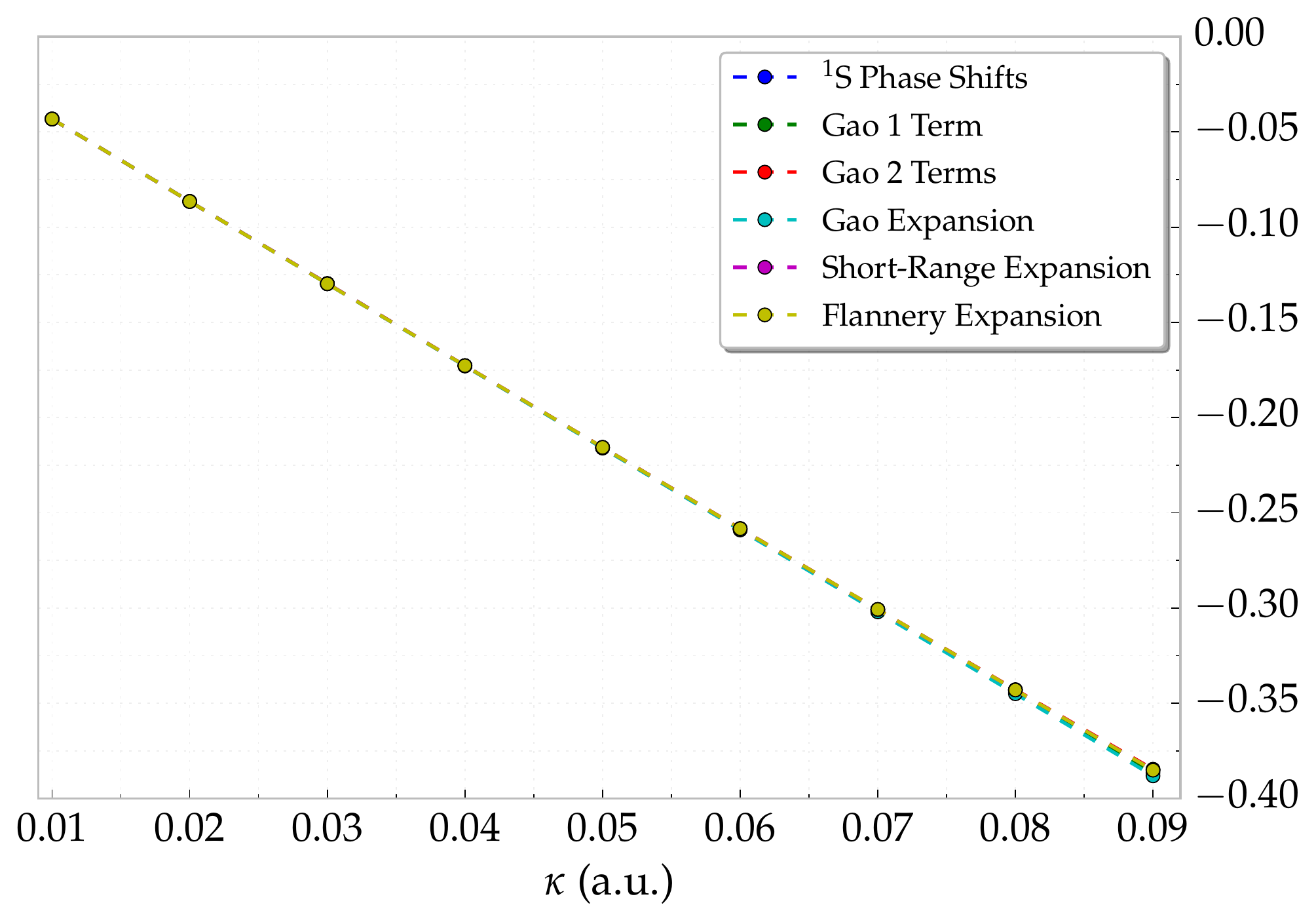} }}%
    \caption[Comparison of results from effective range theories for $^1S$ Ps-H]{Comparison of results from effective range theories for $^1S$ Ps-H. Figure (a) shows a larger set of $\kappa$ values, and (b) shows a range with small $\kappa$ values. The Gao 1 and 2 term results use 1 and 2 terms from \cref{eq:GaoKTaylor}, respectively. The Gao expansion uses the method in Ref.~\cite{Gao1998a}. The short-range and Flannery expansions are given in \cref{sec:EffectiveRangeShort,sec:Flannery}, respectively.}%
    \label{fig:ERT-Comparisons-Singlet}%
\end{figure}

\begin{figure}[H]%
    \centering
    \subfloat[ ]{{\includegraphics[width=3.2in]{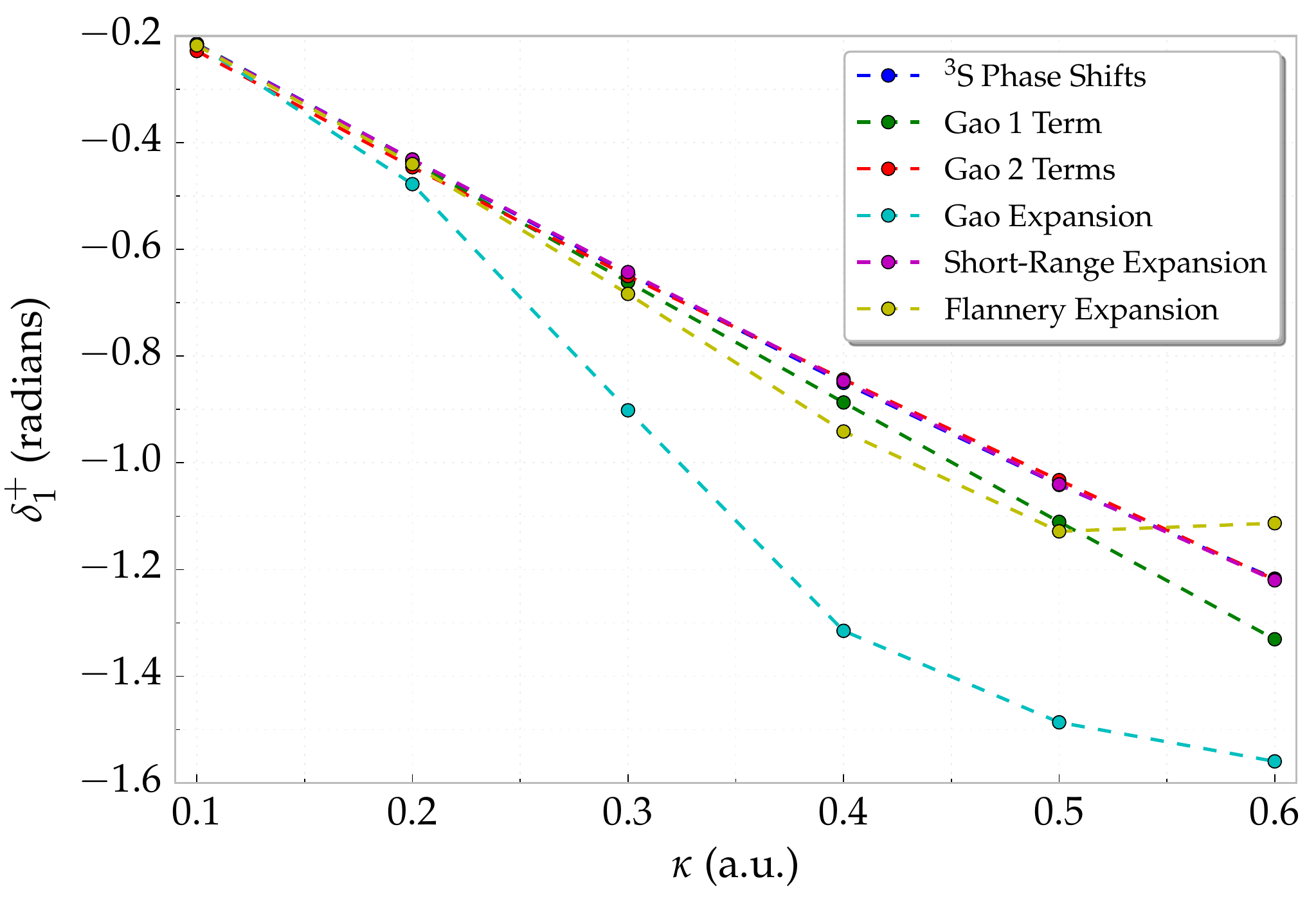} }}%
    \subfloat[ ]{{\includegraphics[width=3.2in]{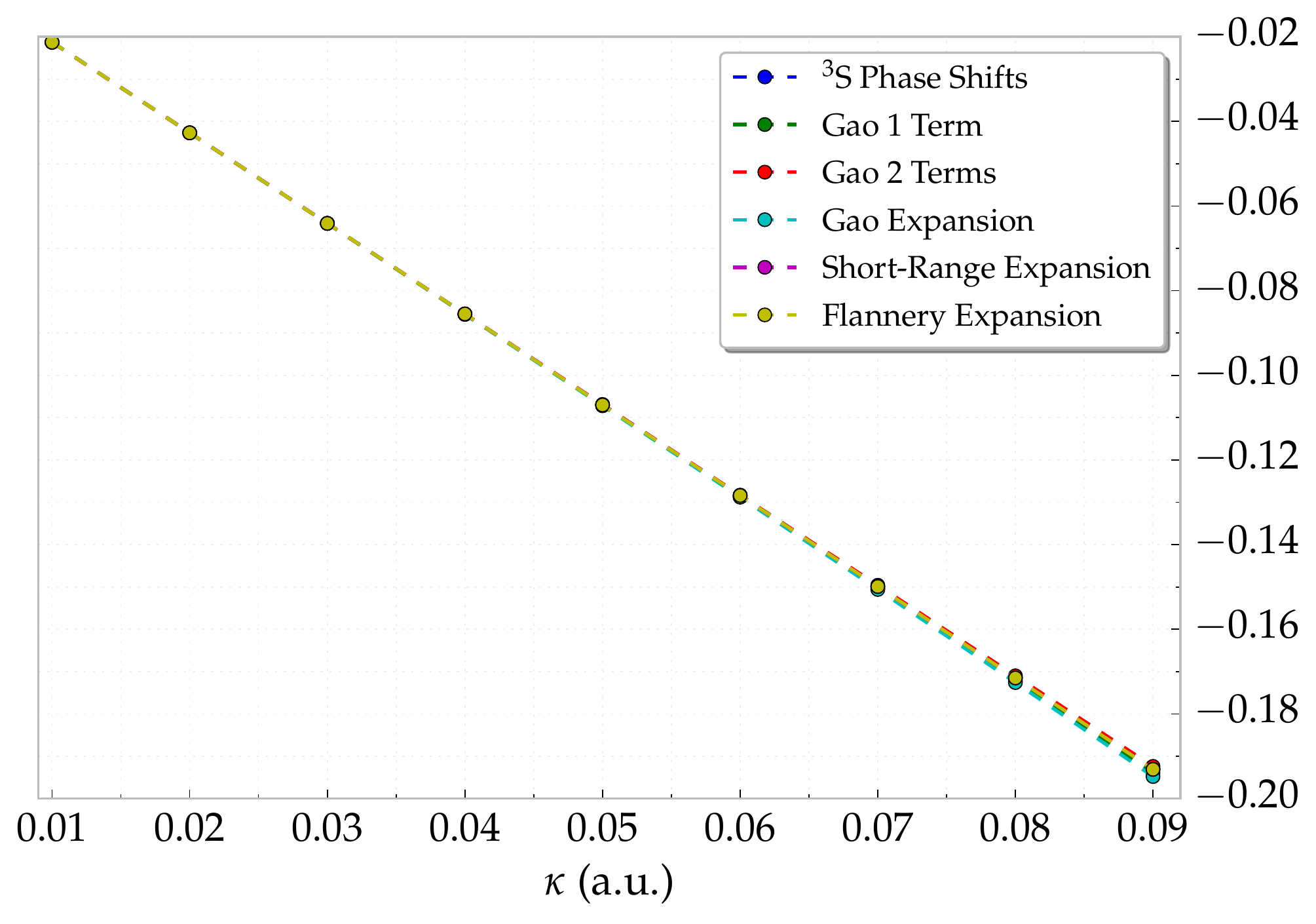} }}%
    \caption[Comparison of results from effective range theories for $^3S$ Ps-H]{Comparison of results from effective range theories for $^3S$ Ps-H. Figure (a) shows a larger set of $\kappa$ values, and (b) shows a range with small $\kappa$ values. The Gao 1 and 2 term results use 1 and 2 terms from \cref{eq:GaoKTaylor}, respectively. The Gao expansion uses the method in Ref.~\cite{Gao1998a}. The short-range and Flannery expansions are given in \cref{sec:EffectiveRangeShort,sec:Flannery}, respectively.}%
    \label{fig:ERT-Comparisons-Triplet}%
\end{figure}

To obtain the plots in
\cref{fig:ERT-Comparisons-Singlet,fig:ERT-Comparisons-Triplet},
after obtaining fits for each of the 
models, we solved the respective equations for the phase shifts.  This 
provides a reliable way to determine how well each model fits the phase shift 
data. When $\kappa$ is small, the different models agree extremely well, as 
seen in Figures \ref{fig:ERT-Comparisons-Singlet}(b) and \ref{fig:ERT-Comparisons-Triplet}(b).
Higher order terms in each of the model equations 
become negligible as $\kappa$ gets smaller.
\Cref{fig:ERT-Comparisons-Singlet,fig:ERT-Comparisons-Triplet}
do not include the Hinckelmann-Spruch results since, as seen in \cref{tab:HinckERT}, 
this expression does not generate particularly good results due to the need to
fit to the $d$ parameter in \cref{eq:HinckelmannEqn}.

\subsubsection{Comparisons with Other Groups' Results}


\begin{table}
\centering
\footnotesize
\centerline{
\begin{tabular}{l c c c c c}
\toprule
Method & $\kappa$ & \multicolumn{1}{c}{$a^+$} & \multicolumn{1}{c}{$r_0^+$} & \multicolumn{1}{c}{$a^-$} & \multicolumn{1}{c}{$r_0^-$}\\
\midrule
Approx. to def. - Eq. (\ref{eq:ScatLenApprox})                 & $0.001$ & $4.331 \pm 0.012$ & --- & $2.137 \pm 0.008$ & --- \\
Extrapolated ($\omega = 4 \to 7)$ - Eq. (\ref{tab:ScatLenDef}) & $0.001$ & $4.319$ & --- & $2.129$ & --- \\
ERT Short - Eq. (\ref{eq:EffectiveRangeShort})                 & $0.001 - 0.009$ & $4.331 \pm 0.012$ & 2.197 & $2.137 \pm 0.008$ & 2.035 \\
ERT Short - Eq. (\ref{eq:EffectiveRangeShort})                 & $0.1 - 0.5$ & $4.308 \pm 0.003$ & 2.275 & $2.162 \pm 0.003$ & 1.343 \\
ERT vdW - Eq. (\ref{eq:EffectiveRangeLongAu})                  & $0.001 - 0.009$ & $4.331 \pm 0.012$ & 2.221 & $2.137 \pm 0.008$ & 2.137 \\
QDT - Eqs. (\ref{eq:GaoZEqn}), (\ref{eq:GaoKTaylor})           & $0.002, 0.003$ & $4.331 \pm 0.012$ & 2.210 & $2.136 \pm 0.008$ & 2.151 \\
QDT expansion - Eq. (6) in Ref.~\cite{Gao1998a}                & $0.001$ & $4.331 \pm 0.012$ & 2.535 & $2.136 \pm 0.008$ & 3.086 \\
Eq. (\ref{eq:BlackwoodERT})                                    & --- & --- & 2.106 & --- & --- \\
\midrule
CC 14Ps14H+H$^-$ (Walters \emph{et al} 2004) \cite{Blackwood2002}       & ---       & 4.327 & ---   & ---   & --- \\
Kohn extrapolated (Van Reeth \emph{et al} 2003) \cite{VanReeth2003}     & ---       & 4.311 & 2.27  & 2.126 & 1.39 \\
Kohn 721 terms (Van Reeth \emph{et al} 2003) \cite{VanReeth2003}        & ---       & 4.334 & ---   & 2.143 & --- \\
Kohn Eq. (\ref{eq:EffectiveRangeShort}) \cite{VanReeth2003}             & up to 0.5 & 4.30  & 2.27  & 2.147 & \,\,--- \\
CC 14Ps14H (Blackwood \emph{et al} 2002) \cite{Blackwood2002}           & ---       & 4.41  & 2.19  & 2.06  & 1.47 \\
SVM (Ivanov \emph{et al} 2002) \cite{Ivanov2002}                        & ---       & 4.34  & 2.39  & 2.22  & 1.29 \\
DMC (Chiesa \emph{et al} 2002) \cite{Chiesa2002}                        & ---       & 4.375 & 2.228 & 2.246 & 1.425 \\
T-matrix (Biswas \emph{et al} 2002) \cite{Biswas2002a}                  & ---       & 3.89  & ---   & ---   & --- \\
SVM (Ivanov \emph{et al} 2001) \cite{Ivanov2001}                        & ---       & 4.3   & ---   & 2.2   & --- \\
Variational basis-set (Adhikari \emph{et al} 2001) \cite{Adhikari2001b} & ---       & 3.49  & ---   & 2.46  & --- \\
6-state CC (Sinha \emph{et al} 2000) \cite{Sinha2000}                   & ---       & 5.90  & 2.73  & 2.32  & 1.29 \\
5-state CC (Adhikari \emph{et al} 1999) \cite{Adhikari1999}             & ---       & 3.72  & 1.67  & ---   & --- \\
22-state CC (Campbell \emph{et al} 1998) \cite{Campbell1998}            & ---       & 5.20  & 2.52  & 2.45  & 1.32 \\
9-state CC (Campbell \emph{et al} 1998) \cite{Campbell1998}             & ---       & 5.51  & 2.63  & 2.45  & 1.33 \\
Stabilization (Drachman \emph{et al} 1976) \cite{Drachman1976}          & ---       & ---   & ---   & 2.36  & 1.31 \\
Stabilization (Drachman \emph{et al} 1975) \cite{Drachman1975}          & ---       & 5.33  & 2.54  & ---   & --- \\
Kohn 35 terms (Page 1976) \cite{Page1976}                               & ---       & 5.844 & 2.90  & 2.319 & --- \\
SE (Hara \emph{et al} 1975) \cite{Hara1975}                             & ---       & 7.275 & ---   & 2.476 & --- \\
\bottomrule
\end{tabular}
}
\caption{$^{1,3}$S-wave scattering lengths and effective ranges}
\label{tab:SWaveScatLenOther}
\end{table}

The first eight entries of \cref{tab:SWaveScatLenOther} show the scattering 
lengths and effective ranges we calculate with the various methods described 
in this chapter. Other than the ERT Short for $\kappa = 0.1 - 0.5$, the
scattering lengths from the different methods are identical. There is much less
agreement in the effective range. For the singlet, $r_0^+ \approx 2.2$, and for
the triplet, $r_0^- \approx 2.0 - 2.1$. The effective range is much more
sensitive to slight variations in the phase shifts. More agreement could likely
be achieved if the wavefunction was fully optimized for this very low energy
range. We also tried a smaller $\kappa$ range of 0.0001 - 0.0009, but the phase
shifts were less numerically stable in that region.

As noted earlier, there is good agreement between the scattering lengths and
effective ranges for recent calculations from other groups and the results we
obtain using the complex Kohn phase shifts. The older 
calculations tend to have higher values for $a^\pm$. The $r_0^-$ values from
other groups lie close to our ERT Short values for $\kappa = 0.1-0.5$, but as
seen in \cref{fig:swave-ERT-short,tab:SWaveScatLenOther}, smaller $\kappa$
values give a much larger effective range.

\section{P-Wave Scattering Lengths}
\label{sec:PWaveScatLen}

From Refs.~\cite{Levy1963,Gao1998a}, the scattering length is only defined for
a partial wave if $2\ell+3 < n$, and the effective range is only defined if
$2\ell+5 < n$. For the van der Waals interaction, $n = 6$. Thus for Ps-H
scattering, the P-wave has a scattering length but not an effective range.
Ref.~\cite{Ivanov2002} calculates a P-wave effective range, but this
does not appear to be physical.

To determine the scattering length, we can use \cref{eq:ScatLenApprox} with $\ell = 1$.
This is the approximation to the definition shown as the first entry in
\cref{tab:PWaveScatLen}. We also extrapolate the scattering lengths from this
for $\omega = 4-7$ using \cref{eq:ScatLenExtrap}.
Additionally, we use the QDT of Gao from \cref{sec:GaoModel}
for the P-wave and the QDT expansion of Gao given by \cite{Gao1998a}
\begin{equation}
\label{eq:GaoScatLenP}
a_1 = -\frac{\uppi}{18[\Gamma(3/4)]^2} \frac{K_{l=1}^0(0) + 1}{K_{l=1}^0(0)} \beta_6^3.
\end{equation}
Each of these methods agree reasonably well for the $^1$P and
$^3$P scattering lengths. The $^1$P scattering length is negative and has a
magnitude much larger than the $^{1,3}$S scattering lengths, while the $^3$P
scattering length is positive and closer to the $^{1,3}$S scattering lengths.

The only other calculation we have found of P-wave scattering lengths is the
SVM calculation \cite{Ivanov2002}. Their $a_1^+$ matches well with the complex Kohn value, but
their $a_1^-$ is much larger. They determine the scattering lengths with an
effective range formula involving a $r_1^\pm$ term, but even if we fit the
complex Kohn phase shifts to this form, the scattering lengths do not change
much. The discrepancy in the $a_1^-$ scattering lengths appears mainly to be
due to the much smaller SVM phase shifts at low $\kappa$.

\begin{table}
	\centering
	\footnotesize
	\begin{tabular}{l l c c}
	\toprule
	Model & $\,\,\kappa$ & $a_1^+$ & $a_1^-$ \\
	\midrule
	Approx. to def. - Eq. (\ref{eq:ScatLenApprox}) & 0.01 & $-22.130 \pm 0.173$ & $1.4530 \pm 0.1104$ \\
	Extrap. approx. to def. ($\omega = 4 \to 7)$ - Eq. (\ref{tab:ScatLenDef}) & $0.01$ & $-22.262$ & $1.378$ \\
	QDT - Eq. (\ref{eq:GaoZEqn}) & 0.01, 0.02 & $-22.200 \pm 0.173$ & $1.4158 \pm 0.1107$ \\
	QDT expansion - Eq. (\ref{eq:GaoScatLenP}) & 0.01 & $-22.198 \pm 0.172$ & $1.4102 \pm 0.1104$ \\
	\midrule
	SVM \cite{Ivanov2002} & --- & $-20.7$ & $6.80$ \\
	\bottomrule
	\end{tabular}
	\caption{$^{1,3}$P scattering lengths summarized and compared}
	\label{tab:PWaveScatLen}
\end{table}

\begin{figure}
	\centering
	\includegraphics[width=\textwidth]{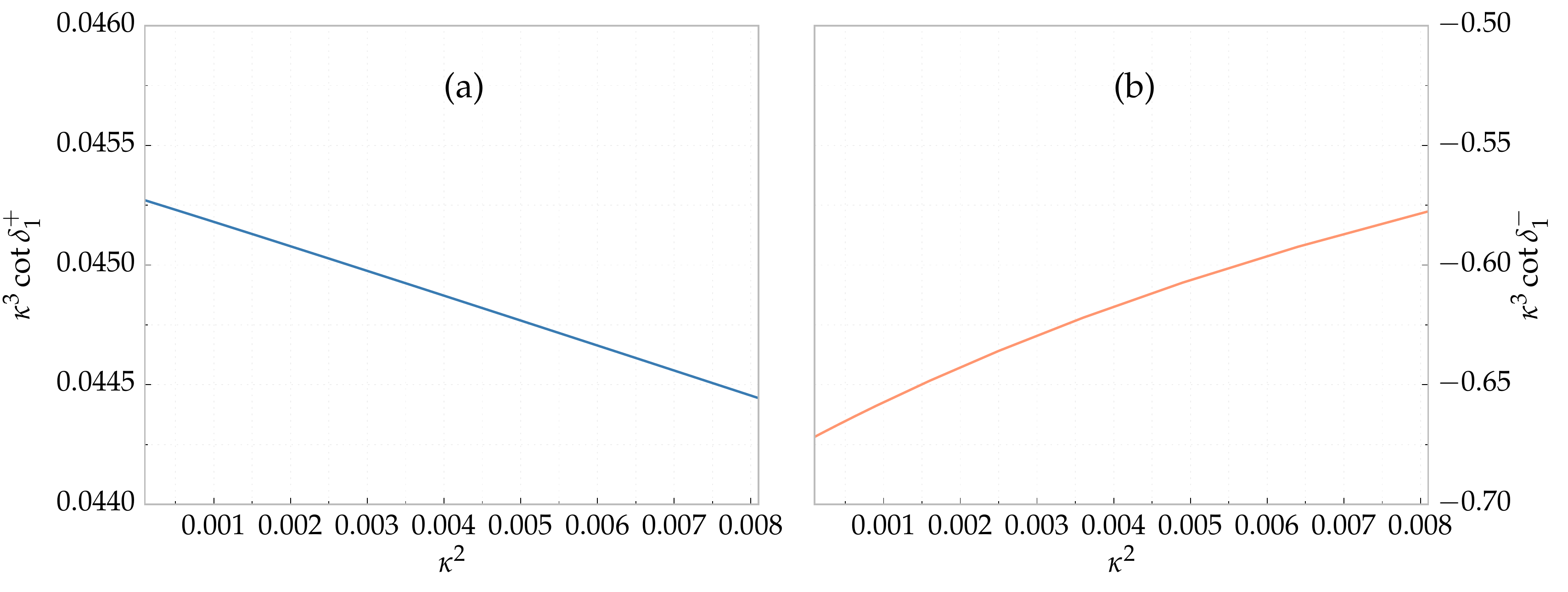}
	\caption[P-wave short-range ERT fitting]{(a) $^1P$ and (b) $^3P$ phase shifts, plotted as
$\kappa^3 \cot \delta_1^\pm$ versus $\kappa^2$.}
	\label{fig:ERT-pwave}
\end{figure}

Despite there not being a P-wave effective range, spurred by the SVM
\cite{Ivanov2002} use of an ERT expression, we performed a similar plot to
\cref{fig:swave-ERT-short}, given in \cref{fig:ERT-pwave}. Similar to the
S-wave, the $^1$P gives a relatively straight line, while the $^3$P curves downward
at low $\kappa$. Negating and inverting the y-intercept gives another way to
calculate the scattering length. Doing this, we get $a_1^+ = -22.101$ and
$a_1^- = 1.474$. These match well with the other values we obtain in
\cref{tab:PWaveScatLen}.

\section{Summary}
\label{sec:SummaryERT}

We calculated the $^{1,3}$S-wave scattering lengths using the approximation to
the definition, yielding similar results to the prior Kohn / inverse Kohn 
calculation \cite{VanReeth2003}. In addition, we calculated the $^{1,3}$S-wave
scattering lengths and effective ranges using the short-range expansion,
several models using the van der Waals interaction, and a QDT model 
incorporating the van der Waals interaction. These models yield nearly 
identical $^1$S and $^3$S scattering lengths when small $\kappa$ values are 
used. The effective ranges vary more between the different models than the
scattering lengths, but most models agree relatively well for several $\kappa$
ranges for the $^1$S-wave, with the exception of the QDT expansion, which 
gives less accurate results. Consequently, we do not report the QDT expansion
results in Ref.~\cite{Woods2015}.

We are able to reproduce approximately the $^3$S effective range reported by
other groups and the earlier Kohn / inverse Kohn calculations
\cite{VanReeth2003} by using $\kappa = 0.1 - 0.5$. As Van Reeth and Humberston
\cite{VanReeth2003} note, the graph of $\kappa \cot\delta^-$ versus $\kappa^2$
is not a straight line but curves down at low $\kappa$. Due to this, we also
used a much smaller $\kappa$ range of $0.001 - 0.009$, giving a more stable
effective range that is higher than previously reported.

The P-wave does not have an effective range but does have a scattering length,
which we calculated using the approximation to the definition and the QDT model.
The scattering lengths between the different methods agreed relatively well,
and the $^1$P scattering length is similar to the SVM result \cite{Ivanov2002}.
However, the $^3$P scattering length between our work and the SVM does not
agree, with the SVM $a_1^-$ being approximately 4.7 times as large, which is
likely due to the smaller SVM $^3$P-wave phase shifts.



\chapter{Conclusions}
\label{chp:Conclusion}

The Kohn variational method and variants of the method have been successful 
at treating Ps-H scattering in this work. We give a general formalism that 
can use any of the variants of the Kohn variational method,
including the inverse Kohn, generalized Kohn, $S$-matrix complex Kohn,
$T$-matrix complex Kohn, generalized $S$-matrix complex Kohn, and generalized
$T$-matrix complex Kohn. We have also developed a general formalism and code
that works for arbitrary $\ell$, allowing us to calculate phase shifts for
multiple partial waves. We have presented results for the
first six partial waves through the H-wave using highly correlated
Hylleraas-type short-range terms and an appropriate choice of long-range
terms. The computational techniques that have enabled to do this work are 
also presented.

We used these short-range terms to calculate the binding energy for $^1$S PsH,
which compares very well to the accurate results of
Refs.~\cite{Yan1999,Bubin2006}. This gave us confidence in using these 
terms for the Ps-H scattering problem. Despite there not being bound states
for other partial waves, we also used these short-range terms to create
stabilization plots to get rough estimates of the resonance positions for
$^1$P and $^1$D.

The S-wave and P-wave phase shifts compare well to other accurate
calculations, including the CC \cite{Walters2004,Blackwood2002}. The current
complex Kohn phase shifts for these two partial waves are highly accurate.
The resonance parameters we have computed for the singlet partial waves 
through the $^1$F-wave generally compare better with the complex rotation
\cite{Yan1999, Yan1998a, Ho1998, Ho2000} than the CC \cite{Walters2004}.

The much larger differences come in with the lower $\kappa$ phase shifts for
the $^{1,3}$D-wave. The complex Kohn $^3$D phase shifts are consistently below
the CC results \cite{Blackwood2002}, and the $^1$D phase shifts are generally
below but become slightly higher than the CC \cite{Walters2004} near the
resonance region. An analysis of the nonlinear parameters improved the
phase shifts slightly, but we believe that some of this discrepancy with the
CC can be explained by our omission of the mixed symmetry terms for the
D-wave (and higher partial waves). These terms are very difficult to work with,
and an analysis \cite{VanReeth2015} of their contributions for e$^+$-H and
e$^-$-H scattering shows that they could be important for Ps-H scattering.
The mixed symmetry terms for the three-body e$^+$-H and e$^-$-H
systems are much easier to use than for the four-body Ps-H system.

However, the multiple cross sections we calculate are not affected much by this
discrepancy in the $^{1,3}$D phase shifts. This is due to the D-wave mainly
contributing only at higher $\kappa$ values, while the S-wave and P-wave are
dominant through much of the lower $\kappa$ range. In this region, the 
complex Kohn phase shifts are better converged and match well with the CC 
phase shifts. The resonance parameters that we calculate for the $^1$D 
resonance also match well with the complex rotation results. The $^{1,3}$F, 
$^{1,3}$G, and $^{1,3}$H partial waves do not contribute significantly to the
elastic integrated or momentum transfer cross sections, but they do 
contribute some to the elastic differential cross section. The differential 
cross section is relatively well converged by the H-wave, so we have 
presented all of these cross sections in this work. The differential cross
section has interesting features, such as the contributions from the singlet
$^1$S through $^1$F resonances. The differential cross section is nearly
isotropic at very low energy and becomes slightly backward peaked at low
energy, then becomes forward peaked at approximately \SI{0.5}{eV},
continuing this way to the Ps(n=2) threshold.

Multiple effective range theories were used to analyze the effective ranges
for $^{1,3}$S and the scattering lengths for $^{1,3}$S and $^{1,3}$P, including
several that incorporate the van der Waals interaction. We found that these all
agreed relatively well at very low $\kappa$ -- even the short-range expansion.
We find that the $^3$S effective range changes significantly depending on the
range of $\kappa$ used. The $r_0^-$ we obtain agrees with the values from other
groups if $\kappa = 0.1 - 0.5$ but increases when we use the much smaller
range of $\kappa = 0.001 - 0.009$. The other S-wave scattering lengths and
effective ranges match up well with those calculated by other groups. The
$^1$P scattering length compares well with the SVM result \cite{Ivanov2002},
but the $^3$P scattering length differs significantly.


\appendix

\chapter{External Angular Integrations}
\label{chp:AngularInt}

\iftoggle{UNT}{We}{\lettrine{\textcolor{startcolor}{W}}{e}}
perform rotations and then integrations over the 3 
external angles to reduce the 9-dimensional integrations to 6 dimensions. The
procedure described here is the same as in Van Reeth's thesis
\cite{VanReethThesis}.

\begin{figure}
	\centering
	\includegraphics[width=4.5in]{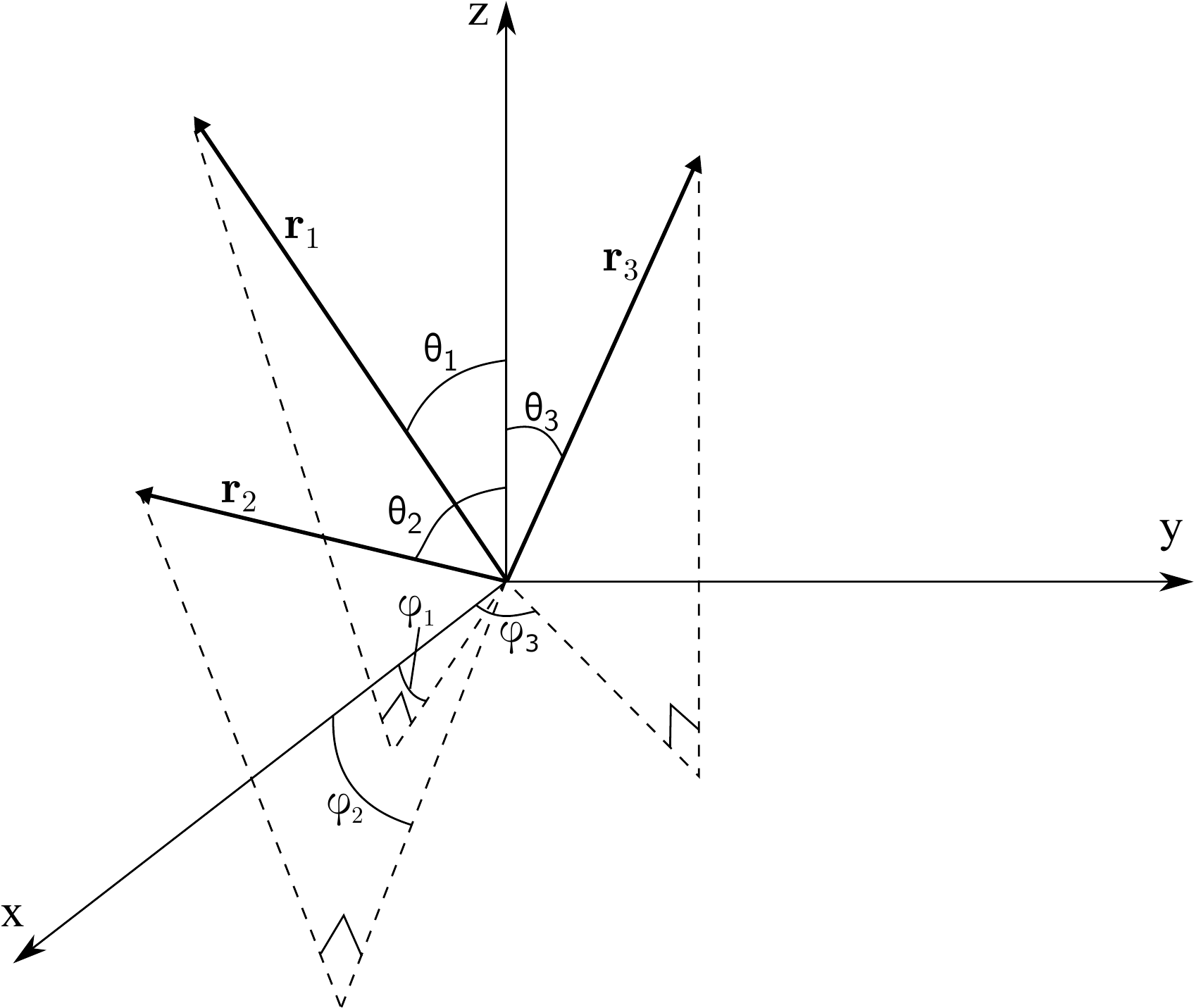}
	\caption[Ps-H original coordinate system]{Ps-H original coordinate system. This figure is the same as in Van Reeth \cite{VanReethThesis}.}
	\label{fig:CoordinateSystemOriginal}
\end{figure}

The integrals needed in
\cref{eq:BoundHFull,eq:GeneralKohnMatrix,eq:FourBody,eq:ShortIntGen} have
volume elements of
\begin{equation}
\label{eq:dTau1}
d\tau = d\mathbf{r}_1 d\mathbf{r}_2 d\mathbf{r}_3.
\end{equation}
In spherical coordinates, this becomes
\begin{equation}
\label{eq:dTau2}
d\tau = r_1^2 dr_1 \sin\theta_1 d\theta_1 d\varphi_1 r_2^2 dr_2 \sin\theta_2
  d\theta_2 d\varphi_2 r_3^2 dr_3 \sin\theta_3 d\theta_3 d\varphi_3.
\end{equation}
These coordinates are given in an arbitrary coordinate system as in
\cref{fig:CoordinateSystemOriginal}.

\begin{figure}
	\centering
	\includegraphics[width=4.5in]{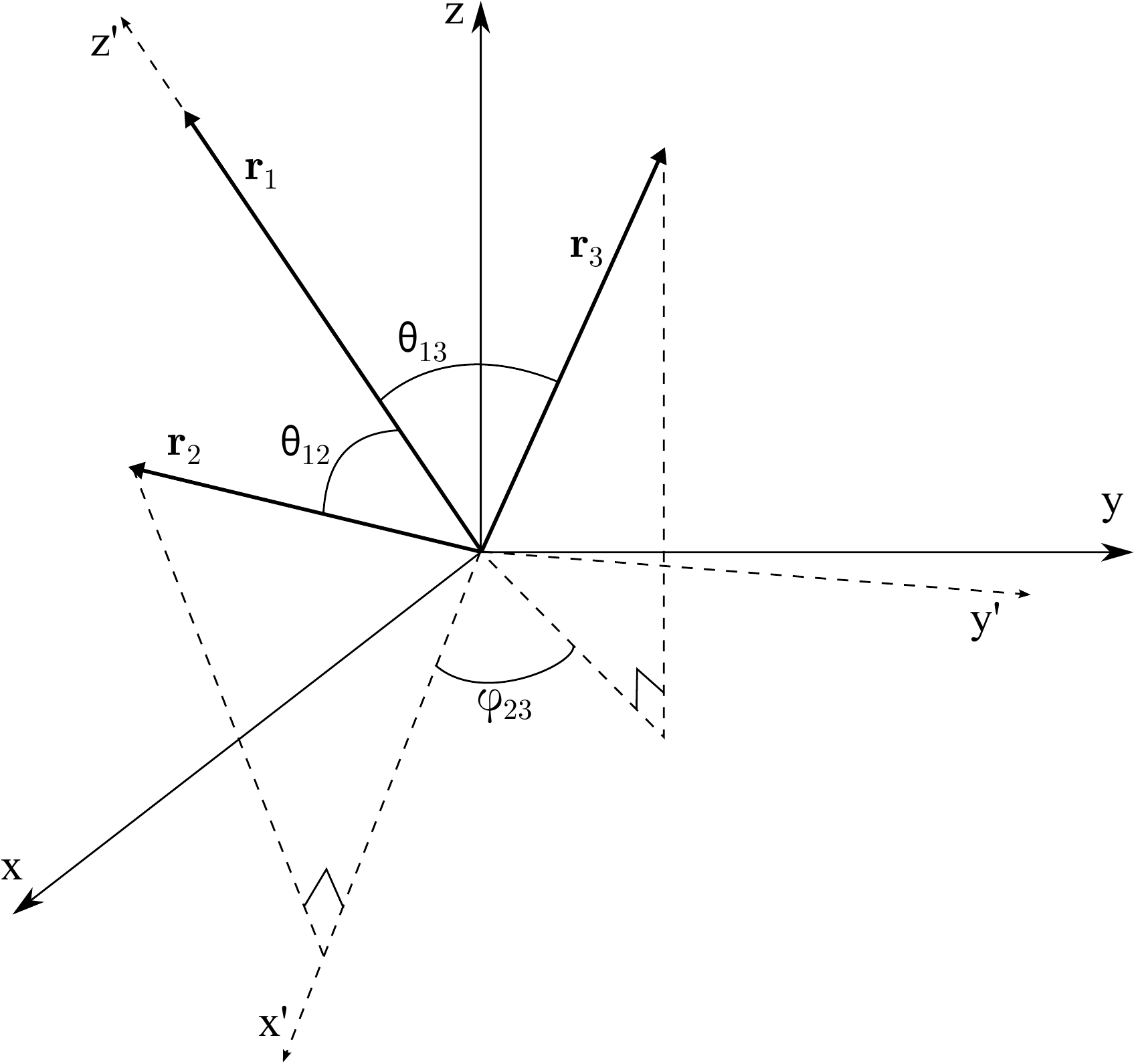}
	\caption[Ps-H rotated coordinate system]{Ps-H rotated coordinate system. This figure is the same as in Van Reeth \cite{VanReethThesis}.}
	\label{fig:CoordinateSystemRotated}
\end{figure}

We can rotate the coordinate system multiple ways in order to integrate over 
the external angles. First consider rotating the coordinate system so that the
z-axis is along $\mathbf{r}_1$. If we then choose to perform another rotation
so that $\mathbf{r}_2$ is in the x$'$-z$'$ plane, from
\cref{fig:CoordinateSystemRotated}, the volume element becomes
\begin{equation}
\label{eq:dTau3}
d\tau = r_1^2 dr_1 \sin\theta_1 d\theta_1 d\varphi_1 r_2^2 dr_2 \sin\theta_{12}
  d\theta_{12} d\phi'_2 r_3^2 dr_3 \sin\theta_{13} d\theta_{13} d\varphi_{23}.
\end{equation}
The angles $\theta_{12}$, $\theta_{13}$, and $\varphi_{23}$ are 
internal angles. The external angle $\phi'_2$ is measured from the x$'$-axis 
before the rotation into $\mathbf{r}_2$. The angles $\theta_1$ and $\varphi_1$
are also external angles.
As Peter Van Reeth points out in his thesis \cite{VanReethThesis} on page 78, 
$\varphi_{23}$ is the angle between the planes of the triangles
$(r_1,r_2,r_{12})$ and $(r_1,r_3,r_{13})$. This angle can range between
$\varphi_{23} = 0$ and $\varphi_{23} = 2 \pi$.

For the PsH bound state and S-wave Ps-H scattering, integrating over the
external angles gives
\begin{equation}
\label{eq:dTauS1}
d\tau = 8 \uppi^2 r_1^2 dr_1 r_2^2 dr_2 \sin\theta_{12} d\theta_{12} r_3^2 dr_3
  \sin\theta_{13} d\theta_{13} d\varphi_{23}.
\end{equation}
We can transform this into integrations over $r_{12}$ and $r_{13}$ instead of
$d\theta_{12}$ and $d\theta_{13}$ by differentiating the following expression
from the law of cosines with respect to $\mathbf{r}_{kl}$:
\begin{equation}
\label{eq:rkl}
r_{kl}^2 = r_k^2 + r_l^2 - 2 r_k r_l \cos\theta_{kl}.
\end{equation}
When differentiated, this gives
\begin{equation}
\label{eq:rklDer}
dr_{kl} r_{kl} = r_k r_l \sin\theta_{kl} d\theta_{kl}.
\end{equation}
Thus, we have
\begin{subequations}
\label{eq:rklDer2}
\begin{align}
dr_{12} r_{12} &= r_1 r_2 \sin\theta_{12} d\theta_{12} \\
dr_{13} r_{13} &= r_1 r_3 \sin\theta_{13} d\theta_{13}.
\end{align}
\end{subequations}
Substituting these into \cref{eq:dTauS1}, we have
\begin{equation}
\label{eq:dTauS23}
d\tau = 8 \uppi^2 dr_1 r_2 dr_2 r_3 dr_3 r_{12} dr_{12} r_{13} dr_{13}
   d\varphi_{23}.
\end{equation}
This is the final form that we use in the short-range and long-range
integrations, described in \cref{sec:CompShort,sec:CompLong}.

If we do the rotations so that the z-axis is pointing in the direction of
$\mathbf{r}_2$, then rotate so that $\mathbf{r}_1$ is in the x$'$-z$'$ plane,
the volume element is instead
\begin{equation}
\label{eq:dTauS13}
d\tau = 8 \uppi^2 r_1 dr_1 dr_2 r_3 dr_3 r_{12} dr_{12} r_{23} dr_{23}
   d\varphi_{13}.
\end{equation}
We use this in the long-range integrations when there is an $r_{23}^{-1}$ term,
as in \cref{sec:LongLongInt,sec:ShortLongInt}.

There are 6 possible orderings to performing these rotations, but we only
consider 3 and only use 2 in the final code. The last one rotates the z-axis
into $\mathbf{r}_3$, then performs another rotation where the x$'$-z$'$ plane
contains $\mathbf{r}_2$, giving a volume element of
\begin{equation}
\label{eq:dTauS12}
d\tau = 8 \uppi^2 r_1 dr_1 r_2 dr_2 dr_3 r_{13} dr_{13} r_{23} dr_{23}
   d\varphi_{12}.
\end{equation}

For the P-wave and higher, there is no spherical symmetry, but there is
azimuthal symmetry for $m = 0$ of the spherical harmonics for all partial 
waves.
The following sections show the results of performing these external angular
integrations for general $\ell$. These are shown in a \emph{Mathematica}
notebook entitled ``General Angular Integrations.nb'' \cite{GitHub}.

\section{Terms with Identical Spherical Harmonics}
\label{sec:AngSame}
Since we are only investigating integrals over external angles in this
Appendix, the orthogonality relations of the
spherical harmonics do not hold. However, it can be seen that if we integrate
over all external angles, $d\tau_{ext}$, any spherical harmonic with itself
will give $2\pi$, i.e.
\begin{equation}
\label{eq:AngSame}
\Int{Y_\ell^0(\theta_i, \varphi_i) Y_\ell^0(\theta_i, \varphi_i)}{\tau_{ext}, \tau_{ext}} = 2 \pi,
\end{equation}
where $i$ = $1$, 2, 3, $\rho$ or $\rho^\prime$.

\section{Terms with \texorpdfstring{$Y_\ell^0(\theta_i, \varphi_i) Y_\ell^0(\theta_j, \varphi_j)$}{Yi-Yj} }
\label{sec:AngRiRj}

These terms are not completely necessary to generalize here, because they 
only appear in the short-short calculations. The derivations and code for the 
S-, P-, and D-wave short-short calculations use these. The general short-short
derivations and code (\cref{sec:GeneralShort}) use a formalism from Drake and
Yan \cite{Yan1997} that does not do these external angular integrations. The 
general code is used for the F-, G-, and H-waves and can be used for 
arbitrary $\ell$. Derivations for all of these can be seen in the
\emph{Mathematica} notebook ``General Angular Integrations.nb''
\cite{GitHub,Wiki}. Note that the
result of $2\pi$ in \cref{sec:AngSame} is recovered if $i=j$. The general 
result is
\begin{equation}
\label{eq:AngRiRj}
\Int{Y_\ell^0(\theta_i, \varphi_i) Y_\ell^0(\theta_j, \varphi_j)}{\tau_{ext}, \tau_{ext}} = 2 \pi \LegendreP{\ell, \cos\theta_{ij}}.
\end{equation}
where $i$ = $1$, 2, 3, $\rho$ or $\rho^\prime$. The results in the
\emph{Mathematica} notebook ``General Angular Integrations.nb'' are equivalent
to this.

These integrations for the first three partial waves are given below:
\begin{subequations}
\begin{align}
\Int{Y_0^0(\theta_i, \varphi_i) Y_0^0(\theta_j, \varphi_j)}{\tau_{ext}, \tau_{ext}} &= 2 \pi \\
\Int{Y_1^0(\theta_i, \varphi_i) Y_1^0(\theta_j, \varphi_j)}{\tau_{ext}, \tau_{ext}} &= 2 \pi \cos\theta_{ij} \\
\Int{Y_2^0(\theta_i, \varphi_i) Y_2^0(\theta_j, \varphi_j)}{\tau_{ext}, \tau_{ext}} &= 2 \pi \left(3 \cos^2 \theta_{ij} - 1 \right).
\end{align}
\end{subequations}

When performing the external angular integrations instead of using the general short-range integrals in \cref{sec:GeneralShort}, the very specific form of the short-short integrals has to be that of the four-body integrals in \cref{eq:FourBody}. From \cref{eq:AngRiRj}, the results here have a $\cos\theta_{ij}$, which does not appear in the four-body integrals. To use these results, we have to use the law of cosines to replace any $\cos\theta_{ij}$ terms by
\begin{equation}
\label{eq:LawCosines}
\cos\theta_{ij} = \frac{r_i^2 + r_j^2 - r_{ij}^2}{2 r_i r_j}.
\end{equation}
This allows us to split these into multiple integrations with only polynomial $r_i$ and $r_{ij}$ terms with the decaying exponentials. Most of the other external angular integrations in this Appendix also end up with $\cos\theta_{ij}$, but those only apply to the long-range code, which does not have this restriction.

\section{Terms with \texorpdfstring{$Y_\ell^0(\theta_1, \varphi_1) Y_\ell^0(\theta_{\rho}, \varphi_{\rho})$}{Y1-Yrho} and \texorpdfstring{$Y_\ell^0(\theta_1, \varphi_1) Y_\ell^0(\theta_{\rho^\prime}, \varphi_{\rho^\prime})$}{Y1-Yrho'}}
\label{sec:AngR1Rho}

After doing these integrations by hand for the P-, D-, and F-wave (the S-wave is just $2\pi$), I realized that it is possible to generalize these. This can be seen in the ``Vector Gaussian Integration.cpp'' file of the general long-range integration code (\cref{chp:Programs}). If we define
\begin{equation}
w_1 = \frac{r_1 + r_2 \cos\theta_{12}}{2 \rho},
\end{equation}
then
\begin{equation}
\Int{Y_\ell^0(\theta_1, \varphi_1) Y_\ell^0(\theta_{\rho}, \varphi_{\rho})}{\tau_{ext}, \tau_{ext}} = 2 \pi \LegendreP{\ell, w_1},
\end{equation}
where $P_\ell$ is the standard Legendre polynomial. This easily lets us calculate these terms in the C++ code. Likewise, for
\begin{equation}
w_1^\prime = \frac{r_1 + r_3 \cos\theta_{13}}{2 \rho^\prime},
\end{equation}
then
\begin{equation}
\Int{Y_\ell^0(\theta_1, \varphi_1) Y_\ell^0(\theta_{\rho^\prime}, \varphi_{\rho^\prime})}{\tau_{ext}, \tau_{ext}} = 2 \pi \LegendreP{\ell, w_1^\prime}.
\end{equation}

\section{Terms with \texorpdfstring{$Y_\ell^0(\theta_2, \varphi_2) Y_\ell^0(\theta_{\rho}, \varphi_{\rho})$}{Y2-Yrho} and \texorpdfstring{$Y_\ell^0(\theta_3, \varphi_3) Y_\ell^0(\theta_{\rho^\prime}, \varphi_{\rho^\prime})$}{Y3-Yrho'}}
\label{sec:AngR2Rho}

Similar to \cref{sec:AngR1Rho}, these integrals are also generalizable and are used in the C++ file ``Vector Gaussian Integration.cpp''. Defining
\begin{equation}
w_2 = \frac{r_2 + r_1 \cos\theta_{12}}{2 \rho},
\end{equation}
then
\begin{equation}
\Int{Y_\ell^0(\theta_2, \varphi_2) Y_\ell^0(\theta_{\rho}, \varphi_{\rho})}{\tau_{ext}, \tau_{ext}} = 2 \pi \LegendreP{\ell, w_2}.
\end{equation}
Since $\rho^\prime$ is constructed from $r_3$, we can do the same method for the next set. We define
\begin{equation}
w_3 = \frac{r_3 + r_1 \cos\theta_{13}}{2 \rho^\prime},
\end{equation}
which gives
\begin{equation}
\Int{Y_\ell^0(\theta_3, \varphi_3) Y_\ell^0(\theta_{\rho^\prime}, \varphi_{\rho^\prime})}{\tau_{ext}, \tau_{ext}} = 2 \pi \LegendreP{\ell, w_3}.
\end{equation}

\section{Terms with \texorpdfstring{$Y_\ell^0(\theta_2, \varphi_2) Y_\ell^0(\theta_{\rho^\prime}, \varphi_{\rho^\prime})$}{Y2-Yrho'} and \texorpdfstring{$Y_\ell^0(\theta_3, \varphi_3) Y_\ell^0(\theta_{\rho}, \varphi_{\rho})$}{Y3-Yrho}}
\label{sec:AngR2Rhop}

This set is more complicated than those in \cref{sec:AngR2Rho} but look similar. We first define
\begin{equation}
w_4 = \frac{r_1 \cos\theta_{12} + r_3 \cos\theta_{23}}{2 \rho^\prime},
\end{equation}
and then
\begin{equation}
\Int{Y_\ell^0(\theta_2, \varphi_2) Y_\ell^0(\theta_{\rho^\prime}, \varphi_{\rho^\prime})}{\tau_{ext}, \tau_{ext}} = 2 \pi \LegendreP{\ell, w_4}.
\end{equation}
For the permuted version of this, we use
\begin{equation}
w_5 = \frac{r_1 \cos\theta_{13} + r_2 \cos\theta_{23}}{2 \rho},
\end{equation}
giving
\begin{equation}
\Int{Y_\ell^0(\theta_3, \varphi_3) Y_\ell^0(\theta_{\rho}, \varphi_{\rho})}{\tau_{ext}, \tau_{ext}} = 2 \pi \LegendreP{\ell, w_5}.
\end{equation}

\section{Terms with \texorpdfstring{$Y_\ell^0(\theta_\rho, \varphi_\rho) Y_\ell^0(\theta_{\rho^\prime}, \varphi_{\rho^\prime})$}{Yrho-Yrho'}}

These are the most difficult angular integrations considered. After 
performing these integrations by hand through the D-wave, I came up with a 
way to compute them in \emph{Mathematica}. These are derived in a
\emph{Mathematica} notebook entitled ``General Angular Integrations.nb''. Using 
this method, the H-wave integral ($\ell = 5$) took approximately 1 hour and 26
minutes to calculate in \emph{Mathematica} using an Intel Q6600 processor-based
desktop computer. 

I discovered a way to generalize these integrals as well, but it was not until recently. Without realizing this relation, these integrals are very difficult to compute. Defining
\begin{equation}
y = \frac{4\left(\rho^2 + {\rho^\prime}^2\right) - r_{23}^2}{8 \rho \rho^\prime},
\end{equation}
these integrals can be generalized to
\begin{equation}
\Int{Y_\ell^0(\theta_\rho, \varphi_\rho) Y_\ell^0(\theta_{\rho^\prime}, \varphi_{\rho^\prime})}{\tau_{ext}, \tau_{ext}} = 2 \pi \LegendreP{\ell, y}.
\end{equation}
The results of these integrals for the first six partial waves follow.

\begin{align}
\Int{&Y_0^0(\theta_\rho, \varphi_\rho) Y_0^0(\theta_{\rho^\prime}, \varphi_{\rho^\prime})}{\tau_{ext}, \tau_{ext}} = 2 \pi
\end{align}

\begin{align}
\Int{&Y_1^0(\theta_\rho, \varphi_\rho) Y_1^0(\theta_{\rho^\prime}, \varphi_{\rho^\prime})}{\tau_{ext}, \tau_{ext}} = \frac{\pi}{4 \rho \rho^\prime}  \left[4 \left(\rho^2+{\rho^\prime}^2\right)-r_{23}^2\right]
\end{align}

\begin{align}
\Int{Y_2^0(\theta_\rho, \varphi_\rho) & Y_2^0(\theta_{\rho^\prime}, \varphi_{\rho^\prime})}{\tau_{ext}, \tau_{ext}} = \frac{\pi}{64 \rho ^2 {\rho^\prime}^2}  \left[16 \left(3 \rho ^4+2 \rho ^2 {\rho^\prime}^2 +3 {\rho^\prime}^4\right)  \right. \nonumber \\
& \left. +3 r_{23}^4-24 r_{23}^2 \left(\rho ^2+{\rho^\prime}^2\right)\right]
\end{align}

\begin{align}
\Int{&Y_3^0(\theta_\rho, \varphi_\rho) Y_3^0(\theta_{\rho^\prime}, \varphi_{\rho^\prime})}{\tau_{ext}, \tau_{ext}} = \frac{\pi}{512 \rho ^3 {\rho^\prime}^3} \left[64 \left(5 \rho ^6+3 \rho ^4 {\rho^\prime}^2+3 \rho ^2 {\rho^\prime}^4+5 {\rho^\prime}^6\right) \right.  \nonumber \\
& \left. -5 r_{23}^6 +60 r_{23}^4 \left(\rho ^2+{\rho^\prime}^2\right)-48 r_{23}^2 \left(5 \rho ^4+6 \rho ^2 {\rho^\prime}^2+5 {\rho^\prime}^4\right)\right]
\end{align}

\begin{align}
\Int{&Y_4^0(\theta_\rho, \varphi_\rho) Y_4^0(\theta_{\rho^\prime}, \varphi_{\rho^\prime})}{\tau_{ext}, \tau_{ext}} = \frac{\pi}{16384 \rho^4 {\rho^\prime}^4}  \left[8960 {\rho^\prime}^8+1280 {\rho^\prime}^6 \left(4 \rho^2-7 r_{23}^2\right) \right.  \nonumber \\
& -80 {\rho^\prime}^2 \left(r_{23}^2-4 \rho ^2\right)^2 \left(7 r_{23}^2-4 \rho ^2\right)+35 \left(r_{23}^2-4 \rho ^2\right)^4 \nonumber \\
& \left. +96 {\rho^\prime}^4 \left(48 \rho ^4+35 r_{23}^4-120 \rho ^2 r_{23}^2\right)\right]
\end{align}

\begin{align}
\Int{&Y_5^0(\theta_\rho, \varphi_\rho) Y_5^0(\theta_{\rho^\prime}, \varphi_{\rho^\prime})}{\tau_{ext}, \tau_{ext}} = \frac{\pi}{131072 \rho ^5 {\rho^\prime}^5} \left[64512 {\rho^\prime}^{10}-8960 {\rho^\prime}^8 \left(9 r_{23}^2-4 \rho ^2\right) \right.  \nonumber \\
& + 140 {\rho^\prime}^2 \left(r_{23}^2-4 \rho ^2\right)^3 \left(9 r_{23}^2-4 \rho ^2\right)-63 \left(r_{23}^2-4 \rho ^2\right)^5  \nonumber \\
& + 1920 {\rho^\prime}^6 \left(16 \rho ^4+21 r_{23}^4-56 \rho^2 r_{23}^2\right)  \nonumber \\
& \left. -480 {\rho^\prime}^4 (r_{23}-2 \rho ) (2 \rho +r_{23}) \left(16 \rho ^4+21 r_{23}^4-56 \rho ^2 r_{23}^2\right)\right]
\end{align}


\chapter{Extra Derivations}
\label{chp:ExtraDer}

\iftoggle{UNT}{This}{\lettrine{\textcolor{startcolor}{T}}{his}}
Appendix mainly contains short derivations and equations that are not 
critical to understanding the main results but are nonetheless needed for 
this work.

\section{Spherical Functions}
\label{sec:SphericalFunc}

This section gives the spherical harmonics (\cref{tab:SphHarm}),
spherical Bessel functions (\cref{tab:SphBess}), and
spherical Neumann functions (\cref{tab:SphNeum})
through $\ell = 5$ for easier reference. These
were all obtained using the appropriate functions in \emph{Mathematica}
\cite{Mathematica}. Most sources \cite{Arfken2005,Abramowitz1965} give
these for $\ell \leq 2$.

{
\renewcommand{\arraystretch}{1.5}
\begin{table}
\centering
\begin{tabular}{l l}
\iftoggle{UNT}{\toprule\\[-0.9cm]}{\toprule\\[-1.2cm]}
Partial Wave & $\SphericalHarmonicY{\ell}{0}{\theta}{\phi}$ \\
\midrule
S-Wave & $\frac{1}{\sqrt{4\pi}}$ \\
P-Wave & $\sqrt{\frac{3}{4\pi}} \cos\theta$ \\
D-Wave & $\sqrt{\frac{5}{16\pi}} (3\cos^2\theta - 1)$ \\
F-Wave & $\sqrt{\frac{7}{16\pi}} \left(5 \cos^3\theta - 3 \cos\theta \right)$ \\
G-Wave & $\sqrt{\frac{9}{256\pi}} \left(35 \cos^4\theta - 30 \cos^2\theta + 3 \right)$ \\
H-Wave & $\sqrt{\frac{11}{256\pi}} \left(63 \cos^5\theta - 70 \cos^3\theta + 15 \cos\theta \right)$ \\
\bottomrule
\end{tabular}
\caption{Spherical harmonics for partial waves $\ell = 0$ through 5}
\label{tab:SphHarm}
\end{table}
}

{
\renewcommand{\arraystretch}{1.5}
\begin{table}
\centering
\begin{tabular}{l l}
\iftoggle{UNT}{\toprule\\[-0.9cm]}{\toprule\\[-1.2cm]}
Partial Wave & $j_\ell(z)$ \\
\midrule
S-Wave & $\frac{\sin(\kappa\rho)}{\kappa\rho}$ \\
P-Wave & $\frac{\sin z}{z^2} - \frac{\cos z}{z}$ \\
D-Wave & $\left(\frac{3}{z^3}-\frac{1}{z}\right)\sin z - \frac{3}{z^2}\cos z$ \\
F-Wave & $\frac{\left(z^2-15\right) \cos z}{z^3}-\frac{3 \left(2 z^2-5\right) \sin z}{z^4}$ \\
G-Wave & $\frac{5 \left(2 z^2-21\right) \cos z}{z^4}+\frac{\left(z^4-45 z^2+105\right) \sin z}{z^5}$ \\
H-Wave & $\frac{15 \left(z^4-28 z^2+63\right) \sin z}{z^6}+\frac{\left(-z^4+105 z^2-945\right) \cos z}{z^5}$ \\
\bottomrule
\end{tabular}
\caption{Spherical Bessel functions for partial waves $\ell = 0$ through 5}
\label{tab:SphBess}
\end{table}
}

{
\renewcommand{\arraystretch}{1.5}
\begin{table}
\centering
\begin{tabular}{l l}
\iftoggle{UNT}{\toprule\\[-0.9cm]}{\toprule\\[-1.2cm]}
Partial Wave & $n_\ell(z)$ \\
\midrule
S-Wave & $-\frac{\cos z}{z}$ \\
P-Wave & $-\frac{\cos z}{z^2} - \frac{\sin z}{z}$ \\
D-Wave & $-\left(\frac{3}{z^3}-\frac{1}{z}\right)\cos z - \frac{3}{z^2}\sin z$ \\
F-Wave & $\frac{3 \left(2 z^2-5\right) \cos z}{z^4}+\frac{\left(z^2-15\right) \sin z}{z^3}$ \\
G-Wave & $\frac{5 \left(2 z^2-21\right) \sin z}{z^4}+\frac{\left(-z^4+45 z^2-105\right) \cos z}{z^5}$ \\
H-Wave & $\frac{\left(-z^4+105 z^2-945\right) \sin z}{z^5}-\frac{15 \left(z^4-28 z^2+63\right) \cos z}{z^6}$ \\
\bottomrule
\end{tabular}
\caption{Spherical Neumann functions for partial waves $\ell = 0$ through 5}
\label{tab:SphNeum}
\end{table}
}

\section{\texorpdfstring{$\rho$ and $\rho'$} {rho and rho'} Definitions}
\label{sec:RhoDef}
From equation 2.1 of Peter Van Reeth's thesis \cite{VanReethThesis} and Armour and Humberston's paper \cite{Armour1991},

\begin{equation}
\bm{\rho} = \frac{1}{2} \left( \bm{r}_1 + \bm{r}_2 \right).
\label{eq:RhoDef1}
\end{equation}
By switching coordinates 2 and 3, we have
\begin{equation}
\bm{\rho}' = \frac{1}{2} \left( \bm{r}_1 + \bm{r}_3 \right).
\label{eq:RhoDef2}
\end{equation}
In the original coordinate system (\cref{fig:CoordinateSystemOriginal}),
\begin{align}
\nonumber \bm{r}_1 &= \left( r_1 \sin \theta_1 \cos \varphi_1, r_1 \sin \theta_1 \sin \varphi_1, r_1 \cos \theta_1 \right) \\
\bm{r}_3 &= \left( r_3 \sin \theta_3 \cos \varphi_3, r_3 \sin \theta_3 \sin \varphi_3, r_3 \cos \theta_3 \right)
\end{align}
In the rotated coordinate system (\cref{fig:CoordinateSystemRotated}),
\begin{align}
\nonumber \bm{r}_1 &= (0, 0, r_1) \\
\bm{r}_3 &= \left( r_3 \sin \theta_{13} \cos \varphi_{13}, r_3 \sin \theta_{13} \sin \varphi_{13}, r_3 \cos \theta_{13} \right)
\end{align}
\begin{align}
\nonumber \left| r_1 + r_3 \right|^2 &= r_3^2 \sin^2 \theta_{13} \cos^2 \varphi_{13} + r_3^2 \sin^2 \theta_{13} \sin^2 \varphi_{13} + (r_1 + r_3 \cos \theta_{13})^2\\
\nonumber &= r_3^2 \sin^2 \theta_{13} + r_1^2 + r_3^2 \cos^2 \theta_{13} + 2 r_1 r_3 \cos \theta_{13} \\
&= r_1^2 + r_3^2 + 2 r_1 r_3 \cos \theta_{13}.
\label{eq:RhoDef3}
\end{align}

Also, using the law of cosines,
\begin{equation}
r_{13}^2 = r_1^2 + r_3^2 - 2 r_1 r_3 \cos \theta_{13}
\label{eq:CosLaw}
\end{equation}
Substituting \cref{eq:CosLaw} into \cref{eq:RhoDef3} gives
\begin{equation}
\left| r_1 + r_3 \right|^2 = 2 \left( r_1^2 + r_3^2 \right) - r_{13}^2.
\label{eq:RhoDef4}
\end{equation}
From \cref{eq:RhoDef2,eq:RhoDef4},
\begin{equation}
\label{eq:RhopRDef}
\rho' = \frac{1}{2} \left[ 2 \left(r_1^2 + r_3^2 \right) - r_{13}^2 \right] ^ \frac{1}{2}.
\end{equation}
Similarly,
\begin{equation}
\label{eq:RhoRDef}
\rho = \frac{1}{2} \left[ 2 \left(r_1^2 + r_2^2 \right) - r_{12}^2 \right] ^ \frac{1}{2}.
\end{equation}

\section{Perimetric Coordinates}
\label{sec:PerimetricCoords}

Perimetric coordinates are used for the long-long integrations in the S-wave
code. If perimetric coordinates are used for $r_1$, $r_2$ and $r_{12}$, then
these are defined by \cite{Armour1991}

\begin{align}
\label{eq:PerimetricCoords1}
\nonumber x &= r_1 + r_2 - r_{12} \\
\nonumber y &= r_2 + r_{12} - r_1 \\
z &= r_{12} + r_1 - r_2.
\end{align}
These can alternately be written as
\begin{align}
\label{eq:PerimetricCoords2}
\nonumber r_1 &= \frac{x+z}{2} \\
\nonumber r_2 &= \frac{x+y}{2} \\
r_{12} &= \frac{y+z}{2}.
\end{align}

From \cref{eq:dTauS23}, the volume element after integration over the external angles is
\begin{equation}
d\tau = 8\pi^2 dr_1 r_2 dr_2 r_3 dr_3 r_{12} dr_{12} r_{13} dr_{13} d\varphi_{23}.
\end{equation}
We need to perform a change of variables to use perimetric coordinates for $r_1$, $r_2$ and $r_{12}$. The Jacobian is
\begin{equation}
\label{eq:PerimetricJacobian}
J(x,y,z) = 
\left| {\begin{array}{ccc}
 \frac{\partial r_1}{\partial x} & \frac{\partial r_1}{\partial y} & \frac{\partial r_1}{\partial z}  \\
 \frac{\partial r_2}{\partial x} & \frac{\partial r_2}{\partial y} & \frac{\partial r_2}{\partial z}  \\
 \frac{\partial r_{12}}{\partial x} & \frac{\partial r_{12}}{\partial y} & \frac{\partial r_{12}}{\partial z}  \\
 \end{array} } \right|
=
\left| {\begin{array}{ccc}
 \frac{1}{2} & 0 & \frac{1}{2} \\
 \frac{1}{2} & \frac{1}{2} & 0 \\
 0 & \frac{1}{2} & \frac{1}{2}
 \end{array} } \right|
=
\frac{1}{4}.
\end{equation}

\noindent This gives a transformed volume element of
\begin{equation}
\label{eq:PerimetricVolEl}
d\tau = 2\pi^2 r_2 r_3 r_{12} r_{13} dx\, dy\, dz\, dr_3\, dr_{13}\, d\varphi_{23}.
\end{equation}

\noindent The limits for each of the perimetric coordinates are 0 to $\infty$.

\section{Spherical Bessel Derivatives}
\label{sec:SphBess}

\subsection{First Derivative}
\label{sec:SphBess1}

From Abramowitz and Stegun \cite[p.437]{Abramowitz1965},
\begin{subequations}
\begin{align}
j_\ell(z) &= z^{-1} \left[ P(\ell+\tfrac{1}{2}, z ) \sin(z-\tfrac{1}{2}\ell\pi) + Q(\ell+\tfrac{1}{2},z) \cos(z-\tfrac{1}{2}\ell\pi) \right] \\
n_\ell(z) &= (-1)^{\ell+1} z^{-1} \left[ P(\ell+\tfrac{1}{2}, z ) \cos(z+\tfrac{1}{2}\ell\pi) - Q(\ell+\tfrac{1}{2},z) \sin(z+\tfrac{1}{2}\ell\pi) \right].
\end{align}
\end{subequations}
where
\begin{subequations}
\begin{align}
P(\ell+\tfrac{1}{2}, z) &= 1 - \frac{\Factorial{\ell+2}}{\Factorial{2} \GammaFunc{\ell-1}} (2z)^{-2} + \ldots \\
Q(\ell+\tfrac{1}{2}, z) &= \frac{\Factorial{\ell+1}}{\Factorial{1} \GammaFunc{\ell}} (2z)^{-1} - \frac{\Factorial{n+3}}{\Factorial{3} \GammaFunc{\ell-2}} (2z)^{-3} + \ldots
\end{align}
\end{subequations}
Using \emph{Mathematica} with these expansions, we get
\begin{subequations}
\begin{align}
\label{eq:SphBesExpan}
j_\ell(z) &= \frac{\sin(z - \tfrac{1}{2}\ell\pi)}{z} + \ldots \\
n_\ell(z) &= \frac{(-1)^{\ell+1} \cos(z + \tfrac{1}{2}\ell\pi)}{z} + \ldots
\end{align}
\end{subequations}
and
\begin{subequations}
\begin{align}
\label{eq:SphBesExpanDer}
{j_\ell}^\prime(z) &= \frac{\cos(z - \tfrac{1}{2}\ell\pi)}{z} + \ldots \\
{n_\ell}^\prime(z) &= \frac{(-1)^{\ell+2} \sin(z + \tfrac{1}{2}\ell\pi)}{z} + \ldots
\end{align}
\end{subequations}

In a more general format than that of Abramowitz and Stegun \cite[p.73]{Abramowitz1965},
\begin{equation}
\cos \left(z-\frac{\ell \pi}{2} \right) = \begin{cases} (-1)^{\ell/2} \cos z & \mbox{if } \ell \mbox{ is even} \\ 
(-1)^{(\ell-1)/2} \sin z & \mbox{if } \ell \mbox{ is odd} \end{cases} 
\end{equation}
and
\begin{equation}
\sin \left(z-\frac{\ell \pi}{2} \right) = \begin{cases} (-1)^{\ell/2} \sin z & \mbox{if } \ell \mbox{ is even} \\ 
(-1)^{(\ell+1)/2} \cos z & \mbox{if } \ell \mbox{ is odd} \end{cases} .
\end{equation}
Also,
\begin{equation}
\cos \left(z+\frac{\ell \pi}{2} \right) = \begin{cases} (-1)^{\ell/2} \cos z & \mbox{if } \ell \mbox{ is even} \\ 
(-1)^{(\ell+1)/2} \sin z & \mbox{if } \ell \mbox{ is odd} \end{cases} 
\end{equation}
and
\begin{equation}
\sin \left(z+\frac{\ell \pi}{2} \right) = \begin{cases} (-1)^{\ell/2} \sin z & \mbox{if } \ell \mbox{ is even} \\ 
(-1)^{(\ell-1)/2} \cos z & \mbox{if } \ell \mbox{ is odd} \end{cases} .
\end{equation}
Using these with \cref{eq:SphBesExpan,eq:SphBesExpanDer}, we see that to first order, there is a relationship between these functions and their derivatives given by
\begin{subequations}
\begin{align}
\label{eq:SphBesDerRel}
{j_\ell}^\prime(z) &\approx -n_\ell(z) \\
{n_\ell}^\prime(z) &\approx j_\ell(z).
\end{align}
\end{subequations}
This allows us to write the gradient of $\widetilde{S}_\ell$ and
$\widetilde{C}_\ell$ for arbitrary $\ell$ to first order in \cref{eq:GradSC}.

\subsection{Second Derivative}
\label{sec:SphBess2}

The output we get from the \emph{Mathematica} notebook ``First Partial Waves LS\\ General.nb'' \cite{GitHub,Wiki} is
\begin{align}
\label{eq:LSGenLapl1}
\frac{\Laplacian_\rho \left[\SphericalHarmonicY{\ell}{0}{\theta_\rho}{\varphi_\rho} j_\ell(\kappa\rho) \right]}{\SphericalHarmonicY{\ell}{0}{\theta_\rho}{\varphi_\rho} j_\ell(\kappa\rho)} = \frac{(n + n^2 - \kappa^2 \rho^2) \LegendreP{\ell, \cos\theta} + 2 \cot\theta \, \AssocLegendreP{\ell}{1}{\cos\theta} + \AssocLegendreP{\ell}{2}{\cos\theta}} {\rho^2 \LegendreP{\ell, \cos\theta}}.
\end{align}
From Ref.~\cite{WolframPnm}, a recurrence relation for the associated Legendre polynomials is
\begin{equation}
\AssocLegendreP{\ell}{\mu+1}{z} - \left[\mu(\mu-1) - \ell(\ell+1)\right] \AssocLegendreP{\ell}{\mu-1}{z} + \frac{2 \mu z}{\sqrt{1-z^2}} \AssocLegendreP{\ell}{\mu}{z} = 0.
\end{equation}
Note that other books \cite{Abramowitz1965,Zwillinger2003} give slightly different forms of this recurrence relation. If we set $\mu = 1$ and $z = \cos\theta$, this becomes
\begin{equation}
\AssocLegendreP{\ell}{2}{\cos\theta} + \left(\ell^2+\ell \right) \AssocLegendreP{\ell}{0}{z} + \frac{2 \cos\theta}{\sin\theta} \AssocLegendreP{\ell}{1}{\cos\theta} = 0.
\end{equation}
Using the definition of $\cot$ and solving for $\left(\ell^2+\ell \right)\AssocLegendreP{\ell}{0}{z} = \left(\ell^2+\ell \right)\LegendreP{\ell, \cos\theta}$,
\begin{equation}
\left(\ell^2+\ell \right)\LegendreP{\ell, \cos\theta} = -\AssocLegendreP{\ell}{2}{\cos\theta} - 2 \cot\theta \, \AssocLegendreP{\ell}{1}{\cos\theta}.
\end{equation}
Substituting this back into \cref{eq:LSGenLapl1}, we get cancellations with the associated Legendre polynomials and see that $\SphericalHarmonicY{\ell}{0}{\theta_\rho}{\varphi_\rho} j_\ell(\kappa\rho)$ is an eigenfunction of $\Laplacian_\rho$ with eigenvalue $-\kappa^2$:
\begin{align}
\label{eq:LSGenLapl2}
\Laplacian_\rho \left[\SphericalHarmonicY{\ell}{0}{\theta_\rho}{\varphi_\rho} j_\ell(\kappa\rho) \right] = \frac{(-\kappa^2 \rho^2) \LegendreP{\ell, \cos\theta} } {\rho^2 \LegendreP{\ell, \cos\theta}}
= -\kappa^2 \, \SphericalHarmonicY{\ell}{0}{\theta_\rho}{\varphi_\rho} j_\ell(\kappa\rho).
\end{align}

\section{\texorpdfstring{$_2F_1$}{Hypergeometric} Recursion Relation}
\label{sec:Hypergeometric}

The backwards recursion relation for the hypergeometric function is used in the
short-range code (\cref{sec:CompShort}). This is given in Refs.~\cite{Drake1995,Frolov2003} as
\begin{equation}
\label{eq:HyperIdentity}
\Hypergeometric{2}{1}{1,a}{c}{z} = 1 + \left( \frac{a}{c} \right) z \,\, \Hypergeometric{2}{1}{1,a+1}{c+1}{z}.
\end{equation}

From Abramowitz and Stegun \cite{Abramowitz1965}, the definition of the hypergeometric function is given by
\begin{equation}
\label{eq:HyperDef}
_2F_1(\alpha,\beta;\gamma;z) = 1 + \sum_{n=1}^{\infty} \frac{(\alpha)_n \cdot (\beta)_n}{(\gamma)_n} \frac{z^n}{n!},
\end{equation}
where $(x)_n$ is the Pochhammer symbol given by \cite{Abramowitz1965}
\begin{equation}
\label{eq:PochhammerDef}
(x)_n \equiv \frac{\Gamma(x+n)}{\Gamma(x)} = x(x+1) \cdots  (x+n-1)
\end{equation}
with $(x)_0 = 1$.  A special case is $(x)_1 = n!$, where $n!$ is the factorial.

Using the above definition of the Pochhammer symbol, we can easily see that
\begin{equation}
\label{eq:PochPlus1}
(x+1)_n = \frac{(x+n)}{(x)} \cdot (x)_n = \frac{(x)_{n+1}}{x}.
\end{equation}
From \cref{eq:HyperDef,eq:PochPlus1}, 
\begin{align}
\nonumber _2F_1(1,a+1;c+1;z) &= 1 + \sum_{n=1}^{\infty} \frac{(1)_n \cdot (a+1)_n}{(c+1)_n} \frac{z^n}{n!} = 1 + \sum_{n=1}^{\infty} \frac{(a+1)_n}{(c+1)_n} z^n \\
&= 1 + \left(\frac{c}{a}\right) \sum_{n=1}^{\infty} \frac{(a)_{n+1}}{(c)_{n+1}} z^n = 1 + \left(\frac{c}{a}\right) \sum_{n=2}^{\infty} \frac{(a)_n}{(c)_n} z^{n-1}.
\end{align}
Multiplying by $\left(\frac{c}{a}\right) z$, we now have
\begin{equation}
\label{eq:Hyper1}
\left(\frac{c}{a}\right) z \,\, _2F_1(1,a+1;c+1;z) = \left(\frac{c}{a}\right) z + \sum_{n=2}^{\infty} \frac{(a)_n}{(c)_n} z^n = \sum_{n=1}^{\infty} \frac{(a)_n}{(c)_n} z^n.
\end{equation}
From the definition in \cref{eq:HyperDef},
\begin{equation}
\label{eq:Hyper2}
_2F_1(1,a;c;z) = 1 + \sum_{n=1}^{\infty} \frac{(a)_n}{(c)_n} z^n,
\end{equation}
which, when combined with \cref{eq:Hyper1}, gives the final result of
\begin{equation}
_2F_1(1,a;c;z) = 1 + \left( \frac{a}{c} \right) z \,\,_2F_1(1,a+1;c+1;z).
\end{equation}

\section{Shielding Function}
\label{sec:ShieldingFunc}

The spherical Neumann functions, $n_\ell(\kappa\rho)$, in $C_\ell$ go to
$-\infty$ at the origin, and we remove this singularity with the shielding
function. The shielding function given by \cref{eq:PartialWaveShielding},
$f_\ell(\rho) = \left[1 - \ee^{-\mu \rho} \left(1+\frac{\mu}{2}\rho\right)\right]^{m_\ell}$,
is slightly different than earlier work \cite{VanReeth2003,VanReeth2004} on
Ps-H scattering, which used $f(\rho) = (1 - \ee^{-\lambda \rho})^3$.

This work is based on notes from Van Reeth \cite{VanReethPrivate}\
for the S-wave shielding function.
We want to have $C_\ell$ behaving similar to $S_\ell$ at the origin. To
accomplish this, we take a series expansion of both to see what the dominant
terms are. To more easily see the behavior, we only take expansions of the
spherical Bessel and Neumann functions for the S-wave. For $S_0$, the series expansion of
$j_0$ is
\begin{equation}
j_0(\kappa\rho) \sim 1 - \frac{\kappa^2 \rho^2}{6} + \frac{\kappa^4 \rho^4}{120} + \mathcal{O}(\rho^5).
\end{equation}
For $C_0$, the series expansion of $n_0$ is
\begin{equation}
n_0(\kappa\rho) \sim -\frac{1}{\kappa\rho} + \frac{\kappa \rho}{2} - \frac{\kappa^3 \rho^3}{24} + \mathcal{O}(\rho^4).
\end{equation}
The singular nature at the origin is easily seen by the first term. To have
$C_0 \sim S_0$ at the origin, the shielding function needs to change the
leading term to a constant. If we choose a shielding function of the form
$\left[1 - \ee^{-\mu \rho} \left(1+a \rho\right)\right]$, the
expansion at the origin of this multiplied by the spherical Neumann function is
\begin{equation}
n_0(\kappa\rho) \left[1 - \ee^{-\mu \rho} \left(1+a \rho\right)\right] \sim
-\frac{\mu-a}{\kappa} - \frac{\left(a\mu - \frac{\mu^2}{2}\right)\rho}{\kappa}
+ \left(\frac{1}{2}\kappa(\mu-a) - \frac{\mu^2 - 3 a \mu^2}{6\kappa}\right) \rho^2 + \ldots
\end{equation}
This is no longer singular, but it has a $\rho$ term, so if we set
$a = \frac{\mu}{2}$, the second term disappears, leaving us with
\begin{equation}
n_0(\kappa\rho) \left[1 - \ee^{-\mu \rho} \left(1+\frac{\mu}{2}\rho\right)\right] \sim
-\frac{\mu}{2\kappa} + \left(\frac{\kappa\mu}{4} + \frac{\mu^3}{12\kappa}\right) \rho^2 + \ldots \ .
\end{equation}
This shows that with the choice of shielding function in \cref{eq:PartialWaveShielding},
we have $C_0$ behaving similar to $S_0$ at the origin.

The \emph{Mathematica} notebook ``Shielding Factor.nb'' found on the GitHub
page \cite{GitHub} shows these expansions and the expansions for the P-wave
and D-wave. We normally choose $m_\ell = (2\ell+1)$, but for the D-wave, we
used $m_\ell = 7$ when we were trying to improve the convergence, which
ultimately did not give improved results over $m_\ell = 5$. This notebook
also shows the first and second derivatives of the shielding function given
in \cref{eq:Shielding1Der,eq:Shielding2Der}.

\begin{figure}
	\centering
	\includegraphics[width=5in]{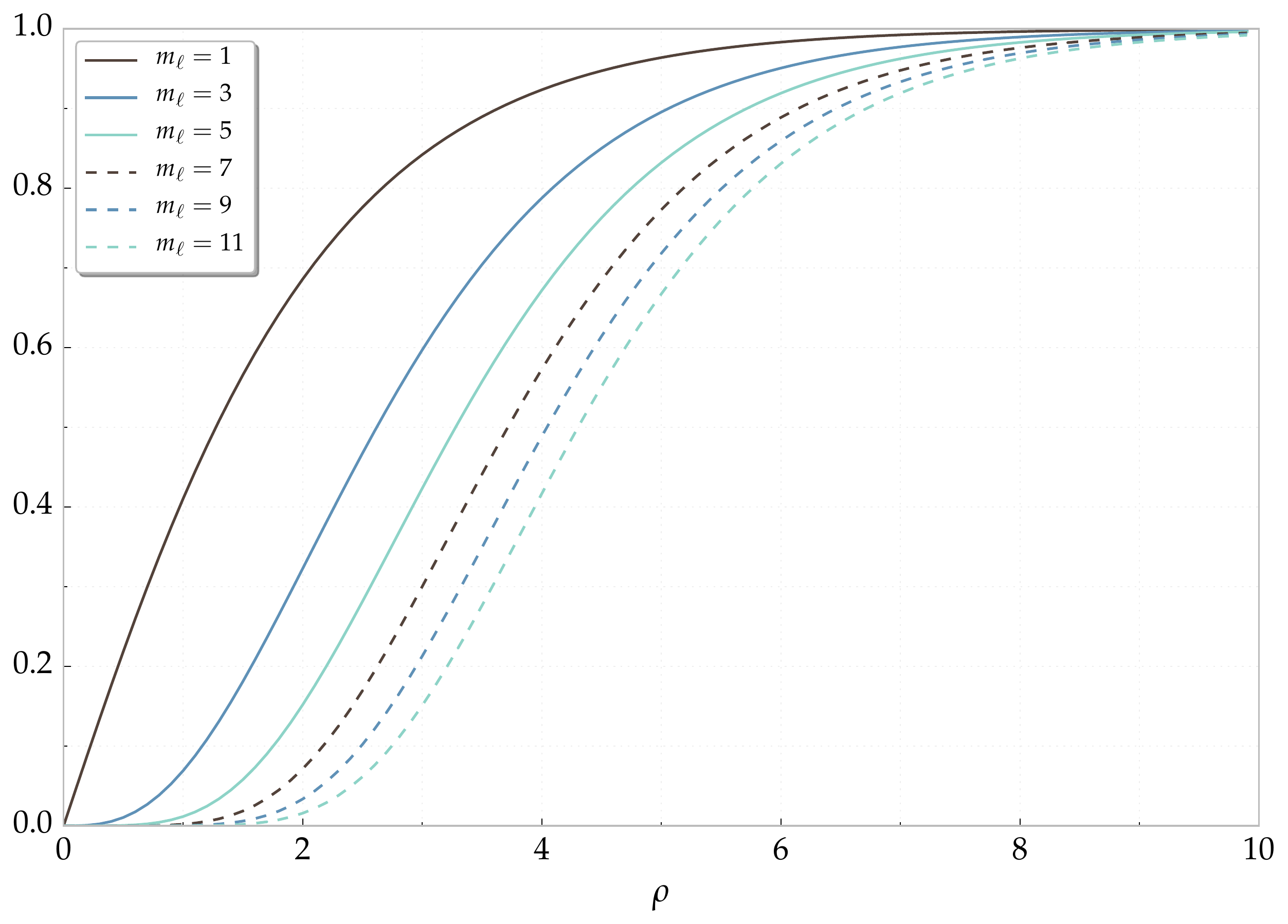}
	\caption[Shielding function $f_\ell$ variation with respect to $\rho$ for multiple values of $m_\ell$]{Shielding function $f_\ell$ variation with respect to $\rho$ for multiple values of $m_\ell$ with $\mu = 0.9$}
	\label{fig:shielding-func-n}
\end{figure}

\begin{figure}
	\centering
	\includegraphics[width=5in]{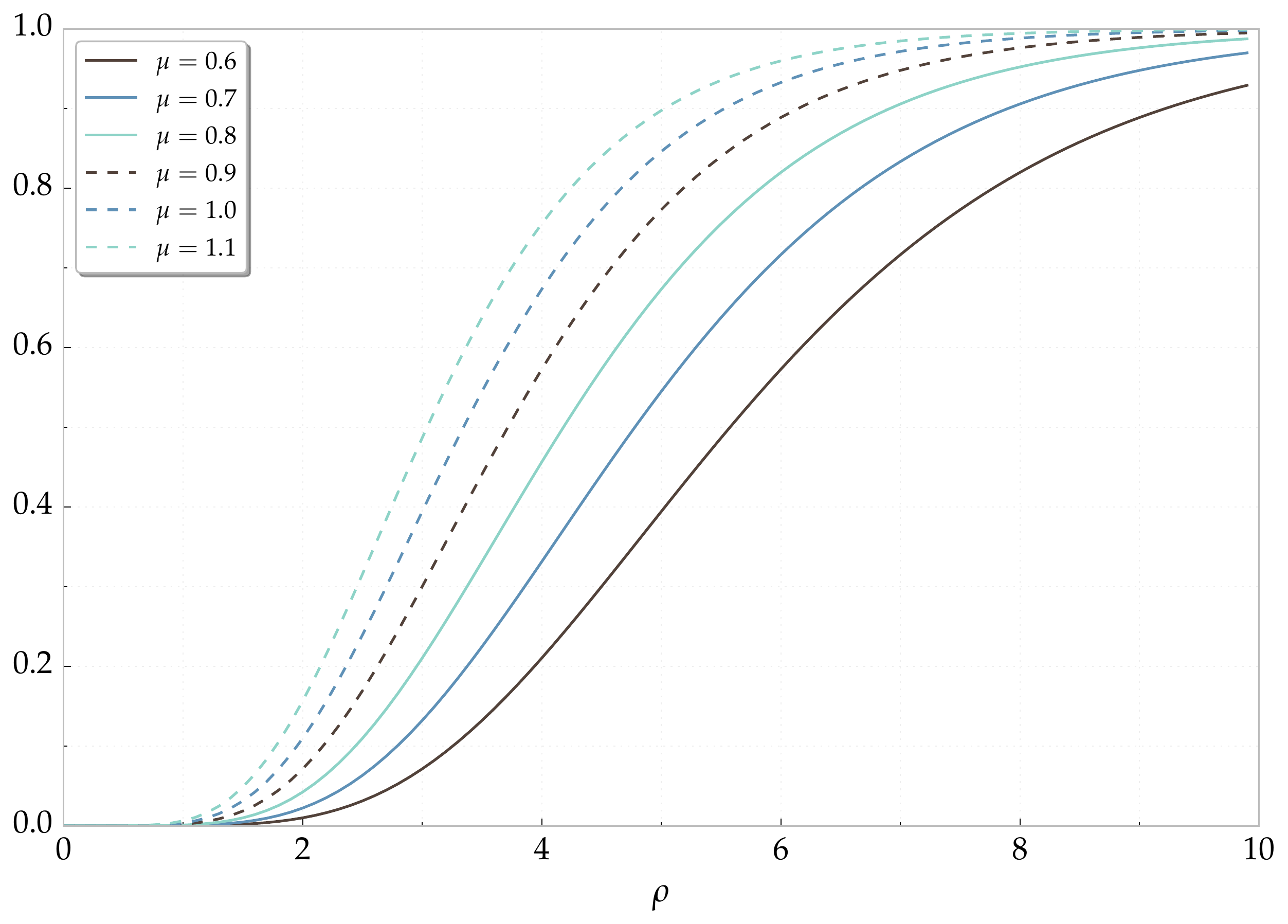}
	\caption[Shielding function $f_\ell$ variation with respect to $\rho$ for multiple values of $\mu$]{Shielding function $f_\ell$ variation with respect to $\rho$ for multiple values of $\mu$ with $m_\ell = 7$}
	\label{fig:shielding-func-mu}
\end{figure}

This notebook also has interactive graphs that show how the shielding function
with and without the spherical Neumann function behaves with differing $m_\ell$,
$\mu$, and $\ell$ values. \Cref{fig:shielding-func-n} shows that as $m_\ell$
increases for fixed $\mu$ of 0.9, it takes a larger $\rho$ before $C_\ell$
becomes significant. \Cref{fig:shielding-func-mu} keeps $m_\ell$ constant at 7
and varies $\mu$. This figure shows that smaller $\mu$ values give a strong
contribution for $C_\ell$. \Cref{tab:Nonlinear} shows the $\mu$ and $m_\ell$
values we use for each partial wave.

\section{D-Wave Mixed Symmetry Terms}
\label{sec:MixedDerivation}

This derivation proves \cref{eq:MixedAngSimple}. Using \cref{eq:MixedAng} and
substituting the appropriate spherical harmonics and Clebsch-Gordan coefficients,
\begin{align}
\label{eq:Psi1}
\psi_{(1,1,2,0)} = &\sum_{m=-1}^{+1} Y_{1,m}(\theta_1,\varphi_1) Y_{1,m}(\theta_2,\varphi_2) \left< 1,m; 1,-m,0 | 2,0 \right> \nonumber \\
	= &-\sqrt{\frac{3}{8\uppi}} \sin\theta_1 \ee^{-\ii \varphi_1} \sqrt{\frac{3}{8\uppi}} \sin\theta_2 \ee^{\ii \varphi_1} \frac{1}{\sqrt{6}} \nonumber \\
	& + \sqrt{\frac{3}{4\uppi}} \cos\theta_1 \sqrt{\frac{3}{4\uppi}} \sin\theta_2 \frac{2}{\sqrt{6}} \nonumber \\
	& -\sqrt{\frac{3}{8\uppi}} \sin\theta_1 \ee^{\ii \varphi_1} \sqrt{\frac{3}{8\uppi}} \sin\theta_2 \ee^{-\ii \varphi_1} \frac{1}{\sqrt{6}}.
\end{align}
Using \cite[p.192]{VanReethThesis}
\begin{equation}
\cos\theta_{12} = \sin\theta_1 \sin\theta_2 \cos(\varphi_1 - \varphi_2) + \cos\theta_1 \cos\theta_2
\end{equation}
in \cref{eq:Psi1}, we obtain \cref{eq:MixedAngSimple}:
\begin{align}
\label{eq:Psi2}
\psi_{(1,1,2,0)} &= -\frac{3}{8\uppi} \frac{1}{\sqrt{6}} \sin\theta_1 \sin\theta_2 \cdot 2 \cos(\varphi_1 - \varphi_2) + \frac{3}{4\uppi} \frac{1}{\sqrt{6}} \cdot 2 \cos\theta_1 \cos\theta_2 \nonumber \\
&= \frac{3}{4\uppi} \frac{1}{\sqrt{6}} \left(3 \cos\theta_1 \cos\theta_2 - \cos\theta_{12} \right).
\end{align}

\section{Miscellaneous}
\label{sec:Misc}

The cosine factors present in many of the matrix element equations are easily 
expressed in terms of $r_i$ and $r_{ij}$ by using the law of cosines
\cite[p.174]{CRC1978}:
\begin{equation}
\label{eq:Cosines}
\cos\theta_{12} = \frac{r_1^2 + r_2^2 - r_{12}^2}{2 r_1 r_2}, \ \ \ \ \cos\theta_{13} = \frac{r_1^2 + r_3^2 - r_{13}^2}{2 r_1 r_3} \ \ \ \text{and}  \ \ \cos\theta_{23} = \frac{r_2^2 + r_3^2 - r_{23}^2}{2 r_2 r_3}.
\end{equation}
This allows us to express all short-short matrix elements in the form needed by
the short-short methods described in \cref{sec:CompShort}.


\chapter{Extra Numerics}
\label{chp:ExtraNumerics}

\iftoggle{UNT}{This}{\lettrine{\textcolor{startcolor}{T}}{his}}
Appendix gives more
details than what is provided in \cref{chp:Computation} on
computations.

\section{Short-Range Code Nonlinear Parameter Optimization}
\label{sec:BoundOptimization}

A powerful property of the Rayleigh-Ritz variational method is the 
ability to systematically improve the wavefunction to lower the upper bound 
on the energy. By either adding terms to the expansion in
\cref{eq:BoundWavefn_psi} or changing the nonlinear parameters $\alpha$,
$\beta$ and $\gamma$, the energy can be reduced and a possible minimum
found. This is a three-dimensional optimization problem, and we tried 
multiple methods for the nonlinear parameter optimization.

\subsection{Broyden's Method}
\label{sec:Broyden}

We used Broyden's method \cite{Sauer2006}, which can 
solve for all three nonlinear parameters simultaneously. This was more stable 
than the 1-D Newton method. The second Broyden's method, sometimes referred 
to as the ``bad Broyden's method'' was used here. As Kvaalen points out, this 
method is perfectly usable and can be faster than the first Broyden's method 
\cite{Kvaalen1991}. \Cref{tab:NonlinearOptimized3SBroyden,tab:BroydenPWaveSingOpt}
show the nonlinear parameters used for $^3$S and $^1$P. All other partial waves
used the simplex method described in \cref{sec:Simplex}. This is because I
already had a number of results for $^3$S and $^1$P, so running everything
again for simplex-optimized nonlinear parameters was unnecessary. For $^3$S,
we were able to use even more terms than we could use for $^1$S \cref{tab:Nonlinear},
and we were able to use the same number of terms for $^1$P and $^3$P.

\setlength{\abovecaptionskip}{6pt}   
\setlength{\belowcaptionskip}{6pt}   
\begin{table}
\centering
\begin{tabular}{c c c c}
\toprule
$\omega$ & $\alpha$ & $\beta$ & $\gamma$ \\ [0.5ex]
\midrule
1 & 0.264440 & 0.831645 & 0.498871 \\
2 & 0.356175 & 0.452426 & 0.829591 \\
3 & 0.347611 & 0.467298 & 0.814971 \\
4 & 0.323300 & 0.333783 & 0.974653 \\
\bottomrule
\end{tabular}
\caption{Broyden optimized $^3$S nonlinear parameters}
\label{tab:NonlinearOptimized3SBroyden}
\end{table}

\begin{table}
\centering
\begin{tabular}{c c c c}
\toprule
$\omega$ & $\alpha$ & $\beta$ & $\gamma$ \\ [0.5ex]
\midrule
1 & 0.47767 & 0.50273 & 0.97498 \\
2 & 0.48253 & 0.49342 & 0.96874 \\
3 & 0.42803 & 0.43099 & 0.98993 \\
4 & 0.39740 & 0.37617 & 0.96205 \\
\bottomrule
\end{tabular}
\caption{Broyden optimized $^1$P nonlinear parameters}
\label{tab:BroydenPWaveSingOpt}
\end{table}

\subsection{Simplex Method}
\label{sec:Simplex}

\begin{table}
\centering
\begin{tabular}{c c c c}
\toprule
$\omega$ & $\alpha$ & $\beta$ & $\gamma$ \\ [0.5ex]
\midrule
0 & 0.30226 & 0.45479 & 1.07962 \\
1 & 0.53592 & 0.59453 & 1.02206 \\
2 & 0.57450 & 0.65222 & 0.98020 \\
3 & 0.58966 & 0.63150 & 0.97397 \\
4 & 0.58493 & 0.60995 & 0.98610 \\
5 & 0.58691 & 0.58045 & 1.03321 \\
\bottomrule
\end{tabular}
\caption{Simplex optimized $^1$S nonlinear parameters}
\label{tab:NonlinearOptimized1SSimplex}
\end{table}

Broyden's method was more stable than Newton's method for this work, but I 
also tried the \texttt{gsl\_multimin\_fminimizer\_nmsimplex} routine from the 
GNU Scientific Library, which is an implementation of the simplex method
\cite{GSL,GSLsimplex}. This was the most stable of the three methods tried to 
optimize $\alpha$, $\beta$, and $\gamma$ simultaneously. I normally stopped 
at $\omega = 5$ for the optimization, and the S-wave singlet runs are shown 
in \cref{tab:NonlinearOptimized1SSimplex}. \Cref{tab:NonlinearOptimizedPD} 
has the optimized nonlinear parameters used for the P-wave and D-wave. Due to 
the slowness of the general short-range code (see \cref{sec:GeneralShort}), 
the F-wave through G-wave just use the parameters $\alpha = 0.5$, $\beta = 0.6$,
and $\gamma = 1.1$.

\begin{table}
\small
\centering
\begin{tabular}{c c c c}
\toprule
Partial Wave & $\alpha$ & $\beta$ & $\gamma$ \\
\midrule
$^3$P & 0.310 & 0.311 & 0.995 \\
$^1$D & 0.359 & 0.368 & 0.976 \\
$^3$D & 0.356 & 0.365 & 0.976 \\
\bottomrule
\end{tabular}
\caption{Simplex optimized nonlinear parameters for the P-wave and D-wave}
\label{tab:NonlinearOptimizedPD}
\end{table}

With the work on the second formalism for the P-wave \cref{sec:PWaveOpt}, it 
was also possible to use the simplex method to optimize all 6 nonlinear 
parameters simultaneously instead of having to optimize each symmetry 
separately.

\subsection{P-Wave Nonlinear Parameter Optimization}
\label{sec:PWaveOpt}

Using the simplex method described in \cref{sec:Simplex}, we optimized the nonlinear parameters for both the first and second formalisms. \Cref{tab:SimplexPWaveSingOpt,tab:SimplexPWaveTripOpt} show the results of these optimizations in the second and third columns. We also let each symmetry have its own set of nonlinear parameters for each symmetry in the fourth and fifth columns. This was inspired by the work of Yan and Ho \cite{Yan1999}, where they used 5 different sets of nonlinear parameters to calculate the PsH ground state energy.

\begin{table}
\footnotesize
\centering
\begin{tabular}{c c c c c}
\toprule
\toprule
$\omega$ & 1st formalism / 1 set & 2nd formalism / 1 set & 1st formalism / 2 sets & 2nd formalism / 2 sets \\
\midrule
\midrule
 &  &  & 0.5341, 0.4536, 1.0139 & 0.5668, 0.4686, 0.9787 \\
1 & 0.4776, 0.5027, 0.9749 & 0.4571, 0.5700, 0.9266 & 0.3792, 0.5455, 0.9816 & 0.6777, 0.8587, 0.4813 \\
 & $\textbf{-0.666819968640}$ & $\textbf{-0.663226610680}$ & $\textbf{-0.670015702237}$ & $\textbf{-0.665355147531}$ \\
\midrule
 &  &  & 0.5270, 0.4394, 1.0036 & 0.4497, 0.5039, 0.9459 \\
2 & 0.4825, 0.4934, 0.9687 & 0.4734, 0.5162, 0.9584 & 0.4087, 0.5213, 0.9704 & 0.3963, 1.0233, 0.4327 \\
 & $\textbf{-0.700681070987}$ & $\textbf{-0.699190666285}$ & $\textbf{-0.701936448530}$ & $\textbf{-0.700245066225}$ \\
\midrule
 &  &  & 0.4632, 0.3918, 1.001 & 0.4653, 0.4512, 0.9905 \\
3 & 0.4297, 0.4337, 0.9808 & 0.4317, 0.4564, 0.9621 & 0.3844, 0.4620, 0.9740 & 0.8745, 0.9796, 0.4957 \\
 & $\textbf{-0.718093418924}$ & $\textbf{-0.717282613790}$ & $\textbf{-0.718548496648}$ & $\textbf{-0.717931026880}$ \\
\midrule
 &  &  & 0.3954, 0.3505, 0.9997 & 0.3744, 0.3746, 0.9537 \\
4 & 0.3740, 0.3744, 0.9898 & 0.3803, 0.3951, 0.9648 & 0.3478, 0.39493, 0.9798 & 0.4078, 0.9010, 0.3351 \\
 & $\textbf{-0.727918723553}$ & $\textbf{-0.727281885394}$ & $\textbf{-0.728067443345}$ & $\textbf{-0.727981667586}$ \\
\midrule
 &  &  & 0.3401, 0.3181, 0.9983 & 0.3373, 0.3390, 0.9635 \\
5 & 0.3293, 0.3289, 0.9939 & 0.3371, 0.3468, 0.9680 & 0.3174, 0.3397, 0.9880 & 0.4299, 0.9452, 0.3023 \\
 & $\textbf{-0.734233160953}$ & $\textbf{-0.733680013812}$ & $\textbf{-0.734264232997}$ & $\textbf{-0.734219573964}$ \\
\bottomrule
\bottomrule
\end{tabular}
\caption{Simplex $^1$P-Wave Short-Range Optimization}
\label{tab:SimplexPWaveSingOpt}
\end{table}

\begin{table}
\footnotesize
\centering
\begin{tabular}{c c c c c}
\toprule
\toprule
$\omega$ & 1st formalism / 1 set & 2nd formalism / 1 set & 1st formalism / 2 sets & 2nd formalism / 2 sets \\
\midrule
\midrule
 &  &  & 0.3832, 0.2911, 0.9894 & 0.3367, 0.3436, 0.9517 \\
1 & 0.3316, 0.3535, 0.9956 & 0.3302, 0.3540, 0.9909 & 0.2854, 0.3959, 1.0112 & 0.7606, 0.6593, 0.2304 \\
 & $\textbf{-0.624190839847}$ & $\textbf{-0.622936676609}$ & $\textbf{-0.629448013148}$ & $\textbf{-0.626442778437}$ \\
\midrule
 &  &  & 0.4056, 0.3294, 0.9799 & 0.3623, 0.3952, 0.9716 \\
2 & 0.3664, 0.3750, 0.9900 & 0.3679, 0.3896, 0.9753 & 0.3226, 0.4097, 1.0078 & 0.1961, 0.7724, 0.7692 \\
 & $\textbf{-0.674819577647}$ & $\textbf{-0.672880514312}$ & $\textbf{-0.676837948726}$ & $\textbf{-0.673438894110}$ \\
\midrule
 &  &  & 0.3874, 0.3314, 0.9823 & 0.3557, 0.3794, 0.9593 \\
3 & 0.3585, 0.3613, 0.9911 & 0.3639, 0.3810, 0.9690 & 0.3257, 0.3881, 1.0036 & 0.2080, 0.5994, 0.7711 \\
 & $\textbf{-0.704764841366}$ & $\textbf{-0.703191879865}$ & $\textbf{-0.705411707053}$ & $\textbf{-0.703452729525}$ \\
\midrule
 &  &  & 0.3539, 0.3192, 0.9879 & 0.3392, 0.3502, 0.9789 \\
4 & 0.3357, 0.3366, 0.9936 & 0.3427, 0.3554, 0.9680 & 0.3162, 0.3539, 0.9989 & 0.1624, 0.6564, 0.9512 \\
 & $\textbf{-0.721696022987}$ & $\textbf{-0.720513417195}$ & $\textbf{-0.721859567659}$ & $\textbf{-0.720596381145}$ \\
\midrule
 &  &  & 0.3184, 0.3038, 0.9935 & 0.3177, 0.3271, 0.9769 \\
5 & 0.3106, 0.3109, 0.9953 & 0.3182, 0.3275, 0.9682 & 0.3025, 0.3186, 0.9966 & 0.1142, 0.4274, 1.1088 \\
 & $\textbf{-0.731463030326}$ & $\textbf{-0.730592482596}$ & $\textbf{-0.731486455300}$ & $\textbf{-0.730615490923}$ \\
\bottomrule
\bottomrule
\end{tabular}
\caption{Simplex $^3$P Short-Range Optimization}
\label{tab:SimplexPWaveTripOpt}
\end{table}

From \cref{tab:SimplexPWaveSingOpt,tab:SimplexPWaveTripOpt}, we see that 
lower energy eigenvalues are always obtained with the first formalism with 1 
set versus the second formalism with 1 set. Likewise, the first formalism has 
lower energy eigenvalues than the second formalism when both have 2 sets. For 
$\omega \geq 2$, the first formalism with 1 set even has lower energy 
eigenvalues than the second symmetry with 2 sets. We had trouble obtaining 
phase shifts with two different sets of nonlinear parameters due to increased 
linear dependence, but for higher $\omega$, we also see that the energy does 
not change much. Using more than one set of nonlinear parameters could be 
explored further in future work, and we have done preliminary investigation 
into this for the S-wave.

\subsection{D-Wave Nonlinear Parameter Optimization}
\label{sec:DWaveNonlinear}

The comparison to the CC results \cite{Walters2004,Blackwood2002}
in \cref{tab:DWaveComparisons,fig:DWavePhase} for the $^1$D-wave is
reasonable, with the CC results below the complex Kohn results at $\kappa = 0.7$.
For $^3$D, the CC results are much higher, as can be seen in the inset
in \cref{fig:DWavePhase}. The $^1$D phase shifts are small, so their overall
contribution to the integrated cross section is small. 
This lead us to investigate whether the phase shifts could be improved by a
better selection of the short-range nonlinear parameters. If the phase shifts
were fully converged, varying the nonlinear parameters should have little
effect on their values. Both Van Reeth \cite{VanReethPrivate} and I
investigated this.

Using the simplex method described in \cref{sec:Simplex}, we obtained a set of
nonlinear parameters for $^1$D and $^3$D in \cref{tab:NonlinearOptimizedPD}.
We realized when calculating the phase shifts however that these were more
sensitive to the values of the nonlinear parameters than the S-wave and P-wave,
especially for $^3$D. This is likely due to the short-range terms trying to
make up for the missing mixed symmetry terms, and for higher partial waves, the
interaction region is more extended. We performed some manual
optimization of the nonlinear parameters for two $\kappa$ values to try to
improve the phase shifts.

For this investigation, we chose $\kappa$ values in two different regions:
one at lower $\kappa$ and another at higher $\kappa$. After optimization with
these, we also checked convergence ratios in the more sensitive resonance
region. The $\kappa = 0.1$ choice
was made, since this is the lowest value that we report. We chose $\kappa = 0.6$
for the higher $\kappa$ region, as this gets closer to the Ps(n=2) threshold
but is far enough away from the $^1$D resonance to avoid sensitivity of the
nonlinear parameters due to the resonance.

For these variations, we kept $\gamma$ constant and used $\omega = 4$. From 
\cref{tab:Nonlinear}, the value of $\gamma$ was found to be near $1$ using 
the simplex method (\cref{sec:Simplex}) for every partial wave for both the 
singlet and triplet. An explanation is that the $r_3$ coordinate represents 
the electron in H, and $\gamma = 1$ gives the short-range terms multiplied by 
the H wavefunction given in \cref{eq:HWave} \cite{VanReethPrivate}. We used the
original nonlinear parameters found by the simplex method and given in
\cref{tab:NonlinearOptimizedPD} as a starting set. For
$^1$D, these are $\alpha = 0.359$, $\beta = 0.368$, and $\gamma = 0.976$. For
$^3$D, these are $\alpha = 0.356$, $\beta = 0.365$, and $\gamma = 0.976$.

\begin{figure}
	\centering
	\includegraphics[width=\textwidth]{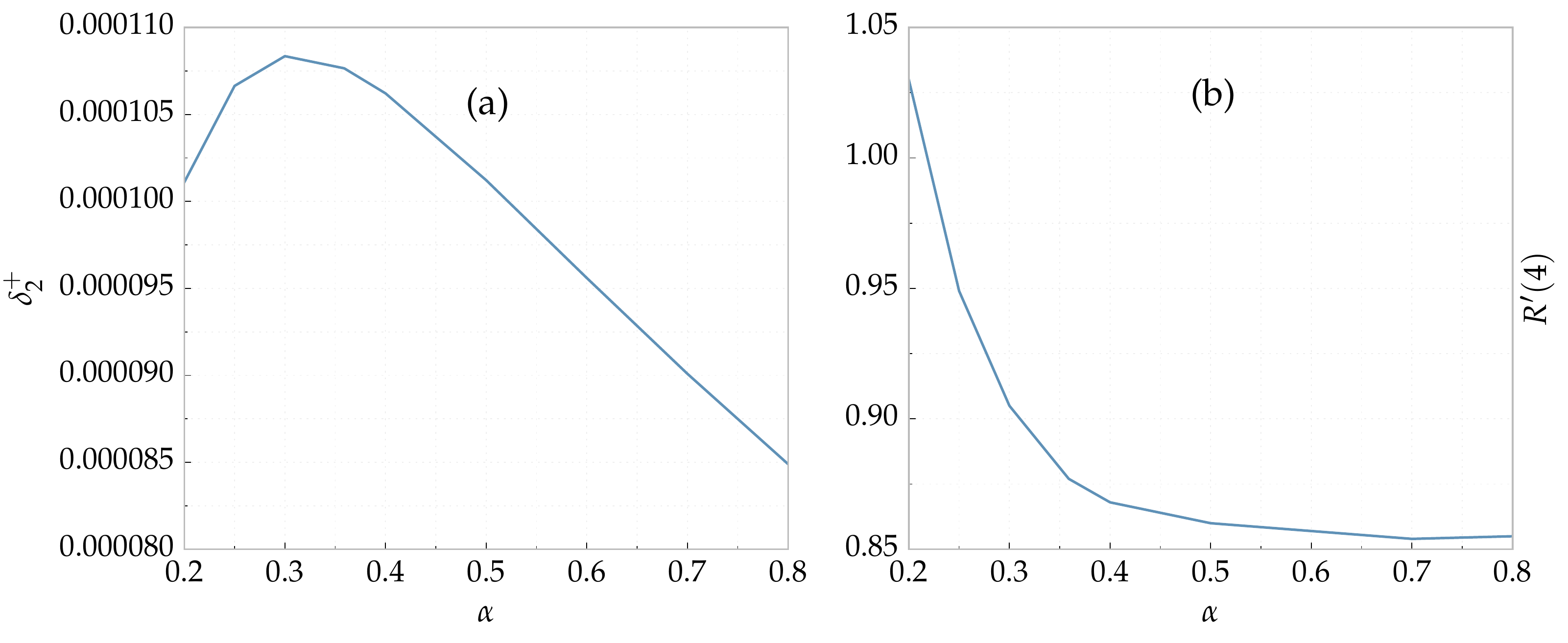}
	\caption[Variation of the nonlinear parameter $\alpha$ for $^{1}$D at $\kappa = 0.1$]{Phase shifts (a) and convergence ratios (b) for variation of the nonlinear parameter $\alpha$ for $^{1}$D at $\kappa = 0.1$}
	\label{fig:dwave-singlet-alpha-k01-variation}
\end{figure}

For the first variation, we investigated the $\alpha$ nonlinear parameter at
$\kappa = 0.1$. We varied $\alpha$ to see its effect on the phase shifts.
\Cref{fig:dwave-singlet-alpha-k01-variation} shows the results of this
variation. There is a maximum in the phase shift in 
\cref{fig:dwave-singlet-alpha-k01-variation}(a) and a large difference between
the phase shifts at low and high $\alpha$. If we decrease
$\alpha$ from its value of 0.359, however, from
\cref{fig:dwave-singlet-alpha-k01-variation}(b),
the convergence ratio $R'(4)$, given by \cref{eq:ConvRatio},
increases drastically. From this analysis, we had hoped for higher phase shifts,
but we have a trade-off between this and reasonable convergence ratios. Due to
this, we have kept the nonlinear parameter $\alpha$ at 0.359, which seems to be
a reasonable compromise between higher phase shifts and better convergence
ratios. This is likely an indication of the amount of numerical instability
we have with small phase shifts.

\begin{figure}
	\centering
	\includegraphics[width=\textwidth]{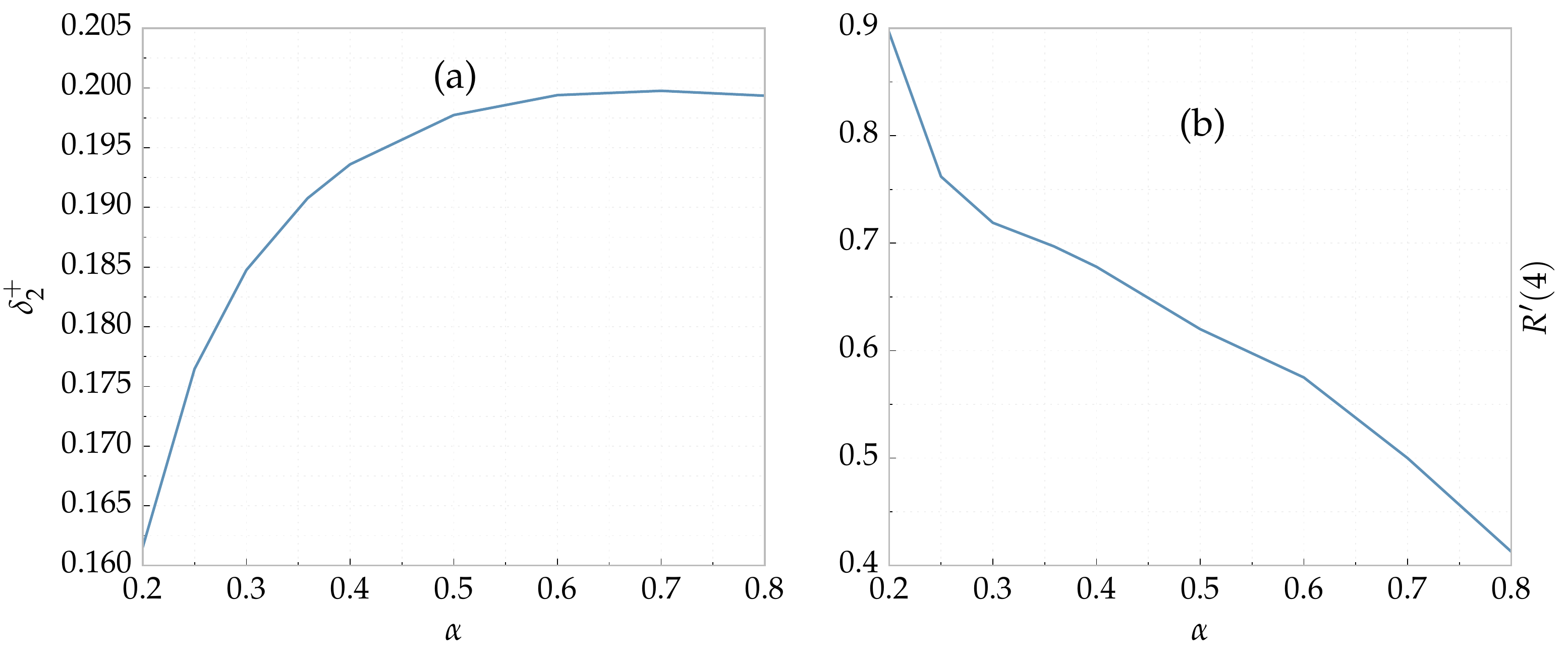}
	\caption[Variation of the nonlinear parameter $\alpha$ for $^{1}$D at $\kappa = 0.6$]{Phase shifts (a) and convergence ratios (b) for variation of the nonlinear parameter $\alpha$ for $^{1}$D at $\kappa = 0.6$}
	\label{fig:dwave-singlet-alpha-k06-variation}
\end{figure}

At the higher $\kappa$ of 0.6, the variation looks very different, as seen in
\cref{fig:dwave-singlet-alpha-k06-variation}. The maximum is at about
$\alpha = 0.6$, and $R'(4)$ is much less than 1. Interestingly,
$R'(4)$ decreases monotonically as $\alpha$ is increased. For
$\kappa = 0.6$, it is clear that choosing $\alpha = 0.6$ is much better than
the original 0.359.

Starting from the original nonlinear parameters, we also varied $\beta$. The
$\beta$ variation looks very similar to the $\alpha$ derivation, but there is
a surprising breakdown of the phase shifts when $\beta > 0.6$. At $\beta = 0.7$,
$\delta_2^+$ increases significantly, and $R'(4) > 5$. For $\beta = 0.8$,
$\delta_2^+ = -2.0694^{-3}$, and $R'(4) = -128.3$, so the phase shifts
for large $\beta$ are obviously not reliable. When $\beta$ is smaller, we see
very similar behavior to that of the $\alpha$ variation in
\cref{fig:dwave-singlet-alpha-k01-variation}. The convergence ratio increases as
$\beta$ gets too low, and the maximum is around $\beta = 0.3$. Similar to the
$\alpha$ variation, we kept the original $\beta = 0.368$ value.

\begin{figure}
	\centering
	\includegraphics[width=\textwidth]{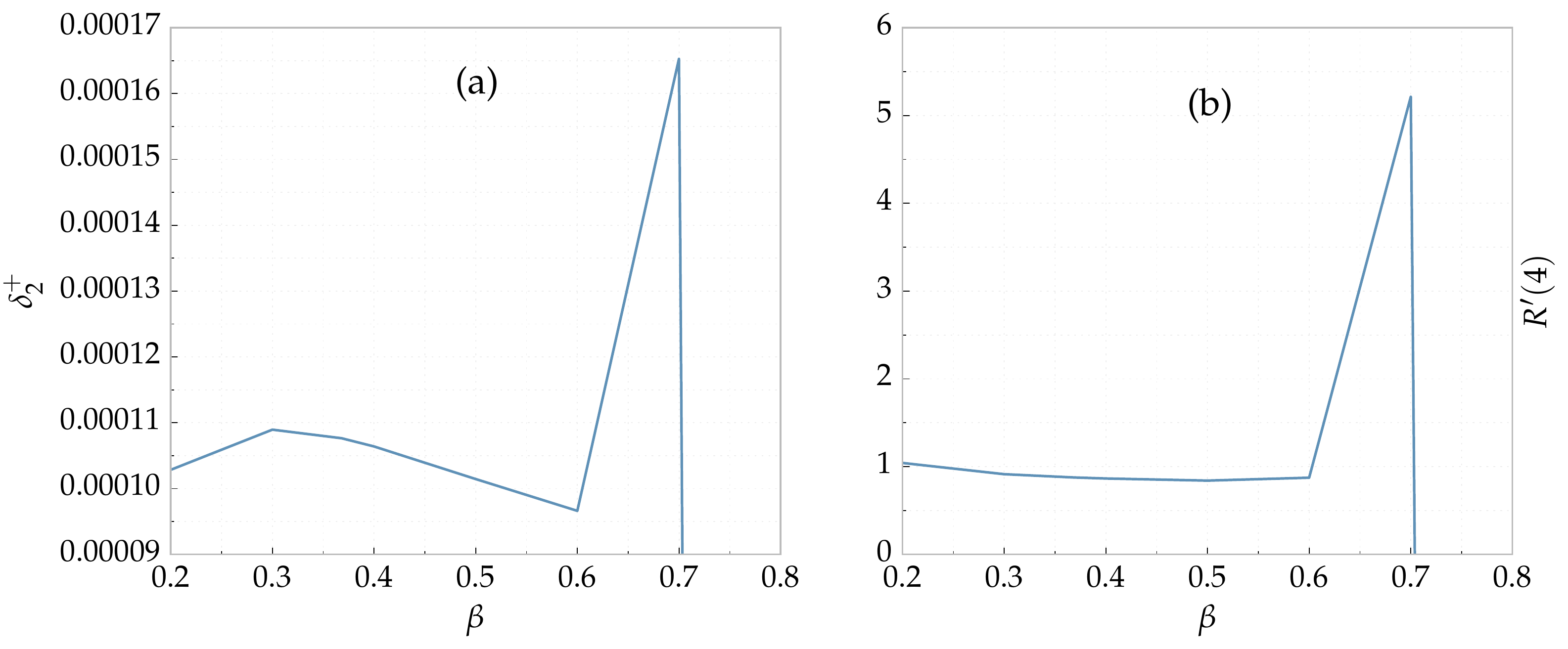}
	\caption[Variation of the nonlinear parameter $\beta$ for $^{1}$D at $\kappa = 0.1$]{Phase shifts (a) and convergence ratios (b) for variation of the nonlinear parameter $\beta$ for $^{1}$D at $\kappa = 0.1$}
	\label{fig:dwave-singlet-beta-k01-variation}
\end{figure}

\begin{figure}
	\centering
	\includegraphics[width=\textwidth]{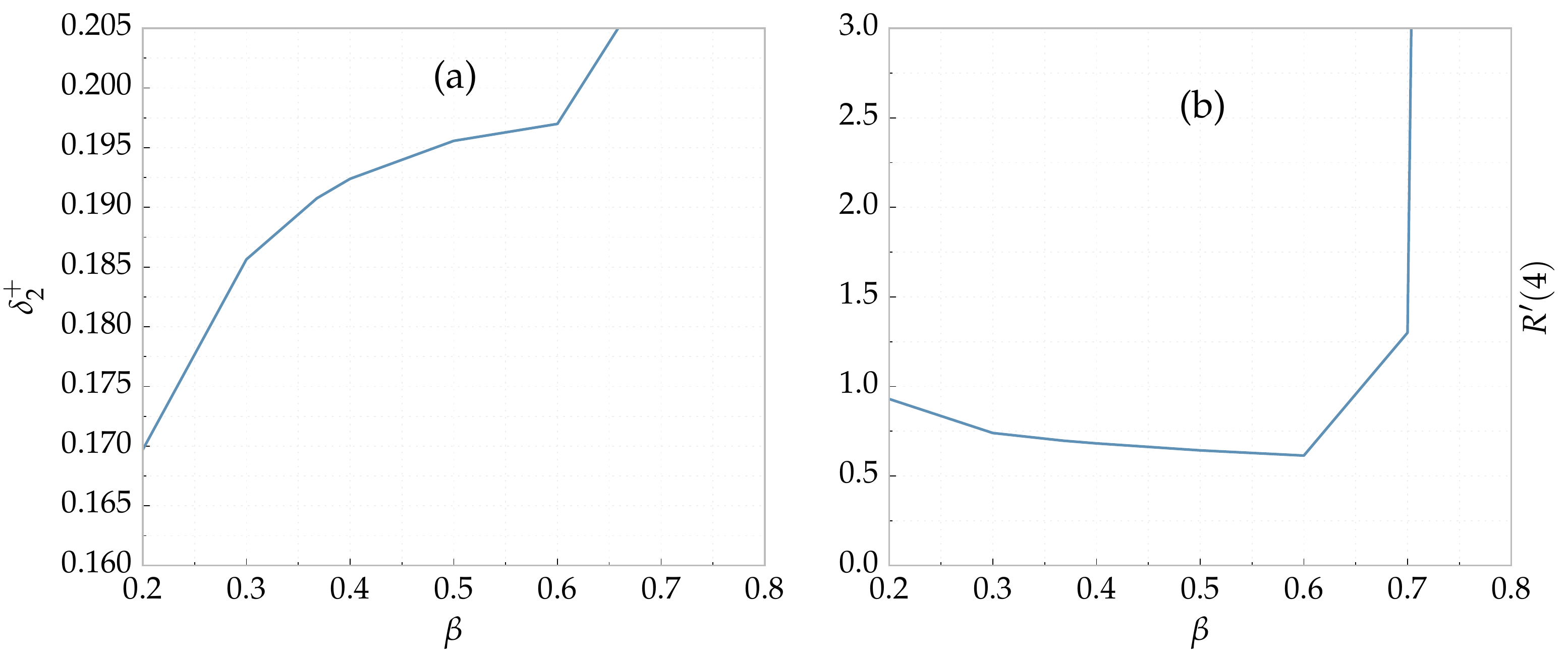}
	\caption[Variation of the nonlinear parameter $\beta$ for $^{1}$D at $\kappa = 0.6$]{Phase shifts (a) and convergence ratios (b) for variation of the nonlinear parameter $\beta$ for $^{1}$D at $\kappa = 0.6$}
	\label{fig:dwave-singlet-beta-k06-variation}
\end{figure}

For $\kappa = 0.6$, we see a breakdown when $\beta$ is large as well. The plot
in \cref{fig:dwave-singlet-beta-k06-variation} shows that we have not hit the maximum
$\delta_2^+$ before the phase shifts exhibit breakdown. The $\alpha$ variation
appears to be more stable than the $\beta$ variation from these runs.

\begin{figure}
	\centering
	\includegraphics[width=\textwidth]{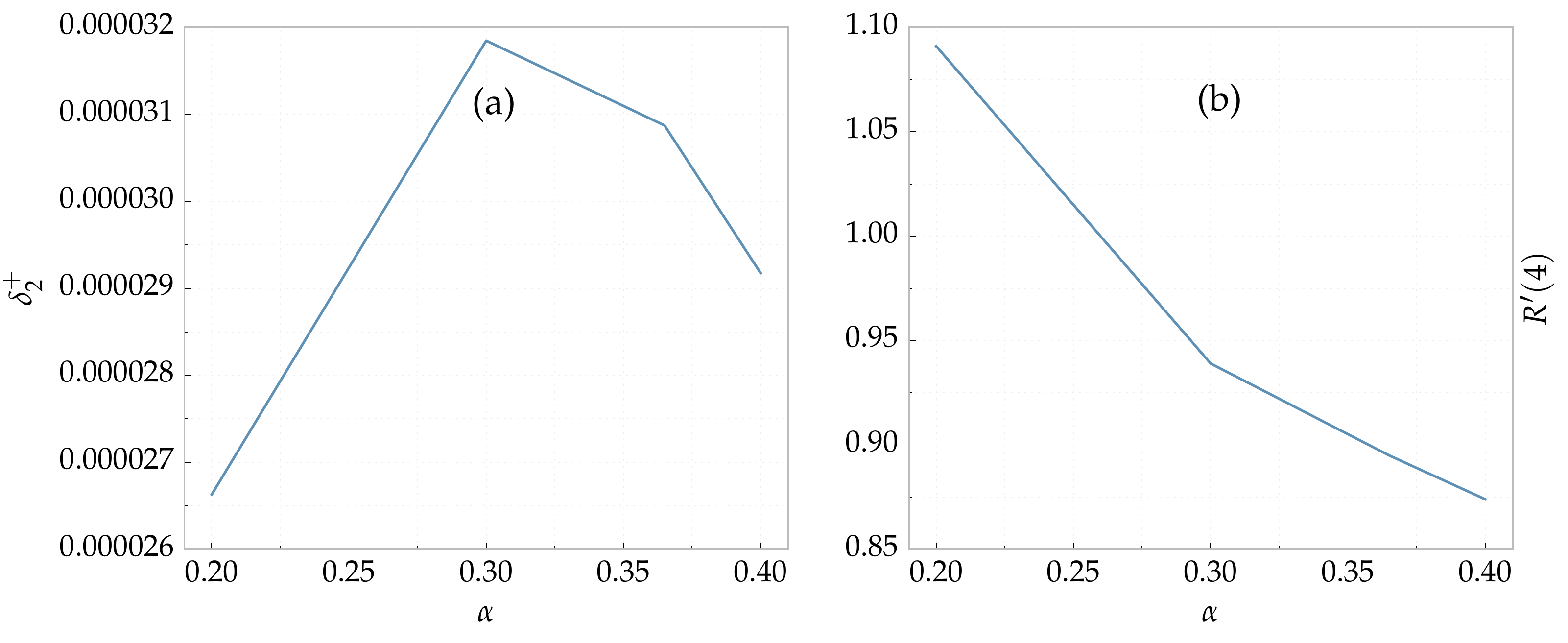}
	\caption[Variation of the nonlinear parameter $\alpha$ for $^{3}$D at $\kappa = 0.1$]{Phase shifts (a) and convergence ratios (b) for variation of the nonlinear parameter $\alpha$ for $^{3}$D at $\kappa = 0.1$}
	\label{fig:dwave-triplet-alpha-k01-variation}
\end{figure}

The variations for $^3$D look similar to that of $^1$D. There are less points
in \cref{fig:dwave-triplet-alpha-k01-variation}, but we can see that as
$\alpha$ is lowered, $R'(4)$ increases, and there is a maximum around
$\alpha = 0.3$. Again, we kept the original $\alpha$ of 0.356.

\begin{figure}
	\centering
	\includegraphics[width=\textwidth]{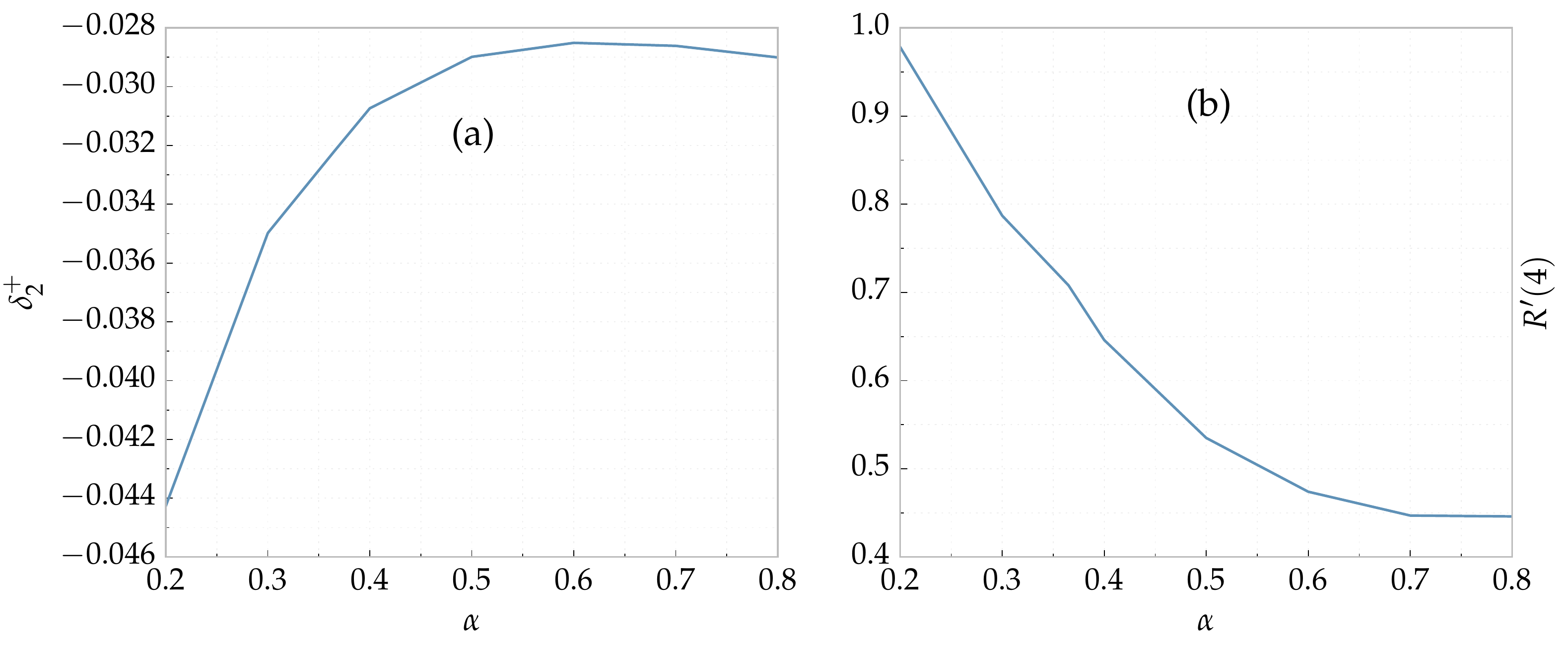}
	\caption[Variation of the nonlinear parameter $\alpha$ for $^{3}$D at $\kappa = 0.6$]{Phase shifts (a) and convergence ratios (b) for variation of the nonlinear parameter $\alpha$ for $^{3}$D at $\kappa = 0.6$}
	\label{fig:dwave-triplet-alpha-k06-variation}
\end{figure}

\begin{figure}
	\centering
	\includegraphics[width=\textwidth]{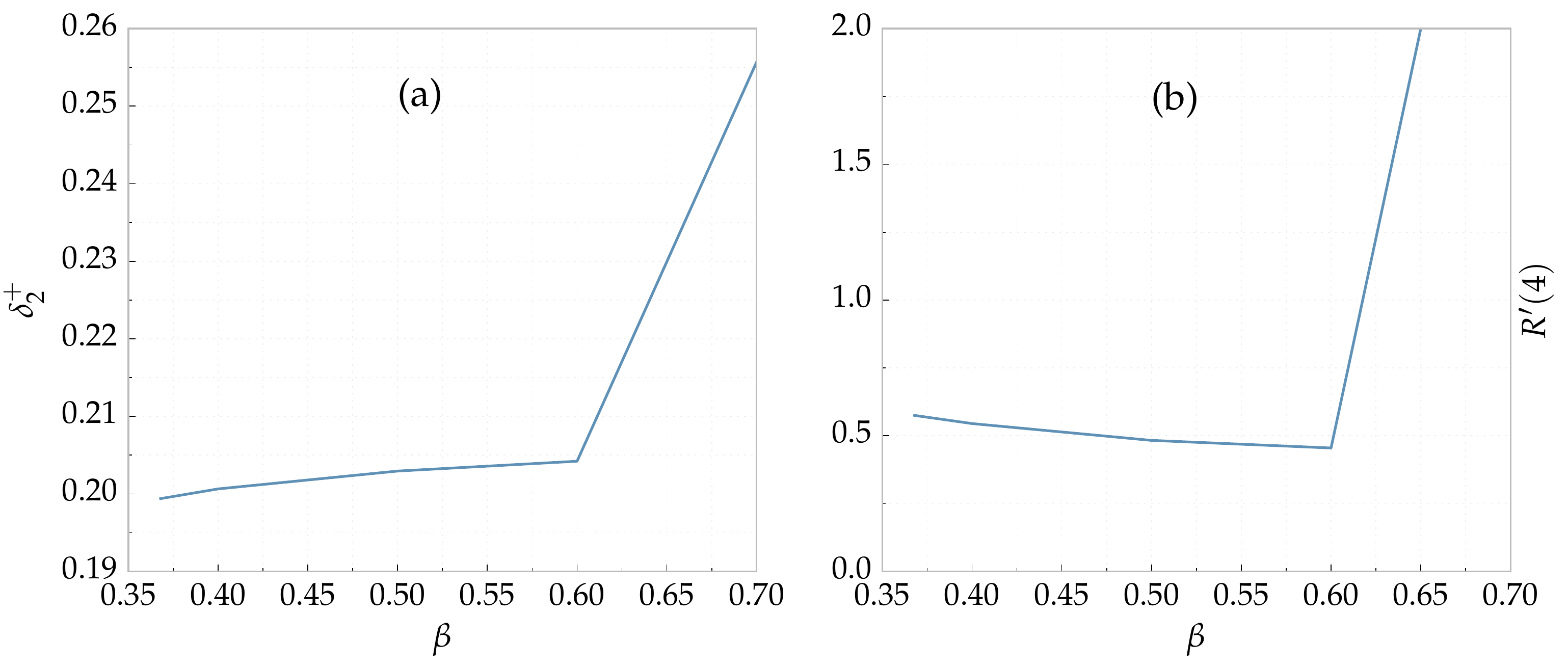}
	\caption[Variation of the nonlinear parameter $\beta$ for $^{1}$D at $\kappa = 0.1$ and $\alpha = 0.6$]{Phase shifts (a) and convergence ratios (b) for variation of the nonlinear parameter $\beta$ for $^{1}$D at $\kappa = 0.1$ and $\alpha = 0.6$}
	\label{fig:dwave-singlet-alphabeta-k06-variation}
\end{figure}

\begin{figure}
	\centering
	\includegraphics[width=\textwidth]{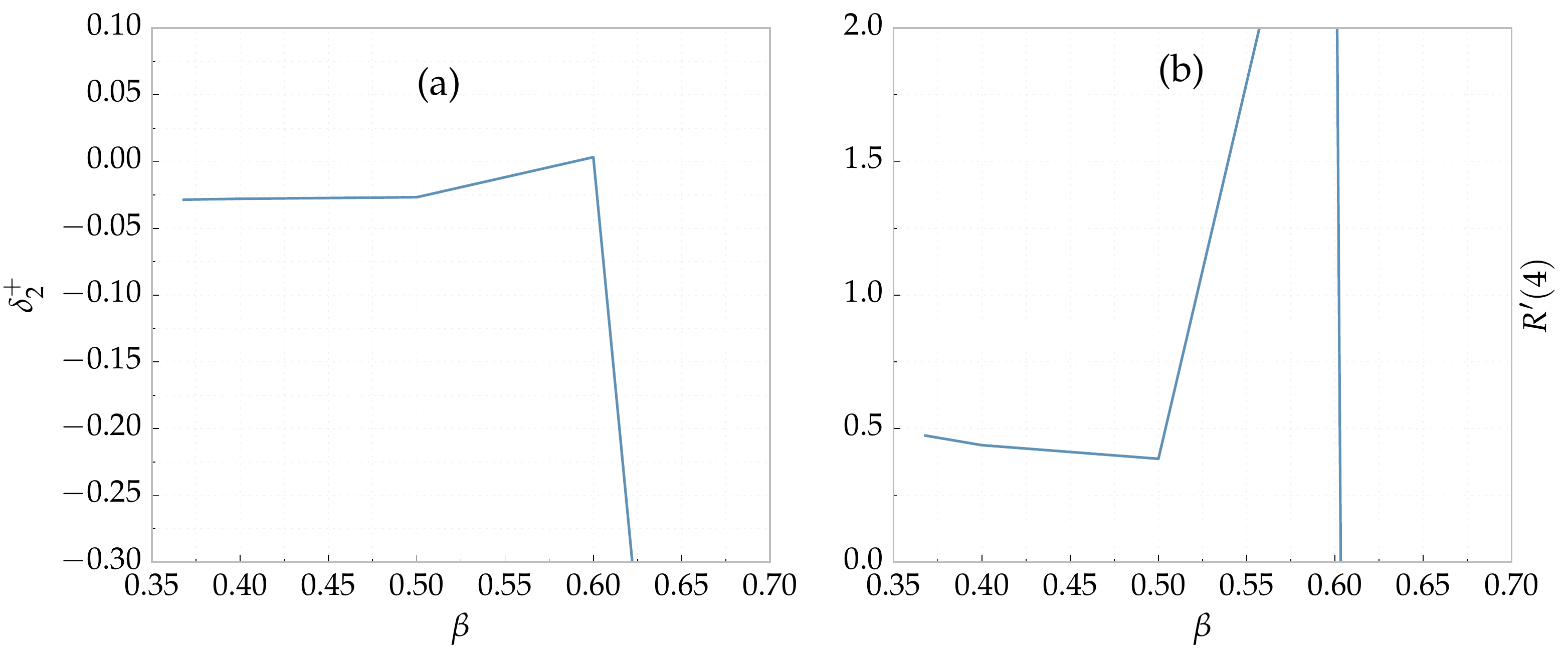}
	\caption[Variation of the nonlinear parameter $\beta$ for $^{3}$D at $\kappa = 0.1$ and $\alpha = 0.6$]{Phase shifts (a) and convergence ratios (b) for variation of the nonlinear parameter $\beta$ for $^{3}$D at $\kappa = 0.1$ and $\alpha = 0.6$}
	\label{fig:dwave-triplet-alphabeta-k06-variation}
\end{figure}

Due to the problems we had with the $\beta$ variation for $^1$D
and the original $\alpha$, we did not pursue this for $^3$D.
With these new choices of $^{1,3}$D nonlinear parameters for higher $\kappa$
that use a higher $\alpha$ of 0.6, we also investigated varying $\beta$, as
shown in \Cref{fig:dwave-singlet-alphabeta-k06-variation} for $^1$D.
Again, the phase shifts are unreliable as $\beta$ is increased much.
The equivalent plots for $^3$D are in \cref{fig:dwave-triplet-alphabeta-k06-variation}.
For this, the phase shifts break down even earlier, starting after $\beta = 0.5$.

We had previously noticed for multiple partial waves that if $\alpha$ and
$\beta$ are equal, linear dependence becomes a large issue. Based on this and
the graphs in \cref{fig:dwave-singlet-alphabeta-k06-variation,fig:dwave-triplet-alphabeta-k06-variation},
we tried the set of nonlinear parameters $\alpha = 0.6$, $\beta = 0.5$,
and $\gamma = 0.976$ for $^{1,3}$D for a full run of $\omega = 6$.
Using the Todd procedure in \cref{sec:ToddBound}, this $^1$D set only uses 844
terms, and the $^3$D set uses 854 terms.

\Cref{tab:D1WaveBetaVar} compares the sets with $\beta = 0.368$ and $0.5$ (using
$\alpha = 0.6$ and $\gamma = 0.976$). We do not use the restricted set described
in \cref{sec:Restricted} for our final calculations, but a comparison with it
can give an idea of how stable the phase shifts are. Comparing the full and
restricted sets for $\beta = 0.368$, the phase shifts do not change much, even
though we are reducing the basis set from 913 to 720 terms. However, comparison
of the full and restricted sets for $\beta = 0.5$ shows a much larger change in
the phase shifts, despite only changing from 844 to 720 terms. This seems to
indicate that the set with $\beta = 0.5$ is not as stable and could potentially
suffer from linear dependence.

\begin{table}[h]
\centering
\begin{tabular}{c....}
\toprule
$\kappa$ & \multicolumn{1} {c}{$\delta_2^-$ full $\beta = 0.368$} & \multicolumn{1} {c}{$\delta_2^-$ restricted $\beta = 0.368$}  & \multicolumn{1} {c}{$\delta_2^-$ full $\beta = 0.5$}  & \multicolumn{1} {c}{$\delta_2^-$ restricted $\beta = 0.5$} \\
\midrule
0.3 & 1.599^{-2} & 1.595^{-2} & 1.693^{-2} & 1.600^{-2} \\
0.4 & 4.978^{-2} & 4.965^{-2} & 5.176^{-2} & 4.987^{-2} \\
0.5 & 1.126^{-1} & 1.124^{-1} & 1.152^{-1} & 1.130^{-1} \\
0.6 & 2.058^{-1} & 2.053^{-1} & 2.083^{-1} & 2.066^{-1} \\
0.7 & 3.275^{-1} & 3.269^{-1} & 3.302^{-1} & 3.293^{-1} \\
\bottomrule
\end{tabular}
\caption[$^1$D phase shifts for varying $\beta$]{$^1$D phase shifts for sets of nonlinear parameters with $\alpha = 0.6$ and $\gamma = 0.976$. The full set for $\beta = 0.368$ has 913 terms, and the full set for $\beta = 0.5$ has 844 terms. The restricted sets have 720 terms.}
\label{tab:D1WaveBetaVar}
\end{table}

To see whether the $\beta = 0.5$ set has linear dependence issues, we compared
the convergence ratios. From \cref{tab:D1Beta368VarConv}, we see that the
convergence ratios are less than 0.5 for $\omega = 6$, which indicates good
convergence. \Cref{tab:D1Beta5VarConv} however shows a problem with linear
dependence for the $\beta = 0.5$ set. Based on this analysis, we have chosen
the set $\alpha = 0.359$, $\beta = 0.368$, and $\gamma = 0.976$ for higher
$\kappa$ for $^1$D.

\begin{table}
\centering
\begin{tabular}{cccccc}
\toprule
$\kappa$ & $R'(2)$ & $R'(3)$ & $R'(4)$ & $R'(5)$ & $R'(6)$ \\
\midrule
0.3 & 1.523 & 0.795 & 0.572 & 0.491 & 0.377 \\
0.4 & 1.286 & 0.680 & 0.475 & 0.412 & 0.429 \\
0.5 & 1.048 & 0.603 & 0.464 & 0.442 & 0.466 \\
0.6 & 0.826 & 0.534 & 0.575 & 0.435 & 0.481 \\
0.7 & 0.471 & 0.874 & 0.472 & 0.521 & 0.465 \\
\bottomrule
\end{tabular}
\caption{Convergence ratios for $^1D$ at multiple $\omega$ values for the full $\beta = 0.368$ set}
\label{tab:D1Beta368VarConv}
\end{table}

\begin{table}
\centering
\begin{tabular}{cccccc}
\toprule
$\kappa$ & $R'(2)$ & $R'(3)$ & $R'(4)$ & $R'(5)$ & $R'(6)$ \\
\midrule
0.3 & 2.475 & 0.352 & 1.611 & 0.316 & 1.628 \\
0.4 & 2.157 & 0.291 & 1.372 & 0.209 & 1.902 \\
0.5 & 1.835 & 0.254 & 1.133 & 0.225 & 1.343 \\
0.6 & 1.539 & 0.237 & 0.979 & 0.299 & 0.667 \\
0.7 & 1.206 & 0.306 & 0.696 & 0.360 & 0.441 \\
\bottomrule
\end{tabular}
\caption{Convergence ratios for $^1D$ at multiple $\omega$ values for the full $\beta = 0.5$ set}
\label{tab:D1Beta5VarConv}
\end{table}

\begin{table}
\centering
\begin{tabular}{c....}
\toprule
$\kappa$ & \multicolumn{1} {c}{$\delta_2^-$ full $\beta = 0.368$} & \multicolumn{1} {c}{$\delta_2^-$ restricted $\beta = 0.368$}  & \multicolumn{1} {c}{$\delta_2^-$ full $\beta = 0.5$}  & \multicolumn{1} {c}{$\delta_2^-$ restricted $\beta = 0.5$} \\
\midrule
0.3  & 1.100^{-3}  & 1.058^{-3}  & 1.977^{-3}  & 1.079^{-3}   \\
0.4  & -1.796^{-3} & -1.898^{-3} & 8.016^{-6}  & -1.751^{-3}  \\
0.5  & -1.070^{-2} & -1.087^{-2} & -8.432^{-3} & -1.052^{-2}  \\
0.6  & -2.544^{-2} & -2.568^{-2} & -2.331^{-2} & -2.501^{-2}  \\
0.7  & -4.281^{-2} & -4.314^{-2} & -4.085^{-2} & -4.191^{-2}  \\
\bottomrule
\end{tabular}
\caption[$^3$D phase shifts for varying $\beta$]{$^3$D phase shifts for sets of nonlinear parameters with $\alpha = 0.6$ and $\gamma = 0.976$. The full set for $\beta = 0.365$ has 913 terms, and the full set for $\beta = 0.5$ has 854 terms. The restricted sets have 720 terms.}
\label{tab:D3WaveBetaVar}
\end{table}

We also looked at the convergence ratios for $^3$D, as given in
\cref{tab:D3Beta365VarConv,tab:D3Beta5VarConv}. Again, the set of nonlinear
parameters with $\beta = 0.5$ is problematic, giving $R'(6) > 1$ for
most $\kappa$ values. From this analysis, we decided on the $^3$D nonlinear
parameters of $\alpha = 0.356$, $\beta = 0.365$, and $\gamma = 0.976$.

\begin{table}
\centering
\begin{tabular}{cccccc}
\toprule
$\kappa$ & $R'(2)$ & $R'(3)$ & $R'(4)$ & $R'(5)$ & $R'(6)$ \\
\midrule
0.3 & 3.950 & 0.932 & 0.564 & 0.512 & 0.432 \\
0.4 & 3.399 & 0.811 & 0.446 & 0.404 & 0.469 \\
0.5 & 2.812 & 0.740 & 0.417 & 0.411 & 0.519 \\
0.6 & 2.261 & 0.730 & 0.466 & 0.400 & 0.527 \\
0.7 & 1.714 & 0.860 & 0.490 & 0.433 & 0.484 \\
\bottomrule
\end{tabular}
\caption{Convergence ratios for $^3D$ at multiple $\omega$ values for the full $\beta = 0.365$ set}
\label{tab:D3Beta365VarConv}
\end{table}

\begin{table}
\centering
\begin{tabular}{cccccc}
\toprule
$\kappa$ & $R'(2)$ & $R'(3)$ & $R'(4)$ & $R'(5)$ & $R'(6)$ \\
\midrule
0.3 & 7.392 & 3.424 & 0.505 & 0.426 & 1.768 \\
0.4 & 6.978 & 3.035 & 0.390 & 0.265 & 2.421 \\
0.5 & 6.435 & 2.625 & 0.345 & 0.229 & 2.235 \\
0.6 & 5.739 & 2.241 & 0.393 & 0.242 & 1.251 \\
0.7 & 5.082 & 2.063 & 0.504 & 0.244 & 0.723 \\
\bottomrule
\end{tabular}
\caption{Convergence ratios for $^3D$ at multiple $\omega$ values for the full $\beta = 0.5$ set}
\label{tab:D3Beta5VarConv}
\end{table}

\section{Nonlinear Parameters and Terms Used in the Scattering Wavefunction}
\label{sec:NonlinParam}

\begin{table}
  \centering
  \begin{tabular}{cclccccc}
	\toprule
	Partial wave & $\omega$ & $N^\prime(\omega)$ & $\alpha$ & $\beta$ & $\gamma$ & $\mu$ & $m_\ell$ \\
	\midrule
	$^1$S                      & 7 & 1505        & 0.568 & 0.580 & 1.093 & 0.9 & 1 \\
	$^3$S                      & 7 & 1633        & 0.323 & 0.334 & 0.975 & 0.9 & 1 \\
	$^1$P                      & 7 & 1000        & 0.397 & 0.376 & 0.962 & 0.9 & 3 \\
	$^3$P                      & 7 & 1000        & 0.310 & 0.311 & 0.995 & 0.9 & 3 \\
	$^1$D $(\kappa < 0.3)$     & 6 & 916         & 0.359 & 0.368 & 0.976 & 0.7 & 7 \\
	$^1$D $(\kappa \geq 0.3)$  & 6 & 913         & 0.600 & 0.368 & 0.976 & 0.7 & 7 \\
	$^3$D $(\kappa < 0.3)$     & 6 & 919         & 0.356 & 0.365 & 0.976 & 0.7 & 7 \\
	$^3$D $(\kappa \geq 0.3)$  & 6 & 913         & 0.600 & 0.365 & 0.976 & 0.7 & 7 \\
	$^1$F $(\kappa < 0.4)$     & 5 & $385^\star$ & 0.359 & 0.368 & 0.976 & 0.7 & 7 \\
	$^1$F $(\kappa \geq 0.4)$  & 5 & 462         & 0.500 & 0.600 & 1.100 & 0.7 & 7 \\
	$^3$F $(\kappa < 0.4)$     & 5 & $385^\star$ & 0.356 & 0.365 & 0.976 & 0.7 & 7 \\
    $^3$F $(\kappa \geq 0.4)$  & 5 & 462         & 0.600 & 0.365 & 0.976 & 0.7 & 7 \\
	$^1$G $(\kappa < 0.45)$    & 5 & 462         & 0.359 & 0.368 & 0.976 & 0.7 & 9 \\
    $^1$G $(\kappa \geq 0.45)$ & 5 & 462         & 0.500 & 0.600 & 1.100 & 0.7 & 9 \\
	$^3$G $(\kappa < 0.45)$    & 5 & 462         & 0.356 & 0.365 & 0.976 & 0.7 & 9 \\
    $^3$G $(\kappa \geq 0.45)$ & 5 & 462         & 0.600 & 0.365 & 0.976 & 0.7 & 9 \\
	$^1$H $(\kappa < 0.5)$     & 5 & 462         & 0.359 & 0.368 & 0.976 & 0.7 & 11 \\
	$^1$H $(\kappa \geq 0.5)$  & 5 & 462         & 0.500 & 0.600 & 1.100 & 0.7 & 11 \\
	$^3$H $(\kappa < 0.45)$    & 5 & 462         & 0.356 & 0.365 & 0.976 & 0.7 & 11 \\
    $^3$H $(\kappa \geq 0.45)$ & 5 & 462         & 0.600 & 0.365 & 0.976 & 0.7 & 11 \\
	\bottomrule
  \end{tabular}
  \caption[Parameters for each partial wave]{Nonlinear parameters $\alpha$, $\beta$, $\gamma$, $\mu$, integer power $m_\ell$
in the shielding function, $\omega$, and the number of terms $N'(\omega)$ of each symmetry
in the wavefunction for each partial wave. Numbers marked with a star indicate the
restriction in the $r_3$ power described in \cref{sec:Restricted}.}
  \label{tab:Nonlinear}
\end{table}

\Cref{tab:Nonlinear} gives the nonlinear parameters and total short-range
terms used for each partial wave. This table is the same as that in our
Ref.~\cite{Woods2015}.

\section{S-Wave Maximum \texorpdfstring{$\mu$}{mu}}
\label{sec:MaxMu}

\begin{figure}
	\centering
	\includegraphics[width=5in]{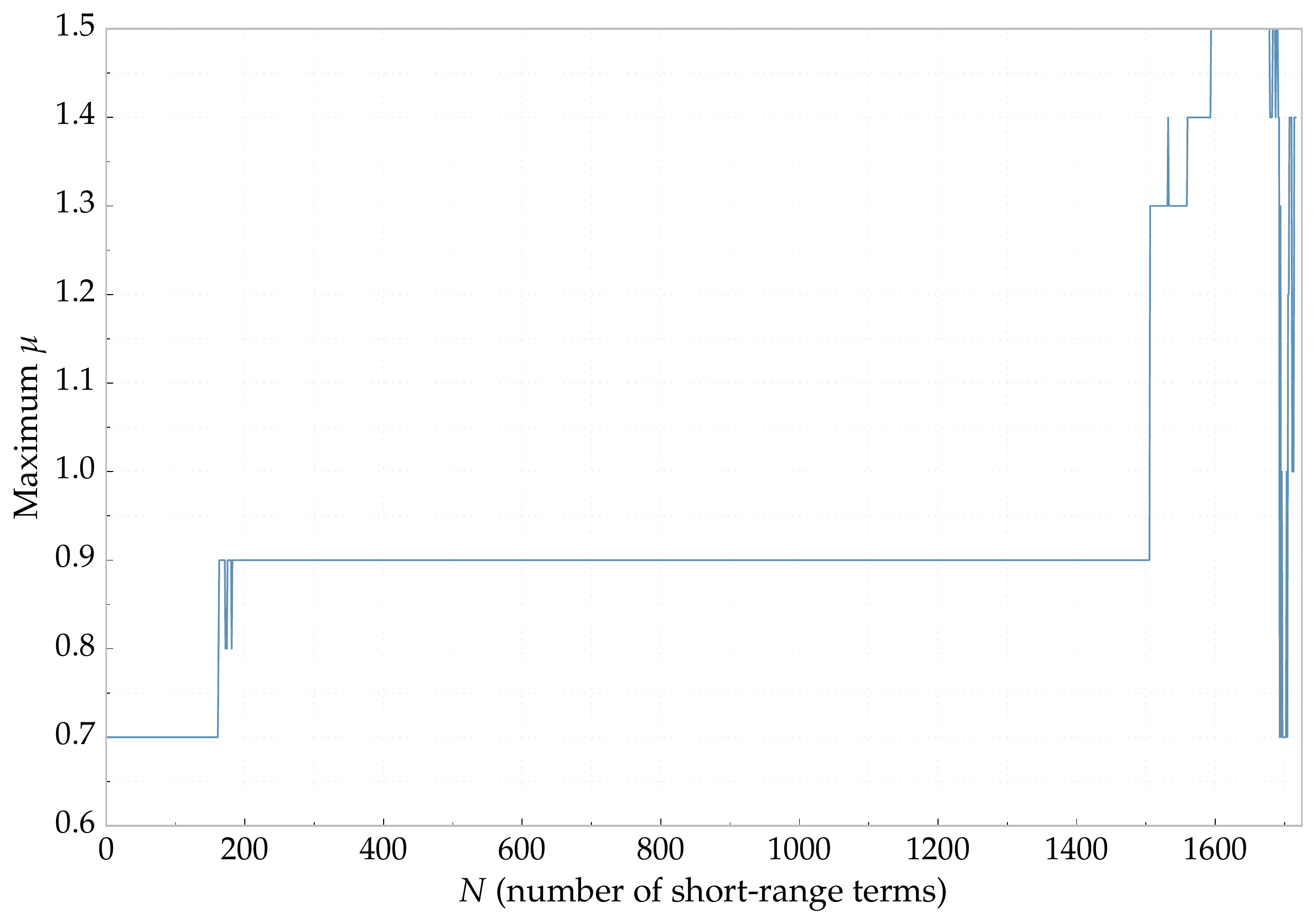}
	\caption[Maximum $\mu$ for $^1$S-wave]{Value of $\mu$ in the shielding function, \cref{eq:PartialWaveShielding}, that gives the largest phase shift versus the number of terms for the $^1$S-wave with $\kappa = 0.1$}
	\label{fig:swave-max-mu}
\end{figure}

For the first three partial waves, we truncated the short-range basis set of
size $N(\omega)$ using the method in \cref{sec:CompPhase}. For $^1$S however,
we found that we could more easily determine the cutoff by performing phase 
shift runs for multiple $\mu$ values and various $\kappa$ to see what the 
optimal value of $\mu$ was that gave the highest phase shift. As shown in
\cref{fig:swave-max-mu}, this value of $\mu$ was fairly stable through most of
the $N$ range, but it suddenly spikes up after 1505 terms and fluctuates
greatly. For some $\kappa$ values, we also saw that when this happened, the
value of $\mu$ that gave the highest phase shift was sometimes unbounded,
definitely indicating linear dependence. Interestingly, this behavior was
not noticed for the $^3$S-wave. We did not try this for other partial waves,
but it gave a good cutoff point for the $^1$S-wave.

\section{Gaussian Quadratures}
\label{sec:GaussQuad}

Gaussian quadratures are used to integrate many classes of integrals. In their
most general form, these quadratures are given by \cite[p.887]{Abramowitz1965}
\beq
\label{eq:GeneralQuadratures}
\int_a^b W(x) f(x) dx \approx \sum_{i=1}^n w_i f(x_i).
\eeq
Gaussian quadratures are particularly attractive, since they give exact 
results for polynomials up to degree $2n-1$. The weight function $W(x)$ can 
be chosen for certain classes of integrals. Three main types of weight 
functions are used in this work. For a discussion on the number of quadrature 
points used, refer to \cref{sec:QuadraturePoints}.

For most integrals over finite intervals, we use the Gauss-Legendre quadrature.
The exception is when we integrate over an interior angle ($\varphi_{ij}$),
where we use the Chebyshev-Gauss quadrature. For semi-infinite integrations,
we use Gauss-Laguerre quadratures. Each of these can be seen in
Ref.~\cite[p.887-890]{Abramowitz1965}.

The Gauss-Laguerre quadrature is typically over the range $(-1,1)$ but can be
modified to arbitrary finite intervals using a standard formula given in
Ref.~\cite[p.887]{Abramowitz1965}. Discussion over adapting the Gauss-Laguerre
quadrature to use an arbitrary lower limit and an extra constant multiplying
the variable in the exponential is found in \cref{sec:GaussLag}. Discussion over
adapting the Chebyshev-Gauss quadrature to the $\varphi_{ij}$ innermost
integrations is given in \cref{sec:ChebyshevGauss}.

\subsection{Gauss-Laguerre Quadrature}
\label{sec:GaussLag}

The Gauss-Legendre quadrature cannot be used on semi-infinite intervals, so 
we use the Gauss-Laguerre quadrature in these cases. The orthogonal 
polynomials in this case are the Laguerre polynomials, $L_n(x)$, and the 
weight function is $W(x) = e^{-x}$. The specific form of
\cref{eq:GeneralQuadratures} is
\beq
\label{eq:GaussLag}
\int_0^\infty e^{-x} f(x) dx \approx \sum_{i=1}^n w_i f(x_i).
\eeq
When the integration is over the interval $(a,\infty)$ instead, \cref{eq:GaussLag} is transformed by
\beq
\label{eq:GaussLagGen1}
\int_a^\infty e^{-x} f(x) dx = \int_0^\infty e^{-(x+a)} f(x+a) dx = e^{-a} \int_0^\infty e^{-x} f(x+a) dx \approx e^{-a} \sum_{i=1}^n w_i f(x_i+a).
\eeq

\noindent A more general form of this is obtained by using a coefficient in the exponential, i.e.
\beq
\label{eq:GaussLagGen2}
\int_a^\infty e^{-m x} f(x) dx = \frac{1}{m} \int_a^\infty e^{-y} f\left(\frac{y}{m}\right) dy,
\eeq
where we have defined $y = m x$.  This allows for \cref{eq:GaussLagGen1} to be generalized to
\begin{align}
\label{eq:GaussLagGen}
\nonumber \int_a^\infty e^{-m x} f(x) dx &= \frac{1}{m} \int_{ma}^\infty e^{-y} f\left(\frac{y}{m}\right) dy = \frac{1}{m} \int_0^\infty e^{-(y+ma)} f\left(\frac{y}{m}+a\right) dy \\
& = \frac{e^{-ma}}{m} \int_0^\infty e^{-y} f\left(\frac{y}{m}+a\right) dy \approx \frac{e^{-ma}}{m} \sum_{i=1}^n w_i f\left(\frac{y_i}{m}+a\right).
\end{align}
The $y_i$ abscissas and $w_i$ weights are the same as 
the standard Gauss-Laguerre quadrature. This more general form
quadrature is what we use for semi-infinite integrations.

\subsection{Chebyshev-Gauss Quadrature}
\label{sec:ChebyshevGauss}

To integrate over the $\varphi_{ij}$ variable, we can use \cite[p.79]{VanReethThesis}
\begin{equation}
\Int{ D(\cos \varphi_{23})}{\varphi_{23},0,2\pi} \approx \frac{2\pi}{n}\sum_{i=1}^n D\left[\cos\left(\frac{2i-1}{2n}\pi\right)\right].
\label{eq:PVRAng}
\end{equation}
To prove this, we start with
\begin{equation}
\Int{ D(\cos \varphi_{23}) }{\varphi_{23},0,\pi} \approx \frac{\pi}{n}\sum_{i=1}^n D\left[\cos\left(\frac{2i-1}{2n}\pi\right)\right].
\label{eq:PVRAngNew}
\end{equation}
Both of these equations are variations on Gaussian quadratures. The
Chebyshev-Gauss quadrature is given by
\cite{Abramowitz1965,MathworldChebyshevGauss}
\begin{equation}
\Int{ \frac{f\left( x \right)}{\sqrt{1 - x^2}} }{x,-1,1} \approx \sum_{i = 1}^n {w_i}f\left( {{x_i}} \right),
\label{eq:ChebyshevGauss}
\end{equation}
with $w_i = \frac{\pi}{n}$ and $x_i = \cos\left(\frac{2i-1}{2n}\pi\right)$.

Starting from the left side of \cref{eq:PVRAngNew}, we have
\begin{align}
\Int{ D\left( {\cos \varphi } \right) }{\varphi,0,\pi} &= \Int{ \frac{1}{{\sqrt {1 - {{\cos }^2}\varphi } }}\sqrt {1 - {{\cos }^2}\varphi } \: D\!\left( {\cos \varphi } \right) }{\varphi,0,\pi} \\
& = \Int{ \frac{1}{{\sqrt {1 - {{\cos }^2}\varphi } }} D\!\left( {\cos \varphi } \right)\sin \varphi \: }{\varphi,0,\pi}.
\label{eq:PVRAngNew2}
\end{align}

Using the substitution $x = \cos \varphi$ and $dx = -\sin \varphi \:d\varphi$
changes the limits to $x_{min} \!\!=\! \cos 0 \! =\! 1$ and
$x_{max} = \cos \pi = -1$. \Cref{eq:PVRAngNew2} becomes
\begin{equation}
\int_0^\pi  D\left( {\cos \varphi } \right){\textrm{d}}\varphi = - \int_1^{ - 1} \frac{1}{{\sqrt {1 - {x^2}} }}D\left( x \right){\textrm{d}}x = \int_{ - 1}^1 \frac{1}{{\sqrt {1 - {x^2}} }}D\left( x \right){\textrm{d}}x.
\label{eq:PVRAngNew3}
\end{equation}
This is the exact form needed for Gauss-Chebyshev quadrature (with f = D):
\begin{equation}
\int_0^\pi  D\left( {\cos \varphi } \right){\textrm{d}}\varphi \approx \frac{\pi }{n} \sum_{i = 1}^n D\!\left[ {\cos \left( {\frac{{2i - 1}}{{2n}}\pi } \right)} \right]
\label{eq:AngQuadrature}
\end{equation}

This proves only form \cref{eq:PVRAngNew}. To prove \cref{eq:PVRAng}, we split up the integration into two parts:
\begin{equation}
\int_0^{2\pi } D\left( {\cos {\varphi_{23}}} \right){\textrm{d}}{\varphi_{23}} = \int_0^\pi  D\left( {\cos {\varphi_{23}}} \right){\textrm{d}}{\varphi_{23}} + \int _\pi ^{2\pi } D\left( {\cos {\varphi _{23}}} \right){\textrm{d}}{\varphi _{23}}.
\label{eq:AngSplit}
\end{equation}
The first integration is just \cref{eq:PVRAngNew}. The only difference between the first and second integration is the limits. Defining $y = \varphi - \pi$ gives
\begin{equation}
\int _\pi ^{2\pi } D\left( {\cos {\varphi}} \right){\textrm{d}}{\varphi} = \int_0^\pi D\left[\cos(\varphi)\right] dy = \int_0^\pi D\left[\cos(y+\pi)\right] dy.
\end{equation}
If we also define $z = \cos \varphi = \cos(y+\pi)$ and $dz = \sin y \,dy$, we get an expression the same as \cref{eq:PVRAngNew3}:
\begin{equation}
\int_0^\pi D\left(\cos\varphi\right) dy = \int_0^\pi \frac{1}{\sqrt{1-z^2}} D(z)\, sin\varphi\, dy = \int_{-1}^1 \frac{1}{\sqrt{1-z^2}} D(z) dz
\end{equation}
Since this is the same as \cref{eq:PVRAngNew3}, we just combine this with \cref{eq:AngSplit,eq:AngQuadrature} to get \cref{eq:PVRAng}, proving the first form.

\subsection{Selection of Quadrature Points}
\label{sec:SelQuadPoints}

The number of quadrature points for each coordinate in the 6-dimensional 
integrations is critical to have fully converged results. In our testing, the 
$r_1$ coordinate (the coordinate of $e^+$) required more integration points 
than any other. The $r_2$ and $r_3$ coordinates required less points, and the 
interparticle terms ($r_{12}$, $r_{13}$ and $r_{23}$) required the least. 
These results only apply to the long-long and long-short terms, as the short-
short terms are integrated using the asymptotic expansion method (see
\cref{sec:MatrixShort}).

To determine this, we held the number of integration points fixed, except for 
one coordinate, which we increased in steps. The difference in the output 
between steps was used to analyze how important each coordinate was. Also, 
some terms are more sensitive to the number of integration points than other. 
For instance, the $(\bar{S}_\ell,\mathcal{L} \bar{S}_\ell)$ term converges 
relatively quickly, while the $(\bar{C}_\ell,\mathcal{L} \bar{C}_\ell)$ 
term requires more integration points.

We also used a sample input file from Van Reeth \cite{VanReethPrivate} for
the number of quadrature points he used in his code as a starting point.
This is shown in \cref{tab:BaseEffectiveCoords}.
Using a comparison program with the Developer's Image Library \cite{DevIL}, we
determined a set of quadrature points that yielded good convergence of
the matrix elements with a reasonable runtime. The final set of quadrature
points that we use for all partial waves is shown in
\cref{tab:OptimalEffectiveCoords}. This set yielded good convergence for
partial waves through the H-wave.

\subsection{Quadrature Points}
\label{sec:QuadraturePoints}

Describing the number of points used in integrating the different coordinates 
in \cref{sec:LongLongInt,sec:ShortLongInt} can be confusing, so we have taken 
to grouping the sets of points as in
\Cref{tab:EffectiveCoords1,tab:EffectiveCoords2}. Each column of these tables 
is referred to as an ``effective coordinate'', as the $r_2$ and $r_3$ 
integrations are split into two parts, as described in \cref{sec:Cusps}.
The ``Lag'' entries use the Gauss-Laguerre quadrature, ``Leg'' uses the
Gauss-Legendre quadrature, and ``Che'' uses Chebyshev-Gauss quadrature. Note
that the only difference between \cref{tab:EffectiveCoords1,tab:EffectiveCoords2}
is that the S-wave long-long matrix elements are integrated using perimetric
coordinates when the $r_{23}^{-1}$ is not present, while the other partial
waves use the same coordinates as the long-short terms.

\begin{table}
\centering
\footnotesize
\begin{tabular}{c c c c c c c c c}
\toprule
 & Coord & Coord & Coord & Coord & Coord & Coord & Coord & Coord \\
Integral & 1 & 2 & 3 & 4 & 5 & 6 & 7 & 8 \\
\midrule
 Long-long, no $r_{23}^{-1}$ & $x$ Lag & $y$ Lag & $z$ Lag & $r_3$ Leg & $r_3$ Leg & $r_{13}$ Leg & & \\
 Long-long, $r_{23}^{-1}$ & $r_1$ Lag & $r_2$ Leg & $r_2$ Lag & $r_3$ Leg & $r_3$ Lag & $\varphi_{12}$ Che & $r_{13}$ Leg & $r_{23}$ Leg \\
\midrule
 Long-short $q_i = 0$, no $r_{23}^{-1}$ & $r_1$ Lag & $r_2$ Leg & $r_2$ Lag & $r_3$ Leg & $r_3$ Lag & $r_{12}$ Leg & $r_{13}$ Leg & \\
 Long-short $q_i = 0$, $r_{23}^{-1}$ & $r_1$ Lag & $r_2$ Leg & $r_2$ Lag & $r_3$ Leg & $r_3$ Lag & $r_{12}$ Leg & $\varphi_{13}$ Che & $r_{23}$ Leg \\
 Long-short $q_i > 0$ & $r_1$ Lag & $r_2$ Leg & $r_2$ Lag & $r_3$ Leg & $r_3$ Lag & $r_{12}$ Leg & $r_{13}$ Leg & $\varphi_{23}$ Che \\
\bottomrule
\end{tabular}
\caption[Description of the values in \cref{tab:BaseEffectiveCoords,tab:OptimalEffectiveCoords} for the S-wave]{Description of the values in \cref{tab:BaseEffectiveCoords,tab:OptimalEffectiveCoords} for the S-wave. Refer to the text for explanation of this table.}
\label{tab:EffectiveCoords1}
\end{table}

\begin{table}
\centering
\footnotesize
\begin{tabular}{c c c c c c c c c}
\toprule
 & Coord & Coord & Coord & Coord & Coord & Coord & Coord & Coord \\
Integral & 1 & 2 & 3 & 4 & 5 & 6 & 7 & 8 \\
\midrule
 Long-long, no $r_{23}^{-1}$ & $r_1$ Lag & $r_2$ Leg & $r_2$ Lag & $r_3$ Leg & $r_3$ Lag & $r_{12}$ Leg & $r_{13}$ Leg & \\
 Long-long, $r_{23}^{-1}$ & $r_1$ Lag & $r_2$ Leg & $r_2$ Lag & $r_3$ Leg & $r_3$ Lag & $\varphi_{12}$ Che & $r_{13}$ Leg & $r_{23}$ Leg \\
\midrule
 Long-short $q_i = 0$, no $r_{23}^{-1}$ & $r_1$ Lag & $r_2$ Leg & $r_2$ Lag & $r_3$ Leg & $r_3$ Lag & $r_{12}$ Leg & $r_{13}$ Leg & \\
 Long-short $q_i = 0$, $r_{23}^{-1}$ & $r_1$ Lag & $r_2$ Leg & $r_2$ Lag & $r_3$ Leg & $r_3$ Lag & $r_{12}$ Leg & $\varphi_{13}$ Che & $r_{23}$ Leg \\
 Long-short $q_i > 0$ & $r_1$ Lag & $r_2$ Leg & $r_2$ Lag & $r_3$ Leg & $r_3$ Lag & $r_{12}$ Leg & $r_{13}$ Leg & $\varphi_{23}$ Che \\
\bottomrule
\end{tabular}
\caption[Description of the values in \cref{tab:BaseEffectiveCoords,tab:OptimalEffectiveCoords} for $\ell > 0$]{Description of the values in \cref{tab:BaseEffectiveCoords,tab:OptimalEffectiveCoords} for $\ell > 0$. Refer to the text for explanation of this table.}
\label{tab:EffectiveCoords2}
\end{table}

\Cref{tab:BaseEffectiveCoords} shows the set of integration points used by Van
Reeth \cite{VanReethPrivate}. This was the starting point for trying to obtain
better convergence of the matrix element integrations, as described in
\cref{sec:SelQuadPoints}. Using a comparison program to investigate this
convergence, we determined a more extensive set of integration points given
in \cref{tab:OptimalEffectiveCoords} that we use for all partial waves.

\begin{table}
\centering
\footnotesize
\begin{tabular}{c c c c c c c c c}
\toprule
 & Coord & Coord & Coord & Coord & Coord & Coord & Coord & Coord\\
Integral & 1 & 2 & 3 & 4 & 5 & 6 & 7 & 8 \\
\midrule
 Long-long, no $r_{23}^{-1}$ & 45 & 35 & 35 & 35 & 28 & 15 & & \\
 Long-long, $r_{23}^{-1}$ & 65 & 35 & 28 & 35 & 28 & 12 & 15 & 15 \\
\midrule
 Long-short $q_i = 0$, no $r_{23}^{-1}$ & 90 & 57 & 34 & 57 & 34 & 30 & 30 & \\
 Long-short $q_i = 0$, $r_{23}^{-1}$ & 90 & 58 & 30 & 55 & 35 & 33 & 33 & 33 \\
 Long-short $q_i > 0$ & 90 & 57 & 34 & 57 & 34 & 30 & 30 & 30 \\
\bottomrule
\end{tabular}
\caption[Base set of effective coordinates for integrations]{Base set of effective coordinates for integrations from Van Reeth~\cite{VanReethPrivate}}
\label{tab:BaseEffectiveCoords}
\end{table}

\begin{table}
\centering
\footnotesize
\begin{tabular}{c c c c c c c c c}
\toprule
 & Coord & Coord & Coord & Coord & Coord & Coord & Coord & Coord\\
Integral & 1 & 2 & 3 & 4 & 5 & 6 & 7 & 8 \\
\midrule
 Long-long, no $r_{23}^{-1}$			&  75 & 40 & 40 & 40 & 40 & 25 & & \\
 Long-long, $r_{23}^{-1}$				&  75 & 40 & 40 & 40 & 40 & 25 & 25 & 25 \\
\midrule
 Long-short $q_i = 0$, no $r_{23}^{-1}$	& 100 & 65 & 45 & 65 & 45 & 45 & 45 & \\
 Long-short $q_i = 0$, $r_{23}^{-1}$	& 115 & 65 & 45 & 65 & 45 & 45 & 45 & 45 \\
 Long-short $q_i > 0$					& 100 & 65 & 45 & 65 & 45 & 45 & 45 & 45 \\
\bottomrule
\end{tabular}
\caption{Final set of effective coordinates for integrations}
\label{tab:OptimalEffectiveCoords}
\end{table}

\section{Resonance Fitting}
\label{sec:ResonanceFit}

To find the resonance parameters (positions and widths), the phase shifts for 
multiple energy values are fitted to \cref{eq:ResonanceCurve}. There have 
been multiple programs described in the literature
\cite{Tennyson1984, Stibbe1998, Sochi2013} to do these fittings.
Some of the difficulty with this type of 
fitting is choosing appropriate guesses for the resonance parameters.
Ref.~\cite{Sochi2013} describes methods that some groups use to try to identify 
resonance parameters.

With the help of Bosca \cite{BoscaPrivate}, I wrote a MATLAB\textsuperscript{\textregistered}
\cite{matlab} script that uses the \texttt{nlinfit} routine of MATLAB. 
\texttt{nlinfit} is specifically designed for fitting to nonlinear functions, 
and its robust option allows for a variety of weighting functions to be used. 
This script does the fitting for all eight of the possible weightings: 
Bisquare, Andrews, Cauchy, Fair, Huber, Logistic, Talwar and Welsch. This 
fitting routine is also not as sensitive to the initial guesses as the 
Mathematica and SciPy routines.

Later on, I adapted this resonance fitting code to be called from IPython 
\cite{ipython} using the mlabwrap \cite{mlabwrap} Python to MATLAB wrapper. 
This allows fits and graphs of the fittings to be in the same IPython 
notebook, including the flexibility of querying a MySQL database for the 
phase shift data for any partial wave at any number of terms and for
any Kohn-type variational method. The mlabwrap package is difficult to install properly, 
requiring a compilation against MATLAB. The mlabwrap-purepy package
\cite{mlabwrappurepy} has been created to simplify this, but I have not tried it 
yet.

\begin{figure}
	\centering
	\includegraphics[width=\textwidth]{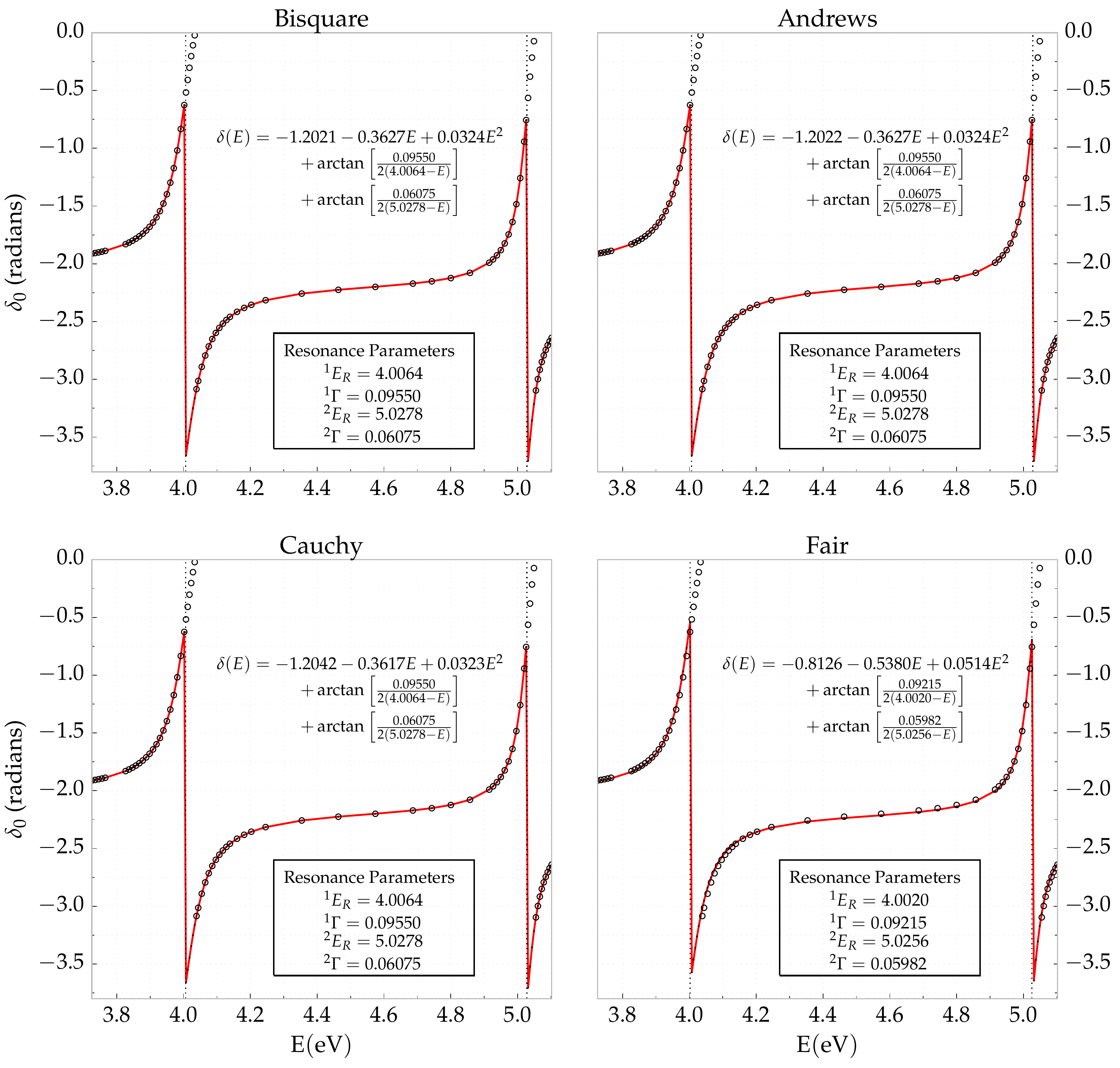}
	\caption{First set of uncorrected resonance fitting graphs for $^1$S at $\omega = 7$}
	\label{fig:swave-resonance-uncorrected-fits1}
\end{figure}

\begin{figure}
	\centering
	\includegraphics[width=\textwidth]{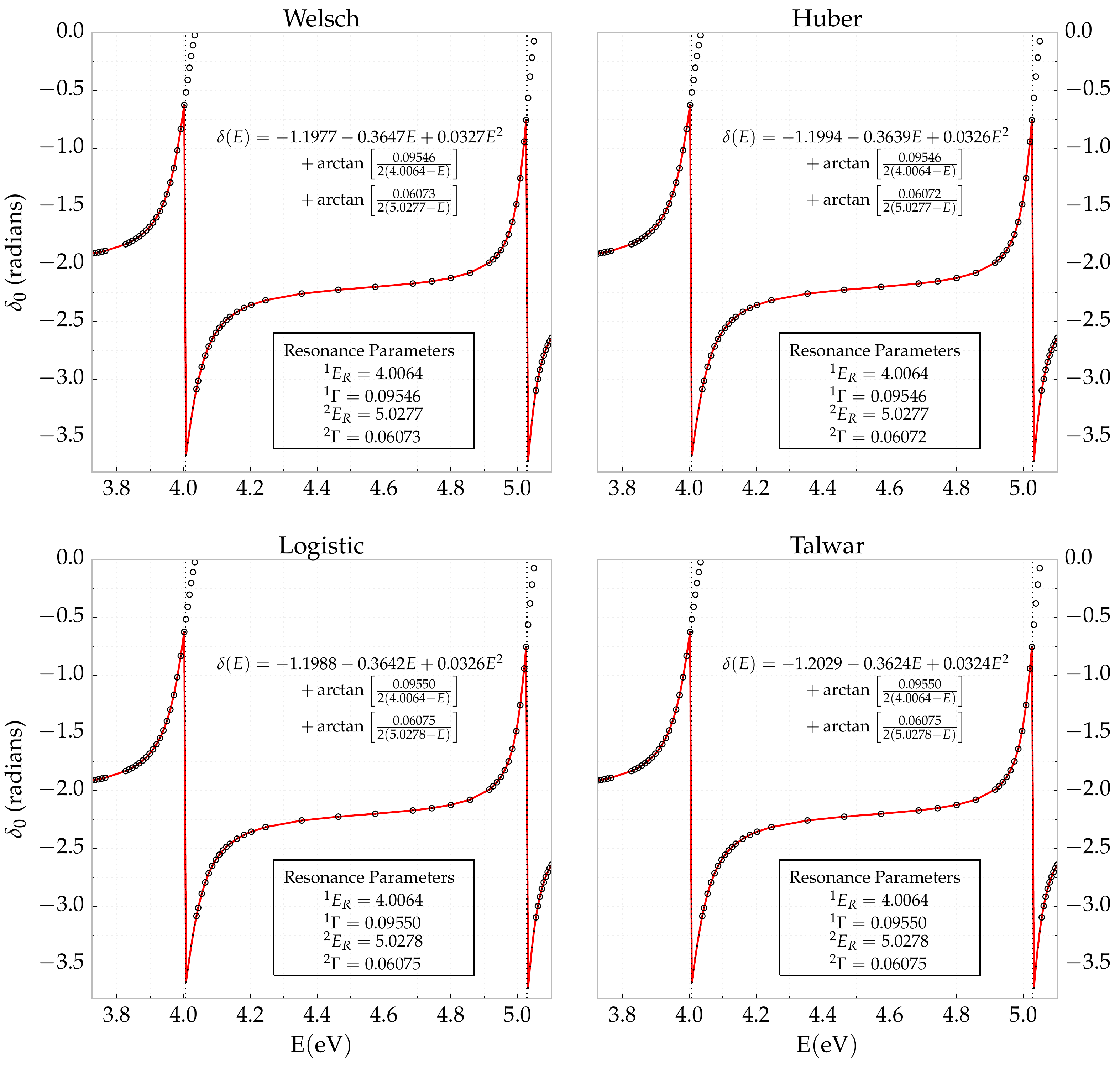}
	\caption{Second set of uncorrected resonance fitting graphs for $^1$S at $\omega = 7$}
	\label{fig:swave-resonance-uncorrected-fits2}
\end{figure}

\begin{figure}
	\centering
	\includegraphics[width=\textwidth]{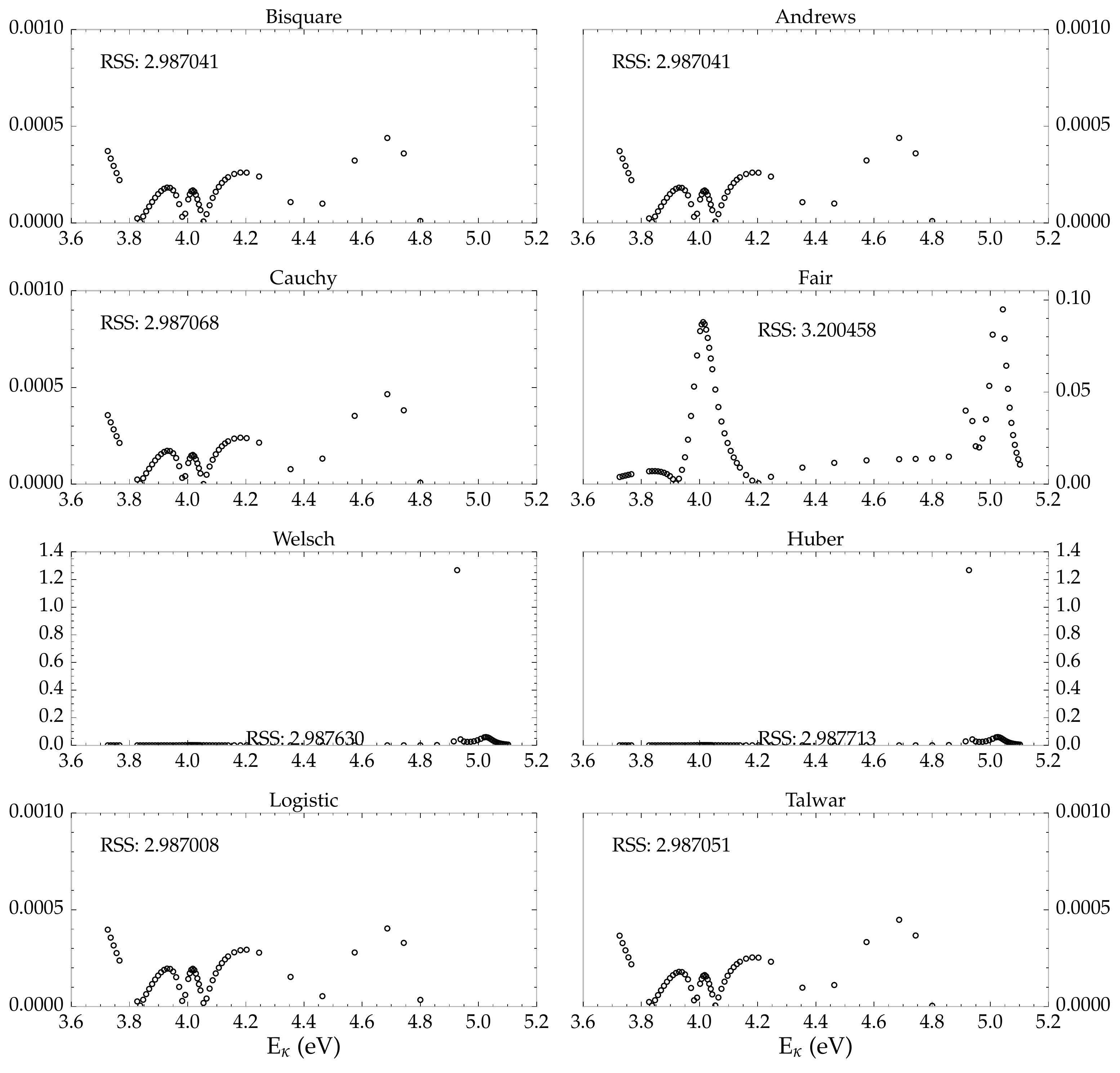}
	\caption{Residuals for uncorrected resonance fitting graphs for $^1$S at $\omega = 7$}
	\label{fig:swave-resonance-uncorrected-res}
\end{figure}

The results of fitting the phase shifts from the $S$ matrix are shown in
\cref{fig:swave-resonance-uncorrected-fits1,fig:swave-resonance-uncorrected-fits2} 
for each of the weighting functions, and the resonance parameters are given 
in each subfigure. There is good agreement between the fits performed with 
each of the weighting functions, and the Fair is the furthest from the 
others. In our testing, the Cauchy is consistently one of the best choices 
for fitting. In addition, \cref{fig:swave-resonance-uncorrected-res} has 
plots of the residuals (the absolute value of the difference between the 
fitted curve and the actual phase shifts). Each graph also has the residual 
sum of squares (RSS) calculated. The RSS gives an easy way to compare the 
performance of the weightings with each other. The Fair has a notably larger 
RSS than all of the other fittings.

One important thing to notice with the fits in
\cref{fig:swave-resonance-uncorrected-fits1,fig:swave-resonance-uncorrected-fits2} is that the
$S$-matrix complex Kohn phase shifts extend above the fitted curve at the resonances before 
matching up again on the right side of the resonance below $-3.0$. The red 
curves still fit to the data well, but the fits can be improved by correcting 
for these. 

In \cref{eq:ResonanceCurve}, the first three polynomial terms form the 
background, and the two $\arctan$ terms represent the resonances. The range 
of $\arctan$ is $(-\frac{\pi}{2},\frac{\pi}{2})$, so the $\arctan$ parts of 
this model cannot bring the phase shift from the background (near 2.0) all 
the way to 0.0, as it can only add up to $\frac{\pi}{2}$ to the background. 
It is important to realize that the Kohn-type variational methods will only return phase 
shifts in a certain range. From \cref{eq:GenKohnL}, we are not finding the 
phase shifts directly but are rather calculating $\tan \delta_\ell$. The 
phase shifts found this way sometimes have to have $\pi$ added or subtracted 
if they are outside this range.

\begin{figure}
	\centering
	\includegraphics[width=5in]{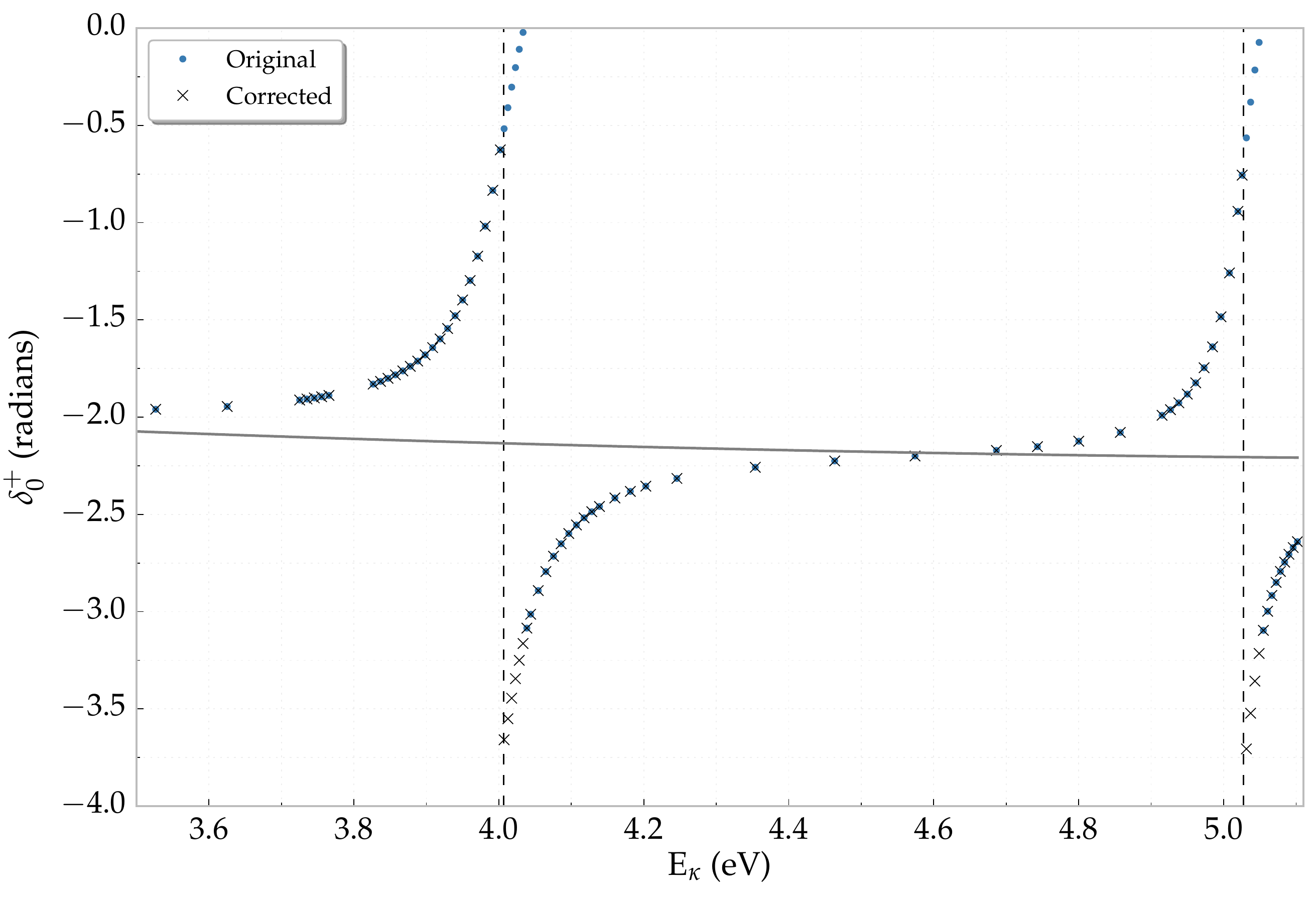}
	\caption[$^1$S resonance data showing correction]{$^1$S resonance data showing raw data from $S$-matrix complex Kohn and the corrected version. The solid gray line shows the polynomial background, and the vertical dashed lines give the calculated resonance positions.}
	\label{fig:swave-phases-reson-pi}
\end{figure}

\Cref{fig:swave-phases-reson-pi} shows the results of subtracting $\pi$ from 
phase shifts that are more than $\frac{\pi}{2}$ above the background. The 
original and the corrected data are both shown on this plot, and the 
background is given as a gray line. Note that the slope of the original data 
changes when crossing over the vertical dashed lines, which gives the 
resonance positions. When corrected, these upper points are moved down to 
their appropriate place, shown as x's on the graph. These match up better 
with the fitting curve. This fitting is an iterative process, because the 
background polynomial has to be determined with \cref{eq:ResonanceCurve} 
before we can do this correction. Then the phase shifts are fitted again 
after performing the correction.

\begin{figure}
	\centering
	\includegraphics[width=\textwidth]{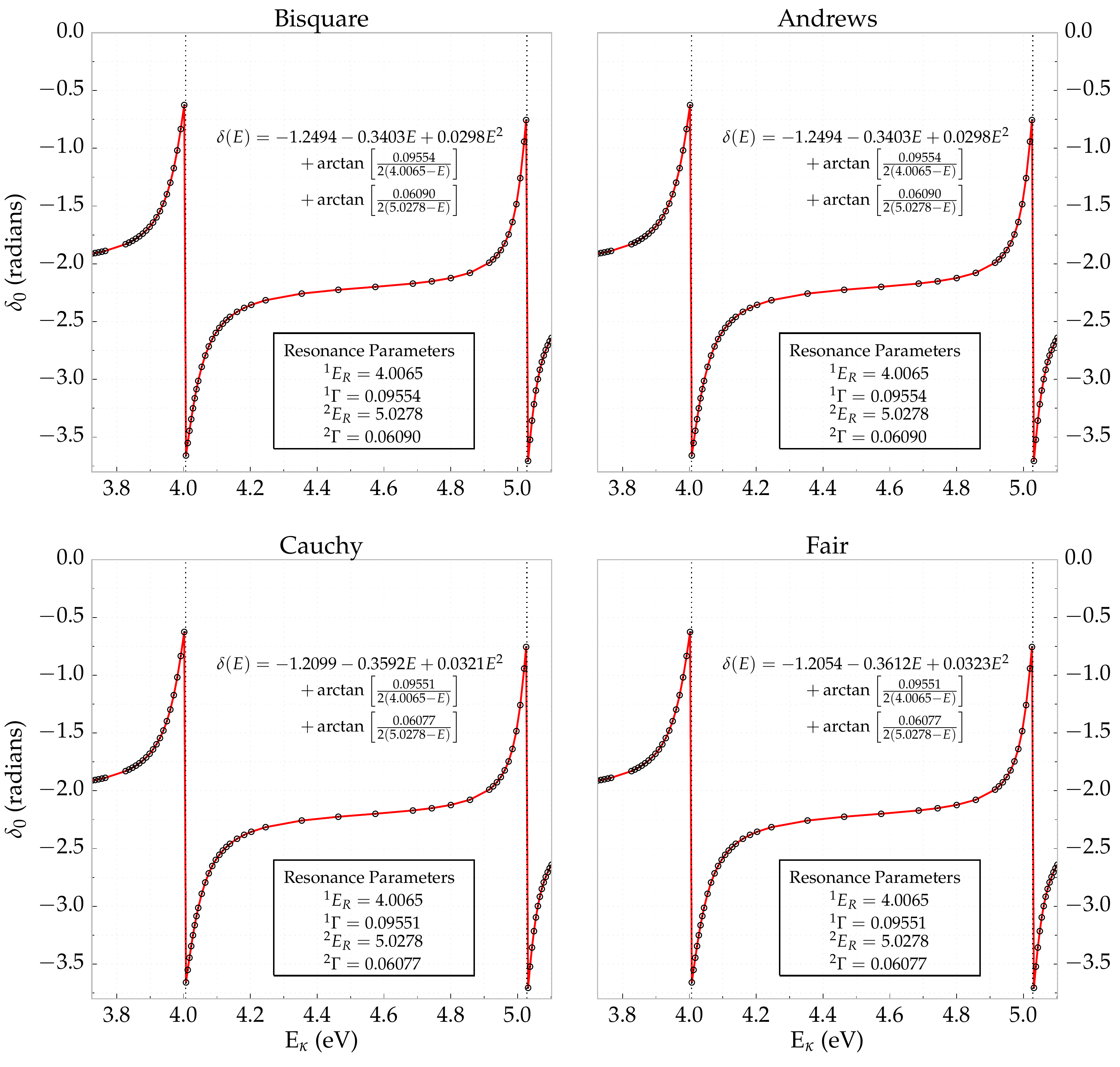}
	\caption{First set of corrected resonance fitting graphs for $^1$S at $\omega = 7$}
	\label{fig:swave-resonance-corrected-fits1}
\end{figure}

\begin{figure}
	\centering
	\includegraphics[width=\textwidth]{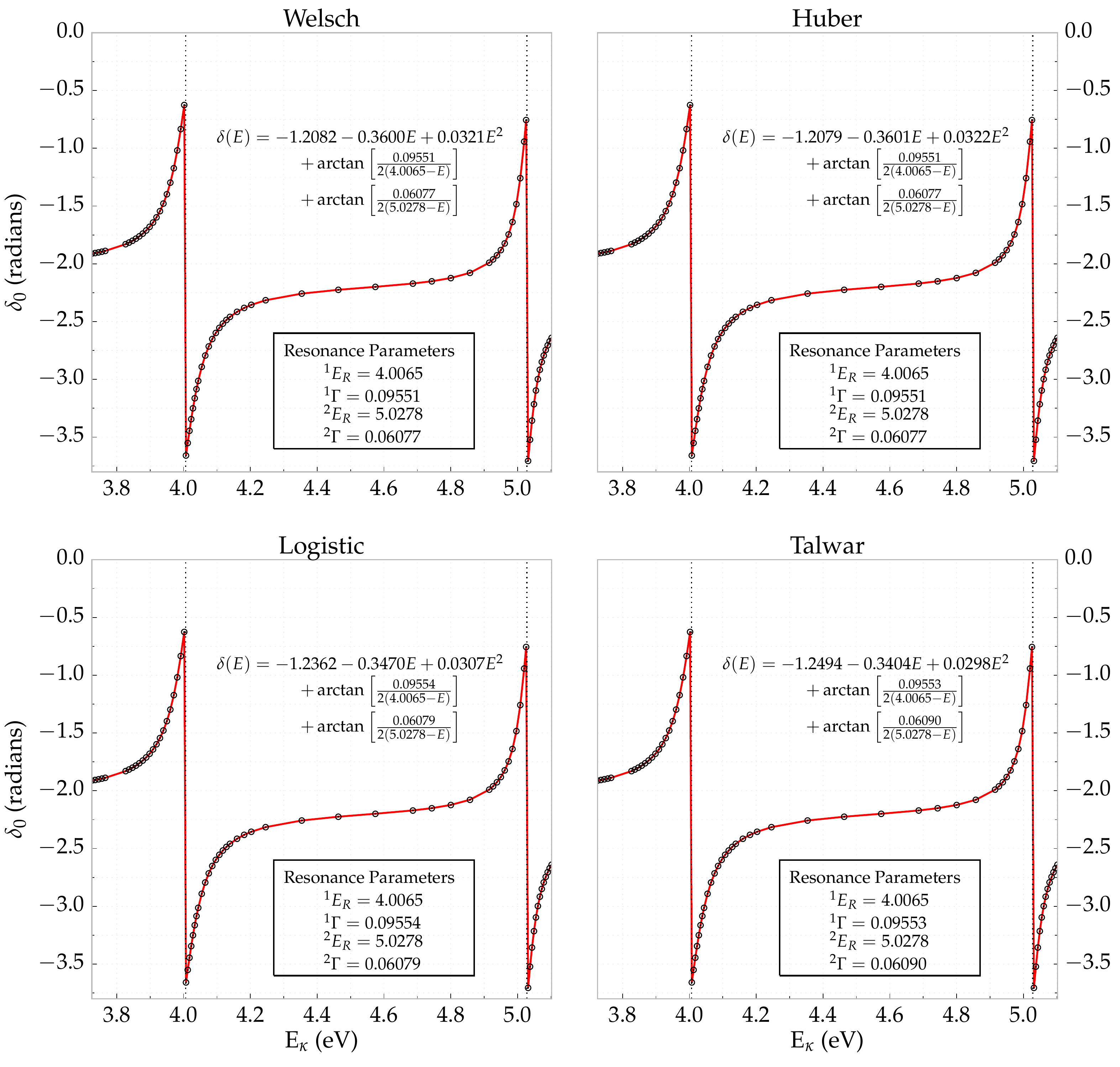}
	\caption{Second set of corrected resonance fitting graphs for $^1$S at $\omega = 7$}
	\label{fig:swave-resonance-corrected-fits2}
\end{figure}

\begin{figure}
	\centering
	\includegraphics[width=\textwidth]{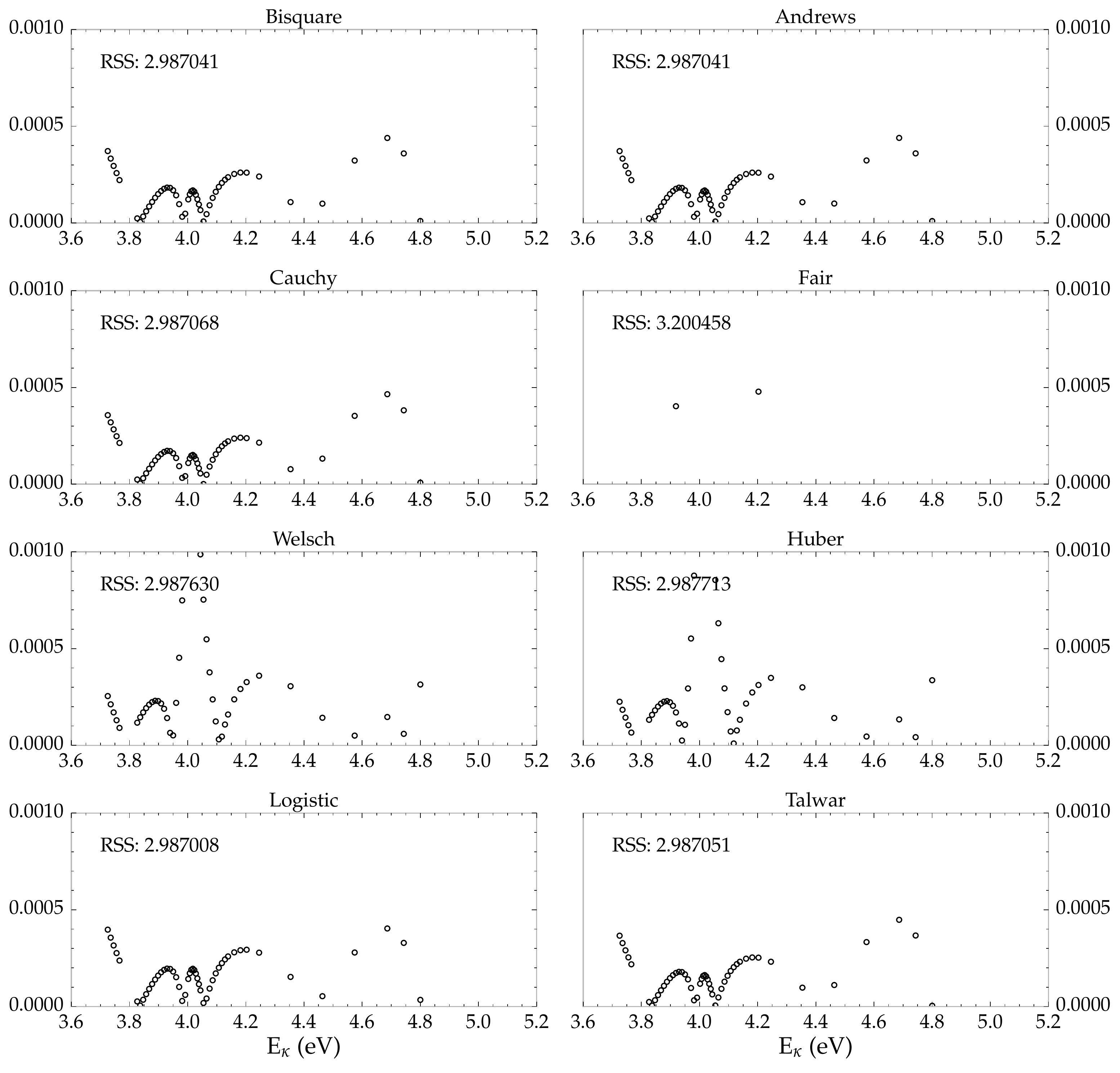}
	\caption{Residuals for corrected resonance fitting graphs for $^1$S at $\omega = 7$}
	\label{fig:swave-resonance-corrected-res}
\end{figure}

Finally, this corrected and re-fitted data is shown in
\cref{fig:swave-resonance-corrected-fits1,fig:swave-resonance-corrected-fits2}. The different 
weighting methods agree extremely well now, with the only notable differences 
being in the $^2\Gamma$ width of the second resonance for the Bisquare, 
Andrews, and Talwar fits. This is still a very small difference, and the 
other resonance parameters are almost all identical.
\Cref{fig:swave-resonance-corrected-res} gives the residuals and RSS for the corrected phase
shifts, similar to \cref{fig:swave-resonance-uncorrected-res}. The Bisquare, Andrews, 
and Talwar weightings have larger RSS values than the other weightings, 
indicating that their fits are not quite as accurate, but they are still 
relatively accurate.

The differences between the different weighting methods gives us one way to 
determine errors for the resonance parameters. Additionally, comparing the 
different Kohn-type variational methods methods gives us an idea of the error. The real-valued 
generalized Kohn variational methods give more disagreement, so we compute the resonance 
parameters for each of these. We also have to take into account 
Schwartz singularities. In \cref{fig:schwartz-singularity}(a) on
\pageref{fig:schwartz-singularity}, the second resonance parameters are not as 
accurate. The fitting routine described here chooses fitting parameters of
$^2E_R = \SI{5.0295}{eV}$ and $^2\Gamma = \SI{0.06011}{eV}$.
In \cref{fig:schwartz-singularity}(b), where there are no Schwartz 
singularities, $^2E_R = \SI{5.0278}{eV}$ and $^2\Gamma = \SI{0.06075}{eV}$. 
This highlights the importance of using multiple Kohn-type variational methods and trying to 
detect Schwartz singularities. After removing obvious Schwartz singularities, 
the mean value of the different Kohn-type variational methods for all weightings is taken for 
each resonance parameter, and these are the results listed for each of the 
partial waves in
\cref{sec:SWaveResonances,sec:PWaveResonances,sec:DWaveResonance,sec:FWaveResonance}.
The errors given in these tables are simply the standard deviation with all 
generalized Kohn variational methods and weightings used for each resonance parameter. For this, we 
do not use the generalized $S$-matrix or $T$-matrix complex Kohn methods,
because using these would decrease the error, as they agree to a significant
precision.



\chapter{Program Descriptions}
\label{chp:Programs}

\begin{figure}
	\centering
	\includegraphics[width=4.5in]{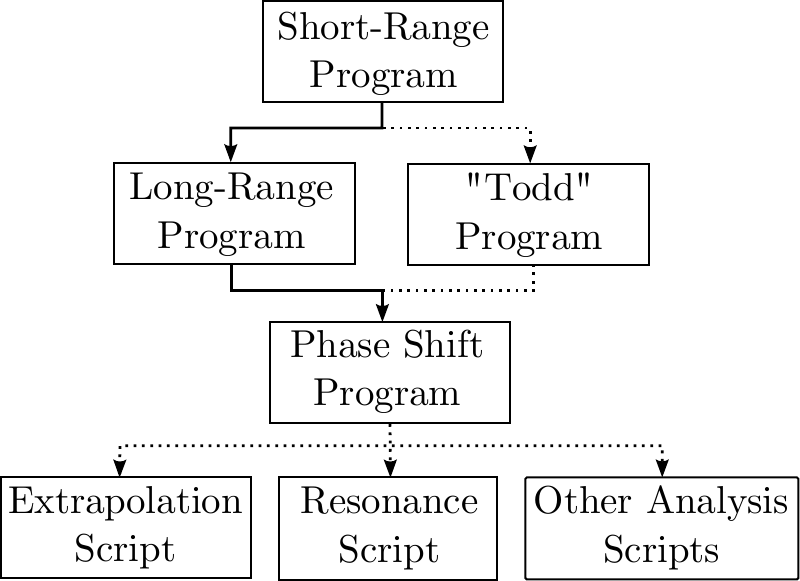}
	\caption{Flowchart for programs and scripts}
	\label{fig:ProgramFlowchart}
\end{figure}

\iftoggle{UNT}{Doing}{\lettrine{\textcolor{startcolor}{D}}{oing}}
this work required writing multiple
codes, and I have made my codes available on GitHub at 
\url{https://github.com/DentonW/Ps-H-Scattering} \cite{GitHub}. Much of the
codes are what is described in \cref{chp:Computation}. A flowchart of the
steps required to calculate phase shifts and any additional quantities, such
as cross sections, is given by \cref{fig:ProgramFlowchart}.

The short-range program typically only needs to be run once for a partial wave,
and those short-range matrix elements are used for all $\kappa$ runs. For
$\ell \geq 2$, we have a low and high $\kappa$ set of nonlinear parameters, so
there are two runs of the short-range program that have to be done. The
long-range program is run for every $\kappa$ value for each partial wave, and
if needed, we restrict the short-range terms using Todd's method from
\cref{sec:ToddBound}. Then the phase shift program uses these inputs to run
many Kohn-type variational methods and generate phase shifts. These phase shifts
can then be further analyzed with several scripts.

\begin{figure}[H]
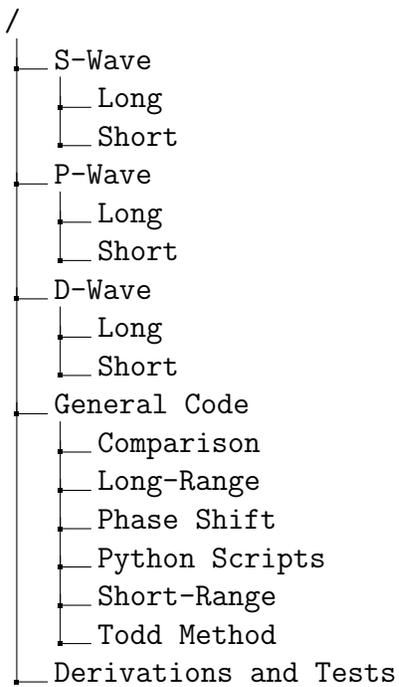

\label{fig:directory}
\begin{flushleft}
\dirtree{%
.1 /.
.2 S-Wave.
.3 Long.
.3 Short.
.2 P-Wave.
.3 Long.
.3 Short.
.2 D-Wave.
.3 Long.
.3 Short.
.2 General Code.
.3 Comparison.
.3 Long-Range.
.3 Phase Shift.
.3 Python Scripts.
.3 Short-Range.
.3 Todd Method.
.2 Derivations and Tests.
}
\end{flushleft}
\caption[Directory tree of code]{Directory tree of code on GitHub \cite{GitHub}}
\label{fig:dirtree}
\end{figure}

As the codes are freely available on GitHub, it is worthwhile to show the
directory structure when this package is downloaded with git or as a zip.
\Cref{fig:dirtree} shows this directory tree. The S-wave, P-wave, and D-wave
codes only work for their respective partial wave, and the code in ``General 
Code'' works for arbitrary $\ell$, as described in \cref{chp:WaveKohn,chp:General}.

\subsection*{S-, P-, and D-wave Codes}
The S-, P-, and D-wave folders each have a Short subfolder (short-short
code) and a Long subfolder (short-long and long-long code). As shown in
the flowchart in \cref{fig:ProgramFlowchart}, the Short code is run first,
and then Long code is run for all $\kappa$ values of interest.
To obtain phase shifts, use the "Phase Shift" code in the "General Code"
folder. Extrapolations, etc.\ are also in the "General Code" folder.

\subsection*{General Short and Long}
The general Short and Long codes act similarly to the S-, P-, and D-wave
codes but for arbitrary $\ell$. It should be noted that because these
codes are general, they are much slower than running the corresponding
code specific to a partial wave if $\ell \leq 2$.

\subsection*{Phase Shifts}
The phase shift code, located in ``General Code/Phase Shift'', uses the
outputs of the Short and Long codes from the S-Wave, P-Wave, D-wave or
``General Code'' folders to calculate the phase shifts. An optional file
describing which short-range terms to use (such as that generated by the
``Todd Code''), can also be used as an input. The output is an XML file
showing phase shifts calculated for every N value for each of the 109
total Kohn-type methods used.

\subsection*{Extrapolation Program}
I wrote a Python \cite{Python} script, located in ``General Code/Python Scripts'',
to extrapolate the phase shifts for a 
run for all Kohn-type methods used, including the 35 values of $\tau$ used 
in each of the generalized Kohn, generalized $S$-matrix Kohn, and generalized
$T$-matrix Kohn. The extrapolation is performed with a least-squares fitting 
using the \texttt{polyfit} function of the SciPy package \cite{SciPy} to the 
form of \cref{eq:PhaseExtrap}. This program can do the extrapolation over any 
interval of $\omega$ values requested. The extrapolations from the 109 total 
Kohn-type methods are compared to see if there is any large discrepancy between 
them, indicating numerical instability. The phase shifts can have a 
singularity in the real-valued Kohn-type variational methods
as seen in \cref{fig:schwartz-singularity}, so 
care must be taken to ensure extrapolations are not taken around this interval.

\subsection*{Derivations and Tests}
The ``General Code/Derivations and Tests'' folder has files referenced
throughout this document that are usually \emph{Mathematica} notebooks.
Examples are the ``Shielding Factor.nb'' notebook that is referenced in
\cref{sec:ShieldingFunc} and notebooks to generate the powers and coefficients
needed for the short-short integrations for each partial wave.


\bibliographystyle{UNTamsplain-modded}

\bibliography{Dissertation}

\end{document}